%% file: main.tex
\documentclass[12pt,notitlepage]{report}

\usepackage{subcaption}
\PassOptionsToPackage{dvipdfmx}{graphicx}

\usepackage[toc,page]{appendix}
\usepackage{etoolbox}

\usepackage[utf8]{inputenc}
\usepackage{graphicx}
\usepackage{intro_macros}
\usepackage{sectors_macros_thesis}
\usepackage{category_macros_thesis}
\usepackage[left=1in, right=1in, top=1in, bottom=1in]{geometry}
\usepackage{dsfont}
\usepackage{amsthm}
\usepackage{amsfonts}
\usepackage{thmtools}
\usepackage{thm-restate}
\usepackage{physics}
\usepackage{amssymb}
\usepackage{amsmath}
\usepackage{xcolor}
\usepackage{graphicx}
\usepackage{xparse}
\usepackage{mathtools}
\usepackage{subcaption}
\PassOptionsToPackage{dvipdfmx}{graphicx}
\usepackage[sectionbib]{chapterbib}
\usepackage{quiver}
\usepackage{minitoc}
\usepackage[toc,page]{appendix}
\usepackage{etoolbox}
\usepackage[most]{tcolorbox}
\usepackage[pdftex, hypertexnames=false]{hyperref}
\usepackage{cleveref}
\usepackage{enumitem}
\usepackage{titlesec}
\usepackage{setspace}

\usepackage{tikz}
\usetikzlibrary{arrows,backgrounds}
\newcommand{\tikzmath}[2][]
{\vcenter{\hbox{\begin{tikzpicture}[#1]#2\end{tikzpicture}}}
}

\usetikzlibrary{decorations.pathmorphing}

\tikzset{snake it/.style={decorate, decoration=snake}}

\definecolor{azure}{HTML}{007fff}
\definecolor{OliveGreen}{HTML}{6D712E}
\definecolor{boysenberry}{HTML}{873260}
\definecolor{violet}{RGB}{148,0,211}
\definecolor{salmon}{HTML}{ff8c69}

\hypersetup{
    colorlinks=true,                            
    linkcolor=azure,                          
    citecolor=OliveGreen,                          
    filecolor=red,                           
    urlcolor=azure                             
}

\newcommand{\nocontentsline}[3]{}
\newcommand{\tocless}[2]{\bgroup\let\addcontentsline=\nocontentsline#1{#2}\egroup}

\theoremstyle{plain}
\newtheorem{thm}{Theorem}[section]

\newtheorem{cor}[thm]{Corollary}
\newtheorem{lem}[thm]{Lemma}
\newtheorem{prop}[thm]{Proposition}

\theoremstyle{definition}
\newtheorem{defn}[thm]{Definition}
\newtheorem{nota}[thm]{Notation}

\newtheorem{facts}[thm]{Facts}
\newtheorem{rem}[thm]{Remark}
\newtheorem{asmp}{Assumption}
\newtheorem*{asmp_recall}{Assumption}

\newenvironment{chapterabstract}
  {\begin{quote}\small\textbf{Abstract.} }
  {\end{quote}\normalsize}

\newtcolorbox{thm_border}[1][]{
    colback=white, 
    colframe=cyan, 
    fonttitle=\bfseries, 
    coltitle=white, 
    title=Theorem, 
    sharp corners, 
    boxrule=1pt, 
    #1, 
}

\newtcolorbox{setting}[1][]{
    colback=white, 
    colframe=cyan, 
    fonttitle=\bfseries, 
    title=Setting, 
    sharp corners, 
    boxrule=1pt, 
    #1, 
}

\newtcbtheorem[number within=thm]{example}{Example}%
{enhanced, breakable,
 colback=white,
 colframe=azure,
 boxrule=0.8pt,
 sharp corners,
 before skip=10pt,
 after skip=10pt}%
{eg}

\titleformat{\chapter}[display]
  {\normalfont\huge\bfseries\centering}  
  {\chaptertitlename\ \thechapter}       
  {1ex}                                  
  {\Huge}                                

\newcommand{\chapterauthors}[1]{%
  \par\vspace{-.75\baselineskip}%
  \begin{center}
    #1
  \end{center}
  \vspace{.5\baselineskip}%
}

\newcommand{\chapterauthor}[2]{%
  {\Large\textbf{#1}}\\[0.2cm]%
  {\normalsize \textbf{#2}}\par\vspace{0.5cm}%
}

\begin{document}

\pagenumbering{roman}   

\begin{center}
  \vspace*{\fill}


  {\Huge\bfseries
    An Operator-Algebraic Framework for Anyons and Defects in Quantum Spin Systems\par}
  \vspace{1.5cm}

  {\Huge Siddharth Vadnerkar\par}
  \vspace{1cm}
  {\large Dissertation\par}
  \vspace{1cm}

  {Submitted in partial satisfaction of the requirements for the degree of DOCTOR OF PHILOSOPHY in Physics in the OFFICE OF GRADUATE STUDIES of the UNIVERSITY OF CALIFORNIA DAVIS\par}

    \vspace{1cm}
  {\large December 10th, 2025 \par}
  {\large University of California, Davis}
  
  \vspace*{\fill}
\end{center}


\clearpage

\begin{center}
  \vspace*{\fill}

  {\bfseries Abstract\par}
  \vspace{0.5cm}

  \begin{minipage}{0.8\textwidth}
    \small
    In this dissertation, we detail an operator algebraic approach to studying topological order in the infinite volume setting. We give a thorough and self-contained review of the DHR-style approach on quantum spin systems, which builds a category $\DHR$ of anyon sectors starting from microscopic lattice spin systems. In general, this category has the structure of a braided $\rmC^*$-tensor category. We will verify in full detail that $\DHR$ is the expected category in Kitaev's Quantum Double model, a paradigmatic model for studying topological order on the lattice. We will then extend the DHR-style analysis to systems in the presence of a global on-site symmetry, and introduce a category of symmetry defects, $\GSec$, and show that it has the structure of a $G$-crossed braided
    $\rmC^*$-tensor category.
  \end{minipage}

  \vspace*{\fill}
\end{center}

\begin{center}
  \vspace*{\fill}

  {\bfseries Funding\par}
  \vspace{0.5cm}

  \begin{minipage}{0.8\textwidth}
    \small
    The author acknowledges funding from NSF grant numbers DMS-2510824, DMS-2108390, DMS–1813149 over the course of this doctorate and writing this dissertation. Separate funding acknowledgements are given in chapters   \ref{chap:quantum double sectors}, \ref{chap:category of quantum double}, \ref{chap:symmetry}.
  \end{minipage}

  \vspace*{\fill}
\end{center}

  \clearpage
  {\centering \bfseries Acknowledgements\par}
\vspace{0.5cm}
    It is an impossible task to put in writing the countless souls who have helped me along in my journey toward this doctorate. I would like to try nonetheless, knowing that any acknowledgment will not fully convey the depths of my gratitude.

    I would like to thank my parents, Arvind and Shalini, who nurtured me to value curiosity and creativity. I would not be in this privileged position without their constant love and support, and an assurance that I will always have a home no matter what I chose to do with my life. I remember strongly how they fought to shield me from the clutches of societal pressures and to let me follow my own path, however perilous it may be. Thank you, aaji, for tolerating my many practical jokes. I'm sorry I couldn't come visit you one last time before your passing. Though I left India to find the best path toward my goal as a researcher, my heart frequently wishes to be close to my family and my roots. 

    I would like to thank my lovely wife, Tamar, who sat through various mock-presentations and gave me excellent academic advice despite not knowing half the words. She's my apostle, cheer-leader, critic, PR-team, confidant. I am grateful to her for giving me plenty to look forward to outside my research, and for walking this path beside me. She should really be getting an honorary PhD for taking on the tough job of rounding out my (many) rough edges. Her constant compassion inspires me to be a better person every day.

    I could not have hoped for a better, more learned advisor to guide me in the quest to obtain my PhD. I would like to thank my advisor Bruno, who has witnessed first-hand my growth as a researcher and as a mathematician. I thank him for patiently handling my initial refusal to learn the beautiful mathematical structures that govern my research. Many of my research headaches would've been avoided if I chose to listen to his well-founded advice. And he certainly gave me plenty of it, which will stay with me well after the completion of my PhD.

    Science is not about individual brilliance, but rather about collaborative efforts that together push the collective frontiers of knowledge. Thank you, Alex, for showing me by example the effort needed to succeed as a researcher. Thank you, Daniel (Wallick), Daniel (Spiegel), Michael, for teaching me how to navigate the complexity of mathematical structures. Thank you, Dominic and Kyle, for strengthening my physical intuition with a myriad of examples. Thank you, Sven, Mukund, Martin and Dominic, for the many research discussions and for guiding me throughout my journey.

    My friends have played an important role in this journey. Thank you Francisco for being one of my closest friends, for your unbiased advice, late-night gaming sessions, conversations about nothing in particular. Thank you Patty, for being a close friend and urging me to spend more time in the great outdoors. You will be deeply missed. Thank you Karthik for our shared bond rooted in good music. Thank you Yash, Michael, Tyler, Jake, Pratik, Patrick, Julie for being my cohort and giving me a sense of community at Davis. Thank you to my friends ``from discord'', Akshay, Shreyas, Haroun, Bhattu, for keeping me virtual company during the isolation of covid. Thank you to my closest friends from Satpura, Utkarsh, Naveen, for not letting the inter-continental divide be a barrier to organize our annual trips.

    Lastly, I would like to thank my furry kids, Charlie and Sir Gregory, for supplying me with a limitless supply of cuddles and adorable pictures.
    
    \begin{center}
    \vspace*{\fill}
        \begin{minipage}{0.25\linewidth}
      \centering
      \includegraphics[width=\linewidth]{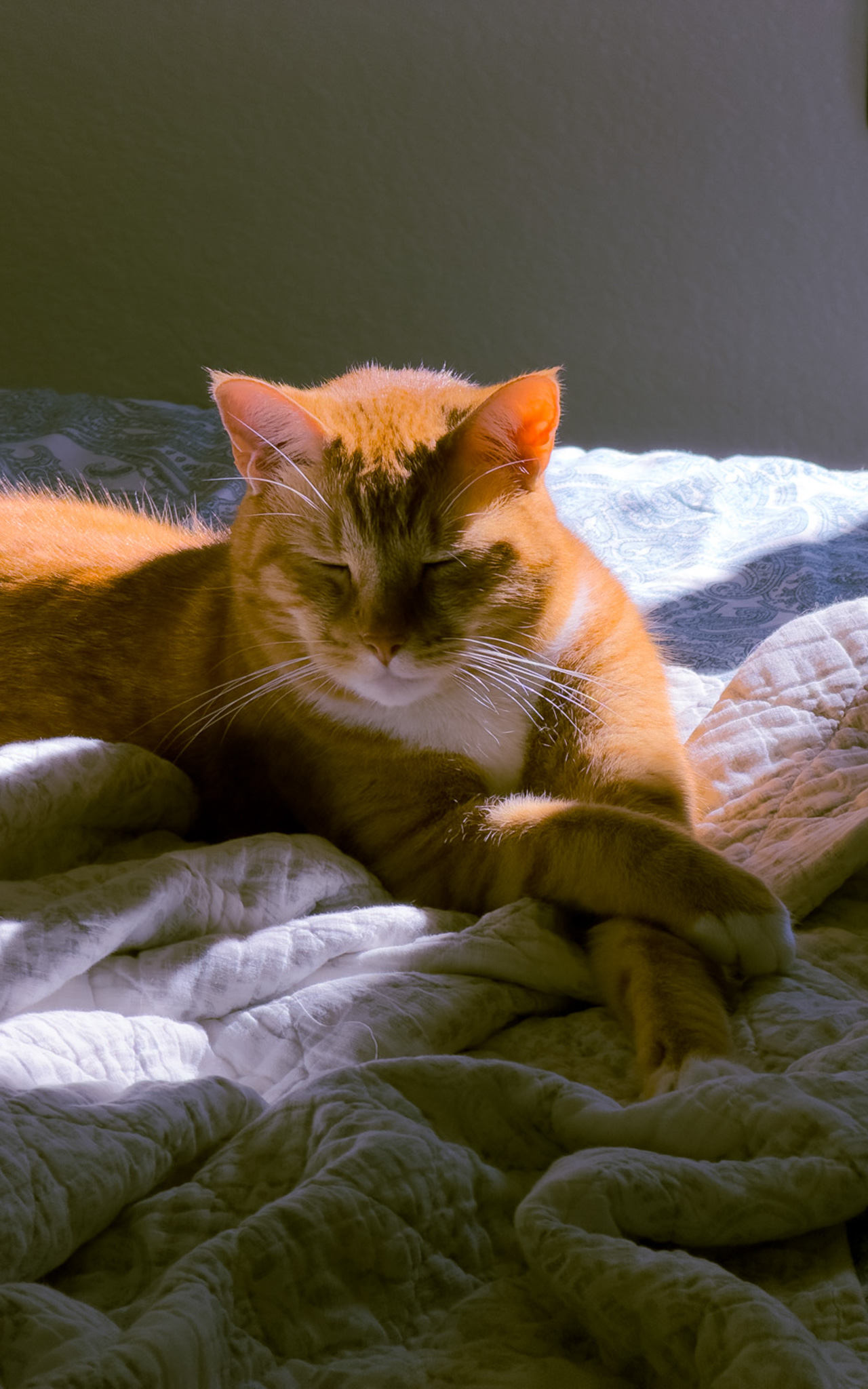}\\[0.5em]
      {\footnotesize Charlie}
    \end{minipage}
    \hspace{2cm}
    \begin{minipage}{0.25\linewidth}
      \centering
      \includegraphics[width=\linewidth]{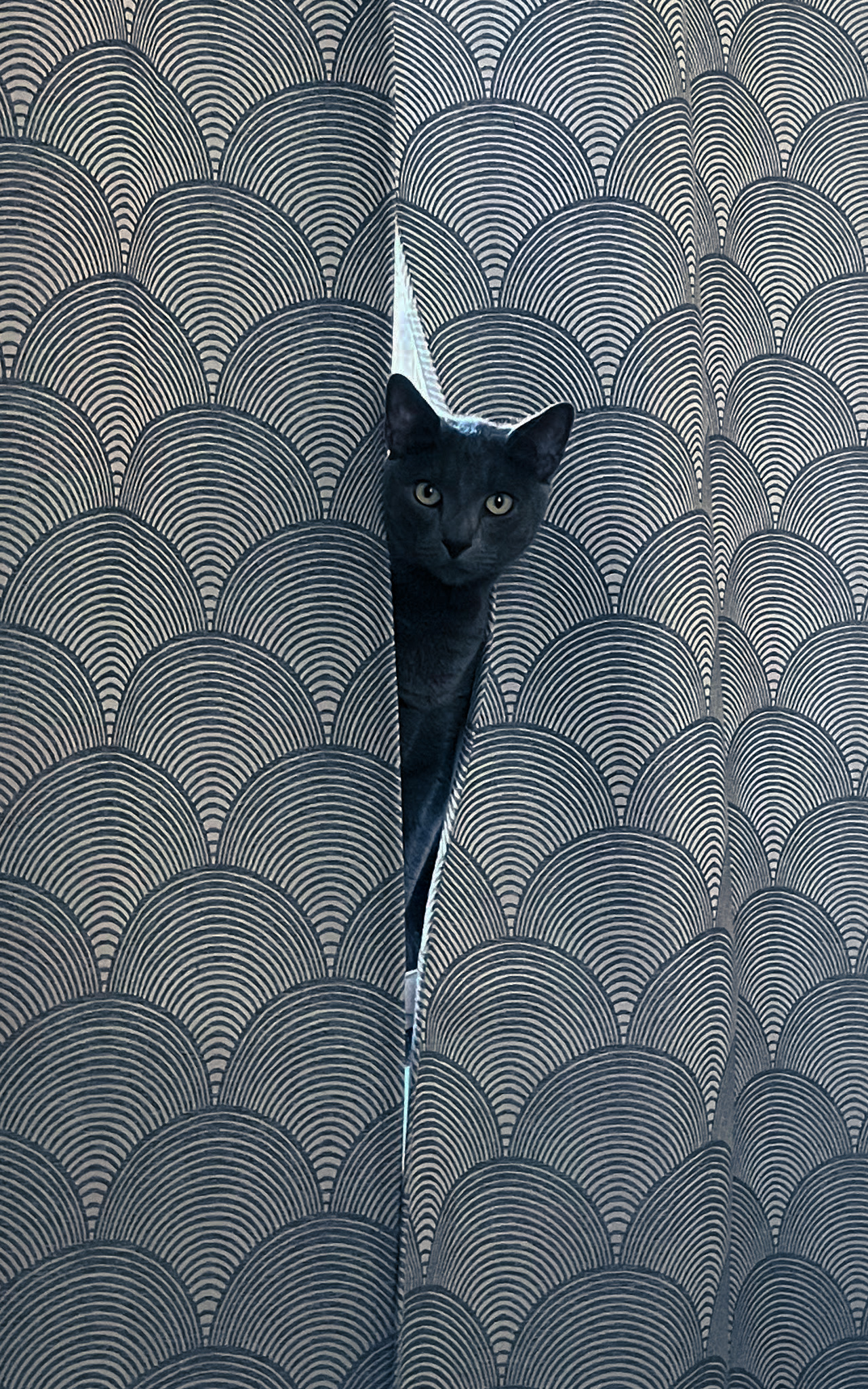}\\[0.5em]
      {\footnotesize Sir Gregory}
    \end{minipage}
    \end{center}

  \vspace*{\fill}

\clearpage

{\centering \bfseries Foreword \par}
\vspace*{0.5cm}
Since I started out as a condensed-matter physicist who was interested in topological phases, I had close to no mathematical background. As you may imagine, operator algebra is a difficult field to parse through, with a rich history and many developments that one must first understand before using them to study physical problems like the classification of topological phases. Many crucial results used to study quantum spin systems are decades old, and many-a-times hidden behind somewhat outdated language/perspectives. In terms of modern and accessible introductions to operator algebras with a focus on use in quantum spin systems, the book by Pieter Naaijkens probably comes the closest, but there have been several developments in this subfield since then, so another perspective is warranted. 

In contrast, while topological phases is a currently a very popular field and has seen many recent developments, by now there are several works that have done a great job at providing modern and accessible introductions to the subject. The same can be said for Kitaev's Quantum Double models, on which 2 whole chapters are dedicated. 

As such, this dissertation is written with the intention of providing a focused operator algebraic introduction with the sole aim of tackling topological phases. It answers and clarifies the many questions that I had when I was starting out. I hope that others will use this resource to more effectively begin the study of topological orders using operator algebras.

Notably, absent is a thorough treatment of the physics of topological phases, the categorical viewpoint of topological phases, and an introduction to Kitaev's Quantum Double models. I felt like there were enough resources which give a great introductory treatment of these topics, and any contribution of mine would be incremental at best.

Chapter \ref{chap:history} provides a thorough account of the many historical developments in the field of topological orders and operator algebraic efforts therein.

Chapter \ref{chap:physics motivation} provides an intuition of anyons, and the type of properties one should expect it to have. Some challenges in passing to infinite volume limits are addressed heuristically.

Chapter \ref{chap:UMTC background} provides a bare-bones introduction to tensor categories, ultimately building up to the concept of a Braided $\rmC^*$-tensor category and a Unitary Modular Tensor Category.

Chapter \ref{chap:operator algebras} has the meat of the introductory chapters. It first introduces the various $\rmC^*$-algebras that arise in quantum spin systems like the quasi-local algebra, cone von Neumann algebras, the auxilliary algebra. Then it discusses the \emph{anyon selection criterion} and the cornerstone result that any anyon sector can be equivalently thought of as a localized, transportable endomorphism of the auxilliary algebra. Finally, the category of anyons is constructed and it is shown that it has the structure of a braided $\rmC^*$-tensor category.

Chapters \ref{chap:quantum double sectors}, \ref{chap:category of quantum double} then switch focus and construct a UMTC from the anyon sectors of Kitaev's quantum double model. The former first classifies the irreducible anyon sectors in this model, and the latter establishes the categorical structure.

Chapter \ref{chap:symmetry} switches back to the general theory of topological orders and defines the concept of a defect sector, which is roughly a global point-like symmetry defect that one can obtain in systems with an onsite symmetry by twice-truncating the symmetry using the Else-Nayak construction. Then it is established that defect sectors have the structure of a $G$-crossed braided $\rmC^*$-tensor category, and some illustrative examples are considered.


\setcounter{secnumdepth}{3}

\dominitoc         

\setcounter{tocdepth}{1}
\tableofcontents



\clearpage               
\pagenumbering{arabic}   
\setcounter{page}{1}

\include{history_introduction}

\include{motivation_v2}


\include{UMTCbackground}

\include{OperatorAlgebrasBackground}

\include{main_sectors_thesis}

\include{main_category_thesis}
\include{main_symmetry_thesis}

\end{document}

%% file: history_introduction.tex
\chapter{History Of Topological Phases}
\label{chap:history}
Mathematics and physics have always been intertwined since their inception. Sometimes physics discovers systems that necessitate the study of whole new mathematical structures; other times mathematics uncovers a beautiful language to describe the physics of these systems. One finds countless examples of this dance throughout history, with the group theoretic structure of symmetries in physics, to differential geometry describing the phenomenon of relativity, to gauge theories and their connections to electrodynamics.

It should come as no surprise, then, that the study of topological phases comes with a similar story. The novelty in this story is perhaps the uncovering of topological structures in physical systems, two very distinct worlds: the messy system-dependent physics of real systems, and the sanitized world of topology where structures are continuously deformable without effect. The collision course of these two worlds is the story of topological phases.

\section{Discovery And Early Understanding}
Let's set the stage with some theoretical predictions. In 3+1D, one may only have bosons and fermions. Bosons can trivially be exchanged with each other without effect. Fermions, when exchanged, get a phase change of $\theta = \pi$. This is due to the deep connections between the space-time Lie group and its connections to the permutation group in 3+1D. In 1977, Leinaas and Myrheim studied the exchange statistics of particles in 2+1D, and pointed out the possibility of \emph{fractional} statistics in 2+1D \cite{Leinaas:1977fm}. The connections of fractional statistics to the braid group were discussed by Wu in 1984 \cite{PhysRevLett.52.2103}. We now understand that this is due to connections between the space-time Lie group and the braid group in 2+1D. 

The discovery of the fractional quantum hall effect (FQHE) in 1982 \cite{PhysRevLett.48.1559} by Tsui, Störmer and Gossard kickstarted the study of topological phases in physics. They plotted the Hall conductance of a 2D electron gas at a $\nu = \frac{1}{3}$ filling fraction of the Landau level, and observed that it was quantized at a plateau just like the integer Hall effect, but at a fractional value. This plateau was incredibly flat and robust to small system doping as confirmed by later experiments, and indicated the incompressible nature of this quantum fluid state at this filling fraction. The history of topological phases begins in trying to explain the reason for this phenomenon, which indicated the presence of entirely new physics hiding in a routine material. 

Soon enough, more plateaus were discovered at other filling fractions. The following year, Robert Laughlin proposed a trial wavefunction to explain this state. This wavefunction successfully explained the incompressible nature of this state, where the electrons were correlated by repulsive interactions, and resisting compression due to the repulsive interactions. However this wavefunction predicted the existence of fractionally charged quasiparticles that carry the charge $e/q$ for the filling fraction $1/q$. This prediction was later confirmed by shot-noise experiments in the 90s, which directly measured the charge quanta of $e/3$ in the $\nu = 1/3$ filling fraction \cite{DEPICCIOTTO1998395}.

In 1984, Halperin argued that exchanging two $1/q$ particles would produce a fractional statistical phase $\theta = \pi /q $ \cite{PhysRevLett.52.1583}. In the same year, Arovas, Schreiffer and Wilczek computed the Berry phase for adiabatically braiding two Laughlin quasiholes and showed that this leads to the phase $e^{i \theta}$ to the resulting state \cite{PhysRevLett.53.722}. These particles thus had an exchange phase that is between that of a boson ($\theta = 0$) and a fermion ($\theta = \pi$). These particles could thus have any rational phase, and were christened \emph{anyons} by Frank Wilczek. Exchange statistics are also independent of the distance between the anyons, and is thus a long-range property of the system. It would thus be undetectable using order-parameters.

Many other filling fraction plateaus could be explained using the hierarchy construction of FQHE systems proposed by Haldane and Halperin: many different FQHE states stacked together to form a new fraction. Another alternative approach was proposed by Jain which explained these fractions due to the presence of quasi-particles called composite fermions: electrons bound to flux quanta.

In 1984, Tao and Wu realized that putting a FQH system on a manifold with a non-trivial topology like a torus yielded a ground-state degeneracy (GSD). The origin of this effect was unclear and there were some misguided explanations involving symmetry-breaking mechanisms in the usual physics tradition (see \cite{PhysRevB.31.8305, PhysRevB.31.3372, PhysRevB.28.2264}). The origin of this GSD was clarified by Wen, Niu \cite{doi:10.1142/S0217979290000139, PhysRevB.41.9377} by noting that it was robust to arbitrary weak perturbations. Thus there was no symmetry involved in this effect and it was beyond Landau's paradigm. Moreover, this GSD seemed to be inherently related to the statistics of anyons present in the system, and was thus a long-range effect. The GSD provided an entirely new quantum invariant; two systems with the same Hall conductance could be distinguished by their GSD.

FQH states provided the first example of ``topologically ordered phases'' as termed by Xiao-Gang Wen \cite{doi:10.1142/S0217979290000139}. These are ground states with long-range quantum entanglement and emergent gauge structure, not characterized by any local order parameter. A defining feature of topological order is the presence of robust ground state degeneracy that depends on the topology of the underlying surface rather than its local geometry.

Parallel to these developments, there were QFT efforts to explain the universal properties of FQH states. In 1989, Zhang, Hansson and Kivelson proposed that the $\nu = 1/3$ FQH state could be described by a $U(1)$ Chern-Simons gauge theory, which successfully reproduced the Hall conductance and anyon braiding statistics via a mechanism that binds fluxes with charges. These charge-flux quasi-particles yield, when braided, an Aharanov Bohm phase of $e^{i \theta}$ which matches the exchange statistics of anyons. Notably, the Chern-Simons gauge theory is a Topological Quantum Field Theory (TQFT), meaning its observables are topological in nature. Wen, Niu \cite{PhysRevB.41.9377} showed that indeed the long wavelength limit of the FQH state is equivalent to a $U(1)$ Chern-Simons TQFT.

This was a profound realization: the macroscopic phenomena of fractional charge, fractional statistics, and ground-state degeneracy could all be understood as emerging from an effective TQFT. In such a TQFT, the anyons correspond to quantized flux-charge tubes, and braiding them corresponds to nontrivial Wilson loop operators in the gauge theory. The success of the Chern–Simons description hinted at a deeper connection between condensed matter anyons and abstract topological invariants, a connection later made explicit by mathematicians studying knot theory and category theory.

By the end of the 1980s, experiments had revealed fractional quantum numbers and anyonic statistics, and theorists had identified topological ground-state degeneracy and field theoretic descriptions. The community began to appreciate that these ``fractional'' quantum Hall states were exemplars of a whole new class of phases that were characterized by topological order and supporting anyonic excitations. 

In hindsight, all of these states were a class of quantum states hosting quasi-particles called \emph{abelian anyons}. These anyons actually correspond to the $1$-dimensional representations of the braid group. The mathematically inclined reader may realize that the braid group also carries higher dimensional representations. There is thus a possibility of particles corresponding to these representations, termed \emph{non-abelian anyons}. The next major breakthrough would be the prediction of non-abelian anyons, which carry even more exotic statistics and possibilities.

In 1987, an even denominator plateau was observed at the $\nu = 5/2$ FQH state \cite{PhysRevLett.59.1776}. This plateau was unexplained by any previously discussed theory. Moore and Read proposed that this plateau was explained by a Pfaffian wavefunction arising from a pairing of the quasi-particles in a BCS-like fashion \cite{moore1991nonabelions}. Crucially, they showed that the quasi-particles would carry \emph{non-abelian} statistics. Meaning, braiding operations between multiple different particles do not commute, and instead depend on the order of the braids. This property was what lead to the connection between the higher dimensional representations of the braid group. The underlying TQFT for the Moore-Read Pffaffian is now understood to be the $SU(2)_2$ Chern-Simons TQFT, or equivalently an Ising-type TQFT. Other filling fractions like the $\nu = 12/5$ state have also been theorized to host non-abelian anyons. The $\nu= 5/2$ FQH state remains a leading contender for a liquid with experimental signatures of non-abelian anyons.

\section{Braided Tensor Categories And TQFTs}
As the variety of models expanded, there was a necessity of a unifying framework. There was obviously a rich framework involved in the braiding of anyons, especially non-abelian anyons. Physicists and mathematicians gradually converged on this issue and realized that topological phases are described by braided tensor categories. There were several interesting developments that led to this understanding. 

In 1989, Moore, Seiberg \cite{moore1989classical} analyzed 1+1D Conformal Field Theories (CFTs) and derived consistency conditions involved in the fusion and braiding of particles in these theories. These consistency conditions, now known as the pentagon and hexagon equations, ensure well-behaved, single valued crossing symmetries in the CFT and encode the same information as anyon fusion and braiding rules. Around the same time, Witten's work revealed deep connections between TQFTs and topological Knot invariants \cite{witten1989quantum}, detailing that the knot invariants could be understood as the braiding of charges in a 2+1D TQFT, making clear the connections between topology and anyons.

When there are many non-abelian anyons, they cannot arbitrarily braid and fuse due to the importance of the order of the braiding. The system should also be consistent for any number of anyons. Thus the braid group representations can't be chosen independently, and must be dependent on the fusion and splitting rules which change the number of anyons in the system. The mathematical solution to these conditions then resulted in the notion of a Braided Tensor Category (BTC). It is an algebraic structure with the objects as anyons, fusion rules, braiding rules, and consistency conditions leading to $F,R$ symbols which ensure consistency in the order of these operations.

Researchers like Kitaev and Freedman, having the quantum computation backgrounds, explicitly formulated anyon theories in category theoretic terms. The paper by Freedman, Kitaev, Larsen, Wang \cite{freedman2002topologicalquantumcomputation} established the connection of anyons to quantum computing and established the notation of \emph{topological quantum computation} (TQC). This paper connected Witten's work to anyon theories and explicitly established the BTC strcture present in anyon systems. The physics community gradually transitioned to this viewpoint in the 90s and 00s, particularly as there was interest in TQC. By the 00s, it became folklore that the mathematical structure describing anyons is a BTC.

\section{Lattice Realizations Of Topological Order}
The FQH state is a continuum state and thus it is natural to expect its behaviour to be characterized by continuum theories like TQFTs. In 1989, Wen proposed a theory of chiral spin liquids \cite{PhysRevB.40.7387} which are spin systems where the spins form a resonating valence bond (RVB) liquid that could potentially harbour a $\bbZ_2$ gauge theory with fractionalized excitations. 

Even before topological order was clearly defined, Kogut and Wegner had already drawn parallels between spin systems and gauge theories. Wegner's 1971 model of an Ising gauge theory on a lattice was essentially a precursor of Kitaev's toric code \cite{Wegner:1971app}. Kogut in his review on lattice gauge theories noted that a $\bbZ_2$ gauge theory can be formulated as a spin-$1/2$ system with a four spin interaction on each plaquette \cite{RevModPhys.51.659}. The idea that long-range ordered spin states could similarly host anyonic excitations gained traction in the 1990s, especially in the context of RVB states. 

In 1997, Kitaev introduced the toric code model and its generalizations, the quantum double models \cite{MR1951039} (See \cite{HamdanThesis} for an operator algebraic introduction or \cite{Bombin2007-uw} for a physical perspective). This kickstarted the study of topological phases using exactly solvable lattice models. These models had the specialty that they are exactly solvable, in the sense that one can has a closed-form expression of the eigenstates of these models. Moreover these models, in contrast to the field theoretic models proposed above, are computationally tractable. This enabled many different avenues for research into topological phases. 

Besides these benefits, the toric code and its generalizations established a link between lattice models and TQC. On a torus, the ground-state is 4-fold degenerate, and this degeneracy is robust to arbitrary weak perturbations. The ground-state space thus forms a logical qubit, and due to the degeneracy, the logical qubit is \emph{fault tolerant} by design, i.e., the faults introduced by arbitrary weak perturbations don't change the state of the logical qubit. Freedman, Kitaev, Larsen, Wang \cite{freedman2002topologicalquantumcomputation} established the connection of anyons to quantum computing. With the end goal of engineering anyons to perform TQC, focus turned to constructing topologically ordered lattice spin systems. 

In 2005, Levin and Wen introduced the string-net models \cite{PhysRevB.71.045110} (see \cite{christian2023lattice, green2024enriched} for a mathematical perspective or \cite{hu2018full} for a physical treatment), which was a profound breakthrough in using lattice models for quantum computation, as it provided an exactly solvable model for realizing a \emph{doubled order}, meaning a gapped topological order associated with the Drinfel'd center of an input fusion category.
In summary, the string-net model provides a toolkit for lattice Hamiltonians: given any desired anyon content without edge modes, one can construct a local spin model that has that topological order. This was a monumental conceptual step. It also firmly cemented the role of category theory in condensed matter. Phrases like ``fusion rules'', ``$F$-symbols'', and ``$R$-symbols'' became part of the working language for characterizing lattice topological phases.

\section{Operator Algebras And Topological Phases}
\label{sec:DHR and lattice}
\subsection{Continuum field theories}
In 1964, Rudolf Haag and Daniel Kastler introduced the Algebraic (or Axiomatic) approach to QFT, called AQFT \cite{haag1964algebraic}. In this approach, one assigns to each region of space-time a von Neumann algebra of local observables. The idea was that instead of quantizing the fields themselves, one could study the algebra of observables and their representations. This framework encapsulated locality and provided a mathematically robust foundation for QFT using $\rmC^*$-algebras and von Neumann algebras. This approach showed that many structural results follow just from this operator-algebraic approach,  like Haag-Ruelle scattering theory \cite{haag1958quantum, ruelle1962asymptotic}, and more relevantly, a general analysis of superselection sectors by Doplicher-Haag-Roberts (DHR) starting in the late 1960s \cite{doplicher1971local, doplicher1974local}. The DHR theory demonstrated that under the algebraic framework in 4d QFT, charges are associated with inequivalent representations of the observable algebra that are localized in space, and these charges must obey either Bose or Fermi statistics with an underlying global gauge symmetry. In fact, Dophlicher and Roberts showed that one can reconstruct a compact gauge group from the properties of these superselection sectors \cite{doplicher1989new}, a profound result establishing a duality between the algebraic description and gauge symmetry. Operator algebra (OA) methods thus solved a conceptual problem: how to classify and combine charges without presuming a gauge group, deriving it instead from representation theory of observable algebras.

By the 1980s OAs had become a standard tool in mathematical physics. They offered a unifying language for quantum fields and many-body systems: local algebras of observables, states as algebraic functional, dynamics as automorphisms, and charges as representations. These ideas set the stage for tackling topological phases, which are subtle quantum orders not characterized by conventional observables.

One of the great successes of OA methods is the classification and analysis of superselection sectors in low-dimensional models, which directly applies to the anyonic excitations in topologically ordered systems. In 1989, Fredenhagen, Rehren, and Schroer extended the DHR superselection theory to $2+1$D, showing that localized excitations in two dimensions can obey braid group statistics rather than ordinary Bose/Fermi exchange \cite{fredenhagen1989superselection}. This seminal work gave a rigorous underpinning to anyons as localized endomorphisms of the observable algebra (which we will elaborate on in Chapter \ref{chap:operator algebras}). This provided a conceptual breakthrough: one could classify anyon types by classifying the representation category of the local observable algebra, a problem amenable to OA techniques.

Another fruitful thread is the use of \emph{subfactor theory} to classify and construct 2D topological orders. Vaughan Jones' discovery in 1983 of the Jones index for subfactors \cite{jones1983index} revealed a surprising quantization: the index $[M:N]$, for subfactors corresponding to extensions $ N \subset M$ of factors, could only take specific values. This led to rich algebraic structures (notably the Temperley-Lieb algebra and planar algebras) and ultimately knot invariants (the Jones polynomial). Jones' work opened a new field of quantum topology by linking OAs to knot theory and low-dimensional topology. Soon after, it was realized that the standard invariants of a subfactor could serve as data for a TQFT. In 1988, Witten’s interpretation of the Jones polynomial via Chern–Simons TQFT \cite{witten1989quantum} gave a physical context to these categories: the Jones representation of the braid group corresponds to anyonic braiding in a $2+1$D TQFT.

Ocneanu and others in the 1990s further developed the connection, showing how to construct state-sum invariants of 3D manifolds using subfactor data. In essence, each subfactor with finite index provides a fusion category, and often a rich structure (like a UMTC) describing some hypothetical anyon system. For example, the even part of the $E_6$ subfactor yields the Ising modular category and the so-called ``Haagerup subfactor'' yields an exotic modular category not obviously realized by any known quantum symmetry. This line of research indicated that OAs could predict new topological orders in principle, by enumerating possible consistent anyon models.

\subsection{Lattice systems}
In parallel with these developments, OA techniques were applied to lattice quantum systems (quantum spins or lattice fermions). The infinite lattice can be treated as an inductive limit of finite-subsystem algebras - often called the quasi-local $\rmC^*$-algebra of the spin system. This approach, systemized in the classic texts by Brratteli and Robinson (1979, 1981), allowed rigorous definitions of phases, symmetry breaking, and locality for infinitely extended systems. For instance, a ferromagnetic phase is described by a state (expectation functional) on the quasi-local algebra that is invariant under the symmetry breaking, and locality for infinitely extended systems. For instance, the ferromagnetic phase is described by a state (expectation functional) on the quasi-local algebra that is invariant under the symmetry but not clustering. Concepts like the \emph{split property} (the fact that in a gapped system, the algebra of a region and its complement can have a tensor product split) were discovered, linking the type of von Neumann algebra to physical properties like correlation length. The lattice algebraic approach was essential to later understand topological order: it provides a language to define a phase as an equivalence class of states on the quasi-local algebra (or of gapped Hamiltonians generating those states), without referring to any particular local order parameter.

Lattice models of topological order, while not Lorentz-invariant QFTs, can be treated with similar operator-algebraic ideas. Pieter Naaijkens in 2011 rigorously studied Kitaev's toric code model using OA methods, and identified the corresponding superselection structure of the anyons \cite{MR2804555}. He found that the resulting superselection structure can be turned into a $\rmC^*$-braided tensor category, and moreover that it is equivalent to the conjectured category $\Rep D(\bbZ_2)$, the representation category of the Drinfel'd center of the category $\Rep(\bbZ_2)$, or equivalently, the representation category of the quantum double $D(\bbZ_2)$ \cite{MR3135456}. Naaijkens \cite{Naaijkens2015} later extended this structure to quantum double models with abelian finite group $G$, and very recently this has been extended to quantum double models with arbitrary finite group $G$ \cite{bols2025classification, bols2025category}.

This was a breakthrough on 2 fronts. First, of mathematical interest, is that there are no additional superselection sectors than the ones that were already conjectured (this is not guaranteed to be the case, and in fact fails spectacularly in higher dimensions for the simplest of models \cite{vadnerkar2023superselection}, though that is a problem of not yet having the correct criterion for higher dimensions). This showed that the OA approach and studying the DHR-style superselection theory was a valid approach to understand topological phases. The other, of relevance to physicists, is that small perturbations of the toric code Hamiltonian do not spontaneously generate new sectors. 

Ogata showed that starting from some axioms like a pure, gapped ground-state, the dynamics satisfying Lieb-Robinson style bounds, and approximate Haag duality (the latter two providing a lattice analogue for light-cone type information propagation), one can impart a braided $\rmC^*$-tensor category (BTC) structure to the superselection sectors with respect to this ground-state. Another major result attributable to the OA approach is that this BTC structure is robust if the Hamiltonian is ``slowly'' perturbed in a way that does not close the gap. This result was shown in various parts in \cite{cha2020stability}. This approach also works in systems with a boundary \cite{jones2023local}, and has recently been shown to work also in SET orders \cite{kawagoe2024operator}. The main technical assumption in this analysis is Haag duality, which has been shown to work in a large class of lattice models \cite{ogata2025haag} including Quantum Double models and the Levin-Wen string-net models, and is expected to hold (at least approximately) in the bulk for all symmetry-enriched topological phases.

A more recent bridge between OAs and topological order on the lattice comes from the study of tensor network states, and in particular, matrix product states (MPS) in 1D and projected entangled-pair states (PEPS) in higher dimensions. These ansätze, popular in computational physics, turned out to have deep OA connections. In 1992, Fannes, Nachtergaele, and Werner \cite{fannes1992finitely} showed that the set of translationally invariant MPS with a given local dimension can be identified with states on a certain AF (approximately finite) C*-algebra, and that the structure of that algebra’s representations gives rise to the “finitely correlated states” classification. In essence, they proved that any 1D gapped ground state with a unique infinite-volume pure state is an MPS, and different phases correspond to different inequivalent representations of the MPS transfer-operator algebra. This area has since seen lots of progress, including the advent of matrix product operator (MPO) algebras that appear as symmetries on the boundary of 2D PEPS \cite{csahinouglu2021characterizing, williamson2016matrix}. From these MPO symmetries one can construct a $\rmC^*$-algebra (sometimes called the annular fusion algebra of the PEPS transfer operator), whose structure of idempotents and simple modules corresponds exactly to the different anyon types in the topological order.

\section{Future}
The rich interplay between Operator algebras, Category theory, TQFTs and topological order continues to be a fascinating collaboration. Physicists are now aware that to classify exotic phases, one often ends up classifying some algebraic invariant. Mathematicians, on the other hand, are using physical intuition to guide the search for new algebraic structures. The math provides clarity and rigor, while the physics provides examples and intuition. This dialogue stands as a shining example of interdisciplinary synergy, one that is sure to continue yielding deep insights into the nature of quantum order.

\bibliographystyle{alpha}
\bibliography{intro_bibliography}

%% file: motivation_v2.tex
\chapter{The Physics Of Anyons}
\label{chap:physics motivation}
\minitoc

This chapter is intentionally light on formalism. The goal is to build intuition for how anyonic excitations are created, moved, braided, and fused in gapped $2+1$D systems. Precise definitions and proofs will appear in later chapters.

\section{Quantum lattice systems}
We imagine a large (but finite) patch of a two–dimensional lattice with identical bosonic microscopic degrees of freedom on each site. The system is governed by a (frustration-free) Hamiltonian consisting of local interaction terms, and has a spectral gap above its ground space. In practice, rigorous treatments carry ``exponentially small tails'' in locality statements, but here we ignore such technicalities unless they matter for intuition.

We can think in the following simple terms:
\begin{itemize}
\item \textbf{Ground states and excitations.} Ground states simultaneously minimize all the local interaction terms, and have eigenvalue $0$. Excited states are orthogonal to the space of ground states. We say that a state has an excitation in some region if it does not minimize all the interaction terms in that region.
\item \textbf{Local operations.} Acting in a small disk changes only what lives there, and observers living in far away regions cannot tell what you did locally.
\item \textbf{Moving excitations.} In many gapped models we can move localized excitations along narrow paths using finite-depth local unitaries (FDQCs). Meaning we can apply a fixed number of successive circuits, each consisting of local unitary operations living in a region no bigger than a circle of fixed radius.
\item \textbf{Adiabatic transport.} Slowly changing the Hamiltonian (or, equivalently, applying a controlled sequence of local unitaries) transports excitations along a thin strip containing chosen path $\gamma$ without creating new excitations in the process.
\end{itemize}

It is often useful to visualize motion in {space–time}. A world-line is the space-time history of a localized excitation. Two different processes are the same for distant observers if their world-lines can be deformed into one another without crossing other world-lines.

\begin{figure}[!ht]
\centering
\includegraphics[width=0.5\linewidth]{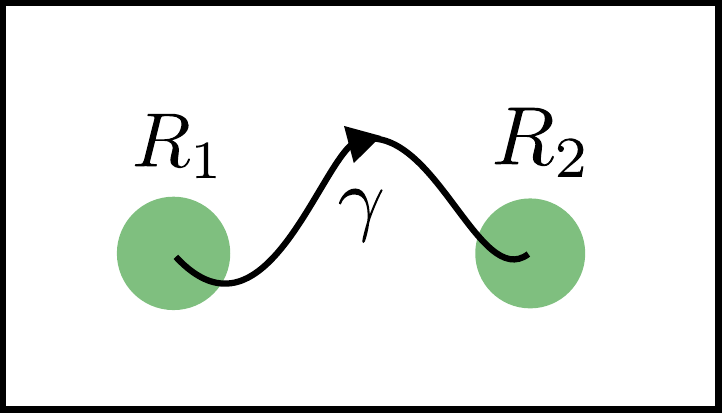}
\caption{Transporting localized excitations from a disk $R_1$ to a disk $R_2$ along a thin strip containing path $\gamma$.}
\label{fig:particle transport}
\end{figure}

\begin{figure}
    \centering
    \includegraphics[width=0.5\linewidth]{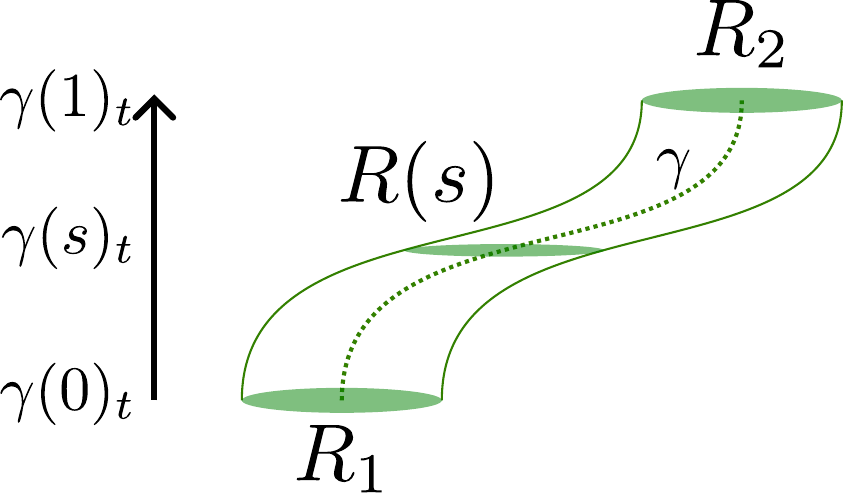}
    \caption{The transport process from Figure \ref{fig:particle transport} viewed as a world-line in space-time.}
    \label{fig:particle transport space-time}
\end{figure}

\begin{figure}[!ht]
\centering
\includegraphics[width=0.5\linewidth]{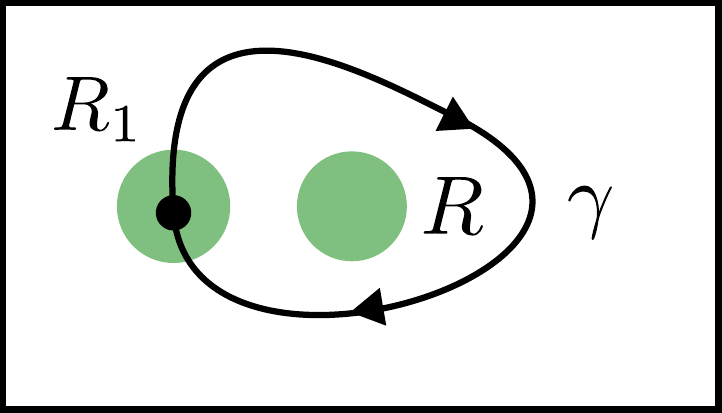}
\caption{For usual excitations, circling one around the other does not change the state at long distances.}
\label{fig:particle transport in circle}
\end{figure}

In a topologically trivial gapped phase (short-range entangled), excitations can be created or annihilated \emph{locally}. If you take one such excitation around another far away and return, nothing measurable changes, i.e., the process is invisible at long distances (Fig.~\ref{fig:particle transport in circle}). We term these excitations \emph{usual excitations}. Bosons\footnote{and after a few modifications to our simple picture, fermions too are examples of these particles. One has to account for non-trivial exchange statistics obeyed by fermions, and mod measurable changes by a possible phase shift of $\pi$.} are examples of usual excitations.

\section{Anyons}
By contrast, anyonic excitations, or simply \emph{anyons}, are special: a single anyon cannot be created or destroyed by a strictly local operation. They must be \emph{pair-created} as an anyon with label $a$ and its conjugate $a^*$. Separating the pair and looking inside a disk only containing the $a$ anyon leaves a ``locally persistent'', nontrivial topological charge: small operations inside a disk cannot change the anyon label $a$. Usual excitations, by virtue of being locally created or destroyed do not leave behind such a charge. We denote the topologically trivial charge by $1$, and call it the \emph{trivial anyon}. 

Due to the principle of homogeneity, and the fact that anyons are ``locally persistent'' topological charges, we would rightfully expect that the anyon labels are actually independent of the size of this disk, or where it is placed. It turns out to be the case when we carry out a rigorious analysis of anyons in the operator algebra setting. We also observe that due to homogeneity, a disk that contains $a$ must also be capable of containing its conjugate $a^*$. Additionally, we may hope that the index set of possible anyons is finite. This fact is certainly true in lattice models that are sufficiently nice, but in general may not hold. In this simple setting we assume it to hold. 

\section{Strings and remote detectability}
\begin{figure}[!ht]
\centering
\includegraphics[width=0.7\linewidth]{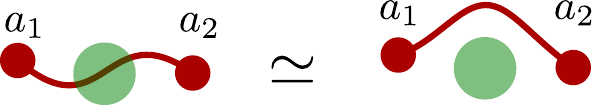}
\caption{One can freely deform a string (in red) as long as an anyon is not crossed. Local probes (drawn in green) supported away from anyons cannot see how the string is drawn.}
\label{fig:strings can be deformed}
\end{figure}

An intuitive way to keep track of pair creation of $a-a^*$ is to draw a thin \emph{string} from $a$ to $a^*$ to remember that they can, in principle, pair re-annihilate. Strings are bookkeeping devices. You can slide and wiggle them freely as long as you do not force them to cross an anyon (See Fig.~\ref{fig:strings can be deformed}).

\begin{figure}[!ht]
\centering
\includegraphics[width=0.3\linewidth]{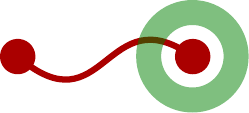}
\caption{Any loop around the hole must intersect the annulus, so loop processes can probe the enclosed charge.}
\label{fig:operators_encircling}
\end{figure}

\begin{figure}[!htbp]
\centering
\begin{subfigure}[b]{0.3\textwidth}
\centering
\includegraphics[width=0.9\linewidth]{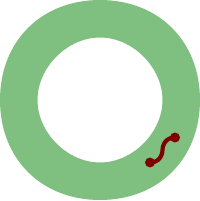}
\caption{Create $a$–$a^*$ in the annulus}
\label{fig:anyon creation in annulus}
\end{subfigure}
\hspace{0.2cm}
\begin{subfigure}[b]{0.3\textwidth}
\centering
\includegraphics[width=0.9\linewidth]{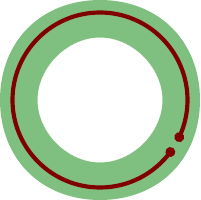}
\caption{Carry $a$ around once}
\label{fig:anyon encirclement in annulus}
\end{subfigure}
\hspace{0.2cm}
\begin{subfigure}[b]{0.3\textwidth}
\centering
\includegraphics[width=0.9\linewidth]{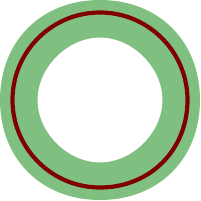}
\caption{Annihilate the pair}
\label{fig:anyon annihilation in annulus}
\end{subfigure}
\caption{A schematic loop operator process.}
\label{fig:loop operator}
\end{figure}

If you place a thin annulus around an anyon and run a string once around that annulus, the string {must} intersect the annulus (Fig.~\ref{fig:operators_encircling}).
This motivates \emph{loop operators}: prepare an $a$–$a^*$ pair in the annulus, carry $a$ once around the hole, and annihilate the pair again (Fig.~\ref{fig:loop operator}).
Because nothing is left inside the annulus at the end, such a loop is insensitive to microscopic details but can still pick up a topological signature of what sits in the hole.

If an anyon is well isolated from other excitations, suitable loop processes in a surrounding annulus can distinguish its label. This is called the \emph{remote detectability principle}, and is not automatically guaranteed. Nevertheless, under some relatively tame assumptions, it is expected that this principle holds.

\section{Fusion}
\begin{figure}[!htbp]
\centering
\begin{subfigure}[b]{0.45\textwidth}
\centering
\includegraphics[width=0.8\linewidth]{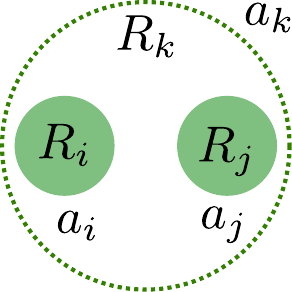}
\caption{Two anyons $a_i,a_j$ inside a larger region $R_k$, behaving as $a_k$ localized in $R_k$. }
\label{fig:fusion of two anyons}
\end{subfigure}
\hspace{0.2cm}
\begin{subfigure}[b]{0.45\textwidth}
\centering
\includegraphics[width=0.8\linewidth]{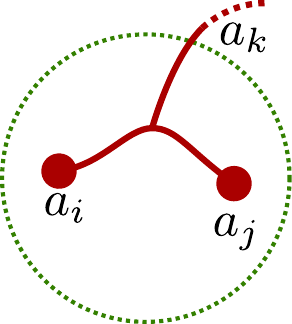}
\caption{String picture: $a_i$ and $a_j$ viewed as a single effective charge $a_k$ localized in $R_k$}
\label{fig:string representation of fusion}
\end{subfigure}
\caption{Fusion viewed as a coarse–graining two well separated anyons into one effective charge inside a larger disk.}
\label{fig:The fusion space of two anyons}
\end{figure}

Place two well separated anyons of labels $a_i$ and $a_j$ inside a larger disk $R_k$ (Fig.~\ref{fig:The fusion space of two anyons}). If we only probe outside $R_k$, all that is visible is a \emph{single} effective topological charge. This can be thought of as zooming out. Different microscopic states of the pair can lead to different effective charges $a_k$; these possibilities are called \emph{fusion channels} of $a_i$ and $a_j$. There may be more than one independent way to obtain the same $a_k$, and the number of such ways is the \emph{multiplicity} for the channel $a_i \otimes a_j\rightarrow a_k$.

We immediately observe that the trivial charge $1$ acts as a unit. Fusing $1$ with any $a$ does nothing. Also, since an anyon can be pair-annihilated, when a $a-a^*$ string-connected pair are brought together in a larger disk, they annihilate to $1$. However if we bring together an unconnected $a$ and conjugate $a^*$, they may not annihilate. 

We can summarize these fusion rules in the equation $$a_i \otimes a_j = \bigoplus_k N^k_{ij} \, \, a_k$$ where $N^k_{ij}$ is the multiplicity and is a positive integer.

\section{Braiding}
\begin{figure}[!ht]
    \centering
    \includegraphics[width=0.2\linewidth]{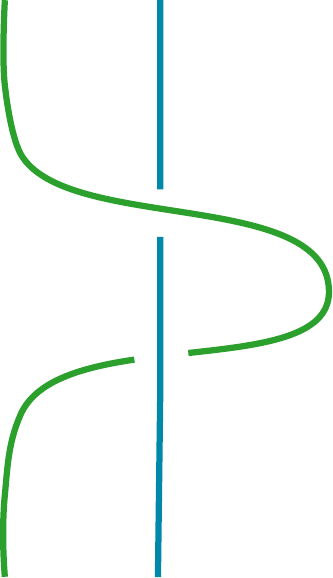}
    \caption{The space-time picture of taking $a_i$ around $a_j$. The world-lines of $a_i, a_j$ (viewed as space-time strings rather than cylinders) can be viewed as braiding around each other. This specific braid is called a double-braid.}
    \label{fig:braiding anyon world-lines}
\end{figure}

Now we again consider the setup of Figure \ref{fig:The fusion space of two anyons}, and take $a_i$ around $a_j$ by moving it along a path well separated and disjoint from $a_j$. When viewed as a process in space-time, this is a braid of their world-lines and thus called a braid (See Figure \ref{fig:braiding anyon world-lines}). This specific braid is called a double-braid. Since nothing happens outside $R_k$, the overall charge seen outside remains whatever it was before. What can change, however, is the internal state of the $a_k$ sector (since the multiplicity of the $a_i \otimes a_j \rightarrow a_k$ channel might be non-trivial). Thus a double braid implements a well-defined linear transformation on the $a_k$ sector. Intuitively, a braid is a unitary that depends only on the anyon labels and the braid, not on microscopic details of the path.

In the string picture of Fig.~\ref{fig:string representation of fusion}, an exchange forces one anyon to cross the other’s string once. That crossing is the source of the nontrivial transformation $R_{ij}^k$ associated with the pair $a_i, a_j$ fusing to $a_k$ and is a matrix of dimension equal to its multiplicity. 


It turns out that the fusion and braiding rules need certain compatibility relations, which ensure that consistency between different order of braiding/fusion operations. For example, the physics should be unchanged if two anyons fuse and then braid around a third anyon or if the two anyons braid around a third anyon and then fuse. Requiring that anyons can be pair-created or pair-annihilated, their movement is unitary etc.~lead to more compatibility conditions. 

When we abstract out this essential physics of anyons, it turns out that the anyon physics is universally described by a structure called braided $\rmC^*$-tensor categories, and in many cases by unitary modular tensor categories, which are particularly nice examples of the former. We will introduce and study these structures in Chapter \ref{chap:UMTC background}.

We now conclude this section by summarizing the string rules.
\begin{itemize}
\item Connect pair-created $a$ and $a^*$ by a directed string labelled $a$.
\item Reversing string orientation switches $a$ with $a^*$.
\item If $a$ and $a^*$ are connected by a single $a$-string, they may re-annihilate to the vacuum $1$.
\item Without the appropriate string connection, nearby $a$ and $a^*$ need not annihilate—history matters.
\item If an anyon crosses another anyon string, then the overall state gets modified by a factor given by the braiding matrix.
\item Two anyon strings in the same orientation can fuse to form a new string with the same rules as the anyons.
\end{itemize}

\section{Making anyons rigorous}
\subsection{Infinite volume}
We've been deliberately working in the finite volume picture so far. Let's explore heuristically what happens in the infinite volume case. In chapter \ref{chap:operator algebras} we will make this discussion rigorous.

In the infinite volume limit (known also as the thermodynamic limit), we work with the quasi-local algebra, which is an algebra of operators that are approximable with local operators. States are no longer vectors but density matrices\footnote{Strictly states are positive linear functionals of the quasi-local algebra, and a subset of states are normal, meaning they can be represented as a density matrix.}. A Hamiltonian is an unbounded operator and does not belong to the quasi-local algebra. However with some clever tricks, dynamics still exist in the infinite volume with the appropriate generalizations. So the concept of ground-states is well-defined with respect to this dynamics. There is no natural notion of a physical Hilbert space, so we have to get clever and use a foundational theorem in operator algebras, called the GNS construction. Given a state on a quasi-local algebra, one can represent this state on a Hilbert space called the GNS Hilbert space. There are vastly many states, and thus many representations. Many of these representations are unphysical: they may have infinite energy, infinitely many excitations, etc. So in practice, it is often best to fix a ``nice'' state like a frustration-free pure ground-state of the dynamics (often this state is also translation-invariant). Let's call this ground-state $\rho_0$.

\subsection{Creating an anyon state}
Now let's see how to create anyons in the infinite volume limit. We can of course create an $a-a^*$ pair using a local operator $V_a$ in the quasi-local algebra. Next we transport the conjugate anyon, $a^*$, $n$-steps away using a local unitary $U_n$. The new state (denoted $\rho_n$) now looks like\footnote{recall that for a density matrix $\rho$, a local operation by $U$ looks like $U \rho U^*$ where $(^*)$ means the adjoint.} $$\rho_n := U_n V_a \rho_0 V_a^* U_n^*$$

Since we always have that two locally different-looking states will be related by a local operation, any locally different anyon shares the same label. The right infinite volume generalization of an anyon contained in a finite disk is to consider an ``infinitely large'' disk containing an anyon, and send the conjugate anyon ``to infinity''. We can create the anyon-state from $\rho_0$ essentially by sending the conjugate $a^*$ outside the infinitely large disk via limits. In other words, the state $$\rho_a := \lim_{n \uparrow \infty} \rho_n$$ should correspond to an anyon state. This essentially means we've sent the conjugate, $a^*$, infinitely far away. This limit ends up working even though by definition there's no local operation that can send an anyon to infinity, because for local observers, ``really far away'' is essentially the same as ``infinitely far away''.

Recall that during pair-creation, we attached a string to the anyon-pair. As we would expect, sending the conjugate $a^*$ to infinity leaves a semi-infinite string attached to the anyon $a$. Since the string can be freely deformed, we enforce that it goes to infinity in a relatively tame way, meaning it stays inside some cone $\Lambda$ containing the end-point of the string (which is also where $a$ is located). 

\subsection{Anyons are localized and transportable}
\begin{figure}
    \centering
    \includegraphics[width=0.4\linewidth]{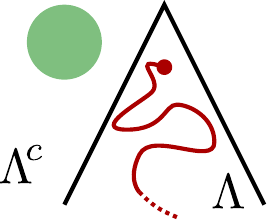}
    \caption{An anyon (in dark red) localized in cone $\Lambda$. A local observer located in $\Lambda^c$ (in green) will only see the ground-state $\rho_0$ around them.}
    \label{fig:localized}
\end{figure}

For local observers situated in $\Lambda^c$ (the complementary region to $\Lambda$), no matter where, $\rho_a$ will \emph{always} look like the ground-state $\rho_0$ because nothing has changed in $\Lambda^c$ (See Figure \ref{fig:localized}). That is, for all operators $A$ in the quasi-local algebra that are localized in $\Lambda^c$ we will have, $$\Tr{\rho_a A} = \Tr{\rho_0 A}$$ If this is true, we say that $\rho_a$ is \emph{localized} in cone $\Lambda$. Let us now switch the notation of $\rho_a$, instead denoting it as $\rho_a^\Lambda$ to signal that it is localized in $\Lambda$.

\begin{figure}[!htbp]
\centering
\begin{subfigure}[b]{0.3\textwidth}
\centering
\includegraphics[width=0.9\linewidth]{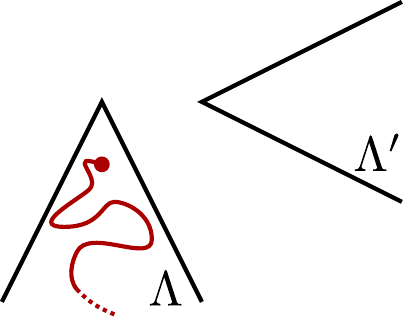}
\caption{Anyon is localized in $\Lambda$}
\label{fig:trans1}
\end{subfigure}
\hspace{0.2cm}
\begin{subfigure}[b]{0.3\textwidth}
\centering
\includegraphics[width=0.9\linewidth]{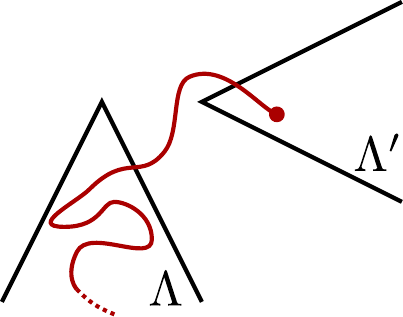}
\caption{Transport anyon to $\Lambda'$}
\label{fig:trans2}
\end{subfigure}
\hspace{0.2cm}
\begin{subfigure}[b]{0.3\textwidth}
\centering
\includegraphics[width=0.9\linewidth]{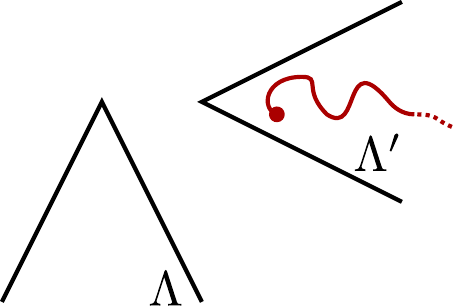}
\caption{Freely deform string to $\Lambda'$}
\label{fig:trans3}
\end{subfigure}
\caption{A schematic for a local unitary that transports anyon localized in cone $\Lambda$ to an anyon localized in cone $\Lambda'$.}
\label{fig:transportable}
\end{figure}

Now consider another arbitrarily chosen cone $\Lambda'$. If $\Lambda$ is disjoint from $\Lambda'$, then we can always use some local unitary $U_{\Lambda, \Lambda'}$ to move the anyon $a$ situated inside cone $\Lambda$ to cone $\Lambda'$, and since the string can be freely deformed and is just a bookkeeping device, move the string freely into $\Lambda'$ as well. The new state looks like $U_{\Lambda, \Lambda'}\rho_a^\Lambda U_{\Lambda, \Lambda'}^*$ (See Figure \ref{fig:transportable} for an algorithmic action of $U_{\Lambda, \Lambda'}$). Now there should be no physical difference between this state and a state $\rho_a^{\Lambda'}$ which contains the anyon $a$ and the string localized in cone $\Lambda'$. We thus have $\rho_a^{\Lambda'} = U_{\Lambda, \Lambda'}\rho_a^\Lambda U_{\Lambda, \Lambda'}^*$. We thus have another physical property on our hands: anyons, strings included, can be freely transported to arbitrary cones. We call an anyon state $\rho_a^\Lambda$ \emph{transportable} if for any arbitrary cone $\Lambda'$, there exists a local unitary $U$ such that $\rho_a^\Lambda = U \rho_a^{\Lambda'} U^*$.

\subsection{Anyon selection criterion}
Recall that for a given quasi-local algebra there are many unphysical states and representations, so we usually want to select the relevant states that exhibit the phenomena that we're interested in capturing. Now that we know that anyon states are localized and transportable, we're in a position to propose the anyon selection criterion: 
\vspace{0.3cm}
\begin{center}
    \textit{An anyon state is defined as a state that is localized and transportable.}
\end{center}
\vspace{0.3cm}
Having proposed this selection criterion, we are at serious risk of allowing \emph{too few states}, i.e., states that should be considered an anyon but are excluded by this criterion. Conversely, we are also at risk of this criterion allowing spurious states, i.e., states that should not be considered anyons, but are included by this criterion. Fortunately, neither worry materializes, and the criterion is well-designed to capture anyon states. The criterion allows for states that have finitely many local anyon-pairs, as well as for states containing two anyons going to infinity in different directions. It also excludes, among others, states with infinitely many anyons.

\section{Symmetry-Enriched Topological (SET) phases}
\begin{figure}[!ht]
    \centering
    \includegraphics[width=0.3\linewidth]{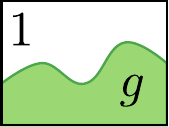}
    \caption{Example of a symmetry domain wall. To local observers deep in the green region it looks like a global symmetry action. To observers deep in the white region, it looks like nothing has changed.}
    \label{fig:symmetry domain wall}
\end{figure}

Consider a quantum lattice system that supports anyon states, and let the Hamiltonian have a global on-site (spontaneously unbroken) symmetry action of some group $G$. The fundamental physics of anyons remains unchanged even after taking into account the action of the symmetry on the anyons. In fact, if the symmetry is ``broken'' such that it only acts on a subset of the entire system (See Figure \ref{fig:symmetry domain wall} for a pictorial example of a symmetry domain wall), the anyon physics is still unchanged.


\begin{figure}[!ht]
    \centering
    \includegraphics[width=0.5\linewidth]{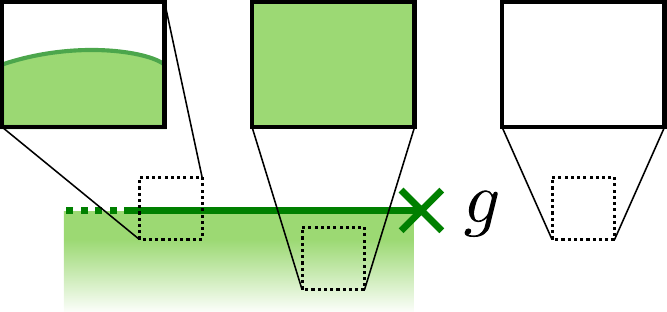}
    \caption{Example of a symmetry defect. Symmetry defects act as end-points for symmetry domains: Close to the defect line, the defect looks like a domain-wall to local observers. Far away in the bulk of the defect, the defect looks like a global symmetry action. Far away from the defect line, the defect looks like the ground-state.}
    \label{fig:symmetry defect}
\end{figure}

\begin{figure}[!ht]
    \centering
    \includegraphics[width=0.4\linewidth]{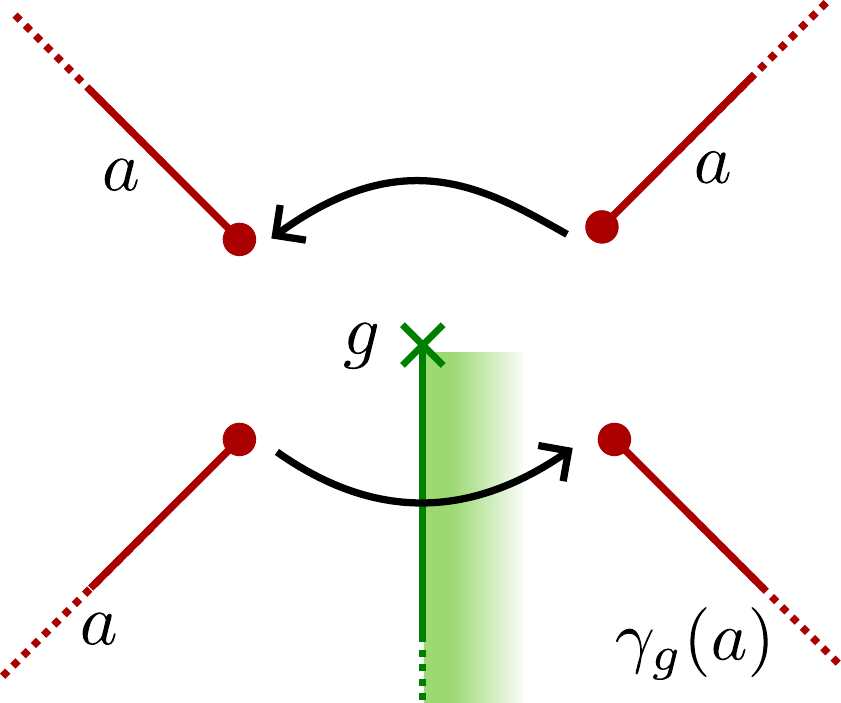}
    \caption{Symmetry defects permute anyon labels. If an anyon is moved across the defect line, there is a possible permutation to another anyon $\gamma_g(a)$. If it is moved far away from the defect, then it retains its label. Thus if we take an anyon (ccw) around the symmetry defect, it changes to the anyon $\gamma_g(a)$.}
    \label{fig:defect anyon permutation}
\end{figure}

It is when the symmetry is further broken into a \emph{symmetry defect} (See Figure \ref{fig:symmetry defect}), that we see that the physics of anyons in the presence of these defects is `enriched'. Taking an anyon around a symmetry defect can permute its label (See Figure \ref{fig:defect anyon permutation}), and symmetry defects can even act as sources and sinks for anyons, allowing them to be individually created or annihilated when brought close to a symmetry defect, an otherwise forbidden feature of anyons. 

The symmetry defects also have their own physics: they are mobile, and can fuse into another symmetry defect, and can ``crossed-braid'' around each other, which is a sort of generalization of braiding. This physics is theoretically described by mathematical structures called $G$-crossed braided $\rmC^*$-tensor categories, and in particularly nice cases, by $G$-crossed unitary modular tensor categories. We will explore these structures in Chapter \ref{chap:UMTC background}, and rigorously understand the microscopic behaviour of symmetry defects in Chapter \ref{chap:symmetry}.


%% file: UMTCbackground.tex
\chapter{Braided Tensor Categories}
\label{chap:UMTC background}
\minitoc

In this chapter we will go over the basics of braided ($\rmC^*$-)tensor categories. Everything covered in this chapter is standard material in tensor categories and we omit references to elementary facts. A definitive treatment can be found in the modern reference \cite{etingof2015tensor} and a fantastic introduction can be found in \cite{kong2022invitationtopologicalorderscategory}. A more general introduction can be found in \cite{mac1998categories}. For a general discussion on Hopf algebras and their representation theory we recommend \cite{kytola2011introduction}, and an exhaustive treatment can be found in \cite{majid2000foundations}. 

In this section we will only touch on finite dimensional categories as examples. To make contact with $\rmC^*$-algebras we direct the reader to \cite{jones2017operator, chen2022q} and references therein. 

We will be extensively relying on \cite{kong2022invitationtopologicalorderscategory, etingof2015tensor} as the backbone for this chapter.

First we review basic category theory, discussing some pedestrian constructions and structures. Next we talk about a particular type of categories called tensor categories, along the way discussing structures like rigidity and unitarity and braiding. These structures represent the universal physics of anyons (cf.~Chapter \ref{chap:physics motivation}). We conclude the section by talking about the Drinfel'd center of a tensor category, and establish connections to the Drinfel'd center of the category of $G$-graded vector spaces and $G$-crossed braided $\rmC^*$-tensor categories via the category of Yetter-Drinfel'd modules over $G$.

\section{Basics of category theory}
A category $\cC$ is defined as 
\begin{itemize}
    \item A set $\cC_0$ of \emph{objects} (we will abuse notation and say $X \in \cC$ if $X \in \cC_0$)
    \item A set $\cC_1(X\rightarrow Y)$ of \emph{morphisms} for all $X,Y \in \cC$ where for all $f \in \cC_1(X \rightarrow Y)$, $X$ is the \emph{source} and $Y$ is the \emph{target} (We will again abuse notation and say $f \in \cC(X \rightarrow Y)$ if $f \in \cC_1(X \rightarrow Y)$)
    \item A \emph{composition} of morphisms $f \circ g \in \cC(X \rightarrow Z)$ for $X,Y,Z \in \cC$ and $g \in \cC(X \rightarrow Y), f \in \cC(Y \rightarrow Z)$
    \item An \emph{identity morphism} $\Id_X \in \cC(X\rightarrow X)$ for all $X \in \cC$
    \item \emph{Associativity}: for objects $A,B,C,D$ and any $f \in \cC(B \rightarrow A), g \in \cC(C \rightarrow B), h \in \cC(D \rightarrow C)$ we have $(f \circ g) \circ h = f \circ  (g \circ h)$
    \item \emph{Unitality}: $\Id_Y \circ f = f = f \circ \Id_X$ for some objects $X,Y \in \cC$ and $f \in \cC(X \rightarrow Y)$
\end{itemize}

\subsection{Commutative diagrams}
\label{sec:commutative diagrams}
Given morphisms in a category, we can graphically represent equations involving morphisms by using commutative diagrams. For example, the following diagram commutes if and only if the equation $k \circ f = j \circ g$ is true for objects $W,X,Y,Z \in \cC$ and morphisms $f \in \cC(W \rightarrow X)$, $g \in \cC(W \rightarrow Y)$, $k \in \cC(X \rightarrow Z)$, $j \in \cC(Y \rightarrow Z)$:

$$\begin{tikzcd}
	W & X \\
	Y & Z
	\arrow["f"', from=1-1, to=1-2]
	\arrow["g", from=1-1, to=2-1]
	\arrow["k"', from=1-2, to=2-2]
	\arrow["j", from=2-1, to=2-2]
\end{tikzcd}$$

\subsection{Construction of new categories}
Given a category $\cC$, we can construct a category $\cC^{\opp}$ whose objects are the same as that in $\cC$ but whose morphism space is given by reversing the morphisms in $\cC$: $$\cC^\opp(X\rightarrow Y) := \cC(Y \rightarrow X)$$

Given categories $\cC, \cD$ we can define a Cartesian product $\cC \times \cD$, whose objects are given by $[\cC\times \cD]_0 := \cC_0 \times \cD_0$ and the morphisms are given by $$[\cC \times \cD]_1(X\times Y \rightarrow X' \times Y') := \cC(X\rightarrow X') \times \cD(Y \rightarrow Y')$$

\begin{defn}
\label{def:subcategory and full subcategory}
    Let $\cC, \cD$ be two categories. Then $\cD$ is called a \emph{subcategory} of $\cC$ if we have $\cD_0 \subseteq \cC_0$ and $\cD(X \rightarrow Y) \subseteq \cC(X \rightarrow Y)$ for every $X,Y \in \cD \subseteq \cC$. $\cD$ is called \emph{full} if $\cD(X\rightarrow Y) = \cC(X\rightarrow Y)$ for all $X,Y \in \cD$.
\end{defn}

\begin{defn}
    \label{def:isomorphism and inverse}
    Let $\cC$ be a category and $X,Y \in \cC$. A morphism $f \in \cC(X\rightarrow Y)$ is called an \emph{isomorphism} if there exists a morphism $g \in \cC(Y \rightarrow X)$ such that $g \circ f = \Id_X$ and $f \circ g = \Id_Y$. Such a morphism $g$, if it exists, is called the \emph{inverse} of $f$ and denoted by $f^{-1}$. Two objects $X,Y \in \cC$ are \emph{isomorphic} if there exists an isomorphism between them. We denote it by $X \sim Y$. 
\end{defn}

\begin{defn}
\label{def:linear category and direct sums}
    A $\bbC$-\emph{linear category} is a category in which each morphism space is equipped with a structure of a vector space over $\bbC$, such that the composition of morphisms is $\bbC$-bilinear.

    Since we only consider the base field $\bbC$ in this thesis, we will henceforth call $\cC$ a \emph{linear category} if $\cC$ is a $\bbC$-linear category.

    Let $\cC$ be a $\bbC$-linear category. A \emph{direct sum}\footnote{We will tacitly assume finite biproducts where we speak of direct sums. In that case the injections/projections satisfy the biproduct identities.} of objects $X_1,\cdots, X_n \in \cC$ is an object $X \in \cC$ equipped with morphisms $\iota_i \in \cC(X_i \rightarrow X)$ and $\pi_i \in \cC(X \rightarrow X_i)$ for $1 \leq i \leq n$, such that the following equations hold:
    \begin{align*}
        &\pi_i \circ \iota_j = \delta_{ij} \Id_{X_j} \qquad \forall i,j \in \{1,\cdots, n\}\\
        &\sum_{j=1}^n \iota_j \circ \pi_j = \Id_X
    \end{align*}
\end{defn}

This direct sum, if it exists, is unique up to a unique isomorphism.

\begin{defn}
\label{def:semisimple and finite semisimple category and set of irreducible objects}
    Let $\cC$ be a linear category. We say that an object $X \in \cC$ is \emph{simple} if $\End_{\cC}(X) := \cC(X \rightarrow X) \simeq \bbC \Id_X$ (as algebras). Two objects $X, Y \in \cC$ are \emph{disjoint} if $$\cC(X\rightarrow Y) = 0 = \cC(Y\rightarrow X)$$
    We say $\cC$ is \emph{semisimple} if (1) the direct sum of finitely many objects in $\cC$ exists and (2) if there exists a set of mutually disjoint simple objects $\{X_i\}_{i \in I}$ such that every object is isomorphic to a finite direct sum of objects in $\{X_i\}_{i \in I}$. We will denote this set by $\Irr(\cC)$. If the index set of $\Irr(\cC)$ is finite and if the morphism spaces are finite dimensional, then we say that $\cC$ is \emph{finite semisimple}. 
\end{defn}

A structure preserving map between categories is called a functor. 

\begin{defn}
\label{def:functor and linear functor}
    Let $\cC, \cD$ be categories. A functor $F: \cC \rightarrow \cD$ consists of the following data:
    \begin{itemize}
        \item A map $F: \cC_0 \rightarrow \cD_0$
        \item A map $F_{X,Y}: \cC(X\rightarrow Y) \rightarrow \cD(F(X) \rightarrow F(Y))$ for each pair of objects $X,Y \in \cC$. We will abuse notation and for every $f \in \cC(X\rightarrow Y)$ we will denote $F_{X,Y}(f)$ by $F(f)$.
        \item Given $f \in \cC(X\rightarrow Y)$ and $g \in \cC(Y\rightarrow Z)$ we have $F(g) \circ F(f) = F(g \circ f)$
        \item For all objects $X \in \cC$, we have $F(\Id_X) = \Id_{F(X)}$
    \end{itemize}

    A \emph{$\bbC$-linear functor} is a functor such that $F_{X,Y}$ is a $\bbC$-linear map for every $X,Y \in \cC$. We again shorten notation by saying $F$ is a \emph{linear functor} if $F$ is a $\bbC$-linear functor.
\end{defn}

\begin{lem}
    Let $F: \cC \rightarrow \cD$ be a functor for linear categories $\cC, \cD$. Then $F$  is linear if and only if it preserves direct sums, i.e., given $(X, \{\pi_i\}, \{\iota_i\})$ is a direct sum of $X_1,\cdots , X_n \in \cC$, then $(F(X), \{F(\pi_i)\}, \{F(\iota_i)\})$ is a direct sum of $F(X_1),\cdots, F(X_n) \in \cD$.
\end{lem}
\begin{proof}
Standard. Can be found for instance in \cite[Ch.~VIII,Prop.~4]{mac1998categories}.
\end{proof}

For a category $\cC$, we can define an \emph{identity functor} $\Id_\cC: \cC \rightarrow \cC$, which leaves both the objects as well as the morphisms in the category unchanged.

For two functors $F: \cC\rightarrow \cD$ and $G: \cD \rightarrow \cE$, we can define the composition functor $G \circ F$ which for every $f \in \cC(X\rightarrow Y)$ and objects $X,Y \in \cC$ gives $G\circ F: f \mapsto G(F(f))$.

Every functor preserves isomorphisms. Indeed, if $f \in \cC(X\rightarrow Y)$ is an isomorphism, then $F(f) \circ F(f^{-1}) = F(f \circ f^{-1}) = F(\Id_X) = \Id_{F(X)}$ and also for the left inverse.

A structure preserving map is called a natural transformation:

\begin{defn}
    Let $\cC, \cD$ be two categories and let $F, G$ be functors from $\cC$ to $\cD$. A \emph{natural transformation} $\alpha: F \Rightarrow G$ is a family of morphisms $\{\alpha_X : F(X) \rightarrow G(X)\}_{X \in \cC}$ in $\cD$ such that the following diagram commutes for any morphism $f \in \cC(X \rightarrow Y)$:
$$\begin{tikzcd}
	{F(X)} & {G(X)} \\
	{F(Y)} & {G(Y)}
	\arrow["{\alpha_X}"', from=1-1, to=1-2]
	\arrow["{F(f)}"', from=1-1, to=2-1]
	\arrow["{G(f)}", from=1-2, to=2-2]
	\arrow["{\alpha_Y}", from=2-1, to=2-2]
\end{tikzcd}$$

A natural transformation $\alpha$ is called a \emph{natural isomorphism} if every morphism $\alpha_X$ is an isomorphism. If there exists a natural isomorphism $\alpha$ between two functors $F,G$ then we say $F$ is \emph{naturally isomorphic} to $G$, and denote it by $F \sim G$.
\end{defn}

\begin{defn}
\label{def:equivalence of categories}
    Let $F: \cC\rightarrow \cD$ be a functor. We say $F$ is an \emph{equivalence} if there exists a functor $G : \cD \rightarrow \cC$ such that $G \circ F \sim \Id_\cC$ and $F \circ G \sim \Id_\cD$. We say that the categories $\cC, \cD$ are \emph{equivalent}, denoted by $\cC \sim \cD$ if there exists an equivalence between them.
\end{defn}

The following definitions are useful to state an important theorem to compute an equivalence between categories.

\begin{defn}
\label{def:fully faithful and essentially surjective}
    Let $F:\cC \rightarrow \cD$  be a functor.
    
    $F$ is called \emph{faithful} if every map $F_{X,Y}$ for objects $X,Y \in \cC$ is injective, and \emph{full} if it is surjective. If $F$ is both full and faithful, we call $F$ as \emph{fully faithful}.

    $F$ is called \emph{essentially surjective} if for every object in $A \in \cD$ there exists an object $X \in \cC$ such that $F(X) \sim A$
\end{defn}

\begin{thm}
    A functor $F : \cC \rightarrow \cD$ is an equivalence if and only if it is fully faithful and essentially surjective. 
\end{thm}
\begin{proof}
Standard. It is shown for instance in \cite[Ch.~IV,~Thm.1.(iii)]{mac1998categories}. 
\end{proof}

\begin{example}{\texorpdfstring{$\Vect$}{Vec}}{vec}
This is the category of finite-dimensional vector spaces. The objects are finite-dimensional vector spaces over $\bbC$. Given $V,W \in \Vect$, the morphisms $\Vect(V\rightarrow W)$ are defined as the set of linear maps from $V$ to $W$. The identity map is the usual identity map on vector spaces, and the composition of morphisms is the usual composition of linear maps between vector spaces. $\Vect$ is a linear category by construction. There is only isomorphism class of simples in $\Vect$, represented by $\bbC$. every object $V \in \Vect$ satisfies $V \sim  \bbC^{\oplus {n}}$ for some $n \in \bbN$, i.e., $V$ is isomorphic to $n$ copies of $\bbC$. Therefore it is a finite semisimple category. 
\end{example}

\begin{example}{\texorpdfstring{$\Grep$}{Rep G}}{rep G}
This is the category of representations of $G$, where $G$ is a group. The objects in this category are pairs $(\rho, V)$ where $\rho : G \rightarrow \GL(V)$ is a representation of $G$ onto the finite dimensional vector space $V$. Given two objects $(\rho, V)$ and $(\sigma, W)$, the space $\Grep((\rho, V) \rightarrow (\sigma, W))$ is defined as the space of intertwining morphisms (or simply intertwiners) $f \in \Vect(V \rightarrow W)$ such that for all $g \in G$ we have $\sigma(g) \circ f = f \circ \rho(g)$. The composition of morphisms is the usual composition of linear maps, and so is the identity map. $\Grep$ is linear by construction, and if $G$ is a finite group, then $\Grep$ is finite semisimple by Maschke's theorem \cite{zbMATH03552764}.

Let us define a functor $F: \Grep \rightarrow \Vect$ which maps $(\rho, V) \mapsto V$ for objects $(\rho, V) \in \Grep$. It acts on the morphisms $f \in \Grep((\rho, V) \rightarrow (\sigma, W))$ as the inclusion map, i.e., $F: f \in \Grep((\rho, V) \rightarrow (\sigma, W)) \mapsto f \in \Vect({V\rightarrow W})$. We call this the \emph{forgetful functor}, since it forgets the action of the representation on the vector spaces. The forgetful functor is a linear, faithful functor. It is full if and only if $G$ is trivial.
\end{example}

\begin{example}{\texorpdfstring{$\Gvec$}{Vec G}}{vec G}
    This is the category of $G$ graded vector spaces, where $G$ is a finite group. The objects in this category are a set $\{V_g \in \Vect \}_{g \in G}$. The (finite)\footnote{There are only finitely many non-zero graded pieces} direct sum $V = \bigoplus_{g \in G}V_g$ is called the \emph{total space}. We also use $V$ to denote this set of graded vector spaces. A morphism $f \in \Gvec(V \rightarrow W)$ for objects $V,W \in \Gvec$ is defined as a degree-preserving linear map $f = \bigoplus_{g \in G} f_g$ for the set of morphisms $\{f_g \in \Vect(V_g\rightarrow W_g)\}_{g\in G}$. The composition of morphisms is given by the usual composition of morphisms for each graded component. The identity morphism is the usual identity map for each graded component. This category is linear by construction. Let $G$ be a finite group. Then $\Gvec$ is a finite semisimple category as every object $V = \{V_g\} \in \Gvec$ satisfies $V \sim \bigoplus_{g \in G}\bbC(g)^{\oplus n(g)}$, where $n(g) \in \bbN$ for every $g \in G$ and $\bbC(g) = \{\bbC(g)_h\}_{h \in G}$ is the $G$-graded vector space with components $\bbC(g)_h := \delta_{g,h} \bbC$. 

We can construct a forgetful functor $F: \Gvec \rightarrow \Vect$ which maps $\{V_g\}_{g \in G} \mapsto V = \bigoplus_{g \in G} V_g$. $F$ is a linear, faithful functor. It is full if and only if $G$ is trivial.
\end{example}

\section{Tensor Categories}
\begin{defn}
\label{def:tensor category or monoidal category}
    A tensor category $(\cC, \otimes, \ds[1], \alpha, \lambda^L, \lambda^R)$ (also called a monoidal category) consists of the following data:
    \begin{itemize}
        \item a category $\cC$
        \item a functor $\otimes: \cC \times \cC \rightarrow \cC$, where $\otimes(X,Y)$ is denoted by $X \otimes Y$ and for morphisms $f \in \cC(X \rightarrow X'),g \in \cC(Y \rightarrow Y')$, $\otimes(f, g): X\otimes Y \rightarrow X' \otimes Y'$ is denoted by $f \otimes g$.
        \item A distinguished object $\ds[1] \in \cC$ called the \emph{tensor unit}
        \item A natural isomorphism $\alpha$ with $\alpha_{X,Y,Z}: (X\otimes Y) \otimes Z \rightarrow X \otimes (Y \otimes Z)$ called the \emph{associator}
        \item Natural isomorphisms $\lambda_X^L: \ds[1] \otimes X \rightarrow X$ called the \emph{left unitor} and $\lambda^R_X: X\otimes \ds[1] \rightarrow X$ called the \emph{right unitor}
        \item These data satisfy the \emph{pentagon equation} given by
\[
\tag{pentagon}
\begin{tikzcd}
	&& {((A \otimes B) \otimes C) \otimes D} \\
	{(A \otimes (B \otimes C)) \otimes D} &&&& {(A \otimes B) \otimes (C \otimes D)} \\
	{A \otimes ((B \otimes C) \otimes D)} &&&& {A \otimes (B \otimes (C \otimes D))}
	\arrow["{\alpha_{A,B,C}\otimes \Id_D}", from=1-3, to=2-1]
	\arrow["{\alpha_{A\otimes B, C, D }}"', from=1-3, to=2-5]
	\arrow["{\alpha_{A, B\otimes C, D}}", from=2-1, to=3-1]
	\arrow["{\alpha_{A, B, C\otimes D}}"', from=2-5, to=3-5]
	\arrow["{\Id_A \otimes \alpha_{B,C,D}}", from=3-1, to=3-5]
\end{tikzcd}\]
    \item These data satisfy the \emph{triangle equation} given by 
\[
\tag{triangle} 
\begin{tikzcd}
	{(X \otimes \ds[1]) \otimes Y} && {X \otimes (\ds[1] \otimes Y)} \\
	& {X \otimes Y}
	\arrow["{\alpha_{X, \ds[1], Y}}"', from=1-1, to=1-3]
	\arrow["{\lambda^R_X \otimes \Id_Y}"', from=1-1, to=2-2]
	\arrow["{\Id_X \otimes \lambda^L_Y}", from=1-3, to=2-2]
\end{tikzcd}\]
\end{itemize}
A $\bbC$-linear tensor category $\cC$ is a linear category and a tensor category such that the functor $\otimes$ is bilinear, i.e., linear in both components.

A tensor category is called \emph{strict} if the associators and unitors are all trivial, i.e., associativity and unitality is automatically satisfied for the tensor operation.

We abuse notation and denote $\cC := (\cC, \otimes, \ds[1], \alpha, \lambda^L, \lambda^R)$ as a tensor category, keeping the tensor structure implicit unless stated.
\end{defn}

\begin{defn}
\label{def:tensor functor}
    Let $\cC, \cD$ be tensor categories. A tensor functor $F : \cC \rightarrow \cD$ consists of the following data:
    \begin{itemize}
        \item The functor $F$
        \item A natural isomorphism $F^2_{X,Y}: F(X) \otimes F(Y) \rightarrow F(X \otimes Y)$ called the \emph{tensorator}
        \item An isomorphism $F^0: \ds[1]_\cD \rightarrow F(\ds[1]_\cC)$ called the \emph{unit isomorphism}
        \item The following diagram commutes for all $X,Y,Z$:

\[\tag{tensor hexagon}
\begin{tikzcd}
	{(F(X) \otimes F(Y)) \otimes F(Z)} && {F(X) \otimes (F(Y) \otimes F(Z))} \\
	{F(X\otimes Y) \otimes F(Z)} && {F(X)\otimes F(Y\otimes Z)} \\
	{F((X\otimes Y) \otimes Z)} && {F(X\otimes(Y\otimes Z))}
	\arrow["{\alpha^\cD_{F(X), F(Y), F(Z)}}", from=1-1, to=1-3]
	\arrow[ld,"{F_{X,Y}^2 \otimes \Id_{F(Z)}}"', from=1-1, to=2-1]
	\arrow["{\Id_{F(X)} \otimes F^2_{Y,Z}}", from=1-3, to=2-3]
	\arrow["{F^2_{X\otimes Y, Z}}"', from=2-1, to=3-1]
	\arrow["{F^2_{X,Y\otimes Z}}", from=2-3, to=3-3]
	\arrow["{F(\alpha^\cC_{X,Y,Z})}", from=3-1, to=3-3]
\end{tikzcd}\]

    \item The following diagram (and its counterpart for $\lambda_X^R$) commutes for all $X \in \cC$:
\[\tag{tensor unit left}
\begin{tikzcd}
	{F(\ds[1]_\cC) \otimes F(X)} && {F(\ds[1]_\cC \otimes X)} \\
	{\ds[1]_\cD \otimes F(X)} && {F(X)}
	\arrow["{F^2_{\ds[1]_\cC, X}}", from=1-1, to=1-3]
	\arrow["{F((\lambda_X^L)^\cC)}", from=1-3, to=2-3]
	\arrow["{F^0 \otimes \Id_{F(X)}}", from=2-1, to=1-1]
	\arrow["{(\lambda^L_{F(X)})^\cD}"', from=2-1, to=2-3]
\end{tikzcd}\]
    \end{itemize}

    A tensor functor that is also an equivalence is called a \emph{tensor equivalence}.
\end{defn}

\begin{defn}
\label{def:tensor natural transformation}
    Let $\cC, \cD$ be tensor categories and $F,G: \cC \rightarrow \cD$ be tensor functors. A \emph{tensor natural transformation} $\alpha: F \implies G$ is a natural transformation satisfying the following commutative diagrams: 
\[\tag{tensor naturality}\begin{tikzcd}
	{F(X) \otimes F(Y)} & {F(X\otimes Y)} \\
	{G(X) \otimes G(Y)} & {G(X\otimes Y)}
	\arrow["{F_{X,Y}^2}", from=1-1, to=1-2]
	\arrow["{\alpha_X \otimes \alpha_Y}"', from=1-1, to=2-1]
	\arrow["{\alpha_{X \otimes Y}}", from=1-2, to=2-2]
	\arrow["{G^2_{X,Y}}"', from=2-1, to=2-2]
\end{tikzcd}\]

\[\tag{tensor natural unit}\begin{tikzcd}
	{\ds[1]_{\cD}} & {F(\ds[1]_{\cC})} \\
	& {G(\ds[1]_{\cC})}
	\arrow["{F^0}", from=1-1, to=1-2]
	\arrow["{G^0}"', from=1-1, to=2-2]
	\arrow["{\alpha_{\ds[1]_\cC}}", from=1-2, to=2-2]
\end{tikzcd}\]

A tensor natural transformation that is also a natural isomorphism is called a \emph{tensor natural isomorphism}.
\end{defn}

\begin{thm}[Mac Lane's coherence theorem]
    Every tensor category is tensor equivalent to a strict tensor category.
\end{thm}
\begin{proof}
    Standard. Can be found for instance in \cite[Thm.~2.9.2]{etingof2015tensor}.
\end{proof}

\subsection{Unitarity}

\begin{defn}
\label{def:*-structure and unitary category}
    Let $\cC$ be a linear category. A $*-$structure (sometimes called the \emph{dagger structure}, especially in physics literature) on $\cC$ is a contravariant, involutive, anti-linear (on morphism spaces) functor $*: \cC^\opp \rightarrow \cC$ which acts on the objects of $\cC$ as the identity. We write $*(f)$ as $f^*$ for some morphism $f \in \cC(X\rightarrow Y)$. 

    Moreover, the $*$-structure is called a \emph{unitary structure} if for any morphism $f \in \cC(X\rightarrow Y)$ we have $f^* \circ f = 0$ if and only if $f =0$. A \emph{unitary category} $\cC$ is a linear category equipped with a unitary structure. 
\end{defn}

\begin{defn}
\label{def:unitary tensor category}
    A \emph{unitary tensor category} is a linear tensor category that is equipped with a unitary structure such that the functor $*$ is a tensor functor, i.e., $f^* \otimes g^* = (f\otimes g)^*$.
\end{defn}

\begin{defn}
\label{def:unitary morphism}
    A morphism $f$ in a unitary category is called \emph{unitary} if it is an isomorphism and satisfies $f^* = f^{-1}$.
\end{defn}

\subsection{\texorpdfstring{$\rmC^*$}{C*}-property}
\begin{defn}
\label{def:C^* category and C^* tensor category}
    A \emph{$\rmC^*$-category} $\cC$ is a category $\cC$ equipped with a $*$-functor, and the spaces $\cC(X\rightarrow Y)$ are Banach spaces and the norms satisfy
    \begin{align*}
        ||g \circ f|| &\leq ||g|| \, ||f|| \tag{contractivity of compositions}\\
        ||f^* \circ f|| &= ||f||^2 \tag{$\rmC^*$-identity}
    \end{align*}
    For any $f \in \cC(X \rightarrow Y)$ and $g \in \cC(Y \rightarrow Z)$. Thus $\cC(X \rightarrow X)$ are $\rmC^*$-algebras for all $X \in \cC$. 

    A $\rmC^*$-tensor category is a $\rmC^*$-category that is also a tensor category, and the $*$-functor is a tensor functor.
\end{defn}

For categories having finite dimensional morphism spaces, the notion of a unitary category is equivalent to the notion of a $\rmC^*$-category \cite[Prop.~2.1]{mueger1999galoistheorybraidedtensor}. A unitary category with finite dimensional morphism spaces is automatically semisimple \cite[Lem.~3.2]{yamagami2004frobenius}.

\begin{defn}
    Let $\cC,\cD$ be $\rmC^*$-tensor categories. A functor $F:\cC \rightarrow \cD$ is called a \emph{$\rmC^*$-functor} if it is linear, $*$-preserving, i.e., for all $f \in \cC(X\rightarrow Y)$ we have $F(f)^* = F(f^*)$.

    A \emph{$\rmC^*$-tensor} functor is a $*$-functor that is also a tensor functor, and the tensorators and identity isomorphism are unitary.

    A unitary natural isomorphism (sometimes called \emph{$\rmC^*$-isomorphism}) is a natural transformation such that each component is unitary.
\end{defn}

\subsection{Rigidity}

\begin{defn}
    \label{def:rigid tensor category and dualizability}
    Let $\cC$ be a tensor category and $X \in \cC$. A \emph{left dual} of $X$ is an object $X^L \in \cC$ equipped with two morphisms $b_X^L \in \cC(\ds[1] \rightarrow X \otimes X^L)$ and $d_X^L \in \cC( X^L \otimes X \rightarrow \ds[1])$ satisfying two \emph{zig-zag} equations:
    \begin{align*}
    \tag{\text{zig-zag L}}
        (\Id_X \otimes d_X^L) \circ (b_X^L \otimes \Id_X) = \Id_X \qquad \qquad (d_X^L \otimes \Id_{X^L}) \circ (\Id_{X^L} \otimes b_X^L) = \Id_{X^L}
    \end{align*}
    Here we ignore the associators and unitors for simplicity. Similarly a \emph{right dual} of $X$ is an object $X^R \in \cC$ equipped with two morphisms $b_X^R \in \cC(\ds[1] \rightarrow X^R \otimes X)$ and $d_X^R \in \cC( X \otimes X^R \rightarrow \ds[1])$ satisfying two \emph{zig-zag} equations:
    \begin{align*}
    \tag{\text{zig-zag R}}
        (d_X^R \otimes \Id_X) \circ ( \Id_X \otimes b_X^R ) = \Id_X \qquad \qquad ( \Id_{X^R} \otimes d_X^R) \circ (b_X^R \otimes \Id_{X^R}) = \Id_{X^R}
    \end{align*}
    An object in $\cC$ is called \emph{dualizable} if it admits both left and right duals. If every object in $\cC$ is dualizable, then we say that $\cC$ is a \emph{rigid tensor category}.
\end{defn}

\begin{defn}
\label{def:unitary fusion category}
    A \emph{unitary fusion category} (UFC) is a linear, rigid, unitary tensor category that is also finite semi-simple and the tensor unit $\ds[1]$ is a simple object.
\end{defn}

\begin{lem}
\label{lem:left and right duals are equal in a rigid, unitary tensor cat}
    Let $\cC$ be a rigid, unitary fusion category. Then for every object $X \in \cC$, $X^L$ (or equivalently $X^R$) is both the left dual as well as the right dual of $X$ (upto a canonical isomorphism).
\end{lem}
\begin{proof}
This is a trivial consequence of there being a canonical spherical structure in a UFC \cite[Prop.~8.23]{etingof2005fusion}. Roughly, this means that for every object the left and right duals coincide, and that there is only one canonical way to make this dual object.
\end{proof}

Following the results of Lemma \ref{lem:left and right duals are equal in a rigid, unitary tensor cat}, in a UFC $\cC$ and an object $X \in \cC$, we denote both the left and the right dual of $X$ as $(\bar X, b_X, d_X)$, where $b_X := b_X^L = (d_X^R)^*$ and $d_X := d_X^L = (b_X^R)^*$. 

\begin{defn}
    \label{quantum dimension}
    In a UFC $\cC$, we define the \emph{trace} of a morphism as follows. Choose a morphism $f \in \cC(X \rightarrow X)$ for some $X \in \cC$. Then $$\Tr{f} := b_X^* \circ (f \otimes \Id_{\bar X}) \circ b_X = d_X \circ (\Id_{\bar X} \otimes f) \circ d_X^*$$

    In a generic rigid tensor category we're only able to define the left-trace (first equality) and separately the right-trace (second equality). But since our category is unitary, due to the canonical spherical structure in any UFC, the left-trace and right-trace agree.

    Moreover, we define the \emph{quantum dimension} of an object $X \in \cC$ as $\dim(X):= \Tr{\Id_X}$. We define the total quantum dimension of $\cC$ as $$\dim{(\cC)} := \abs{\sqrt{\sum_{X \in \Irr(\cC)} [\dim(X)]^2}}$$ (see Def \ref{def:semisimple and finite semisimple category and set of irreducible objects} for the definition of $\Irr(\cC)$).
\end{defn}

Let $\cC$ be a UFC. Let $X \in \cC$ be a non-zero object. Then it follows that $\dim(X) = b_X^* \circ b_X = d_X \circ d_X^* \geq 0$ due to the unitary structure and the fact that $b_X, d_X$ are non-zero, and thus $\dim(\cC) {\geq 0}$.

\subsection{Braiding}

\begin{defn}
\label{def:braided tensor category}
    A \emph{braided tensor category} consists of the following data:
    \begin{itemize}
        \item a tensor category $\cC = (\cC, \otimes , \ds[1], \alpha, \lambda^L, \lambda^R)$
        \item A natural isomorphism $c_{X,Y}: X\otimes Y \rightarrow Y\otimes X$ for all $X,Y \in \cC$ satisfying the \emph{naturality square}: 
        \[
        \tag{braiding naturality}
        \begin{tikzcd} X\otimes Y \arrow[r,"c_{X,Y}"] \arrow[d,"f\otimes g"'] & Y\otimes X \arrow[d,"g\otimes f"]\\ X'\otimes Y' \arrow[r,"c_{X',Y'}"'] & Y'\otimes X' \end{tikzcd}
        \]
        \item These data satisfy the following commutative diagrams (called \emph{hexagon equations}):
\[\begin{tikzcd}
\tag{hexagon 1}
	& {X \otimes (Y\otimes Z)} && {(Y\otimes Z) \otimes X} \\
	{(X\otimes Y) \otimes Z} &&&& {Y\otimes (Z \otimes X)} \\
	& {(Y \otimes X) \otimes Z} && {Y \otimes (X \otimes Z)}
	\arrow["{c_{X,Y\otimes Z}}", from=1-2, to=1-4]
	\arrow["{\alpha_{Y,Z,X}}"{pos=0.5}, from=1-4, to=2-5]
	\arrow["{\alpha_{X,Y,Z}}"{pos=0.5}, from=2-1, to=1-2]
	\arrow["{c_{X,Y}\otimes \Id_Z}"'{pos=0.5}, from=2-1, to=3-2]
	\arrow["{\alpha_{Y,X,Z}}"', from=3-2, to=3-4]
	\arrow["{\Id_{Y}\otimes c_{X,Z}}"'{pos=0.5}, from=3-4, to=2-5]
\end{tikzcd}\]
        \[\begin{tikzcd}
        \tag{hexagon 2}
        	& {(X\otimes Y) \otimes Z} && {Z\otimes(X\otimes Y)} \\
        	{X \otimes (Y\otimes Z)} &&&& { (Z \otimes X)\otimes Y} \\
        	& {X\otimes (Z \otimes Y)} && { (X \otimes Z)\otimes Y}
        	\arrow["{c_{X\otimes Y, Z}}", from=1-2, to=1-4]
        	\arrow["{\alpha_{Z,X,Y}^{-1}}"{pos=0.5}, from=1-4, to=2-5]
        	\arrow["{\alpha^{-1}_{X,Y,Z}}"{pos=0.5}, from=2-1, to=1-2]
        	\arrow["{\Id_X\otimes c_{Y,Z}}"'{pos=0.5}, from=2-1, to=3-2]
        	\arrow["{\alpha_{X,Z,Y}^{-1}}"', from=3-2, to=3-4]
        	\arrow["{ c_{X, Z} \otimes \Id_Y}"'{pos=0.5}, from=3-4, to=2-5]
        \end{tikzcd}\]
    \end{itemize}
\end{defn}

\begin{defn}
    Let $\cC, \cD$ be braided tensor categories. A \emph{braided tensor functor} (or simply a \emph{braided functor}) $F: \cC \rightarrow \cD$ is a tensor functor $F: \cC \rightarrow \cD$ such that the following diagram commutes for any $X,Y \in \cC$:
\[\tag{braided functor}\begin{tikzcd}
	{F(X) \otimes F(Y)} & {F(X\otimes Y)} \\
	{F(Y) \otimes F(X)} & {F(Y\otimes X)}
	\arrow["{F^2_{X,Y}}", from=1-1, to=1-2]
	\arrow["{c^\cD_{F(X), F(Y)}}"', from=1-1, to=2-1]
	\arrow["{F(c^\cC_{X,Y})}", from=1-2, to=2-2]
	\arrow["{F^2_{Y,X}}", from=2-1, to=2-2]
\end{tikzcd}\]

A braided tensor functor that is also an equivalence is called a braided equivalence.
\end{defn}

\begin{defn}
    A braided tensor category $\cC$ equipped with the $*$-functor is called a \emph{unitary braided tensor category} if the $*$-functor is a braided functor and the braiding isomorphisms are unitary. 

    A unitary braided tensor category $\cC$ that is also a unitary fusion category is called a \emph{unitary braided fusion category}.
\end{defn}

\subsection{Modularity}
\begin{defn}
\label{def:unitary modular tensor category}
    Let $\cC$ be a unitary braided fusion category. Then $\cC$ is called a \emph{unitary modular tensor category} (UMTC) if for any simple object $X \in \cC$ for all objects $Y \in \cC$ the identity $c_{Y,X} \circ c_{X,Y} = \Id_{X \otimes Y}$  implies $X \simeq \ds[1]$ \footnote{The set of all such objects is known as the Müger center, and this condition is equivalent to the Müger center being trivial. See \cite{muger2003structure}.}.
\end{defn}

This principle arises from the remote detectability principle, which is related to anomalies in topological phases.

\subsection{Routine categories are UFCs}
\label{sec:routine categories are UFC}
In this section we elaborate on the structures possessed by the routine categories discussed in Examples \ref{eg:vec}, \ref{eg:rep G}, \ref{eg:vec G}.

First we note that since $\Vect$ has objects as finite dimensional Hilbert spaces, one can always endow it with an extra structure of an inner product for free. Thus there exists an equivalence $\Vect \simeq \mathsf{Hilb}$ where $\mathsf{Hilb}$ is the category of finite dimensional Hilbert spaces. Henceforth we assume that $\Vect$ is the category of finite dimensional Hilbert spaces, and $\Gvec$ is the category of finite dimensional $G$-graded Hilbert spaces. As a consequence, $\Vect$ is a $\rmC^*$-category with the operator norm and $*$-functor being the usual vector space adjoint.

In a similar vein, every $G$-representation on a finite dimensional Hilbert space is equivalent to a unitary $G$-representation. Thus WLOG we assume $\Grep$ to be the category of finite dimensional unitary representations.

\begin{example}{\texorpdfstring{$\Vect$}{Vec}}{vec UFC}
We can define the tensor functor $\otimes$ in the usual way as the tensor product of finite dimensional Hilbert spaces making $\Vect$ into a strict tensor category. The associator $\alpha$ and unitors are given by the canonical unitary isomorphisms. By Mac Lane's coherence theorem, we may strictify and treat $\Vect$ as a strict tensor category. By above discussion, $\Vect$ is $\rmC^*$-tensor category (after the straightforward verification of $f^* \otimes g^* = (f\otimes g)^*$) with finite dimensional morphism spaces, making it a unitary tensor category.

We define the evaluation map $d_V : \bar V \otimes V \rightarrow \bbC$ (here $\bar V$ is the conjugate space to $V$) with $d_V : \bar v \otimes w \mapsto \inner{v}{w}$, while the coevaluation map $b_X: \bbC \rightarrow V \otimes \bar V$ is given by $1 \mapsto \sum_i e_i \otimes \bar e_i$, where $e_i$ are the ONB vectors of $V$. It is straightforward to check the zig-zag equations, giving us that $\Vect$ is indeed a UFC.
\end{example}

\begin{example}{\texorpdfstring{$\Grep$}{Rep G}}{rep G UFC}
$\Grep$ is a strict tensor category when endowed with the usual algebraic (associative) tensor product of $G$-representations. It is a $\rmC^*$-category with $*$ the usual adjoint operation compatible with $\otimes$ as in $\Vect$, making it a $\rmC^*$-tensor category. Since the morphism spaces are finite dimensional, $\Grep$ is a unitary tensor category. 

For any representation $(\pi,V)$, we define the dual representation $(\bar \pi, \bar V)$\footnote{Here $\bar V$ is the conjugate Hilbert space corresponding to $V$.} as $\pi(g) \bar v := \overline{\pi(g)v} $ for all $\bar v \in \bar V$. The evaluation, coevaluation maps are inherited from $\Vect$. The inherited evaluation map $d_V$ is an intertwiner between $(\bar \pi,\bar V)\otimes ( \pi,  V)$ and $(\pi_1, \bbC)$, and similarly the inherited coevaluation map is an intertwiner between $(\pi_1, \bbC)$ and $(\pi,V)\otimes (\bar \pi, \bar V)$. Thus $\Grep$ is a rigid category. Combining these structures we have that $\Grep$ is a UFC.
\end{example}

\begin{example}{\texorpdfstring{$\Gvec$}{Vec G}}{vec G UFC}
The morphisms preserve grading of the objects. The tensor product is defined as $$(V \otimes W)_g := \bigoplus_{jk = g} V_j \otimes W_k$$ where the tensor product is inherited from $\Vect$ viewing $V_j, W_k$ as objects in $\Vect$. The usual $*$-functor inherited from $\Vect$ preserves the gradings, making $\Gvec$ into a $\rmC^*$-category. Moreover $*$-functor is compatible with the tensor product of morphisms since it preserves gradings. Thus $\Gvec$ is a (finite dimensional) $\rmC^*$-tensor category and hence a unitary tensor category.

For a Hilbert space $V$ we set the dual Hilbert space $V^* = \oplus \bar V_{g^{-1}}$. The evaluation, coevaluation maps are the graded versions of the ones in $\Vect$, and thus $\Gvec$ is a UFC.
\end{example}

\section{Drinfel'd center}
\begin{defn}
    
Let $\cC$ be a $\rmC^*$-tensor category with associator $\alpha$. The \emph{Drinfel'd center} (or simply \emph{center}) of $\cC$ is a category $\cZ(\cC)$ defined as follows:
\begin{itemize}
    \item The objects in $\cZ(\cC)$ are pairs $(X,\sigma_{X,(\cdot)})$ where $X \in \cC$ and the \emph{half-braid} $\sigma_{X,(\cdot)}$ such that $\sigma_{X,Y}: X\otimes Y \rightarrow Y \otimes X$ is a unitary natural (in $Y$) isomorphism for all $Y \in \cC$.

    \item The morphisms in $\cZ(\cC)$, $f:(X,\sigma_X)\rightarrow (X',\sigma_{X'})$ are morphisms $f:X\rightarrow X'$ in $\cC$ such that for all $Y\in\cC$, $$(\Id_Y\otimes f)\circ \sigma_{X,Y}=\sigma_{X',Y}\circ (f\otimes \Id_Y)$$

    \item $\sigma_{X,Z}$ satisfies the following commutative diagram for all $Z \in \cC$:

        \[\tag{Half-braiding hexagon}\begin{tikzcd}
        	& {(X\otimes Y) \otimes Z} && {Z\otimes(X\otimes Y)} \\
        	{X \otimes (Y\otimes Z)} &&&& { (Z \otimes X)\otimes Y} \\
        	& {X\otimes (Z \otimes Y)} && { (X \otimes Z)\otimes Y}
        	\arrow["{\sigma_{X\otimes Y, Z}}", from=1-2, to=1-4]
        	\arrow["{\alpha_{Z,X,Y}^{-1}}"{pos=0.5}, from=1-4, to=2-5]
        	\arrow["{\alpha^{-1}_{X,Y,Z}}"{pos=0.5}, from=2-1, to=1-2]
        	\arrow["{\Id_X\otimes \sigma_{Y,Z}}"'{pos=0.5}, from=2-1, to=3-2]
        	\arrow["{\alpha_{X,Z,Y}^{-1}}"', from=3-2, to=3-4]
        	\arrow["{ \sigma_{X, Z} \otimes \Id_Y}"'{pos=0.5}, from=3-4, to=2-5]
        \end{tikzcd}\]
        \end{itemize}
\end{defn}

Notice that the commutative diagram is the same as the hexagon 2 commutative diagram, which should give some intuition behind the naming of $\sigma_{X,(\cdot)}$.

$\cZ(\cC)$ is in fact a braided $\rmC^*$-tensor category with the tensor product structure given by $$(X_1, \sigma_{X_1,(\cdot)}) \otimes (X_2, \sigma_{X_2,(\cdot)}) := (X_1 \otimes X_2, \sigma_{X_1 \otimes X_2,(\cdot)})$$ while the $*$-functor, associator $\alpha$, unitors are inherited from $\cC$. Here $\sigma_{X_1 \otimes X_2, X}: (X_1 \otimes X_2) \otimes Y   \rightarrow Y \otimes (X_1 \otimes X_2)$ is defined by $$\sigma_{X_1\otimes X_2,\,Y}:=\alpha_{Y,X_1,X_2}\circ (\sigma_{X_1,Y}\otimes \Id_{X_2})\circ \alpha_{X_1,Y,X_2}^{-1}\circ (\Id_{X_1}\otimes \sigma_{X_2,Y})\circ \alpha_{X_1,X_2,Y}$$
The \emph{braiding} in $\cZ(\cC)$ is given by $$c_{(X,\sigma_X),(Y,\sigma_Y)} := \sigma_{X,Y}: X\otimes Y \rightarrow Y\otimes X$$

\begin{rem}
\label{rem:physics of drinfeld center}
    The construction of Drinfel'd center has a physical motivation. Consider a UFC $\cC$. Since there is no braiding isomorphism in $\cC$, there is a priori no notion of ``swapping'' of tensor factors. However, objects in $\cZ(\cC)$ have a half-braiding, so given an object $(Z,\sigma_{Z,(\cdot)}) \in \cZ(\cC)$ and $X \in \cC$, indeed there is a notion of ``swapping'' the tensor factor $Z\otimes X$, which is achieved using the half-braid isomorphism $\sigma_{Z,X}$. Physically, this can be interpreted as the object $Z$ ``crossing over'' the object $X$ with the help of $\sigma_{Z,X}$.
\end{rem}

\begin{prop}[\cite{muger2003structure, turaev2016quantum}]
\label{prop:center of UFC is UMTC}
    If $\cC$ is a UFC, then $\cZ(\cC)$ is a UMTC.
\end{prop}

\begin{rem}
    We note that by Proposition \ref{prop:center of UFC is UMTC} and the discussion in Section \ref{sec:routine categories are UFC}, the Drinfel'd center $\cZ(\Vect), \cZ(\Gvec), \cZ(\Grep)$ is a UMTC. In fact, $\cZ(\Vect)$ is a trivial UMTC since $\Vect$ (with the braid isomorphism being the swap isomorphism) is a trivial UMTC and the Drinfel'd center of $\Vect$ is itself, with the half-braiding being the swap.
\end{rem}

\begin{example}{$\cZ(\Gvec)$}{Z vec G}
As a useful example, we construct the Drinfel'd center for the category $\Gvec$.

The \emph{Drinfel'd center} $\cZ(\Gvec)$ is the category with,
\begin{itemize}
  \item \emph{Objects:} pairs $(V,\sigma_{V,(\cdot)})$ with $V\in\Gvec$.
  \item \emph{Morphisms:} $f:(V,\sigma_{V, (\cdot)})\rightarrow (V',\sigma'_{V', (\cdot)})$ are maps $f:V\rightarrow V'$ in $\Gvec$ such that  $$(\Id_W\otimes f)\circ \sigma_{V,W} = \sigma'_{V',W}\circ (f\otimes \Id_W)\qquad \forall \, W\in\Gvec$$
\end{itemize}

The tensor product is
\begin{align*}
    (V,\sigma_{V, (\cdot)})\otimes (W,\delta_{W, (\cdot)})&:=\bigl(V\otimes W,\ \sigma_{V\otimes W,(\cdot)}\bigr)\\
    \sigma_{V\otimes W,X} &:= \alpha_{X,V,W} \circ (\sigma_{V,X} \otimes \Id_W) \circ \alpha^{-1}_{V,X,W} \circ (\Id_V \otimes \delta_{W,X}) \circ \alpha_{V,W,X},
\end{align*}

with unit object $(\ds[1],\sigma_{\ds[1],(\cdot)})$ given by the unitors. The \emph{braiding} is given by $$c_{(V,\sigma_{V, (\cdot)}), (W,\delta_{W, (\cdot)})} := \sigma_{V,W}: V\otimes W \rightarrow W\otimes V$$
\end{example}

\begin{rem}
There is an obvious forgetful functor $F: \caZ(\caC) \rightarrow \caC $ given by $$F((X, \sigma_{X,(\cdot)})) \mapsto X$$ It is easy to show that $F$ is a $\rmC^*$-tensor functor as it acts trivially on the $\rmC^*$-tensor structure of $\caZ(C)$.
\end{rem}

\section{G-crossed braided \texorpdfstring{$\rmC^*$}{C*}-tensor category}

\begin{defn}{\cite[Def.~8.24.1]{etingof2015tensor}}
\label{def:G crossed BTC}
A \emph{$G$-crossed braided $\rmC^*$-tensor category} for a finite group $G$ consists of:
\begin{enumerate}
\item A (not necessarily faithful) $G$–grading $$\cC = \bigoplus_{g\in G} \cC_g,\qquad \cC_g\otimes \cC_h \subset \cC_{gh},\quad \ds[1]\in \cC_e$$ where $\cC$ is a $\rmC^*$–tensor category and all structural isomorphisms are unitary.

\item A unitary tensor $G$–action $\gamma: G \rightarrow \Aut (\cC)$, $g\mapsto \gamma_g$ with $$\gamma_g(\cC_h)\subset \cC_{g h g^{-1}}$$ together with unitary tensor structures $$\mu_g^{X,Y}: \gamma_g(X)\otimes \gamma_g(Y)\rightarrow \gamma_g(X\otimes Y) \quad X,Y \in \cC, \qquad \mu_g^0: \ds[1]\rightarrow \gamma_g(\ds[1])$$ and unitary tensor natural isomorphisms $$\eta_{g,h}: \gamma_g\circ \gamma_h \Rightarrow \gamma_{gh}$$ satisfying the standard pentagon for $\eta_{g,h}$ and the compatibility of $\mu,\gamma$. This guarantees that $\gamma: G \rightarrow \Aut(\cC)$ is a tensor functor.

\item A \emph{$G$–braiding}: for $X\in\cC_g$ and $Y\in\cC$,
a unitary natural isomorphism $$c_{X,Y}: X\otimes Y \rightarrow \gamma_g(Y)\otimes X$$ which natural in both variables, satisfying $$c_{X,\ds[1]} = \Id_X  \text{ (via } \mu_g^0) \qquad \qquad c_{\ds[1], Y} = \Id_Y$$ and the following coherence conditions \cite[eqns.~(8.105)–(8.107)]{etingof2015tensor}.

\item \emph{Equivariance under $\gamma_g$:}
For all $g,h\in G$, $X\in\cC_h$, $Y\in\cC$, the diagram
\[\tag{\shortstack{\text{$G$–equivariant} \\\text{braiding}}}
\begin{tikzcd}[column sep=huge,row sep=large]
\gamma_g(X)\otimes \gamma_g(Y)
  \arrow[r,"{c_{\gamma_g(X),\gamma_g(Y)}}"]
  \arrow[d,swap,"{(\mu_g^{X,Y})}"]
&
\gamma_{g h g^{-1}}\!\bigl(\gamma_g(Y)\bigr)\otimes \gamma_g(X)
  \arrow[d,"{(\eta_{g h g^{-1},g})_Y\otimes\Id_{\gamma_g(X)}}"]
\\
\gamma_g(X\otimes Y)
  \arrow[d,swap,"{\gamma_g(c_{X,Y})}"]
&
\gamma_{g h}(Y)\otimes \gamma_g(X)
\\
\gamma_g\!\bigl(\gamma_h(Y)\otimes X\bigr)
  \arrow[r,swap,"{(\mu_g^{\gamma_h(Y),X})^{-1}}"]
&
\gamma_g\!\bigl(\gamma_h(Y)\bigr)\otimes \gamma_g(X)
  \arrow[u,swap,"{(\eta_{g,h})_Y\otimes\Id_{\gamma_g(X)}}"]
\end{tikzcd}
\]

\item For $g\in G$,
$X\in\cC_g$, $Y,Z\in\cC$, the following diagram commutes:
\[\tag{heptagon 1}
\begin{tikzcd}
	& {(X \otimes Y )\otimes Z} \\
	{X \otimes (Y \otimes Z)} && {(\gamma_g(Y) \otimes X)\otimes Z} \\
	{\gamma_g(Y\otimes Z) \otimes X} && {\gamma_g(Y) \otimes (X\otimes Z)} \\
	{(\gamma_g(Y) \otimes \gamma_g(Z))\otimes X} && {\gamma_g(Y) \otimes (\gamma_g(Z)\otimes X)}
	\arrow["{\alpha_{X,Y,Z}}"', from=1-2, to=2-1]
	\arrow["{c_{X,Y}\otimes \Id_Z}", from=1-2, to=2-3]
	\arrow["{c_{X, Y \otimes Z}}"', from=2-1, to=3-1]
	\arrow["{\alpha_{\gamma_g(Y),X,Z}}", from=2-3, to=3-3]
	\arrow["{(\mu_g^{Y,Z})^{-1} \otimes \Id_X}"', from=3-1, to=4-1]
	\arrow["{\Id_{\gamma_g(Y)} \otimes c_{X,Z}}", from=3-3, to=4-3]
	\arrow["{\alpha_{\gamma_g(Y), \gamma_g(Z), X}}"', from=4-1, to=4-3]
\end{tikzcd}\]

\item 
For all $g,h\in G$, $X\in\cC_g$, $Y\in\cC_h$, $Z\in\cC$, the following diagram commutes:

\[\tag{heptagon 2}
\begin{tikzcd}
	& {X \otimes (Y \otimes Z)} \\
	{(X \otimes Y) \otimes Z} && {X\otimes (\gamma_h(Z) \otimes Y)} \\
	{\gamma_{gh}(Z)\otimes (X\otimes Y)} && {(X \otimes \gamma_h(Z)) \otimes Y} \\
	{\gamma_g(\gamma_h(Z)) \otimes (X \otimes Y)} && {(\gamma_g(\gamma_h(Z)) \otimes X) \otimes Y}
	\arrow["{\Id_X \otimes c_{Y,Z}}", from=1-2, to=2-3]
	\arrow["{\alpha_{X,Y,Z}}", from=2-1, to=1-2]
	\arrow["{\alpha^{-1}_{X,\gamma_h(Z),Y}}", from=2-3, to=3-3]
	\arrow["{c^{-1}_{X\otimes Y,Z}}", from=3-1, to=2-1]
	\arrow["{c_{X, \gamma_h(Z)}\otimes \Id_Y}", from=3-3, to=4-3]
	\arrow["{\eta_{g,h} \otimes \Id_{X\otimes Y}}", from=4-1, to=3-1]
	\arrow["{\alpha^{-1}_{\gamma_g( \gamma_h(Z)), X,Y}}"', from=4-1, to=4-3]
\end{tikzcd}\]

\end{enumerate}
\end{defn}

\section{\texorpdfstring{$\Rep D(G)$}{Rep D(G)} and relationship to \texorpdfstring{$\cZ(\Gvec)$}{Z(Vec(G))}}
\label{subsec:center-pointed-double}
In this section we show that $\cZ(\Gvec)$ is braided $\rmC^*$-tensor equivalent to $\Rep D(G)$, the category of representations of the Quantum Double algebra (defined below). For this result, we will need $\YD_G$, the Yetter-Drinfel'd category.

The reader may also treat this section as a basic introduction to the Quantum Double category, a crucial object studied in this thesis.

\begin{defn}
We denote the category of Yetter-Drinfel'd modules $\YD_G$ as the $\rmC^*$–tensor category whose objects are finite $G$–graded Hilbert spaces $$V=\bigoplus_{g\in G} V_g$$ equipped with a unitary action $\rho:G\rightarrow\Aut(V)$ satisfying the conjugation covariance $$\rho(h)(V_g)\subset V_{hgh^{-1}}\qquad(\forall\,g,h\in G)$$ Morphisms preserve both the grading and the action. The tensor structure is the graded tensor product $$(V\otimes W)_x=\bigoplus_{ab=x} V_a\otimes W_b,\qquad \rho(h)(v\otimes w):=\rho(h)v\otimes \rho(h)w$$ and the $\rmC^*$–structure is inherited componentwise.

The braiding isomorphism is given by $$ c^{\YD}_{V,W}(v_g\otimes w)=(\rho(g) w)\otimes v_g\qquad v_g\in V_g, $$ extended by linearity.
\end{defn}

\subsection{The Drinfel'd double $D(G)$ and category $\Rep D(G)$.}{\cite[Chap.~7]{majid2000foundations}}
\label{sec:quantum double category}
Let $\bbC G$ be the group algebra and $\bbC^G$ the function algebra with basis $\{\delta_s\}_{s\in G}$. The quantum double is the crossed (bicrossed) Hopf algebra $$D(G)=\bbC^G\bowtie \bbC G\quad\text{(as a vector space } \bbC^G\otimes \bbC G)$$ with:
\begin{itemize}
    \item \textit{Cross relation:} $\delta_x g = g \delta_{g^{-1}xg}\qquad(x,g\in G),$
    \item \textit{Multiplication:} $ (\delta_s\otimes g)(\delta_t\otimes h)=\delta_{s,g t g^{-1}}\ (\delta_s\otimes gh),$
    \item \textit{$*$-structure:} $\delta_g^* = \delta_g ,\qquad g^* = g^{-1}, \qquad (\delta_h \otimes g)^* = \delta_{ghg^{-1}} \otimes g^{-1},$
    \item \textit{Hopf structure:} $$\Delta(\delta_s\otimes g)=\sum_{ab=s} (\delta_a\otimes g)\otimes (\delta_b\otimes g),\qquad \epsilon(\delta_s\otimes g)=\delta_{s,e},\qquad S(\delta_s\otimes g)=\delta_{g^{-1}s^{-1}g}\otimes g^{-1}. $$
\end{itemize}

The universal $R$–matrix\footnote{Roughly, it is a braiding isomorphism on the Hopf algebra} is $$R = \sum_{g\in G} (\delta_g\otimes 1)\ \otimes\ (1\otimes g)\ \in D(G)\otimes D(G)$$ which satisfies the quasitriangular identities $$(\Delta\otimes\Id)(R)=R_{13}R_{23},\quad(\Id\otimes \Delta)(R)=R_{13}R_{12},\quad\Delta^{\opp}(x)=R\Delta(x)R^{-1} (\forall x\in D(G))$$

\begin{defn}
\label{def:RepDG}
The category $\Rep D(G)$ is defined as follows.
\begin{itemize}
  \item \emph{Objects:} finite–dimensional $D(G)$ $*$-representations $(V,\pi_V)$.
  \item \emph{Morphisms:} Bounded linear maps $f:V\rightarrow W$ such that, such that $f\pi_V(x)=\pi_W(x)f$ for all $x\in D(G)$.
\end{itemize}
It is a linear $\rmC^*$-tensor category with tensor product $$(V,\pi_V)\otimes (W, \pi_W) := (V \otimes W, \pi_{V\otimes W}) \qquad \qquad \pi_{V\otimes W}(x):=(\pi_V\otimes \pi_W)\bigl(\Delta(x)\bigr) \qquad x \in D(G)$$ unit object $(\bbC,\epsilon(x) \Id_\bbC)$ and (strict) associativity/unitality inherited from vector spaces.

The \emph{braiding}
on $\Rep D(G)$ is $$c^{D(G)}_{V,W}:V\otimes W \longrightarrow W\otimes V,\qquad c^{D(G)}_{V,W}=\tau\circ (\pi_V\otimes \pi_W)(R)$$ where $\tau$ is the flip $v\otimes w\mapsto w\otimes v$. This makes $\Rep D(G)$ a braided $\rmC^*$-tensor category.

The category is rigid with dual $V^*$ carrying the action $$\pi_{V^*}(x)\phi:=\phi\circ \pi_V\bigl(S(x)\bigr)\qquad (x\in D(G),\ \phi\in V^*)$$ and the usual evaluation/coevaluation maps.

\end{defn}

In \cite{gould1993quantum}, it was shown that the irreducible objects in $\Rep D(G)$ are classified by the pairs $(C, \pi_{Z_C})$ where $C \subset G$ is a conjugacy class of $G$, $Z_C \subset G$ is the centralizer of a representative element $g_C \in C$ and $\pi_{Z_C}$ is an irreducible representation of $Z_C$. This fact is crucially used in Kitaev's Quantum Double model to construct representative states and string operators corresponding to every irreducible representation of $D(G)$ (see Chapter \ref{chap:quantum double sectors}, and in particular Sections \ref{subsection:irreps and Wigner projections}, \ref{sec:ribbon operators and limiting maps} for more detail of this standard fact).

In \cite{bols2025classification} it was shown that the irreducible anyon representations are classified by the irreducible objects of $\Rep D(G)$. This was the cornerstone result enabling the analysis of \cite{bols2025category}, which built the category of anyons and showed that it is braided $\rmC^*$-equivalent to $\Rep D(G)$.

\subsection{$\cZ(\Gvec) \simeq \YD_G$.}
The reference \cite[Prop.~7.15.3]{etingof2015tensor} sketches the ($\rmC^*$-tensor) equivalence for a more general setting. Here we fill in some basic details and promote it to braided equivalence. Let $\bbC(h)\in\Gvec$ denote the one–dimensional graded object concentrated in degree $h\in G$, and fix unit vectors $e_h\in\bbC(h)$ once and for all.

Define a unitary action $\rho^\gamma:G\rightarrow \Aut(V)$ by $$\lambda^L_V\ \circ\bigl(\Id_{\bbC(h)}\otimes \rho^\gamma(h)\bigr):=\gamma_{V,\bbC(h)}\ \circ\ \bigl(\Id_V\otimes (\lambda^R_{\bbC(h)})^{-1}\bigr)$$ equivalently, $$e_h\otimes \rho^\gamma(h)v := \gamma_{V,\bbC(h)}(v\otimes e_h)\qquad (h\in G,\ v\in V)$$ Set $$\Phi(V,\gamma):=(V,\rho^\gamma)\ \in\ \YD_G,\qquad\Phi(f):=f\ \text{ on morphisms.}$$

\begin{prop}\label{prop:Phi-braided-Cstar-tensor}
$\Phi$ is a well-defined braided $\rmC^*$–tensor equivalence.
\end{prop}

\begin{proof}
Naturality of $\gamma_{V,(\cdot)}$ with respect to the inclusions $\bbC(h)\hookrightarrow W$ for any $W\in\Gvec$ implies the group law $\rho^\gamma(hk)=\rho^\gamma(h)\rho^\gamma(k)$ and the covariance $\rho^\gamma(h)(V_g)\subset V_{hgh^{-1}}$.

Unitarity of $\gamma$ implies unitarity of each $\rho^\gamma(h)$, so $\Phi$ is $\rmC^*$–functor.

The half–braiding on $(V,\gamma)\otimes(W,\delta)$ equals

$\gamma_{V\otimes W,(\cdot)}=(\Id\otimes \alpha)\circ (\delta_{W,(\cdot)}\otimes \Id)\circ (\Id\otimes \gamma_{V,(\cdot)})\circ \alpha^{-1}$, hence $\rho^{\gamma\otimes\delta}(h)=\rho^\gamma(h)\otimes \rho^\delta(h)$. The tensorator of $\Phi$ is thus the identity on $V\otimes W$, and $\Phi$ is a $\rmC^*$-tensor functor.

For homogeneous $v\in V_g$, $w\in W$, $$c^{\cZ(\Gvec)}_{(V,\gamma_{V,(\cdot)}),(W,\delta_{W,(\cdot)})}(v\otimes w)=\gamma_{V,W}(v\otimes w)=\rho^\gamma(g)(w)\otimes v=c^{\YD}_{\Phi(V,\gamma_{V,(\cdot)}),\Phi(W,\delta_{W,(\cdot)})}(v\otimes w)$$

Since $\Phi$ is the identity on underlying linear maps, it is faithful. If $f:V\rightarrow W$ is a morphism in $\YD_G$, then $f\rho^\gamma(h)=\rho^\delta(h)f$ for all $h\in G$, which is equivalent to $$(\Id_{\bbC(h)}\otimes f)\gamma_{V,\bbC(h)} =\gamma_{W,\bbC(h)}(f\otimes\Id_{\bbC(h)})$$ As the simples $\bbC(h)$ generate $\Gvec$ under finite direct sums and tensor products, this is exactly the naturality condition defining morphisms in $\cZ(\Gvec)$. Thus $${\cZ(\Gvec)}((V,\gamma),(W,\delta)) ={\YD_G}(\Phi(V,\gamma),\Phi(W,\delta))$$ so $\Phi$ is full and faithful.

Let $(V,\rho,\bigoplus_{g\in G}V_g)\in\YD_G$. Define $\gamma_{V,\bbC(h)}:V\otimes\bbC(h)\rightarrow\bbC(h)\otimes V$ by $$ \gamma_{V,\bbC(h)}(v\otimes e_h):=e_h\otimes \rho(h)v, $$ and extend uniquely to all $X\in\Gvec$ by additivity and multiplicativity in $X$ (the objects $\bbC(h)$ generate $\Gvec$). The Yetter–Drinfel'd conditions are equivalent to the center hexagon axioms, so this yields a half–braiding $\gamma$ making $(V,\gamma)\in\cZ(\Gvec)$; by construction $\Phi(V,\gamma)=(V,\rho)$. This gives essential surjectivity.
\end{proof}

\begin{rem}
The forgetful tensor functor $U:\YD_G\rightarrow\Gvec$, $(V,\rho)\mapsto V$, corresponds under $\Phi$ to the canonical forgetful $F:\cZ(\Gvec)\rightarrow\Gvec$. A fully faithful tensor embedding $\Gvec\hookrightarrow \YD_G$ exists exactly on the full subcategory supported on $Z(G)$ (trivial action is YD–compatible only there).
\end{rem}

\subsection{$\YD_G \simeq \Rep D(G)$}

Define the functor $$\Psi: \YD_G\longrightarrow \Rep D(G),\qquad(V,\rho,\textstyle\bigoplus_{g\in G}V_g)\ \longmapsto\ \pi_V$$ defined on generators by $$\pi_V(\delta_g\otimes 1):=P_g \qquad \pi_V(1\otimes g):=\rho(g)$$ and extended multiplicatively. Here $P_g$ is the projection $P_g : V \rightarrow V_g$. It acts on the morphisms as identity.

\begin{prop}
\label{prop:Psi-braided-Cstar-tensor} 
The functor $\Psi$ is a well-defined braided $\rmC^*$–tensor equivalence.
\end{prop}

\begin{proof}
Note that by definition, $\rho(g) V_h \subset V_{g h g^{-1}}$, we have $\rho(g) P_h = P_{g h g^{-1}} \rho(g)$, so $\pi_V$ respects the cross relation of $D(G)$. It is straightforwardly verified that $\pi_V$ is a $*$-homomorphism of $D(G)$. The functor $\Psi$ also preserves the grading and action.

It is straightforward that $\Psi$ is a $\rmC^*$-functor. Moreover, by defining $\Psi^0:\ds[1]\rightarrow \Psi(\ds[1])$ to be the identity on the trivial module and the tensorator $\Psi^2_{V,W}:\Psi(V)\otimes \Psi(W)\rightarrow \Psi(V\otimes W)$ to be the identity on $V\otimes W$, we can show that $\Psi$ is a $\rmC^*$-tensor functor. Indeed, one verifies the identity $$\pi_{V\otimes W}(x)=(\pi_V\otimes \pi_W)\bigl(\Delta(x)\bigr) \qquad x \in D(G)$$ by verifying it on the generators of $D(G)$.

Thus $\Psi^2_{V,W}$ is trivially a $D(G)$–intertwiner, natural in $V,W$, and the coherence with associator/unitors is immediate since all structure maps are identities on the underlying spaces.

The braiding in $\Rep D(G)$ is given by $c^{D(G)}_{\Psi(V),\Psi(W)}=\tau\circ (\pi_V\otimes \pi_W)(R)$. Let $v_h \in V_h$ be homogenous and $w \in W$. We confirm using the definition of $\pi_V,\pi_W$ and $R$ that  $$(\pi_V\otimes \pi_W)(R)(v_h\otimes w)=\sum_{g\in G} (P_g v_h)\otimes \rho_W(g)w =v_h\otimes \rho_W(h)w$$ so $$c^{D(G)}_{\Psi(V),\Psi(W)}(v_h\otimes w)= \rho_W(h)w\otimes v_h$$ which is precisely the braiding $c_{V,W}^{\YD}$. It follows that the braiding square commutes and that $\Psi$ is braided.

All remaining unit/naturality axioms are immediate from the definitions. Therefore $\Psi$ is a braided $\rmC^*$–tensor functor.

The functor $\Psi$ is full and faithful by design, and for any $(\pi,V) \in \Rep D(G)$ we define the grading on $V$ by $V_g := \pi(\delta_g \otimes 1) V$ and set $\rho: G \rightarrow \Aut(V)$ as $\rho(g) v \mapsto \pi(1 \otimes g) v$. Thus $(V, \rho, \oplus_{g \in G} V_g)\in \YD_G$ and satisfies $\Psi(V, \rho_{\pi}, \oplus_{g \in G} V_g) = (\pi_V, V)$, showing the essential surjectivity of $\Psi$. 

We thus get that $\Psi$ is a braided $\rmC^*$-tensor equivalence.
\end{proof}

Combining Propositions \ref{prop:Phi-braided-Cstar-tensor} and \ref{prop:Psi-braided-Cstar-tensor}, we have shown,
\begin{thm}{\cite[Sec.~7.14,~8.5]{etingof2015tensor}}
\label{thm:center-pointed-equals-Drinfeld-double-untwisted}
There are braided $\rmC^*$–tensor equivalences $$\cZ(\Gvec)\ \simeq\ \YD_G\ \simeq\ \Rep D(G)$$ In particular, $\Rep D(G), \YD_G$ are UMTCs.
\end{thm}
\begin{rem}
    The proof given in \cite{etingof2015tensor} is as follows. By \cite[Prop.~7.14.16, 7.14.18(iii)]{etingof2015tensor} we have $\cZ(\Gvec) \simeq \Rep D(G)$. By \cite[Prop.~7.15.3]{etingof2015tensor} we have $\YD_G \simeq \cZ(\Gvec)$. Here the equivalence is actually braided equivalence, as noted in \cite[Ex.~8.5.5, 8.5.6]{etingof2015tensor}. Promoting it to a braided $\rmC^*$-equivalence is trivial.
\end{rem}

\begin{rem}
    This theorem has deep connections to the ubiquitous concepts of condensation and gauging in physics. Roughly, it states that one can start with a system whose degrees of freedom are $G$-graded vector spaces with a compatible $G$-action, and understand the anyons in the system by gauging the $G$-action. Equivalently, one can start with a system hosting anyons corresponding to the quantum double $D(G)$ phase, and condense the anyons to get a $G$-action on the system.
\end{rem}

\begin{rem}
    This theorem is highly generalizable, and in fact holds if one replaces $G$ with an arbitrary weak (not necessarily finite dimensional) Hopf-$*$ algebra, with the appropriate generalizations to the definitions of $\Gvec, \YD_G$.
\end{rem}

\begin{rem}
    We note the physical importance of the above theorem in Kitaev's Quantum Double model. The braided equivalence $\cZ(\Gvec) \simeq \Rep D(G)$ tells us that in fact the objects of $\Rep D(G)$ are capable of ``crossing over'' the objects of $\Gvec$ (cf.~remark \ref{rem:physics of drinfeld center}). Since each edge of QD is assigned a $G$-graded vector space\footnote{This is strictly speaking untrue in the original model proposed by Kitaev, but holds in an equivalent model called the Levin-Wen model with input category $\Gvec$. These two models are in the same phase.}, categorically modelled by $\Gvec$, we see that objects of $\Rep D(G)$ carry a natural interpretation of anyons, as they are able to freely ``cross over'' the objects of $\Gvec$ on each edge with the help of the half-braid isomorphism.
\end{rem}

\bibliographystyle{alpha}
\bibliography{intro_bibliography}

%% file: OperatorAlgebrasBackground.tex
\chapter{Operator Algebras \& anyons}
\label{chap:operator algebras}

\minitoc

While in the domain of this thesis operator algebras are standard-issue, we take a moment to highlight why one should care about operator algebras in physics.

Operator algebras let us talk about infinite quantum systems without hand-waving. They encode locality, limits, and symmetries in a single analytic language. By operator algebras, we primarily mean $\rmC^*$-algebras. Roughly, $\rmC^*$-algebras are vector spaces with an underlying multiplication structure, a notion of adjoint, and also some notion of being a closed set in some topology. This means there is some notion of distance and neighborhoods, strength of an operator etc. Additionally, the strength of an observable (self-adjoint operator) in a $\rmC^*$-algebra is equal to the largest magnitude of a measurement outcome if you prepare the ``worst-case'' state. The set of (approximately) local operators, all too common in physics, forms a $\rmC^*$-algebra called the local algebra. In fact, in physics there's a special reason to care about operators that can be approximated by local operators, since labs are usually localized in finite regions. This set of operators forms the quasi-local algebra. The set of $n\times n$ complex matrices also forms a $\rmC^*$-algebra called the ($n$-dimensional) matrix algebra. In fact, if the $\rmC^*$-algebra is ``finite dimensional'' (meaning there are finitely many basis elements), then it is isomorphic to finitely many copies of matrix algebras. Results like this already expose why it is useful to study operator algebras.

Ubiquitous in physics are questions related to the thermodynamic limit of a quantum system and its stability under small perturbations. Here we argue that $\rmC^*$-algebras make for an excellent tool to study these two questions. Regarding thermodynamic limits, since the algebras already have a topological structure, the notion of limits is built into the structure of an $\rmC^*$-algebra. So questions like ``what happens if I take a limit of this operator" have an automatic (if usually hard to compute) answer. Stability questions can also be studied elegantly using the $\rmC^*$-algebra structure. Instead of studying the stability under a specific kind of perturbation (for example a small magnetic field in some direction), in $\rmC^*$-algebras one imposes a bound on the ``strength'' of the operators corresponding to perturbations, and then studies the stability of the spectral gap. 

Operator algebras have a rich and long history, and have as such been extensively studied in many, many works. Here is a list of sources ranging from introductory (\cite{hunter2001applied, MR3617688}) to authoritative treatments (\cite{bratteli2012, bratteli2013, takesaki2003theory1, takesaki2003theory2, kadison1986fundamentals, MR1468230}). Some of these treatments are more mathematical, while others have many applications to physics, and others like \cite{kadison1986fundamentals, MR1468230} have helpful exercises. Treatments like \cite{jones2017operator, chen2022q} categorify many of the fundamental results in operator algebras. In this introduction, we will primarily follow \cite{MR3617688, naaijkens2012anyons}.

The introduction is structured as follows. First we introduce algebras, and in particular $\rmC^*$-algebras, which will be our main object of study. We will briefly explore important structures and properties of these algebras such as tensor products and direct sums. Then we will define the notion of states in infinite volume, which will unlock a lot more structure of these $\rmC^*$-algebras since it will allow us to use the {GNS construction} to build a Hilbert space. Next, we will move on to study von Neumann algebras, which will be especially important for the works in this thesis.

We then switch gears to talk about $\rmC^*$-algebras on a lattice, and in particular the quasi-local algebra, cone algebra, and auxiliary algebra. Finally we conclude by discussing the notion of an anyon sector, the category of anyon sectors, and showing that it is a braided $\rmC^*$-tensor category. 

Much of what we do in the latter half of this introduction is the lattice version of the DHR-style AQFT analysis \cite{haag1964algebraic, doplicher1989new,doplicher1971local,doplicher1974local, doplicher1969fields}, first adapted to the lattice setting by Naaijkens \cite{MR2804555} and then further developed in \cite{MR4362722, MR4426734, MR3426207, bols2025classification, bols2025category, kawagoe2024operator, ogata2025haag, bachmann2024tensor, Naaijkens2015}. Many results will mirror the AQFT side, which is absolutely by design. For an introduction to the AQFT literature, the reader may wish to read \cite{fredenhagen2015introduction} for a modern introduction to the topic. 

An important omission from this introduction is an in-depth treatment of dynamics of spin systems. For the works in this thesis, the only concepts that are necessary will be the existence of ``nice" dynamics, the existence of a gap in the spectrum, and a ground-state. Thus we will leave the treatment of these concepts to works that will do justice to these important concepts (\cite{hunter2001applied, MR3617688}).

In order to curtail the length of this introduction, we will assume familiarity with vector spaces and inner products, norms, direct sums, tensor products, as well as basic linear algebra results. A basic familiarity with topological spaces is also assumed, though we will review them.

\section{\texorpdfstring{$\rmC^*$}{C*}-Algebras}
An \emph{algebra} $\cA$ is a vector space $\cA$ (over $\bbC$) equipped with a multiplication operation $\cA \times \cA \rightarrow \cA$ satisfying:
\begin{itemize}
    \item $(xy)z = x(yz)$ for all $x,y,z \in \cA$
    \item $x(y+z) = xy + xz$ and $(x+y)z = xz + yz$ for all $x,y,z \in \cA$
    \item $c(xy) = (cx)y = x(cy)$ for all $x,y \in \cA$
\end{itemize}
An algebra $\cA$ is \emph{unital} if there exists $1 \in \cA$ satisfying $1 x = x 1 = x$ for all $x \in \cA$. 

A \emph{$\rmC^*$-algebra} $\cA$ is an algebra $\cA$ equipped with a norm $|| \cdot || : \cA \rightarrow [0,\infty)$ and is complete with respect to this norm\footnote{Completeness means that $\cA$ contains the limit of all Cauchy sequences $(a_n)_{n \in \bbN}$ with respect to the norm.}, equipped with an involution $*$ (called the adjoint) that is also an anti-homomorphism, i.e.,
\begin{itemize}
    \item $(x + y)^* = x^* + y^*$ for all $x,y \in \cA$
    \item $(c x)^* = \bar c x^*$ for all $x \in \cA$
    \item $(x^*)^* = x$ for all $x \in \cA$
    \item $(xy)^* = y^* x^*$ for all $x,y \in \cA$
\end{itemize} and in addition satisfies for all $x,y \in \cA$ the $\rmC^*$-identity $||x^* x|| = ||x||^2$, from which it follows that $||xy|| \leq ||x|| \, ||y||$ and $||x|| = ||x^*||$.

\begin{example}{Bounded operators on a Hilbert space}{bdd ops on H}
\label{example:bounded ops on a Hilbert space}
Given a Hilbert space $\cH$ we define a \emph{bounded linear map} $x : \cH \rightarrow \cH$ as a map $x$ for which the \emph{operator norm}, defined as $$||x|| := \underset{||\xi|| = 1}{\sup} || x \xi||,$$ is finite. Define $\cB(\cH)$ as the set of bounded linear maps over $\cH$. We notice that for two bounded linear maps $x,y \in \cB(\cH)$, we can define another linear map $x \circ y : \cH \rightarrow \cH$. We observe $$||x \circ y|| = \underset{||\xi|| = 1}{\sup} ||x (y \xi)|| \leq ||x|| \underset{||\xi|| = 1}{\sup}  || y \xi|| \leq ||x|| \,||y||$$ and thus $x \circ y$ is also another bounded linear operator. $\cB(\cH)$ is thus an algebra with composition as the multiplication and the identity map as the unit (denoted $\Id$). We observe also that the operator norm is aptly named, in the sense that it is a norm. In the metric induced by the operator norm, a Cauchy sequence $(x_n)$ converges in $\cB(\cH)$
to some bounded operator $x$, and moreover $\|x_n-x\|\rightarrow 0$. In particular $\cB(\cH)$ is complete.

\begin{prop}
    For every $x \in \cB(\cH)$ there exists a unique element $y \in \cB(\cH)$ satisfying for all $\xi,\eta \in \cH$, $$\inner{\xi}{x \eta} = \inner{y \xi}{\eta}$$ Moreover, we have $||y|| = ||x||$.
\end{prop}
\begin{proof}
    Standard.
\end{proof}

Using the above proposition we may define $*: x \mapsto x^*$ as our adjoint operation, where $x^*$ is defined as the unique element corresponding to $x$ afforded to us by the above proposition. It is trivial to check that $*$-operation is indeed an adjoint by using standard properties of inner products.

Finally we will show $||x^* x || = ||x||^2$. We note that $||x^*x|| \leq ||x^*|| \, ||x|| = ||x||^2$ by above proposition. We also note that $x^* x$ is positive, so $\inner{x^* x \xi}{\xi} \geq 0$. Now by definition of $||x||$ for every $\epsilon > 0$ there exists a unit vector $\xi \in \cH$ such that $||x\xi || > ||x|| -\epsilon$. Thus, $$||x^*x|| \geq |\inner{x^*x \xi}{\xi}| = \inner{x^*x \xi}{\xi} = ||x \xi||^2 \geq (||x|| - \epsilon)^2 $$ where the first inequality is Cauchy-Schwarz. Letting $\epsilon \downarrow 0$ we get $||x^* x|| \geq ||x||^2$, and thus $||x^*x|| = ||x||^2$, giving us that $\cB(\cH)$ is a $\rmC^*$-algebra.
\end{example}

\begin{example}{}{}
The set of $n\times n$ complex matrices, $M_n(\bbC)$, is an example of $\cB(\cH)$ with $\cH = \bbC^n$. The usual matrix multiplication is the multiplication operation and the identity matrix is the unit. The usual matrix adjoint is a $*$-operation.
\end{example}

Notice that $M_n(\bbC)$ can be thought of as the set of (trivially bounded) linear maps $M : \bbC^n \rightarrow \bbC^n$ and is thus an elementary example of $\cB(\cH)$ with $\cH = \bbC^n$.

Let $\cA, \cB$ be algebras. A (algebra) \emph{homomorphism} $\phi: \cA \rightarrow \cB$ is a linear map preserving the algebra structure, i.e., $\phi(a)\phi(b) = \phi(ab)$ for all $a, b \in \cA$. If $\cA, \cB$ are unital, then we call $\phi$ a \emph{unital} homomorphism if $\phi(1_{\cA}) = 1_{\cB}$. If $\cA, \cB$ are $\rmC^*$-algebras, then $\phi$ is called a \emph{$*$-homomorphism} if it commutes with the $*$ operation (i.e., $\phi(a)^* = \phi(a^*)$).

It is well known that any $*$-homomorphism of $\rmC^*$-algebras $\phi: \cA \rightarrow \cB$ is automatically continuous\footnote{It uses the $\rmC^*$-property, an \emph{algebra} homomorphism needn't be continuous.} (with respect to the norm topologies), which follows from $\phi$ being contractive ($||\phi (a)|| \leq ||a||$).

If we have two algebras $\cA, \cB$, we say that $\cA \simeq \cB$ if there exists a $*$-isomorphism $\phi:\cA \rightarrow \cB$. That is, a $*$-homomorphism that is also an invertible map.

\begin{defn}
    Given some $x \in \cB(\cH)$ we say that $x$ is
    \begin{itemize}
        \item \textbf{(self-adjoint)} if $x = x^*$
        \item \textbf{(normal)} if $xx^* = x^* x$
        \item \textbf{(unitary)} if $x x^* = x^* x = 1$
        \item \textbf{(projection)} if $x = x^* = x^2$
        \item \textbf{(isometry)} if $x^* x = 1$
        \item \textbf{(partial isometry)} if $x = x x^* x$ (equivalently, if $xx^*$, hence also $x^*x$, is a projection)
        \item \textbf{(invertible)} if there exists $y \in \cB(\cH)$ such that $xy = yx = 1$
    \end{itemize}
\end{defn}
We note that self-adjoint operators, unitaries, and projections are all normal.  A unitary is precisely a normal isometry;  equivalently,  a  unitary  is  an  invertible  isometry.   Isometries,  unitaries,  and  projections  are  all partial isometries.

\begin{lem}
Let $\cH,\cK$ be Hilbert spaces. An isometry $x : \cH \rightarrow \cK$ is unitary if and only if it has dense range.
\end{lem}

\begin{proof}
If $x$ is unitary, then by definition it is surjective, hence has dense range. Conversely, suppose $x$ is an isometry with dense range. Then $x$ is bounded and has an adjoint $x^*$. For all $\xi,\eta \in \cH$, we have $$\inner{x^*x\xi}{\eta} = \inner{x\xi}{x\eta} = \inner{\xi}{\eta},$$ and thus $x^*x = 1_{\cH}$. 

Let $P := xx^*$. For any $\zeta \in \cK$ and any $\xi \in \cH$, $$\inner{(1-P)\zeta}{x\xi} = \inner{\zeta}{x\xi} - \inner{xx^*\zeta}{x\xi} = \inner{\zeta}{x\xi} - \inner{x^*\zeta}{x^*x\xi} = \inner{\zeta}{x\xi} - \inner{x^*\zeta}{\xi} = 0$$ Thus $(1-P)\zeta$ is orthogonal to $\Ran x$, which is dense, so $(1-P)\zeta = 0$ for all $\zeta$, hence $P = 1_{\cK}$. Therefore $x^*x = 1_{\cH}$ and $xx^* = 1_{\cK}$, so $x$ is unitary.
\end{proof}

\subsection{Tensor products and direct sums of \texorpdfstring{$\rmC^*$}{C*}-algebras}
\label{sec:tensors and sums of C^* algebras}
Let $\cA,\cB$ be $\rmC^*$-algebras. Their \emph{algebraic tensor product} $\cA\odot\cB$ is the complex vector space spanned by simple tensors $a\otimes b$ with multiplication and $*$-operation defined on simple tensors by $$(a\otimes b)(c\otimes d)=ac\otimes bd,\qquad (a\otimes b)^*=a^*\otimes b^*,$$and extended by bilinearity and conjugate linearity, respectively. To obtain a $\rmC^*$-algebra, we equip $\cA \odot \cB$ with the \emph{minimal} (spatial) $\rmC^*$-tensor norm and complete. The result is denoted by $\cA\otimes\cB$\footnote{Interestingly, the definition of tensor product allows for the definition of more than one $\rmC^*$-norm. Two canonical choices for the completion are the \emph{minimal} (spatial) tensor product completion $\otimes_{\min}$, defined via faithful representations $(\pi, \cH), (\rho, \cK)$ and completion inside $B(\cH\otimes \cK)$, and the \emph{maximal} tensor product $\otimes_{\max}$ completion, defined by the universal $\rmC^*$-norm. If either $\cA$ or $\cB$ is \emph{nuclear}, then $\otimes_{\min}$ and $\otimes_{\max}$ coincide. Here we suppress this subtlety.}. 

For every $\rmC^*$-algebra $\cA$ there is a canonical $*$-isomorphism $$\cA\otimes M_n(\C)\simeq M_n(\cA),$$ given by $a\otimes E_{ij}\mapsto a E_{ij}$, where $E_{ij}$ is the matrix with $1$ in entry $(i,j)$ and zeros elsewhere. In particular, for a Hilbert space $\cH$, $$B(\cH)\otimes M_n(\C)\simeq B(\cH\otimes \C^n)\simeq M_n(B(\cH))$$

If $\phi_1:\cA_1\rightarrow\cB_1$ and $\phi_2:\cA_2\rightarrow\cB_2$ are $*$-homomorphisms, their algebraic tensor $\phi_1\odot\phi_2$ on $\cA_1\odot\cA_2$ extends uniquely by continuity to a $*$-homomorphism $$\phi_1\otimes \phi_2:\ \cA_1\otimes \cA_2\ \longrightarrow\ \cB_1\otimes \cB_2 $$

\begin{example}{Matrix amplifications}{amp}
Given a linear map $\phi: \cA \rightarrow \cB$, we can define an amplification as follows. Define an identity map $I \in M_n(\bbC)$. Then we have $\phi \otimes I: \cA \otimes M_n(\bbC) \rightarrow \cB \otimes M_n(\bbC)$. By above, since we have $\cA \otimes M_n(\bbC) \simeq M_n(\cA)$ and $\cB \otimes M_n(\bbC) \simeq M_n(\cB)$, we see that $\phi \otimes I$ acts on $M_n(\cA)$ by applying $\phi$ to each component. This is called a \emph{matrix amplification} of $\phi$, and sometimes denoted $\phi^{(n)}$.
\end{example}

Given $\rmC^*$-algebras $\cA,\cB$, their \emph{direct sum} is $$\cA\oplus \cB:=\{(a,b):a\in\cA,\ b\in\cB\},$$ with pointwise operations, involution $(a,b)^*=(a^*,b^*)$, and norm $\|(a,b)\|=\max\{\|a\|,\|b\|\}$.

\subsection{Representations}
An important type of $*$-homomorphism from a $\rmC^*$-algebra $\cA$ is given by $\pi: \cA \rightarrow \cB(\cH)$ where $\cH$ is some Hilbert space. The $*$-homomorphism $\pi$ is called a \emph{representation} of $\cA$. We write it as $(\pi, \cH)$. 

\begin{defn}
    A representation $(\pi, \cH)$ of a $\rmC^*$-algebra is called:
    \begin{itemize}
        \item \emph{non-degenerate} if $\pi(\cA) \cH := \{ \pi(a)\xi: a \in \cA, \xi \in \cH\}$ is dense in $\cH$. If $\cA$ is unital, then this is equivalent to $\pi(1) = I$. We will only consider non-degenerate representations in this thesis.
        \item \emph{faithful} if $\pi$ is an injective $*$-homomorphism.
        \item \emph{cyclic} if there exists $\Omega \in \cH$ such that $\pi(\cA) \Omega := \{\pi(a) \Omega: a \in \cA\}$ is dense in $\cH$. $\Omega$ is called a \emph{cyclic vector}.
        \item \emph{irreducible} if it leaves no non-trivial closed subspace of $\cH$ invariant. Otherwise, $\pi$ is \emph{reducible}.
    \end{itemize}
\end{defn}

We have a $\rmC^*$-version of Schur's lemma for representations of $\rmC^*$-algebras:
\begin{lem}
    A non-degenerate representation $\pi$ is irreducible if and only if its commutant, defined as $$\pi(\cA)' := \{x \in \cB(\cH): x \pi(a) = \pi(a) x \text{ for all }a \in \cA\},$$ satisfies $\pi(\cA)' = \bbC I$.
\end{lem}

Two representations $(\pi_1, \cH_1)$ and $(\pi_2, \cH_2)$ (of a $\rmC^*$-algebra $\cA$) are unitarily equivalent (or simply equivalent) if there exists a unitary map $U: \cH_1 \rightarrow \cH_2$ such that we have $$U \pi_1(a) U^* = \pi_2(a) \qquad \qquad \text{for all $a \in \cA$}$$ We denote this by $\pi_1 \simeq \pi_2$. 

Consider representations $(\pi_1,\cH_1), (\pi_2, \cH_2)$. We can make a new representation $(\pi_1 \oplus \pi_2, \cH_1 \oplus \cH_2)$ called the \emph{direct sum} of representations. It acts as $$(\pi_1 \oplus \pi_2)(a) (\xi_1 \oplus \xi_2) := \pi_1(a) \xi_1 \oplus \pi_2 (a) \xi_2$$ for $\xi_i \in \cH_i$. Similarly we can make a representation $(\pi_1 \otimes \pi_2, \cH_1 \otimes \cH_2)$ which acts on the simple tensors $\xi_1 \otimes \xi_2$ of $\cH_1 \otimes \cH_2$ as $$(\pi_1\otimes \pi_2)(a_1 \otimes a_2)(\xi_1 \otimes \xi_2 ) := \pi_1(a_1) \xi_1 \otimes \pi_2 (a_2) \xi_2$$ For some representation $(\pi, \cH)$ if there exists a (non-trivial, non-zero) projection $P \in \cB(\cH)$ such that $P \cH$ is reducing under $\pi$ (meaning invariant under both $\pi(\cA)$ as well as $\pi(\cA)^*$), then we have that $\pi|_{P \cH} : \cA \rightarrow \cB(P\cH)$ and $\pi|_{(1-P)\cH}: \cA \rightarrow \cB{((1-P) \cH)}$ are representations, such that $\pi \simeq \pi_{P\cH}\oplus \pi_{(1-P) \cH}$. $\pi_{P \cH}, \pi_{(1-P)\cH}$ are then called (non-zero) \emph{subrepresentations} of $\pi$.

\subsection{States}
A state $\omega: \cA \rightarrow \bbC$ on a $\rmC^*$-algebra $\cA$ is a positive\footnote{positivity means $\omega(a^*a) \geq 0$ for all $a \in \cA$} linear functional of norm $1$. We denote a state on $\cA$ by $(\omega, \cA)$. By positivity, it is easily shown that the Cauchy-Schwarz inequality holds: $$|\omega(b^* a)|^2 \leq \omega(a^*a) \omega(b^*b) \qquad \qquad  a,b \in \cA$$

A state $\omega$ is called \emph{pure} if for every pair of states $\omega_1, \omega_2$ such that $$\omega(a) = \lambda \omega_1 (a) + (1-\lambda)\omega_2(a) \qquad \qquad a \in \cA, \, \lambda \in (0,1)$$ we have $\omega_1 = \omega_2 = \omega$, i.e., it cannot be written as a convex combination of two distinct states. Pure states are the extreme points of all states of a $\rmC^*$-algebra $\cA$. We denote the set of all states on $\cA$ by $\cS(\cA)$. This set is non-empty by the Hahn-Banach theorem, and is convex with pure states as its extremal points by the Krein-Milmann theorem.

We note that for a representation $(\pi,\cH)$ and unit vector $\Omega \in \cH$, the map $\phi: \cA \rightarrow \bbC$ given by $\phi: a \mapsto \inner{\Omega}{\pi(a) \Omega}$ defines a state. Such a state is called a \emph{vector state}.

The following theorem lets us associate to any state $(\omega, \cA)$ a vector state on some representation $(\pi, \cH)$. 

\begin{thm}[GNS construction]
\label{thm:GNS construction}
    For every state $(\omega, \cA)$ there exists a triple $(\pi, \cH, \Omega)$ such that $\pi: \cA  \rightarrow \cB(\cH)$ is a cyclic representation with cyclic vector $\Omega \in \cH$, and $$\omega(a) = \inner{\Omega}{\pi(a) \Omega} \qquad \qquad a \in \cA$$
    This triple is unique up to unitary equivalence of representations, i.e., if there exists another triple $(\pi', \cH', \Omega')$ then there exists a unitary $U: \cH \rightarrow \cH'$ such that $\pi'(a) = U \pi(a) U^*$ for all $a \in \cA$ and $U \Omega = \Omega'$.
\end{thm}

The GNS construction is very useful for many reasons. First it allows us to talk about Hilbert spaces where there may not be a natural notion of a Hilbert space to begin with. With this we can talk about vectors and bounded linear operators on this Hilbert space, which is a particularly nice kind of $\rmC^*$-algebra.

\begin{lem}
    A state $(\omega, \cA)$ is pure if and only if its GNS representation $(\pi, \cH)$ is irreducible.
\end{lem}
\begin{proof}
    $(\impliedby)$: We argue by contradiction. Assume $\omega$ is not pure, and has an irreducible GNS representation $\pi$. Consider a convex decomposition of $\omega$ given by $\lambda\omega_1 + (1-\lambda) \omega_2$. Let $(\pi_i, \cH_i, \Omega_i)$ be the GNS triple for $\omega_i$, and $(\pi, \cH, \Omega)$ be the GNS triple of $\omega$. We design a map $U: \cH \rightarrow \cH_1 \oplus \cH_2$ given by $$U: \pi(a) \Omega \mapsto \sqrt{\lambda} \pi_1(a) \Omega_1\oplus \sqrt{1-\lambda} \pi_2(a) \Omega_2,$$
    Check that $U$ is indeed a unitary, and note
    \begin{align*}
        U \pi(a) (\pi(c)\Omega) &= \sqrt{\lambda} \pi_1(a c) \Omega_1 \oplus \sqrt{1-\lambda} \pi_2(a c) \Omega_2 = (\pi_1(a) \oplus \pi_2(a)) U (\pi(c)\Omega)
    \end{align*}
    Since $\pi(\cA)\Omega$ is dense in $\cH$, we thus have that $\pi \simeq \pi_1 \oplus \pi_2$, showing $\pi$ is reducible contradicting the assumed irreducibility of $\pi$. The state $\omega$ is thus pure.

    $(\implies)$: We again show a contradition. Consider a reducible GNS representation $(\pi,\cH)$ for a pure state $\omega$. Then by definition, there exists an invariant subspace $\cK \subset \cH$. Let $p \in \cB(\cH)$ be the projection onto this subspace. Note that since $p$ is a (non-trivial, non-zero) projection onto an invariant subspace, we can define $t := \inner{\Omega}{p \Omega}$, giving us $0 < t < 1$. Now we define $$\omega_1 (a) := \inner{\Omega}{p \pi(a) \Omega}/t \qquad \qquad \omega_2(a) := \inner{\Omega}{(1-p) \pi(a) \Omega}/(1-t)$$ which are well-defined states since $0 < t < 1$. Clearly $\omega(a) = t \omega_1(a) + (1-t) \omega_2(a)$ and $\omega$ is thus not pure, giving us a contradiction to the assumed purity of $\omega$. Thus $\pi$ must be irreducible.
\end{proof}

The following lemma is useful in Chapter \ref{chap:symmetry} and follows immediately from the uniqueness of the GNS construction. 

\begin{lem}
\label{lem:invariance of state under a symmetry}
    Let $\omega$ be a state and $\alpha \in \Aut{(\cA)}$ be an automorphism. If $\omega = \omega \circ \alpha$, then we have $U \in \cB(\cH)$ such that $$\pi \circ \alpha(a) = U \pi (a) U^*, \qquad U \Omega = \Omega, \qquad \qquad a \in \cA$$ i.e., $\alpha$ is implemented by a unitary $U \in \cB(\cH)$, where $(\pi, \cH, \Omega)$ is the GNS triple of $\omega$.
\end{lem}
\begin{proof}
    We notice that $\omega \circ \alpha (a) = \inner{\Omega}{\pi \circ \alpha(a) \Omega}$, and thus $(\pi \circ \alpha, \cH, \Omega)$ is a GNS triple of $\omega \circ \alpha$. By Theorem \ref{thm:GNS construction}, we may take this triple to be the GNS triple of $\omega \circ \alpha$. Now we note that $\omega = \omega \circ \alpha$. Again by Theorem \ref{thm:GNS construction}, we have $\pi \circ \alpha \simeq \pi$. Let $U: \cH \rightarrow \cH$ be the unitary implementing the equivalence. We then have the required result, $$U \pi(a) U^* = \pi \circ \alpha(a), \qquad U\Omega = \Omega \qquad a \in \cA $$\qedhere
\end{proof}

Another important consequence of the GNS construction is that every $\rmC^*$-algebra $\cA$ can be represented as a $\rmC^*$-subalgebra of $\cB(\cH)$ for some Hilbert space $\cH$: 

\begin{thm}[Gel'fand-Naimark]
    Let $\cA$ be a $\rmC^*$-algebra. Then there exists a faithful isometric representation $\phi: \cA \rightarrow \cB(\cH)$ to a norm-closed self-adjoint $*$-subalgebra $\cB \subset \cB(\cH)$ for some Hilbert space $\cH$.
\end{thm}

\section{Topologies on \texorpdfstring{$\cB(\cH)$}{B(H)}}
\subsection{Topological spaces}
We recall that a topological space is the pair $(X,\tau)$ where $X$ is a set and $\tau$ is a collection of subsets of $X$ called \emph{open sets} of $X$ satisfying:
\begin{itemize}
    \item $\emptyset, X \in \tau$
    \item If $U_i \in \tau$ with $i\in I$ an (possibly infinite) index set, then $\bigcup_{i \in I} U_i \in \tau$
    \item If $U_1, \cdots, U_n \in \tau$ then $\bigcap_{i =1}^n U_i \in \tau$
\end{itemize}
Unless stated otherwise, we assume that our topological spaces are Hausdorff. A subset $N_x \subseteq X$ corresponding to a point $x \in X$ is called a \emph{neighbourhood} of $x$ if there exists an open set $U \in \tau$ with $x \in U \subseteq N_x$.

We notice that a normed vector space is a particularly nice kind of topological space. Thus a Hilbert space, $\rmC^*$-algebra, $\bbR^n$, etc. are all examples of a topological space.

Consider two different topologies $(X,\tau_s), (X, \tau_w)$ on the same space $X$. We say that $\tau_s$ is a \emph{stronger/finer} topology than $\tau_w$ if $\tau_w \subset \tau_s$, i.e., if $\tau_s$ contains more open sets.

In general a Cauchy sequence is not a well-defined concept in a topological space, and requires a metric (e.g.\ a norm). However we may still talk about convergence on topological spaces using nets. 

\begin{defn}
    A \emph{directed set} is a set $I$ equipped with an operation $\leq$ that satisfies:
    \begin{itemize}
        \item $i \leq i $ for all $i \in I$
        \item if $i \leq j$ and $j \leq k$ then $i \leq k$ for all $i,j,k \in I$
        \item for any $i,j \in I$ there exists a $k \in I$ such that $i,j \leq k$
    \end{itemize}
\end{defn}

\begin{defn}
    A \emph{net} in a topological space is a map $x: I\rightarrow X$ where $I$ is a directed set. For some set $S \subset X$, a net is usually denoted as $(x_i)_{i \in I} \subset S$  with $x_i \in S$ (we sometimes drop ($\subset S$) from the notation when it is clear from the context). A net $(x_i)_{i \in I}$ \emph{converges} to $x \in X$ in topology $\tau$ if for any chosen $U_x \in \tau$ containing $x$, there exists some $i_0 \in I$ such that for all $i \geq i_0$, $x_i \in U_x$. In this case we call $x$ the limit of $(x_i)_{i \in I}$ and write $x = \lim_i x_i$. A set $\S \subseteq X$ is \emph{closed} in topology $\tau$ if for any convergent net $(x_i)_{i \in I} \subset S$ in $\tau$ we have that $\lim_i x_i \in S$.
\end{defn}

We observe when $I = \bbN$, convergence of a net is the familiar notion of the convergence of a sequence.

As a trivial result, the ambient set $X$ is by definition closed with respect to any topology we put on it.

\begin{lem}
\label{lem:convergence in stronger topology implies convergence in weaker topology}
    Consider $(X, \tau_s), (X,\tau_w)$ as topological spaces. The following are equivalent:
    \begin{enumerate}
        \item The topology $\tau_s$ is stronger than $\tau_w$ (i.e., $\tau_w \subset \tau_s$).
        \item If a net $(x_i)_{i\in I} \subset X$ converges to $x \in X$ in $\tau_s$, then it converges to $x \in X$ in $\tau_w$. 
    \end{enumerate}
\end{lem}
\begin{proof}
    Standard. We show $(1 \implies 2)$ since we use it later. Consider a net $(x_i)_{i \in I} \subset X$ converging to $x \in X$ in $\tau_s$. Thus for any open set $U_x \in \tau_s$ containing $x$ there exists some $i_0 \in I$ such that for all $i \geq i_0$ we have $x_i \in U_x$. Now take any open set $V_x \in \tau_w$ containing $x$. Since $\tau_w \subset \tau_s$ we have $V_x \in \tau_s$. Thus by above, there exists some $i_0$ such that for all $i \geq i_0$ we have $x_i \in V_x$, and thus $(x_i)_{i \in I}$ converges to $x$ in $\tau_w$. 
\end{proof}

\begin{lem}
\label{lem:closure in weaker topology implies closure in stronger topology}
    Consider two topologies $(X, \tau_s), (X, \tau_w)$ with $\tau_s$ stronger than $\tau_w$. Now consider $S \subseteq X$ such that $S$ is closed in $\tau_w$. Then $S$ is closed in $\tau_s$.
\end{lem}
\begin{proof}
    Consider a net $(x_i)_{i \in I} \in S$ convergent to $x \in X$ in $\tau_s$. Then by Lemma \ref{lem:convergence in stronger topology implies convergence in weaker topology}, we have that $(x_i)_{i\in I}$ is convergent to $x \in X$ in $\tau_w$. But since $S$ is closed with respect to $\tau_w$, we have that $x \in S$, and thus $S$ is also closed with respect to $\tau_s$.
\end{proof}

A map $\Phi: X \rightarrow Y$ between topological sets $(X,\tau_x), (Y, \tau_y)$ is continuous if for every net $(x_i) \subset X$ converging in $\tau_x$ to $x \in X$, the net $\Phi(x_i) \subset Y$ converges in $\tau_y$ to $\Phi(x) \in Y$. Equivalently, $\Phi$ is continuous if the pre-image of an open set is open.

\begin{lem}
    \label{lem:continuity in stronger topology implies continuity in weaker topology}
    Consider a map $\Phi: X \rightarrow Y$ between topological sets $(X, \tau_x^s), (Y, \tau_y^s)$. Consider also a weaker topology $\tau^w_x \subseteq \tau_x^s$ on $X$ and $\tau^w_y \subseteq \tau^s_y$ on $Y$. The following statements hold:
    \begin{itemize}
        \item If $\Phi:(X, \tau_x^w) \rightarrow (Y, \tau_y^s)$ is continuous, then $\Phi:(X, \tau_x^s) \rightarrow (Y, \tau_y^s)$ is continuous. 
        \item If $\Phi:(X, \tau_x^s) \rightarrow (Y, \tau_y^s)$ is continuous, then $\Phi:(X, \tau_x^s) \rightarrow (Y, \tau_y^w)$ is continuous. 
    \end{itemize}
\end{lem}
\begin{proof}
    Consider an open set $U \in \tau_y^s$. Then $\Phi^{-1}(U) \in \tau^w_x$ by definition of continuity. But since $\tau^w_x \subseteq \tau_x^s$, we have that $\Phi^{-1}(U) \in \tau^s_x$ is open. 

    Now consider an open set $U \in \tau^w_y$ Since $\tau^w_y \subseteq \tau^s_y$, we have $U \in \tau^s_y$. Then since $\Phi$ is continuous, we have $\Phi^{-1}(U) \in \tau^s_x$. 
\end{proof}

\subsection{Norm topology, SOT, WOT on \texorpdfstring{$\cB(\cH)$}{B(H)}}
We may now talk about the different topologies on $\cB(\cH)$.

\begin{defn}
Fix an operator $x_0 \in \cB(\cH)$. $\cB(\cH)$ has various topologies:
    \begin{itemize}
        \item \emph{\textbf{Norm Topology}}: The operator norm imbues $\cB(\cH)$ with a topology, called the {norm topology}. The basic neighborhoods are given by $$N(x_0, \epsilon) := \{x \in \cB(\cH): ||x - x_0|| < \epsilon \}$$
        A net $(x_i)_{i \in I} \subset \cB(\cH)$ converges to $x \in \cB(\cH)$ in the norm topology if and only if for all $\epsilon >0$ there exists some $i_0 \in I$ such that for all $i \geq i_0$ we have $||x_i - x|| < \epsilon$.
        \item \emph{\textbf{Strong Operator Topology (SOT)}}: This is the topology of pointwise convergence on $\cH$. The basic neighborhoods are given by $$N(x_0, \{\xi_i\}, \epsilon) := \{x \in \cB(\cH): ||(x - x_0)\xi_i|| < \epsilon; \, i \in \{1, \cdots, n\};\, \xi_i \in \cH \}$$
        A net $(x_i)_{i \in I} \subset \cB(\cH)$ converges to $x \in \cB(\cH)$ in the strong operator topology if and only if for all $\epsilon > 0$ and each fixed $\xi \in \cH$ there exists some $i_0 \in I$ such that for all $i \geq i_0$ we have $||(x_i - x)\xi|| < \epsilon$.
        \item \emph{\textbf{Weak Operator Topology (WOT)}}: This is the weakest topology making all maps $x\mapsto \langle x\xi,\eta\rangle$ continuous for all $\xi,\eta\in\cH$. The basic neighborhoods are given by $$N(x_0, \{\xi_i\}, \{\eta_i\}, \epsilon) := \{x \in \cB(\cH): |\inner{\xi_i}{(x - x_0) \eta_i}| < \epsilon; \, i \in \{1, \cdots, n\};\, \xi_i, \eta_i \in \cH\}$$
        A net $(x_i)_{i \in I} \subset \cB(\cH)$ converges to $x \in \cB(\cH)$ in the weak operator topology if and only if for all $\epsilon > 0$ and each fixed $\xi , \eta \in \cH$ there exists some $i_0 \in I$ such that for all $i \geq i_0$ we have $|\inner{\eta}{(x_i - x) \xi}| < \epsilon$.
    \end{itemize}
\end{defn}
There are many different topologies on $\cB(\cH)$ other than the ones listed above: for example the ultra-strong topology (= $\sigma$-strong topology), ultra-strong$^*$ topology (= $\sigma$-strong$^*$ topology), strong-$^*$ topology (also called SOT$^*$), ultraweak topology (= $\sigma$-weak and weak$^*$ topologies). These will not be necessary for our works and so we avoid addressing them.

If $\cH$ is finite dimensional, we have that these topologies are all equal. 

\begin{lem}
\label{lem:ordering of topologies on bounded ops on H}
    The following ordering statements are true on the various topologies of $\cB(\cH)$:
    \begin{itemize}
        \item The norm topology is stronger than SOT, which in turn is stronger than WOT. 
        \item Norm convergence implies SOT convergence, which in turn implies WOT convergence.
        \item WOT closure implies SOT closure, which in turn implies norm closure.
    \end{itemize}    
\end{lem}
\begin{proof}
    By definition of operator norm, $||y|| \geq ||y\xi|| /||\xi||$ for all $\xi \in \cH$. Thus if we have $(x_i)$ converging to $x \in \cB(\cH)$ in norm topology, then $||(x-x_i)\xi|| \leq ||x-x_i|| \rightarrow 0$ for all $\xi \in \cH$, so norm convergence implies SOT convergence. By Lemma \ref{lem:convergence in stronger topology implies convergence in weaker topology} norm topology is stronger than SOT. 

    Similarly, Consider a net $(x_i)_{i \in I}$ converging to $x \in \cB(\cH)$ in SOT. Then by definition, for any chosen $\xi \in \cH$, $||(x - x_i) \xi|| \rightarrow 0$. But by Cauchy-Schwarz inequality, for any chosen $\eta,\xi \in \cH$ we have $|\langle \eta,(x_i-x)\xi\rangle|\le \|\eta\|\,\|(x_i-x)\xi\|\rightarrow 0$ and thus $(x_i)$ converges to $x$ in WOT. So SOT convergence implies WOT convergence. By Lemma \ref{lem:convergence in stronger topology implies convergence in weaker topology} we have that SOT is stronger than WOT.
    
    Now Lemma \ref{lem:closure in weaker topology implies closure in stronger topology} implies the required result for closures.
\end{proof}

If $\cH$ is infinite dimensional, the containment of the above Lemma is strict.

\begin{lem}
    \label{lem:adjoint action of a unitary is continuous}
    Consider a unitary $U \in \cB(\cH)$. Then the map $\alpha: \cB(\cH) \rightarrow \cB(\cH)$ given by $\alpha: a \mapsto U a U^*$ is continuous in WOT. 
\end{lem}
\begin{proof}
    Consider a net $(a_i)_{i \in I} \subset \cB(\cH)$ convergent in WOT to some $a \in \cB(\cH)$. Then we have for any $\xi, \eta \in \cH$, $$|\inner{\eta}{\alpha(a - a_i)\xi}| = |\inner{U^*\eta}{(a - a_i)U\xi}| \rightarrow 0$$ by definition of convergence in WOT. 
\end{proof}

\section{Von Neumann algebras}
Consider a $\rmC^*$-subalgebra $\cA \subseteq \cB(\cH)$ for some Hilbert space $\cH$. We define $\cA'$, called the \emph{commutant} of $\cA$, as the set $$\cA' := \{x \in \cB(\cH): [x, a] = 0 \text{ for all }a\in \cA\}$$
Notice that $I \in \cA'$. If $a,b \in \cA'$ then we also have $ab \in \cA'$. Similarly, if $a \in \cA'$ then $a^* \in \cA'$. $\cA'$ carries the norm-topology inherited from $\cB(\cH)$. Finally, if $(x_i)_{i \in I} \in \cA'$ is a net converging to $x \in \cB(\cH)$ in norm then $x \in \cA'$ as well. Thus $\cA'$ is a unital $\rmC^*$-algebra.

\begin{lem}
\label{lem:commutant is closed in WOT}
    For any subset $\cS \subseteq \cB(\cH)$, we have that $\cS'$ is closed in  WOT.
\end{lem}
\begin{proof}
    Consider a net $(x_i) \subset \cS'$ converging to $x \in \cB(\cH)$ in WOT. We note that left and right multiplication by a fixed element is continuous in WOT. So we have for all $a \in \cS$ and $\eta, \xi \in \cH$ and all chosen $\epsilon$ there exists $i_0$ such that for all $i \geq i_0$ we have $|\inner{\eta}{(x a - x_i a) \xi}| < \epsilon$ and $|\inner{\eta}{(a x - a x_i ) \xi}| < \epsilon$. Now we observe,
    \begin{align*}
        |\inner{\eta}{(x a - a x) \xi}| &\leq |\inner{\eta}{(x a - x_i a) \xi}| +  |\inner{\eta}{(x_i a - a x_i ) \xi}| + |\inner{\eta}{( a x_i - a x) \xi}|\\
        &= |\inner{\eta}{(x a - x_i a) \xi}| +  |\inner{\eta}{( a x_i - a x) \xi}| < 2 \epsilon
    \end{align*}
    where we have used that $[x_i, a ] = 0$. The result follows.
\end{proof}

\begin{defn}
    Consider a $\rmC^*$-subalgebra $\cM \subseteq \cB(\cH)$ for some Hilbert space $\cH$. We say $\cM$ is a \emph{von Neumann algebra} (we will shorten it to vN algebra) if it satisfies $$\cM = \cM'' := (\cM')'$$
    Here $(\cdot)''$ is called the bicommutant.
\end{defn}

If a subset $\cS \subseteq \cB(\cH)$ is closed under taking adjoints, then it is easily shown that $\cS'$ is a vN algebra. Of course, taking the commutant again, $\cS''$ is again a vN algebra. Obviously $\cS \subseteq \cS''$ by definition. In fact, $\cS''$ is the smallest vN algebra containing $\cS$.

Let $\cS_1, \cS_2 \subseteq \cB(\cH)$ be two $*$-subalgebras. Then we write $\cS_1 \vee \cS_2 := (\cS_1 \cup \cS_2)''$ and $\cS_1 \wedge \cS_2 := (\cS_1 \cap \cS_2)''$. The following are useful identities: $$(\cS_1 \cup \cS_2)' = \cS_1 ' \cap \cS_2 ' \qquad \qquad (\cS_1 \cap \cS_2)' \supset \cS_1' \vee \cS_2'$$

A core result in the study of vN algebras is the following theorem due to von Neumann, bridging the above algebraic definition with a more analytic one:

\begin{thm}[von Neumann bicommutant theorem]
\label{thm:von Neumann bicommutant theorem}
    The following statements are equivalent for a non-degenerate $*$-subalgebra $\cM \subseteq \cB(\cH)$: 
    \begin{enumerate}
        \item $\cM = \cM''$
        \item $\cM$ is SOT closed
        \item $\cM$ is WOT closed
    \end{enumerate}
\end{thm}

In particular, by Lemma \ref{lem:ordering of topologies on bounded ops on H}, $\cM$ is norm-closed and is thus a $\rmC^*$-algebra. This means that by demanding $\cM = \cM''$ we could have relaxed the condition that $\cM$ is a $\rmC^*$-algebra, and derived it instead. 

Theorem \ref{thm:von Neumann bicommutant theorem} and the above observation now indicate that we may equivalently define a vN algebra as a unital, WOT-closed $\rmC^*$-subalgebra $\cM \subset \cB(\cH)$.

As a special case, we have the following lemma for $\cB(\cH)$.

\begin{lem}
$\cB(\cH)$ is a vN algebra for any Hilbert space $\cH$.
\end{lem}
\begin{proof}
    We note that $\cB(\cH)$ is the whole space so by definition closed under any topology. By Theorem \ref{thm:von Neumann bicommutant theorem}, we have that $\cB(\cH) = \cB(\cH)''$ and $\cB(\cH)$ is thus a vN algebra.
\end{proof}

\begin{lem}
    For any set $\cS \subseteq \cB(\cH)$ that's closed under taking adjoints, we have $\cS''' := (\cS')'' = \cS'$
\end{lem}
\begin{proof}
    The set $\cS'$ is weakly closed (Lemma \ref{lem:commutant is closed in WOT}). It is thus a vN algebra. By theorem \ref{thm:von Neumann bicommutant theorem}, it implies that $\cS' = (\cS')''$.
\end{proof}

\begin{lem}
\label{lem:cone quasi-local algebra is dense in the cone vN algebra}
    For a non-degenerate $*$-subalgebra $\cA \subseteq \cB(\cH)$ we have that $\cA$ is WOT-dense in $\cA''$.
\end{lem}
\begin{proof}
    Consider the WOT closure $\overline{\cA}^{w}$ of $\cA$. By Theorem \ref{thm:von Neumann bicommutant theorem}, $\overline{\cA}^{w}$ is a vN algebra, i.e., $\overline{\cA}^{w} = \cA''$. Since $\cA$ is dense in $\overline{\cA}^{w}$ by construction, the result follows.
\end{proof}

\begin{thm}[Kaplansky density theorem]{\cite[Thm.~5.3.5]{kadison1986fundamentals}}
\label{thm:Kaplansky's density thm}
    Let $\cA \subset \cB(\cH)$ be a $\rmC^*$-subalgebra and $\cM = \overline{\cA}^{SOT}$ be the vN algebra it generates. Then the unit ball in $\cA$ is SOT-dense in the unit ball in $\cM$.

    In particular, for every element $x \in \cM$ there exists a bounded net $(a_i) \subset \cA$ converging in SOT to $x$ such that $\sup_i ||a_i|| \leq ||x||$. If $x \geq 0$, the net can be chosen such that each $a_i \geq 0$.
\end{thm}

vN algebras admit a canonical predual (hence ultraweak/ultrastrong topologies), have a dense set of projections, and support normal states and maps. These features make them especially convenient for infinite-volume quantum systems.

\subsection{Factors and classification}

Suppose $\cM$ is a vN algebra. It can be shown that $\cM$ is generated by a set of projections in $\cM$. Suppose $P, Q \in \cM$ are two projections. Then $P,Q$ are \emph{Murray-von Neumann} equivalent, $P \sim Q$, if there exists a partial isometry $V \in \cM$ with $V^* V = P$ and $VV^* = Q$. A projection $P$ is a subprojection of $Q$, written $P \leq Q$, if the range of $P$ is contained in the range of $Q$. Equivalently, if $P,Q$ satisfy $PQ = QP = P$.

\begin{defn}
    Let $P \in \cM$ be a projection. Then $P$ is called:
    \begin{itemize}
        \item \emph{finite} if for a projection $Q \in \cM$, $Q \leq P$ and $P \sim Q$ implies $P = Q$.
        \item \emph{infinite} if $P$ is not finite.
        \item \emph{properly infinite} if there is no finite projection $Q \in \cM$  with $Q \leq P$. 
        \item \emph{minimal} if requiring $0 \neq Q \leq P$ for some projection $Q \in \cM$ implies $Q=P$.
        \item \emph{central} if $P \in \cM \cap \cM'$. We call $\cM \cap \cM'$ as the \emph{center} of $\cM$.
    \end{itemize}
\end{defn}

\begin{defn}
    A non-degenerate vN algebra $\cM \subseteq \cB(\cH)$ is called:
    \begin{itemize}
        \item a \emph{factor} if $\cM \cap \cM' = \bbC I$. 
        \item \emph{finite/infinite/properly infinite} if the identity $I \in \cM$ is finite/infinite/properly infinite.
    \end{itemize}
\end{defn}

\begin{defn}
A factor $\cM$ is said to be:
\begin{itemize}
    \item type $I$ if there exist minimal projections in $\cM$ (equivalently $\cM \simeq \cB(\cH)$ for some Hilbert space $\cH$)
    \item type $II$ if there are no minimal projections but there exists at least one finite non-zero projection.
    \subitem type $II_1$ if the unit is finite.
    \subitem type $II_\infty$ if unit is infinite and $\cM \simeq \cN \otimes \cB(\cH)$ where $\cN$ is a type $II_1$ factor.
    \item type $III$ if every (non-zero) projection is properly infinite.
\end{itemize}
\end{defn}

Every factor $\cM$ is one of type $I/ II_1/ II_\infty/ III$. Every vN algebra $\cM$ can be uniquely written in the form $$\cM = Z_{I} \cM_1 Z_{I}\oplus Z_{II_1} \cM  Z_{II_1}\oplus Z_{II_\infty} \cM Z_{II_\infty} \oplus Z_{III} \cM Z_{III}$$ where $Z_{I},Z_{II_1},Z_{II_\infty},Z_{III} \in \cM$ are central projections adding up to the identity and somewhat suggestively, $Z_{II_\infty} \cM Z_{II_\infty}$ is a factor of type $II_\infty$ and similarly for the others.

\begin{rem}
    The classification of factors is very useful for endowing the category of anyons with the $\bbC$-linear structure (i.e., every object has subobjects and a direct sum of two objects is another object in the category.). In particular it turns out that on the Quantum Double models, the cone algebras are of type $II_\infty$, which allows us to 'fold a cone algebra into itself finitely many times'.
\end{rem}

\section{\texorpdfstring{$\rmC^*$-algebras on a lattice}{C*-algebras on a lattice}}
So far we've avoided talking about any geometry in our $\rmC^*$-algebras. But for applications in lattice systems, there is a definite notion of spatial and temporal locality. This will enable us to in particular talk about the quasi-local algebra, and cone vN algebra, which are central objects in the works considered in this thesis. We begin by talking about inductive limits. 

\begin{defn}
Consider a directed set $\Gamma$. An \emph{inductive system} is the set $$(\cA_\Gamma) :=\{(\cA_{\Lambda_1}, \iota_{\Lambda_1, \Lambda_2}): \Lambda_1, \Lambda_2 \subset \Gamma; \Lambda_1 \leq \Lambda_2\}$$ Here $\cA_{\Lambda_i}$ is a $\rmC^*$-algebra and $\iota_{\Lambda_1, \Lambda_2}: \cA_{\Lambda_1} \hookrightarrow \cA_{\Lambda_2}$ is an injective $*$-homomorphism satisfying $\iota_{\Lambda_1, \Lambda_2} \circ \iota_{\Lambda_2, \Lambda_3} = \iota_{\Lambda_1, \Lambda_3}$, $\iota_{\Lambda, \Lambda} = \Id$.
\end{defn}

We can take an inductive limit of this inductive system to obtain a $\rmC^*$-algebra $\cA$ as follows. First we take the algebraic direct sums of this system: $$\bigoplus_{\Lambda_i} \cA_{\Lambda_i} := \{(a_i)_{i \in I}: a_i \in \cA_{\Lambda_i}; a_i =0 \text{ for all but finitely many $i$}\}$$ with component-wise addition, multiplication, and adjoint. This turns $\bigoplus_{\Lambda_i} \cA_{\Lambda_i}$ into a $*$-algebra. Define for any $\Lambda_j$ the map $$\theta_j: \bigoplus_{\Lambda_i} \cA_{\Lambda_i} \rightarrow \cA_{\Lambda_j} \qquad \qquad \theta_j : (a_i)_i \mapsto \sum_{\Lambda_i \leq \Lambda_j} \iota_{\Lambda_i, \Lambda_j}(a_i)$$

We can define the seminorm\footnote{A seminorm is a norm but without the non-degeneracy condition} for any $x = (a_i)_i \in \bigoplus_{\Lambda_i} \cA_{\Lambda_i}$ as $p(x) := \sup_{j \in I} ||\theta_j(x)||$ which is well defined because for any $j$, we have $||\theta_j(x)|| \leq \sum_i ||a_i|| $. Taking the quotient $\cQ := \bigoplus_{\Lambda_i} \cA_{\Lambda_i} / N$ with respect to $N := \{x: p(x) = 0\}$, turns $p$ into a true norm on $\cQ$ and allows us to complete $\cQ$ into $\overline{\cQ}$ with respect to $p$, giving us a $\rmC^*$-algebra. 

This procedure satisfies certain universality properties, in the sense that different ways of defining $\theta_j$ are all equivalent.

We now construct a \emph{local net}, which is an inductive system imbued with (spatial) locality: if $\Lambda_1 \cap \Lambda_2 = \emptyset$ then every element of $\cA_{\Lambda_1}$ commutes with every element of $\cA_{\Lambda_2}$.

\subsection{Quasi-local algebra}
A \emph{Quantum Spin System (QSS)} is defined as follows. Consider a graph $\Gamma$ embedded in $\bbR^2$ (i.e., a 2d graph) consisting of edges, vertices, faces. For simplicity, the reader may want to keep on hand a simple graph like $\bbZ^2$. Subsets of $\Gamma$ are called \emph{regions}. We generically refer to an element of $\Gamma$ as a \emph{site}. On each site of $\Gamma$, we place a finite dimensional Hilbert space $\cH_s$. Let $\Lambda \subset_f \Gamma$ denote that $\Lambda$ is a finite subset of $\Gamma$. Regions $\Lambda \subset_f \Gamma$ are called \emph{local} regions. For each $\Lambda \subset_f \Gamma$, we define $\cH_{\Lambda} := \bigotimes_{s \in \Lambda} \cH_s$. We can also define $\cA_\Lambda := \cB(\cH_\Lambda)$ as the $\rmC^*$-algebra of local operators supported on region $\Lambda$. 

We define $(\Lambda) \subseteq \Gamma$ to be a set of increasingly bigger finite subsets of $\Gamma$ such that every finite subset of $\Gamma$ is eventually contained in an element of $(\Lambda)$. The set $(\Lambda)$ is directed with the $\leq$ operation being defined by inclusion: for any two subsets $\Lambda_1, \Lambda_2 \subseteq (\Lambda)$ we may define $\Lambda_1 \leq \Lambda_2$ if $\Lambda_1 \subseteq \Lambda_2$. Defining for every $\Lambda_1 \subseteq \Lambda_2$ the map $\iota_{\Lambda_1, \Lambda_2}: \cA_{\Lambda_1} \hookrightarrow \cA_{\Lambda_2}$ as the canonical embedding map $\cA_{\Lambda_1} \mapsto \cA_{\Lambda_1} \otimes 1_{\Lambda_2 \setminus \Lambda_1}$ (which is clearly a $*$-homomorphism satisfying $\iota_{\Lambda_1, \Lambda_2} \circ \iota_{\Lambda_2, \Lambda_3} = \iota_{\Lambda_1, \Lambda_3}$) enables us to define an inductive system. We can define the \emph{local algebra} $$\cA^{\loc}_\Gamma := \bigcup_{\Lambda \subset_f \Gamma} \cA_\Lambda$$ by identifying $\cA_{\Lambda_1} \subset \cA_{\Lambda_2}$ for $\Lambda_1 \leq \Lambda_2$ using the embedding map. $\cA^{\loc}_\Gamma$ is a $*$-algebra as we have not yet completed it. Completing it with respect to the norm of the inductive system defined as above will result in a $\rmC^*$-algebra $\cA_\Gamma$, called the \emph{quasi-local} algebra\footnote{In fact, since our embedding maps are injective (hence isometric), the semi-norm equals the operator norm on the nose and there is no need for taking the quotient. }. We write $$\cA_\Gamma := \overline{\cA^{\loc}_\Gamma}^{||\cdot||}$$

 For brevity, we denote $\cA := \cA_\Gamma$ and $\cA^{\loc} := \cA_\Gamma^{\loc}$. 

We can also take limits over a possibly infinite subgraph $\Omega \subseteq \Gamma$ in a similar manner, by defining an inductive system over $\Omega$ and following the above procedure. We call the resulting $*$-algebra the \emph{local algebra over $\Omega$}, denoted $\cA^{\loc}_\Omega$, and complete it in norm to obtain the \emph{quasi-local algebra over $\Omega$}, $\cA_\Omega$. We have $\cA_\Omega \subseteq \cA_\Gamma$, and in fact $\cA_\Gamma \simeq \cA_{\Omega} \otimes \cA_{\Gamma \setminus \Omega}$\footnote{To be precise, $\otimes$ here denotes $\otimes_{min}$ (c.f Section \ref{sec:tensors and sums of C^* algebras}).}.

An operator $a \in \cA$ has \emph{finite support} if there exists some $\Lambda \subset_f \Gamma$ such that for all $\Lambda' \subset \Gamma$ disjoint from $\Lambda$ and any $b \in \cA_{\Lambda'}$, we have $[a,b] = 0$. In this case, we say $\Lambda$ is the support of $a$ (or $a$ is supported in $\Lambda$) if $\Lambda$ is the smallest such subset.

\begin{rem}
\label{rem:norm completion or inductive limit}
    Instead of taking the inductive limit to obtain $\cA$, we could equivalently have defined $\cA$ as the completion of $\cA^{\loc}$ with respect to the operator norm in the usual way as the notation suggests. It turns out that the two definitions of $\cA$ are $*$-isomorphic, so the precise route of obtaining $\cA$ is unimportant.
\end{rem}

\subsection{Interactions, Hamiltonians, Dynamics}
Consider a $\rmC^*$-algebra $\cA$. We define \emph{dynamics} on $\cA$ as the pair $(\cA, \alpha_t)$ where $\alpha : \bbR \rightarrow \Aut(\cA)$ is a strongly continuous one-parameter group of automorphisms, i.e., $$\alpha_t \circ \alpha_s = \alpha_{t + s}, \qquad \alpha_0 = \Id, \qquad \lim_{t \rightarrow 0}||\alpha_t(A) - A || \rightarrow 0 \text{ for all } A \in \cA$$

A \emph{derivation} $\delta$ (called the \emph{generator} of $\alpha$) is defined on a suitable domain (i.e., where the norm limit exists) as $$\delta(a) := \lim_{t \rightarrow 0} \frac{\alpha_t (a) - a}{t} \qquad \qquad a \in \text{Dom}(\delta)$$ and satisfies $$\delta(ab) = a \delta(b) + \delta(a) b \qquad \qquad \delta(a)^* = \delta(a^*)$$ we note that $\delta$ is densely-defined on $\cA$, closed, and $\alpha_t$-invariant.

Usually in physics one talks about Heisenberg dynamics on the operators, which results from a Hamiltonian on a system dictating how the system behaves under time-evolution. Hamiltonians are ill-defined on the quasi-local algebra since they are unbounded operators. However, under suitable assumptions on the interaction terms of the Hamiltonian, it is still possible to derive well-defined dynamics on the entire quasi-local algebra. We now explore these connections by specializing to the case of the quasi-local algebra.

A (uniformly bounded) finite-range\footnote{This assumption can be relaxed for the following discussion to having a suitably ``nice'' decay with the size of the region.} interaction is defined as a map $\Phi: \cP_f(\Gamma) \rightarrow \cA$ where $\cP_f(\Gamma)$ is the set of finite subsets of $\Gamma$ and such that for each $\Lambda \subset \cP_f(\Gamma)$, $\Phi(\Lambda) \in \cA_\Lambda$ satisfies $\Phi \geq 0$ and $0$ if $\text{diam}(\Lambda) > r$\footnote{The diameter of any finite region $\Lambda$, {diam}($\Lambda$) is defined in the usual geometric way, i.e., as the size of the minimal ball that contains $\Lambda$ viewed as a subset of points in $\bbR^2$.}\footnote{Since $\Phi$ is uniformly finite-range, $r$ does not depend on $\Lambda$ and is the uniform upper bound.} for some $r \in \bbR_{>0}$. A (frustration-free) \emph{Hamiltonian} on some $\Lambda \subset_f \Gamma$ is defined as $$H_\Lambda := \sum_{X \in \cP_f(\Lambda)} \Phi(X)$$

A Hamiltonian $H_\Lambda$, due to its self-adjointness, defines dynamics $\alpha_t$ of $\cA_\Lambda$ by setting $$\alpha_t^\Lambda(\cdot) := e^{i t H_\Lambda} (\cdot) e^{ - i t H_\Lambda}$$ which is called Heisenberg dynamics in physics. For finite-range interactions, the limit $\lim_{\Lambda \uparrow \infty} \alpha_t^\Lambda$ is Cauchy uniformly for $t$ in compact intervals as $\Lambda \uparrow \infty$, and thus there exists a unique strongly-continuous dynamics $\alpha_t^\Phi$ on $\cA_{\loc}$ corresponding to interactions $\Phi$, which extends to $\cA$ by continuity. The generator for $\alpha_t^\Phi$ is then the derivation $\delta: \cA_{\loc} \rightarrow \cA_{\loc}$ given by $$\delta(A) := i \sum_{\Lambda \subset_f \Gamma}[\Phi(\Lambda), a]$$ in finite volume, this is exactly the Heisenberg equation $\dot{a} = i [H_\Lambda, a]$ whose solution for finite $t$ is the time evolution of $a$ given by $a(t) = e^{i t H_\Lambda} a e^{ - i t H_\Lambda}$ and exactly the same as the dynamics $\alpha_t^\Lambda(a)$.

Now we investigate the effect of dynamics $(\cA, \alpha_t)$ on an $\alpha_t$ invariant state $(\omega, \cA)$. Consider the GNS triple $(\pi_\omega, \cH_\omega, \Omega_\omega)$ of the state $\omega$. We notice that since $\omega$ is invariant under the action of $\alpha_t$, we have $\omega \circ \alpha_t = \omega$. By Lemma \ref{lem:invariance of state under a symmetry} there exists a unitary $U_t \in \cB(\cH_\omega)$\footnote{It can be shown that $U_t \in \pi_\omega(\cA)'' \subseteq \cB(\cH_\omega)$.} implementing $\alpha_t$, i.e., $\pi \circ \alpha_t (\cdot) = U_t \pi(\cdot) U_t^*$ with $U_t \Omega_\omega = \Omega_\omega$. By Stone's theorem \cite[Sec.~3.1]{bratteli2012}, there exists a (generally unbounded) self-adjoint operator $H_\omega$ on $\cH_\omega$\footnote{$H_\omega$ is affiliated with $\pi(\cA)''$, i.e., for every Borel set $B \subset \bbR$, the spectral projection $E^{H_\omega}(B)$ lies in $\pi(\cA)''$.} such that $U_t = e^{i t H_\omega}$. $H_\omega$ is the infinite-volume equivalent of the usual Hamiltonian in physics. We call $H_\omega$ the \emph{GNS Hamiltonian}.

A \emph{ground-state} of dynamics $(\cA,\alpha_t^\Phi)$ is a state $\omega$ such that its GNS Hamiltonian $H_\omega$ satisfies $H_\omega \geq 0$, $H_\omega \Omega_\omega = 0$. Equivalently, $\omega$ is a ground-state if and only if it satisfies the inequality $$-i \omega(a^* \delta_\Phi(a)) \geq 0 \qquad \qquad a \in \text{Dom}(\delta_\Phi)$$ Here $\delta_\Phi$ is the derivation generating $\alpha_t^\Phi$. The former condition is closer to the physics-level idea that the ground-state has minimal energy.

A ground-state $\omega$ is \emph{gapped} if $H_\omega$ has a spectral gap. Equivalently, $\omega$ is gapped if and only if it satisfies the Poincaré inequality $$-i \omega(a^* \delta_\Phi(a)) \geq \gamma (\omega(a^* a) - |\omega(a)|^2) \qquad \qquad a \in \text{Dom}(\delta_\Phi)$$

A ground-state $\omega$ for interactions $\Phi$ is \emph{frustration-free} if $\omega(\Phi(\Lambda)) = 0$ for all $\Lambda \in \cP_f(\Gamma)$.

\subsection{Symmetries}
Symmetries play an important role in physics. For a $\rmC^*$-algebra $\cA$, we define a symmetry $\beta : G \rightarrow \Aut(\cA)$ as $\beta: g \mapsto \beta_g$, with $\beta_1 = \Id$ and $\beta_g \circ \beta_h = \beta_{gh}$. We will also assume that $\beta$ is a faithful representation of $G$. 

Let $\omega \in \cS(\cA)$ be a state. The action of $g$ on $\omega$ is defined as precomposition by $\beta_{\bar g}$ where $\bar g$ is the inverse of $g$. A state $\omega$ is invariant if for all $g \in G$ we have $\omega \circ \beta_g = \omega$. By Lemma \ref{lem:invariance of state under a symmetry} we have that $\beta_g$ is implemented by a unitary $U_g$.

\begin{rem}
    We observe the dynamics $(\alpha_t, \cA)$ is a type of group action, called time-translation. 
\end{rem}

We ignore the anti-unitary symmetries like time-reversal to unburden ourselves from unnecessary complexity as they are not required for the works presented in this thesis.

We now specialize to the case of the quasi-local algebra $\cA$. For each site $s \in \Gamma$ we define $\beta_{g,s}$ as the symmetry action of $G$ on the algebra $\cA_s$ of site $s$, and $\beta_{g, \Lambda} := \bigotimes_{s \in \Lambda} \beta_{g,s} \in \Aut{(\cA_\Lambda)}$ for $\Lambda \subset_f \Gamma$. Since $\beta_{g,\Lambda}$ are consistent under inclusions of $\Lambda$, they extend to a symmetry action $\beta_g: G \rightarrow \cA$ on the entire quasi-local algebra $\cA$ defined on each $\cA_\Lambda$ as $\beta_g |_{\cA_\Lambda} := \beta_{g, \Lambda}$ called the \emph{on-site} action of $g$. Because the action of $\beta_g$ is on-site, we have that $\beta_g(\cA_\Lambda) = \cA_\Lambda$ for all $\Lambda \subset \Gamma$. 

An interaction $\Phi$ (hence the dynamics) on $\cA$ is invariant (under the action of $G$) if we have $\beta_g(\Phi(\Lambda)) = \Phi(\Lambda)$ for all $g \in G$ and $\Lambda \in \cP_f(\Gamma)$.

\subsection{Cone algebras}
A \emph{cone} $\Lambda(\theta_1, \theta_2, x_0) \subset \bbR^2$ with angles $\theta_1, \theta_2 \in [0,2\pi)$ satisfying $\theta_1 \neq \theta_2$, and point $x_0 \in \bbR^2$, is defined as the set $$\Lambda(\theta_1, \theta_2, x_0):= \{x \in \bbR^2 : \text{angle}(x - x_0) \in (\theta_1,\theta_2)\}$$ Here $(\theta_1, \theta_2)$ is understood to mean the interval of angles going counter-clockwise from $\theta_1$ to $\theta_2$. We simply write $\Lambda := \Lambda(\theta_1, \theta_2, x_0)$ for brevity. For a cone $\Lambda$ we denote by $\Lambda^c := \bbR^2 \setminus \overline\Lambda$ as its complement cone, where $\overline\Lambda$ is the closure of $\Lambda$ in the topology of $\bbR^2$.

A cone $\overline\Lambda \subset \Gamma$ corresponding to $\Lambda \subset \bbR^2$ is defined as follows. Notice that $\Gamma$ is a subset of $\bbR^2$ when each edge is assigned a point in $\bbR^2$ corresponding to the center of the edge, and similarly for a face and its corresponding geometric center. Then $\overline{\Lambda}$ is the subset of sites in $\Gamma$ that lie in $\Lambda$ when $\Gamma$ is taken as a subset of $\bbR^2$. We abuse notation for the sake of simplicity and denote $\Lambda \subset \Gamma$ when we mean $\overline{\Lambda}$.

We notice that $\cP_f(\Lambda)$ is a directed set with the $\leq$ operation once again being inclusion. As above, we take the inductive limit of the QSS defined on the cone $\Lambda$ and denote it by $\cA_\Lambda$\footnote{Here by $\cA_\Lambda$ we mean that the graph $\Gamma$ embedded in $\bbR^2$ we consider is the cone $\Lambda$ instead of the usual $\bbZ^2$.}. Consider now a representation $\pi: \cA_\Lambda \rightarrow \cB(\cH)$ (it may well be the GNS representation of some state $\omega$ on $\cA_\Lambda$). Since $\pi(\cA)$ defines a $*$-subalgebra of $\cB(\cH)$, we may define the vN algebra $$\cR^\pi_\Lambda := \pi(\cA_\Lambda)''$$ called the \emph{cone algebra}. If $\pi$ is non-degenerate, then by Theorem \ref{thm:von Neumann bicommutant theorem} we have $\cR^\pi_\Lambda$ is WOT-closed (and hence SOT-closed, norm-closed). By Lemma \ref{lem:cone quasi-local algebra is dense in the cone vN algebra} we have that $\pi(\cA_\Lambda)$ is WOT-dense in $\cR_\Lambda^\pi$.

\begin{rem}
    Cone algebras contain many useful operators not contained in the quasi-local algebra (for example the unitaries that implement equivalences of `anyons' (see Section \ref{sec:anyon selection rules}), or projections to a particular anyon sector) and are crucial to define a fusion categorical structure on the category of anyons (to be defined in Section \ref{sec:category of anyons}). 
\end{rem}

\begin{lem}{\cite[Lem.~5.3,3.5]{MR4362722}}
\label{lem:infinite factor}
    Let $\omega$ be a pure gapped ground-state of dynamics given by finite-range interactions with a uniform bound, and let $(\pi, \cH, \Omega)$ be its GNS triple. Then for any cone $\Lambda$, $\cR_\Lambda^\pi$ does not admit a normal tracial state. In particular, $\cR_\Lambda^\pi$ is either of type $II_\infty$ or type $III$, and is thus a properly infinite factor.
\end{lem}

\section{Anyon selection criterion}
\label{sec:anyon selection rules}
In principle there are many faithful representations of a given $\rmC^*$-algebra like the quasi-local algebra. Many of these representations are unphysical. For instance, there are representations with ``infinite energy'' or with infinitely many ``excitations''. We thus select physical representations by imposing suitable criteria to select for the physical phenomena that we wish to filter out. In this case, since we would like to describe an anyon (see Chapter \ref{chap:physics motivation} for physical characteristics of anyons), we impose the superselection criterion.

In the following definitions, we will frequently be talking about a reference state. This state is usually physically significant and is typically chosen to be a pure ground state. In case of ``nice models'' like the Quantum Double models, it is chosen to be the unique frustration free ground-state. 

Fix a reference state $\omega_0$ and denote its GNS triple as $(\pi_0, \cH_0, \Omega_0)$. Also denote the cone algebras as $\cR_\Lambda := \pi_0(\cA_\Lambda)''$ for some cone $\Lambda$. Since $\pi_0$ is faithful, we will identify $\cA$ with its image $\pi_0(\cA) \subset \cB(\cH_0)$ and $\cA_\Lambda$ with its image $\pi_0(\cA_\Lambda) \subset \cB(\cH_0)$ to avoid clutter whenever the context is clear.

\begin{defn}
\label{def:anyon sector}
A representation $\pi:\cA\rightarrow\cB(\cH)$ satisfies the \emph{anyon selection criterion} (with respect to GNS $(\pi_0,\cH_0,\Omega_0)$ of $\omega_0$) if for every cone $\Lambda$ there exists a unitary $U_\Lambda:\cH\rightarrow\cH_0$ such that $$U_\Lambda\,\pi(a)\,U_\Lambda^* = \pi_0(a) \qquad\text{for all } a\in \cA_{\Lambda^c}$$
Such a representation $\pi$ is called an \emph{anyon representation}, and its unitary equivalence class is an \emph{anyon sector}. 
\end{defn}

\begin{rem}
    A common rephrasing of Definition \ref{def:anyon sector} is that for all cones $\Lambda$ it satisfies $$\pi |_{\cA_{\Lambda^c}} \simeq \pi_0 |_{\cA_{\Lambda^c}}$$ where $\pi|_{\cA_\Lambda}$ is the restriction of $\pi$ to $\cA_\Lambda$.
\end{rem}

\begin{rem}
    In \cite{bols2025classification} the definition of an anyon sector additionally includes irreducibility. This was done primarily for the purposes of classification of all irreducible sectors. However, when building a general category of anyon sectors, such a requirement of ireducibility is detrimental as one cannot construct sub-objects in such a category.
\end{rem}

\begin{rem}
    The idea for this criterion is to select for representations that can `hide a half-infinite string operator', for which cones are very useful.
\end{rem}

\begin{rem}
    The anyon selection criterion has been traditionally called the superselection criterion in the literature. However we elect to use the above terminology to help clarify the etymology with respect to the `defect selection criterion' proposed in one of the works in this thesis and elaborated on below.
\end{rem}

\begin{lem}
\label{lem:anyon reps are unital}
    Any anyon representation $\pi$ is unital, hence non-degenerate.
\end{lem}
\begin{proof}
    Fix any cone $\Lambda$. By definition of an anyon representation, there exists a $U: \cH \rightarrow \cH_0$ such that $\pi(a) = U^* \pi_0(a) U$ for all $a \in \cA_{\Lambda^c}$. But the unit $1 \in \cA_{\Lambda^c}$ is shared with $1 \in \cA$, since the inclusion $\cA_{\Lambda^c}\hookrightarrow \cA$ is unital. Thus we have, $\pi(1) = U^* \pi_0(1) U$. Noting that $\pi_0$ is unital, we get the required result.
\end{proof}

In the definition of anyon selection criterion we didn't really need to consider cones, rather any poset satisfying certain axioms would do, as discussed in \cite{bhardwaj2024superselectionsectorsposetsvon}.

A very important technical condition necessary for the works in this thesis is called Haag duality.

\begin{defn}
\label{def:Haag duality}
    We say that \emph{Haag duality} holds for a reference state $\omega_0$ if for any cone $\Lambda$, the cone algebras satisfy $$\cR_\Lambda = \cR_{\Lambda^c}'$$
\end{defn}

This condition can be seen as the space-time equivalent of the locality principle, i.e., two observables outside each other's light cone commute. It can also be relaxed significantly, as we will see in Chapter \ref{chap:symmetry} or an even weaker property called approximate Haag duality \cite{MR4362722}.

\begin{defn}
    We define the \emph{auxiliary algebra}\footnote{Sometimes also called the \emph{allowed algebra}.} for the reference state $\omega$ and corresponding cone algebras $\cR_\Lambda$ as follows. Choose a forbidden direction $\theta \in [0,2\pi)$. We say a cone $\Lambda$ is \emph{allowed} if $\theta \notin (\theta_1, \theta_2)$ where $\theta_1, \theta_2$ are the two bounding angles of $\Lambda$. Let the set of all allowed cones be denoted as $\cL$. Then we define the auxiliary algebra as:

    $$\cA^a := \overline{\bigcup_{\Lambda \in \cL}\cR_\Lambda}^{||\cdot||}$$
\end{defn}

\begin{rem}
    In defining $\cA^a$ we took the norm completion. But equivalently we could have carried out the following procedure (c.f. Remark \ref{rem:norm completion or inductive limit}): we define an increasing set $(\cL)\subset \cL$ directed by inclusion with $(\cL) \uparrow \cL$. This gives us an inductive system with the canonical embedding maps $\iota_{\Lambda_1, \Lambda_2}: \cR_{\Lambda_1} \hookrightarrow \cR_{\Lambda_2}$ if $\Lambda_1\leq \Lambda_2$. We take the inductive limit to obtain the auxiliary algebra $\cA^a$.
\end{rem}

\begin{rem}
    In the above definition of auxilliary algebra, we made the choice of a forbidden direction. Readers uncomfortable with the axiom of choice may note that it is not necessary to choose a forbidden direction \cite{benini2026c}. The resulting structure is still braided $\rmC^*$-tensor category.
\end{rem}

\begin{lem}
\label{lem:aux alg is cstar, contains quasiloc alg, is subset of bounded ops}
    $\cA^a$ contains the quasi-local algebra $\cA$ and is a $\rmC^*$-subalgebra of $\cB(\cH_0)$.
\end{lem}
\begin{proof}
    Let us first show $\cA \subset \cA^a$. Consider some $a \in \cA$ with finite support. There exists an allowed cone $\Lambda$ which contains the support of $a$, so $a \in \cA_\Lambda$. Since $\cR_\Lambda$ by definition contains $\cA_\Lambda$, we have $a \in \cR_\Lambda$. Now consider a norm-convergent net $(a_i)_{i \in I} \subset \cA_{\loc}$ converging to $a \in \cA$ ($\cA_{\loc}$ is norm-dense in $\cA$). By above, there exists $\Lambda_i$ such that $a_i \in \cR_{\Lambda_i}$, and thus $a_i \in \cA^a$. By definition, $\cA^a$ is norm-closed, and thus $a \in \cA^a$.

    Now we show $\cA^a$ is a $\rmC^*$-subalgebra of $\cB(\cH_0)$. By definition, $\cR_\Lambda \subset \cB(\cH_0)$, so $\bigcup_{\Lambda \in \cL} \cR_\Lambda \subset \cB(\cH_0)$. Since $\cB(\cH_0)$ is norm-closed, we have $\cA^a \subset \cB(\cH_0)$. Since $\cA^a$ is norm-closed it is a Banach space. It is obviously a $\rmC^*$-algebra, with the $*$-algebra structure and the $\rmC^*$-property inherited from $\cB(\cH_0)$, the result follows.
\end{proof}

\begin{rem}
     Instead of choosing a forbidden direction, we may choose a forbidden cone $\Lambda_0$ such that $\cL$ consists of all cones $\Lambda$ such that there exists an $x \in \bbZ^2$ for which $\Lambda$ is disjoint from $\Lambda_0 +x$. The allowed algebra $\cA^0$ is then defined in the same way as $\cA^a$ but with the appropriate definition of $\cL$ (i.e., by replacing a forbidden direction with an interval of forbidden directions). 
 \end{rem}

We may also define another algebra $\cB^0$ as follows: $$\cB^0 := \overline{\bigcup_{x \in \bbZ^2} \cR_{\Lambda_0^c+x}}^{||\cdot||}$$ Many earlier works in the DHR-style AQFT analyses adopted the above alternate definition. However this definition is exactly the same as the auxiliary algebra definition, as the following lemma shows.

 \begin{lem}
     We have $\cB^0 = \cA^0$.
     \label{lem:two equivalent definitions of aux algebra}
 \end{lem}
 \begin{proof}
     Clearly $\Lambda_0^c \in \cL$ since $\Lambda_0^c \cap \Lambda_0 = \emptyset$. It follows that $\Lambda_0^c + x \in \cL$. Thus we have $\cB^0 \subset \cA^0$. On the other hand, consider an allowed cone $\Lambda \in \cL$. By definition, there exists some $x$ such that $\Lambda \subset \Lambda_0^c +x$. Thus $\cA^0 \subset \cB^0$. The lemma follows. 
 \end{proof}

\section{Localized, transportable endomorphisms}
\label{sec:localized, transportable endomorphisms}

In this section we will understand that anyon representations live as a special type of endomorphism of $\cA^a$. As a reminder, we've identified $\cA$ with $\pi_0(\cA)$ and $\cA_\Lambda$ with $\pi_0(\cA_\Lambda)$ due to the faithfulness of $\pi_0$.

\begin{rem}
\label{rem:why-normal}
In what follows we will occasionally need to extend $*$-homomorphisms from
$\cA_\Lambda$ to  $\cR_\Lambda$ and
to justify taking limits inside such extensions. For this it is convenient to
use the \emph{ultraweak} ($\sigma$-weak) topology on $\cB(\cH_0)$ and the associated
notion of \emph{normal} maps. Concretely, a linear map between vN algebras is
called \emph{normal} if it is ultraweakly continuous. We will only use the two
standard facts recorded below. We otherwise avoid developing the general theory.
\end{rem}

\begin{lem}
\label{lem:normal-extension-to-cone}
Let $\Lambda\subset \Gamma$ be a cone and let
$\psi:\cA_\Lambda \rightarrow \cB(\cH_0)$ be a unital bounded $*$-homomorphism.
Then $\psi$ extends uniquely to a normal $*$-homomorphism
$\overline\psi:\cR_\Lambda\rightarrow \cB(\cH_0)$.
\end{lem}
\begin{proof}
Standard. \cite[Lem.~2.2]{takesaki2003theory1} shows this result in a more general setting of enveloping vN algebras.
\end{proof}

\begin{lem}
    \label{lem:uw and WOT coincide on unit ball}
    On a vN algebra $\cM \subset \cB(\cH)$, the ultraweak topology and WOT coincide on the closed unit ball. Equivalently, also on any norm-bounded set.
\end{lem}
\begin{proof}
    Standard. Can be found in \cite[Lem.~2.5]{takesaki2003theory1} in a more general setting.
\end{proof}

\begin{lem}
\label{lem:normal maps are WOT cts on norm-bounded nets}
    A normal map is WOT-continuous on norm-bounded nets.
\end{lem}
\begin{proof}
    A normal map is by definition ultraweakly continuous. By Lemma \ref{lem:uw and WOT coincide on unit ball} the result follows.
\end{proof}

\begin{defn}
    A $*$-endomorphism $\rho \in \End(\cA^a)$ is called \emph{localized} in cone $\Lambda$ if for all $a \in \cA_{\Lambda^c} \subset \cA^a$ we have $\rho(a)= a$. $\rho$ is said to be \emph{transportable} if for all chosen cones $\Lambda'$, there exists an $*$-endomorphism $\sigma \in \End(\cA^a)$ localized in $\Lambda'$ such that $\sigma \simeq \rho$\footnote{By $\sigma \simeq \rho$ we mean that there exists a unitary $u \in \cA^a$ such that $\sigma = \Ad (u) \circ \rho$.}.
\end{defn}

\begin{thm}
\label{thm:there is a localized transportable endo for every anyon rep}
Assume Haag duality for cones. Let
$\pi:\cA\rightarrow\cB(\cH)$ be an anyon representation. Then there exists a unital $*$-endomorphism $\rho\in\End(\cA^a)$ such that $\rho\circ\pi_0 \simeq \pi$.
\end{thm}

\begin{proof}
\textbf{Step 1: Construction of $\rho_0$ on $\cA$.}
We keep $\pi_0$ explicit for this step to avoid ambiguity. By the anyon selection criterion, for every cone $\Lambda$ there is a unitary $U_\Lambda:\cH\rightarrow\cH_0$ such that $$\pi(a) = U_\Lambda^*\pi_0(a)U_\Lambda \qquad\text{for all }a\in\cA_{\Lambda^c}$$

We fix once and for all an allowed cone $\Lambda_0\in\cL$ and write $U := U_{\Lambda_0}$. Define $$\phi:\cA\rightarrow\cB(\cH_0), \qquad \phi(a) := U \pi(a) U^*$$
Then $\phi$ is a non-degenerate $*$-representation. Moreover, for all $a\in\cA_{\Lambda_0^c}$ we have $$\phi(a) = U\pi(a)U^* = \pi_0(a)$$ by the anyon selection criterion for $\Lambda_0$. We now define a $*$-homomorphism  $$\rho_0:\pi_0(\cA)\rightarrow\cB(\cH_0),\qquad \rho_0(\pi_0(a)) := \phi(a) = U\pi(a)U^*$$
This map is unital and contractive. 

\textbf{Step 2: Extension to cone algebras.}
We now restore the identification of $\cA,\cA_\Lambda$ with its image under $\pi_0$.
Fix any cone $\Lambda\in\cL$. With the standing identification of $\cA$ with its image $\pi_0(\cA)$, we will write simply
$\rho_0(a)=U\pi(a)U^*$ when no confusion can arise. Consider the restriction $$\rho_0|_{\cA_\Lambda}:\cA_\Lambda\rightarrow\cB(\cH_0),$$ which is a bounded $*$-homomorphism between $\rmC^*$-algebras, hence norm-continuous. Lemma \ref{lem:normal-extension-to-cone} now yields a unique normal $*$-homomorphism $$\rho_\Lambda:\cR_\Lambda\rightarrow\cB(\cH_0)$$ such that $$\rho_\Lambda(a) = \rho_0(a) \qquad\forall a\in\cA_\Lambda$$

Now let $\Lambda,\Sigma\in\cL$ with $\Lambda\subset\Sigma$. Then $\cA_\Lambda\subset\cA_\Sigma$ and $\cR_\Lambda\subset\cR_\Sigma$. On the common dense $*$-subalgebra $\cA_\Lambda \subset\cR_\Lambda$ we have $$\rho_\Lambda(a) = \rho_0(a) = \rho_\Sigma(a) \qquad\forall a\in\cA_\Lambda$$

By uniqueness of $\rho_\Lambda$ (Lemma \ref{lem:normal-extension-to-cone}), it follows that $$\rho_\Sigma|_{\cR_\Lambda} = \rho_\Lambda$$ Thus the family $\{\rho_\Lambda\}_{\Lambda \in \cL}$ is compatible along the inclusions $\cR_\Lambda\hookrightarrow\cR_\Sigma$ whenever $\Lambda\subset\Sigma$.

\textbf{Step 3: Definition of $\rho$ on $\cA^a$.}
For $x\in \bigcup_{\Lambda\in\cL}\cR_\Lambda$, choose $\Lambda$ with $x\in\cR_\Lambda$ and set $\rho(x):=\rho_\Lambda(x)$. This is well-defined, since if $x\in\cR_\Lambda\cap\cR_\Sigma$, we can choose $\Theta\in\cL$ with $\Lambda\cup\Sigma\subset\Theta$. Then by the compatibility $\rho_\Theta|_{\cR_\Lambda}=\rho_\Lambda$ and $\rho_\Theta|_{\cR_\Sigma}=\rho_\Sigma$, so $\rho_\Lambda(x)=\rho_\Theta(x)=\rho_\Sigma(x)$. Thus $\rho$ is a unital $*$-homomorphism on $\bigcup_{\Lambda\in\cL}\cR_\Lambda$, hence norm-bounded, and it extends uniquely by continuity to a unital $*$-homomorphism $\rho:\cA^a\rightarrow \cB(\cH_0)$.

\textbf{Step 4: The range of $\rho$ lies in $\cA^a$.}
We next prove that for each allowed cone $\Lambda\in\cL$, the map $\rho_\Lambda$ actually has range in some cone algebra $\cR_\Sigma$ for a suitable \emph{allowed} cone $\Sigma\in\cL$. This will imply $\rho(\cA^a)\subset\cA^a$.

Fix $\Lambda\in\cL$. Since $\cL$ is directed under inclusion, we can choose $\Sigma\in\cL$ such that $\Lambda_0\cup\Lambda \subset \Sigma$. Let $b\in\cA_\Lambda$ and $a\in\cA_{\Sigma^c}$. Then $\Sigma^c\subset\Lambda_0^c$, so $a\in\cA_{\Lambda_0^c}$ and hence $\rho_0(a) = \phi(a) = a$. Since $a,b$ have disjoint supports, we have $ab=ba$. Using that $\phi$ is a $*$-representation and $\phi(a)=a$, $\phi(b)=\rho_0(b)$, we obtain
$$a\rho_0(b) = \phi(a)\phi(b)= \phi(ab) = \phi(ba) = \phi(b)\phi(a) = \rho_0(b)a$$ 
So $\rho_0(b)$ commutes with $a$ for all $a\in\cA_{\Sigma^c}$, and therefore commutes with the vN algebra $\cR_{\Sigma^c}$. In other words, $\rho_0(b)\in\cR_{\Sigma^c}'$. By Haag duality for cones, $\cR_{\Sigma^c}'=\cR_\Sigma$, so $\rho_0(b)\in\cR_\Sigma$ for all $ b\in\cA_\Lambda$.

Since $\rho_\Lambda$ extends $\rho_0$ on $\cA_\Lambda$, we have \(\rho_\Lambda(b)=\rho_0(b)\in\cR_\Sigma\).

Now let $x\in\cR_\Lambda$. Choose a norm-bounded net $b_i\in\cA_\Lambda$ converging in SOT to $x$ by Theorem \ref{thm:Kaplansky's density thm}, hence also in WOT (Lemma \ref{lem:convergence in stronger topology implies convergence in weaker topology}). By normality of $\rho_\Lambda$, we get WOT-continuity of $\rho_\Lambda$ (Lemma \ref{lem:normal maps are WOT cts on norm-bounded nets}). Moreover $\cR_\Sigma$ is WOT-closed. Combining these facts we get, $$\rho_\Lambda(x) = \rho_\Lambda\bigl(\lim_i b_i\bigr) = \lim_i \rho_\Lambda( b_i) = \lim_i \rho_0(b_i) \in \cR_\Sigma$$ Thus for each $\Lambda\in\cL$ there is an allowed cone $\Sigma=\Sigma(\Lambda)\in\cL$ with $\rho_\Lambda(\cR_\Lambda)\subset\cR_\Sigma\subset\cA^a$.

Since $\cA^a$ is the $\rmC^*$-closure of the union of the images $\iota_\Lambda(\cR_\Lambda)$ and $\rho\circ\iota_\Lambda=\rho_\Lambda$ maps each $\cR_\Lambda$ into some cone algebra inside $\cA^a$, it follows that $\rho(\cA^a)\subset\cA^a$. Hence $\rho$ is indeed a unital $*$\nobreakdash-endomorphism of $\cA^a$ in the usual sense.

\textbf{Step 5: $\rho\circ\pi_0\simeq\pi$.}
Finally, for every $a\in\cA$ we have $$(\rho\circ\pi_0)(a) = \rho_0(\pi_0(a)) = \phi(a) = U \pi(a) U^*$$ Thus $$U^* (\rho\circ\pi_0)(a) U = \pi(a) \qquad\forall a\in\cA,$$ so $\pi$ and $\rho\circ\pi_0$ are unitarily equivalent as representations, with intertwiner $U^*$.
\end{proof}

\begin{prop}
\label{lem:anyon rep endo is localized and trans}
Assume Haag duality for cones and let $\pi$ and $\rho\in\End(\cA^a)$ be as in Theorem~\ref{thm:there is a localized transportable endo for every anyon rep}, constructed from the anyon representation $\pi$ using the distinguished cone $\Lambda_0$ and unitary $U_{\Lambda_0}$. Then $\rho$ is localized in $\Lambda_0$ and transportable.
\end{prop}

\begin{proof}
\textbf{Localization in $\Lambda_0$.}
By construction of $\rho_0$ and $\phi$, for $a\in\cA_{\Lambda_0^c}$ we have $\rho_0(a) = \phi(a) = a$. Moreover, for local $a$ supported in $\Lambda_0^c$, we can choose $\Lambda\in\cL$ with $\mathrm{supp}(a)\subset\Lambda$, so that $a\in\cR_\Lambda$ and $$\rho(a) = \rho_\Lambda(a) = \rho_0(a) = a$$ By norm-approximation of arbitrary $a\in\cA_{\Lambda_0^c}$ by such local observables and continuity of $\rho$, the equality $\rho(a)=a$ extends to all $a\in\cA_{\Lambda_0^c}$, showing localization in $\Lambda_0$.

\textbf{Transportability.}
Let $\Theta\in\cL$ be any other allowed cone. By the anyon selection criterion there exists a unitary $U_\Theta:\cH\rightarrow\cH_0$ such that $$\pi(a) = U_\Theta^* \pi_0(a) U_\Theta \qquad\forall a\in\cA_{\Theta^c}$$

Repeating the construction of Theorem~\ref{thm:there is a localized transportable endo for every anyon rep} with $U_\Theta$ in place of $U_{\Lambda_0}$, and repeating the range argument of Step~4 (with $\Theta$ in place of $\Lambda_0$), we obtain a $*$-endomorphism $\rho^{(\Theta)}\in\End(\cA^a)$, localized in $\Theta$ satisfying for all $\Lambda \in \cL$, $$\rho^{(\Theta)}_\Lambda(a) = U_\Theta \pi(a) U_\Theta^* \qquad\forall a\in\cA_\Lambda$$

Define $W_\Theta := U_\Theta U_{\Lambda_0}^* \in\cB(\cH_0)$. Then for every $a\in\cA$,
$$\rho^{(\Theta)}(a) = U_\Theta\pi(a)U_\Theta^* = W_\Theta  U_{\Lambda_0}\pi(a)U_{\Lambda_0}^* W_\Theta^* = W_\Theta \rho(a) W_\Theta^*$$

Moreover, by the anyon selection criterion and Haag duality for cones, the unitary $W_\Theta$ is localized in a cone $\Sigma \in \cL$ with $\Lambda_0 \cup \Theta \subset \Sigma$, so that $W_\Theta \in \cR_\Sigma \subset \cA^a$. 

Now fix a cone $\Lambda\in\cL$ and let $x\in\cR_\Lambda$. Choose a bounded net $a_i \in \cA_\Lambda$ converging in SOT to $x$. Using normality of $\rho_\Lambda$ and $\rho^{(\Theta)}_\Lambda$ on $\cR_\Lambda$, Lemma \ref{lem:normal maps are WOT cts on norm-bounded nets} and SOT-continuity of $\Ad(W_\Theta)$, we obtain
$$\rho^{(\Theta)}_\Lambda(x) = \lim_i \rho^{(\Theta)}_\Lambda(a_i) = \lim_i W_\Theta \rho_\Lambda(a_i) W_\Theta^* = W_\Theta \rho_\Lambda(x) W_\Theta^*$$
Thus, $\rho^{(\Theta)}_\Lambda = \Ad(W_\Theta)\circ\rho_\Lambda \text{ on } \cR_\Lambda\text{ for all }\Lambda\in\cL$.

By the norm-extension to $\rho$ and $\rho^{(\Theta)}$, this implies $$\rho^{(\Theta)} = \Ad(W_\Theta)\circ\rho \quad\text{on all of }\cA^a$$

In particular, $\rho^{(\Theta)}$ is unitarily equivalent to $\rho$, and it is localized in $\Theta$ by construction, showing transportability. 
\end{proof}

\begin{lem}
\label{lem:there is an anyon rep for every loc, trans endo of aux algebra}
    For every localized, transportable $\rho \in \End(\cA^a)$ there exists an anyon representation $\pi:\cA \rightarrow \cB(\cH_0)$ such that $\rho \circ \pi_0 \simeq \pi$.
\end{lem}
\begin{proof}
    For every cone $\Lambda$, since $\rho$ is transportable, there exists $\rho_\Lambda \simeq \rho$ with $U_\Lambda \in \cA^a$ implementing the equivalence, such that $\rho_\Lambda$ is localized in $\Lambda$.

    Since $\cA_{\Lambda^c} \subset \cA \subset \cA^a$, we have for all $a \in \cA_{\Lambda^c}$, $$\Ad(U_\Lambda) \circ \rho(a) = \rho_{\Lambda}(a) = a$$
    Thus the representation\footnote{the fact that $\rho \circ \pi_0$ is a representation of $\cA$ is easily verified by noting that $\rho|_{\cA}$ is a $*$-homomorphism to $\cA^a \subset \cB(\cH_0)$.} $ \pi := \rho \circ \pi_0: \cA \rightarrow \cB(\cH_0)$ is an anyon representation (trivially equivalent to $\rho \circ \pi_0$).
\end{proof}

\begin{cor}
\label{cor:for every anyon rep, there is an anyon rep to vacuum Hilbert space}
    For every anyon representation $\pi: \cA \rightarrow \cB(\cH)$ there exists an anyon representation $\sigma: \cA \rightarrow \cB(\cH_0)$ such that $\pi \simeq \sigma$.
\end{cor}
\begin{proof}
    We apply Theorem \ref{thm:there is a localized transportable endo for every anyon rep} for $\pi$ to get $\rho \in \End(\cA^a)$ such that $\pi \simeq \rho \circ \pi_0$. Then using Lemma \ref{lem:anyon rep endo is localized and trans} we get that $\rho$ is localized, transportable. Now we apply Lemma \ref{lem:there is an anyon rep for every loc, trans endo of aux algebra} to get an anyon representation $\sigma: \cA \rightarrow \cB(\cH_0)$ corresponding to $\rho$ such that $\sigma \simeq \rho \circ \pi_0$. Putting these results together, we have $\pi \simeq \sigma$. 
\end{proof}

Having shown in Theorem \ref{thm:there is a localized transportable endo for every anyon rep} that for every anyon representation we have the existence of a localized, transportable $*$-endomorphism of $\cA^a$ that essentially implements $\pi:\cA \rightarrow \cB(\cH)$ on $\cB(\cH_0)$, and in Lemma \ref{lem:there is an anyon rep for every loc, trans endo of aux algebra} that there is an anyon representation for every localized, transportable $*$-endomorphism of $\cA^a$, we get the fascinating result of Corollary \ref{cor:for every anyon rep, there is an anyon rep to vacuum Hilbert space}. Quite remarkably, the precise selection criterion was unimportant\footnote{i.e., the choice of starting framework (representations vs endomorphisms, different ambient algebras)}! We could have chosen arbitrary representations $\pi:\cA\rightarrow \cB(\cH)$, or fixed the Hilbert space to $\cH_0$ and considered only representation $\pi:\cA\rightarrow \cB(\cH_0)$. 

In fact, as the Theorem quite often uses, we could also have changed the $\rmC^*$-algebra in question! Instead of starting from the quasi-local algebra $\cA$, we could have considered representations of $\cA_\Lambda$, or of $\cR_\Lambda$, or of $\cA^a$ and it would still have given us exactly the same selection criterion.

As if this wasn't the end of the series of remarkable fairytale-esque coincidences, the paper \cite{bhardwaj2024superselectionsectorsposetsvon} shows that even conic regions are unimportant, and that one would have obtained the same anyon sectors had one started from a very general poset satisfying general geometric conditions.

\section{Category of anyons}
\label{sec:category of anyons}
Building on the discussion at the end of last section, we've essentially shown that if we want to study the category of anyon representations, we may equivalently study the category of localized, transportable $*$-endomorphisms of $\cA^a$. Going forward, the latter will be our working definition of a category of anyons, as expounded below.

We now identify $\cA$ with its image $\pi_0(\cA)$ to avoid notational clutter. We want to arrive at the result that the category of anyons is a braided $\rmC^*$-tensor category. Before proceeding, we make explicit the assumptions used in the construction of this category. 

We consider a reference representation $\pi_0: \cA \rightarrow \cB(\cH_0)$ arising from a state $\omega_0$ which is a gapped ground-state of dynamics given by finite-range interactions with a uniform bound. We assume also that Haag duality holds for $\pi_0$.

\begin{defn}
    We define the category $\DHR$ to have objects as $*$-endomorphisms of $\cA^a$ which are localized in some allowed cone and are transportable. For $\eta, \sigma \in \DHR$, the intertwiner space $$(\eta, \sigma) := \{T \in \cB(\cH_0): T \eta(A) = \sigma(A) T \,\, \forall \,\, A \in \cA^a\}$$

    We define the category $\DHR(\Lambda)$ to be the full subcategory of $\DHR$ having objects localized in an allowed cone $\Lambda \in \cL$.
\end{defn}

\begin{rem}
    We will show that the definition of $\DHR(\Lambda)$ does not depend on the particular chosen cone $\Lambda$ (upto $\rmC^*$-tensor equivalence, see Corollary \ref{cor:cone of DHR is irrelevant}). While all our proofs and constructions work with $\DHR$, we use $\DHR(\Lambda)$ to simplify proofs without having to explicitly address the cone of localization every time.
\end{rem}

\begin{rem}
    Importantly, we do not restrict transportability to just allowed cones $\Lambda$. This is because (cf.~discussion at the end of Lemma \ref{lem:anyon rep endo is localized and trans}) for any disallowed cone $\Lambda$ there exists an allowed cone $\Lambda'$ such that $\Lambda' \subset \Lambda$.
\end{rem}

Since $\eta, \sigma$ are localized in $\Lambda$, we have for all $a \in \cR_{\Lambda^c}$, $$T a = T \eta(a) = \sigma(a) T = a T$$ By Haag duality, it follows that $T \in \cR_{\Lambda} \subset \cA^a$. Hence for any $\eta,\sigma \in \DHR(\Lambda)$, the morphism space satisfies $(\eta, \sigma) \subset \cR_\Lambda$. In particular, for any $\eta \in \DHR(\Lambda)$, $(\eta, \eta)$ is a $\rmC^*$-subalgebra of $\cR_\Lambda$ (cf.~Lemma \ref{lem:DHR is a strict tensor cat}).

Recall Definition \ref{def:linear category and direct sums}. By definition $(\eta, \rho)$ is obviously a vector space for all $\eta,\rho \in \DHR(\Lambda)$, and the composition of morphisms is bilinear. Thus $\DHR(\Lambda)$ is a linear category. We will show in Lemma \ref{lem:DHR has direct sums and subobjects} that indeed we also have existence of direct sums.

\subsection{$\rmC^*$-tensor structure}
\begin{defn}
\label{def:DHR tensor product}
    We define $\otimes: \DHR(\Lambda) \times \DHR(\Lambda) \rightarrow \DHR(\Lambda)$, as follows. 
    \begin{align*}
        \otimes &: \eta, \sigma \mapsto \eta \circ \sigma \qquad \text{(denoted $\eta \otimes \sigma$)}\\ 
        \otimes &: U \in (\eta, \eta'),V \in (\sigma, \sigma') \mapsto U \eta(V) \in (\eta \otimes \sigma, \eta' \otimes \sigma')
    \end{align*}
\end{defn}

It is easily checked that if $\rho, \sigma \in \DHR(\Lambda)$ then $\rho \circ \sigma \in \DHR(\Lambda)$. Indeed, it is obviously a $*$-endomorphism of $\cA^a$. For any $a \in \cA_{\Lambda^c}$ we have $\rho \circ \sigma( \pi_0(a)) = \rho (\pi_0(a)) = \pi_0(a)$ so $\rho \circ \sigma$ is localized in $\Lambda$. For any cone $\Lambda'$, there exists $\rho',\sigma' \in \DHR(\Lambda')$ such that $\rho \simeq \rho'$ and $\sigma \simeq \sigma'$. Let $V,W$ be the intertwining unitaries respectively. Then $V \rho(W)$ is a unitary and intertwines $\rho \circ \sigma$ with $\rho' \circ \sigma'$. Thus $\rho \circ \sigma$ is transportable.

\begin{rem}
    In the above definition the reader may be slightly puzzled at the inclusion of $\eta$ in the definition of $U\otimes T$. The definition ensure that $U \otimes V$ indeed intertwines $\eta \circ \sigma$ with $\eta' \circ \sigma'$: Consider some $a \in \cA^a$. 
    $$U\eta(V) \eta(\sigma(a)) = U \eta(V\sigma(a)) = U\eta(\sigma'(a)V) = U\eta(\sigma'(a)) \eta(V) = \eta'(\sigma'(a) )U \eta(V)$$
    So indeed we have $U\eta(V) \in (\eta \otimes \sigma, \eta' \otimes \sigma')$.
\end{rem}
\begin{lem}
\label{lem:alternate definition of tensor of intertwiners}
    We have $U\eta(V) = \eta'(V)U$ where $U \in (\eta, \eta')$ and $V \in (\sigma, \sigma')$. In particular, this gives us an alternate definition of a tensor product of intertwiners: $U\otimes V:= \eta'(V) U$
\end{lem}
\begin{proof}
     Since for all $a \in \cA^a$ we have $U\eta(a) = \eta'(a)U$, we set $a = V$ (which is allowed since $V \in \cR_{\Lambda} \subset \cA^a$), giving us the result.
\end{proof}

\begin{lem}
\label{lem:DHR is a strict tensor cat}
    Recall Definition \ref{def:tensor category or monoidal category}.
    The tuple $(\DHR(\Lambda), \circ , \Id, \alpha, \lambda^L, \lambda^R)$ forms a (strict) tensor category. Here $\Id \in \End(\cA^a)$ is the tensor unit, $\alpha$ is the trivial associator since composition is associative.
\end{lem}
\begin{proof}
    The pentagon and triangle equations are trivially satisfied since $\alpha$ is the trivial associator (which also makes $\DHR(\Lambda)$ strict), and for any $\rho \in \DHR(\Lambda)$, the maps $\lambda^L_\rho, \lambda^R_\rho$ are also trivial since $\rho \circ \Id = \rho = \Id \circ \rho$. 
\end{proof}

Since after taking $\otimes$-functor to mean composition the rest of the data in the tuple of $\DHR(\Lambda)$ is trivial, we henceforth suppress it.

\begin{defn}
    We define the functor $*: \DHR(\Lambda) \rightarrow \DHR(\Lambda)$ as follows:
    \begin{align*}
        * &: \eta \mapsto \eta\\
        * &: T \in (\eta, \eta') \mapsto T^* \in (\eta', \eta)
    \end{align*}
    Here the $*$-operation is the usual adjoint action on $\cA^a$ inherited from $\cB(\cH_0)$.
\end{defn}

\begin{rem}
    The $*$-operation is really a $*$-functor on $\DHR(\Lambda)$. It is trivially checked that if $T \in (\eta,\eta')$ then $T^* \in (\eta', \eta)$, so $*: \DHR(\Lambda) \rightarrow \DHR(\Lambda)^{op}$. Moreover, the $*$-operation is contravariant, involutive, anti-linear (on morphism spaces), so it is indeed a $*$-functor on $\DHR(\Lambda)$. 
\end{rem}

\begin{lem}
\label{lem:DHR is a BTC}
    Recall Definition \ref{def:C^* category and C^* tensor category}. The category $\DHR(\Lambda)$, equipped with the $*$-functor, is a (strict) $\rmC^*$-tensor category.
\end{lem}
\begin{proof}
    By Lemma \ref{lem:DHR is a strict tensor cat} we only need to show that $\DHR(\Lambda)$ is a $\rmC^*$-category, and that $*$-functor is monoidal. 

    For any $\rho,\sigma \in \DHR(\Lambda)$, consider a map $\Phi_a: \cB(\cH_0) \rightarrow \cB(\cH_0)$ given by $\Phi: T \mapsto T\rho(a) - \sigma(a)T$ with $a \in \cA^a$. Then $(\rho,\sigma) =\bigcap_{a \in \cA^a}\ker{\Phi_a}$. Since $\ker{\Phi}$ is a closed set and $(\rho,\sigma)$ is an intersection of closed sets, $(\rho,\sigma)$ is a Banach space. In particular, for any $\rho\in \DHR(\Lambda)$,  $(\rho,\rho)\subset \cR_\Lambda$ is a $\rmC^*$-subalgebra. Also, we have that the norm inherited from $\cB(\cH_0)$ satisfies the $\rmC^*$-identity and composition is automatically contractive (recall Example \ref{eg:bdd ops on H}). Thus $\DHR(\Lambda)$ is a $\rmC^*$-category.

    Moreover, for $T \in (\eta,\eta')$ and $S \in (\sigma, \sigma')$ we have $$(T\otimes S)^* = (T\eta(S))^* = (\eta(S))^* T^* = \eta(S^*) T^* = T^* \eta'(S^*) = T^* \otimes S^*$$ where in the last equality we used Lemma \ref{lem:alternate definition of tensor of intertwiners}. So the $*$-functor is monoidal. This gives us the required result.
\end{proof}

\begin{lem}
\label{lem:DHR has direct sums and subobjects}
    The category $\DHR(\Lambda)$ has subobjects and direct sums.
\end{lem}
\begin{proof}
Shown in \cite{MR4362722}. We write the proof for convenience.

    \textbf{Direct sums}: Since $\cR_\Lambda$ is a properly infinite factor, there exist projections $p \in \cR_\Lambda$ such that $p \leq 1$ and $p \sim (1 - p) \sim 1$ where $\sim$ is Murray-von Neumann equivalence \cite[Lemma 6.3.3]{MR1468230}. It follows that one can find isometries $v, w \in \cR_\Lambda$ such that $v^*v = w^*w = 1$, $v^*w = w^*v = 0$ and $vv^* = p, ww^* = (1-p)$.
    
    We then construct a direct-sum as follows: for $\mu, \sigma \in \DHR(\Lambda)$ choose the isometries $v,w$ as above. Then, $$(\mu \oplus \sigma) (a) := v \mu(a) v^* + w \sigma(a) w^* \qquad a \in \cA^a$$
    It is easily verified that $\mu \oplus \sigma \in \DHR(\Lambda)$. In particular, $\mu \oplus \sigma$ is a unital $*$-endomorphism of $\cA^a$. Indeed for all $a \in \cA^a$ we have:
    \begin{align*}
        (\mu \oplus \sigma)(a) (\mu \oplus \sigma)(b) &= (v \mu(a) v^* + w \sigma(a) w^*)(v \mu(b) v^* + w \sigma(b) w^*)\\
        &= v\mu(ab)v^* + w\sigma(ab)w^* = (\mu \oplus \sigma)(ab)
    \end{align*}
    Unitality, $*$-property are immediate.

    The direct sum is well-defined because for a different set of mutually orthogonal isometries $v',w' \in \cR_\Lambda$ resulting in the direct sum $\oplus'$, it is easily shown that $\mu \oplus \sigma \simeq \mu \oplus' \sigma$ with the unitary $v' v^* + w' w^*$ implementing the equivalence.

    $\mu \oplus \sigma$ is localized in $\Lambda$ using the localization of $\mu,\sigma$, the fact that $v,w \in \cR_\Lambda = \cR_{\Lambda^c}'$ (by Haag duality) and that $vv^* + ww^* =1$. It is also transportable. Indeed, for a cone $\Lambda'$, there exist $\mu', \sigma' \in \DHR(\Lambda')$ with unitaries $V: \mu \rightarrow \mu'$ and $W: \sigma \rightarrow \sigma'$ implementing the equivalence respectively. Then we define $\mu' \oplus \sigma'$ using the tuple of mutually orthogonal isometries $\{v',w' \in \cR_{\Lambda'}\}$. The unitary $U = v' V v^* + w' W w^*$ then intertwines $\mu \oplus \sigma$ with $\mu' \oplus \sigma'$, which is a straightforward check.

     \textbf{Subobjects}: Let $\mu \in \DHR(\Lambda)$ and let $p \in (\mu,\mu)$ be any non-zero projection. Then by \cite[Lem.~5.8]{MR4362722} we have that $ p \sim 1 \in \cR_\Lambda$. Thus there exists an isometry $v \in \cR_\Lambda$ such that $vv^* = p$. We now construct the object $\sigma (\cdot) = v^* \mu(\cdot) v$. 

    The map $\sigma$ is straightforwardly a $*$-endomorphism of $\cA^a$. Let us check for example the morphism property. For any $a,b \in \cA^a$, $$\sigma(a)\sigma(b) = v^* \mu(a) v v^* \mu(b) v = v^* \mu(a) p \mu(b) v = v^* p \mu(a) \mu(b) v = v^*p \mu(ab) v = v^* \mu(ab) v = \sigma(ab)$$ Here we have used that $p \in (\mu,\mu)$ to obtain $p\mu(a) = \mu(a) p$, and the identity $v^*p = v^* v v^* = v^*$. 

    We verify that $\sigma$ is localized in $\Lambda$. Consider $a \in \cA_{\Lambda^c}$. Then\footnote{recall the identification of $\cA_\Lambda$ with its image under $\pi_0$}, $$\sigma (a) = v^* \mu(a) v = v^* \pi_0(a) v = a$$
    where we used that $v \in \cR_\Lambda = \cR_{\Lambda^c}'$ and $\cA_{\Lambda^c} \subset \cR_{\Lambda^c}$ to get $a v = v a$.

    We verify that $\sigma$ is transportable. We observe $\mu$ is transportable. Consider a cone $\Lambda'$ and some $\mu' \simeq \mu$ localized in $\Lambda'$ and the unitary $U: \mu \rightarrow \mu'$ implementing the equivalence. Define $\sigma' (\cdot) := U \sigma(\cdot) U^*$. By definition, $\sigma' \simeq \sigma$. We note that $U v U^* \in \cR_{\Lambda'}$ by observing that $UvU^* \in (\mu',\mu')\subset \cR_{\Lambda'}$. 
    
    Now for all $a \in \cA_{(\Lambda')^c}$ we have,
    \begin{align*}
        \sigma'(a) &= U v \mu(a) v^* U^* = U v U^* U \mu(a) U^* U v^* U^*\\
        &= U v U^* \mu'(a) Uv^* U^* = UvU^* a U v^* U^* = a
    \end{align*}
    where we have used $U vU^* \subset \cR_{\Lambda'}$ to get $[UvU^* , a] = 0$.

    Since $\sigma$ as constructed is a localized (in $\Lambda$), transportable endomorphism of $\cA^a$, it follows that $\sigma \in \DHR(\Lambda)$.

    Now we verify that $\sigma$ is indeed a subobject of $\mu$. Indeed for all $a \in \cA^a$, $$ v \sigma(a) = vv^* \mu(a) v = p \mu(a) v = \mu(a) p v = \mu(a) v$$ Where we used that $p \in (\mu, \mu)$, and $pv = vv^* v = v$. Thus $v \in (\sigma, \mu)$, but since $v$ is an isometric intertwiner, $\sigma$ is indeed a subobject of $\mu$, completing the proof.
\end{proof}

\subsection{The full subcategory $\DHR_0(\Lambda)$}
It is standard that one can always construct a ($\rmC^*$-)tensor equivalent full skeletal subcategory starting from a ($\rmC^*$-)tensor category (see e.g. \cite[Exercise~2.8.8]{etingof2015tensor}). We now explicitly spell out this construction for the case of $\DHR(\Lambda)$. 
\begin{defn}
    We define $\DHR_0(\Lambda)$ to be a full skeletal subcategory of $\DHR(\Lambda)$ by fixing a representative for each equivalence class of objects in $\DHR(\Lambda)$ and importing the hom spaces from $\DHR(\Lambda)$. In particular, if $\rho_0, \rho_0' \in \DHR_0(\Lambda)$ and $\rho_0 \simeq \rho_0'$, then $\rho_0 = \rho_0'$.
\end{defn}

We now fix the unitary intertwiners $T_\rho: \rho \rightarrow \rho_0$ for each object $\rho \in \DHR(\Lambda)$ and the representative $\rho_0\in \DHR_0(\Lambda)$ for the equivalence class $[\rho]$. We observe $T_{\Id} = 1$ up to a complex phase since by definition, $\Id \in \DHR(\Lambda)$ is localized in $\Lambda$. Now since $(\Id, \Id)$ is the set of all operators $T \in \cR_\Lambda$ that commute with all $a \in \cR_\Lambda$ and $\cR_\Lambda$ is assumed to be a factor, it follows that $(\Id, \Id) = \bbC 1$. Since $T$ was assumed unitary, it follows that $T = c1$ with $|c| = 1$. We now fix $c = 1$ for all $T_{\rho_0}$ with $\rho_0 \in \DHR_0(\Lambda)$. In particular, we also fix $T_{\Id} = 1$.

We define the tensor functor $\otimes_0$ on $\DHR_0(\Lambda)$ by setting for all $\rho_0,\sigma_0 \in \DHR_0(\Lambda)$, $$\rho_0 \otimes_0 \sigma_0 := T_{\rho_0 \otimes \sigma_0}  ({\rho_0 \circ \sigma_0}) T_{\rho_0 \otimes \sigma_0}^*$$ 

Now since $\DHR_0(\Lambda)$ was built from $\DHR(\Lambda)$ by fixing a representative of each equivalence class, for all $V \in (\rho_0,\rho_0')$, $W \in (\sigma_0,\sigma_0')$ setting $$V \otimes_0 W := T_{\rho_0' \otimes \sigma_0'} (V \rho_0(W)) T_{\rho_0 \otimes \sigma_0}^*$$ gives the tensor product on the morphisms.

We notice that $\rho_0 \otimes_0 \Id = \Ad[T_{\rho_0 \otimes \Id}] (\rho_0 \otimes \Id) = \Ad[T_{\rho_0}] (\rho_0) = \rho_0$, implying that the right unitor is always trivial. Similarly for the left unitor.

The $*$-functor on $\DHR_0(\Lambda)$ is inherited from $\DHR(\Lambda)$ since $\DHR_0(\Lambda)$ is full and $*$-functor only acts non-trivially on morphisms.

\begin{lem}
\label{lem:DHR0 is a BTC}
    The tuple $(\DHR_0(\Lambda), \otimes_0, \Id, \alpha^0, \lambda^L, \lambda^R)$ with the $*$-functor\ is a $\rmC^*$-tensor category. Here the associator $\alpha^0$ is given by $$\alpha^0_{\rho_0,\sigma_0,\mu_0} = T_{\rho_0\otimes(\sigma_0\otimes_0 \mu_0)} \rho_0(T_{\sigma_0 \otimes \mu_0}) (T_{\rho_0 \otimes \sigma_0}\otimes \Id_{\mu_0})^* T_{(\rho_0\otimes_0\sigma_0)\otimes \mu_0}^*,$$ and $\lambda^L, \lambda^R$ are the trivial unitors.
\end{lem}
\begin{proof}
    First we verify that $\alpha^0$ is indeed an associator:
        \begin{align*}
        & \alpha^0_{\rho_0, \sigma_0, \mu_0} \left((\rho_0\otimes_0 \sigma_0) \otimes_0 \mu_0\right)\\
        &=T_{\rho_0\otimes(\sigma_0\otimes_0 \mu_0)} \rho_0(T_{\sigma_0 \otimes \mu_0}) (T_{\rho_0 \otimes \sigma_0}\otimes 1_{\mu_0})^* T_{(\rho_0\otimes_0\sigma_0)\otimes \mu_0}^* \left(\Ad[T_{(\rho_0\otimes_0\sigma_0)\otimes \mu_0}](\Ad[T_{\rho_0 \otimes \sigma_0}](\rho_0 \circ \sigma_0) \circ \mu_0) \right)\\
        &=T_{\rho_0\otimes(\sigma_0\otimes_0 \mu_0)} \rho_0(T_{\sigma_0 \otimes \mu_0}) T_{\rho_0 \otimes \sigma_0}^*  (\Ad[T_{\rho_0 \otimes \sigma_0}](\rho_0 \circ \sigma_0) \circ \mu_0) T_{(\rho_0\otimes_0\sigma_0)\otimes \mu_0}^* \\
        &=T_{\rho_0\otimes(\sigma_0\otimes_0 \mu_0)} \rho_0(T_{\sigma_0 \otimes \mu_0})  ((\rho_0 \circ \sigma_0) \circ \mu_0)  T_{\rho_0 \otimes \sigma_0}^* T_{(\rho_0\otimes_0\sigma_0)\otimes \mu_0}^* \\
        &=T_{\rho_0\otimes(\sigma_0\otimes_0 \mu_0)} \rho_0(T_{\sigma_0 \otimes \mu_0})  (\rho_0 \circ (\sigma_0 \circ \mu_0))  T_{\rho_0 \otimes \sigma_0}^* T_{(\rho_0\otimes_0\sigma_0)\otimes \mu_0}^* \\
        &=T_{\rho_0\otimes(\sigma_0\otimes_0 \mu_0)}  (\rho_0 \circ \Ad[T_{\sigma_0 \otimes \mu_0}](\sigma_0 \circ \mu_0)) \rho_0(T_{\sigma_0 \otimes \mu_0})  T_{\rho_0 \otimes \sigma_0}^* T_{(\rho_0\otimes_0\sigma_0)\otimes \mu_0}^* \\
        &=\left(\Ad[T_{\rho_0\otimes(\sigma_0\otimes_0 \mu_0)}]  (\rho_0 \circ \Ad[T_{\sigma_0 \otimes \mu_0} ](\sigma_0 \circ \mu_0)) \right)T_{\rho_0\otimes(\sigma_0\otimes_0 \mu_0)} \rho_0(T_{\sigma_0 \otimes \mu_0})  (T_{\rho_0 \otimes \sigma_0}\otimes 1_{\mu_0})^* T_{(\rho_0\otimes_0\sigma_0)\otimes \mu_0}^*\\
        &=  \left(\rho_0\otimes_0 (\sigma_0 \otimes_0 \mu_0)\right) \alpha^0_{\rho_0, \sigma_0, \mu_0}
    \end{align*}
    Since $\alpha$ is trivial, it follows that the pentagon equation is satisfied by the coherence of $T$'s. Since the unitors are trivial by design, the triangle equation is trivially satisfied. 
    
    We now observe that the $*$-functor satisfies for $V \in (\rho_0,\rho_0')$ and $W \in (\sigma_0,\sigma_0')$,
    \begin{align*}
        (V \otimes_0 W)^* &= \left(T_{\rho_0'\otimes \sigma_0'}(V\rho_0(W))T_{\rho_0\otimes \sigma_0}^*\right)^* = T_{\rho_0\otimes \sigma_0}(\rho_0(W^*)V^*)T_{\rho_0'\otimes \sigma_0'}^* \\
        &=T_{\rho_0\otimes \sigma_0}(V^* \rho_0'(W^*))T_{\rho_0'\otimes \sigma_0'}^* = V^* \otimes_0 W^*
    \end{align*}
    Showing that the $*$-functor is indeed monoidal. The result thus follows. 
\end{proof}

\begin{lem}
\label{lem:tensor equivalence of DHR and DHR0}
    $\DHR_0(\Lambda)$ is $\rmC^*$-tensor equivalent to $\DHR(\Lambda)$.
\end{lem}
\begin{proof}
    To prove equivalence, we consider the inclusion functor $F: \DHR_0(\Lambda)\rightarrow \DHR(\Lambda)$. The functor $F$ is fully faithful by design and essentially surjective since every object in $\DHR(\Lambda)$ is equivalent to its representative in $\DHR_0(\Lambda)$, which makes it an equivalence.

Define the natural transformations $$\phi_{\rho_0,\sigma_0}:F(\rho_0)\otimes F(\sigma_0) \rightarrow F(\rho_0\otimes_0 \sigma_0),\qquad \phi_0: \Id \rightarrow F(\Id),$$ by assigning for all objects $\rho_0,\sigma_0$ of $\DHR_0(\Lambda)$, $$\phi_{\rho_0,\sigma_0} = T_{\rho_0\otimes \sigma_0},\qquad \phi_0 = 1$$

Since $T_{\rho_0 \otimes \sigma_0}$ is chosen to be a unitary equivalence, it is readily verified that $\phi_{\rho_0,\sigma_0}$ is a natural isomorphism. And $\phi_0$ is is trivially an isomorphism.

We verify that the hexagon equation holds:
\begin{align*}
F(\alpha^0_{\rho_0,\sigma_0,\mu_0})\phi_{\rho_0\otimes_0 \sigma_0,\mu_0}(\phi_{\rho_0,\sigma_0} \otimes 1_{\mu_0}) &= \alpha_{\rho_0,\sigma_0,\mu_0}\phi_{\rho_0\otimes_0 \sigma_0,\mu_0}\phi_{\rho_0,\sigma_0}\\
&= T_{\rho_0\otimes(\sigma_0\otimes_0 \mu_0)} \rho_0(T_{\sigma_0 \otimes \mu_0}) T_{\rho_0 \otimes \sigma_0}^* T_{(\rho_0\otimes_0\sigma_0)\otimes \mu_0}^* T_{(\rho_0 \otimes_0 \sigma_0)\otimes \mu_0} T_{\rho_0 \otimes \sigma_0}\\
&= T_{\rho_0\otimes(\sigma_0\otimes_0 \mu_0)} \rho_0(T_{\sigma_0 \otimes \mu_0})\\
&= T_{\rho_0\otimes(\sigma_0\otimes_0 \mu_0)} (1_{\rho_0} \circ \rho_0(T_{\sigma_0 \otimes \mu_0}))\\
&= \phi_{\rho_0, \sigma_0 \otimes_0 \mu_0} (1_{\rho_0} \otimes \phi_{\sigma_0,\mu_0}) = \phi_{\rho_0, \sigma_0 \otimes_0 \mu_0} (1_{\rho_0} \otimes \phi_{\sigma_0,\mu_0}) 
\end{align*}
Which is the correct equation to show since $\alpha$ is trivial.

Showing that $F$ is indeed a tensor functor. Since $F$ is also an equivalence, it follows that $F$ is a tensor equivalence. Thus we have that $\DHR(\Lambda)$ is tensor equivalent to $\DHR_0(\Lambda)$.

Fullness of $\DHR_0(\Lambda)$ gives the same $*$-functor and identical morphism spaces as $\DHR(\Lambda)$. Hence the inclusion functor $F$ is compatible with the $*$-functor, i.e., $F(V)^* = F(V^*)$ for all $V \in (\rho_0,\rho_0)$ and is isometric on the morphism spaces (hence positive). Since the tensorator and unit isomorphism are unitary maps due to the unitarity of T maps, $F$ is actually a $\rmC^*$-tensor equivalence. It follows that $F$ is a $\rmC^*$-tensor equivalence.
\end{proof}

\begin{cor}
\label{cor:cone of DHR is irrelevant}
    For cones $\Lambda_1, \Lambda_2$ with $\Lambda_1 \subset \Lambda_2$, we have $\DHR(\Lambda_1)$ is $\rmC^*$-tensor equivalent to $\DHR(\Lambda_2)$. 
\end{cor}
\begin{proof}
    We use Lemma \ref{lem:tensor equivalence of DHR and DHR0} twice (once for $\DHR(\Lambda_1)$, once for $\DHR(\Lambda_2)$) to obtain the required result. 
\end{proof}

\subsection{Braided structure}
\label{sec:braiding of anyons}
We can imbue $\DHR(\Lambda)$ with a braiding. This style of braiding was first done in the AQFT literature by Doplicher, Haag, Roberts. A braiding isomorphism $c: \DHR(\Lambda) \times \DHR(\Lambda) \rightarrow \DHR(\Lambda) \times \DHR(\Lambda)$ is a $\rmC^*$-natural isomorphism that swaps the tensor factors: $\otimes \Rightarrow \otimes^{\opp}$. In many treatments the tensor functor is usually a geometric object, with a well-defined notion of left/right given by the position in the tensor factor. However in the DHR treatment, since the tensor is given by composition, there is a natural ``temporal'' geometry associated with it. The object in the first tensor factor acts after the object in the second tensor factor. Let us define this ``temporal braiding'' after observing the following useful result (which is the lattice analogue of ``locality''):

\begin{lem}
    \label{lem:switch order of composition for disjoint cones}
    Let $\rho \in \DHR(\Lambda)$ and $\sigma \in \DHR(\Lambda')$ where $\Lambda' $ is disjoint from $\Lambda$. Then we have, $$\rho \otimes \sigma = \sigma \otimes \rho$$
\end{lem}

\begin{proof}
Fix some allowed cone $\Sigma \in \cL$. For every finite region $X \subset \Sigma$, we define 
    $X = X_1 \sqcup X_2 \sqcup X_3$, 
  where $X_1 := X\cap \Lambda$, $X_2 := X\cap \Lambda'$, and
  $X_3 := X\setminus (X_1\cup X_2)$.
  Then we have a canonical tensor product decomposition of the local algebra
    $\cA_X \simeq \cA_{X_1} \otimes \cA_{X_2} \otimes \cA_{X_3}$. Every $a\in\cA_X$ can therefore be written as a finite sum
    $a = \sum_{j=1}^n a_j^{(1)} a_j^{(2)} a_j^{(3)}$ with $a_j^{(k)} \in \cA_{X_k}$.

  By localization, $\rho$ acts trivially on any observable supported in
  $\Lambda^c$, and $\sigma$ acts trivially on any observable supported in
  $(\Lambda')^c$. Since $\Lambda$ and $\Lambda'$ are disjoint, we have $X_1 \subset (\Lambda')^c, X_2 \subset \Lambda^c$,
  and hence
  \begin{align*}
    \sigma(b) = b \quad \text{for all } b\in \cA_{X_1},\qquad \rho(c)  = c \quad \text{for all } c\in \cA_{X_2}.
  \end{align*}
  Moreover, both $\rho$ and $\sigma$ act trivially on $\cA_{X_3}$, since
  $X_3 \subset \Lambda^c\cap (\Lambda')^c$.

  Now let us compute $\rho\circ\sigma$ and $\sigma\circ\rho$ on a simple
  tensor $a_j^{(1)} a_j^{(2)} a_j^{(3)}$. We get
  \begin{align*}
    \rho\circ\sigma(a_j^{(1)} a_j^{(2)} a_j^{(3)})
      &= \rho\big( \sigma(a_j^{(1)}) \sigma(a_j^{(2)}) \sigma(a_j^{(3)}) \big) = \rho\big( a_j^{(1)} \sigma(a_j^{(2)}) a_j^{(3)} \big) \\
      &= \rho(a_j^{(1)}) \rho(\sigma(a_j^{(2)})) \rho(a_j^{(3)}) = \rho(a_j^{(1)}) \sigma(a_j^{(2)}) a_j^{(3)},
  \end{align*}
  where we used that $\rho$ acts trivially on $a_j^{(3)}$, and that
$\sigma(a_j^{(2)}) \in \cR_{\Lambda'} \subset \cR_{\Lambda^c}$ while $\rho$
acts as the identity on $\cR_{\Lambda^c}$, so $\rho(\sigma(a_j^{(2)})) = \sigma(a_j^{(2)})$. Similarly,
  \begin{align*}
    \sigma\circ\rho(a_j^{(1)} a_j^{(2)} a_j^{(3)})
      &= \sigma\big( \rho(a_j^{(1)}) \rho(a_j^{(2)}) \rho(a_j^{(3)}) \big) = \sigma\big( \rho(a_j^{(1)}) a_j^{(2)} a_j^{(3)} \big) \\
      &= \sigma(\rho(a_j^{(1)})) \sigma(a_j^{(2)}) \sigma(a_j^{(3)}) = \rho(a_j^{(1)}) \sigma(a_j^{(2)}) a_j^{(3)}.
  \end{align*}
  Thus $\rho\circ\sigma$ and $\sigma\circ\rho$ agree on each simple tensor
  $a_j^{(1)} a_j^{(2)} a_j^{(3)}$ and hence, by linearity, on all of
  $\cA_X$, and by norm-continuity on all of $\cA_\Sigma$.
  
We observe that the restrictions of $\rho$ and $\sigma$ to each $\cR_\Sigma$ are normal $*$-endomorphisms (hence SOT-continuous) and $\cA_\Sigma$ is strongly dense in $\cR_\Sigma$. Thus $\rho \circ \sigma$ and $\sigma \circ \rho$ agree on $\cR_\Sigma$.

By definition, $\cA^a$ is the norm-closure of the $*$-algebra $\bigcup_{\Sigma \in \cL} \cR_\Sigma$. Since $\rho\circ\sigma$ and $\sigma\circ\rho$ agree on each $\cR_\Sigma$, they agree on $\bigcup_{\Sigma \in \cL} \cR_\Sigma$. Since $\rho\circ\sigma$ and $\sigma\circ\rho$ are norm-bounded linear maps on $\cA^a$, the claim follows.
\end{proof}

Categorically the interpretation of Lemma \ref{lem:switch order of composition for disjoint cones} is the guaranteed existence of the trivial morphism $\SWAP := 1 \in (\rho \otimes \sigma, \sigma \otimes \rho)$ for $\rho,\sigma \in \DHR(\Lambda_a)$ localized in disjoint cones $\Lambda, \Lambda' \subset \Lambda_a$ respectively.

\begin{figure}[!ht]
    \centering
    \includegraphics[width=0.4\linewidth]{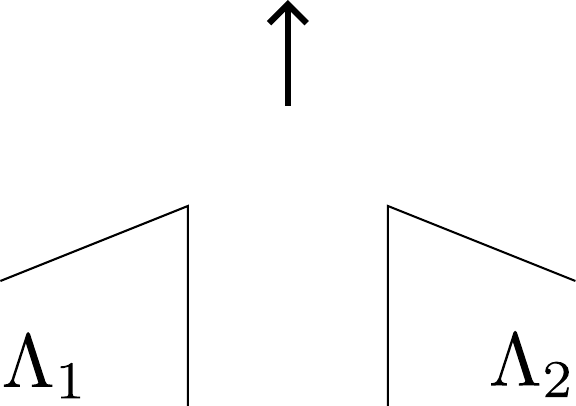}
    \caption{An example figure of cone $\Lambda_1$ lying to the left of cone $\Lambda_2$ (equivalently, $\Lambda_2$ lies to the right of $\Lambda_1$). Here the arrow represents the forbidden direction $\theta$.}
    \label{fig:left right for cones}
\end{figure}

\begin{defn}
    Recall that the definition of $\cA^a$ was reliant on a forbidden direction $\theta$. We define for two allowed, mutually disjoint cones $\Lambda(\theta_1,\theta_2, x), \Lambda'(\theta_1',\theta_2',x')$ a canonical notion of left/right. We say $\Lambda$ is \emph{to the left of} $\Lambda'$ if $\theta_1' > \theta_1 > \theta$. Otherwise we say $\Lambda$ is \emph{to the right of} $\Lambda'$. See Figure \ref{fig:left right for cones} for an example.
\end{defn}

We pick $\rho,\sigma \in \DHR(\Lambda)$ { for some allowed cone $\Lambda \in \cL$} and define braiding procedurally as follows:
\begin{itemize}
    \item Initially our object is $\rho \circ \sigma$, with $\rho,\sigma$ localized in cone $\Lambda$.
    \item First we pick an allowed cone $\Lambda_L \in \cL$ lying to the left of $\Lambda$. Since $\rho$ is transportable, there exists a $\rho_L$ localized in $\Lambda_L$ such that $\rho \simeq \rho_L$. Let $U:\rho \rightarrow \rho_L$ be the unitary implementing the equivalence. In categorical notation, we have $(U \otimes 1) \in (\rho \otimes \sigma, \rho_L \otimes \sigma)$. 
    \item Now since $\rho_L$ is localized in $\Lambda_L$ and $\sigma$ is localized in $\Lambda$ with $\Lambda$ disjoint from $\Lambda_L$, we're able to use Lemma \ref{lem:switch order of composition for disjoint cones} to trivially obtain $\rho_L \circ \sigma = \sigma \circ \rho_L$. While algebraically this is a trivial operation, categorically, we accomplish this by the morphism $\SWAP \in (\rho_L \otimes \sigma, \sigma \otimes \rho_L)$.
    \item Now we again use the transportability of $\rho_L$ to get back to $\rho$ localized in $\Lambda$. But this time, $\rho_L$ is temporally acting first, i.e., it is in the second tensor factor. Therefore the correct morphism is $(1 \otimes U^*) \in (\sigma \otimes \rho_L, \sigma \otimes \rho)$.
    \item We have successfully exchanged $\rho,\sigma$ temporally. We define $c_{\rho,\sigma} \in (\rho \otimes \sigma, \sigma \otimes \rho)$ for any $\rho,\sigma \in \DHR(\Lambda)$ by combining these morphisms.
\end{itemize}

\begin{defn}
    Consider $\rho, \sigma \in\DHR(\Lambda)$, and a unitary $U:(\rho, \rho_L)$ with $\Lambda_L$ lying to the left of $\Lambda$ and $\rho_L \in \DHR(\Lambda_L)$ satisfying $\rho \simeq \rho_L$. We define the natural isomorphism (c.f.~Lemma \ref{lem:naturality of braiding iso}) called the \emph{braiding isomorphism} by $$c_{\rho,\sigma}:= (1\otimes U^*) \circ  \SWAP \circ (U\otimes 1) = \sigma(U^*) U$$

    We call the isomorphism $S(\rho,\sigma) := c_{\sigma,\rho}c_{\rho,\sigma} \in (\rho\otimes \sigma, \rho \otimes \sigma)$ the \emph{double braiding}.
\end{defn}

\begin{lem}
\label{lem:braiding iso independence of chosen cone or unitary}
    The braiding isomorphism $c_{\rho,\sigma}$ is independent of choice of cone $\Lambda'$ and of the unitary used to transport to $\Lambda'$
\end{lem}
\begin{proof}
    Consider another unitary $V :\rho \rightarrow \rho'$. We notice that $VU^* \in (\rho', \rho') \subset \cR_{\Lambda'}$. Since $\rho,\sigma \in \DHR(\Lambda)$ and $\Lambda$ is disjoint from $\Lambda'$, we obtain $$\sigma(U^*) U = \sigma(U^*) UV^* V = \sigma(U^*)\sigma(UV^*) V = \sigma(V^*)V$$ showing the independence of the particular unitary chosen in defining $c_{\rho,\sigma}$.

    Now let $\Lambda'$ and $\Lambda''$ be two allowed cones to the left of $\Lambda$, and choose unitaries $U:\rho\rightarrow\rho'$ and $V:\rho\rightarrow\rho''$ with $\rho',\rho''$ localized in $\Lambda',\Lambda''$ respectively. Choose a cone $\Lambda_a\in\cL$ containing both $\Lambda'$ and $\Lambda''$, and transport further to a common $\Lambda_a$-localized endomorphism $\tilde\rho$ via unitaries $W:\rho'\rightarrow\tilde\rho$ and $W':\rho''\rightarrow\tilde\rho$. Then $WU$ and $W'V$ are two unitaries from $\rho$ to $\tilde\rho$, and by the first part the resulting braidings coincide. Hence $c_{\rho,\sigma}$ does not depend on the choice of left cone.
\end{proof}

\begin{lem}
    \label{lem:naturality of braiding iso}
    The family of braiding isomorphisms $\{c_{\rho,\sigma}\}$ defines a natural isomorphism.
\end{lem}
\begin{proof}
We verify naturality of $c_{\rho,\sigma}$ by exhibiting it as a composition of
natural transformations. Fix for each $\rho\in\DHR(\Lambda)$ a
charge $\rho_L \in \DHR(\Lambda_L)$ localized in $\Lambda_L$ to the left of $\Lambda$ and a unitary intertwiner $U_\rho\in (\rho,\rho_L)$. By
Lemma~\ref{lem:braiding iso independence of chosen cone or unitary}, the
resulting braiding $$c_{\rho,\sigma}
  :=
  (1_\sigma\otimes U_\rho^*)\circ \SWAP \circ (U_\rho\otimes 1_\sigma)$$ is independent of all these choices. 

For fixed $\sigma$, we can define the functors $F_\sigma$ acting as $F_\sigma(\rho):= \rho \otimes \sigma$ and $F_\sigma(f) := f \otimes 1_\sigma \in (\rho \otimes \sigma, \rho' \otimes \sigma)$, and $G_\sigma$ acting as $G_\sigma(\rho) := \rho_L \otimes \sigma$, and $G_\sigma(f) := U_{\rho'} f U_\rho^* \otimes 1_\sigma \in (\rho_L \otimes \sigma, \rho_{L'}\otimes \sigma)$. Now it is easily verified that the family $\{\eta_\rho: U_\rho \otimes 1_\sigma\}$ implements a natural transformation $F_\sigma \Rightarrow G_\sigma$.  

Since $\rho_L$ is localized to the left of $\sigma$,  $\SWAP:\rho_L\otimes\sigma\rightarrow\sigma\otimes\rho_L$ is trivially natural in both variables, because both $\rho_L,\sigma$ are localized in disjoint cones, and thus $\SWAP$ is just the identity as an operator, and thus the identity natural transformation. Thus $c_{(\cdot),\sigma}$ is obtained by conjugating a natural transformation ($\SWAP$) by natural isomorphisms ($\{\eta_{\rho}\}$), and it follows that $c_{\rho,\sigma}$ is therefore natural in $\rho$. A symmetric argument gives naturality in $\sigma$. Since the tensor product is bifunctorial, this yields the full naturality condition $(g\otimes f)c_{\rho,\sigma}=c_{\rho',\sigma'}(f\otimes g)$ for all intertwiners $f:\rho\rightarrow\rho'$ and $g:\sigma\rightarrow\sigma'$.
\end{proof}

\begin{rem}
    The definition of the braiding isomorphism $c_{\rho,\sigma}$ was reliant on picking a disjoint allowed cone $\Lambda'$ lying to the left of $\Lambda$. If instead, $\Lambda'$ lies to the right of $\Lambda$, then we would get another braided isomorphism $d_{\rho,\sigma}$. In fact, these two isomorphisms are actually inverses of each other. 
\end{rem}

We equip $\DHR(\Lambda)$ with the braided isomorphism $c_{\rho,\sigma}$, and equip $\DHR_0(\Lambda)$ with the braided isomorphism $c_{\rho_0,\sigma_0}^0 := T_{\sigma_0\otimes \rho_0} c_{\rho_0,\sigma_0} T_{\rho_0\otimes \sigma_0}^*$.

\begin{lem}
    The categories $\DHR(\Lambda)$ and $\DHR_0(\Lambda)$ are braided $\rmC^*$-tensor categories. Moreover, they are braided $\rmC^*$-tensor equivalent.
\end{lem}
\begin{proof}
    We already have from Lemmas \ref{lem:DHR is a BTC}, \ref{lem:DHR0 is a BTC} that the categories $\DHR(\Lambda), \DHR_0(\Lambda)$ are $\rmC^*$-tensor categories. To establish that they are braided $\rmC^*$-tensor categories, we must establish that the hexagon equations are satisfied.

    We show $\DHR(\Lambda)$ is a braided $\rmC^*$-tensor category: We verify the first hexagon equation. Since the associators are trivial, we have for $U: \rho\rightarrow \rho'$ (by Lemma \ref{lem:braiding iso independence of chosen cone or unitary} the precise unitary used or cone of localization does not matter), 
    $$c_{\rho, \sigma \otimes \mu} = \sigma( \mu(U^*))U = \sigma(\mu(U^*) U) \sigma(U^*) U = \sigma(c_{\rho,\mu}) c_{\rho,\sigma} = (1_{\sigma}\otimes c_{\rho,\mu})\circ(c_{\rho,\sigma}\otimes 1_{\mu})$$
    The second hexagon equation follows similarly. Thus $\DHR(\Lambda)$ is a braided $\rmC^*$-tensor category.

    We show $\DHR_0(\Lambda)$ is a braided $\rmC^*$-tensor category as follows. Since $c_{\rho_0,\sigma_0}$ satisfies the hexagon equations for $\DHR(\Lambda)$, and $c_{\rho_0,\sigma_0}^0, \rho_0 \otimes_0\sigma_0$ are built from $c_{\rho_0,\sigma_0},\rho_0\otimes\sigma_0$ using conjugation by the same $T$ maps, it immediately follows that $c^0_{\rho_0,\sigma_0}$ will satisfy the hexagon equations for $\DHR_0(\Lambda)$.

    To show $F:\DHR_0(\Lambda) \rightarrow \DHR(\Lambda)$ is a braided $\rmC^*$-tensor equivalence, we observe
    \begin{align*}
        F(c_{\rho_0,\sigma_0}^0) \phi_{\rho_0,\sigma_0} &= c_{\rho_0,\sigma_0}^0 \phi_{\rho_0,\sigma_0}= T_{\sigma_0 \otimes \rho_0} c_{\rho_0,\sigma_0} T_{\rho_0 \otimes \sigma_0}^* T_{\rho_0 \otimes \sigma_0} \\
        &=  T_{\sigma_0 \otimes \rho_0}c_{\rho_0,\sigma_0}= \phi_{\sigma_0,\rho_0} c_{\rho_0,\sigma_0} = \phi_{\sigma_0,\rho_0} c_{F(\rho_0),F(\sigma_0)}
    \end{align*}
    which completes the proof.
\end{proof}

\subsection{The category of anyon representations}
With the discussion in Section \ref{sec:localized, transportable endomorphisms}, specifically \ref{thm:there is a localized transportable endo for every anyon rep}, \ref{lem:anyon rep endo is localized and trans}, we see that the set of anyon representations is in bijection with the set of localized transportable endomorphisms of $\cA^a$. In fact, this statement can be categorified to show that the anyon representation category (denoted $\DHR_\pi$) is equivalent to the category of anyons $\DHR(\Lambda)$. But since $\DHR(\Lambda)$ has additional structure of being a braided $\rmC^*$-tensor category, which $\DHR_\pi$ is missing, we can promote $\DHR_\pi$ into a braided $\rmC^*$-tensor category using this equivalence.

\subsection{A rigid subcategory of \texorpdfstring{$\DHR(\Lambda)$}{DHR(L)}}

\begin{defn}
\label{def:rigid subcategory}
    We denote by $\DHR_f(\Lambda)$ the full subcategory of $\DHR(\Lambda)$ consisting of objects having a dual (see Definition \ref{def:rigid tensor category and dualizability}). That is, $\DHR_f(\Lambda)$ has objects $\rho \in \DHR(\Lambda)$ such that there exists an object $\bar \rho \in \DHR(\Lambda)$ and a distinguished morphism $R \in (\Id, \bar \rho \otimes  \rho)$ and $\bar R \in (\Id, \rho \otimes \bar \rho)$ satisfying the \emph{zig-zag} equations, i.e.,
    \begin{align*}
        (\bar R^* \otimes 1_\rho) \circ (1_\rho \otimes R) = 1_\rho \qquad ( R^* \otimes 1_{\bar\rho}) \circ (1_{\bar\rho} \otimes \bar R) = 1_{\bar\rho}
    \end{align*}
    where $1_\rho \in (\rho, \rho)$ is the identity morphism, and similarly for $1_{\bar \rho}$. We call $\bar \rho$ the dual of $\rho$, and we say that $R, \bar R$ satisfy the dual equations.
\end{defn}

\begin{rem}
    Definition \ref{def:rigid subcategory} is symmetric for $\rho, \bar \rho$ so if $\rho \in \DHR_f(\Lambda)$ then so is $\bar \rho$.
\end{rem}

Consider some $T \in (\rho, \rho)$ for some $\rho \in \DHR_f(\Lambda)$ with dual $\bar \rho$ and $R, \bar R$ satisfying the duality equations. Then we define\footnote{Throughout we suppress associators and unit constraints in tensor products.} $$\bar T :=  (R^*\otimes 1_{\bar\rho})\circ (1_{\bar\rho}\otimes T \otimes 1_{\bar\rho})\circ (1_{\bar\rho}\otimes \bar R) \in(\bar\rho,\bar\rho)$$ Then $T\mapsto \bar T$ is a unital $*$-anti-isomorphism $(\rho,\rho)\rightarrow (\bar\rho,\bar\rho)$.

\begin{prop}
$\DHR_f(\Lambda)$ is a rigid, braided $\rmC^*$-tensor subcategory of $\DHR(\Lambda)$
\end{prop}
\begin{proof}
    The $\rmC^*$-structure on $\DHR_f(\Lambda)$ is inherited from $\DHR(\Lambda)$. Rigidity follows by definition. The tensor unit $\Id\in\DHR(\Lambda)$ is dualizable (take $\bar{\Id}=\Id$ and $R=\bar R=1_{\Id}$), hence $\Id\in\DHR_f(\Lambda)$. We must show that $\DHR_f(\Lambda)$ is closed under direct sums, subobjects, $\otimes$, braided in order to get the required result.

    \textbf{Closure under $\otimes$.} Consider $\rho_1,\rho_2 \in \DHR_f(\Lambda)$. We explicitly define the dual for $\rho_1 \otimes \rho_2$ as follows. Let $R_i \in (\Id, \bar \rho_i \otimes \rho_i)$ and $\bar R_i \in (\Id, \rho_i \otimes \bar \rho_i)$ solve the dual equations for $i = 1,2$. Define
    \begin{align*}
        R := (1_{\bar \rho_2} \otimes R_1 \otimes 1_{\rho_2}) \circ R_2 \qquad \bar R := (1_{\rho_1} \otimes \bar R_2 \otimes 1_{\bar \rho_1}) \circ \bar R_1
    \end{align*}
    
    Then $R \in (\Id, (\bar \rho_2 \otimes \bar \rho_1) \otimes (\rho_1 \otimes \rho_2))$ and $\bar R \in (\Id, (\rho_1 \otimes \rho_2) \otimes (\bar \rho_2 \otimes \bar \rho_1))$ and it can be checked that $R, \bar R$ satisfy the zig-zag equations, thus defining the dual $\bar \rho_2 \otimes \bar \rho_1$ for $\rho_1 \otimes \rho_2$. Therefore the tensor product of $\DHR(\Lambda)$ restricts to $\DHR_f(\Lambda)$ on objects, and on morphisms it restricts because the subcategory is full. Thus $\DHR_f(\Lambda)$ is a tensor subcategory.

    \textbf{Closure under direct sums.}
    Consider $\rho_i \in \DHR_f(\Lambda)$ and consider the corresponding $R_i, \bar R_i$ as above. We construct the dual for $\rho_1 \oplus \rho_2$ as follows. Consider isometries $W_i \in (\rho_i , \rho_1 \oplus \rho_2)$ and $\bar W_i \in (\bar \rho_i, \bar \rho_1 \oplus \bar \rho_2)$ satisfying $W_i^*W_j=\delta_{ij}1_{\rho_i}$ and $\sum_i W_iW_i^*=1_{\rho_1\oplus\rho_2}$, and similarly for $\bar W_i$. Then setting
    \begin{align*}
        R := \sum_i (\bar W_i \otimes W_i ) \circ R_i \qquad \bar R := \sum_i (W_i \otimes \bar W_i ) \circ \bar R_i
    \end{align*}
    We get that $R \in (\Id, (\bar \rho_1 \oplus \bar \rho_2) \otimes  (\rho_1 \oplus \rho_2))$ and $\bar R \in (\Id,  (\rho_1 \oplus \rho_2) \otimes (\bar \rho_1 \oplus \bar \rho_2))$. It is easily checked that $R, \bar R$ satisfy the zig-zag equations, giving us the dual $(\bar \rho_1 \oplus \bar \rho_2)$ to $(\rho_1 \oplus \rho_2)$. 

    \textbf{Closure under subobjects.}
    Consider some object $\rho \in \DHR_f(\Lambda)$ and let $\bar \rho$ be its dual, with $R, \bar R$ solving the dual equations. Let there be a (non-zero) projection $P \in (\rho, \rho)$. Then set $\sigma \in \DHR(\Lambda)$ to be the corresponding subobject using the isometry $v \in (\sigma, \rho)$ satisfying $vv^* = P$ (see Lemma \ref{lem:DHR has direct sums and subobjects}).

    Notice that $\bar P \in (\bar \rho, \bar \rho)$\footnote{Here $\bar P$ is constructed as above using $R, \bar R$} and in particular $\bar P$ is a non-zero projection. Then we define another object $\bar \sigma \in \DHR(\Lambda)$ using $\bar P$ and the corresponding isometry $\bar v \in (\bar \sigma, \bar \rho)$ satisfying $\bar v \bar v^* = \bar P$. We define 
    \begin{align*}
        S := (\bar v^* \otimes v^*) \circ R \qquad \bar S := (v^* \otimes \bar v^*) \circ \bar R
    \end{align*}
    with $S \in (\Id, \bar \sigma \otimes \sigma), \bar S \in (\Id, \sigma \otimes \bar \sigma)$ and are easily shown to satisfy the conjugate equations, and thus define a dual $\bar \sigma$, showing that $\sigma$ (hence $\bar \sigma$) belongs to $\DHR_f(\Lambda)$.

    \textbf{Braiding.}
    Let $c_{\rho,\sigma}\in(\rho\otimes\sigma,\sigma\otimes\rho)$ denote the braiding of $\DHR(\Lambda)$. For $\rho,\sigma\in\DHR_f(\Lambda)$, we have already shown $\rho\otimes\sigma,\sigma\otimes\rho\in\DHR_f(\Lambda)$, and since the subcategory is full, $c_{\rho,\sigma}$ is a morphism in $\DHR_f(\Lambda)$. Naturality and the hexagon identities hold in $\DHR(\Lambda)$, hence they hold after restriction. Thus $\DHR_f(\Lambda)$ is braided.

    Collecting the above facts we get the required result.
\end{proof}

Any two conjugates of $\rho$ are unitarily equivalent. Moreover, once $\bar\rho$ is fixed, different solutions $(R,\bar R)$ of the conjugate equations lead to conjugation maps $T\mapsto \bar T$ that differ by inner conjugacy on $(\bar\rho,\bar\rho)$. We will therefore fix a \emph{standard} (normalized) solution.

In fact, one can show the following powerful result.

\begin{thm}
    Every object $\rho \in \DHR_f(\Lambda)$ is finite, i.e, it is a direct-sum of finitely many copies of finitely many irreducible objects in $\DHR_f(\Lambda)$. In particular, $(\rho,\rho)$ is finite dimensional for all $\rho \in \DHR_f(\Lambda)$. 
\end{thm}
\begin{proof}
    The proof is given in \cite[Lem.~3.2]{longo1996theory}.
\end{proof}

\begin{rem}
If $\rho\in\DHR_f(\Lambda)$, then there are natural isomorphisms $$(\rho\sigma,\tau)\simeq(\sigma,\bar\rho\tau),\qquad (\sigma\rho,\tau)\simeq(\sigma,\tau\bar\rho)$$ implemented by $(R,\bar R)$. This is called Frobenius reciprocity.
\end{rem}

Actually, one can define a notion of dimension in $\DHR_f(\Lambda)$ given by $$\dim \rho := R^* \circ R = \bar R^* \circ \bar R =: \dim \bar \rho$$ called the \emph{statistical dimension}. One gets the following very powerful result:

\begin{thm}[\cite{longo1996theory}]
    For any $\rho \in \DHR(\Lambda)$, $\dim \rho < \infty$ if and only if there exists a corresponding dual $\bar \rho$ for $\rho$.
\end{thm}

\subsection{Is $\DHR(\Lambda)$ a UMTC?}
The short answer is \textbf{probably not}. We comment that to obtain a UMTC (See Definition \ref{def:unitary modular tensor category}), one would require finite semisimplicity, rigidity, and non-degeneracy of braiding. As explained in the previous section, the problem of establishing rigidity is equivalent to the statistical dimension being finite \cite{longo1996theory}, which is an index that arises from subfactor theory. It is unlikely that a generic anyon category will be a UMTC, and in-fact has been shown to be false in the AQFT setting \cite{fredenhagen1994superselection}. However, additional assumptions such as translation invariance and a mass-gap lead to the existence of conjugate sectors in the AQFT setting \cite{fredenhagen1981existence}. It is an open problem to establish the necessary assumptions (if any) that lead to the reduction to a UMTC. 

\begin{rem}
    The results of \cite{fredenhagen1981existence} to establish the existence of conjugate sectors can perhaps be ported over to the lattice using \cite{MR3510469}, but with the additional assumption that the mass-gap is ``regular'' or ``pseudo-relativistic''. As noted in the discussion of that paper, exactly solvable models like Toric Code fail to satisfy these assumptions due to having a flat spectrum. However, there are perturbative variations of these models like \cite{bachmann2023dynamical, bachmann2025anyons} which do. It is unclear if every topologically ordered gapped frustration-free model can be perturbatively made to satisfy these spectral assumptions.
\end{rem}

\begin{rem}
    The recent work \cite{bachmann2025stacking} discusses the difficulty of reducing $\DHR(\Lambda)$ to a finite-semisimple category. In particular, every sector can be decomposed in terms of a direct integral of irreducible representations, which (apart from possibly a measure $0$ set) are all anyon sectors. However when the anyon sector satisfies the ``type $I$ property'' (that is, the reference representation in the anyon selection criterion is approximately split), and the Hilbert space is separable, then the direct integral can be reduced to countably-many direct sums of irreducible anyon sectors. However the general problem of addressing the necessary and sufficient conditions under which $\DHR(\Lambda)$ reduces to a finite-semisimple category is still open.
\end{rem}

\begin{rem}
    Non-degeneracy of braiding is another open problem, even when $\DHR(\Lambda)$ is established to be a finite-semisimple, rigid category. One of the key results of the original AQFT analysis \cite{doplicher1989new} was that the Müger center of $\DHR(\Lambda)$ is a compact group. This group is not a-priori related to a symmetry of a quantum spin system, nor to any gauge freedom. So the problem of establishing non-degeneracy is equivalent to the problem of showing that this group is in-fact trivial. 
\end{rem}

\bibliographystyle{alpha}
\bibliography{intro_bibliography}

%% file: main_sectors_thesis.tex
\chapter{Classification Of The Anyon Sectors Of Kitaev’s Quantum Double Model}
\chapterauthors{
\chapterauthor{Alex Bols}{Institute for Theoretical Physics, ETH Z{\"u}rich}
\chapterauthor{Siddharth Vadnerkar}{Department of Physics, University of California, Davis}
}
\label{chap:quantum double sectors}

This chapter is taken verbatim from \cite{bols2025classification} and is published in Communications in Mathematical Physics. Reprinted with the permission of Alex Bols and Siddharth Vadnerkar. Redistribution is allowed under the \href{https://s100.copyright.com/AppDispatchServlet?title=Classification%20of%20the%20Anyon%20Sectors%20of%20Kitaev%E2%80%99s%20Quantum%20Double%20Model&author=Alex%20Bols%20et%20al&contentID=10.1007%2Fs00220-025-05363-w&copyright=The%20Author%28s%29&publication=0010-3616&publicationDate=2025-07-02&publisherName=SpringerNature&orderBeanReset=true&oa=CC%20BY}{copyright terms of this article} (\href{https://creativecommons.org/licenses/}{Creative Commons CC BY license}). First we take a moment to discuss the scope and context of this work. 

At the time of publishing, there already existed general framework of anyon sectors and their physical structure on a lattice \cite{Ogata2022-wp, naaijkens2022split}, where in particular, Ogata showed that on lattice spin systems satisfying a technical condition called Haag duality\footnote{Technically Ogata uses an approximate version of Haag duality, which is a weaker condition.} (Definition \ref{def:Haag duality}) the anyon sectors (Definition \ref{def:anyon sector}) form a Braided $\rmC^*$-tensor category (Definition \ref{def:braided tensor category}). 

It was also standard lore at this point that a ``nice'' gapped topological phase in 2+1D should be described mathematically by a Unitary Modular Tensor Category (Definition \ref{def:unitary modular tensor category}). By the discussion in Chapter \ref{chap:UMTC background}, a Braided $\rmC^*$-tensor category is a more general concept than a UMTC. In particular, the anyon sectors need to be finite in number. So it is imperative to study concrete lattice models where one would expect a UMTC structure for anyon sectors rather than the more general structure. 

The natural family of models where one would first look to verify the UMTC structure is Kitaev's Quantum Double models \cite{Kitaev2003-qr} and the Levin-Wen string-net models \cite{levin2005string}. At the time of publishing, the most exhaustively studied family of models was the abelian Quantum Double models analyzed by Naaijkens \cite{Naaijkens2010-aq, Naaijkens2015-xj}. But the abelian Quantum Double model only hosts abelian anyon sectors due to the structural property of anyon-pair creation/annihilation operators being unitary that mutually commute up to a phase. This considerably simplifies the sector classification process. Categorically, the anyon category for abelian Quantum Double models is pointed, and thus braiding two abelian anyons only results in a phase and abelian anyons always fuse to a unique abelian anyon. In this sense abelian anyons are `boring'. It is well known in the quantum computing literature that abelian anyons cannot be used for universal fault-tolerant quantum computation, so non-abelian anyons are a natural next step to investigate.

The non-abelian Quantum Double models were paradigmatic in this respect, as they are capable of universal quantum computation \cite{chen2025universal}.
The expectation is that the anyon sectors in these models would carry the full representation theory of the Drinfel'd Double of a finite group $G$, denoted $D(G)$, which is categorically a UMTC. So the category of anyon sectors would also be a UMTC, and there would be a braided $\rmC^*$-tensor equivalence between the two categories. We set out to clarify this lore in this work.

Naaijkens had some formalism developed in \cite{Naaijkens2015-xj}, in particular he had explored a $*$-homomorphism of the quasi-local algebra called \emph{amplimorphisms} and had a conjectured list of candidates for anyon-sectors. However the full classification was missing many key ingredients, including an upper bound on the number of sectors, and a proof that these candidates were indeed anyon sectors. 

However, passing from abelian to non-abelian quantum doubles within the operator-algebraic, infinite-volume setting is not a routine upgrade. We highlight \cite{Bombin2007-uw} for a physical introduction and \cite{HamdanThesis} for an introduction following the operator algebraic perspective. Technically, several of the tools that work cleanly for abelian $G$ become fragile. For instance, while abelian closed-ribbon operators give commuting families of projectors that neatly decompose charge types, in the non-abelian case the ribbon-operator algebra no longer reduces to commuting projectors, and the bookkeeping of charge and flux constraints (and their transport) is subtler. In short, the ``obvious'' adaptation of the abelian ribbon-projector machinery does not directly deliver the desired non-abelian sector classification. 

Beyond these technicalities lies a more conceptual obstacle: the classification problem for anyon sectors is not merely about producing some sectors, but rather about proving that a chosen selection criterion yields exactly the physically correct anyon sectors, and in fact it yields all of them. It is straightforward to see that if one prepares an anyon pair and sends one partner to infinity, the remaining localized excitation defines a sector satisfying the anyon selection criterion (Definition \ref{def:anyon sector}). But apriori this only shows existence, not completeness. Depending on how the selection criterion is phrased, one risks admitting spurious sectors, or conversely, one risks missing legitimate ones. The DHR literature emphasizes exactly these pitfalls: localization regions matter, transportability is essential, and Haag duality/split properties control whether the representation theory truly reflects the intrinsic particle content. Lattice analogues inherit the same hazards, so a proof of completeness is non-trivial.

This is the juncture at which our work sits. We give a complete classification of the anyon sectors of Kitaev’s Quantum Double model for arbitrary finite groups $G$, thereby covering the non-abelian case that had remained open in this operator-algebraic setting. Concretely, we prove that the anyon sectors are in one-to-one correspondence with the irreducible representations of the quantum double $D(G)$.  Establishing this identification puts the non-abelian models on the same footing as their abelian cousins.

Our methodology is as follows. We first construct a family of pure states for every irreducible representation $D$ of $D(G)$, which are special ground-states of the Quantum Double model, containing only a single excitation corresponding to $D$ localized at some fixed site. Conversely, We also show that all possible states that have only one excitation localized at some fixed site must have this form. We then show that these states lie in the `anyon sector candidate' representations of Naaijkens'. Finally we prove completeness by showing that every anyon sector \emph{must} contain one of these special ground-states.

A crucial ingredient in our proof is that each anyon sector contains states with only finitely many excitations. We prove that any representation satisfying the anyon selection criterion admits a pure state that is gauge-invariant and has trivial flux outside a finite region, i.e, the state has only finitely many excitations. Coupled with the well-known fact that in the Quantum Double model we can sweep finitely many excitations onto a single site, and that we've classified all such states, we're able to say that we've captured every possible anyon sector.

As a concluding remark, we comment that this anyon sector Quantum Double correspondence can be categorized. Meaning the anyon sector category constructed in Section \ref{sec:category of anyons} is braided $\rmC^*$-equivalent to the category of representations of $D(G)$ constructed in Section \ref{sec:quantum double category}. The result of this work is crucial to make this correspondence explicit, and is the subject of the follow-up work in chapter \ref{chap:category of quantum double}.

\setcounter{tocdepth}{2}
\appto\appendix{\addtocontents{toc}{\protect\setcounter{tocdepth}{1}}}

\appto\listoffigures{\addtocontents{lof}{\protect\setcounter{tocdepth}{1}}}
\appto\listoftables{\addtocontents{lot}{\protect\setcounter{tocdepth}{1}}}



\begin{chapterabstract}
    We give a complete classification of the anyon sectors of Kitaev's quantum double model on the infinite triangular lattice and for finite gauge group $G$, including the non-abelian case. As conjectured, the anyon sectors of the model correspond precisely to equivalence classes of irreducible representations of the quantum double algebra of $G$.
\end{chapterabstract}

\minitoc

\subsection*{Acknowledgements}
A.B. was supported by the Simons Foundation for part of this project. S.V. was supported by the NSF grant DMS-2108390. We thank Yoshiko Ogata for pointing out a serious error in the first version of the manuscript. We also thank Gian Michele Graf, Pieter Naaijkens, and Bruno Nachtergaele for useful conversations.

\subsection*{Data availability and conflict of interest}
Data availability is not applicable to this article as no new data were created or analyzed in this study. There are no known conflicts of interest.

\input{sectors/intro}

\input{sectors/excited_states}

\input{sectors/anyon_representations}

\input{sectors/completeness}

\input{sectors/conclusions}

\begin{subappendices}
\input{sectors/ribbonprops}

\input{sectors/string_nets}
\end{subappendices}

\bibliographystyle{alpha}
\bibliography{sectors/sector_bibliography}

%% file: sectors/intro.tex
\section{Introduction}

Over the decades since the discovery of the integer quantum Hall effect, the notion of topological phases of matter has come to be a central paradigm in condensed matter physics. In contrast to the conventional Landau-Ginsburg paradigm of spontaneous symmetry breaking, topological phases of matter are not distinguished by any local order parameter. Instead they are characterised by a remarkably wide variety of topological properties, ranging from toplogically non-trivial Bloch bands to topological ground-state degeneracy. What all these topological materials seem to have in common is that they are characterised by robust patterns in the entanglement structure of their ground states \cite{Li2008-le, Fidkowski2010-gj}.

Within this zoo of topological phases, the \emph{topologically ordered} phases in two dimensions have received a great deal of attention. The reason for this is in part because of their possible applications to quantum computation \cite{Kitaev2003-qr, Freedman1998-tz, Nayak2008-ef}. Topologically ordered materials exhibit robust ground state degeneracy depending on the genus of the surface on which they sit, and they support anyonic excitations which have braiding statistics that differs from that of bosons or fermions.

With the ever increasing experimental control of quantum many-body systems in the lab in mind, it is desirable to understand topological order from a microscopic point of view. On the one hand, an important role is played in this endeavor by exactly solvable quantum spin models that exhibit topologcial order, such as Kitaev's quantum double models \cite{Kitaev2003-qr} and, more generally, the Levin-Wen models \cite{levin2005string}. On the other hand, one wants to obtain a good understanding of the mathematical structures involved in characterising topological orders in general models \cite{Kitaev2006-ts, 
Shi2019-tl, Kawagoe2020-vt}. The latter problem has proven to be a rich challenge for mathematical physics \cite{Naaijkens2010-aq, Naaijkens2012-fh, Cha2018-ke, Cha2020-rz, Ogata2022-wp}. These works have yielded a rigorous, albeit still incomplete, description of topological order in gapped quantum spin systems in two dimensions. They provide robust definitions of anyon types, their fusion rules, and their braiding statistics, as well as a rigorous understanding of how these data fit together in a braided $C^*$-tensor category.

In this paper we study Kitaev's quantum double models from this mathematical physics point of view. The quantum double models can be thought of as discrete gauge theories with a finite gauge group $G$. These models are of particular interest because for non-abelian $G$, they are paradigmatic examples of models that support \emph{non-abelian anyons} \cite{Kitaev2003-qr}. We take a first step towards integrating the quantum double models for general $G$ into the mathematical framework referred to above. In particular, we classify all the anyon types of these models.

Roughly, an anyon type corresponds to a superselection sector, \ie a unitary equivalence class of representations of the observable algebra that are unitarily equivalent to the ground state representation when restricted to the complement of any cone-like region of the plane. We call such sectors \emph{anyon sectors}. Intuitively an anyon sector contains states that can be made to look like the ground state locally by moving the anyon somewhere else, but globally, the anyon is always detectable by braiding operations.

In order to completely classify the anyon sectors of the quantum double model we construct states $\omega_{s}^{RC;u}$ labeled by an irreducible representation $RC$ of the quantum double algebra $\caD(G)$, a site $s$, and additional microscopic data $u$. These states look like the ground state when evaluated on any local observable whose support does not contain or encircle the site $s$. We characterise these states by showing that they are the unique states that satisfy certain local constraints depending on the site $s$ and the data $RC$ and $u$. In particular, the states $\omega_s^{RC;u}$ are pure. In the particular case where $RC$ corresponds to the trivial representation of the quantum double algebra, the state $\omega_s^{RC;u}$ is the frustration free ground state, so we get existence and uniqueness of the frustration free ground state as a corollary, a result which was first proven in \cite{naaijkens2012anyons}.

We continue by showing that the pure states $\omega_{s}^{RC;u}$ belong to different superselection sectors if and only if they differ in their $RC$ label. It follows that the GNS representations of the states $\omega_{s}^{RC;u}$ give us a collection of pairwise disjoint irreducible representations $\pi^{RC}$ labeled by irreducible representations of the quantum double algebra. By relating the representations $\pi^{RC}$ to so-called amplimorphism representations \cite{Naaijkens2015-xj, vecsernyes1994quantum, fuchs1994quantum, nill1997quantum}, we show that these representations do in fact belong to anyon sectors. Finally, we show that any anyon sector must contain one of the states $\omega_s^{RC;u}$, thus showing that all anyon sectors contain one of the $\pi^{RC}$.

The paper is structured as follows. In Section \ref{sec:results} we set up the problem and state our main results. In Section \ref{sec:excited states} we construct the states $\omega_{s}^{RC;u}$ that `contain an anyon' at site $s$ and prove that these states are pure. Section \ref{sec:anyon representations} is devoted to constructing for each irreducible representation $RC$ of the quantum double algebra a representation $\pi^{RC}$ of the observable algebra that contains the states $\{\omega_s^{RC;u}\}_{s, u}$, and proving that these representations are disjoint. Finally, in section \ref{sec:completeness} we show that any anyon sector contains one of the $\pi^{RC}$, thus showing that the $\pi^{RC}$ exhaust all anyon sectors of the model.

\section{Setup and main results}
\label{sec:results}

\subsection{Algebra of observables}

\begin{figure}
    \centering
    \includegraphics[ width=0.4\textwidth]{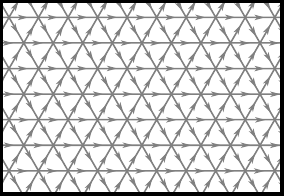}
    \caption{A snapshot of the triangular lattice with all edges oriented towards the right.}
    \label{fig:oriented lattice snapshot}
\end{figure}

Let $\Gamma$ be the regular triangular lattice (see Figure \ref{fig:oriented lattice snapshot}) whose set of vertices $\latticevert$ we regard as a subset of the plane $\R^2$ such that nearest neighbouring vertices are separated by unit distance.

The set of oriented edges of $\Gamma$ is identified with the set of ordered pairs of neighbouring vertices:
$$\orientededges := \{ (v_0, v_1) \in \latticevert \times \latticevert \, : \, v_0 \, \text{and} \, v_1 \, \text{are nearest neighbours}  \}.$$
We let $\latticeedge \subset \orientededges$ consist of the oriented edges pointing from left to right as in Figure \ref{fig:oriented lattice snapshot}. Note that $\latticeedge$ contains exactly one oriented edge for every edge of $\Gamma$. We denote the set of faces of $\Gamma$ by $\latticeface$.

Any oriented edge $e = (v_0, v_1)$ has an initial vertex $\partial_0 e = v_0$, a final vertex $\partial_1 e = v_1$, and an opposite oriented edge $\bar e = (v_1, v_0)$. The vertices $\latticevert$ are equipped with the graph distance $\mathrm{dist}( \cdot, \cdot )$ and similarly for the faces (regarded as elements of the dual graph). \\

We fix a finite group $G$ and associate to each edge $e \in \latticeedge$ a Hilbert space $\hilb_e = \mathbb{C}^{|G|}$ and a matrix algebra $\cstar[e] = \mathrm{End}(\hilb_e)$. For any finite $S \subset \latticeedge$  we have a Hilbert space $\hilb_S = \otimes_{e \in S} \hilb_e$ and the algebra of operators $\cstar[S] = \mathrm{End}(\hilb_S)$ on this space.

\begin{figure}
    \centering
    \includegraphics[width=0.8\textwidth]{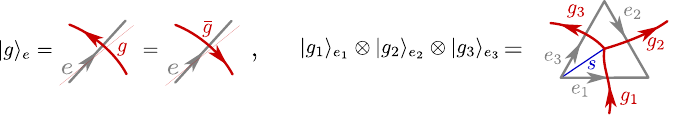}
    \caption{A graphical representation of the basis vector $\ket{g}_e \in \hilb_e$ and of the tensor product vector $\ket{g_1}_{e_1} \otimes \ket{g_2}_{e_2} \otimes \ket{g_3}_{e_3} \in \hilb_{\{e_1, e_2, e_3\}}$ for edges $e_1, e_2, e_3$ belonging to a single face $f$. We have also indicated a site $s$ with $f(s) = f$ in blue (see Section \ref{sec:excited states}).}
    \label{fig:graphical_representation}
\end{figure}

We employ the following graphical representation of states $\ket{\alpha}$. For any edge $e$, the basis state $\ket{g}_{e}$ of $\hilb_e$ is represented by the edge $e$ being crossed from right to left by an oriented \emph{string} labeled $g$. An equivalent representation of $\ket{g}_e$ is the edge $e$ being crossed from left to right by a string labeled $\bar g$, see Figure \ref{fig:graphical_representation}. The basis element $\ket{1}_e$ is represented by the edge $e$ not being crossed by any string at all. A tensor product of several of such basis states is represented by a figure where each participating edge is crossed by a labeled oriented string by the rules just described. See Figure \ref{fig:graphical_representation} for an example.

Let $S_1, S_2 \subset \latticeedge$ be finite sets of edges such that $S_1 \subset S_2$, then there is a natural embedding $\iota_{S_1, S_2}: \cstar[S_1] \hookrightarrow \cstar[S_2]$ given by tensoring with the identity, i.e.
$$\iota_{S_1, S_2}(O) = O \otimes \mathds{1}_{\cstar[{S_2 \setminus S_1}]}$$
for all $O \in \cstar[S_1]$. With these embeddings the algebras $\cstar[S]$ for finite $S \subset \latticeedge$ form a directed system of matrix algebras. Their direct limit is called the \emph{local algebra}, and is denoted by $\cstar[loc]$. The norm closure of the local algebra
$$\cstar = \overline{\cstar[loc]}^{||\cdot||}$$
is called the \emph{quasi-local algebra} or \emph{observable algebra}.

Similarly, for any (possibly infinite) $S \subset \latticeedge$ we have the algebra $\cstar[S] \subset \cstar$ of quasi-local observables supported on $S$.\\

A \emph{state} on $\cstar$ is a positive linear functional $\omega: \cstar \rightarrow \mathbb{C}$ with $\omega(\mathds{1}) =1$. Given a state $\omega$ on $\cstar$ there is a representation $\pi_{\omega} : \cstar \rightarrow \mathcal{B}({\hilb_{\omega}})$ for some separable Hilbert space $\hilb_{\omega}$ containing a unit vector $\ket{\Omega}$ that is cyclic for the representation $\pi_{\omega}$ and such that $\omega(O) = \langle \Omega, \pi_{\omega} (O) \Omega\rangle$ for all $O \in \cstar$. The triple $(\pi_{\omega}, \hilb_{\omega}, \ket{\Omega})$ satisfying these properties is unique up to unitary equivalence, and is called the GNS triple of the state $\omega$.\\

Throughout this paper, we will use the word `projector' to mean a self-adjoint operator that squares to itself. A collection of projectors is called \emph{orthogonal} if the product of each pair of projectors in the set vanishes. A set of projectors is called \emph{commuting} if each pair of projectors in the set commutes with each other.

\subsection{The quantum double Hamiltonian and its frustration free ground state}
\label{sec:model}

We say an edge $e$ belongs to a face $f$ and write $e \in f$ when $e$ is an edge on the boundary of $f$. Similarly, we say a vertex $v$ belongs to $f$ and write $v \in f$ if $v$ neighbours $f$, and we say a vertex $v$ belongs to an edge $e$ and write $v \in e$ if $v$ is the origin or endpoint of $e$.\\

We fix for each edge $e$ an orthonormal basis $\{ \ket{g} \}_{g \in G}$ for $\hilb_e$ labeled by elements of the group $G$. For $g \in G$ we denote its inverse by $\overline{g}$, and we define the left group action $L_e^h := \sum_{g \in G} \ket{hg}\bra{g}$, the right group action $R^h_e := \sum_{g \in G} \ket{g \bar h} \bra{g}$, and the projectors $T_e^g := | g \rangle \langle g |$.

For each vertex $v$ and edge $e$ such that $v$ belongs to $e$ we set $L^h(e, v) = L^h_e$ if $\partial_0 e = v$ and $L^h(e, v) = R^{h}_e$ if $\partial_1 e = v$. For each $h \in G$ we define a unitary $A_v^h := \prod_{e \in v} L^h(e, v)$. These are called the \emph{gauge transformations at $v$}. Graphically,
\begin{center}
    \includegraphics[width=0.6\textwidth]{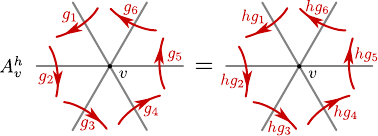}
\end{center}
We then define the \emph{gauge constraint} $A_v := \frac{1}{\abs{G}} \sum_{h \in G} A_v^h$, which is the projector enforcing gauge invariance at the site $v$. Similarly, for each face $f$ and edge $e \in f$ we set $T^h(e, f) = T_e^h$ if $f$ is to the left of $e$, and $T^h(e, f) = T_e^{\bar h}$ if $f$ is to the right of $e$. If the face $f$ has bounding edges $(e_1, e_2, e_2)$ ordered counterclockwise (with arbitrary initial edge), then we define the \emph{flat gauge projector} $B_f := \sum_{ \substack{ g_1, g_2, g_3 \in G : \\ g_1 g_2 g_3 = 1 }} T^{g_1}(e_1, f) \, T^{g_2}(e_2, f) \, T^{g_3}(e_3, f)$ which is also a projector. Note that this expression does not depend on which edge goes first in the triple $(e_1, e_2, e_3)$. Graphically,
\begin{center}
    \includegraphics[width=0.6\textwidth]{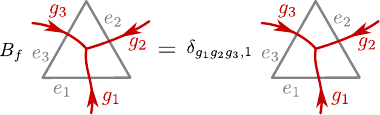}
\end{center}
 The set of projectors $\{ A_v \}_{v \in \latticevert} \cup \{ B_f \}_{f \in \latticeface}$ all commute with each other.\\

The quantum double Hamiltonian is the formal sum (interaction) of commuting projectors 
\begin{equation*}
    H = \sum_{v \in \latticevert} ( \mathds{1} - A_v ) + \sum_{f \in \latticeface} ( \mathds{1} - B_f ).
\end{equation*}

\begin{defn} \label{def:ffgs}
    A state $\omega$ on $\cstar$ is a \ffgs{} of $H$ if $\omega(A_v) = \omega(B_f) = 1$ for all $v \in \latticevert$ and all $f \in \latticeface$.
\end{defn}

If a state $\omega$ satisfies $\omega(A_v) = 1$ then we say it is \emph{gauge-invariant} at $v$, and if $\omega(B_f) = 1$ then we say it is \emph{flat} at $f$. A \ffgs{} is a state that is gauge invariant and flat everywhere.

The following Proposition was first proven for the Toric code (the case $G = \Z_2$) in \cite{alicki2007statistical}, and for general $G$ in \cite{naaijkens2012anyons}. See also \cite{chuah2024boundary} which uses general results on commuting projector Hamiltonians from \cite{jones2023local}. We obtain a new proof of this Proposition as a Corollary to Proposition \ref{prop:characterisation of S^RCu}.
\begin{prop}[\cite{naaijkens2012anyons}]
\label{prop:ffgsunique}
    The quantum double Hamiltonian has a unique frustration free ground state.
\end{prop}
We will denote the unique \ffgs{} by $\omega_0$, and let $(\pi_0, \hilb_0, \ket{\Omega_0})$ be its GNS triple. Note that since $\omega_0$ is the \emph{unique} \ffgs{} of the quantum double model it is a pure state, and therefore $\pi_0$ is an irreducible representation.

\subsection{Classification of anyon sectors}

In the context of infinite volume quantum spin systems or field theories, types of anyonic excitations over a ground state $\omega_0$ have a very nice mathematical characterisation. They correspond to the irreducible representations of the observable algebra that satisfy a certain superselection criterion w.r.t. the GNS representation $\pi_0$ of the ground state (\cite{Doplicher1971-jd}, \cite{Doplicher1974-hb}, \cite{fredenhagen1989superselection}, \cite{fredenhagen1992superselection}, \cite{frohlich1990braid}). In our setting of quantum spin systems, the appropriate superselection criterion was first formulated in \cite{Naaijkens2010-aq} in the special case of the Toric code.\\

The cone with apex at $a \in \R^2$, axis $\hat v \in \R^2$ of unit length, and opening angle $\theta \in (0, 2\pi)$ is the open subset of $\R^2$ given by
\begin{equation*}
 	\Lambda_{a, \hat v, \theta} := \{ x \in \R^2 \, : \, (x - a) \cdot \hat v > \norm{x-a} \cos (\theta/2)   \}. \quad\quad\quad\quad \includegraphics[width=2.5cm]{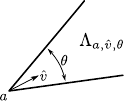}
\end{equation*}
Any subset $\Lambda \subset \R^2$ of this form will be called a cone.

For any subset $S \subset \R^2$ of the plane we denote by $\overline S \subset \latticeedge$ the set of edges whose midpoints lie in $S$, and write $\cstar[S] := \cstar[\overline S]$ for the algebra of observables supported on $\overline S$. With this definition we have $\overline S \cup \overline{ S^c } = \latticeedge$ for any $S \subset \R^2$.

\begin{defn}
\label{def:anyonsector}
    An irreducible representation $\pi: \cstar \rightarrow \mathcal{B}(\hilb)$ is said to satisfy the superselection criterion w.r.t. $\pi_0$ if for any cone $\Lambda$, there is a unitary $U_\Lambda : \hilb_0 \rightarrow \hilb$ such that
    $$\pi(A) = U_\Lambda \pi_0 (A) U^*_\Lambda$$
    for all $A\in \cstar[\Lambda]$. We will call such a representation an \emph{anyon representation} for $\omega_0$. A unitary equivalence class of anyon representations we call an \emph{anyon sector}.
\end{defn}

Let us denote by $\caD(G)$ the quantum double algebra of $G$. The irreducible representations of $\caD(G)$ are, up to isomorphism, uniquely labeled by a conjugacy class $C$ of $G$ together with an irreducible representation $R$ of the centralizer $N_C$ of $C$ (see for example \cite{Gould1993-bt}). We denote the irreducible representation of $\caD(G)$ corresponding to conjugacy class $C$ and irreducible representation $R$ by $RC$.

Our main result is the complete classification of the anyon sectors of $\omega_0$ in terms of the irreducible representations of $\caD(G)$. This result was first obtained for the Toric code in \cite{naaijkens2013kosaki} using very different methods. See also \cite{Fiedler2015-na} where it is indicated how the methods of \cite{naaijkens2013kosaki} can be applied to quantum double models for abelian $G$.
\begin{thm} \label{thm:main theorem}
    For each irreducible representation $RC$ of $\caD(G)$ there is an anyon representation $\pi^{RC}$. The representations $\{ \pi^{RC} \}_{RC}$ are pairwise disjoint, and any anyon representation is unitarily equivalent to one of them.
\end{thm}
This Theorem is restated and proven in Section \ref{sec:completeness}, Theorem \ref{thm:classification theorem}.

%% file: sectors/excited_states.tex
\section{Anyon states}
\label{sec:excited states}

In this section we construct states containing a single anyonic excitation of arbitrary type, and show that these states are pure. These are states that satisfy the frustration free ground state constraints everywhere except at a fixed site $\site$, where instead they are constrained by some Wigner projector onto an irreducible representation for the quantum double action on that site (see Remark \ref{rem:quantum double at s}). We completely classify the states satisfying such constraints by first classifying all states that satisfy appropriate local versions of these constraints. We note that the methods presented in this section are sufficient to establish that anyon types of the quantum double model are in one-to-one correspondence with the irreducible representations of the quantum double algebra $\caD(G)$ in the context of the entanglement bootstrap program \cite{Shi2019-tl}.\\

\subsection{Ribbon operators, gauge configurations, and gauge transformations} \label{subsec:preliminary notions}

In this subsection we introduce sites, triangles, and ribbons in order to then describe the ribbon operators introduced by \cite{Kitaev2003-qr}. These ribbon operators will play a crucial role from Section \ref{sec:anyon representations} onward. We then introduce the notion of local gauge transformations.

Let $\Gamma^*$ be the dual lattice to $\Gamma$. To each oriented edge $e \in \orientededges$ we associate a unique oriented dual edge $\de \in (\vec{\Gamma^*})^E$ with orientation such that along $e$, the dual edge $\de$ passes from right to left. Here
$$(\vec{\Gamma^*})^E = \{ (f_0, f_1) \in \latticeface \times \latticeface \, : \, f_0 \, \text{and} \, f_1 \, \text{are neighbouring faces}  \}$$
is the set of oriented dual edges of $\Gamma$.

\subsubsection{Sites and triangles}

A \emph{site} $s$ is a pair $s = (v, f)$ of a vertex and a face such that $v$ is on the boundary of $f$. We write $v(s) = v$ for the vertex of $s$ and $f(s) = f$ for the face of $s$. We represent a site graphically by a line from the site's vertex to the center of its face.

\begin{figure}
    \centering
    \includegraphics[ width=0.4\textwidth]{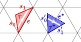}
    \caption{A direct triangle $\tau = (s_0, s_1, e)$ in red, and a dual triangle $\tau^* = (s'_0, s'_1, e')$ in blue. The dual edge $e^*$ associated to $e'$ is indicated as a dotted blue line. Note  that for this particular dual triangle, $e^*$ is oriented opposite to the white arrow, which instead follows the orientation of $(e_{\tau^*})^*$.}
    \label{fig:direct and dual triangles}
\end{figure}

A direct triangle $\tau = (s_0, s_1, e)$ consists of a pair of sites $s_0, s_1$ that share the same face, and the edge $e \in \latticeedge$ that connects the vertices of $s_0$ and $s_1$. We write $\partial_0 \tau = s_0$ and $\partial_1 \tau = s_1$ for the initial and final sites of the direct triangle $\tau$, and $e_{\tau} = (v(s_0), v(s_1))$ for the oriented edge associated to $\tau$, see Figure \ref{fig:direct and dual triangles}. Note that $e$ and $e_{\tau}$ need not be the same, $e$ always has the left to right orientation used in the definition of $\latticeedge$ while $e_{\tau}$ is oriented in the direction of the direct triangle. The opposite triangle to $\tau$ is the direct triangle $\bar \tau = (s_1, s_0, e)$. A direct triangle $\tau = (s_0, s_1, e)$ is \emph{positive} if the face $f = f(s_0) = f(s_1)$ lies to the left of $e_{\tau}$ and \emph{negative} otherwise.

Similarly, a dual triangle $\tau = (s_0, s_1, e)$ consists of a pair of sites $s_0, s_1$ that share the same vertex, and the edge $e \in \latticeedge$ whose associated dual edge $\de$ connects the faces of $s_0$ and $s_1$. We write again $\partial_0 \tau = s_0$ and $\partial_1 \tau = s_1$ and write $e^*_{\tau} = (f(s_0), f(s_1))$ for the oriented dual edge associated to $\tau$. We also write $e_{\tau}$ for the oriented edge whose dual is $e_{\tau}^*$. Note again that $e^*$ and $e_{\tau}^*$ need not be the same. The orientation of $e^*$ is determined by the left to right orientation of $e \in \latticeedge$ while $e^*_{\tau}$ is oriented in the direction of the dual triangle. We define the opposite dual triangle by $\bar \tau = (s_1, s_0, e)$. A dual triangle $\tau = (s_0, s_1, e)$ is called \emph{positive} if the vertex $v = v(s_0) = v(s_1)$ lies to the right of $e^*_{\tau}$ and $\emph{negative}$ otherwise.

To each dual triangle $\tau = (s_0, s_1, e)$ we associate unitaries $L^h_{\tau}$ supported on the edge $e$. The way $L^h_{\tau}$ acts depends on whether the edge $e^*$ dual to $e$ satisfies $e^* = (f(s_0), f(s_1))$ or $e^* = (f(s_1), f(s_0))$, and on whether $v(s_0) = \partial_0 e$ or $v(s_0) = \partial_1 e$ as follows. If $e^* = (f(s_0), f(s_1))$ and $v(s_0) = \partial_0 e$ then $L^h_{\tau} := L_e^h$. If $e^* = (f(s_0), f(s_1))$ and $v(s_0) = \partial_1 e$ then $L^h_{\tau} := R_e^{\bar h}$. If $e^* = (f(s_1), f(s_0))$ and $v(s_0) = \partial_0 e$ then $L^h_{\tau} := L_e^{\bar h}$ ($L^h_e, R^h_e$ were defined in section \ref{sec:model}). Finally, If $e^* = (f(s_1), f(s_0))$ and $v(s_0) = \partial_1 e$ then $L^h_{\tau} := R_e^h$. Similarly, to each direct triangle $\tau = (s_0, s_1, e)$ we associate projectors $T^g_\tau := T^g_{e}$ if $e = (v(s_0), v(s_1))$ and $T^g_{\tau} := T_e^{\bar g}$ if $e = (v(s_1), v(s_0))$.

\subsubsection{Ribbons}

\begin{figure}
    \centering
    \includegraphics[ width=0.6\textwidth]{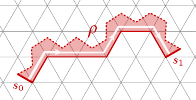}
    \caption{A positive finite ribbon $\rho$ with $\partial_0 \rho = s_0$ and $\partial_1 \rho = s_1$. The direct path of $\rho$ consists of the solid red edges.}
    \label{fig:finite ribbon}
\end{figure}

We define a finite ribbon $\rho:= \{\tau_i\}_{i = 1}^l$ to be an ordered tuple of triangles such that $\partial_1 \tau_i = \partial_0 \tau_{i+1}$ for all $i = 1, \cdots, l-1$, and such that for each edge $e \in \latticeedge$ there is at most one triangle $\tau_i$ for which $\tau_i = (\partial_0 \tau_i, \partial_1 \tau_i, e)$. The empty ribbon is denoted by $\ep$. For non-empty ribbons $\rho$ we write $\partial_0 \rho := \partial_0 \tau_1$ for the initial site of the ribbon and $\partial_1 \rho := \partial_1 \tau_n$ for the final site of the ribbon. See Figure \ref{fig:finite ribbon}. If all triangles belonging to a ribbon $\rho$ are direct, we say that $\rho$ is a direct ribbon, and if all triangles belonging to a ribbon $\rho$ are dual, we say that $\rho$ is a dual ribbon.

A ribbon is said to be positive if all of its triangles are positive, and negative if all of its triangles are negative. All non-empty ribbons are either positive or negative.

If we have two tuples $\rho_1 = \{ \tau_i \}_{i=1}^{l_1}$ and $\rho_2 = \{ \tau_i \}_{i = l_1 + 1}^{l_2}$ then we can concatenate them to form a tuple $\rho = \{ \tau_i \}_{i = 1}^{l_1 + l_2}$. We denote this concatenated tuple by $\rho = \rho_1 \rho_2$. Note that if $\rho$ is a finite ribbon and $\rho = \rho_1 \rho_2$, then $\rho_1$ and $\rho_2$ are automatically finite ribbons and (if $\rho_1$ and $\rho_2$ are non-empty), $\partial_0 \rho = \partial_0 \rho_1$, $\partial_1 \rho = \partial_1 \rho_2$ and $\partial_1 \rho_1 = \partial_0 \rho_2$.

The orientation reversal of a ribbon $\rho = \{ \tau_i \}_{i = 1}^l$ is the ribbon $\bar \rho = \bar\tau_l \cdots \bar \tau_1$. We say a finite ribbon $\rho$ is \emph{closed} if $\partial_0 \rho = \partial_1 \rho$.

The support of a ribbon $\rho = \{ \tau_i = (s_0^{(i)}, s_1^{(i)}, e_i)  \}_{i=1}^l$ is $\supp(\rho) := \{ e_i \}_{i = 1}^l.$ If $\supp(\rho) \subseteq S \subseteq \latticeedge$ we say $\rho$ is supported on $S$.

\subsubsection{Direct paths}
\label{sec:direct paths}
A direct path $\gamma = \{ e_i \}_{i=1}^l$ is an ordered tuple  of oriented edges $e_i \in \orientededges$ such that $\partial_1 e_i = \partial_0 e_{i+1}$ for $i = 1, \cdots, l-1$. We write $\partial_0 \gamma = \partial_0 e_1$ for the initial vertex, and $\partial_1 \gamma = \partial_1 e_l$ for the final vertex of $\gamma$. Given two direct paths $\gamma_1 = \{ e_i \}_{i = 1}^{l_1}$ and $\gamma_2 = \{ e_i \}_{i= l_1 + 1}^{l_1 + l_2}$ such that $\partial_1 \gamma_1 = \partial_0 \gamma_2$ we can concatenate them to form a new direct path $\gamma = \{ e_i \}_{i = 1}^{l_1+l_2}$. We denote the concatenated path by $\gamma = \gamma_1 \gamma_2$. The orientation reversal of a direct path $\gamma = \{ e_i \}_{i=1}^l$ is the direct path $\bar \gamma = \bar e_l \cdots \bar e_1$. The support of a direct path $\gamma = \{ e_i \}_{i=1}^l$ is
$$\supp(\gamma) := \{ e \in \latticeedge \, : \, e = e_i \,\, \text{or} \,\, \bar e = e_i \,\, \text{for some} \,\, i = 1, \cdots, l  \}.$$
If $\supp(\gamma) \subseteq S \subseteq \latticeedge$ we say $\gamma$ is supported on $S$.

To each ribbon $\rho$ we can associate a direct path as follows. Let $\rho = \{ \tau_1, \cdots, \tau_l  \}$ be a finite ribbon, and let $J = \{ j_1, \cdots, j_{l'} \} \subset \{1, \cdots, l\}$ be the ordered subset such that $\tau_{j}$ is a direct triangle if and only if $j \in J$. Then $\rho^{dir} := \{ e_{\tau_j} \,: \, j \in J  \}$ is the direct path of $\rho$. To see that this is indeed a direct path, take indices $j_{\nu}, j_{\nu+1} \in J$ and suppose $j_{\nu+1} = j_{\nu} + m$. Then we want to show that $\partial_1 e_{\tau_{j_{\nu}}} = \partial_0 e_{\tau_{j_{\nu+1}}}$. By construction all triangles $\tau_{j_{\nu} + n}$ for $n = 1, \cdots, m-1$ are dual and therefore $v = v(\partial_0 \tau_{j_{\nu}+n}) = v(\partial_1 \tau_{j_{\nu}+n})$ are equal for all these $n$. We therefore have $\partial_1 e_{\tau_{j_{\nu}}} = v(\partial_1 \tau_{j_{\nu}}) = v$ and $\partial_0 e_{\tau_{j_{\nu}+m}} = v(\partial_0 \tau_{j_{\nu+1}}) = v$ as required. We have $\bar \rho^{dir} = \overline{ \rho^{dir} }$, and if $\rho$ is supported in $S \subseteq \latticeedge$ then $\rho^{dir}$ is also supported in $S$.

\subsubsection{Ribbon operators}

Here we describe the ribbon operators introduced by \cite{Kitaev2003-qr}, and state some of their elementary properties. For proofs and many more properties, see Appendix \ref{app:ribbonprops} of this paper or appendices B and C of \cite{Bombin2007-uw}. To each ribbon $\rho$ we associate a ribbon operator $F^{h,g}_\rho$ as follows. If $\epsilon$ is the trivial ribbon, then we set $F_{\epsilon}^{h, g} = \delta_{1, g} \mathds{1}$. For ribbons composed of a single direct triangle $\tau$ we put $F^{h,g}_\tau =  T_\tau^g$. For ribbons composed of a single dual triangle $\tau$, we put $F^{h,g}_{\tau} = \delta_{g,1} L_{\tau}^h$. For longer ribbons the ribbon operators are defined inductively as
\begin{equation} \label{eq:F inductive def}
    F^{h,g}_\rho = \sum_{k \in G} F^{h, k}_{\rho_1} F^{\overline{k}hk,\overline{k}g}_{\rho_2}
\end{equation}
for $\rho = \rho_1 \rho_2$. It follows from the discussion at the beginning of appendix \ref{app:ribbonprops} that this definition is independent of the way $\rho$ is split into two smaller ribbons. By construction, the ribbon operator $F_{\rho}^{h, g}$ is supported on $\supp(\rho)$. Let us define
$$T^g_\rho := F^{1,g}_\rho, \qquad L^h_\rho := \sum_{g \in G} F^{h,g}_\rho$$
so that $F^{h,g}_\rho = L^h_\rho T^g_\rho = T^g_\rho L^h_\rho$ (Lemma \ref{eq:F breaks into LT}).\\

\begin{figure}
    \centering
    \includegraphics[ width=0.4\textwidth]{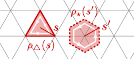}
    \caption{The elementary direct ribbon $\rho_{\triangle}(s)$ associated to the site $s$, and the elementary dual ribbon $\rho_{\star}(s')$ associated to the site $s'$.}
    \label{fig:elementary direct and dual ribbons}
\end{figure}

We define gauge transformations $A_s^h$ and flux projectors $B_s^g$ at site $s$ in terms of the ribbon operators as follows:
$$A_s^h := F^{h,1}_{\rho_{\star}(s)}, \qquad B_s^g := F^{1, g}_{\rho_{\triangle}(s)}$$
where $\rho_{\triangle}(s)$ (resp. $\rho_{\star}(s)$) is the unique counterclockwise closed direct (dual) ribbon with end sites at $s$, see Figure \ref{fig:elementary direct and dual ribbons}. It is easily verified that $A_s^{h_1} A_s^{h_2} = A_s^{h_1 h_2}$ for all $h_1, h_2 \in G$, so the gauge transformations at $s$ form a representation of $G$. Similarly, one verifies that $B_s^{g_1} B_s^{g_2} = \delta_{g_1, g_2} B_s^{g_1}$ for all $g_1, g_2 \in G$. We further note that the gauge transformations $A_s^h$ depend only on the vertex $v(s)$, so we may put $A_v^h := A_s^h$ for any site $s$ such that $v = v(s)$ and speak of the gauge transformations at the vertex $v$. Similarly, the projectors $B_s^1$ onto trivial flux depend only on the face $f(s)$ so we may put $B_f^1 = B_{s}^1$ for any site $s$ such that $f = f(s)$.

\begin{rem} \label{rem:quantum double at s}
    For each site $s$ the operators $A_s^h$ and $B_s^g$ generate a realisation of the \emph{quantum double algebra} of $G$. This fact justifies the name of the model, and will be central to our analysis.
\end{rem}

The projectors $A_v, B_f$ appearing in the quantum double Hamiltonian can now be written as follows:
$$ A_v = \frac{1}{\abs{G}} \sum_h A_v^{h}, \quad B_f = B_f^1.$$
They are the projectors onto states that are gauge invariant at $v$, and that have trivial flux at $f$, respectively.

\subsubsection{Gauge configurations and gauge transformations}

It is very helpful to think of the frustration free ground state of the quantum double model as a string-net condensate, see \cite{levin2005string}. In what follows we establish the language of string-nets, which in this case correspond to gauge configurations. For any $S \subseteq \latticeedge$ we denote by $\gc[S]$ the set of maps $\alpha : S \rightarrow G$. We will denote by $\alpha_e$ the evaluation of $\alpha$ on an edge $e \in S$. We call such maps \emph{gauge configurations} on $S$. Let us write $\orientededges[S] = \{ e \in \orientededges \, : \, e \in S \,\, \text{or} \,\, \bar e \in S \}$ for the set of oriented edges corresponding to $S$. Any gauge configuration $\alpha$ on $S \subseteq \latticeedge$ extends to a function $\alpha : \orientededges[S] \rightarrow G$ on oriented edges by setting $\alpha_{\bar e} = \bar \alpha_{e}$. The meaning of $\alpha_e$ is the parallel transport of a discrete gauge field as one traverses the edge $e$.\\ 

For any finite $V \subset \latticevert$ we define $\gauge[V]$ to be the group of unitaries generated by $\{ A_{v}^{g_v} \, : \, v \in V , \,\,  g_v \in G  \}$. Since $A_v^g$ and $A_{v'}^{g'}$ commute whenever $v \neq v'$, any element $U \in \gauge[V]$ is uniquely determined by an assignment $V \ni v \mapsto g_v \in G$ of a group element to each vertex in $V$ so that
$$U = U[\{ g_v \}] = \prod_{v \in V} \, A_v^{g_v}.$$
We call $\gauge[V]$ the group of gauge transformations on $V$.\\

If each $U \in \gauge[V]$ is supported on a set $S \subseteq \latticeedge$ then the gauge group $\gauge[V]$ acts on the gauge configurations $\gc[S]$ as follows. The gauge transformation $U = U[\{g_v\}] \in \gauge[V]$ acts on a gauge configuration $\al \in \gc[S]$, yielding a new gauge configuration $\al' := U(\al) \in \gc[S]$ given by $\al'_e = g_{\partial_0 e} \, \al_e \, \bar g_{\partial_1 e}$, where we set $g_v = 1$ whenever $v \not\in V$.\\

If $S \subset \latticeedge$ is finite then we let $\caH_{S} := \bigotimes_{e \in S} \caH_e$. The set of gauge configurations $\gc[S]$ then labels an orthonormal basis of $\caH_{S}$ given by $\ket{\al} := \bigotimes_{e \in S} \ket{\al_e}$. If the gauge transformations $\gauge[V]$ for some finite $V \subset \latticevert$ are supported in $S$ then these gauge transformations act on the Hilbert space $\caH_{S}$ as $U \ket{\al} = \ket{ U(\al) }$, i.e. Gauge transformations map basis states to basis states.\\

\subsection{Local gauge configurations and boundary conditions} \label{subsec:local gauge configurations and boundary conditions}

\begin{figure}
    \centering
    \includegraphics[ width=0.4\textwidth]{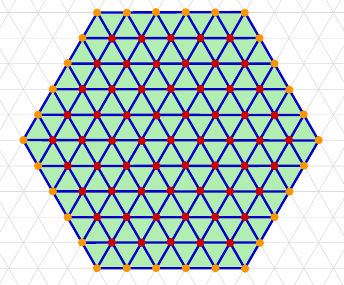}
    \caption{The sets $\vregion(\site)$ (red), $\fregion(\site)$ (green), $\eregion(\site)$ (blue), and $\partial \vregion(\site)$ (orange) are depicted for a site $\site = (v_0, f_0)$. The vertex $v_0$ sits in the center of the figure. The set $\partial \eregion$ consists of the blue edges on the boundary of the figure.}
    \label{fig:regions}
\end{figure}

Recall that $\mathrm{dist}(\cdot, \cdot)$ is the graph distance on $\latticevert$. We fix an arbitrary site $\site = (v_0, f_0)$ and define (see figure \ref{fig:regions}):
\begin{align*}
\vregion(\site) &:= \{ v \in \latticevert \, : \,  \mathrm{dist}(v,v_0) \leq n\},\\
\fregion(\site) &:= \{f \in \latticeface \, : \, \exists v \in f \text{ such that } v \in \vregion \}, \\  
\eregion(\site) &:= \{e \in \latticeedge  \, : \, \exists f \in \fregion \text{ such that } e \in f \},\\
\partial \eregion(\site) &:= \{e \in \eregion \, : \,  \exists! f \in \fregion  \text{ with } e \in f \}, \\
\partial \vregion(\site) &:=  \vregion[n+1] \setminus \vregion.
\end{align*}
Note that these regions depend on the choice of an origin $\site$. Throughout this paper, we will want to consider different sites as the origin. In order to unburden the notation we will nevertheless drop $\site$ from the notation and simply write $\vregion, \fregion, \eregion$ and $\partial \eregion$ whenever it is clear from context which site is to serve as the origin.\\

For the remainder of this section, we fix a site $\site$ as our origin. We write $\gc := \gc[\eregion]$ for the gauge configurations on $\eregion$ and let
$$ \hilb_n := \hilb_{\eregion} = \bigotimes_{e \in \eregion} \hilb_e $$
be the Hilbert space associated to the region $\eregion$. The set of gauge configurations $\gc$ then labels an orthonormal basis of $\hilb_n$, given by $\ket{\alpha} = \bigotimes_{e \in \eregion} \ket{\alpha_e}$ for all $\alpha \in \gc$.\\

\begin{figure}
    \centering
    \includegraphics[width = 0.6 \textwidth]{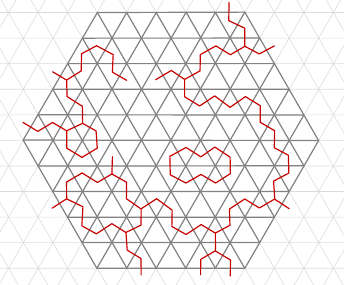}
    \caption{An example of a gauge configuration on $\eregion[5]$. The dark grey edges belong to $\eregion[5]$, they are crossed by red oriented strings that carry group labels. For edges that are not crossed by any red string, the gauge degree of freedom takes the value $1 \in G$. The orientations and group labels of the red strings are not shown. This picture corresponds to a definite basis state $\ket{\alpha} \in \hilb_n$ for some $\alpha \in \gc$.}
    \label{fig:string_net_example}
\end{figure}

For any $\alpha \in \gc$, the corresponding basis state $\ket{\alpha} \in \hilb_n$ has a graphical representation, see Figure \ref{fig:string_net_example} for a schematic example.

\begin{defn} \label{def:flux thourgh ribbon}
    For a gauge configuration $\alpha$, the \emph{flux of $\alpha$ through a ribbon $\rho$} is defined as
    $$\phi_\rho(\alpha) := \prod_{e \in \rho^{dir}} \alpha_e$$
    where the product is ordered by the order of $\rho^{dir}$, the direct path of the ribbon $\rho$ as defined in section \ref{sec:direct paths}. 
\end{defn}

We will be interested in gauge configurations that satisfy certain constraints. Recall that to any site $s$ we can associate the elementary closed direct ribbon $\rho_{\triangle}(s)$ that starts and ends at $s$ and circles $f(s)$ in a counterclockwise direction. Let $\alpha$ be a gauge configuration on a region that contains all edges of $f(s)$ and define
$$\phi_s(\alpha) := \phi_{\rho_{\triangle}(s)}(\alpha)$$
to be the flux of $\alpha$ at $s$. By construction, we have $B_s^{g} \ket{\alpha} = \delta_{g, \phi_s(\alpha)} \ket{\alpha}$. For example, the flux at $s$ for the gauge configuration $\alpha$ depicted in Figure \ref{fig:graphical_representation} is $\phi_s(\alpha) = g_1 \bar g_2 \bar g_3$.\\

Let $\bc := \gc[\partial \eregion]$ be the set of gauge configurations on $\partial \eregion$. We call its elements $b : \partial \eregion \rightarrow G$ \emph{boundary conditions}. For any gauge configuration $\alpha \in \gc$ we denote by $b(\alpha) = \alpha|_{\partial \eregion}$ the \emph{boundary condition of $\alpha$} given by restriction of $\alpha$ to the boundary $\partial \eregion$. We write $b = \emptyset$ for the trivial boundary condition $\emptyset_e = 1 \in G$ for all $e \in \partial \eregion$.\\

Having fixed a site $\site = (v_0, f_0)$ we can regard $v_0$ as the origin of the plane and define unit vectors in $\R^2$ as follows. We let $\hat y$ be the unit vector with base at $v_0$ pointing towards the center of the face $f_0$, and we let $\hat x$ be the unit vector with base at $v_0$, perpendicular to $\hat y$ and such that $\hat x \times \hat y = 1$, \ie $(\hat x, \hat y)$ is a positive basis for $\R^2$. Let us now set $\hat l_1 = \hat x$ and $\hat l_2 = \cos(\pi/3) \hat x + \sin(\pi/3) \hat y$. Then each vertex $v \in \latticevert$ can be identified with its coordinate $(n_1, n_2)$ relative to $v_0$, \ie $v = v_0 + n_1 \hat l_1 + n_2 \hat l_2$. Using these coordinates, let $v_i = (i, 0)$ for $i \in \Z$ and consider the direct path $\nu_n^{dir} = ( (v_0, v_1), (v_1, v_2), \cdots, (v_{n-1}, v_n) )$.

\begin{figure}
    \centering
    \includegraphics[width=0.8\textwidth]{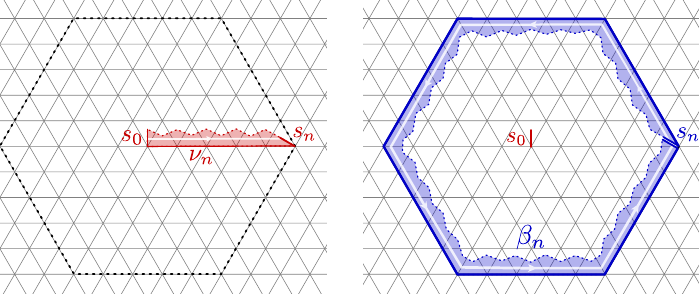}
    \caption{The fiducial ribbon $\fidu$ in red (left) and the boundary ribbon $\bdy$ in blue (right) for a given site $\site$ and $n = 5$. For the fiducial ribbon $\fidu$, we have $\start[ \fidu] = \site$ and $\en[ \fidu] = s_n$. For the boundary ribbon $\bdy$, we have $\start[\bdy] = \en[\bdy] = s_n$, and a counterclockwise orientation around $\partial \eregion$. Any site $s$ is related to the site $\site$ by lattice rotation and translation. We define the fiducial ribbons and boundary ribbons for arbitrary $s$ by the corresponding rotations and translations of the fiducial and boundary ribbons of $\site$.}
    \label{fig:fiducial_and_boundary_ribbons}
\end{figure}

We define the \emph{fiducial ribbon} $\fidu$ to be the unique positive ribbon such that $\partial_0 \fidu = \site$, such that $\nu_n^{dir}$ is the direct path of $\fidu$, and such that the final triangle of $\fidu$ is direct. We let $s_n = \partial_1 \fidu$ denote the final site of $\fidu$. See Figure \ref{fig:fiducial_and_boundary_ribbons}.

We define the \emph{boundary ribbon} $\bdy$ to be the unique closed positive ribbon starting and ending at $s_n$ such that its direct path consists of the edges in $\partial \eregion$, oriented counterclockwise around $\eregion$. See Figure \ref{fig:fiducial_and_boundary_ribbons}.

\begin{defn} \label{def:boundary condition projector}
    For any boundary condition $b \in \bc$ we define a projector $P_b \in \cstar[\eregion]$ given by
    $$P_b = \prod_{\{\tau_e \in \bdy| \tau_e \text{ direct}\}} T^{b_e}_{\tau_e}.$$
\end{defn}

\begin{defn}
    We call $\phi_{\bdy}(\alpha)$ the \emph{boundary flux} of the gauge configuration $\alpha \in \gc$.
\end{defn}

\begin{defn}
    For any boundary condition $b \in \bc$ we write $\phi_{\bdy}(b)$ for the associated \emph{boundary flux} as measured through the boundary ribbon $\bdy$.
\end{defn}

\subsection{Irreducible representations of \texorpdfstring{$\caD(G)$}{G(G)}, Wigner projectors, and local constraints} \label{subsection:irreps and Wigner projections}

As mentioned above, we have at each site $s$ a realisation of the quantum double algebra $\caD(G)$ generated by the gauge transformations and flux projectors at $s$.

Let us introduce some terminology and conventions that will allow us to analyse representations of the quantum double algebra. Denote by $(G)_{cj}$ the set of conjugacy classes of $G$. For each conjugacy class $C \in (G)_{cj}$ let $C = \{c_i\}_{i = 1}^{|C|}$ be a labeling of its elements. Any $g \in C$ has $g = c_i$ for a definite label $i$, and we define the label function $i := i(g)$. Pick an arbitrary \emph{representative element} $r_C \in C$. All elements of $C$ are conjugate to the chosen representative $r_C$ so we can fix group elements $q_i$ such that for all $c_i \in C$ we have $c_i = q_i r_C \dash{q}_i$. We let $Q_C:= \{q_i\}_{i = 1}^{|C|}$ be the \emph{iterator set} of $C$. Let $N_C:=\{n\in G| n r_C = r_C n \}$ be the commutant of $r_C$ in $G$. Note that the group structure of $N_C$ does not depend on the choice of $r_C$, it is a realization of the centralizer of $C$. Denote by $(N_C)_{irr}$ the collection of irreducible representations of $N_C$.

As mentioned in the setup, the irreducible representations of the quantum double algebra of $G$ are in one-to-one correspondence with pairs $RC$ where $C \in (G)_{cj}$ is a conjugacy class and $R \in (N_C)_{irr}$ is an irreducible representation of the group $N_C$.

For each $R \in (N_C)_{irr}$ we fix a concrete unitary matrix representation $N_C \ni m \mapsto R(m) \in \caM_{\dimR}(\C)$ with components $R^{j j'}(m)$.

In what follows we will often consider a label $i \in \{1, \cdots, \abs{C}\}$ for $C$ together with a label $j \in \{ 1, \cdots, \dimR \}$. We define $I_{RC} := \{1, \cdots, \abs{C}\} \times \{ 1, \cdots, \dimR \}$ so that $(i, j) \in I_{RC}$.

\begin{defn} \label{def:Wigner projectors}
    Let us define the Wigner projector to $RC$ at site $s$ by
    $$\qd := \frac{\dimR}{|N_C|} \sum_{m\in N_C} \chi_R(m)^* \sum_{q \in Q_C} A_{s} ^{q  m \dash{q}} B_{s}^{q r_C \dash{q}}.$$
    $\qd$ decomposes as a sum of commuting projectors $\{ \qd[RC;u] \}_{u \in I_{RC}}$ (Lemma \ref{lem:DRC decomposes into DRCu}). For $u = (i, j)$ these are given by
    $$\qd[RC;u] := \frac{\dimR}{|N_C|} \sum_{m \in N_C} R^{jj}(m)^* A_{s} ^{q_i m\dash{q}_i} B_{s}^{c_i}.$$ 
\end{defn}

Fix a site $\site = (v_0, f_0)$ and introduce the following notations
$$\dlatticeface := \latticeface \setminus \{f_0\} \qquad \dlatticevert := \latticevert \setminus \{v_0\}.$$
The site $s_0$ will be fixed throughout this section, and will therefore often not be made explicit in the notation. \\

Let us define the following sets of states.

\begin{defn} \label{def:state spaces}
    Let $\overline{\S}_{s_0}$ be the set of states $\omega$ on $\cstar$ that satisfy
    \begin{equation}
        \label{eq:cons1}
        \omega(A_v) = \omega(B_f) = 1 \quad  \text{ for all } \,\,\,  v \in \dlatticevert, \quad f \in \dlatticeface.
    \end{equation}
    Similarly, we denote by $\S_{s_0}^{RC}$ the set of states $\omega$ on $\cstar$ that in addition to \eqref{eq:cons1} also satisfy
    \begin{equation}
        \label{eq:cons2}
        \omega(D^{RC}_{s_0}) = 1,
    \end{equation}
    and by $\S_{s_0}^{RC;u}$ the set of states that in addition to \eqref{eq:cons1} also satisfy
    \begin{equation}
        \label{eq:cons3}
        \omega(\qdsu) = 1.
    \end{equation}
\end{defn}

In this section we prove that the set $\S_{\site}^{RC;u}$ contains a single pure state. Considering the case where $C = C_1 := \{1\}$ is the trivial conjugacy class and $R$ is the trivial representation of $N_{\{1\}} = G$, we see that $\S_{\site}^{RC_1}$ is precisely the set of frustration free ground states, so we get in particular a new proof of Proposition \ref{prop:ffgsunique}.

\subsection{Local constraints}

We will characterise the state spaces $\overline{\S}_{\site}, \S_{\site}^{RC}$ and $\S_{\site}^{RC;u}$ by investigating the restrictions of states belonging to these spaces to finite volumes $\eregion$. These restrictions correspond to density matrices acting on $\caH_n$ that are supported on subspaces of $\caH_n$ defined by local versions of the constraints \eqref{eq:cons1}, \eqref{eq:cons2} and \eqref{eq:cons3}. Here we introduce these subspaces.

Let us write
$$\dfregion := \fregion \setminus \{f_0\} \qquad \dvregion := \vregion \setminus \{v_0\}.$$

\begin{defn} \label{def:local constraints}
    Let $\overline{\V} \subset \hilb_n$ be the subspace consisting of vectors $\ket{\psi} \in \hilb_n$ that satisfy
    \begin{equation}
        \label{eq:lcons1}
        A_v \ket{\psi} = B_f \ket{\psi} = \ket{\psi} \quad \text{for all} \,\, v \in \dvregion, f \in \dfregion.
    \end{equation}
    Let $\VRC \subset \overline{\V}$ be the subspace consisting  of vectors $\ket{\psi} \in \overline{\V}$ that in addition to \eqref{eq:lcons1} also satisfy
    \begin{equation}
        \label{eq:lcons2}
        D_{\site}^{RC} \ket{\psi} = \ket{\psi},
    \end{equation}
    and let $\VRCu \subset \VRC$ be the subspace consisting of vectors $\ket{\psi} \in \VRC$ that in addition to \eqref{eq:lcons1} also satisfy
    \begin{equation}
        \label{eq:lcons3}
        D_{\site}^{RC;u} \ket{\psi} = \ket{\psi}.
    \end{equation}
\end{defn}

Note that since $\mathds{1} = \sum_{RC} D_{\site}^{RC}$ (Lemma \ref{lem:DRCprops}) and  $D_{\site}^{RC} = \sum_{u} D_{\site}^{RC;u}$ (Lemma \ref{lem:DRC decomposes into DRCu}) we have orthogonal decompositions
$$\overline{\V} = \bigoplus_{RC} \VRC \quad \text{and} \quad \VRC = \bigoplus_u \VRCu$$

\subsection{Imposing flux constraints and boundary conditions}

For each $RC$ and $u = (i, j) \in I_{RC}$, and each site $s$, we define
\begin{equation*}
    A_s^{RC;u} := \frac{\dimR}{|N_C|} \, \sum_{m \in N_C} \, R^{jj}(m)^* \, A_s^{q_i m \bar q_i},
\end{equation*}
so $D_s^{RC;u} = A_s^{RC;u} B_s^{c_i}$. From Lemma \ref{lem:a and B commute with A_v and B_f} we have that the $A_{s}^{RC;u}$ are projectors that commute with $B_{s}^{c_i}$. In other words, the projector $D_{s_0}^{RC;u}$ really imposes two independent constraints, namely a flux constraint $B_{\site}^{c_i}$ and a gauge constraint $A_{\site}^{RC;u}$.\\

Throughout this section we will find the following Lemma useful. Recall the projectors $P_b$ from Definition \ref{def:boundary condition projector} that project on the boundary condition $b \in \bc$.
\begin{lem} \label{lem:commuting local constraints}
    For any $C \in (G)_{cj}$, any $R \in (N_C)_{irr}$, any $u = (i, j) \in I_{RC}$, and any boundary condition $b \in \bc$, the set
    $$\{ B_f  \}_{f \in \dfregion} \cup \{ A_v \}_{v \in \dvregion} \cup \{ B_{\site}^{c_i}, \, A_{\site}^{RC;u}, \, P_b \}$$
    is a set of commuting projectors.
\end{lem}

\begin{proof}
    The set
    $$\{ B_f  \}_{f \in \dfregion} \cup \{ A_v \}_{v \in \dvregion} \cup \{ B_{\site}^{c_i}, \, A_{\site}^{RC;u}\}$$
    is a set of commuting projectors by Eq.  \eqref{eq:ABcommproj} and Lemma \ref{lem:a and B commute with A_v and B_f}. The projectors $\{ A_v \}_{v \in \dvregion} \cup \{ B_{\site}^{c_i}, \, A_{\site}^{RC;u}\}$ are all supported on $\eregion \setminus \partial \eregion$ while $P_b$ is supported on $\partial \eregion$, so these projectors commute with $P_b$. The projectors $\{ B_f  \}_{f \in \dfregion}$ and $P_b$ are all diagonal in the basis of gauge configurations, so they also commute.
\end{proof}

We first investigate the space of vectors in $\caH_n$ that satisfy flux constraints.

\begin{defn} \label{def:WCi}
    Let $\WCi \subset \caH_n$ be the subspace consisting of vectors $| \psi \rangle \in \caH_n$ such that
    \begin{equation*}
        \ket{\psi} = B_f \ket{\psi} = B_{s_0}^{c_i} \ket{\psi}
    \end{equation*}
    for all $f \in \dfregion$.
\end{defn}

The space $\WCi$ is spanned by vectors $| \al \rangle$ for certain $\al \in \gc$ that satisfy these constraints.

\begin{defn} \label{def:string nets Ci}
    For any conjugacy class $C$ and any $i = 1, \cdots, \abs{C}$ we define
    $$\packi :=\{\alpha \in \gc \, : \, \ket{\al} = B_{\site}^{c_i} \ket{\al} = B_f \ket{\al} \quad \text{for all} \,\,\, f \in \dfregion \}. $$
\end{defn}

\begin{figure}
    \centering
    \includegraphics[width=0.6\textwidth]{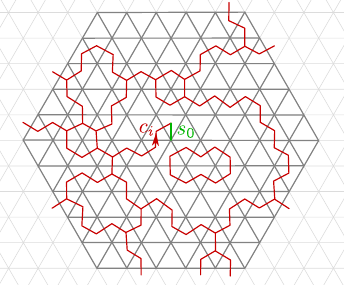}
    \caption{The graphical representation of a string net $\alpha \in \packC[C;i]$. The orientations and labels of the red strings are not shown, except for the piece that ensures that the constraint $B_{\site}^{c_i} \ket{\al} = \ket{\al}$ is satisfied.}
    \label{fig:defect string net examlpe}
\end{figure}

See Figure \ref{fig:defect string net examlpe} for an example of a string net $\al \in \packi$.

\begin{lem} \label{lem:WCi spanned by string nets}
    We have
    $$\WCi = \Span{\ket{\al} \, : \, \al \in \packi }.$$
\end{lem}

\begin{proof}
    That $\ket{\al} \in \WCi$ if $\al \in \packi$ is immediate from the definitions. Conversely, since $\{ B_f \}_{f \in \dfregion} \cup \{ B_{s_0}^{c_i} \}$ is a set of commuting projectors we have
    $$\WCi = \left( \prod_{f \in \dfregion}  B_f \right) \, B_{s_0}^{c_i} \, \caH_n.$$
    Now note that the vectors $\ket{\al}$ for $\al \in \gc$ from an orthonormal basis for $\caH_n$ and that
    $$\left( \prod_{f \in \dfregion}  B_f \right) \, B_{s_0}^{c_i} \ket{\al} = \begin{cases} \ket{\al} \quad &\text{if } \,\,\, \al \in \packi \\
    0 &\text{otherwise}.\end{cases}$$
    The claim follows.
\end{proof}

We can further refine the spaces $\WCi$ by specifying boundary conditions.

\begin{defn} \label{def:compatible boundary conditions}
    We say a boundary condition $b \in \bc$ is compatible with the conjugacy class $C$ if $\phi_{\bdy}(b) \in C$. We denote by $\bc[C]$ the set of boundary conditions compatible with $C$. For $b \in \bc[C]$ we have $\phi_{\bdy}(b) = c_i \in C$ for a definite index $i \in \{1, \cdots, \abs{C}\}$. We write $i = i(b)$.
\end{defn}

\begin{defn} \label{def:WCib}
    For any conjugacy class $C \in (G)_{cj}$, any $i = 1, \cdots, \abs{C}$, and any boundary condition $b \in \bc$ we let $\WCib \subset \caH_n$ be the space of vectors $\ket{\psi} \in \caH_n$ that satisfy
    $$\ket{\psi} = B_f \ket{\psi} = B_{\site}^{c_i} \ket{\psi} = P_b \ket{\psi}$$
    for all $f \in \dfregion$. Here $P_b$ is the projector on the boundary condition $b$ from Definition \ref{def:boundary condition projector}.
\end{defn}

\begin{defn} \label{def:string nets Cib}
    For any boundary condition $b \in \bc$ we let
    $$\packib := \{  \al \in \packi \, : \, b(\al) = b \}.$$
\end{defn}

\begin{lem} \label{lem:WCib spanned by string nets}
    We have
    $$\WCib := \Span{ \ket{\al} \, : \, \al \in \packib }.$$
\end{lem}

\begin{proof}
    This follows from $\WCi = \Span{ \ket{\al} \, : \, \al \in \packi  }$ (Lemma \ref{lem:WCi spanned by string nets}), the fact that $\al \in \packib$ if and only if $\al \in \packi$ and $P_b \ket{\al} = \ket{\al}$, and the fact (Lemma \ref{lem:commuting local constraints}) that $\{B_f\}_{f \in \dfregion} \cup \{ B_{\site}^{c_i}, \, P_b \}$ is a set of commuting projectors.
\end{proof}

\begin{lem} \label{lem:decomposition of WCi according to boundary conditions}
    We have a disjoint union
    $$\packi = \bigsqcup_{b \in \bc[C]} \packib$$
    and an orthogonal decomposition
    $$\WCi = \bigoplus_{b \in \bc[C]}  \WCib.$$

    In particular, $\packib$ is empty if $b$ is not compatible with $C$.
\end{lem}

\begin{proof}
    To show the first claim, it is sufficient to show that $\packib$ is empty if $b$ is not compatible with $C$. This follows from Lemma \ref{lem:bcinC}.
    The second claim then follows immediately from Lemma \ref{lem:WCib spanned by string nets} and Lemma \ref{lem:WCi spanned by string nets}.
\end{proof}

\subsection{Fiducial flux}

The fiducial flux, which is measured by the projectors $T_{\fidu}^g$, remains unconstrained by the projectors defining the spaces $\WCib$, we can therefore further decompose the spaces $\WCib$ according to the fiducial flux.

\begin{lem} \label{lem:commuting constraints with fiducial flux}
    For any $C \in (G)_{cj}$, any $i = 1, \cdots, \abs{C}$, any $g \in G$, and any boundary condition $b \in \bc$, the set
    $$\{ B_f  \}_{f \in \dfregion} \cup \{ A_v \}_{v \in \dvregion} \cup \{ B_{\site}^{c_i}, \, P_b, \, T_{\fidu}^g \}$$
    is a set of commuting projectors.
\end{lem}

\begin{proof}
    That 
    $$\{ B_f  \}_{f \in \dfregion} \cup \{ A_v \}_{v \in \dvregion} \cup \{ B_{\site}^{c_i}, \, P_b\}$$
    is a set of commuting projectors follows from Lemma \ref{lem:commuting local constraints}. To prove the lemma, we must show that $T_{\fidu}^g$ commutes with all the projectors in this set. From Lemma \ref{lem:flux change} we get that $T_{\fidu}^g$ commutes with all $A_v$ for $v \in \dvregion$, all $B_f$ for $f \in \dfregion$ and with $B_{\site}^{c_i}$. To see that $T_{\fidu}^g$ commutes with $P_b$, simply note that $P_b$ is supported on $\partial \eregion$ while $T_{\fidu}^{g}$ is supported on the ribbon $\fidu$, whose support does not contain any edges of $\partial \eregion$.
\end{proof}

Note that by Lemma \ref{lem:inNC} we have for any $\al \in \packib$ that $\bar q_i \phi_{\fidu}(\al) q_{i(b)} \in N_C$. This motivates the following Definition.
\begin{defn} \label{def:WCib(m)}
    For any conjugacy class $C \in (G)_{cj}$, any $i = 1, \cdots, \abs{C}$, any boundary condition $b \in \bc$, and any $m \in N_C$ we let $\WCib(m) \subset \caH_n$ be the space of vectors $\ket{\psi} \in \caH_n$ that satisfy
    $$\ket{\psi} = B_f \ket{\psi} = B_{\site}^{c_i} \ket{\psi} = P_b \ket{\psi} = T_{\fidu}^{q_i m \bar q_{i(b)}} \ket{\psi}$$
    for all $f \in \dfregion$.
\end{defn}

These spaces are again spanned by certain string-net states that have a definite fiducial flux.
\begin{defn} \label{def:string nets Cib(m)}
    For any $m \in N_C$ we define
    $$ \packib(m) := \{  \al \in \packib \, : \, \bar q_i \phi_{\fidu}(\alpha) q_{i(b)} = m  \}.$$
\end{defn}

\begin{lem} \label{lem:WCib(m) spanned by string nets}
    We have
    $$ \WCib(m) = \Span{ \ket{\al} \, : \, \al \in \packib(m) }. $$
\end{lem}

\begin{proof}
    This follows from Lemma \ref{lem:commuting constraints with fiducial flux} and the fact that
    $$ T_{\fidu}^{g} \ket{\al} = \delta_{\phi_{\fidu}(\al), g} \ket{\al}, $$
    see Lemma \ref{lem:flux projector}.
\end{proof}

\subsection{Imposing gauge invariance on \texorpdfstring{$\dvregion$}{V}}

\begin{defn} \label{def:VCib(m)}
    For any conjugacy class $C$, any $i = 1, \cdots, \abs{C}$, any boundary condition $b \in \bc[C]$, and any $m \in N_C$ we define $\VCib(m) \subset \caH_n$ to be the space of vectors $\ket{\psi} \in \caH_n$ that satisfy
    $$ \ket{\psi} = B_f \ket{\psi} = B_{\site}^{c_i} \ket{\psi} = P_b \ket{\psi} = T_{\fidu}^{q_i m \bar q_{i(b)}} \ket{\psi} = A_v \ket{\psi} $$
    for all $f \in \dfregion$ and all $v \in \dvregion$.
\end{defn}

We will show that the spaces $\VCib(m)$ are one-dimensional. To this end we introduce the following group of gauge transformations.
\begin{defn} \label{def:gauge group on dvregion}
    We let $\dgauge := \gauge[\dvregion]$ be the group of gauge transformations on $\dvregion$. All of its elements are unitaries of the form
    $$ U[\{ g_v\}] = \prod_{v \in \dvregion}  \, A_v^{g_v}$$
    for some $\{g_v\} \in G^{|\dvregion|}$.
\end{defn}

Just like for the average over the gauge group $\dgauge$ we have

$$ \frac{1}{\abs{\dgauge}} \, \sum_{U \in \dgauge} U = \frac{1}{\abs{G}^{\abs{\dvregion}}} \sum_{\{ g_v \} \in G^{|\dvregion|} } \, \prod_{v \in \dvregion} A_v^{g_v} = \prod_{v \in \dvregion} \,\left( \frac{1}{\abs{G}} \, \sum_{g_v \in G} A_v^{g_v} \right) = \prod_{v \in \dvregion} A_v. $$

\begin{lem} \label{lem:VCib(m) is dgauge projection of WCib(m)}
    We have
    $$ \VCib(m) = \left( \frac{1}{\abs{\dgauge}} \sum_{U \in \dgauge} U \right) \, \WCib(m).$$
\end{lem}

\begin{proof}
    Since $\frac{1}{\abs{\dgauge}} \sum_{U \in \dgauge} U = \prod_{v \in \dvregion} A_v$ it is sufficient to show that
    $$ \VCib(m) = \bigg( \prod_{v \in \dvregion} A_v \bigg) \, \WCib(m). $$
    This follows immediately from the Definition \ref{def:WCib(m)} and Lemma \ref{lem:commuting constraints with fiducial flux}.
\end{proof}

We now define unit vectors which, as we shall see, span the spaces $\VCib(m)$.
\begin{defn} \label{def:eta^Cib(m)}
    For all $n > 1$, all conjugacy classes $C \in (G)_{cj}$, all labels $i = 1, \cdots, \abs{C}$, all boundary conditions $b \in \bc[C]$, and all $m \in N_C$ we define the unit vector
    $$ \ket{\eta_n^{C;ib}(m)} := \frac{1}{\abs{\packib(m)}^{1/2}} \, \sum_{\al \in \packib(m)} \, \ket{\al}. $$
\end{defn}

We can now use the fact that $\WCib(m)$ is spanned by vectors $\ket{\al}$ with $\al \in \packib(m)$ and that the gauge group $\dgauge$ acts freely and transitively on $\packib(m)$ to show
\begin{lem} \label{lem:orthonormality of the etaCib(m)}
    The space $\VCib(m) \subset \caH_n$ is one-dimensional and spanned by the vector $\ket{ \eta_n^{C;ib}(m) }$. In particular, the vectors $\biggl\{\ket{ \eta_n^{C;ib}(m) }\bigg\}_{C, i, b, m}$ form an orthonormal family.
\end{lem}

\begin{proof}
    By Lemma \ref{lem:WCib(m) spanned by string nets}, we have $\WCib(m) = \Span{ \ket{\al} \, : \, \al \in \packib(m) }$. By Lemma \ref{lem:VCib(m) is dgauge projection of WCib(m)} it is sufficient to show that
    $$  \sum_{U \in \dgauge} U \ket{\al} \propto \ket{\eta_n^{C;ib}(m)}$$
    for all $\al \in \packib(m)$. This follows immediately from Lemma \ref{lem:transitive bulk action} which states that $\dgauge$ acts freely and transitively on $\packib(m)$:
    $$ \sum_{U \in \dgauge} U \ket{\al} = \sum_{\al' \in \packib(m)} \ket{\al'} = \abs{\packib(m)}^{1/2} \ket{\eta_n^{C;ib(m)}}. $$
\end{proof}

\subsection{Action of \texorpdfstring{$N_C$}{N C} on fiducial flux and irreducible subspaces}

The gauge transformations $A_{v_0}^{q_i m \bar q_i}$ realise a left group action of $N_C$ on the vectors $\ket{\eta_n^{C;ib}(m)}$.
\begin{lem} \label{lem:N_C action on VCib}
    For any $m_1, m_2 \in N_C$ we have
    $$ A_{\site}^{q_i m_1 \bar q_i} \ket{ \eta_{n}^{C;ib}(m_2) } = \ket{\eta_n^{C;ib}(m_1 m_2)}. $$
\end{lem}

\begin{proof}
    If $\al \in \packib(m_2)$, then by definition $\phi_{\fidu}(\al) = q_i m_2 \bar q_{i(b)}$. The gauge transformation $A_{\site}^{q_i m_1 \bar q_i}$ acts on such a string-net to yield $\ket{\al'} = A_{\site}^{q_i m_1 \bar q_i} \ket{\al}$ for a new string net $\al'$. Since $A_{\site}^{q_i m_1 \bar q_i}$ commutes with the projectors $\{ B_{\site}^{c_i}, P_b \} \cup \{ B_f \}_{f \in \dfregion}$ (Lemma \ref{lem:commuting local constraints}) and we have $\phi_{\fidu}(\al') = q_i m_1 m_2 \bar q_{i(b)}$ by Lemma \ref{lem:flux change}, we find that $\al' \in \packib(m_1 m_2)$. Since $A_{\site}^{q_i m_2 \bar q_i}$ acts invertibly on string nets, we see that it yields a bijection from $\packib(m_1)$ to $\packib(m_1 m_2)$. In particular, these sets have the same cardinality and
    \begin{align*}
        A_{\site}^{q_i m_1 \bar q_i} \ket{\eta_n^{C;ib}(m_2)} &= \frac{1}{\abs{\packib(m_2)}^{1/2}}  \sum_{\al \in \packib(m_2)} \, {q_i m_1 \bar q_i} \ket{\al} \\
                &= \frac{1}{\abs{\packib(m_1 m_2)}^{1/2}}  \sum_{\al \in \packib(m_1 m_2)} \, \ket{\al} \\  
                &= \ket{\eta_n^{C;ib}(m_1 m_2)}.
    \end{align*}
\end{proof}

The space spanned by the vectors $ \Bigl\{ \ket{\eta_n^{C;ib}(m)}  \Bigl\}_{m \in N_C}$ therefore carries the regular representation of $N_C$, with a left group action provided by the gauge transformations $A_{\site}^{q_i m \bar q_i}$ for $m \in N_C$, which are supported near the site $\site$. It turns out that this space also carries a natural right action of $N_C$ provided by unitaries supported near the boundary of $\eregion$, see Lemma \ref{lem:right action on VCib}.

We can characterise this space as follows. Let us define
\begin{defn} \label{def:VCib}
    $\VCib \subset \caH_n$ is the subspace consisting of vectors $\ket{\psi} \in \caH_n$ such that
    $$\ket{\psi} = A_v \ket{\psi} = B_f \ket{\psi} = B_{\site}^{c_i} \ket{\psi} = P_b \ket{\psi}$$
    for all $v \in \dvregion$ and all $f \in \dfregion$.
\end{defn}
Then
\begin{lem} \label{lem:VCib as direct sum}
    $$\VCib = \bigoplus_{m \in N_C} \, \VCib(m) = \Span{  \ket{\eta_n^{C;ib}(m)} \, : \, m \in N_C }.$$
\end{lem}

\begin{proof}
    The spaces $\VCib(m)$ are defined by the same constraints as the space $\VCib$, plus the constraint $T_{\fidu}^{q_i m \bar q_{i(b)}}$ on the fiducial flux (Definitions \ref{def:VCib(m)} and \ref{def:VCib}).
    
    Since $\VCib \subset \WCib$ (cf. Definition \ref{def:WCib}) is spanned by vectors $\ket{\al}$ for $\al \in \packib$ (Lemma \ref{lem:WCib spanned by string nets}) which satisfy $q_i \phi_{\fidu}(\al) \bar q_{i(b)} \in N_C$ (Lemma \ref{lem:inNC}), we have (using Lemma \ref{lem:flux projector}) $\sum_{m \in N_C} T_{\fidu}^{q_i m \bar q_{i(b)}} \, \ket{\phi} = \ket{\phi}$ for all $\ket{\phi} \in \VCib$.

    Using Lemma \ref{lem:commuting constraints with fiducial flux}, it follows that
    $$ \VCib = \sum_{m \in N_C} \, T_{\fidu}^{q_i m \bar q_{{i(b)}}} \, \VCib = \bigoplus_{m \in N_C} \, \VCib(m)$$
    as required.

    The second equality in the claim now follows immediately from Lemma \ref{lem:orthonormality of the etaCib(m)}.
\end{proof}

Since $\VCib$ carries the regular representation of $N_C$, we can construct an orthonormal basis of $\VCib$ that respects the irreducible subspaces of $\VCib$ for both the left and the right action of $N_C$.
\begin{defn} \label{def:representation basis}
    For any conjugacy class $C \in (G)_{cj}$, any irreducible representation $R \in (N_C)_{irr}$, any label $u = (i, j) \in I_{RC}$, any boundary condition $b \in \bc[C]$, and any label $j' = 1, \cdots, \dimR$, and writing $v = (b, j')$, we define a vector
    $$ \qdpure := \left(\frac{\dimR}{\abs{N_C}} \right)^{1/2} \, \sum_{m \in N_C} \, R^{j j'}(m)^*  \ket{\eta_n^{C;ib}(m)}.$$
\end{defn}

We will write $I'_{RC} := \bc[C] \times \{ 1, \cdots, \dimR \}$ for the possible values of the label $v = (b, j')$.

\begin{lem} \label{lem:etas form ONB}
    The vectors $\{ \qdpure \}$ form an orthonormal family, i.e.
    $$ \inner{\eta_n^{R_1C_1;u_1 v_1}}{ \eta_n^{R_2 C_2 ; u_2 v_2} } = \delta_{R_1 C_1, R_2 C_2} \delta_{u_1, u_2} \delta_{v_1, v_2}. $$
\end{lem}

\begin{proof}
    Let $u_1 = (i_1, j_1)$, $u_2 = (i_2, j_2)$, $v_1 = (b_1, j'_1)$ and $v_2 = (b_2, j'_2)$. Then
    $$ \inner{\eta_n^{R_1C_1;u_1 v_1}}{ \eta_n^{R_2 C_2 ; u_2 v_2} } = \frac{(\dim(R_1) \dim(R_2))^{1/2}}{\abs{N_{C_1}}} \, \delta_{C_1, C_2} \, \delta_{i_1, i_2} \delta_{b_1, b_2} \, \sum_{m \in N_{C_1}} \, R_1^{j_1 j'_1}(m) \, R_2^{j_2 j'_2}(m)^* $$
    where we used that the vectors $\ket{\eta_{n}^{C;ib}(m)}$ are orthonormal (Lemma \ref{lem:orthonormality of the etaCib(m)}). Now using the Schur orthogonality relation \eqref{eq:Schur} gives us the required result.
\end{proof}

The $\qdpure$ were in fact obtained by a unitary rotation of the states $\ket{\eta_n^{C;ib}(m)}$, and this rotation can be reversed.
\begin{lem} \label{lem:inverse rotation}
    We have for all $C \in (G)_{cj}$, all $i = 1, \cdots, \abs{C}$, all $b \in \bc[C]$, and all $m \in N_C$ that
    $$ \ket{\eta_n^{C;ib}(m)} = \sum_{R \in (N_C)_{irr}} \, \left( \frac{\dimR}{\abs{N_C}} \right)^{1/2} \, \sum_{j, j'} \, R^{j j'}(m) \, \ket{ \eta_n^{RC;(i, j)(b, j')} }. $$
\end{lem}

\begin{proof}
    We have
    \begin{align*}
        \sum_{R \in (N_C)_{irr}} \, & \left( \frac{\dimR}{\abs{N_C}} \right)^{1/2} \, \sum_{j, j'} \, R^{j j'}(m) \, \ket{ \eta_n^{RC;(i, j)(b, j')} } \\
        &= \sum_{R \in (N_C)_{irr}} \, \frac{\dimR}{\abs{N_C}}  \, \sum_{j, j'} \, \sum_{m' \in N_C} \, R^{j j'}(m) R^{j j'}(m')^* \, \ket{\eta_n^{C;ib}(m')} \\
        &=  \, \sum_{R \in (N_C)_{irr}} \, \frac{1}{\abs{N_C}} \sum_{m' \in N_C} \, \chi_R(m \bar m') \chi_R(1)^* \, \ket{\eta_n^{C;ib}(m')} \\
        &= \ket{\eta_n^{C;ib}(m)}
    \end{align*}
    where we used $\chi_R(1) = \dimR$ and the Schur orthogonality relation \eqref{eq:Schur2} for irreducible characters.
\end{proof}

\subsection{Characterisation of the spaces \texorpdfstring{$\overline{\V}$}{V}, \texorpdfstring{$\VRC$}{V RC}, and \texorpdfstring{$\VRCu$}{V RC u}}

We can now describe the spaces $\overline{\V}, \VRC$ and $\VRCu$ from Definition \ref{def:local constraints} in terms of the vectors $\qdpure$.

\begin{prop} \label{prop:local spaces spanned by etas}
    We have
    \begin{align*}
        \overline{\V} &= \Span{  \qdpure  \, : \, C \in (G)_{cj}, R \in (N_C)_{irr}, u \in I_{RC}, v \in I'_{RC} }, \\
        \VRC &= \Span{  \qdpure \, : \, u \in I_{RC}, v \in I'_{RC} }, \\
        \VRCu &= \Span{  \qdpure \,: \, v \in I'_{RC} }.
    \end{align*}
\end{prop}

\begin{proof}
    We first note that it follows from Lemma \ref{lem:decomposition of WCi according to boundary conditions}, Definition \ref{def:WCib}, and Lemma \ref{lem:commuting local constraints} that
    $$ B_{\site}^{c_i} \bigg( \prod_{f \in \dfregion}  B_f \bigg)  \, P_b  = 0$$
    whenever $b \in \bc$ is not compatible with $C$. Using this, the fact that $\sum_{C, i} B_{\site}^{c_i} = \sum_{b \in \bc} P_b = \I$, and Lemma \ref{lem:commuting local constraints} we find
    \begin{align*}
        \overline{\V} &= \bigg( \prod_{v \in \dvregion} A_v \prod_{f \in \dfregion} B_f  \bigg)  \, \caH_n = \sum_{C \in (G)_{cj}} \sum_{i = 1}^{\abs{C}} \sum_{b \in \bc}  \bigg( \prod_{v \in \dvregion} A_v \bigg) \, B_{\site}^{c_i} \bigg( \prod_{f \in \dfregion} B_f  \bigg)  \, P_b \, \caH_n \\
        &= \sum_{C \in (G)_{cj}} \sum_{i = 1}^{\abs{C}} \sum_{b \in \bc[C]} \, \bigg( \prod_{v \in \dvregion} A_v \bigg) \, B_{\site}^{c_i} \bigg( \prod_{f \in \dfregion} B_f  \bigg)  \, P_b \, \caH_n \\
        &= \bigoplus_{C \in (G)_{cj}} \bigoplus_{i = 1}^{\abs{C}} \bigoplus_{b \in \bc[C]} \, \VCib,
    \end{align*}
    where we used the Definition \ref{def:VCib} of the spaces $\VCib$. From Lemma \ref{lem:VCib as direct sum} it then follows that
    $$ \overline{\V} = \Span{ \ket{\eta_n^{C;ib}(m)} \, : \, C \in (G)_{cj}, i = 1, \cdots, \abs{C}, b \in \bc[C], m \in N_C  }. $$
    Together with Definition \ref{def:representation basis} and Lemma \ref{lem:inverse rotation} this yields the first claim.
    
    To show the second claim we note that $\VRC = D_{\site}^{RC} \overline{\V}$, and for any $C_1, C_2 \in (G)_{cj}$, $R_1 \in (N_{C_1})_{irr}$, $R_2 \in (N_{C_2})_{irr}$, $u = (i, j) \in I_{R_2 C_2}$, and $v = (b, j') \in I'_{R_2 C_2}$ we have (Lemma \ref{lem:action of Wigner projecitons on string-net condensates})
    $$ D_{\site}^{R_1 C_1} \ket{ \eta_n^{R_2 C_2 ; u v} } = \delta_{R_1 C_1, R_2 C_2}  \, \ket{\eta_n^{R_2 C_2 ; u v}}.$$
    The second claim of the Proposition then follows from the fact that $\overline{\V}$ is spanned by the vectors $\ket{\eta_n^{RC;uv}}$ for arbitrary $RC$ and $u \in I_{RC}$, $v \in I'_{RC}$.

    To show the final claim we note that we have $\VRCu = D_{\site}^{RC;u} \VRC$, and for any $u_1, u_2 \in I_{RC}$ and any $v \in I'_{RC}$ we have (Lemma \ref{lem:action of Wigner sub-projector on string-net condensates})
    $$ D_{\site}^{RC;u_1} \, \ket{ \eta_n^{RC;u_2 v} } = \delta_{u_1, u_2} \, \ket{\eta_n^{RC;u_2 v}}.$$
    The final claim then follows from the fact that $\VRC$ is spanned by the vectors $\ket{\eta_n^{RC;uv}}$ for $u \in I_{RC}$ and $v \in I'_{RC}$.
\end{proof}

\subsection{The bulk is independent of boundary conditions}

Let us define the following operators
\begin{defn} \label{def:label changers}
    For any site $s$, any $n$, any $u_1 = (i_1, j_1), u_2 = (i_2, j_2) \in I_{RC}$, and any $v_1 = (b_1, j_1'), v_2 = (b_2, j_2') \in I'_{RC}$ we define
    \begin{align*}
        A_{s}^{RC; u_2 u_1} &:= \frac{\dimR}{|N_C|}\sum_{m \in N_C} R^{j_2j_1}(m)^* A_{s}^{q_{i_2} m \dash{q}_{i_1}}, \\
        \tilde{A}_n^{RC; v_2 v_1} &:=\frac{\dimR}{|N_C|} \sum_{m \in N_C} R^{j'_2j'_1}(m)  U_{b_2 b_1} L_{\bdy}^{{q_{i(b_1)} \dash{m} \,\dash{q}_{i(b_1)}}}
    \end{align*}
    where $\bdy$ is the boundary ribbon and $U_{b_2 b_1}$ is a boundary unitary provided by Lemma \ref{lem:transitive boundary action} which we choose such that $U_{b_2 b_1} = U_{b_1 b_2}^*$. These boundary unitaries satisfy the following: for any $\alpha \in \packC[C;ib_1]$ we have $U_{b_2 b_1} \ket{\alpha} = \ket{\alpha'}$ where $\alpha' \in \packC[C;i b_2]$, and $\alpha'_e = \alpha_e$ for all $e \in \eregion[n-1]$ and $b(\alpha') = b_2$.
\end{defn}

Note that the $\tilde{A}_n^{RC; v_2 v_1}$ are supported on $\eregion \setminus \eregion[n-1]$ and $A_{\site}^{RC; u_2 u_1}$ is supported on $\eregion[1]$. From Lemma \ref{lem:aconverter} we have for any $u, u_1, u_2 \in I_{RC}$ and any $v, v_1, v_2 \in I'_{RC}$ that
$$A_{\site}^{RC; u_2 u_1} \qdpure[RC;u_1 v] = \qdpure[RC;u_2 v], \quad \quad \tilde{A}_n^{RC; v_2 v_1} \qdpure[RC;u v_1] = \qdpure[RC; u, v_2]$$
as well as
$$(A_{\site}^{RC; u_1 u_2})^* \qdpure[RC;u_1 v] = \qdpure[RC;u_2 v], \quad \quad (\tilde{A}_n^{RC; v_1 v_2})^* \qdpure[RC;u v_1] = \qdpure[RC; u, v_2]$$
\ie these operators change the labels $u$ and $v$ when acting on the states $\qdpure$. We can use these `label changers' to show that expectation values of operators supported on $\eregion[n-1]$ in the state $\qdpure$ are independent of the boundary label $v$.

\begin{lem}
\label{lem:purerest}
    We have for all $O \in \cstar[{\eregion[n-1]}]$, all $u, u' \in I_{RC}$, and all $v, v', v'' \in I'_{RC}$ that
    $$\inner{\eta_n^{uv}}{O \eta_n^{u'v'}} = \delta_{vv'} \inner{\eta_n^{u'v''}}{O \, \eta_n^{u'v''}}.$$
    In particular, $\inner{\qdstpure}{O \, \qdstpure}$ is independent of $v$. 
\end{lem}

\begin{proof}
Using Lemma \ref{lem:aconverter} and the fact that any $O \in \cstar[{\eregion[n-1]}]$ commutes with $\tilde{A}_n^{RC;v v'}$ we find that if $v = (b,j)$ and $v' = (b',j')$ then 
\begin{align*}
    \inner{\eta_n^{uv}}{O \eta_n^{u'v'}} &= \inner{\eta_n^{uv}}{P_{b}O P_{b'} \eta_n^{u'v'}} = \delta_{bb'} \inner{\eta_n^{uv}}{O \eta_n^{u'v'}}\\ 
    &= \inner{\qdstpure[RC;u v'']}{O \, (\tilde{A}_n^{RC;v v''})^{*} \tilde{A}_n^{RC; v' v''} \qdstpure[RC;u' v'']} = \delta_{vv'} \inner{\eta_n^{u'v''}}{O \, \eta_n^{u'v''}},
\end{align*}
where in the last equality we have used the fact that if $v,v'$ have the same boundary $b$ then from lemma \ref{lem:action of two different label changers on eta}, $(\tilde{A}_n^{RC;v' v''})^{*} \tilde{A}_n^{RC; v v''} \ket{\qdstpure[RC;u' v'']}= \delta_{vv'}\ket{\qdstpure[RC;u' v'']}$.
\end{proof}

This Lemma shows that the following is well-defined.
\begin{defn} \label{def:uniform string net superpositions}
    For any $n$ we define the states $\qdstrest$ on $\cstar[{\eregion[n]}]$ by
    $$ \qdstrest(O) := \inner{\eta_{n+1}^{RC;uv}}{ O \, \eta_{n+1}^{RC;uv} } $$
    for any $O \in \cstar[{\eregion}]$ and any boundary label $v$. The choice of boundary label does not matter due to Lemma \ref{lem:purerest}.
\end{defn}

\subsection{Construction of the states \texorpdfstring{$\qdstate$}{omega RC} and proof of their purity}

The following basic Lemma will be useful throughout the paper.
\begin{lem}
    \label{lem:pure components projector Lemma}
    Let $\omega = \sum_{\kappa} \lambda_{\kappa} \omega^{(\kappa)}$ a state on $\cstar[\eregion]$ expressed as a finite convex combination of pure states $\omega^{(\kappa)}$ with positive coefficients $\lambda_{\kappa} > 0$. If $P \in \cstar[\eregion]$ is a projector and $\omega(P) = 1$, then $\omega^{(\kappa)}(P) = 1$ for all $\kappa$.
    
    Moreover, if $\ket{\Omega^{(\kappa)}} \in \hilb_n$ is a unit vector such that $\omega^{(\kappa)}(O) = \inner{\Omega^{(\kappa)}}{ O \, \Omega^{(\kappa)}}$ for all $O \in \cstar[\eregion]$, then $P \ket{\Omega^{(\kappa)}} = \ket{\Omega^{(\kappa)}}$.
\end{lem}

\begin{proof}
       Since $\omega^{(\kappa)}(P) \leq 1$ and the positive numbers $\lambda_{\kappa} > 0$ sum to one, the equality
       $$1 = \omega(P) = \sum_{\kappa} \lambda_{\kappa} \omega^{(\kappa)}(P)$$ 
       can only be satisfied if $\omega^{(\kappa)}(P) = 1$ for all $\kappa$. If $\omega^{(\kappa)}(\cdot) = \inner{\Omega^{(\kappa)}}{ \cdot \, \Omega^{(\kappa)}}$ for a unit vector $\ket{\Omega^{(\kappa)}} \in \hilb_n$ then in particular
       $$1 = \omega^{(\kappa)}(P) = \inner{\Omega^{(\kappa)}}{\, P \, \Omega^{(\kappa)}}.$$
       Since $P$ is an orthogonal projector, this implies $P | \Omega^{(\kappa)} \rangle = \ket{\Omega^{(\kappa)}}$.
\end{proof}

Let us define the following sets of states on $\cstar[\eregion]$.
\begin{defn} \label{def:local RCu constraints for states}
    The set $\mathcal{S}_n^{RC;u}$ consists of states $\omega$ on $\cstar[\eregion]$ such that
    $$ 1 = \omega( D_{\site}^{RC;u} ) = \omega(A_v) = \omega(B_f) $$
    for all $v \in \dvregion$ and all $f \in \dfregion$.
\end{defn}

\begin{lem}
\label{lem:restriction yields eta^RCu}
    Let $1 \leq m < n$. If $\omega \in \mathcal{S}_n^{RC;u}$, and $\omega_m$ is its restriction to $\cstar[{\eregion[m]}]$, then $\omega_m = \qdstrest[m]$.
\end{lem}
\begin{proof}
    Let $\omega_{m+1}$ be the restriction of $\omega$ to $\cstar[{\eregion[m+1]}]$, then $\omega_{m+1} \in \mathcal{S}_{m+1}^{RC;u}$. Let $\omega_{m+1} = \sum_{\kappa} \lambda_{\kappa} \omega_{m+1}^{(\kappa)}$ be the convex decomposition of $\omega_{m+1}$ into finitely many pure components $\omega_{m+1}^{(\kappa)}$. Let $\ket{\Omega_{m+1}^{(\kappa)}} \in \hilb_{m+1}$ be unit vectors corresponding to these pure states. We conclude from Lemma \ref{lem:pure components projector Lemma} that 
    $$A_v \ket{\Omega_{m+1}^{(\kappa)}} = B_f \ket{\Omega_{m+1}^{(\kappa)}} = \qdsu \ket{\Omega_{m+1}^{(\kappa)}} = \ket{\Omega_{m+1}^{(\kappa)}}$$
    for all $\kappa$, all $v \in \dvregion[m+1]$, and all $f \in \dfregion[m+1]$. By Definition \ref{def:local constraints} this means that $\ket{\Omega_{m+1}^{(\kappa)}} \in \mathcal{V}_{m+1}^{RC;u}$ for all $\kappa$. From Proposition \ref{prop:local spaces spanned by etas} it follows that the unit vectors $\ket{\Omega_{m+1}^{(\kappa)}}$ are linear combinations of the $| \eta_{m+1}^{RC;uv} \rangle$ for $v \in I'_{RC}$. Using Lemma \ref{lem:purerest} and Definition \ref{def:uniform string net superpositions} we then have that for any $O \in \eregion[m]$
    \begin{align*}
        \omega_{m+1}^{(\kappa)}(O) &= \inner{ \Omega_{m+1}^{(\kappa)}}{ O \, \Omega_{m+1}^{(\kappa)}} = \eta_{m}^{RC;u}(O)
    \end{align*}
    independently of $\kappa$. The claim follows.
\end{proof}

We define extensions of the states $\qdstrest[n]$ to the whole observable algebra.
\begin{defn}
\label{def:excitation extension}
    We let $\tilde \eta_n^{RC;u}$ be the following extension of $\qdstrest[n]$ to the whole observable algebra. For each $e \in \latticeedge$, let $\zeta_e$ be the pure state on $\cstar[e]$ corresponding to the vector $| 1_e \rangle \in \hilb_e$, and put
    $$\tilde \eta_n^{RC;u} := \qdstrest[n] \otimes \left( \bigotimes_{e \in \latticeedge \setminus \eregion} \zeta_e \right).$$
\end{defn}

Recall the space of states $\S_{s_0}^{RC;u}$ from Definition \ref{def:state spaces}.

\begin{lem}
\label{lem:qdconvergence}
    The sequence of states $\tilde \eta_n^{RC;u}$ converges in the weak-$^*$ topology to a state $\qdstate \in \S_{s_0}^{RC;u}$.
\end{lem}

\begin{proof}
    If $O \in \cstar[{\eregion[m]}]$ then $\tilde \eta_n^{RC;u}(O) = \eta_n^{RC;u}(O)$ for all $n > m$ by construction. Since $\eta_n^{RC;u} \in \mathcal{S}_n^{RC;u}$ we have from Lemma \ref{lem:restriction yields eta^RCu} that $\tilde \eta_n^{RC;u}|_m = \qdstrest[m]$. It follows that $\tilde \eta_n^{RC;u}(O)$ is constant for all $n > m$ and hence converges. Since $m$ was chosen arbitrarily, $\tilde \eta_n^{RC;u}$ converges for any local observable $O \in \cstar[loc]$. Since $\cstar[loc]$ is dense in $\cstar$, the states $\omegaRCu$ converge in the weak-$^*$ topology to some state $\qdstate$ that satisfies the constraints \eqref{eq:cons1} and $\eqref{eq:cons3}$, i.e. $\qdstate \in \S_{\site}^{RC;u}$.
\end{proof}

\begin{lem}
\label{lem:qdunique}
    $\qdstate$ is the unique state in $\S_{\site}^{RC;u}$. It is therefore a pure state.
\end{lem}

\begin{proof}
    Consider any other state $\omega' \in \S_{\site}^{RC;u}$. Then its restriction $\omega'_n$ to $\cstar[\eregion]$ is a state in $\mathcal{S}^{RC;u}_n$. By Lemma \ref{lem:restriction yields eta^RCu} we have $\omega'(O) = \omega'_n(O) = \qdstrest[m](O) = \omegaRCu(O)$ for all $m < n$ and all $O \in \cstar[{\eregion[m]}]$. It follows that $\omega'$ agrees with $\qdstate$ on all local observables and therefore must be the same state.

    To see that this implies that $\qdstate$ is pure, suppose $\qdstate = \lambda \omega' + (1 - \lambda) \omega''$ can be written as a convex combination of states $\omega'$ and $\omega''$. Then for any $v \in \dlatticevert$ we have $1 = \omega_0(A_v) = \lambda \omega'(A_v) + (1 - \lambda) \omega''(A_v)$. Since $A_v$ is a projector we have $\abs{\omega'(A_v)}, \abs{\omega''(A_v)} \leq 1$, so the previous equality can only be satisfied if $1 = \omega'(A_v) = \omega''(A_v)$. By the same reasoning, $1 = \omega'(B_f) = \omega''(B_f)$ for all $f \in \dlatticeface$, and similarly for the projector $D_{s_0}^{RC;u}$. We conclude that $\omega'$ and $\omega''$ both belong to $\S_{\site}^{RC;u}$ and are therefore equal to $\qdstate$. Thus $\qdstate$ is pure.
\end{proof}

Since the site $\site$ was arbitrary, we have in particular shown
\begin{prop} \label{prop:characterisation of S^RCu}
    For any site $s_0$, any irreducible representation $RC$ of $\caD(G)$, and any label $u$, the space of states $\S_{s_0}^{RC;u}$ of Definition \ref{def:state spaces} consists of a single pure state $\omega_{s_0}^{RC;u}$.
\end{prop}

%% file: sectors/anyon_representations.tex
\section{Construction of anyon representations} \label{sec:anyon representations}

In this section we show that the pure states $\omega_{s_1}^{R_1C_1;u_1}, \omega_{s_2}^{R_2C_2;u_2}$ constructed in the previous section are equivalent to each other whenever $R_1C_1 = R_2C_2$. The collection of pure states $\{\omega_{s}^{RC;u}\}_{s, u}$ for fixed $RC$ therefore belong to the same irreducible representation $\pi^{RC}$ of the observable algebra. We will show that the irreducible representations $\{\pi^{RC}\}_{RC}$ are pairwise disjoint. In other words, we show that different $RC$ label different superselection sectors. Finally, we will show that the representations $\pi^{RC}$ are anyon representations by relating them to the so-called amplimorphism representations of \cite{Naaijkens2015-xj}.

\subsection{Ribbon operators and their limiting maps} \label{sec:ribbon operators and limiting maps}

From this point onward, the \emph{ribbon operators} introduced in Section \ref{subsec:preliminary notions} will play an increasingly important role in the analysis. By taking certain linear combinations of these ribbon operators, we construct new ribbon operators that can produce, transport, and detect anyonic excitations above the frustration free ground state. \\

Recall from  section \ref{subsec:preliminary notions} that we can associate to any finite ribbon $\rho$ some ribbon operators $F_{\rho}^{h, g}$. The following linear combinations of these ribbon operators are designed so that when acting on the ground state, they produce excitations that sit in irreducible representations for the action of the quantum double algebra at the endpoints of $\rho$.

\begin{defn}[\cite{Bombin2007-uw}] \label{def:RC ribbons}
    For each irreducible representation $RC$ of the quantum double we define
    \begin{align*}
        \frcuv :=\frac{\dimR}{|N_C|} \sum_{m \in N_C} {R}^{jj'}(m)^* F^{\dash{c}_i, q_i m \dash{q}_{i'}}_\rho 
    \end{align*}
    where $u = (i,j) \in I_{RC}$ and $v = (i',j') \in I_{RC}$.    
\end{defn}

\begin{defn}[\cite{naaijkens2012anyons}, \cite{Naaijkens2015-xj}] \label{def:finite mu}
    For any finite ribbon $\rho$, any $RC$, and any $u_1, u_2 \in I_{RC}$ we define a linear map from $\cstar$ to itself by
    $$ \mu^{RC; u_1 u_2}_{\rho}(O) := \bigg(\frac{|N_C|}{\dimR} \bigg)^2 \sum_{v} \, \big( F_{\rho}^{RC; u_1 v} \big)^* \, O \, F_{\rho}^{RC; u_2 v} $$
    for any $O \in \cstar$.
\end{defn}

We define a half-infinite ribbon to be a sequence $\rho = \{ \tau_i \}_{i \in \N}$ of triangles such that $\partial_1 \tau_i = \partial_0 \tau_{i+1}$ for all $i \in \N$, and such that for each edge $e \in \latticeedge$, there is at most one triangle $\tau_i$ for which $\tau_i = (\partial_0 \tau_i, \partial_1 \tau_i, e)$. We denote by $\partial_0 \rho = \partial_0 \tau_1$ the initial site of the half-infinite ribbon, and by $\rho_n = \{\tau_i\}_{i = 1}^n$ the finite ribbon consisting of the first $n$ triangles of $\rho$. A half-infinite ribbon is positive if all of its triangles are positive, and negative if all of its triangles are negative. Any half-infinite ribbon is either positive or negative.\\

The following Proposition due to \cite{Naaijkens2015-xj} says that we can define $\enRC[RC;u_1 u_2]$ as limits of $\mu_{\rho_n}^{RC;u_1 u_2}$, and states some properties of these limiting maps.

\begin{prop}[Lemma 5.2 in \cite{Naaijkens2015-xj}] \label{prop:ampli properties}
    Let $\rho$ be a half-infinite ribbon. The limit
    $$\enRC[RC;u_1 u_2](O) := \lim_{n \rightarrow \infty} \mu_{\rho_n}^{RC;u_1 u_2} (O)$$
    exists for all $O \in \cstar$ and all $u_1, u_2 \in I_{RC}$, and defines a linear map from $\cstar$ to itself. Moreover, the maps $\mu_{\rho}^{RC;uu}$ are positive, and
    \begin{enumerate}
        \item if $O \in \cstar[loc]$ then there is a finite $n_0$ such that $\mu_{\rho}^{RC;u_1 u_2}(O) = \mu_{\rho_n}^{RC;u_1 u_2}(O)$ for all $n \geq n_0$,
        \item $\mu^{RC; u_1 u_2}_{\rho}(\I) = \delta_{u_1, u_2} \I$.
		\item $\mu^{RC; u_1 u_2}_{\rho}(O) = \delta_{u_1, u_2} O$ if the support of $O$ is disjoint from the support of $\rho$.
		\item $\mu^{RC; u_1 u_2}_{\rho}(OO') = \sum_{u_3 \in I_{RC}} \mu^{RC; u_1 u_3}_{\rho}(O) \mu^{RC; u_3 u_2}_{\rho}(O')$.
		\item $\mu_{\rho}^{RC; u_1 u_2}(O)^* = \mu_{\rho}^{RC; u_2 u_1}(O^*)$.
    \end{enumerate}
\end{prop}

\begin{proof}
    The only thing that is not coming directly from \cite{Naaijkens2015-xj}'s Lemma 5.2 is the claim that the maps $\mu_{\rho}^{RC;uu}$ are positive. To see this, simply note that for any $O \in \cstar$ and using items 4 and 5 we have
    $$\mu_{\rho}^{RC;uu}(O^* O) = \sum_{v} \, \mu_{\rho}^{vu}(O)^* \, \mu_{\rho}^{vu}(O) \geq 0.$$
\end{proof}

Let $\omega_0$ be the frustration free ground state and $(\pi_0, \caH_0, | \Omega_0 \rangle)$ its GNS triple. We write $\chi_{\rho}^{RC; u v} := \pi_0 \circ \mu_{\rho}^{RC; u v} : \cstar \rightarrow \caB(\caH_0)$.

\begin{lem} \label{lem:qdstate from mu action}
    Let $\rho$ be a half-infinite ribbon with $\partial_0 \rho = \site$ for any site $\site$. For any $RC$ and any $u \in I_{RC}$ we have
    $$\omega_{\site}^{RC; u} = \omega_0 \circ \mu_{\rho}^{RC;uu}.$$
\end{lem}

\begin{proof}
    $\omega_0 \circ \mu_{\rho}^{RC;uu}$ is a positive linear functional by Proposition \ref{prop:ampli properties}. Normalisation follows from item 2 of Proposition \ref{prop:ampli properties}.

    Since $\omega_{\site}^{RC;u}$ is completely characterised by
    $$ 1 = \omega_{\site}^{RC;u}(A_v) = \omega_{\site}^{RC;u}(B_f) = \omega_{\site}^{RC;u}(D_{\site}^{RC;u})$$
    for all $v \in \dlatticevert$ and all $f \in \dlatticeface$ (Proposition \ref{prop:characterisation of S^RCu}), it is sufficient to show that $\omega_0 \circ \mu_{\rho}^{RC;uu}$ also satisfies these constraints.
    
    Since for any observable $O \in \cstar$ we have
    $$ (\omega_0 \circ \mu_{\rho}^{RC;uu})(O) = \langle \Omega_0, \chi_{\rho}^{RC;uu}(O) \, \Omega_0 \rangle,$$
    this follows immediately from Lemmas \ref{lem:ampli label projector} and \ref{lem:chi preserves constraints}.
\end{proof}

\begin{lem} \label{lem:transport of anyons}
    For any two sites $s, s'$, any $RC$, and any two labels $u, u' \in I_{RC}$ there is a local operator $T \in \cstar[loc]$ such that
    $$ \omega_{s'}^{RC;u'}(O) = \omega_s^{RC;u}(T O T^*)  $$
    for all $O \in \cstar.$
\end{lem}

\begin{proof}
    Let $\rho$ be a half-infinite ribbon having $\partial_0 \rho = s$, and a half-infinite subribbon $\rho'$ with $\partial_0 \rho' = s'$. Then $\rho = \rho_1 \rho'$ for a finite ribbon $\rho_1$. Let $O \in \cstar$. Using Lemma \ref{lem:qdstate from mu action} we now compute
    \begin{align*}
        \omega_{s}^{RC;u}(O) &= (\omega_0 \circ \mu_{\rho}^{RC;uu})(O) \\
        &= \lim_{n \uparrow \infty} \, \bigg(\frac{|N_C|}{\dimR} \bigg)^2 \,  \langle \Omega_0, \, \sum_{v} \, (F_{\rho_1 \rho'_n}^{RC;uv})^* \,O \, F_{\rho_1 \rho'_n}^{RC;uv} \, \Omega_0 \rangle \\
        \intertext{using Lemma \ref{lem:decomposition of F}}
        &= \lim_{n \uparrow \infty} \, \bigg(\frac{|N_C|}{\dimR} \bigg)^4 \sum_{v, w_1, w_2}  \, \langle \Omega_0, \, (F_{\rho'_n}^{RC;w_1v})^* (F_{\rho_1}^{RC;uw_1})^* \, O \, F_{\rho_1}^{RC;uw_2} \, F_{\rho'_n}^{RC;w_2 v} \, \Omega_0 \rangle \\
        \intertext{then using Lemma \ref{lem:change ribbon operator label}}
        &= \lim_{n \uparrow \infty} \, \bigg(\frac{|N_C|}{\dimR} \bigg)^4 \sum_{v, w_1, w_2}  \, \langle \Omega_0, \, (F_{\rho'_n}^{RC;u'v})^*  (A_{s'}^{RC;w_1 u'})^*  (F_{\rho_1}^{RC;uw_1})^* \\
        & \quad\quad\quad\quad\quad\quad\quad\quad\quad\quad\quad\quad \times \, O \, F_{\rho_1}^{RC;uw_2} \, A_{s'}^{RC;w_2 u'} F_{\rho'_n}^{RC;u'v} \, \Omega_0 \rangle \\
        &= (\omega_0 \circ \mu_{\rho'}^{RC;u'u'}) \bigg(  \bigg(\frac{|N_C|}{\dimR} \bigg)^2 \sum_{w_1, w_2} \, (A_{s'}^{RC;w_1 u'})^*  (F_{\rho_1}^{RC;uw_1})^* \, O \, F_{\rho_1}^{RC;uw_2} \, A_{s'}^{RC;w_2 u'} \bigg)
    \end{align*}
    which proves the claim with
    $$T =  \bigg(\frac{|N_C|}{\dimR} \bigg) \, \sum_w \, (A_{s'}^{RC;w u'})^*  (F_{\rho_1}^{RC;uw})^*.$$
\end{proof}

\subsection{Anyon representations labeled by \texorpdfstring{$RC$}{RC}} \label{sec:construction of piRC}

We define the following GNS representations.
\begin{defn} \label{def:piRC}
    Fix a site $\bsite$. For each $RC$, let $(\pi^{RC}, \hilb^{RC}, \ket{\Omega^{RC; (1, 1)}_{\bsite}})$ be the GNS triple for the pure state $\omega_{\bsite}^{RC;(1, 1)}$.
\end{defn}
Note that $\omega_{\bsite}^{\trivRC;(1, 1)} = \omega_0$ is the \ffgs, so $\pi^{\trivRC} = \pi_0$ is the ground state representation.

In this Section we will show that the representations $\{\pi^{RC}\}_{RC}$ are pairwise disjoint anyon representations with respect to the ground state representation $\pi_0$. In Section \ref{sec:completeness} we will show that any anyon representation is unitarily equivalent to one of the $\pi^{RC}$.

\begin{defn}
\label{def:state belongs to a representation}
    We say a state $\psi$ on $\cstar$ \emph{belongs} to a representation $\pi : \cstar \rightarrow \Bhilb$ of the observable algebra if there is a density operator $\rho \in \Bhilb$ such that
    $$\psi(O) = \Tr \lbrace \rho \pi(O) \rbrace$$
    for all $O \in \cstar$ (This notion is called being \emph{$\pi$-normal} in the operator algebra literature). If $\psi$ is pure and belongs to an irreducible representation $\pi$, then the corresponding density operator is a rank one projector, i.e. $\psi$ has a vector representative in the representation $\pi$. In this case we say $\psi$ is a vector state of $\pi$. Conversely, if $\psi$ belongs to a representation $\pi$, then we say $\pi$ \emph{contains} the state $\psi$.
\end{defn}

We first note that the representation $\pi^{RC}$ contains all the pure states $\{\omega_s^{RC;u}\}_{s, u}$. Since $\pi_{RC}$ is irreducible, it follows that all these states are equivalent to each other.

\begin{lem} \label{lem:statesbelongingtopiRC}
    The pure states $\omega^{RC;u}_{s}$ are vector states of $\pi^{RC}$ for all sites $s$ and all $u \in I_{RC}$.
\end{lem}

\begin{proof}
    This follows immediately from Lemma \ref{lem:transport of anyons}.
\end{proof}

We choose representative vectors for the states $\omega_{s}^{RC;(1, 1)}$ as follows.

\begin{defn}
    For all sites $s \neq \bsite$ we choose unit vectors $\ket{\Omega_s^{RC;(1, 1)}} \in \hilb^{RC}$ such that
    $$ \omega_s^{RC;(1, 1)}(O) = \inner{\Omega_{s}^{RC;(1, 1)}}{\pi^{RC}(O) \,\Omega_{s}^{RC;(1, 1)}} $$
    for all $O \in \cstar$. Such vectors exist by Lemma \ref{lem:statesbelongingtopiRC}, (note that the corresponding vector $\ket{\Omega_{\bsite}^{RC;(1, 1)}}$ for the site $\bsite$ was already fixed in Definition \ref{def:piRC}.)
\end{defn}

\subsection{Disjointness of the representations \texorpdfstring{$\pi^{RC}$}{pi RC}}

We prove that $\pi^{RC}$ and $\pi^{R'C'}$ are disjoint whenever $RC \neq R'C'$.

Let us first show the following basic Lemma, which is due to  \cite{alicki2007statistical}.
\begin{lem} \label{lem:absorption of satisfied projectors}
    Let $\omega$ be a state on $\cstar$ and $P \in \cstar$ an orthogonal projector satisfying $\omega(P) = 1$. Then, $\omega(P O) = \omega(O P) = \omega(O)$ for all $O \in \cstar$.
\end{lem}

\begin{proof}
    Using the Cauchy-Schwarz inequality,
    $$ \abs{\omega(O - PO)}^2 = \abs{\omega(O (\I - P) (\I - P))}^2 \leq \omega( O (\I - P) O^* ) \omega(\I - P) = 0  $$
    which show that $\omega(O) = \omega(PO)$. The equality $\omega(O) = \omega(OP)$ is shown in the same way.
\end{proof}

\begin{defn} (\cite[Eq. (B75)]{Bombin2007-uw}) \label{def:charge detectors}
    For any closed ribbon $\sigma$ we put
    $$\knRC[\sigma] := \dimchbasis \sum_{m \in N_C} \chi_R(m)^* \sum_{q \in Q_C} F_\sigma^{q m \dash{q}, q r_C \dash{q}}.$$
\end{defn}

\begin{lem} \label{lem:projector Lemma}
    If $\omega \in \overline{\S}_{\site}$ then $\omega( O \qds ) = \omega(O K_{\bdy}^{RC}) = \omega(\qds O) = \omega(K_{\bdy}^{RC} O )$ for all $RC$, all $n>1$, and all $O \in \cstar[{\eregion[n]}]$. In fact, $\omega(O D_{\site}^{RC}) = \omega(D_{\site}^{RC} O)$ holds for all $O \in \cstar$.
\end{lem}

\begin{proof}
    The restriction $\omega_n$ of $\omega$ to $\cstar[\eregion]$ satisfies
    $$ 1 = \omega_n(A_v) = \omega_n(B_f) $$
    for all $v \in \dvregion$ and all $f \in \dfregion$. Let $\omega_n = \sum_{\kappa} \lambda_{\kappa} \, \omega_n^{(\kappa)}$ be the convex decomposition of $\omega_n$ into its pure components $\omega_n^{(\kappa)}$, and let $\ket{\Omega_n^{(\kappa)}} \in \caH_n$ be unit vectors such that
    $$\omega_n^{(\kappa)}(O') = \inner{\Omega_n^{(\kappa)}}{O' \, \Omega_n^{(\kappa)}}$$
    for all $O' \in \cstar[\eregion]$. From Lemma \ref{lem:pure components projector Lemma} we find that
    $$ 1 = \omega_n^{(\kappa)}(A_v) = \omega_n^{(\kappa)}(B_f)$$
    and
    $$ \ket{\Omega_n^{(\kappa)}} = A_v \ket{\Omega_n^{(\kappa)}} = B_f \ket{\Omega_n^{(\kappa)}}$$
    for all $v \in \dvregion$ and all $f \in \dfregion$.

    Consider for each $RC$ and each $\kappa$ the vector $\ket{\Omega_n^{(RC, \kappa)}} := D_{\site}^{RC} \ket{\Omega_n^{(\kappa)}}$. Since the $A_v, B_f$ commute with $\qds$ for $v \in \dvregion$ and $f \in \dfregion$ (Lemma \ref{lem:qdcommute}), we have
    $$ \ket{\Omega_n^{(RC, \kappa)}} = A_v \ket{\Omega_n^{(RC,\kappa)}} = B_f \ket{\Omega_n^{(RC,\kappa)}} = D_{\site}^{RC} \ket{\Omega_n^{(RC, \kappa)}} $$
    for all $RC$, $\kappa$, $v \in \dvregion$ and $f \in \dfregion$. \ie we have $\ket{\Omega_n^{(RC,\kappa)}} \in \VRC$ (cf. Definition \ref{def:local constraints}). It then follows from Proposition \ref{prop:local spaces spanned by etas} that
    $$ \ket{\Omega_n^{(RC,\kappa)}} = \sum_{uv} c^{RC, \kappa}_{uv} \qdpure $$
    for some coefficients $c^{RC, \kappa}_{uv} \in \C$. Since $\sum_{RC} \qds = \mathds{1}$ (Lemma \ref{lem:DRCprops}) it follows that
    $$ \ket{\Omega_n^{(\kappa)}} =  \sum_{RC}  \ket{\Omega_n^{(RC,\kappa)}} = \sum_{RC} \sum_{uv} c^{RC, \kappa}_{uv} \qdpure. $$

    We now use Lemma \ref{lem:topological charge detector} to obtain
    $$K_{\bdy}^{R' C'} \ket{\Omega_n^{(\kappa)}} = \sum_{RC} \sum_{uv} c_{uv}^{RC, \kappa} \, K_{\bdy}^{R'C'} \qdpure = \sum_{uv} c_{uv}^{R'C', \kappa} | \eta_n^{R'C';uv} \rangle = D_{\site}^{R'C'} \ket{\Omega_n^{(\kappa)}}$$
    from which it follows that
    $$\omega(O D_{\site}^{RC}) = \sum_{\kappa} \, \lambda_{\kappa} \inner{ \Omega_n^{(\kappa)} }{ O D_{\site}^{RC} \, \Omega_n^{(\kappa)} } = \sum_{\kappa} \, \lambda_{\kappa} \inner{ \Omega_n^{(\kappa)} }{ O K_{\bdy}^{RC} \, \Omega_n^{(\kappa)}} = \omega(O K_{\bdy}^{RC})$$
    for all $O \in \cstar[\eregion]$. Using that the $K_{\sigma}^{RC}$ are hermitian (Lemma \ref{lem:basic properties of K}) and the elementary fact that $\omega(A^* B) = \overline{\omega(B^* A)}$ for all $A, B \in \cstar$ we also get
    $$\omega(D_{\site}^{RC} O) = \omega( K_{\bdy}^{RC} O)$$
    for all $O \in \cstar[\eregion]$.
    
    To show the second claim, note that for any $O \in \cstar[loc]$ we can take $n$ large enough so that $O \in \cstar[{\eregion[n-1]}]$. Then $[K_{\bdy}^{RC}, O] = 0$, so $\omega(O K_{\bdy}^{RC}) = \omega(K_{\bdy}^{RC} O)$. Using the results $\omega(K_{\bdy}^{RC} O) = \omega(D_{\site}^{RC} O)$ and $\omega(O K_{\bdy}^{RC}) = \omega(O D_{\site}^{RC})$ obtained above, we get
    $$ \omega(D_{\site}^{RC} O) =  \omega(O D_{\site}^{RC})$$
    for any $O \in \cstar[loc]$. This result extends to all $O \in \cstar$ by continuity.
\end{proof}

\begin{lem} 
\label{lem:detection Lemma}
    If $\omega \in \S_{\site}^{RC}$ then
    $$ \omega( K^{R'C'}_{\bdy}) = \delta_{RC, R'C'}$$
    for all $n > 1$.
\end{lem}

\begin{proof}
    By definition $\omega(D_{\site}^{RC}) = 1$ and so by Lemma \ref{lem:absorption of satisfied projectors} we have $\omega(O) = \omega(O D_{\site}^{RC}) = \omega(D_{\site}^{RC} O)$ for all $O \in \cstar$. We take $O = K^{R'C'}_{\bdy}$ and use Lemmas \ref{lem:projector Lemma} and \ref{lem:DRCprops} to obtain
    $$ \omega(K^{R'C'}_{\bdy}) = \omega( D_{\site}^{RC} K^{R'C'}_{\bdy}) = \omega( D_{\site}^{RC} D_{\site}^{R'C'} ) = \delta_{RC, R'C'} \omega(D_{\site}^{RC}) = \delta_{RC, R'C'}$$
    as required.
\end{proof}

\begin{lem}
\label{lem:piRCaredisjoint}
    The representations $\pi^{RC}$ and $\pi^{R'C'}$ are disjoint whenever $RC \neq R'C'$
\end{lem}

\begin{proof}
    We have $\omega_{\site}^{RC;(1, 1)} \in \S_{\site}^{RC}$ so Lemma \ref{lem:detection Lemma} says
    $$ \omega_{\bsite}^{RC, (1, 1)}( K_{\bdy}^{R'C'} ) = \delta_{RC,R'C'}$$
    for all $n > 1$. Noting that for any finite region $S \subset \latticeedge$ we can take $n$ large enough such that the projectors $K_{\beta_n}^{RC} \in \cstar[loc]$ are supported outside $S$, the claim follows from Corollary 2.6.11 of \cite{Bratteli2012-gd}.
\end{proof}

\subsection{Construction of amplimorphism representations}

In order to show that the representations $\pi^{RC}$ are anyon representations we first show that they are unitarily equivalent to so-called amplimorphism representations. These are representations which can be obtained from the ground state representation $\pi_0$ by composing $\pi_0 \otimes \id_{\abs{I_{RC}}}$ with an \emph{amplimorphism} $\cstar \rightarrow \mathcal{M}_{\abs{I_{RC}}}(\cstar)$, whose components are given by the maps $\mu_{\rho}^{RC;u_1 u_2}$ for a fixed half-infinite ribbon $\rho$. By the properties listed in Proposition \ref{prop:ampli properties}, this amplimorphism is a homomorphism of $C^*$-algebras, and the composition with $\pi_0$ yields a representation of $\cstar$. The fact that the amplimorphism acts non-trivially only near the ribbon $\rho$ will allow us to establish in Section \ref{sec:pi^RC are anyons representations} that the representations $\pi^{RC}$ are anyons representations.

Amplimorphisms, specifically in the context of non-abelian quantum double models, were first introduced in \cite{Naaijkens2015-xj}. Our presentation here is essentially a completion of the arguments sketched in that work. Amplimorphisms were originally introduced as a tool to investigate generalized symmetries in lattice spin models and quantum field theory, see \cite{szlachanyi1993quantum, vecsernyes1994quantum, fuchs1994quantum, nill1997quantum}.\\

Recall that $(\pi_0, \caH_0, \ket{\Omega_0})$ is the GNS triple of the frustration free ground state $\omega_0$ and we put $\chi_{\rho}^{RC; u v} := \pi_0 \circ \mu_{\rho}^{RC; u v} : \cstar \rightarrow \caB(\caH_0)$. For the remainder of this section we will often write $O$ instead of $\pi_0(O)$ when we are working in the faithful representation $\pi_0$.

We now define the \emph{amplimorphism representation}.
\begin{defn} \label{def:amplimorphism representation}
	We set
	\begin{equation*} 
		\chi_{\rho}^{RC} : \cstar \rightarrow \caB(\caH_0) \otimes M_N(\C) : O \mapsto \begin{bmatrix}
		\chi_{\rho}^{RC; u_1 u_1}(O) & \cdots & \chi_{\rho}^{RC; u_1 u_N}(O) \\
		\vdots & \ddots & \vdots \\
		\chi_{\rho}^{RC; u_N u_1}(O) & \cdots & \chi_{\rho}^{RC; u_N u_N}(O) 
	\end{bmatrix}
	\end{equation*}
	where $N = \abs{I_{RC}} = \abs{C} \dimR$ is the number of distinct values that the label $u \in I_{RC}$ can take.
\end{defn}
Using the properties listed in Proposition \ref{prop:ampli properties}, one can easily check that this is a unital *-representation of the quasi-local algebra.

The amplimorphism representation is carried by the Hilbert space $\caH := \caH_0 \otimes \C^N$. We choose a basis $\{ | u \rangle \}_{u \in I_{RC}}$ of $\C^N$ such that
\begin{equation*}
	\langle \Phi \otimes u, \chi^{RC}_{\rho}(O) \, \Psi \otimes v \rangle = \langle \Phi, \chi_{\rho}^{RC;u v}(O) \, \Psi \rangle
\end{equation*}
for all $| \Phi \rangle, | \Psi \rangle \in \caH_0$.

\subsection{Unitary equivalence of \texorpdfstring{$\chi_{\rho}^{RC}$}{Chi RC} and \texorpdfstring{$\pi^{RC}$}{pi RC}}

Let us fix a half-infinite ribbon $\rho$ with $\partial_0 \rho = \site$.

\begin{lem} \label{lem:omega^u is vector state}
	For any $u \in I_{RC}$, the vector $| \Omega_0 \otimes u \rangle$ represents the pure state $\omega_{\site}^{RC; u}$ in the representation $\chi_{\rho}^{RC}$.
\end{lem}

\begin{proof}
	For any $O \in \cstar$ we have
	\begin{equation*}
		\langle \Omega_0 \otimes u, \chi_{\rho}^{RC}(O) \, \Omega_0 \otimes u \rangle = \langle \Omega_0, \chi_{\rho}^{RC; u u}(O) \, \Omega_0 \rangle = \omega_0 \circ \mu_{\rho}^{RC; u u}(O) = \omega_{s_0}^{RC; u}(O)
	\end{equation*}
	where the last step is by Lemma \ref{lem:qdstate from mu action}.
\end{proof}
\ie the amplimorphism representation $\chi_{\rho}^{RC}$ contains the state $\omega^{RC; u}$, and therefore has a subrepresentation that is unitarily equivalent to $\pi^{RC}$. In fact, we will show that $\chi_{\rho}^{RC}$ is unitarily equivalent to $\pi^{RC}$. To do this, we must show that $| \Omega_0 \otimes u \rangle$ is a cyclic vector for $\chi_{\rho}^{RC}$.

\begin{prop} \label{prop:Omega^u is cyclic}
	For any $u \in I_{RC}$, the vector $| \Omega_0 \otimes u \rangle$ is cyclic for $\chi_{\rho}^{RC}$.
\end{prop}

\begin{proof}
    Let
    \begin{equation*}
    \caH_u := \overline{ \chi_{\rho}^{RC}(\cstar) | \Omega_0 \otimes u \rangle } \subseteq \caH.
    \end{equation*}
    We show that actually $\caH_u = \caH$.

    Consider the subspace
    \begin{equation*}
        \caV := \left\lbrace \sum_{v \in I_{RC}} \,  | \Psi_v \otimes v \rangle \, : \,  |\Psi_v\rangle \in \cstar[loc] | \Omega_0 \rangle \right\rbrace \subset \caH.
    \end{equation*}
    This space is dense in $\caH$.
    
    Take any vector $| \Psi \rangle = \sum_v | \Psi_v \otimes v \rangle \in \caV$ such that $| \Psi_v \rangle = O_v | \Omega_0 \rangle$ with $O_v \in \cstar[loc]$ for each $v \in I_{RC}$.
    
    We want to show that if $|\Psi \rangle$ is approximately orthogonal to $\caH_u$, then $\norm{|\Psi\rangle}$ is small. So let $P_u$ be the orthogonal projector onto $\caH_u$ and suppose that $\norm{ P_u | \Psi \rangle }^2 < \ep$.
    
    We now use the maps $t_{\rho_n}^{RC;vu}$ from Definition \ref{def:magic map}. For any $n \in \N$, any $RC$, and any $u, v \in I_{RC}$, these maps are given by
    $$ t_{\rho_n}^{RC;u v}( O ) :=  \bigg(\frac{\dimR}{|N_C|} \bigg)^2 \, \sum_{w, u'} \, F_{\rho_n}^{RC;u w}  \, O \, \big( F_{\rho_n}^{RC;u' w} \big)^* A_{\site}^{RC;u' v} D_{\site}^{RC;v}$$
    for any $O \in \cstar$. Here $a_{\site}^{RC;u' u}$ is the label changer of Definition \ref{def:label changers}, and $D_{\site}^{RC;u}$ is the projector of Definition \ref{def:Wigner projectors}. For any $O \in \cstar[loc]$, Lemma \ref{lem:magic map} says that for $n$ large enough
    $$\chi_{\rho}^{RC;u_1 v_1} \big( t_{\rho_n}^{RC;u_2 v_2}(O) \big) | \Omega_0 \rangle = \delta_{u_1 u_2} \delta_{v_1 v_2} \, O | \Omega_0 \rangle.$$
    We can therefore take $n$ large enough such that
    \begin{align*}
		\chi^{RC}_\rho \big( t^{RC;v u}_{\rho_n}( O_v ) \big) | \Omega_0 \otimes u \rangle &= \sum_{w} \, | \chi^{RC;w u}_\rho( t_{\rho_n}^{RC; v u}(O_v) ) \Omega_0 \otimes w \rangle \\
      &= \sum_w \, \delta_{w, v} \, | O_v \Omega_0 \otimes w \rangle
      = | \Psi_v \otimes v \rangle \in \caH_u
    \end{align*}
    
    It now follows from our assumption $\norm{ P_u | \Psi \rangle }^2 < \ep$ that
    \begin{equation*}
	 \norm{ | \Psi_v \rangle}^2 = \abs{ \langle \Psi, \chi^{RC}_\rho \big( t^{RC;v u}_{\rho_n}( O_v ) \big) \, \Omega_0 \otimes u \rangle } < \ep
    \end{equation*}
    for all $v \in I_{RC}$ and therefore
    \begin{equation*}
		\norm{ |\Psi \rangle }^2 = \sum_v \, \norm{ | \Psi_v \rangle}^2 < N \ep.
    \end{equation*}

    Now take $| \Psi \rangle \in \caH$ and suppose that $| \Psi \rangle$ is orthogonal to $\caH_u$. Since $\caV$ is dense in $\caH$ there is a sequence of vectors $| \Psi_i \rangle \in \caV$ that converges to $| \Psi \rangle$ in norm. For any $\ep > 0$ we can find $i_0$ such that
    \begin{equation*}
		\norm{ P_u | \Psi_i \rangle  }^2 = \norm{P_u( | \Psi_i \rangle - | \Psi \rangle )}^2 < \ep
    \end{equation*}
    for all $i \geq i_0$. From the above, we conclude that
    \begin{equation*}
		\norm{ | \Psi_i \rangle }^2 < N \ep
    \end{equation*}
    for all $i \geq i_0$. We see that the sequence converges to zero, so $| \Psi \rangle = 0$.

    Since any vector in $\caH$ that is orthogonal to $\caH_u$ must vanish, and since $\caH_u \subseteq \caH$, we find that $\caH_u = \caH$. This shows that $| \Omega_0 \otimes u \rangle$ is a cyclic vector for the representation $\chi_{\rho}^{RC}$.
\end{proof}

\begin{prop} \label{prop:equivalence to pi^RC}
    For any half-infinite ribbon $\rho$ with initial site $s = \partial_0 \rho$, any $RC$, and any $u \in I_{RC}$, the amplimorphism representation $\chi_{\rho}^{RC}$ is unitarily equivalent to the GNS representation of the pure state $\omega_{s}^{RC;u}$. In particular, the amplimorphism representation $\chi_{\rho}^{RC}$ is irreducible and unitarily equivalent to $\pi^{RC}$.
\end{prop}

\begin{proof}
    Unitary equivalence to the GNS representation of $\omega_{s}^{RC;u}$ follows immediately from Lemma \ref{lem:omega^u is vector state} and Proposition \ref{prop:Omega^u is cyclic}. Since $\omega_{s}^{RC;u}$ is a vector state in the irreducible representation $\pi^{RC}$ (Lemma \ref{lem:statesbelongingtopiRC}) we find that $\chi_{\rho}^{RC}$ is unitarily equivalent to $\pi^{RC}$ and in particular irreducible. 
\end{proof}

\subsection{The representations \texorpdfstring{$\pi^{RC}$}{pi RC} are anyon representations} \label{sec:pi^RC are anyons representations}

For any cone $\Lambda$, let $\caR(\Lambda) := \pi_0(\cstar[\Lambda])'' \subset \caB(\caH_0)$ be the von Neumann algebra generated by $\cstar[\Lambda]$ in the ground state representation.\\

The following Proposition is a slight adaptation of part of Theorem 5.4 of \cite{Naaijkens2015-xj}. 
\begin{prop}[\cite{Naaijkens2015-xj}] \label{prop:reduction to H_0}
    If $\rho$ is a half-infinite ribbon whose support is contained in a cone $\Lambda$, then there is a representation $\nu_{\Lambda}^{RC} : \cstar \rightarrow \caB(\caH_0)$ which is unitarily equivalent to the amplimorphism representation $\chi_{\rho}^{RC}$ and satisfies
    \begin{equation*}
		\nu_{\Lambda}^{RC}(O) = \pi_0(O)
    \end{equation*}
    for all $O \in \cstar[\Lambda^c]$.
\end{prop}

The proof is exactly the same as that of Theorem 5.4 of \cite{Naaijkens2015-xj} and we simply point out that the Haag duality assumed in that Theorem is not needed for this part of the statement.

We can now show
\begin{prop} \label{prop:the pi^RC are anyon representations}
    The representations $\pi^{RC}$ are anyon representations.
\end{prop}

\begin{proof}
    Fix any cone $\Lambda$. We want to show that
    \begin{equation*}
		\pi^{RC}|_{\cstar[\Lambda]} \simeq \pi_0|_{\cstar[\Lambda]}.
    \end{equation*}
    To this end, let $\rho$ be a half-infinite ribbon supported in a cone $\widetilde \Lambda$ that is disjoint from $\Lambda$. Since $\pi^{RC}$ is unitarily equivalent to the amplimorphism representation $\chi_{\rho}^{RC}$ (Proposition \ref{prop:equivalence to pi^RC}), we get from Proposition \ref{prop:reduction to H_0} a representation $\nu_{\widetilde \Lambda}^{RC} : \cstar \rightarrow \caB(\caH_0)$ that is unitarily equivalent to $\pi^{RC}$ and satisfies $\nu_{\widetilde \Lambda}^{RC}(O) = \pi_0(O)$ for all $O \in \cstar[\Lambda] \subset \cstar[\widetilde \Lambda^c]$. By the unitary equivalence, there is a unitary $U_{\Lambda} \in \caB(\caH_0)$ such that
    \begin{equation*}
		\pi^{RC}(O) = U_{\Lambda} \, \nu_{\widetilde \Lambda}^{RC}(O) \, U_{\Lambda}^*
    \end{equation*}
    for all $O \in \cstar$. For $O \in \cstar[\Lambda]$ we therefore find
    \begin{equation*}
		\pi^{RC}(O) = U_{\Lambda} \, \pi_0(O) \, U_{\Lambda}^*.
    \end{equation*}
    Since the cone $\Lambda$ was arbitrary, this proves the proposition.
\end{proof}

%% file: sectors/completeness.tex
\section{Completeness}
\label{sec:completeness}

In order to show that all anyon representations are unitarily equivalent to one of the representations $\pi^{RC}$, we prove that any anyon representation $\pi$ contains a pure state $\qdstate$. This we do as follows. In subsection \ref{sec:equivtofinite} we  show that any anyon representation $\pi$ contains a pure state that is gauge invariant and has trivial flux everywhere outside of a finite region. In subsection \ref{sec:sweep}, we show that such a state is unitarily equivalent to a pure state satisfying \eqref{eq:cons1}. Lastly in subsection \ref{sec:classify} we will show that any pure state satisfying \eqref{eq:cons1} is equivalent to some $\qdstate$ and therefore belongs to a definite anyon representation $\pi^{RC}$. Combining these results with the results of the previous Section, we find that the anyon sectors are in one-to-one correspondence with equivalence classes of irreducible representations of the quantum double of $G$.

\subsection{Any anyon representation contains a state that is gauge invariant and has trivial flux outside of a finite region}
\label{sec:equivtofinite}

Let $\pi : \cstar \rightarrow \mathcal{\hilb}$ be an anyon representation.

For any $S \subset \R^2$ let $S^V := \{ v \in \latticevert \, : \, A_v \in \cstar[S] \}$ and $S^F := \{ f \in \latticeface \, : \, B_f \in \cstar[S] \}$. We also write $S^{VF} = S^V \cup S^F$. If $S \subset \R^2$ is bounded then we define
$$P_{S} := \prod_{v \in S^V} A_v \prod_{f \in S^F} B_f$$
which is a projector in $\cstar[S]$. It projects onto states that are gauge invariant and flat on $S$.

We want to define analogous projectors for infinite regions, but clearly such projectors cannot exists in the quasi-local algebra $\cstar$. Instead, we will construct them in the von Neumann algebra $\pi(\cstar)''$.

\begin{defn}
    A non-decreasing sequence $\{S_n\}$ of sets $S_n \subset \R^2$ is said to converge to $S \subset \R^2$ if $\bigcup_{n} S_n = S$.   
\end{defn}

Let $\{S_n\}$ be a non-decreasing sequence of bounded subsets of $\R^2$ converging to a possibly infinite $S \subset \R^2$. Then the sequence of orthogonal projectors $\{\pi(P_{S_n})\}$ is non-increasing and therefore converges in the strong operator topology to the orthogonal projector $p_S$ onto the intersection of the ranges of the $\pi(P_{S_n})$ \cite[Thm 4.32(a)]{Weidmann}. In particular, the limit $p_S$ does not depend on the particular sequence $\{S_n\}$. If $S$ is finite, then $p_S = \pi(P_S)$.

For any $r \geq 0$, let $B_r = \{ x \in \R^2 \, : \norm{x}_2 \leq r\}$ be the closed ball of radius $r$. 

\begin{prop} \label{prop:equivtofinite}
    For any anyon representation $\pi : \cstar \rightarrow \mathcal{B}(\hilb)$ there is an $n > 0$ and a pure state $\omega$ belonging to $\pi$ such that
    $$\omega(A_v) = \omega(B_f) = 1$$
    for all $v$ and $f$ such that $A_v, B_f \in \cstar[B_n^c]$.
\end{prop}

\begin{proof}
    Take two cones $\Lambda_1, \Lambda_2$ such that any $A_v$ and any $B_f$ is supported in (at least) one of them. In particular, $\Lambda_1 \cup \Lambda_2 = \R^2$. Since $\pi$ is an anyon representation, we have unitaries $U_i : \hilb_0 \rightarrow \hilb$ for $i = 1,2$ that satisfy
    $$\pi(O) = U_i \pi_0(O) U_i^* \qquad \forall O \in \cstar[\Lambda_i].$$
    It follows that
    $$\omega_0(O) = \inner{\Omega_0}{\pi_0(O) \Omega_0} = \inner{U_i\Omega_0}{\pi(O) U_i \Omega_0} \quad \forall \, O \in \cstar[\Lambda_i].$$
    Define pure states $\omega_i$ on $\cstar$ given by $\omega_i(O) := \inner{\Omega_i}{\pi(O) \Omega_i}$ where $\ket{\Omega_i} = U_i\ket{\Omega_0} \in \hilb$. The states $\omega_i$ belong to $\pi$ and satisfy $\omega_i(O) = \omega_0(O)$ for all $O \in \cstar[\Lambda_i]$.

    Let $\Lambda_i^{> n} := \Lambda_i \setminus B_n$ and $\Lambda_i^{n, n+m} := \Lambda_i^{>n} \setminus \Lambda_i^{>n+m}$ for all $m, n \in \mathbb{N}$. Then the sequence $m \mapsto \Lambda_i^{n, n+m}$ is a non-decreasing sequence of bounded sets converging to $\Lambda_i^{> n}$. We have
    $$1 = \omega_0 \big( P_{\Lambda_i^{n, n+m}} \big) = \omega_i \big( P_{\Lambda_i^{n, n+m}} \big) = \inner{\Omega_i}{ \pi \big( P_{\Lambda_i^{n, n+m}} \big)  \Omega_i},$$
    where we used that all these projectors are supported in $\Lambda_i$. We now find that
    $$\inner{\Omega_i}{p_{\Lambda_i^{> n}} \Omega_i} = 1$$
    where $p_{\Lambda_i^{> n}}$ is the strong limit of the sequence of projectors $\{ \pi \big( P_{\Lambda_i^{n, n+m}} \big) \}_{m > n}$.

    The pure states $\omega_1$ and $\omega_2$ are unitary equivalent since they are both vector states in the irreducible representation $\pi$. It follows from Corollary 2.6.11 of \cite{Bratteli2012-gd} that for any $\epsilon > 0$ there is an $n(\epsilon) \in \mathbb{N}$ such that
    $$|\omega_1(O) - \omega_2 (O)| \leq \epsilon \norm{O}$$
    for all $O \in \cstar[loc] \cap \cstar[ B_{n(\ep)}^c ]$.

    This gives us
    $$|\inner{\Omega_1}{\pi(P_{\Lambda_i^{n, n+m}}) \Omega_1} - \inner{\Omega_2}{\pi(P_{\Lambda_i^{n, n+m}}) \Omega_2}| < \epsilon$$
    for all $m$ and $n \geq n(\epsilon)$ for some $n(\epsilon) \in \mathbb{N}$. After taking the strong limit we get
    $$\inner{\Omega_1}{p_{\Lambda_2^{>n}} \Omega_1} >1 - \epsilon$$
    where we have used $\inner{\Omega_2}{\pi(P_{\Lambda_2^{n, n+m}}) {\Omega_2}} = 1$.

    Using the fact that $p_{\Lambda_1^{>n}}$ is a projector and $\inner{\Omega_1}{p_{\Lambda_1^{>n}} \Omega_1} = 1$ for all $n \geq n(\epsilon)$, we also have that $p_{\Lambda_1^{>n}} \ket{\Omega_1} = \ket{\Omega_1}$ for all such $n$.

    Now we use the fact that $p_{\Lambda_2^{>n}} p_{\Lambda_1^{>n}}$ projects onto a subspace of the range of $p_{B_{n+1}^c}$ to obtain
    \begin{align*}
        \inner{\Omega_1}{p_{B_{n+1}^c} \Omega_1} \geq  \inner{\Omega_1}{p_{\Lambda_2^{>n}} p_{\Lambda_1^{>n}} \Omega_1} = \inner{\Omega_1}{p_{\Lambda_2^{>n}}  \Omega_1} > 1-\epsilon
    \end{align*}
    for all $n > n(\epsilon)$.

    It follows that for $\epsilon < 1$ we have $p_{B_{n+1}^c} \ket{\Omega_1} \neq 0$ for all $n \geq n(\epsilon)$. Let us therefore fix some $\epsilon < 1$ and $n \geq n(\epsilon)$, and define a normalized vector
    $$\ket{\Omega} := \frac{p_{B_{n+1}^c} \ket{\Omega_1}}{||p_{B_{n+1}^c} \ket{\Omega_1}||} \in \hilb.$$ 

    The vector $\ket{\Omega}$ defines a pure state by $\omega(O):=\inner{\Omega}{\pi(O)\Omega}$ for all $O \in \cstar$. This state belongs to the anyon representation $\pi$.
    
    To finish the proof, we verify that $\omega(A_v) = \omega(B_f) = 1$ whenever $A_v ,B_f \in \cstar[B_{n+1}^c]$. We have
    $$
        \omega(A) = \frac{\inner{\Omega_1}{p_{B_{n+1}^c} \pi(A_v) p_{B_{n+1}^c}\Omega_1}}{||p_{B_{n+1}^c} \ket{\Omega_1}||^2}
        =\frac{\inner{\Omega_1}{p_{B_{n+1}^c} p_{\{v\}} p_{B_{n+1}^c}\Omega_1}}{||p_{B_{n+1}^c} \ket{\Omega_1}||^2} = \frac{\inner{\Omega_1}{p_{B_{n+1}^c} p_{B_{n+1}^c}\Omega_1}}{||p_{B_{n+1}^c} \ket{\Omega_1}||^2} = 1
    $$
    where we used $p_{B_{n+1}^c} \pi(A_v) = p_{B_{n+1}^c}$. The proof that $\omega_\pi(B_f) = 1$ whenever $B_f \in \cstar[B_{n+1}^c]$, is identical. This concludes the proof of the Proposition.
\end{proof}

\subsection{Finite violations of ground state constraints can be swept onto a single site} \label{sec:sweep}

Let $\omega$ be a pure state on $\cstar$ and let $(\pi, \hilb, |\Omega \rangle)$ be its GNS triple. Since $\cstar$ is a simple algebra, the representation $\pi$ is faithful and we can identify $\cstar$ with its image $\pi(\cstar)$. For any $O \in \cstar$ we will write $O$ instead of $\pi(O)$ in the remainder of this Section.

For the remainder of this Section we fix an arbitrary site $s_* = (v_*, f_*)$. We will show that we can move any finite number of violations of the ground state constraints onto the site $s_*$.

The following Lemma will prove useful in achieving this:

\begin{lem}
\label{lem:sweep identities}
    (\cite{Bombin2007-uw}, Eq. B46, B47) Let $\rho$ be a ribbon such that $v = v(\start)$ or $v = v(\en)$ and $v(\start) \neq v(\en)$, and $ f = f(\start) $ or $f = f(\en)$ and $f(\start) \neq f(\en)$. Then we have the following identities:
    \begin{equation}
    \label{eq:sweep}
        {|G|}\sum_{g \in G} T^g_\rho A_v T^g_\rho = 1, \qquad \qquad \sum_{h \in G}L^{\bar{h}}_\rho B_f L^h_\rho = 1
    \end{equation}
\end{lem}

For any vector $\ket{\Psi} \in \caH$, let us define
$$ V_{\Psi} := \{ v \in \latticevert \, : \, A_v \ket{\Psi} \neq \ket{\Psi}  \} \setminus \{ v_* \} $$
and
$$ F_{\Psi} := \{ f \in \latticeface \, : \, B_f \ket{\Psi} \neq \ket{\Psi} \} \setminus \{ f_* \}. $$
The set $V_{\Psi}$ consists of all vertices except $v_*$ where the gauge constraint $A_v$ is violated in the state $\ket{\Psi}$, and $F_{\Psi}$ is the set of all faces except $f_*$ where the flux constraint $B_f$ is violated in the state $\ket{\Psi}$.

The following Lemma says that we can remove a single violation of a gauge constraint.

\begin{lem} \label{lem:remove star violation}
    Let $\ket{\Psi} \in \caH$ be a unit vector and let $v \in \latticevert \setminus \{v_*\}$ be a vertex. Then there is a unit vector $\ket{\Psi'} \in \caH$ such that $V_{\Psi'} \subseteq V_{\Psi} \setminus \{v\}$ and $F_{\Psi'} \subseteq F_{\Psi}$.
\end{lem}

\begin{proof}
    Let $\rho$ be a ribbon such that $v(\partial_0 \rho) = v_*$ and $v(\partial_1 \rho) = v$. For any $g \in G$ we define the vector
    $$ \ket{\Psi_g} := A_v T_{\rho}^g \ket{\Psi}. $$
    It follows immediately from the definition that $A_v \ket{\Psi_g} = \ket{\Psi_g}$, so $v \not\in V_{\Psi_g}$. We now show that all vertices that were not in $V_{\Psi} \setminus \{v\}$ are also not in $V_{\Psi_g}$, \ie $V_{\Psi_g} \subseteq V_{\Psi} \setminus \{ v \}$. We have just shown this for $v$ itself, and it is true by definition for $v_*$. It remains to show it for any $v' \not\in V_{\Psi}$ such that $v' \neq v, v_*$. Then $A_{v'} \ket{\Psi} = \ket{\Psi}$ and since $A_{v'}$ commutes with $A_v$ (Eq. \eqref{eq:ABcommproj}) and with $T_{\rho}^g$ (Lemma \ref{lem:A,B comm T,L except endpt}), we find $A_{v'} \ket{\Psi_g} = \ket{\Psi_g}$. \ie $v' \not\in V_{\Psi_g}$.

    Similarly, if $f \not\in F_{\Psi}$, \ie $B_f \ket{\Psi} = \ket{\Psi}$, then also $B_f \ket{\Psi_g} = \ket{\Psi_g}$ because $B_f$ commutes with $A_v$ (Eq. \eqref{eq:ABcommproj}) and with $T_{\rho}^g$ (Lemma \ref{lem:A,B comm L,T always}). This shows that $F_{\Psi_{g}} \subseteq F_{\Psi}$.

    It remains to show that at least one of the $\ket{\Psi_{g}}$ is non-zero.
    
    Using the first identity of Lemma \ref{lem:sweep identities} we find
    $$ \ket{\Psi} = \abs{G} \, \sum_{g \in G} T_{\rho}^g A_v T_{\rho}^g \ket{\Psi} = \abs{G} \sum_{g \in G} T_{\rho}^g \ket{\Psi_g}. $$
    Since $\ket{\Psi}$ is a unit vector, there must be at least one $g \in G$ such that $\ket{\Psi_g} \neq 0$. Let $g \in G$ be such that $\ket{\Psi_g} \neq 0$. We can normalise this vector, obtaining a unit vector
    $$ \ket{\Psi'} = \frac{\ket{\Psi_g}}{\norm{ \ket{\Psi_g} }}. $$
    The property that $V_{\Psi'} \subseteq V_{\Psi} \setminus \{v\}$ and $F_{\Psi'} \subseteq F_{\Psi}$ follows immediately from the fact that $\ket{\Psi_g}$ satisfies this property, as shown above. This proves the Lemma.
\end{proof}

Similarly, we can remove a single violation of a flux constraint. The proof is essentially identical to the proof of the previous Lemma and is therefore omitted.

\begin{lem} \label{lem:remove face violation}
    Let $\ket{\Psi} \in \caH$ be a unit vector and let $f \in \latticeface \setminus \{f_*\}$ be a face. Then there is a unit vector $\ket{\Psi'} \in \caH$ such that $V_{\Psi'} \subseteq V_{\Psi}$ and $F_{\Psi'} \subseteq F_{\Psi} \setminus \{ f \}$.
\end{lem}

\begin{prop}
\label{prop:sweeping}
    If $\omega$ is a pure state on $\cstar$ and there is an $n$ such that $\omega(A_v) = \omega(B_f) = 1$ for all $A_v, B_f \in \cstar[B_{n}^c]$, then for any site $s_*$ there is a pure state $\psi \in \overline{\S}_{s_*}$ that is unitarily equivalent to $\omega$.
\end{prop}

\begin{proof}
    Let $s_* = (v_*, f_*)$ and let $(\pi, \caH, \ket{\Omega})$ be the GNS triple for $\omega$. From Lemma \ref{lem:omega_0(O) = 1} we find that
    $$ A_v \ket{\Omega} = B_f \ket{\Omega} = \ket{\Omega} $$
    for all $A_v, B_f \in \cstar[B_{n}^c]$. In particular, the sets
    $$ V_{\Omega} = \{ v \in \latticevert \, : \, A_v \ket{\Omega} \neq \ket{\Omega} \} \setminus \{ s_* \} $$
    and
    $$ F_{\Omega} = \{ f \in \latticeface \,: \, B_g \ket{\Omega} \neq \ket{\Omega} \} \setminus \{ f_* \} $$
    are finite.

    We can therefore apply Lemmas \ref{lem:remove star violation} and $\ref{lem:remove face violation}$ a finite number of times to obtain a unit vector $\ket{\Psi} \in \caH$ for which $V_{\Psi} = \emptyset$ and $F_{\Psi} = \emptyset$. In other words, the vector $\ket{\Psi}$ satisfies
    $$  A_v \ket{\Psi} = B_f \ket{\Psi} = \ket{\Psi} $$
    for all $v \neq v_*$ and all $f \neq f_*$.

    The vector $\ket{\Psi}$ corresponds to a pure state $\psi$ given by
    $$\psi(O) = \langle \Psi, O \, \Psi \rangle $$
    for all $O \in \cstar$.

    Since $\omega$ and $\psi$ are vector states in the same representation $\pi$ of $\cstar$, these states are unitarily equivalent. Moreover,
    $$ \psi(A_v) = \langle \Psi, A_v \, \Psi \rangle = 1, \quad \text{and} \quad \psi(B_f) = \langle \Psi, B_f \, \Psi \rangle = 1 $$
    for all $v \neq v_*$ and all $f \neq f_*$, so $\psi \in \overline{\S}_{s_*}$ (Definition \ref{def:state spaces}) as required.
\end{proof}

\subsection{Decomposition of states in \texorpdfstring{$\overline{\mathcal{S}}_{\site}$}{S site}}
\label{sec:classify}

Recall the Definition \ref{def:state spaces} of the set of states $\overline{\S}_{\site}$. We now prove that any state in $\overline{\S}_{\site}$ decomposes into states belonging to the anyon representations $\pi^{RC}$. Let us fix a state $\omega \in \overline{\S}_{\site}$.

\begin{defn} \label{def:omegaRC components defined}
    For each $RC$, define a positive linear functional $\tomegaRC[]$ by
    $$\tomegaRC[](O) := \omega(\qds O \qds)$$
    for all $O \in \cstar$, and a non-negative number
    $$\lamRC := \omega( D^{RC}_{s_0} ) = \tomegaRC[](\mathds{1}) \geq 0.$$
\end{defn}

\begin{lem} \label{lem:tomega vanishes iff it vanishes on 1}
    We have $\tomegaRC[] \equiv 0$ if and only if $\lamRC = 0$
\end{lem}

\begin{proof}
    If $\tomegaRC[](\mathds{1}) = 0$ then $\omega(\qds) = 0$ so by Cauchy-Schwarz
    $$\abs{\tomegaRC[](O)}^2 = \abs{\omega( \qds O \qds )}^2 \leq \omega(\qds) \omega(\qds O^* O \qds ) = 0$$
    for any $O \in \cstar$. i.e. $\tomegaRC[] \equiv 0$ if and only if $\tomegaRC[](\mathds{1}) = 0$. But $\tomegaRC[](\mathds{1}) = \omega( D_{s_0}^{RC} ) = \lamRC$, which yields the claim.
\end{proof}

\begin{lem} \label{lem:omegaRC belongs to S^RC}
    If $\lamRC >0$, we define a linear functional $\omegaRC[]$ by
    $$\omegaRC[](O) := \lamRC^{-1} \,\, \tomegaRC[](O) \qquad \qquad \forall \,\, O \in \cstar.$$
    Then $\omegaRC[](O)$ is a state in $\S_{s_0}^{RC}$.
\end{lem}

\begin{proof}
    We first show that $\omegaRC[]$ is a state. The linear functional $\tomegaRC[]$ is positive by construction and
    $$\omegaRC[](\mathds{1}) = \lamRC^{-1} \, \tomegaRC[](\mathds{1}) = 1,$$
    so $\omegaRC[]$ is normalized. We conclude that $\omegaRC[]$ is indeed a state.

    Let us now check that $\omega^{RC} \in \S_{s_0}^{RC}$. First we note that
    $$\omegaRC[](D_{s_0}^{RC}) = \lamRC^{-1} \, \tomegaRC[](D_{s_0}^{RC}) = 1$$
    because $D_{s_0}^{RC}$ is an orthogonal projector. Furthermore, since the projectors $A_v$ commute with $D_{s_0}^{RC}$ for all $v \in \dvregion$ (Lemma \ref{lem:qdcommute}), we have
    $$\omegaRC[](A_v) = \lamRC^{-1} \, \tomegaRC[](A_v) = \lamRC^{-1} \, \omega\big( D_{s_0}^{RC} A_v D_{s_0}^{RC} \big) = \lamRC^{-1} \, \omega( D_{s_0}^{RC} A_v ) = \lamRC^{-1} \omega(D_{s_0}^{RC}) = 1$$
    where we used $\omega(A_v) = 1$ and Lemma \ref{lem:absorption of satisfied projectors}. In the same way we can show that $\omegaRC[](B_f) = 1$ for all $f \in \dfregion$. We conclude that $\omegaRC[] \in \S_{s_0}^{RC}$, as required.
\end{proof}

\begin{lem} \label{lem:decomposition of omega}
    For any state $\omega \in \overline{\mathcal{S}}_{\site}$, we have 
    $$\omega = \quad \sum_{\mathclap{RC \, : \, \lamRC > 0}} \lamRC \, \omegaRC[]$$
    where the states $\omegaRC[]$ are those defined in Lemma \ref{lem:omegaRC belongs to S^RC}.
\end{lem}

\begin{proof}
    Using the decomposition $\mathds{1} = \sum_{RC} D_{s_0}^{RC}$ (Lemma \ref{lem:DRCprops}) and Lemma \ref{lem:projector Lemma}, we find
    $$
        \omega(O) = \sum_{RC} \omega \big( \, O \, D_{\site}^{RC}  \big) = \sum_{RC} \omega \big( D_{s_0}^{RC} \, O \, D_{s_0}^{RC} \big) = \sum_{RC \, : \, \lamRC > 0} \lamRC \, \omegaRC[](O)
    $$
    where in the last step we used the Definition of $\omegaRC[]$ and Lemma \ref{lem:tomega vanishes iff it vanishes on 1}.
\end{proof}

It remains to show that the states $\omegaRC[]$ belong to $\pi^{RC}$.

\begin{lem} \label{lem:omega in SRC belongs to piRC}
    Any state $\omega \in \S_{\site}^{RC}$ belongs to $\pi^{RC}$
\end{lem}
\begin{proof}
    The restriction $\omega_{n}$ of $\omega$ to $\cstar[{\eregion}]$ satisfies
    $$ 1 = \omega_{n}(A_v) = \omega_{n}(B_f) = \omega_{n}(D_{\site}^{RC}) $$ 
    for all $v \in \dregion$ and all $f \in \fregion$. Let $\omega_{n} = \sum_{\kappa} \lambda_{\kappa} \omega_{n}^{(\kappa)}$ be the convex decomposition of $\omega_{n}$ into its pure components $\omega_{n}^{(\kappa)}$, and let $\ket{\Omega_{n}^{(\kappa)}} \in \caH_{n}$ be unit vectors such that
    $$ \omega_{n}^{(\kappa)}(O) = \langle \Omega_{n}^{(\kappa)}, O \, \Omega_{n}^{(\kappa)} \rangle $$
    for all $O \in \cstar[{\eregion}]$. Then Lemma \ref{lem:pure components projector Lemma} yields
    $$ \ket{\Omega_{n}^{(\kappa)}} = A_v \ket{\Omega_{n}^{(\kappa)}} = B_f \ket{\Omega_{n}^{(\kappa)}} = D_{\site}^{RC} \ket{\Omega_{n}^{(\kappa)}} $$
    for all $v \in \dvregion$ and all $f \in \dfregion$. By Definition \ref{def:local constraints} of the space $\VRC$ we then have $\ket{\Omega_{n}^{(\kappa)}} \in \VRC$ for all $\kappa$. By Proposition \ref{prop:local spaces spanned by etas} it follows that
    $$ \ket{\Omega_{n}^{(\kappa)}} = \sum_{u, v} \, c^{(\kappa)}_{uv} \, \ket{ \eta_{n}^{RC;uv} } $$
    for some coefficients $c_{uv}^{(\kappa)} \in \C$. It follows that
    $$ \omega_{n}(O) = \Tr \lbrace \rho O \rbrace $$
    for any $O \in \cstar[\eregion]$, where $\rho$ is the density matrix on $\caH_n$ given by
    $$\rho = \sum_{u_1 v_1 ; u_2 v_2} \rho_{u_1 v_1 ; u_2 v_2} {\ket{\qdstpure[RC;u_1 v_1]}} \bra{\qdstpure[RC;u_2 v_2]}$$
    with $\rho_{u_1 v_1 ; u_2 v_2} = \sum_{\kappa} \, \lambda_{\kappa} \, c_{u_1 v_1}^{(\kappa)} ( c_{u_2 v_2}^{(\kappa)} )^*$.

    Let $O \in \cstar[{\eregion[n-1]}]$. From Lemma \ref{lem:purerest}, we have that
    $$\inner{\qdstpure[RC;u_1 v_1]}{O \qdstpure[RC;u_2 v_2]} = \delta_{v_1, v_2} \inner{\qdstpure[RC;u_1 v_1]}{O \qdstpure[RC;u_2 v_1]}$$
    is independent of $v_1$, so for any choice $v_0$ we have
    $$ \Tr \lbrace \rho O \rbrace = \sum_{u_1 u_2} \, \left( \sum_v  \rho_{u_1 v;u_2 v}\right) \langle \eta_{n}^{RC;u_1 v_0}, O \, \eta_{n}^{RC;u_2 v_0} \rangle.$$
    Now the numbers
    $$\tilde \rho_{u_1 u_2} := \sum_v \rho_{u_1 v;u_2 v} $$
    are the components of a density matrix $\tilde \rho$ which is a partial trace of $\rho_{u_1 v_1;u_2 v_2}$ over the boundary labels $v_1, v_2$. It follows that there is a basis in which the density matrix $\tilde \rho$ is diagonal, i.e. there is a unitary matrix $U$ with components $U_{uu'} \in \mathbb{C}$ such that
    $$\sum_{u_1 u_2}  (U^*)_{u'_1 u_1} \, \tilde \rho_{u_1 u_2}  \, U_{u_2 u_2'}  = \mu_{u_1'}^{(n)} \delta_{u_1' u_2'}$$
    for non-negative numbers $\mu^{(n)}_{u'}$ that sum to one. 

    We find
    $$\omega_n(O) = \Tr \lbrace \rho O \rbrace = \sum_{u'}  \mu^{(n)}_{u'} \, \tilde \eta_{n}^{RC;u' v_0}(O) $$
    where the $\tilde \eta_{n}^{RC;u' v_0}$ are pure states given by
    $$ \tilde \eta_{n}^{RC;u' v_0}(O) = \langle \tilde \eta_{n}^{RC;u' v_0}, O \, \tilde \eta_{n}^{RC;u'v_0} \rangle $$
    with
    $$ | \tilde \eta_{n}^{RC;u' v_0} \rangle :=  \sum_{u} (U^*)_{u' u} \, | \eta_{n}^{RC;u v_0} \rangle.$$

    Using Lemma \ref{lem:restriction yields eta^RCu}, Definition \ref{def:uniform string net superpositions}, and Lemma \ref{lem:aconverter} we find for any $u'' \in I_{RC}$ that
    $$
    \tilde \eta_{n}^{RC;u' v_0} \big( \sum_{w_1, w_2} D_{s_0}^{RC;w_1} \, (U^*)_{u'' \, w_1} \,  U_{w_2 \, u''} \, D_{s_0}^{RC;w_2} \big) = \delta_{u', u''}.
    $$
    Since the $D_{s_0}^{RC;w}$ are supported in $\cstar[{\eregion[n-1]}]$ for all $n \geq 3$ we therefore find that $\mu_{u''}^{(n)} = \mu_{u''}^{(m)}$ for all $u'' \in I_{RC}$ and all $n, m \geq 3$, \ie the numbers $\mu_u^{(n)}$ do not depend on $n$ if $n \geq 3$. Let us write $\mu_u = \mu_u^{(n)}$ for all $n \geq 3$.

    For any $u \in I_{RC}$, let $A^u := A_{\site}^{RC;u (1, 1)}$ be the label changer operator from Definition \ref{def:label changers}, and define vectors
    $$\ket{\Omega_{\site}^{RC;u}} := \pi^{RC} \big( A^u \big) \ket{\Omega_{\site}^{RC;(1, 1)}}.$$
    These are unit vectors representing the states $\omega_{\site}^{RC;u}$ in the representation $\pi^{RC}$. Define pure states $\tilde \omega^{RC;u'}$ on $\cstar$ by
    $$\tilde \omega^{RC;u'}(O) := \langle \widetilde \Omega^{RC;u'}, \pi^{RC}(O) \, \widetilde \Omega^{RC;u'} \rangle$$
    corresponding to the GNS vectors
    $$|\widetilde \Omega^{RC;u'} \rangle := \sum_{u} (U^*)_{u' u} \, | \Omega^{RC;u} \rangle \in \hilb^{RC}.$$
    
    Then we find
    $$\tilde \omega^{RC;u'}(O) = \omega_{\site}^{RC;(1, 1)} \big( \sum_{u_1, u_2}  U_{u_1 u'} (U^*)_{u' u_2} \, (A^{u_1})^* O A^{u_2} \big)$$
    for any $O \in \cstar$. Using Lemma \ref{lem:restriction yields eta^RCu}, Definition \ref{def:uniform string net superpositions} Lemma \ref{lem:aconverter}, and the fact that the $A^u$'s are supported near $\site$, for any $O$ supported on $\eregion[n-1]$ we get:
    $$ \tilde \omega^{RC;u'}(O) = \eta_{n}^{RC;u' v_0} \big( \sum_{u_1, u_2}  U_{u_1 u'} (U^*)_{u' u_2} \, (A^{u_1})^* O A^{u_2} \big) = \tilde \eta_{n}^{RC;u'v_0}(O). $$
    
    We conclude that
    $$\omega(O) = \Tr \lbrace \rho O \rbrace = \sum_{u'} \, \mu_{u'} \, \tilde \omega^{RC;u'}(O)$$
    for all $n \geq 3$ and all $O \in \cstar[{\eregion[n-1]}]$. It follows that we have an equality of states
    $$\omega = \sum_{u'} \mu_{u'} \tilde \omega^{RC;u'},$$
    which expresses $\omega$ as a finite mixture of pure states belonging to $\pi^{RC}$. It follows that $\omega$ also belongs to $\pi^{RC}$, as was to be shown.
\end{proof}

The results obtained above combine to prove the following proposition.
\begin{prop}
\label{prop:omegadecomp} 
    Let $\omega \in \overline{\S}_{\site}$ for some site $\site$. Then $\omega$ has a convex decomposition
    $$\omega = \sum_{RC} \, \lambda_{RC} \,\, \omegaRC[]$$
    into states $\omegaRC[]$ that belong to the representation $\pi^{RC}$. In particular, if $\omega$ is pure then $\omega$ belongs to $\pi^{RC}$ for some definite $RC$.
\end{prop}

\begin{proof}
    Lemma \ref{lem:decomposition of omega} provides the decomposition
    $$\omega = \sum_{RC : \lambda_{RC} > 0} \lambda_{RC} \, \omega^{RC}$$
    with positive $\lambda_{RC}$ and states $\omega^{RC}$ defined in Definition \ref{def:omegaRC components defined} and Lemma \ref{lem:omegaRC belongs to S^RC}. Moreover, Lemma \ref{lem:omegaRC belongs to S^RC} states that $\omega^{RC} \in \S_{\site}^{RC}$ for each $RC$ with $\lambda_{RC} > 0$. It then follows from Lemma \ref{lem:omega in SRC belongs to piRC} that the states $\omega^{RC}$ belong to $\pi^{RC}$. This concludes the proof.
\end{proof}

\subsection{Classification of anyon sectors}
We finally put together the results obtained above to prove our main result, Theorem \ref{thm:main theorem}, which we restate here for convenience.

\begin{thm} \label{thm:classification theorem}
    For each irreducible representation $RC$ of $\caD(G)$ there is an anyon representation $\pi^{RC}$. The representations $\{ \pi^{RC} \}_{RC}$ are pairwise disjoint, and any anyon representation is unitarily equivalent to one of them. 
\end{thm}

\begin{proof}
    The existence of the pairwise disjoint anyon representations follows immediately from Lemma \ref{lem:piRCaredisjoint} and Proposition \ref{prop:the pi^RC are anyon representations}.

    Let $\pi$ be an anyon representation. By Proposition \ref{prop:equivtofinite}, there is $n \in \N$ such that the anyon representation $\pi$ contains a pure state $\omega$ that satisfies $\omega(A_v) = \omega(B_f) = 1$ for all $A_v, B_f \in \cstar[B_{n}^c]$. For any site $\site$, Proposition \ref{prop:sweeping} then gives us a pure state $\psi \in \overline{\S}_{\site}$ that belongs to $\pi$. Proposition \ref{prop:omegadecomp} shows that this state belongs to an anyon representation $\pi^{RC}$ for some definite $RC$. Since the irreducible representations $\pi$ and $\pi^{RC}$ contain the same pure state, they are unitarily equivalent. This proves the Theorem.
\end{proof}

%% file: sectors/conclusions.tex
\section{Discussion and outlook}
\label{sec:conclusion}

In this paper we have fully classified the anyon sectors of Kitaev's quantum double model for an arbitrary finite gauge group $G$ in the infinite volume setting. The proof of the classification contained several ingredients. First, we constructed for each irreducible representation $RC$ of the quantum double algebra a set of pure states $\{ \omega_{s}^{RC;u} \}_{s, u}$ that are all unitarily equivalent to each other. We then showed that the corresponding irreducible GNS representations $\{\pi^{RC}\}_{RC}$ are a collection of disjoint anyon representations. The proof that these representations are anyon representations crucially relied on their identification with `amplimorphism representations'. To show completeness, we proved that any anyon representation of the quantum double models contains one of the states $\omega_s^{RC;u}$, so that any anyon representation is unitarily equivalent to one of the $\pi^{RC}$.

This result is a first step towards integrating the non-abelian quantum double models in the mathematical framework for topological order developed in \cite{Naaijkens2010-aq, Naaijkens2012-fh, Cha2018-ke, Cha2020-rz, Ogata2022-wp}. To complete the story, we would like to work out the fusion rules and braiding statistics of the anyon types we identified. The amplimorphisms of Section \ref{sec:anyon representations} promise to be an ideal tool to carry out this task, see \cite{Naaijkens2015-xj, szlachanyi1993quantum}. We leave this analysis to an upcoming work. A crucial assumption in the general theory of topological order for infinite quantum spin systems is (approximate) Haag duality for cones. This property has been established for abelian quantum double models \cite{Fiedler2015-na, Naaijkens2012-fh}, but not yet for the non-abelian case.

The proofs of purity of the \ffgs{} and of the states $\omega_{s}^{RC;u}$ use the string-net condensation picture \cite{levin2005string}. The same methods can be used to show purity of ground states of other commuting projector models (for example \cite{Bols2023-qg} shows the purity of the double semion \ffgs{} and \cite{Vadnerkar2023-mm} shows purity of the \ffgs{} of the 3d Toric Code). In principle, the same techniques can be used to show that Levin-Wen models \cite{levin2005string} and quantum double models based on (weak) Hopf algebras have a unique frustration free ground state in infinite volume (this has already been done for Levin-Wen models using different techniques in \cite{jones2023local}).

%% file: sectors/ribbonprops.tex
\section{Ribbon operators, Wigner projections, and amplimorphisms}
\label{app:ribbonprops}

\subsection{Basic properties of ribbon operators, gauge transformations, and flux projectors}

Recall from Section \ref{subsec:preliminary notions} the definitions of ribbons and ribbon operators, as well as the definitions of the gauge transformations $A_s^h$ and flux projectors $B_s^g$, all originally due to Kitaev \cite{Kitaev2003-qr}.

We have the following basic properties of ribbon operators, which can be easily verified. See also Appendix B of \cite{Bombin2007-uw}.

\begin{align}
\label{eq:F elementary props}
    F_\rho^{h,g} F_\rho^{h',g'} = \begin{cases} \delta_{g,g'} F^{h'h,g}_\rho \quad &\text{if  } \, \rho \, \text{ is positive} \\ \delta_{g,g'} F^{hh',g}_\rho \quad &\text{if } \, \rho \, \text{ is negative}  \end{cases} && (F^{h,g}_\rho )^* = F^{\dash{h},g}_\rho.
\end{align}

Let $\rho$ be such that its end sites $s_i = \partial_i \rho = (v_i, f_i)$ satisfy $v_0 \neq v_1$ and $f_0 \neq f_1$. Then if $\rho$ is positive we have
\begin{equation}
    \begin{aligned}
\label{eq:[A,F] and [B,F]}    
    A_{s_0}^k F^{h,g}_\rho = F^{kh\dash{k}, kg}_\rho A_{s_0}^k && \qquad A_{s_1}^k F^{h,g}_\rho = F^{h, g\dash{k}}_\rho A_{s_1}^k \\
    B_{s_0}^k F^{h,g}_\rho = F^{h,g}_\rho B_{s_0}^{hk} && \qquad  B_{s_1}^k F^{h,g}_\rho = F^{h,g}_\rho B_{s_1}^{k\dash{g}\dash{h}g}
    \end{aligned}
\end{equation}
and if $\rho$ is negative
\begin{equation}
    \begin{aligned}
\label{eq:[A,F] and [B,F] negative case}    
    A_{s_0}^k F^{h,g}_\rho = F^{kh\dash{k}, kg}_\rho A_{s_0}^k && \qquad A_{s_1}^k F^{h,g}_\rho = F^{h, g\dash{k}}_\rho A_{s_1}^k \\
    B_{s_0}^k F^{h,g}_\rho = F^{h,g}_\rho B_{s_0}^{kh} && \qquad  B_{s_1}^k F^{h,g}_\rho = F^{h,g}_\rho B_{s_1}^{\dash{g}\dash{h}gk}.
    \end{aligned}
\end{equation}

Importantly, the ribbon is invisible to the gauge transformations and flux projectors away from its endpoints. That is, if $\rho$ is a finite ribbon with $\partial_0 \rho = s_0$ and $\partial_1 \rho = s_1$, and $s = (v, f)$ is such that $v \neq v(s_0), v(s_1)$ while $s' = (v', f')$ is such that $f' \neq f(s_0), f(s_1)$, then
\begin{align}
\label{eq:A,B commute with F except at endpts}
    [F^{h,g}_\rho, A_{s}^k] = 0 = [F^{h,g}_\rho, B_{s'}^l]
\end{align}
for all $g, h, k, l \in G$.

The following properties of the gauge transformations and flux projectors follow immediately from the properties of the ribbon operators listed above.

\begin{align}
\label{eq:Aelementary}
    &A^1_s = \mathds{1}& (A_s^h)^*= A_s^{\dash{h}} && A_s^h A_s^{h'} = A_s^{hh'}\\
    \label{eq:Belementary}
    &\sum_{g \in G} B_s^g = 1 &(B_s^g)^* = B_s^g  && B_s^g B_s^{g'} = \delta_{g,g'} B_s^g\\
    \label{eq:ABonsamesite}
    && A_s^h B_s^g = B^{hg\dash{h}}_s A_s^h &&
\end{align}

If $s \neq s'$ then for any $h,h',g,g' \in G$,
\begin{align}
\label{eq:ABcommuteondiffsites}
    [A_s^h, B_{s'}^g] = [A_s^h, A_s^{h'}] = [B_s^g, B_{s'}^{g'}] = 0
\end{align}

Recall from Section \ref{subsec:preliminary notions} that the projectors $A_v := \frac{1}{|G|} \sum_{h \in G} A_s^h$ and $B_f := B_s^1$ where $v$ is the vertex of $s$ and $f$ is the face of $f$ are well defined. For any vertices $v, v'$ and any faces $f, f'$, we have (\cite{Kitaev2003-qr})
\begin{equation} \label{eq:ABcommproj}
    [A_v, B_f] = [A_v, A_{v'}] = [B_f, B_{f'}] = 0.
\end{equation}

\subsection{Decomposition of \texorpdfstring{$F^{h,g}_\rho$}{F rho} into \texorpdfstring{$L^h_\rho, T^g_\rho$}{L rho T rho}}

\subsubsection{Basic properties}

Recall from Section \ref{subsec:preliminary notions} the definitions $T^g_\rho := F^{1,g}_\rho$ and $L^h_\rho := \sum_{g \in G} F^{h,g}_\rho$.

\begin{lem}
    \label{eq:F breaks into LT}
    $F^{h,g}_\rho = L^h_\rho T^g_\rho =  T^g_\rho L^h_\rho$
\end{lem}

\begin{proof}
    Using Eq. \eqref{eq:F elementary props} we have
    \begin{align*}
        L^h_\rho T^g_\rho = \sum_{g'} F^{h,g'}_\rho F^{1,g}_\rho = F^{h,g}_\rho =  F^{1,g}_\rho \sum_{g'} F^{h,g'}_\rho = T^g_\rho L^h_\rho
    \end{align*}
\end{proof}

\begin{lem} \label{lem:flux change}
    Let $\rho$ be a finite ribbon such that $v(\start) \neq v(\en)$. Then if $s_0,s_1$ are sites such that $v(s_0) = \start$ and $ v(s_1) = \en$, we have
    $$ A^{h}_{s_0} \,  T_{\rho}^g = T_{\rho}^{hg} \, A_{s_0}^h \quad \text{and} \quad A_{s_1}^{h} \, T_{\rho}^g = T_{\rho}^{g \bar h} \, A_{s_1}^h $$
    while for sites $s$ such that $v(s) \neq v(s_0), v(s_1)$, we have
    $$ [A_s^h, T_{\rho}^g] = 0$$
    for all $g, h \in G$. Moreover,
    $$ [B_s^{k}, T_{\rho}^g] = 0 $$
    for all $k, g \in G$ and any site $s$.
\end{lem}

\begin{proof}
    This follows immediately from $T_{\rho}^g = F_{\rho}^{1, g}$ and Eqs. \eqref{eq:[A,F] and [B,F]}, \eqref{eq:[A,F] and [B,F] negative case} and \eqref{eq:A,B commute with F except at endpts}.
\end{proof}

\begin{lem}
\label{lem:A,B comm T,L except endpt}
    Let $\rho$ be a finite ribbon. For all $v \neq v(\partial_i \rho)$ and $f \neq f(\partial_i \rho)$, we have
    $$[A_v,T_\rho^g] = 0 = [B_f, L^h_\rho].$$
\end{lem}
\begin{proof}
    This is a trivial consequence of Eq. \eqref{eq:A,B commute with F except at endpts}.
\end{proof}

\begin{lem}
    \label{lem:A,B comm L,T always}
    Let $\rho$ be a finite ribbon. For all $v$ such that $v\neq v(\start)$ and all $f$ such that $f \neq f(\start)$ we have
    $$[A_v,L_\rho^h] = 0 = [B_f, T^g_\rho].$$
\end{lem}

\begin{proof}
    If we have $v \neq v(\partial_i \rho)$ or $f \neq f(\partial_i \rho)$ for $i = 0, 1$ then the claim follows immediately from Eq. \eqref{eq:A,B commute with F except at endpts}. Now let $v = v(\partial_1 \rho)$, $f = f(\en)$, and let $s,s'$ be sites such that $v(s) = v$ and $f(s') = f$. We then have,
    \begin{align*}
        A_v L_\rho^h &= \sum_k A_s^k L^{h}_\rho = \sum_{g,k} A_s^k F^{h,g}_\rho = \sum_{g,k} F^{h, g\dash{k}}_\rho A_{s}^k = \sum_k L^h_\rho A_s^k = L^h_\rho A_v\\
        B_f T^g_\rho &= B_{s'}^1 F^{1,g}_\rho = F^{1,g}_\rho B_{s'}^1 = T^g_\rho B_f
    \end{align*}
    Which proves the claim.
\end{proof}

\subsubsection{Alternating decomposition of $L_\rho^h$}

In this section we express the operators $L_\rho^h$ in terms of the decomposition of $\rho$ into its alternating direct and dual sub-ribbons. This result will be useful in Section \ref{sec:action of T and L on string nets}.

\begin{lem} \label{lem:L with initial direct triangle}
    Let $\rho = \tau \rho'$ be a finite ribbon whose initial triangle $\tau$ is a direct triangle. Then $$L_{\rho}^h = \sum_{k \in G} T_{\tau}^{k} \, L_{\rho'}^{\bar k h k}.$$
\end{lem}

\begin{proof}
    By definition, $L_{\rho}^h = \sum_g \, F_{\rho}^{h, g}$. Using Eq. \eqref{eq:F inductive def}, this becomes
    $$L_{\rho}^{h} = \sum_g \sum_k F_{\tau}^{h, k} F_{\rho'}^{\bar k h k, g} = \sum_k T_{\tau}^k \, L_{\rho'}^{\bar k h k},$$
    where we used $F_{\tau}^{h, k} = T_{\tau}^k$ because $\tau$ is a direct triangle.
\end{proof}

\begin{lem} \label{lem:L with initial dual triangle}
    Let $\rho = \tau \rho'$ be a finite ribbon such that its initial triangle $\tau$ is a dual triangle. Then $$L_{\rho}^h = L_{\tau}^{h} \, L_{\rho'}.$$
\end{lem}

\begin{proof}
    By definition, $L_{\rho}^h = \sum_g \, F_{\rho}^{h, g}$. Using Eq. \eqref{eq:F inductive def}, this becomes
    $$L_{\rho}^{h} = \sum_g \sum_k F_{\tau}^{h, k} F_{\rho'}^{\bar k h k, g} = L_{\tau}^h \, L_{\rho'}^{h},$$
    where we used $F_{\tau}^{h, k} = \delta_{k, 1} L_{\tau}^h$ because $\tau$ is a dual triangle.
\end{proof}

\begin{lem} \label{lem:unpacking of T for direct ribbon}
    If $\rho = \{\tau_i\}_{i=1}^N$ is a direct ribbon, then
    $$T^g_{\rho} = \sum_{k_1 \cdots k_N = g} \,\,\, \prod_{i = 1}^N \,\,\, T_{\tau_i}^{k_i}.$$
\end{lem}

\begin{proof}
    By definition, $T_{\rho}^g = F_{\rho}^{1, g}$. Using Eq. \eqref{eq:F inductive def} we find
    $$T_{\rho}^g = F_{\rho}^{1, g} = \sum_{k_1} \, F_{\tau_1}^{1, k_1} F_{\rho \setminus \{\tau_1\}}^{1, \bar k_1 g} = \sum_{k_1} T_{\tau_1}^{k_1}  T_{\rho \setminus \{\tau_1\}}^{\bar k_1 g}.$$
    We can apply this result inductively to find
    $$T_{\rho}^g = \sum_{k_1, \cdots, k_N}  \prod_{i= 1}^N  T_{\tau_i}^{k_i} \, F_{\epsilon}^{1, \bar k_N \cdots \bar k_1 g} = \delta_{k_1 \cdots k_N, g} \, \sum_{k_1, \cdots, k_N} \, \prod_{i=1}^N \, T_{\tau_i}^{k_i}.$$
    This proves the claim.
\end{proof}

\begin{lem} \label{lem:unpacking of L for dual ribbon}
    If $\rho = \{\tau_i\}_{i=1}^N$ is a dual ribbon, then
    $$L^h_{\rho} = \prod_{i = 1}^N \,\,\, L_{\tau_i}^{h}.$$
\end{lem}

\begin{proof}
    This follows immediately from a repeated application of Lemma \ref{lem:L with initial dual triangle}.
\end{proof}

Any ribbon decomposes into subribbons that are alternatingly direct and dual.
\begin{defn} \label{def:alternating decomposition}
    Any finite ribbon $\rho$ has a unique decomposition into ribbons $\{I_a, J_a\}_{a = 1, \cdots, n}$ such that the $I_a$ are direct, the $J_a$ are dual and
    $$\rho = I_1 J_1 \cdots I_n J_n.$$
    (possibly, $I_1$ and/or $J_n$) are empty. We call this the alternating decomposition of $\rho$. 
\end{defn}

\begin{lem} \label{lem:L decomposition}
    Let $\rho$ be a finite ribbon with alternating decomposition $\rho = I_1 J_1 \cdots I_n J_n$. We have
    $$L_{\rho}^h = \sum_{k_1, \cdots, k_n \in G} \, \prod_{i = 1}^n \, T_{I_i}^{k_i}  L_{J_i}^{ \bar K_i h K_i }$$
    where $K_i = k_1 k_2 \cdots k_i$.
\end{lem}

\begin{proof}
    The first sub ribbon $I_1 = \{ \tau_1, \cdots, \tau_m\}$ consists entirely of direct triangles. A repeated application of Lemma \ref{lem:L with initial direct triangle} yields
    $$L_{\rho}^h = \sum_{l_1, \cdots, l_m} \prod_{i= 1}^m \,T_{\tau_i}^{l_i} \,\, L_{\rho \setminus I_1}^{\bar k_1 h k_1}$$
    where $k_1 = l_1 \cdots l_m$. Using Lemma \ref{lem:unpacking of T for direct ribbon} we can rewrite this as
    $$L_{\rho}^h = \sum_{k_1} \, T_{I_1}^{k_1} \, L_{\rho \setminus I_1}^{\bar k_1 h k_1}.$$
    Let us now write $\rho' = \rho \setminus I_1$ and let $J_1 = \{\sigma_1, \cdots, \sigma_{m'}\}$ be the first sub-ribbon of $\rho'$ that consists entirely of dual triangles. A repeated application of Lemma \ref{lem:L with initial dual triangle} yields
    $$L_{\rho'}^{\bar k_1 h k_1} = \prod_{i = 1}^{m'} \, L_{\sigma_i}^{\bar k_1 h k_1} \, L_{\rho' \setminus J_1}^{\bar k_1 h k_1} = L_{J_1}^{\bar k_1 h k_1} L_{\rho' \setminus J_1}^{\bar k_1 h k_1}$$
    where we used Lemma \ref{lem:unpacking of L for dual ribbon} in the last step.

    Putting the above results together, we obtain
    $$L_{\rho}^h = \sum_{k_1} \, T_{I_1}^{k_1} \, L_{J_1}^{\bar k_1 h k_1} \, L_{\rho \setminus {I_1 J_1}}^{\bar k_1 h k_2}.$$

    Repeating the same argument for the ribbon $\rho \setminus \{ I_1 J_1 \} = I_2 J_2 \cdots I_n J_n$ we get
    $$L_{\rho}^h = \sum_{k_1, k_2} T_{I_1}^{k_1} L_{J_1}^{\bar K_1 h K_1} \, T_{I_2}^{k_2} L_{J_2}^{\bar K_2 h K_2} \,\, L_{I_3 J_3 \cdots I_n J_n}^{\bar K_2 h K_2}.$$

    Repeating the argument $n-2$ more times yields the claim.
\end{proof}

\subsection{Wigner projectors and their decompositions} \label{sec:Wigner projectors}

\subsubsection{Basic tools}

We provide some facts that will be used in calculations involving irreducible representations of $\caD(G)$ throughout the paper.

\begin{lem}
\label{lem:unique g=qn}
    Let $C \in (G)_{cj}$, then each element $g \in G$ can be written as $g = q n$ with $q \in Q_C$ and $n \in N_C$ in a unique way.
\end{lem}

\begin{proof}
    We have $g r_C \bar g = q r_C \bar q$ for some $q \in Q_C$. So $\bar q g = n \in N_C$, \ie we have $g = q n$.
    
    As for uniqueness, suppose $q_1 n_1 = q_2 n_2$ with $q_1, q_2 \in Q_C$ and $n_1, n_2 \in N_C$. Then $\bar q_2 q_1 = n_2 \bar n_1 \in N_C$, so $r_C = \bar q_2 q_1 r_C \bar q_1 q_2$ from which it follows that $q_2 r_C \bar q_2 = q_1 r_C \bar q_1$. By construction of the iterator set $Q_C$, this is only possible if $q_1 = q_2$, and therefore also $n_1 = n_2$.
\end{proof}

We will often have to use the Schur orthogonality relations, which we state here for reference. Let $H$ be a finite group and $R_1, R_2 \in (H)_{irr}$ irreducible representations of $H$ with matrix realisations $M_{R_1}$ and $M_{R_2}$ respectively. Then
\begin{equation} 
    \label{eq:Schur}
    \sum_{h \in H} M_{R_1}^{jk}(h) M_{R_2}^{lm}(h)^* = \delta_{R_1,R_2} \delta_{j,l} \delta_{k,m} \frac{|H|}{\dim R_1}.
\end{equation}
If $\chi_R$ is the character of the irreducible representation $R$, then we have,
\begin{equation} 
    \label{eq:Schur2}
    \sum_{R \in (H)_{irr}} \, \chi_R(h_1) \chi_R(h_2)^* = \begin{cases} \abs{ \mathcal{Z}_{h_1} } \,\,\,\, & \text{if} \,\, h_1, h_2 \,\, \text{belong to the same} \,\,  C \in (H)_{cj} \\
    0 \, & \text{otherwise} \end{cases}
\end{equation}
where $\mathcal{Z}_{h_1}$ is the commutant of $h_1$ in $H$.

\subsubsection{Wigner projectors}

Recall the Wigner projectors $D_s^{RC}$ and $D_{s}^{RC;u}$ (Definition \ref{def:Wigner projectors}), and the label changers $A_{s}^{RC;u_2 u_1}$ (Definition \ref{def:label changers}). Note also that $\qd[RC;u] = A_s^{RC;u} B_{s}^{c_i}$ with
\begin{equation*}
    A_s^{RC;u} := \frac{\dimR}{|N_C|} \sum_{m \in N_C} \, R^{jj}(m)^* A_s^{q_i m \bar q_i}.
\end{equation*}

\begin{lem}
\label{lem:a and B commute with A_v and B_f}
    Let $s_0 = (v_0, f_0)$ be a site. Then $A_\site^{RC;(i,j)}$ and $B_{\site}^{c_i}$ are commuting projectors that also commute with $A_v$ and $B_f$ for all $v \neq v_0$ and all $f \neq f_0$.
\end{lem}

\begin{proof}
    First we check that the $A_\site^{RC;(i,j)}$ are projectors.
    \begin{align*}
        A_\site^{RC;(i,j)} A_\site^{RC;(i,j)} &= \bigg(\frac{\dimR}{|N_C|} \bigg)^2 \sum_{m, m' \in N_C} R^{jj}(m)^* R^{jj}(m')^*  A_{s_0} ^{q_i m \dash{q}_i} A_{s_0} ^{q_i m' \dash{q}_i}\\
        &= \bigg(\frac{\dimR}{|N_C|} \bigg)^2 \sum_{m, m' \in N_C} R^{jj}(m)^* R^{jj}(m')^* A_{s_0}^{q_i m m' \dash{q}_i}\\
        \intertext{Relabeling $M = m m'$ and using the Schur orthogonality relation Eq. \eqref{eq:Schur} we get}
        &= \bigg(\frac{\dimR}{|N_C|} \bigg) \sum_{M \in N_C} R^{jj}(M)^* \, A_{s_0}^{q_i M \dash{q}_i} = A_\site^{RC;(i,j)}
    \end{align*}
    Showing that $(A_\site^{RC;(i,j)})^* = A_\site^{RC;(i,j)}$ is a straightforward application of Eq. \eqref{eq:Aelementary}.

    $A_\site^{RC;u}$ trivially commutes with $A_v$ for all $v \in \dvregion$ and $B_\site^{c_i}$ trivially commutes with $B_f$ for $f \in \dfregion$ using Eq. \eqref{eq:ABcommuteondiffsites}. Using the same equation, we also have $[A_\site^{RC;u}, B_f] = 0 = [B_\site^{c_i}, A_v]$ for all $f \in \dfregion, v \in \dvregion$.

    It remains to show that $[a_\site^{RC;u}, B_\site^{c_i}] =0$. This follows from Eq. \eqref{eq:ABonsamesite} and the fact that $q_i m \dash{q}_i$ commutes with $c_i$ for all $m \in N_C$. This implies $[A^{q_i m \dash{q}_i}_\site, B^{c_i}_\site] = 0$.
\end{proof}

\begin{lem}
\label{lem:DRC decomposes into DRCu}
    The $\{  D^{RC;u}_s \}_{u \in I_{RC}}$ are a set of commuting projectors such that $\qd = \sum_u D^{RC;u}_s$. In particular, $\qd$ is a projector.
\end{lem}
\begin{proof}
    That $D_s^{RC;u}$ is a projector follows from $D_s^{RC;u} = A_s^{RC;u} B_s^{c_i}$ and the fact that $A_s^{RC;u}$ and $B_s^{c_1}$ are commuting projectors (Lemma \ref{lem:a and B commute with A_v and B_f}).
    
    Now to prove commutativity, let $u_1 = (i_1, j_1), u_2 = (i_2, j_2)$. Then,
    \begin{align*}
        D^{RC;u_1}_s D^{RC;u_2}_s &= \bigg(\frac{\dimR}{|N_C|}\bigg)^2 \sum_{m_1, m_2 \in N_C} R^{j_1 j_1}(m_1)^* A_{s} ^{q_{i_1} m_1 \dash{q}_{i_1}} B_{s}^{c_{i_1}} R^{j_2 j_2}(m_2)^* A_{s}^{q_{i_2} m_2 \dash{q}_{i_2}} B_{s}^{c_{i_2}} \\
        \intertext{Now we use Eqs. \eqref{eq:Belementary}, \eqref{eq:ABonsamesite} to get:}
        &= \delta_{i_1, i_2} \bigg(\frac{\dimR}{|N_C|}\bigg)^2 \sum_{m_1, m_2 \in N_C} \, R^{j_1 j_1}(m_1)^* A_{s}^{q_{i_1} m_1 m_2 \dash{q}_{i_1}} R^{j_2 j_2}(m_2)^* B_{s}^{c_{i_1}} \\
        \intertext{relabelling $m = m_1 m_2$ and using the Schur orthogonality relation Eq. \eqref{eq:Schur} this becomes}
        &= \delta_{u_1, u_2} D^{RC;u_1}_s.
    \end{align*}
    Finally, to show they sum up to $\qd$,
    \begin{align*}
        \sum_u D^{RC;u}_s &= \sum_{i,j} \frac{\dimR}{|N_C|} \sum_{m \in N_C} R^{jj}(m)^* A_{s} ^{q_i m \dash{q}_i} B_{s}^{c_i} = \sum_{i} \frac{\dimR}{|N_C|} \sum_{m \in N_C} \chi_R(m)^* A_{s}^{q_i m \dash{q}_i} B_{s}^{c_i}\\
        &= \frac{\dimR}{|N_C|} \sum_{m \in N_C} \chi_R(d)^* \sum_{q_i \in Q_C}  A_{s}^{q_i m \dash{q}_i} B_{s}^{c_i} = \qd.
    \end{align*}
\end{proof}

The projectors $D_s^{RC}$ satisfy the following properties.
\begin{lem}
\label{lem:DRCprops}
    The $\qd$ are orthogonal projectors and
    $$\qd[RC] \qd[R'C'] = \delta_{RC, R'C'} \qd[RC], \quad \sum_{RC} \, D_s^{RC} = \I.$$
\end{lem}

\begin{proof}
    This follows immediately from Proposition 21 and Eq. (B77) of \cite{Bombin2007-uw}.
\end{proof}

\begin{lem} \label{lem:qdcommute}
    Let $s_0 = (v_0, f_0)$ be a site, then $D^{RC}_\site$ commutes with $A_{v}, B_{f}$ for all $v \neq v_0$ and all $f \neq f_0$.
\end{lem}
\begin{proof}
    Noting that $D_{\site}^{RC} = \sum_u \, D_{\site}^{RC;u}$ (Lemma \ref{lem:DRC decomposes into DRCu}) and $D_{\site}^{RC;(i, j)} = A_{\site}^{RC;(i, j)} B_{\site}^{c_i}$, the claim follows immediately from Lemma \ref{lem:a and B commute with A_v and B_f}.
\end{proof}

\subsection{Representation basis for ribbon operators} \label{sec:RC basis for ribbon operators}

Recall Definition \ref{def:RC ribbons} of the ribbon operators $\frcuv$. These ribbon operators satisfy the following basic properties.
\begin{lem} (\cite[Lemma~4.11]{Naaijkens2015-xj})
    \label{lem:decomposition of F}
    If $\rho = \rho_1 \rho_2$ then
    $$F^{RC;uw}_{\rho} = \bigg( \frac{\abs{N_C}}{\dimR} \bigg) \sum_{v} F^{RC;uv}_{\rho_1} F^{RC;vw}_{\rho_2}.$$
\end{lem}

\begin{lem}(\cite[Eq. (5.1)]{Naaijkens2015-xj})
\label{lem:FdaggerF identity}
    We have
    $$\sum_v (F^{RC;u_1 v}_\rho)^* F^{RC;u_2 v}_\rho = \delta_{u_1,u_2} \bigg(\frac{\dimR}{|N_C|}\bigg)^2 \mathds{1} \quad \text{and} \quad \sum_v F^{RC;u_1 v}_\rho (F^{RC;u_2 v}_\rho)^* = \delta_{u_1,u_2} \bigg(\frac{\dimR}{|N_C|}\bigg)^2 \mathds{1}$$
\end{lem}

\subsection{Detectors of topological charge} \label{sec:topological charge detectors}

Recall Definition \ref{def:charge detectors} of the `charge detectors' $\knRC[\sigma]$. These satisfy the following basic properties.

\begin{lem}(\cite[Eq. (B77)]{Bombin2007-uw}) \label{lem:basic properties of K}
    The $\knRC[\sigma]$ are orthogonal projectors and
    \begin{equation*}
        \knRC[\sigma] K^{R'C'}_\sigma = \delta_{RC, R'C'} \knRC[\sigma], \quad 
        \sum_{RC} \knRC[\sigma] = \I.
    \end{equation*}
\end{lem}

\subsection{Actions on the \ffgs} \label{sec:actions on FFGS}

In this subsection we consider several ways in which the ribbon operators $F_{\rho}^{RC;uv}$, the projectors $D_{s}^{RC;u}$ (Definition \ref{def:Wigner projectors}), and the label changers $A_{s}^{RC;u_2 u_1}$ (Definition \ref{def:label changers}) act on the frustration free ground state. We will work in the GNS representation $(\pi_0, \caH_0, \ket{\Omega_0})$ of the frustration free ground state $\omega_0$, and will in the remainder of this section drop $\pi_0$ from the notation. \ie for any $O \in \cstar$ we simply write $O$ instead of $\pi_0(O)$.

Let us first note the following.
\begin{lem} \label{lem:omega_0(O) = 1}
    If $O$ is a unitary or a projector and $\omega_0(O) = 1$, then $O \ket{\Omega_0} = \ket{\Omega_0}$.
\end{lem}

\begin{proof}
    We have
    $$ 1 = \omega_0(O) = \langle \Omega_0, O \, \Omega_0 \rangle.$$
    since $\norm{O} \leq 1$ this is only possible if $O \ket{\Omega_0} = \ket{\Omega_0}$.
\end{proof}

\begin{lem} \label{lem:change ribbon operator label}
    Let $\rho$ be a finite ribbon with $\partial_0 \rho = \site$. Then
    \begin{equation*}
		A^{RC;u_2 u_1}_{\site} F^{RC;u_1 v}_{\rho} | \Omega_0 \rangle = F^{RC;u_2 v}_{\rho} | \Omega_0 \rangle.
  \end{equation*}
\end{lem}

\begin{proof}
    Let $u_1 = (i_1, j_2)$, $u_2 = (i_2, j_2)$ and $v = (i', j')$, then
    \begin{align*}
		A^{RC;u_2 u_1}_{\site} F^{RC;u_1 v} &= \left( \frac{\dimR}{|N_C|} \right)^2 \sum_{m, n} \, R^{j_2 j_1}(m)^* R^{j_1 j'}(n)^* A_{s_0}^{q_{i_2} m \bar q_{i_1}} F_{\rho}^{\bar c_{i_2}, q_{i_2} n \bar q_{i'}} \\
		&= \left( \frac{\dimR}{|N_C|}  \right)^2 \sum_{m, n} \,  R^{j_2 j_1}(m)^* R^{j_1 j'}(n)^* \, F_{\rho}^{\bar c_{i_1}, q_{i_1} mn \bar q_{i'}} \, A_{s_0}^{q_{i_2} m \bar q_{i_1}}.\\
 	\intertext{where we used Eq. \eqref{eq:[A,F] and [B,F]}. Since $A_{s_0}^h | \Omega_0 \rangle = | \Omega_0 \rangle$ (Lemma \ref{lem:omega_0(O) = 1}) for all $h \in G$ we then find}
		A^{RC;u_2 u_1}_{\site} F^{RC;u_1 v}_{\rho} | \Omega_0 \rangle &= \left( \frac{\dimR}{|N_C|} \right)^2 \sum_{m, n}  \,  R^{j_2 j_1}(m)^* R^{j_1 j'}(n)^* \, F_{\rho}^{\bar c_{i_1}, q_{i_1} mn \bar q_{i'}} \, | \Omega_0 \rangle \\
      &= \left( \frac{\dimR}{|N_C|} \right)^2 \sum_{m, m'} \sum_l \, R^{j_2 j_1}(m)^* R^{j_1 l}(\bar m)^* R^{l j'}(m')^* \, F_{\rho}^{\bar c_{i_2}, q_{i_2} m' \bar q_{i'}} | \Omega_0 \rangle \\
      &= \left( \frac{\dimR}{|N_C|} \right) \sum_{m} \sum_l R^{j_2 j_1}(m)^* R^{l j_1}(m) \, F_{\rho}^{RC;(i_2, l) v}| \Omega_0 \rangle \\
      &= F_{\rho}^{RC;u_2, v} | \Omega_0 \rangle
    \end{align*}
    where we substituted $m' = mn$ in the second line, and used Eq. \eqref{eq:Schur} in the last step.
\end{proof}

\begin{lem} \label{lem:D kills the ground state}
	We have
	\begin{equation*}
		D^{RC; u}_s | \Omega_0 \rangle = \delta_{RC, R_1 C_1} | \Omega_0 \rangle
	\end{equation*}
	where $R_1 C_1$ is the trivial representation. \ie $C_1 = \{ 1 \}$, so $N_{C_1} = G$, and $R_1$ is the trivial representation of $G$.
\end{lem}

\begin{proof}
	 Note that $D_{s}^{R_1 C_1} = A_{v(s)} B_{f(s)}$, so the frustration free ground state satisfies $\omega_0(D_{s}^{R_1 C_1}) = 1$ which proves the claim in the case that $RC = R_1 C_1$. Using Lemma \ref{lem:pure components projector Lemma} and Lemma \ref{lem:DRCprops} we find $\omega_0(D^{RC;u}_s) = \omega_0(D^{RC;u}_s D^{R_1C_1}_s) = 0$. Finally, since $D^{RC;u}_s$ is a projector and $| \Omega_0 \rangle$ is the GNS vector of $\omega_0$, it follows that $D^{RC;u}_s | \Omega_0 \rangle = 0$ (Lemma \ref{lem:omega_0(O) = 1}).
\end{proof}

\begin{lem} \label{lem:ribbon label projector}
	Let $\rho$ be a finite ribbon with $\partial_0 \rho = \site$. Then
	\begin{equation*}
		D^{RC;u_1}_{\site} F^{RC;u_2 v}_{\rho} | \Omega_0 \rangle = \delta_{u_1, u_2} F^{RC;u_1 v}_{\rho} | \Omega_0 \rangle.
	\end{equation*}
\end{lem}

\begin{proof}
    The proof is a computation using the basic commutation rules of the $A_{s_0}^h$ and $B_{s_0}^g$ with the ribbon operators (Eq. \eqref{eq:[A,F] and [B,F]}) and the fact that $B_{s_0}^g | \Omega_0 \rangle = \delta_{g, e} | \Omega_0 \rangle$ and $A_{s_0}^h | \Omega_0 \rangle = | \Omega_0 \rangle$ for all $h, g \in G$ (Lemma \ref{lem:omega_0(O) = 1}).
    
    Let $u_1 = (i_1, j_2)$, $u_2 = (i_2, j_2)$ and $v = (i', j')$, then
    \begin{align*}
    D^{RC;u_1}_{\site} F^{RC;u_2 v}_{\rho} | \Omega_0 \rangle  &= \left( \frac{\dimR}{|N_C|} \right)^2 \sum_{m, n \in N_C} \, R^{j_1 j_1}(m)^* R^{j_2 j'}(n)^* A_{s_0}^{q_{i_1} m \bar q_{i_1}} \, B_{s_0}^{c_{i_1}} \, F^{\bar c_{i_2}, q_{i_2} n \bar q_{i'}}_{\rho} \, | \Omega_0 \rangle \\
      &= \left( \frac{\dimR}{|N_C|} \right)^2 \delta_{i_1, i_2} \, \sum_{m, n \in N_C} \, R^{j_1 j_1}(m)^* R^{j_2 j'}(n)^* F^{\bar c_{i_1}, q_{i_1} m n \bar q_{i'}}_{\rho} \, A_{s_0}^{q_{i_1} m \bar q_{i_1}} \, | \Omega_0 \rangle \\
      &= \delta_{i_1, i_2} \, \left( \frac{\dimR}{|N_C|} \right)^2 \sum_{n, n' \in N_C} \, \sum_{l} R^{j_1 l}(n')^* R^{l j_1}(\bar n)^* R^{j_2 j'}(n)^* \, F^{\bar c_{i_1}, q_{i_1} n' \bar q_{i'}}_{\rho} \, |\Omega_0 \rangle \\
      &= \delta_{i_1, i_2} \delta_{j_1, j_2} \, \left( \frac{\dimR}{|N_C|} \right) \sum_{n'} R^{j_1 j'}(n')^* F^{\bar c_{i_1}, q_{i_1} n' \bar q_{i'}}_{\rho} | \Omega_0 \rangle = \delta_{u_1, u_2} \, F^{RC;u_1 v}_{\rho} \, | \Omega_0 \rangle
    \end{align*}
    where we used Schur orthogonality to get the last line.
\end{proof}

\subsection{Properties of \texorpdfstring{$\mu_{\rho}^{RC;uv}$}{mu rho RC} and \texorpdfstring{$\chi_{\rho}^{RC;uv}$}{chi rho RC}} \label{sec:properties of mu and chi}

\subsubsection{Various actions on the frustration free ground state}

Recall that we defined $\chi_{\rho}^{RC;u_1 u_2} = \pi_0 \circ \mu_{\rho}^{RC;u_1 u_2} : \cstar \rightarrow \caB{(\caH_0)}$ where $(\pi_0, \caH_0, \ket{\Omega_0})$ is the GNS triple of the frustration free ground state $\omega_0$. In the following we will drop $\pi_0$ from the notation. \ie for any $O \in \cstar$ we simply write $O$ instead of $\pi_0(O)$.

\begin{lem} \label{lem:change ampli label}
	Let $\rho$ be a half-infinite ribbon with $\partial_0 \rho = \site$. For any $O \in \cstar$ we have
	\begin{equation*}
		\chi^{RC;u_1 u_2}_{\rho}( O A^{RC;u_3 u_2}_{\site}  ) | \Omega_0 \rangle = \chi^{RC;u_1 u_3}_{\rho}(O) | \Omega_0 \rangle.
	\end{equation*}
	In particular,
	\begin{equation*}
		\chi^{RC;u_1 u_2}_{\rho}( A^{RC;u_3 u_2}_{\site} ) | \Omega_0 \rangle = \chi^{RC;u_1 u_3}_{\rho}(\I) | \Omega_0 \rangle = \delta_{u_1 u_3} | \Omega_0 \rangle.
	\end{equation*}
\end{lem}

\begin{proof}
	By definition
	\begin{equation*}
		\chi^{RC;u_1 u_2}_{\rho}( O A_s^{RC;u_3 u_2}) = \lim_{n \uparrow \infty}  \, \bigg(\frac{|N_C|}{\dimR} \bigg)^2 \,  \sum_v  \big( F_{\rho_n}^{RC;u_1 v} \big)^*  \, O A_{\site}^{RC;u_3 u_2} \, F_{\rho_n}^{RC;u_2 v}
	\end{equation*}
	so using Lemma \ref{lem:change ribbon operator label} we get
	\begin{align*}
		\chi^{RC;u_1 u_2}_{\rho}( O A_{\site}^{RC;u_3 u_2}) | \Omega_0 \rangle &= \lim_{n \uparrow \infty} \, \bigg(\frac{|N_C|}{\dimR} \bigg)^2 \,  \sum_v \, \big( F_{\rho_n}^{RC;u_1 v} \big)^*  \, O \,  F_{\rho_n}^{RC;u_3 v} | \Omega_0 \rangle \\
                &= \chi_{\rho}^{RC;u_1 u_3}(O)
	\end{align*}
	as required.

	The last claim follows immediately from the first and item 2 of Lemma \ref{prop:ampli properties}.
\end{proof}

It follows that
\begin{lem} \label{lem:ampli label projector}
	Let $\rho$ be a half-infinite ribbon with $\partial_0 \rho = \site$. For any $O \in \cstar$ we have
	\begin{equation*}
		\chi^{RC;u_2 u_1}_{\rho}( O D_{\site}^{RC;u_3} ) | \Omega_0 \rangle = \delta_{u_1, u_3} \chi_{\rho}^{RC;u_2 u_1}(O) \, | \Omega_0 \rangle.
	\end{equation*}
\end{lem}

\begin{proof}
	We have
	\begin{align*}
		\chi^{RC;u_2 u_1}_{\rho}(O D_{\site}^{RC;u_3}) | \Omega_0 \rangle &= \lim_{n \uparrow \infty} \, \bigg(\frac{|N_C|}{\dimR} \bigg)^2 \,  \sum_{v} \big(F_{\rho_n}^{RC;u_2 v} \big)^* \, O D_{\site}^{RC;u_3} \, F_{\rho_n}^{RC;u_1 v} \, | \Omega_0 \rangle \\
		\intertext{Using Lemma \ref{lem:ribbon label projector} this becomes}
							     &= \delta_{u_1, u_3} \, \lim_{n \uparrow \infty} \, \bigg(\frac{|N_C|}{\dimR} \bigg)^2 \, \sum_v \, \big( F_{\rho_n}^{RC;u_2 v} \big) \, O \, F_{\rho_n}^{RC;u_1 v} | \Omega_0 \rangle\\
							     &= \delta_{u_1, u_3} \, \chi_{\rho}^{RC;u_2 u_1}(O) | \Omega_0 \rangle,
	\end{align*}
	where in the last step we used $\chi_{\rho}^{RC;u_1 u_2} = \pi_0 \circ \mu_{\rho}^{RC;u_1 u_2}$ and the definition of $\mu_{\rho}^{RC;u_1 u_2}$ given in Proposition \ref{prop:ampli properties}.
\end{proof}

\begin{lem} \label{lem:chi preserves constraints}
    Let $\rho$ be a half-infinite ribbon with $\partial_0 \rho = \site$. For any vertex $v \neq v(\site)$ and any face $f \neq v(\site)$ we have
    $$ \chi_{\rho}^{RC;uu}( A_v ) \ket{\Omega_0} = \chi_{\rho}^{RC;uu}(B_f) \ket{\Omega_0} = \ket{\Omega_0}. $$
\end{lem}

\begin{proof}
    This follows from the definition of $F^{RC;uv}_\rho$, Eq. \eqref{eq:[A,F] and [B,F]}, and the fact that $A_v \ket{\Omega_0} = B_f \ket{\Omega_0} = \ket{\Omega_0}$ for any $v \in \latticevert$ and any $f \in \latticeface$.
\end{proof}

\subsubsection{A tool to prove non-degeneracy of the amplimorphism representation}

We continue to work in the GNS representation $(\pi_0, \caH_0, \ket{\Omega_0})$ of the frustration free ground state $\omega_0$ and again drop $\pi_0$ from the notation.

\begin{defn} \label{def:magic map}
	For any finite ribbon $\rho$ with $\partial_0 \rho = \site$ and any $RC$ and $u, v \in I_{RC}$ define a linear map $t_{\rho}^{RC;u v} : \cstar \rightarrow \cstar$ by
	\begin{equation}
		t_{\rho}^{RC;uv}( O ) :=  \bigg(\frac{\dimR}{|N_C|} \bigg)^2 \, \sum_{w, z} \, F_{\rho}^{RC;u w}  \, O \, \big( F_{\rho}^{RC;z w} \big)^* A_{\site}^{RC;z v} D_{\site}^{RC;v}.
	\end{equation}
\end{defn}

Recall that for a half-infinite ribbon $\rho$ we write $\rho_n$ for the finite ribbons consisting of the first $n$ triangles of $\rho$. We have the following remarkable property:
\begin{lem} \label{lem:magic map}
	Let $\rho$ be a half-infinite ribbon with $\partial_0 \rho = \site$. For any local $O$ whose support does not intersect $\rho \setminus \rho_n$ we have
	\begin{equation}
		\chi_{\rho}^{RC;u_1 v_1} \big( t_{\rho_n}^{RC;u_2 v_2}(O) \big) | \Omega_0 \rangle = \delta_{u_1 u_2} \delta_{v_1 v_2} \, O | \Omega_0 \rangle.
	\end{equation}
\end{lem}

\begin{proof}
	We compute
	\begin{align*}
		\chi_{\rho}^{RC;u_1 v_1} \big( t_{\rho_n}^{RC;u_2 v_2}(O) \big) | \Omega_0 \rangle &= \bigg(\frac{\dimR}{|N_C|} \bigg)^2 \, \sum_{w, u}  \chi^{RC;u_1 v_1}_\rho \left(  F_{\rho_n}^{RC;u_2 w} \, O \, \big( F_{\rho_n}^{RC;u w} \big)^* \, A^{RC;u v_2}_\site D^{RC;v_2}_\site  \right) | \Omega_0 \rangle \\
		\intertext{Using Lemma \ref{lem:change ampli label} and Lemma \ref{lem:ampli label projector}, this becomes: }
      &= \delta_{v_1, v_2} \, \bigg(\frac{\dimR}{|N_C|} \bigg)^2 \, \sum_{w, u} \, \chi^{RC;u_1 u}_\rho \left( F_{\rho_n}^{RC;u_2 w} \, O \, \big( F_{\rho_n}^{RC;u w} \big)^* \right) | \Omega_0 \rangle \\
      \intertext{Taking $m > n$ large enough we get}
      &= \delta_{v_1, v_2} \, \sum_{u, v, w}  \, \big( F_{\rho_m}^{RC;u_1 v} \big)^* F_{\rho_n}^{RC;u_2 w} \, O \, \big( F_{\rho_n}^{RC;u w} \big)^* F_{\rho_m}^{RC;u v} | \Omega_0 \rangle \\
      \intertext{Decomposing $\rho_m = \rho_n \rho'$ we get, using Lemma \ref{lem:decomposition of F}}
      &= \delta_{v_1, v_2} \, \bigg(\frac{\abs{N_C}}{\dimR} \bigg)^2 \,  \sum_{u, v, w} \, \sum_{y, z} \, \big(  F_{\rho_n}^{RC;u_1 y} F_{\rho'}^{RC;y v}  \big)^* \, F_{\rho_n}^{RC;u_2 w} \, O  \\
        & \quad\quad\quad\quad\quad \times \, \big( F_{\rho_n}^{RC;u w} \big)^* \, F_{\rho_n}^{RC;u z} F_{\rho'}^{RC;z v} \, | \Omega_0 \rangle \\
        \intertext{Since the support of $O$ does not intersect $\rho \setminus \rho_n \supset \rho'$ we have $[F_{\rho'}^{RC;z v}, O] = 0$. We also have $[F_{\rho'}^{RC;z v},F_{\rho_n}^{RC;u',w'}] = 0$ for all $u',w' \in I_{RC}$ since $\rho'$ and $\rho_n$ are disjoint. We can therefore commute $F_{\rho'}^{RC;z v}$ to the left and get,}
        &= \delta_{v_1, v_2} \, \bigg(\frac{\abs{N_C}}{\dimR} \bigg)^2 \,  \sum_{w} \, \sum_{y, z} \, \bigg(   \sum_{v} \big(F_{\rho'}^{RC;y v}\big)^*\, F_{\rho'}^{RC;z v} \bigg)  \,  \\
        & \quad\quad\quad\quad\quad \times \, \big( F_{\rho_n}^{RC;u_1 y}\big)^* \, F_{\rho_n}^{RC;u_2 w}  O \bigg( \, \sum_u \, \big( F_{\rho_n}^{RC;u w} \big)^* \, F_{\rho_n}^{RC;u z} \bigg) \, | \Omega_0 \rangle \\
      \intertext{Using Lemma \ref{lem:FdaggerF identity}, the sum over $u$ yields a $\delta_{w, z}$ and the sum over $v$ yields a $\delta_{y, z}$ so}
      &= \delta_{v_1, v_2} \,\bigg(\frac{\dimR}{|N_C|} \bigg)^2 \,  \sum_{w} \, \big(  F_{\rho_n}^{RC;u_1 w} \big)^* \, F_{\rho_n}^{RC;u_2 w} \, O \, | \Omega_0 \rangle  \\
      &= \delta_{u_1, u_2} \delta_{v_1, v_2} \, O | \Omega_0 \rangle
	\end{align*}
	as required.
\end{proof}

%% file: sectors/String_nets.tex
\section{Properties of string nets}
\label{sec:properties of string nets}

\subsection{Direct paths and flux}

Recall the definitions of direct paths and the direct path of a ribbon from Section \ref{subsec:preliminary notions}.

If a direct path $\gamma$ is supported in a region $S \subset \latticeedge$ and $\al \in \gc[S]$ is a gauge configuration on $S$ then we define the flux of $\al$ through $\gamma$ to be
$$ \phi_{\gamma}(\al) := \prod_{e \in \gamma} \, \al_e, $$
where the product is ordered according to the order of $\gamma$. We have $\phi_{\dash{\gamma}}(\alpha) = \dash{\phi_{{\gamma}}(\alpha)}$ and if $\gamma = \gamma_1 \gamma_2$ then we have $\phi_\gamma(\alpha) = \phi_{\gamma_1} (\alpha) \phi_{\gamma_2} (\alpha)$.

Similarly, we say a finite ribbon $\rho = \{ \tau_i \}_i^l$ is supported in $S \subset \latticeedge$ if for all $i = 1, \cdots, l$ we have $e_{\tau_i} \in S$ or $\bar e_{\tau_i} \in S$. In that case the direct path $\rho^{dir}$ is supported in $S$ and we put $\phi_{\rho}(\al) := \phi_{\rho^{dir}}(\al)$. This is consistent with Definition \ref{def:flux thourgh ribbon}.

\subsection{Fluxes of string-nets}

\begin{defn} \label{def:face-move}
    We say two direct paths $\gamma_1$ and $\gamma_2$ are related by a face-move over $f \in \latticeface$ if $\gamma_1 = \gamma' \gamma^{f}_1 \gamma''$ and $\gamma_2 = \gamma' \gamma^{f}_2 \gamma''$ for direct paths $\gamma', \gamma'', \gamma_1^f$ and $\gamma_2^f$ such that $\gamma_1^f \bar \gamma_2^f$ is a closed direct path consisting of three edges circling the face $f$.
\end{defn}

\begin{lem} \label{lem:trivial face-move}
    Let $\gamma_1$, $\gamma_2$ be direct paths in $\eregion$ that are related by a face-move over $f \in \fregion$, and let $\al \in \gc$ be such that $B_f \ket{\al} = \ket{\al}$, then
    $$\phi_{\gamma_1}(\al) = \phi_{\gamma_2}(\al).$$
\end{lem}

\begin{proof}
    From the definition, we have $\gamma_1 = \gamma' \gamma^{f}_1 \gamma''$ and $\gamma_2 = \gamma' \gamma^{f}_2 \gamma''$ for direct paths $\gamma', \gamma'', \gamma_1^f$ and $\gamma_2^f$ such that $\gamma_1^f \bar\gamma_2^f$ is a closed direct path consisting of three edges circling the face $f$.

    It follows from $B_f \ket{\al} = \ket{\al}$ that $\phi_{\gamma_1^f \bar\gamma_2^f }(\al) = 1$, or $\phi_{\gamma^f_1}(\al) = \phi_{\gamma^f_2}(\al)$. It follows that
    $$ \phi_{\gamma_1}(\al) =  \phi_{\gamma'}(\al) \phi_{\gamma^f_1}(\al) \phi_{\gamma''}(\al) = \phi_{\gamma'}(\al) \phi_{\gamma_2^f}(\al) \phi_{\gamma''}(\al) = \phi_{\gamma_2}(\al) $$
    as required.
\end{proof}

Recall the fiducial ribbons $\nu_n$ and boundary ribbons $\beta_n$ defined in Section \ref{subsec:local gauge configurations and boundary conditions}, see Figure \ref{fig:fiducial_and_boundary_ribbons}.
\begin{lem} \label{lem:bcinC}
    Let $C \in (G)_{cj}$ and $i = 1, \cdots, \abs{C}$. If $\alpha \in \packi$ then we have
    $$\phi_{\bdy}(\al) = \dash{ \phi_{\fidu}(\al)  } \, c_i \, \phi_{\fidu}(\al) \in C.$$
\end{lem}

\begin{proof}
    Let $\gamma_{\site}$ be the direct path of $\rho_{\triangle}(\site)$ so $\phi_{\gamma_{\site}}(\alpha) = c_i \in C$. Let $\gamma_{\fidu}$ be the direct path of $\fidu$, and $\gamma_{\bdy}$ the direct path of the boundary ribbon $\bdy$. Consider the direct path $\gamma = \gamma_{\site} \gamma_{\fidu} \dash{\gamma_{\bdy}} \, \dash{\gamma_{\fidu}}$.

    Since $\al \in \packi$ we have by definition that $B_f \ket{\al} = \ket{\al}$ for all $f \in \dfregion$. Since $\gamma$ can be shrunk to the empty ribbon through a sequence of face-moves over faces $f \in \dfregion$, it follows from Lemma \ref{lem:trivial face-move} that $\phi_{\gamma}(\al) = \phi_{\emptyset}(\al) = 1$, which is equivalent to
    $$ \phi_{\bdy}(\al) = \phi_{\gamma_{\bdy}}(\al) = \dash{\phi_{\gamma_{\fidu}}(\al)} \, \phi_{\gamma_{\site}}(\al) \,\ \phi_{\gamma_{\fidu}}(\al) = \dash{\phi_{\fidu}(\al)} \, c_i \, \phi_{\fidu}(\al) $$
    as required.
\end{proof}

\begin{lem} \label{lem:inNC}
    Let $\alpha \in \packib$, then $\bar q_i \phi_{\fidu}(\alpha) q_{i(b)} \in N_C$.
\end{lem}

\begin{proof}
    From Lemma \ref{lem:bcinC} we have
    $$ \phi_{\bdy}(\al) = \dash{ \phi_{\fidu}(\al)  } \, c_i \, \phi_{\fidu}(\al) = \dash{ \phi_{\fidu}(\al)  } \,q_i \, r_C \, \bar q_i \, \phi_{\fidu}(\al), $$
    in particular, $\phi_{\bdy}(\al) = \phi_{\bdy}(b(\al)) \in C$, so we have a unique label $i(b) \in \{ 1, \cdots, \abs{C} \}$ such that $\phi_{\bdy}(\al) = q_{i(b)} r_C \bar q_{i(b)}$. Usign this we obtain
    $$  r_C = \bar q_{i(b)}\dash{ \phi_{\fidu}(\al)  } \,q_i \, r_C \, \bar q_i \, \phi_{\fidu}(\al) \, q_{i(b)}. $$
    This shows that $\bar q_i \phi_{\fidu}(\al) \, q_{i(b)} \in N_C$, as required.
\end{proof}

\subsection{The action of gauge groups on string nets}
\label{app:Uisfaithful}

Recall the group of gauge transformations $\gauge$ consisting of unitaries of the form $\mathcal{U}(\{g_v\}) = \prod_{v \in \vregion} \, A_v^{g_v}$ with $g_v \in G$ for each $v \in \vregion$. These gauge transformations act in the bulk of $\eregion$, they are all supported on $\eregion \setminus \partial \eregion$.

We define \emph{boundary gauge transformations} acting on $\hilb_n$ in a similar way.

\begin{defn} \label{def:boundary gauge transformations}
    Recall $\partial \vregion = \vregion[n+1] \setminus \vregion$ and let $\partial \gauge$ be the group of unitaries of the form $\mathcal{U}(\{g_v\}) = \prod_{v \in \vertex(\partial \eregion)} \tilde{A}_v^{g_v}$ with $g_v \in G$ for each $v \in \partial \vregion$. Here $\tilde{A}_{v}^g$ is the restriction of $A_v^g$ to $\hilb_n$. We call $\partial \gauge$ the group of boundary gauge transformations.
\end{defn}

Note that the boundary gauge transformations are supported on $\eregion \setminus \eregion[n-1]$.

\begin{figure}
    \centering
    \includegraphics[width=0.5\textwidth]{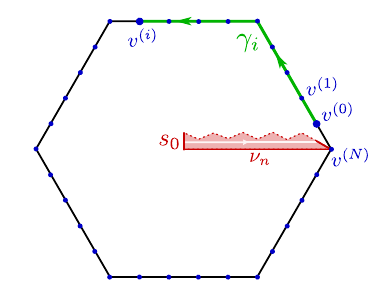}
    \caption{Vertices $v^{(i)}$ and direct paths $\sigma_i$ on the boundary of the region $\eregion$, defined relative to the fiducial ribbon $\fidu$. The region $\eregion$ is indicated by the black hexagon. The individual edges of $\eregion$ are not shown.}
    \label{fig:boundary gauge construction}
\end{figure}

Before proving the Lemma on the free and transitive action of the gauge group $\gauge$, we show the following result, which will help us prove uniqueness statements for gauge transformations.

\begin{lem} \label{lem:unique gauge tranformations}
    If $\alpha \in \gc$ is any gauge configuration on $\eregion$ and $U \in \gauge$ is such that $U \ket{\alpha} = \ket{\alpha}$, then $U = \mathds{1}$.
\end{lem}

\begin{proof}
    Since $U \in \gauge$ it is of the form $U = \prod_{v \in \vregion} \, A_v^{g_v}$ for group elements $g_v \in G$. For any edge $e = (v, v')$ with $v \in \partial \vregion$ and $v' \in \vregion$ we have $\alpha_e = \alpha_e \bar g_{v'}$, so $g_v' = 1$. This shows that in fact $U \in \mathcal{G}_{n-1}$. Proceeding inductively, we find that $U \in \mathcal{G}_0$, i.e. $U$ is of the form $U = A_{v_0}^{g_{v_0}}$. Finally, by considering any edge $e = (v_0, v)$ we find that $\alpha_e = g_{v_0} \alpha_e$ so $g_{v_0} = 1$ and $U = \mathds{1}$.
\end{proof}

Recall the boundary gauge transformations $\partial \gauge$ of Definition \ref{def:boundary gauge transformations}.

\begin{lem} \label{lem:transitive boundary action}
    For any pair of boundary conditions $b, b' \in \bc[C]$ that are compatible with conjugacy class $C$ there is a boundary gauge transformation $U_{b' b} \in \partial \gauge$ such that for any $\alpha \in \packC[C;ib]$ we have $U_{b' b} \ket{\alpha} = \ket{\alpha'}$ for an $\alpha' \in \packC[C;ib']$ that satisfies $\alpha_e = \alpha'_e$ for all $e \in \eregion[n-1]$.
\end{lem}

\begin{figure}[t!]
    \centering
    \begin{subfigure}[t]{0.49\textwidth}
    \centering
    \includegraphics[width=\textwidth]{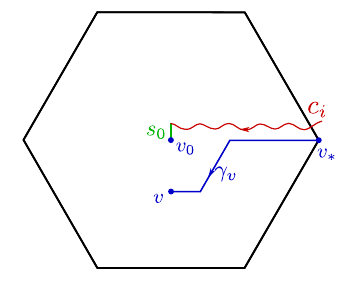}
    \caption{A simple string net $\alpha^{(0)} \in \packC[C;i]$ that is non-trivial only on the dual part of the fiducial ribbon $\fidu$. The corresponding boundary condition $b_0$ is trivial everywhere except at one edge.}
    \label{fig:simple_string_net}      
    \end{subfigure}
    \hfill
    \begin{subfigure}[t]{0.49\textwidth}
    \centering
    \includegraphics[width=\textwidth]{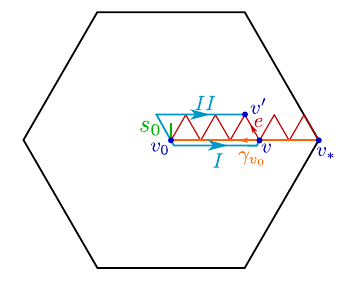}
    \caption{Paths used in the proof of Lemma \ref{lem:transitive bulk action}.}
    \label{fig:paths_I_II}   
    \end{subfigure}
    \caption{}
\end{figure}

\begin{proof}
    Fix a conjugacy class $C$ and a flux $c_i \in C$. We will first prove the claim for the simple boundary condition $b_0$ corresponding to the string net of Figure \ref{fig:simple_string_net}. i.e. let $b$ be an arbitrary boundary condition compatible with $C$. We will construct a boundary gauge transformation $U_{b_0 b} \in \partial \gauge$ such that for any $\alpha \in \packC[C;ib]$ we have $U \ket{\alpha} = \ket{\alpha'}$ with $\alpha' \in \packC[C;ib_0]$ such that $\alpha_e = \alpha'_e$ for all $e \in \eregion[n-1].$

    To that end, let $\partial \vregion = \{v^{(0)}, v^{(1)}, \cdots, v^{(N)}\}$ be a labeling of the vertices in $\partial \vregion$ as in Figure \ref{fig:boundary gauge construction}. For $i = 1, \cdots, N$, let $\sigma_i$ be the direct path proceeding counterclockwise around $\partial \eregion$ from $v^{(0)}$ to $v^{(i)}$ as in the Figure. Let $\sigma_{N+1}$ be the direct path that circles $\partial \eregion$ in a counterclockwise direction starting and ending at $v^{(0)}$.

    Since $\phi_{\sigma_{N+1}}(b) \in C$ it can be written as $\phi_{\sigma_{N+1}}(b) = c_{i'} = q c_i \bar q$ for some $q \in G$. Set $g_{v^{(i)}} := \bar q \, \phi_{\sigma_i}(b)$ and $U_{b_0 b} = \prod_{i = 1}^N \, \tilde A_{v^{(i)}}^{g_{v^{(i)}}} \in \partial \gauge$. Take $\alpha \in \packC[C;ib]$ and let $\alpha'$ be the unique string net such that $U \ket{\alpha} = \ket{\alpha'}$. Then for any $i = 1, \cdots, N-1$ we set $e = (v^{(i)}, v^{(i+1)}) \in \partial \eregion$, and find $$b(\alpha')_e = \alpha'_e = g_{v^{(i)}} \, \alpha_e \, g_{v^{(i+1)}} = \bar q \phi_{\sigma_i}(b)  b_e \bar \phi_{\sigma_{i+1}}(b) q = 1.$$
    Furthermore, for the final boundary edge $e = (v^{(N)}, v^{(0)})$ we have
    $$b(\alpha')_e = \alpha'_e = g_{v^{(N)}} \alpha_e \bar g_{v^{(0)}} = \bar q \phi_{\sigma_{N+1}}(b) q = c_i.$$
    We conclude that $b(\alpha') = b$. Moreover, since $U_{b_0 b}$ is supported on $\eregion \setminus \eregion[n-1]$ we have $\alpha_e = \alpha'_e$ for all $e \in \eregion[n-1]$. This proves the existence part of the claim is this special case.

    Using the same arguments one can show that if $\alpha \in \packC[C;ib_0]$, then $U_{b_0 b}^* \ket{\alpha} = \ket{\alpha'}$ for a string net $\alpha' \in \packC[C;ib_0]$ such that $\alpha_e = \alpha'_e$ for all $e \in \eregion[n-1]$.

    Let us now prove the claim for general boundary conditions $b, b'$ compatible with $C$.
    
    We set $U_{b' b} = U_{b_0 b'}^* U_{b_0 b} \in \partial \gauge$ where the boundary gauge transformations $U_{b_0 b}$ and $U_{b_0 b'}$ are as constructed above. then for any $\alpha \in \packC[C;ib]$ we have $U_{b' b} \ket{\alpha} = \ket{\alpha'}$ for a string net $\alpha' \in \packC[C;ib']$ such that $\alpha_e = \alpha'_e$ for all $e \in \eregion[n-1]$. This proves the general case.
\end{proof}

Recall the Definition \ref{def:string nets Cib(m)} of the collections of string nets $\packC[C;i b](m)$:

Fix a conjugacy class $C$, a boundary condition $b$ compatible with $C$ and a label $i = 1, \cdots, |C|$. For any $m \in N_C$ we have
$$\packib(m) := \{\alpha \in \packib \, : \, \phi_\fidu(\alpha) = q_i m \dash q_{i'}\}$$
where $i' = i(b)$.

\begin{lem} \label{lem:transitive bulk action}
    For any two $\alpha, \alpha' \in \packib(m)$ there is a unique gauge transformation $U \in \dgauge$ such that $U \ket{\alpha} = \ket{\alpha'}$. Moreover, if $\alpha \in \packib(m)$ and $U \in \dgauge$ then $U \ket{\alpha} = \ket{\alpha'}$ with $\alpha' \in \packib(m)$. i.e. $\dgauge$ acts freely and transitively on $\packib(m)$.
\end{lem}

\begin{proof}
    Fix a conjugacy class $C$ and a flux $c_i \in C$. We will first prove the claim for the simple boundary condition $b_0$ corresponding to the string net of Figure \ref{fig:simple_string_net}. Denote by $\alpha^{(0)} \in \packC[C;i b_0](1)$ the string net depicted in that figure. It has trivial gauge configuration everywhere except at the red edges. We will first show that for any $\alpha \in \packC[C;ib_0]$ there is a $U \in \dgauge$ such that $U \ket{\alpha} = \ket{\alpha^{(0)}}$.

    Let $v_* \in \partial \vregion$ be the vertex as defined in Figure \ref{fig:simple_string_net}. For any site $v \in \vregion \cup \partial \vregion$, let $\gamma_v$ be a direct path from $v_*$ to $v$ that does not contain any of the red edges (Lemma \ref{lem:trivial face-move}). i.e. $\gamma_v$ is forbidden from crossing the fiducial ribbon. See Figure \ref{fig:simple_string_net} for an example. Define $g_v := \phi_{\gamma_v}(\alpha)$ for all $v \in \vregion \cup \partial \vregion$. Note that since $\alpha$ satisfies the flat gauge condition for all faces except for $f_0$, the group elements $g_v$ are independent of the choice of path $\gamma_v$, as long as we stick to paths that do not include red edges. (This strip acts as a branch cut.) Note further that since the boundary condition is trivial everywhere except on the red boundary edge, we have $g_v = 1$ for all $v \in \partial \vregion$. Moreover, $g_{v_0} = 1$ because we can take $\gamma_{v_0}$ to run along the direct part of the fiducial ribbon as in Figure \ref{fig:paths_I_II}. Since $\alpha \in \packC[C;i b_0](1)$, we have $\phi_{\fidu}(\alpha) = 1$, therefore $g_{v_0} = 1$.

    Let
    $$U = \prod_{v \in \vregion} \, A_v^{g_v} \,\, \prod_{v \in \partial \vregion} \, \tilde A_v^{g_v} = \prod_{v \in \dvregion} \, A_v^{g_b}$$
    where we used that $g_v = 1$ for all $v \in \partial \vregion$ and for $v = v_0$. i.e. we have $U \in \dgauge$.

    We now let $\alpha' \in \packC[C;i b_0]$ be the unique string net such that $U \ket{\alpha} = \ket{\alpha'}$. We will show that $\alpha' = \alpha^{(0)}$.

    Let $e = (v, v')$ be an edge that is not red. Then $\alpha'_e = g_{v} \alpha_e \bar g_{v'} = 1$ because $g_v \alpha_e$ is the flux of $\alpha$ through $\gamma_{v} e$, which is a path from $v^{(N)}$ to $v'$ that does not involve red edges. As noted before, that implies $g_v \alpha_e = g_{v'}$. We see that $\alpha'_e = 1$ for all edges $e$ except possibly the red edges.

    Let us now consider a red edge $e = (v, v')$ which we take to be oriented upwards so that $\alpha^{(0)}_e = c_i$, see Figure \ref{fig:paths_I_II}. Let $I$ be the path from $v_0$ to $v$ and $II$ the path from $v_0$ to $v'$ as shown in Figure \ref{fig:paths_I_II}. Then $\gamma_{v_0} I$ is a path from $v_*$ to $v$ and since $g_{v_0} = 1$ we have $g_{v} = \phi_I(\alpha)$. Similarly, we have $g_{v'} = \phi_{II}$. Let $\gamma_{s_0}$ be the direct path which starts and ends at $v_0$ and circles $f_0$ in a counterclockwise direction. The closed loop $I e \overline{II}$ can be shrunk to $\gamma_{s_0}$ by a sequence of face-moves (Definition \ref{def:face-move}) over faces $f \in \dfregion$. Since $\alpha \in \packC[C;ib_0]$ we have $B_f \ket{\al} = \ket{\al}$ for all $f \in \dfregion$ so it follows from Lemma \ref{lem:trivial face-move} that $c_i = \phi_{\gamma_{s_0}}(\al) = \phi_{I e \overline{II}}(\al) = g_v \al_{e} \bar g_{v'} = \al_e'$.
    
    Let now $\alpha_1, \alpha_2 \in \packC[C;i b_0](1)$ be arbitrary. We have just shown that there are gauge transformations $U_1, U_2 \in \dgauge$ such that $U_1 \ket{\alpha_1} = U_2 \ket{\alpha_2} = \ket{\alpha^{(0)}}$. It follows that the gauge transformation $U = U_2^* U_1 \in \dgauge$ satisfies $U \ket{\alpha} = \ket{\alpha'}$. i.e. we have shown the existence claim in the special case of $\packC[C;i b_0](1)$.

    Let us now generalise to $\alpha_1, \alpha_2 \in \packC[C;i b_0](m)$ for arbitrary $m \in N_C$. i.e. these string nets satisfy $\phi_{\fidu}(\alpha_1) = \phi_{\fidu}(\alpha_2) = q_i m \bar q_{i}$ where we noted that $i(b_0) = i$.

    Acting with the gauge transformation $U_{v_0} = A_{v_0}^{q_i \bar m \bar q_i}$ yields $U_{v_0} \ket{\alpha_1} = \ket{\alpha'_1}$ and $U_{v_0} \ket{\alpha_2} = \ket{\alpha'_2}$ for string nets $\alpha'_1, \alpha'_2 \in \packC[C;i](1)$. Here the flux $c_i$ at $s_0$ was preserved because $q_i \bar m \bar q_i$ commutes with $c_1$, and the action of $A_{v_0}^{q_i \bar m \bar q_i}$ multiplies the flux through $\fidu$ from the left by $q_i \bar m \bar q_i$, thus trivializing it.

    Applying the above result, we have a  gauge transformation $U \in \dgauge$ such that $U \ket{\alpha'_1} = \ket{\alpha'_2}$. Since $U$ commutes with $U^*_{v_0}$ we then find
    $$U \ket{\alpha_1} = U U_{v_0}^* \ket{\alpha'_1} = U_{v_0}^* \ket{\alpha'_2} = \ket{\alpha_2}.$$
    This proves the existence claim in the case of $\packC[C;ib_0](m)$ for arbitrary $m \in N_C$.

    Let us now consider a general boundary condition $b$ that is compatible with $C$. Take $\alpha_1, \alpha_2 \in \packC[C;ib](m) \subset \packC[C;ib]$ for some $m \in N_C$. Lemma \ref{lem:transitive boundary action} provides a boundary gauge transformation $U_{b_0 b} \in \partial \gauge$ such that $U_{b_0 b} \ket{\alpha_1} = \ket{\alpha'_1}$ and $U_{b_0 b} \ket{\alpha_2} = \ket{\alpha'_2}$ for string nets $\alpha'_1, \alpha'_2 \in \packC[C;ib_0](m')$ for some $m' \in N_C$. Here we noted that since $\alpha_1$ and $\alpha_2$ have the same flux through the fiducial ribbon $\fidu$ and both are acted on by the same boundary gauge transformation $U_{b_0 b}$, the resulting string nets $\alpha'_1$ and $\alpha'_2$ also have the same flux through $\fidu$ (though possibly different from the fluxes of $\alpha_1$ and $\alpha_2$).

    Using the result obtained above, we have a gauge transformation $U \in \dgauge$ such that $U \ket{\alpha'_1} = \ket{\alpha'_2}$. since $U$ commutes with $U_{b_0 b}$ we find
    $$U \ket{\alpha_1} = U U_{b_0 b}^* \ket{\alpha'_1} = U_{b_0 b}^* U \ket{\alpha'_1} = U_{b_0 b}^* \ket{\alpha'_2} = \ket{\alpha_2}.$$
    This proves the existence part of the claim in full generality.

    As for uniqueness, take $\alpha_1, \alpha_2 \in \packC[C;ib](m)$ and suppose that $U, U' \in \dgauge$ both satisfy $U \ket{\alpha_1} = U' \ket{\alpha_1} = \ket{\alpha_2}$. Then $U' U^* \ket{\alpha_2} = \ket{\alpha_2}$ and it follows from Lemma \ref{lem:unique gauge tranformations} that $U' = U$.

    It remains to show that if $\alpha \in \packib(m)$ and $U \in \dgauge$, then $U \ket{\alpha} = \ket{\alpha'}$ for an $\alpha' \in \packib(m)$. To see this it is sufficient to note that $U$ is supported on $\eregion \setminus \partial \eregion$ and therefore it cannot change the boundary condition. Further, By Lemma \ref{lem:a and B commute with A_v and B_f} any gauge transformation not supported on $v_0$ commutes with the projectors $B_{\site}^{c_i}$, so $U \in \dgauge$ cannot change the label $i$. Finally, if $v \in \dvregion$ then either no edges incident on $v$ belong to the direct path of the fiducial ribbon, in which case $\phi_{\fidu}(\alpha') = \phi_{\fidu}(\alpha)$ is obvious. Or, precisely two edges incident on $v$ are part of the fiducial ribbon, say $e_i$ and $e_{i+1}$ where we have labeled the direct edges of the fiducial ribbon $\{e_1, \cdots, e_n \}$ along the orientation of $\fidu$. In that case, if $\ket{\alpha'} = A_v^h \ket{\alpha}$, then $\alpha'_{e_i} = \alpha_{e_i} \bar h$ and $\alpha'_{e_{i+1}} = h \alpha_{e_{i+1}}$, and $\alpha$ and $\alpha'$ agree on all other edges. It follows that
    $$\phi_{\fidu}(\alpha') = \prod_{j = 1}^n \alpha'_{e_j} = \prod_{j = 1}^{i-1} \alpha_{e_j}  \times \alpha_{e_i} \bar h \, h \, \alpha_{e_{i+1}} \times \prod_{j = i+2}^n \, \alpha_{e_j} = \phi_{\fidu}(\alpha).$$
    We see that no $A_v^h \in \dgauge$ changes the flux through the fiducial ribbon. Since $\dgauge$ is generated by these on-site gauge transformations, we get the required result. 
\end{proof}

\subsection{Action of ribbon operators on string-net states}
\label{sec:action of T and L on string nets}

\begin{lem} \label{lem:flux projector}
    Suppose $\rho$ is a finite ribbon supported within $S \subset \latticeedge$ and $\al \in \gc[S]$. Then
    $$ T_{\rho}^g \ket{\al} = \delta_{\phi_{\rho}(\al), g} \ket{\al}  $$
    for any $g \in G$. In particular, $[T_{\rho}^g, T_{\rho'}^{g'}] = 0$ for all ribbons $\rho, \rho'$ and any $g, g' \in G$.
\end{lem}

\begin{proof}
    We prove the Lemma by induction. If $\ep$ is the empty ribbon then $T_{\ep}^g = \delta_{1, g} \I$, which says that the flux through the empty ribbon is always trivial. If $\rho = \{\tau \}$ consists of a single dual triangle then $T_{\rho}^g = F_{\rho}^{1, g} = \delta_{1, g} \I$ which says that the flux through $\rho^{dir} = \emptyset$ is always trivial. If $\rho = \{ \tau \}$ consists of a single direct triangle then $T_{\rho}^g = F_{\rho}^{1, g} = T_{\tau}^g$ which acts on the string net as $T_{\tau}^g \ket{\al} = \delta_{\al_{e_{\tau}}, g} \ket{\al} = \delta_{\phi_{\tau}(\al), g} \ket{\al}$, as required.

    Now suppose $\rho = \rho' \tau$ and suppose the claim is true for the ribbon $\rho'$. Then $T_{\rho}^g = \sum_k \, T_{\rho'}^{k} T_{\tau}^{\bar k g}$. Using the above we get
    $$ T_{\rho}^g \, \ket{\al} = \sum_{k \in G}  \delta_{\phi_{\rho'}(\al), k} \delta_{\phi_{\tau}(\al), \bar k g} \ket{\al} = \delta_{\phi_{\rho}(\al), g} \ket{\al},$$
    as required.

    The commutativity can now be shown as follows. Let $S \subset \latticeedge$ be finite and such that $S$ contains the supports of $T_{\rho}^g$ and $T_{\rho'}^{g'}$. Then for any $\al \in \gc[S]$ we have
    $$ T_{\rho}^g T_{\rho'}^{g'} \ket{\al} = \delta_{\phi_{\rho}(\al), g} \delta_{\phi_{\rho'}(\al), g'} \ket{\al} = T_{\rho'}^{g'} T_{\rho}^g \ket{\al}. $$
    Since $\ket{\al}$ for $\al \in \gc[S]$ is an orthonormal basis for $\caH_S$, the claim follows.
\end{proof}

\begin{figure}
    \centering
    \includegraphics[width=0.5\textwidth]{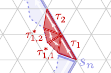}
    \caption{Examples of direct triangles $\tau_i$ and dual triangles $\tau^*_{i, 1}, \tau^*_{i, 2}$ that make up the boundary ribbon $\bdy$. The direct parts $I_1 = \{\tau_1\}$ and $I_2 = \{\tau_2\}$ as well as the first dual part $J_1 = \{ \tau^*_{1, 1}, \tau^*_{1, 2}\}$ of $\bdy$ are depicted.}
    \label{fig:boundary triangles}
\end{figure}

\begin{figure}
    \centering
    \includegraphics[width=0.5\textwidth]{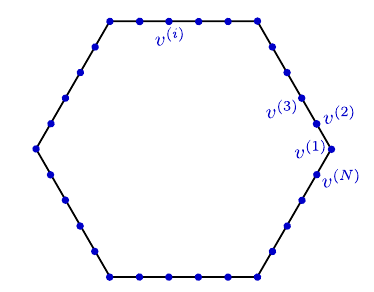}
    \caption{A counterclockwise labeling of the boundary vertices $\partial \vregion$.}
    \label{fig:counterclockwise boundary labeling}
\end{figure}

Let us now consider the boundary ribbon $\bdy$. Its alternating decomposition (cf. Definition \ref{def:alternating decomposition}) $\bdy = I_1J_1 \cdots I_N J_N$ has the direct parts $I_i = \{\tau_i\}$ consisting of a single triangle with $e_{\tau_i} \in \partial \eregion$. The dual parts $J_i$ for $i = 1, \cdots, N-1$ consist of one or two dual triangles each, corresponding to the edges of $\eregion \setminus \partial \eregion$ attached to each boundary vertex in $\partial \vregion$. See Figure \ref{fig:boundary triangles}. For each boundary vertex $v$, let us write $J_v$ for the corresponding dual ribbon. Let us moreover order the boundary vertices $\partial \vregion = \{ v^{(1)}, \cdots, v^{(N)}\}$ counterclockwise as in Figure \ref{fig:counterclockwise boundary labeling}.

Let $\alpha \in \gc$ be a gauge configuration on $\eregion$. With the notations just established, it follows from Lemma \ref{lem:L decomposition} that
\begin{equation} \label{eq:boundary L on gauge config}
    L_{\bdy}^h \ket{\alpha} = \prod_{i = 1}^N  L_{J_{v^{(i)}}}^{\bar K_i h K_i}  \ket{\alpha}
\end{equation}
where
$$K_i = \prod_{j = 1}^i \phi_{\tau_i}(\alpha)$$
is the flux of $\alpha$ through $\tau_1 \cdots \tau_i$. Note that $K_N = \phi_{\bdy}(b)$.

We can now prove
\begin{lem} \label{lem:boundary L on string nets}
    Let $\alpha \in \packC[C;ib]$ and $h \in G$ that commutes with $\phi_{\bdy}(b)$. Then $L_{\bdy}^{h} \ket{\alpha} = \ket{\alpha'}$ for a string net $\alpha' \in \packC[C;ib]$ such that $\phi_{\fidu}(\alpha') = \phi_{\fidu}(\alpha) h$.
\end{lem}

\begin{figure}
    \centering
    \includegraphics[width=0.2\textwidth]{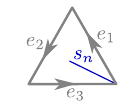}
    \caption{A labeling of the edges of the face $f(s_n)$.}
    \label{fig:final face}
\end{figure}

\begin{proof}
    Using Eq. \eqref{eq:boundary L on gauge config} one easily checks that for each face, except possibly the final one that contains the site $s_n$ (Figure \ref{fig:boundary triangles}), the action of $L_{\bdy}^h$ preserves the trivial flux constraints.

    For that final face, label its edges as in Figure \ref{fig:final face}. From Eq. \eqref{eq:boundary L on gauge config} we see that the operator $L_{\bdy}^h$ acts on the edge degrees of freedom of this triangle as $L_{e_2}^{\bar K_1 \bar h K_1} R_{e_3}^{\bar K_N \bar h K_N}$. On the string net state $\alpha$ this becomes
    $$L_{e_2}^{\bar K_1 \bar h K_1} R_{e_3}^{\bar K_N \bar h K_N} \, \ket{ \alpha_{e_1} } \otimes \ket{ \alpha_{e_2} } \otimes \ket{ \alpha_{e_3} }  = \ket{ \alpha_{e_1} } \otimes \ket{ \bar K_1 \bar h K_1 \alpha_{e_2} } \otimes \ket{ \alpha_{e_3} \bar K_N h K_N }.$$
    
    Noting that $K_N = \phi_{\bdy}(b) $ and $K_1 = \al_{e_1}$ we see that the resulting flux measured at $s_n$ is
    $$ \bar h \al_{e_1} \al_{e_2} \al_{e_3}  \dash{\phi_{\bdy}(b)} \, h \phi_{\bdy}(b) = \bar h \dash{\phi_{\bdy}(b)} \, h \phi_{\bdy}(b)$$
    where we used that $\alpha$ satisfies the trivial flux constraint $\alpha_{e_1} \alpha_{e_2} \alpha_{e_3} = 1$. We now use that $h$ commutes with the boundary flux $\phi_{\bdy}(b)$ to see that the trivial flux condition is also maintained in the final face.

    As already noted, $L_{\bdy}^h$ acts on the degree of freedom at the edge $e_3$ as $R_{e_3}^{\bar K_N \bar h K_N}$, using again that $K_N = \phi_{\bdy}(b)$ and $h$ commutes with $\phi_{\bdy}(b)$ this is the same as $R_{e_3}^{\bar h}$.  since $e_2$ is the final direct edge of the fiducial ribbon $\fidu$, this immediately implies the final claim.
\end{proof}

Recall from Definition \ref{def:eta^Cib(m)} the unit vectors
$$ \ket{\eta_n^{C;ib}(m)} = \frac{1}{\abs{ \packib(m) }^{1/2}} \, \sum_{\al \in \packib(m)} \, \ket{\al}. $$

\begin{lem} \label{lem:action of T_bdy on eta^C;ib(m)}
    For any $g \in G$, $C \in (G)_{cj}, i = 1, \cdots, \abs{C}, b \in \bc[C]$, and $m \in N_C$ we have
    $$T_{\bdy}^g \ket{\eta_n^{C;ib}(m)} = \delta_{g, q_{i(b)} r_C \, \bar q_{i(b)}} \ket{\eta_n^{C;ib}(m)}. $$
\end{lem}

\begin{proof}
    By definition, $\ket{\eta_n^{C;ib(m)}}$ is a linear combination of states $\ket{\al}$ with $\al \in \packib(m)$ hence $\phi_{\fidu}(\al) = q_i m \bar q_{i(b)}$ (cf. Definition \ref{def:string nets Cib(m)}) and $\phi_{f_0}(\al) = c_i$. It follows from Lemmas \ref{lem:bcinC} and \ref{lem:inNC} that for all these string nets we have $\phi_{\bdy}(\al) = \bar \phi_{\fidu(\al)} c_i \phi_{\fidu}(\al) = q_{i(b)} r_C \, \bar q_{i(b)}$. The result now follows immediately from Lemma \ref{lem:flux projector}.
\end{proof}

Recall from Definition \ref{def:representation basis} the unit vectors
$$  \qdpure = \left( \frac{\dimR}{\abs{N_C}} \right)^{1/2} \, \sum_{m \in N_C} \, R^{j j'}(m) \, \ket{\eta_n^{C;ib}(m)}$$
where $u = (i, j) \in I_{RC}$ and $v = (b, j') \in I'_{RC}$.

\begin{lem} \label{lem:topological charge detector}
    We have
    $$ K_{\bdy}^{R_1 C_1} \ket{ \eta_n^{R_2 C_2 ;u v} } = \delta_{R_1 C_1, R_2 C_2}\, \ket{\eta_n^{R_2 C_2; uv}}.  $$
\end{lem}

\begin{proof}
    Let $u = (i, j) \in I_{R_2 C_2}$ and $v = (b, j') \in I'_{R_2 C_2}$, then
    \begin{align*}
        K_{\bdy}^{R_1 C_1} \, \ket{\eta_n^{R_2 C_2; u v}} &= \sum_{\substack{m_1 \in N_{C_1} \\ m_2 \in N_{C_2}}} \, \chi_{R_1}(m_1)^* \, R_2^{j j'}(m_2)^* \, \sum_{q \in Q_{C_1}} \, L_{\bdy}^{q m_1 \bar q} \, T_{\bdy}^{q r_{C_1} \bar q} \, \ket{\eta_n^{C_2;ib}(m_2)} \\
        \intertext{from Lemma \ref{lem:action of T_bdy on eta^C;ib(m)} and noting that $q r_{C_1} \bar q = q_{i(b)} r_{C_2} \bar q_{i(b)}$ implies $C_1 = C_2$ and $q = q_{i(b)}$ we get}
        &= \delta_{C_1, C_2} \, \sum_{m_1, m_2 \in N_{C_2}} \, \chi_{R_1}(m_1)^* \, R_{2}^{j j'}(m_2)^* \, L_{\bdy}^{q_{b(i)} m_1 \bar q_{i(b)}} \, \ket{ \eta_n^{C_2;ib}(m_2) } \\
        \intertext{noting that $\ket{ \eta_n^{C_2;ib}(m_2) }$ is a linear combination of $\ket{\al}$ for $\al \in \packib(m_2)$, and for each such $\al$ we have $\phi_{\bdy} = q_{i(b)} r_{C_2} \bar q_{i(b)}$, it follows from Lemma \ref{lem:boundary L on string nets} that}
        &= \delta_{C_1, C_2} \, \sum_{m_1, m_2 \in N_{C_2}} \, \chi_{R_1}(m_1)^* \, R_{2}^{j j'}(m_2)^* \, \ket{ \eta_n^{C_2;ib}(m_2 m_1) } \\
        \intertext{changing variables to $M = m_2 m_1$ and $m = m_1$ and writing the character as a trace this becomes}
        &= \delta_{C_1, C_2} \, \sum_{M, m \in N_{C_2}} \, \sum_{l, l'} \, R_1^{l l}(m)^* \, R_2^{j l'}(M)^* \, R_2^{j' l'}(m) \, \ket{\eta_n^{C_2;ib}(M)} \\
        \intertext{Finally, applying Schur orthogonality we get}
        &= \delta_{R_1 C_1, R_2 C_2} \, \ket{\eta_n^{R_2 C_2;uv}},
    \end{align*}
    finishing the proof.
\end{proof}

\subsection{Action of Wigner projectors and label changers on string-net states}

Recall the unit vectors (Definitions \ref{def:eta^Cib(m)}) and \ref{def:representation basis})
$$ \ket{\eta_n^{C;ib}(m)} = \frac{1}{\abs{ \packib(m) }^{1/2}} \, \sum_{\al \in \packib(m)} \, \ket{\al} $$
and
$$  \qdpure = \left( \frac{\dimR}{\abs{N_C}} \right)^{1/2} \, \sum_{m \in N_C} \, R^{j j'}(m)^* \, \ket{\eta_n^{C;ib}(m)}$$
where $u = (i, j) \in I_{RC}$ and $v = (b, j') \in I'_{RC}$.

Recall from Definition \ref{def:Wigner projectors} the Wigner projectors
$$\qd := \frac{\dimR}{|N_C|} \sum_{m\in N_C} \chi_R(m)^* \sum_{q \in Q_C} A_{s} ^{q  m \dash{q}} B_{s}^{q r_C \dash{q}}$$
and for each $u = (i, j) \in I_{RC}$
$$\qd[RC;u] := \frac{\dimR}{|N_C|} \sum_{m \in N_C} R^{jj}(m)^* A_{s} ^{q_i m\dash{q}_i} B_{s}^{c_i}.$$

\begin{lem} \label{lem:action of Wigner projecitons on string-net condensates}
    We have
    $$ D_{\site}^{R_1C_1} \ket{ \eta_n^{R_2C_2;uv} } = \delta_{R_1C_1, R_2C_2} \ket{\eta_n^{R_2C_2;uv}}. $$
\end{lem}

\begin{proof}
    Let $u = (i, j) \in I_{RC}$ and $v = (b, j') \in I'_{RC}$. Then
    \begin{align*}
        D_{\site}^{R_1C_1} \ket{\eta_n^{R_2C_2;uv}} &= \bigg( \frac{\dim R_1}{\abs{N_{C_1}}} \bigg)\bigg( \frac{\dim R_2}{ \abs{N_{C_2}} } \bigg)^{1/2} \, \sum_{\substack{ m_1 \in N_{C_1} \\ m_2 \in N_{C_2} }} \, \chi_{R_1}(m_1)^*  R_2^{j j'}(m_2)^* \, \\
        & \quad\quad\quad \times\sum_{q \in Q_{C_1}} \, A_{\site}^{q m_1 \bar q} B_{\site}^{q r_{C_1} \bar q} \, \ket{\eta_n^{C_2;ib}(m_2)} \\
        &= \delta_{C_1, C_2} \, \frac{ \dim R_1 (\dim R_2)^{1/2} }{\abs{N_{C_1}}^{3/2}} \,\sum_{m_1, m_2 \in N_{C_1}} \, \chi_{R_1}(m_1)^*  R_2^{j j'}(m_2)^* \, A_{\site}^{q_i m_1 \bar q_i} \, \ket{\eta_n^{C_1;ib}(m_2)} \\
        \intertext{using Lemma \ref{lem:N_C action on VCib} this becomes}
        &= \delta_{C_1, C_2} \, \frac{ \dim R_1 (\dim R_2)^{1/2} }{\abs{N_{C_1}}^{3/2}} \,\sum_{m_1, m_2 \in N_{C_1}} \, \chi_{R_1}(m_1)^*  R_2^{j j'}(m_2)^*  \ket{\eta_n^{C_1;ib}(m_1m_2)} \\
        \intertext{changing variables to $m = m_1$ and $M = m_1 m_2$, and using using the Schur orthogonality relation \eqref{eq:Schur} we get}
        &= \delta_{R_1C_1, R_2C_2} \, \bigg( \frac{\dim R_1}{\abs{N_{C_1}}} \bigg)^{1/2} \, \sum_{M \in N_{C_1}} \, R_2^{j j'}(M)^*  \ket{\eta_n^{C_1;ib}(M)} \\
        &= \delta_{R_1C_1, R_2C_2} \, \ket{\eta_n^{R_1C_1;uv}}.
    \end{align*}
\end{proof}

\begin{lem} \label{lem:action of Wigner sub-projector on string-net condensates}
    We have
    $$ D_{\site}^{RC;u_1} \, \ket{\eta_n^{RC;u_2 v}} = \delta_{u_1, u_2} \ket{\eta_n^{RC;u_1 v}}. $$
\end{lem}

\begin{proof}
    Let $u_1 = (i_1, j_1)$, $u_2 = (i_2, j_2)$ and $v = (b, j')$. Then
    \begin{align*}
        D_{\site}^{RC;u_1} \ket{\eta_n^{RC;u_2 v}} &= \bigg( \frac{\dim R}{\abs{N_C}} \bigg)^{3/2} \, \sum_{m_1, m_2 \in N_C} \, R^{j_1 j_1}(m_1)^* R^{j_2 j'}(m_2)^* \, A_{\site}^{q_{i_1} m_1 \bar q_{i_1}} B_{\site}^{c_{i_1}} \, \ket{\eta_n^{C;i_2 b}(m_2)} \\
        \intertext{noting that $B_{\site}^{c_{i_1}} \, \ket{\eta_n^{C;i_2 b}(m_2)} = \delta_{i_1, i_2}$ and using Lemma \ref{lem:N_C action on VCib} this becomes}
        &= \delta_{i_1, i_2} \, \bigg( \frac{\dim R}{\abs{N_C}} \bigg)^{3/2} \, \sum_{m_1, m_2 \in N_C} \, R^{j_1 j_1}(m_1)^* R^{j_2 j'}(m_2)^* \, \ket{\eta_n^{C;i_2 b}(m_1m_2)} \\
        \intertext{changing variables to $m = m_1$ and $M = m_1 m_2$, and using Schur orthogonality \eqref{eq:Schur} we get}
        &= \delta_{u_1, u_2} \, \bigg( \frac{\dim R}{\abs{N_C}} \bigg)^{1/2} \, \sum_{M \in N_C}  R^{j_2 j'}(M)^* \, \ket{\eta_n^{C;i_2 b}(M)} = \delta_{u_1, u_2} \, \ket{\eta_n^{RC;u_2 v}}.
    \end{align*}
\end{proof}

Recall the operators from Definition \ref{def:label changers}: 
$$  A_{s}^{RC; u_2 u_1} := \frac{\dimR}{|N_C|}\sum_{m \in N_C} R^{j_2j_1}(m)^* A_{s}^{q_{i_2} m \dash{q}_{i_1}} \qquad  u_1 = (i_1,j_1) \quad u_2 = (i_2,j_2)  $$ 
and 
$$  \tilde{A}_n^{RC; v_2 v_1} :=\frac{\dimR}{|N_C|} \sum_{m \in N_C} R^{j'_2j'_1}(m)  U_{b_2 b_1} L_{\bdy}^{{q_{i(b_1)} \dash{m} \,  \dash{q}_{i(b_1)}}} \qquad v_1 = (b_1,j'_1) \quad u_2 = (b_2,j'_2)  $$
where $U_{b_2 b_1}$ is a unitary provided by Lemma \ref{lem:transitive boundary action}, which we choose such that $U_{b_2 b_1} = (U_{b_1 b_2})^*$. It follows from Lemma \ref{lem:transitive boundary action} that the unitary $U_{b_2 b_1}$ yields a bijection between $\packbc[C;ib_1]$ and $\packbc[C;ib_2]$ whenever $b_1, b_2 \in \bc[C]$.

It was shown in Lemma \ref{lem:N_C action on VCib} that the gauge transformations $A_{\site}^{q_i m \bar q_i}$ for $m \in N_C$ yield a left group action of $N_C$ on the vectors $\ket{\eta_n^{C;ib}(m)}$. We show now that the operators $L_{\bdy}^{q_{i(b)} \bar m \bar q_{i(b)}}$ for $m \in N_C$ yield a right action of $N_C$ on these vectors.

\begin{lem} \label{lem:right action on VCib}
    For any $m_1, m_2 \in N_C$ we have
    $$ L_{\bdy}^{q_{i(b)} \bar m_1 \bar q_{i(b)} } \, \ket{\eta_n^{C;ib}(m_2)} = \ket{\eta_n^{C;ib}(m_2 \bar m_1)}. $$
\end{lem}

\begin{proof}
    From Lemma \ref{lem:boundary L on string nets} and the fact that $L_{\bdy}^{q_{i(b)} \bar m_1 \bar q_{i(b)}}$ is unitary, we see that this operator yields a bijection from $\packib(m_2)$ to $\packib(m_2 \bar m_1)$. It follows that
    \begin{align*}
        L_{\bdy}^{q_{i(b)} \bar m_1 \bar q_{i(b)} } \, &\ket{\eta_n^{C;ib}(m_2)} = \frac{1}{\abs{\packib(m_2)}^{1/2}} \, \sum_{\al \in \packib(m_2)} \, L_{\bdy}^{q_{i(b)} \bar m_1 \bar q_{i(b)} }\, \ket{\al} \\
        &= \frac{1}{\abs{\packib(m_2 \bar m_1)}^{1/2}} \, \sum_{\al \in \packib(m_2 \bar m_1)} \, \ket{\al} = \ket{\eta_n^{C;ib}(m_2 \bar m_1)}.
    \end{align*}
\end{proof}

We can now show
\begin{lem} \label{lem:aconverter}
    For any $u, u_1, u_2 \in I_{RC}$ and any $v, v_1, v_2 \in I'_{RC}$ we have
    $$A_{\site}^{RC; u_2 u_1} \qdpure[RC;u_1 v] = \qdpure[RC;u_2 v], \quad \quad \tilde{A}_n^{RC; v_2 v_1} \qdpure[RC;u v_1] = \qdpure[RC; u, v_2]$$
    as well as
    $$(A_{\site}^{RC; u_1 u_2})^* \qdpure[RC;u_1 v] = \qdpure[RC;u_2 v], \quad \quad (\tilde{A}_n^{RC; v_1 v_2})^* \qdpure[RC;u v_1] = \qdpure[RC; u, v_2].$$
\end{lem}

\begin{proof}
    We prove the claim about the action of $\tilde A_n^{RC;v_2 v_1}$. The claim about $(\tilde{A}_n^{RC; v_1 v_2})^*$ is proven in exactly the same way, and the claims about $A_{\site}^{RC; u_2 u_1}$ and its hermitian conjugate have similar but simpler proofs. Let $u = (i, j)$, $v_1 = (b_1, j'_1)$ and $v_2 = (b_2, j'_2)$, then
    \begin{align*}
        \tilde{A}_n^{RC; v_2 v_1} \qdpure[RC;u v_1] &= \left( \frac{\dimR}{\abs{N_C}} \right)^{3/2} \, \sum_{m_1, m_2 \in N_C} \, R^{j'_2 j'_1}(m_2) \, R^{j j'_1}(m_1)^* \, U_{b_2 b_1} \, L_{\bdy}^{q_{i(b_1)} \bar m_2 \bar q_{i(b_1)}} \, \ket{\eta_n^{C;ib_1}(m_1)} \\
        \intertext{using Lemma \ref{lem:right action on VCib} and the basic properties of $U_{b_2 b_1}$}
        &= \left( \frac{\dimR}{\abs{N_C}} \right)^{3/2} \, \sum_{m_1, m_2 \in N_C} \, R^{j'_2 j'_1}(m_2) \, R^{j j'_1}(m_1)^* \, \ket{\eta_n^{C;ib_2}(m_1 \bar m_2)} \\
        \intertext{letting $M = m_1 \bar m_2$ and $m = m_2$, and using Schur orthogonality, this becomes}
        &= \left( \frac{\dimR}{\abs{N_C}} \right)^{1/2} \, \sum_{M \in N_C}  \, R^{j j'_2}(M)^* \, \ket{\eta_n^{C;ib_2}(M)} = \qdpure[RC;u v_2].
    \end{align*}
\end{proof}

This Lemma tells us that $u$ is a ``bulk" label, as the operator that changes $u_1$ to $u_2$ is $A_{\site}^{RC; u_2 u_1} \in \cstar[{\eregion[1]}]$. We also see that $v$ is a ``boundary" label, as the operator that changes $v_1$ to $v_2$ is $\tilde{A}_n^{RC; v_2 v_1} \in \cstar[{\eregion \setminus \eregion[n-1]}]$.

We can also detect the boundary data by operators supported on $\eregion \setminus \eregion[n-1]$. Recall from Definition \ref{def:boundary condition projector} the projectors $P_b$ supported on $\partial \eregion$ that project onto states with boundary condition $b \in \bc$.

\begin{lem}
    \label{lem:action of two different label changers on eta}
    For any $v_1,v_2, v \in I'_{RC}$ such that $v_1 = (b_0, j'_1)$ and $v_2 = (b_0, j'_2)$ (\ie they have the same boundary label $b_0$), we have $$(\tilde{A}_n^{RC;v_2 v})^{*} \tilde{A}_n^{RC; v_1 v} \ket{\qdstpure[RC;u v]}= \delta_{v_1v_2}\ket{\qdstpure[RC;u v]}$$
\end{lem}
\begin{proof}

    Using lemma \ref{lem:aconverter} we get,
    \begin{align*}
        (\tilde{A}_n^{RC;v v_2})^{*} \tilde{A}_n^{RC; v_1 v} \ket{\qdstpure[RC;u v]} &= (\tilde{A}_n^{RC;v_2 v})^{*}  \ket{\qdstpure[RC;u v_1]}\\
        &= \left( \frac{\dimR}{\abs{N_C}} \right)^{3/2} \sum_{m,m' \in N_C} R^{j'_2 j'}(m')^* R^{jj'_1}(m)^*  \ket{\eta_n^{C;ib}(mm')}\\
        \intertext{Now we relabel $mm' = M$ and use Schur orthogonality to get}
        &=  \left( \frac{\dimR}{\abs{N_C}} \right)^{1/2} \sum_{M \in N_C} \sum_{ j_3'} \delta_{j_3' j} \delta_{j'_2 j_1'}  R^{j'_3 j'}(M)^*  \ket{\eta_n^{C;ib}(M)}\\
        &= \delta_{j_1', j'_2}  \ket{\qdstpure[RC;u v]} = \delta_{v_1, v_2}  \ket{\qdstpure[RC;u v]}
    \end{align*}
\end{proof}

%% file: main_category_thesis.tex
\chapter{The Category Of Anyon Sectors For Non-Abelian Quantum Double Models}
\label{chap:category of quantum double}
\chapterauthors{
\chapterauthor{Alex Bols}{Institute for Theoretical Physics, ETH Z{\"u}rich, Switzerland}
\chapterauthor{Mahdie Hamdan}{School of Mathematics, Cardiff University, United Kingdom}
\chapterauthor{Pieter Naaijkens}{School of Mathematics, Cardiff University, United Kingdom}
\chapterauthor{Siddharth Vadnerkar}{Department of Physics, University of California, Davis, CA, USA}
}

This chapter is taken verbatim from \cite{bols2025category} and published in Communications in Mathematical Physics. Reprinted with the permission of Alex Bols, Mahdie Hamdan, Pieter Naaijkens, Siddharth Vadnerkar. Redistribution is allowed under the \href{https://s100.copyright.com/AppDispatchServlet?title=The%20Category%20of%20Anyon%20Sectors%20for%20Non-Abelian%20Quantum%20Double%20Models&author=Alex%20Bols%20et%20al&contentID=10.1007%2Fs00220-025-05492-2&copyright=The%20Author%28s%29&publication=0010-3616&publicationDate=2025-12-05&publisherName=SpringerNature&orderBeanReset=true&oa=CC%20BY}{copyright terms of this article} (\href{https://creativecommons.org/licenses/}{Creative Commons CC BY license}). This work was born as a direct follow up to the results of \cite{bols2023classificationanyonsectorskitaevs}. In the introduction of Chapter \ref{chap:quantum double sectors}, it was noticed that the bijection between anyon sectors and irreducible representations of the Quantum Double of $G$, denoted $D(G)$, can be categorified. From the discussion in Sections \ref{sec:category of anyons} and \ref{sec:quantum double category}, we already have a category on both sides of this correspondance, and the achievement of \cite{bols2023classificationanyonsectorskitaevs} was to establish the bijection between the irreducible objects in these categories. So it is the next natural step to consider this question. This paper supplies the rest of the ingredients for a braided $\rmC^*$-tensor equivalence between the categories.

We comment that there are several practical advantages to categorifying this correspondance. In \cite{Ogata2021} it was shown that the categorical structure of anyon sectors is actually an invariant of the phase. So if we're able to show a braided $\rmC^*$-tensor equivalence for the Quantum Double models, it will also be true for the rest of the phase. In other words, this correspondance is stable under suitably small perturbations. Another reason is that the category $\Rep D(G)$ has a very different-looking braided structure (Definition \ref{sec:quantum double category}) than the traditional braiding (Section \ref{sec:braiding of anyons}) of the anyon category. So a braided equivalence actually assures us that the definitions are consistent. 

Additionally, the paper takes a slight detour and considers an interesting question, ``is the selection criterion special?''. The answer is in the negative. One can actually take the anyon sector category, or the localized transportable endomorphism (or amplimorphism) category, drop unitality from the list of requirements, and still obtain an equivalent category. It shows that the category of anyons (Definition \ref{sec:category of anyons}) can be equivalently defined in terms of localized transportable amplimorphisms. In addition it also proves that unitality of endomorphisms (as well as amplimorphisms) is an unnecessary condition, and that every non-unital endomorphism (resp.~amplimorphism) is equivalent to a unital endomorphism. This allows us to considerably simplify the definition of subobjects.


The main technical assumption used in the paper, Haag duality for the Quantum Double models, was already established for a wide class of models (including Quantum Double models) before the publishing of this paper \cite{HaagDuality}, and is thus no longer a required assumption.

\begin{chapterabstract}
We study Kitaev's quantum double model for arbitrary finite gauge group in infinite volume, using an operator-algebraic approach.
The quantum double model hosts anyonic excitations which can be identified with equivalence classes of `localized and transportable endomorphisms', which produce anyonic excitations from the ground state. 
Following the Doplicher--Haag--Roberts (DHR) sector theory from AQFT, we organize these endomorphisms into a braided monoidal category capturing the fusion and braiding properties of the anyons. We show that this category is equivalent to $\Rep_f \caD(G)$, the representation category of the quantum double of $G$. This establishes for the first time the full DHR structure for a class of 2d quantum lattice models with non-abelian anyons.
\end{chapterabstract}

\setcounter{tocdepth}{2}
\minitoc


\section{Introduction}
Kitaev's quantum double model~\cite{kitaev2003fault} is the prototypical example of a topologically ordered quantum spin system with long-range entanglement (see~\cite{MR3929747} for an introduction).
Such models host quasi-particle excitations with non-trivial braid statistics called anyons.
The physical properties of such anyons (such as their behavior under exchange or fusion) can be described algebraically by braided (and often even modular) tensor categories~\cite{Kitaev2006,Wang2010}.
In this paper we show that for the quantum double model for a finite gauge group $G$, defined on the plane, this braided  tensor category can be recovered from the unique frustration-free ground state of the model (under some mild technical assumption), and is given by $\Rep_f \caD(G)$, the category of finite dimensional unitary representations of the quantum double algebra of $G$. 

Our approach is motivated by the Doplicher--Haag--Roberts (DHR) theory of superselection sectors (see~\cite{HaagLQP} for an overview).
Mathematically, we can identify the anyons with certain equivalence classes of irreducible representations of the (quasi-local) observable algebra $\alg{A}$.

The relevant representations are those whose vector states approximately agree with the model's ground state on observables supported far away from some fixed point (which we can take as the origin), and whose support does not encircle this point.
The latter condition is to exclude observables corresponding to braiding other anyons around the fixed point, which are able to distinguish non-trivial anyon states from states in the ground state sector.

This intuition is conveniently captured by the \emph{superselection criterion}. Namely, a representation $\pi$
 satisfies the superselection criterion if
\begin{equation}
    \label{eq:sselect}
    \pi | \alg{A}_{\Lambda^c} \cong \pi_0 | \alg{A}_{\Lambda^c},
\end{equation}
where $\Lambda$ is any cone (a notion which we will make more precise later) and $\Lambda^c$ is its complement,  $\pi_0$ is the GNS representation of the (unique) frustration free ground state of the quantum double model, and $\alg{A}_{\Lambda^c}$ is the ${\rm C}^*$-algebra generated by all local observables localized in $\Lambda^c$.
That is, we consider representations that, outside \emph{any} cone, are unitarily equivalent to the ground state representation.
A superselection sector (or simply anyon sector) is an equivalence class of such representations.

The key insight of Doplicher, Haag and Roberts is that the superselection sectors are naturally endowed with a monoidal product (`fusion') and a symmetry describing the exchange of bosonic/fermionic sectors. This was later extended to describe braiding statistics~\cite{frs1,frs2}, yielding a braided monoidal category.
These categories precisely capture the physical properties of anyon sectors, including their braiding and fusion rules.
The essential technical step is that, using a technical property called Haag duality, one can pass from representations to endomorphisms of the quasi-local algebra which are localized (i.e., they act non-trivially only in the localization region) and transportable (the localization region can be moved around with unitaries).
See~\cite{HalvorsonMueger} for an overview of this construction in the language of $\rm{C}^*$-tensor categories.
This theory was initially developed in the context of relativistic quantum field theories. 
The construction has later been adapted to quantum spin systems, see e.g.~\cite{Naaijkens2011,Fiedler2014,Ogata2021}.
For a recent completely axiomatic approach towards anyon sector theory, see~\cite{Bhardwaj2024}.

In this paper we study the anyon sector theory, including fusion and braiding rules, of the quantum double model for arbitrary finite gauge group $G$~\cite{kitaev2003fault}, extending previous results obtained for abelian $G$~\cite{Fiedler2014}. 
In particular, our main result can be paraphrased as follows:

\begin{thm}[Informal]
\notag
Let $\pi_0$ be the GNS representation of the frustration free ground state of the quantum double model for a finite group $G$ defined on the plane and assume that it satisfies Haag duality.
Then the category of representations satisfying~\eqref{eq:sselect} is braided monoidally equivalent to $\Rep_f \caD(G)$, the category of finite dimensional unitary representations of the quantum double algebra $\caD(G)$.
\end{thm}

We will give a precise statement of our main result (including our assumptions) later when we have introduced the necessary terminology, but remark that Haag duality for cones is a technical property that holds for the abelian quantum double model~\cite{Fiedler2014}, and one can still construct a category of anyon sectors without it (or with a weaker version thereof).
A proof of Haag duality for a large class of models has recently been announced~\cite{HaagDuality}.
See Remark~\ref{rem:Haag 2} below for more details.
We also note that since $\Rep_f \caD(G)$ is a unitary modular tensor category, the category of anyon sectors is as well.

As mentioned earlier, our assumptions imply that there is a braided $\rm{C}^*$-category of superselection sectors~\cite{Fiedler2014,Ogata2021}.
Our main contribution in this paper is to construct this category explicitly for the quantum double model for all finite groups $G$.
The main idea is as follows.
For each irreducible represention of $\caD(G)$, examples of representations $\pi$ satisfying the superselection criterion~\eqref{eq:sselect} were constructed in~\cite{Naaijkens2015}.
It was then shown in~\cite{bols2023classificationanyonsectorskitaevs} that these representations are irreducible, and in fact form a complete set of representatives of irreducible representations satisfying~\eqref{eq:sselect}.
These irreducible anyon sectors correspond to the simple objects (i.e., the anyon types) in our category.
Because we have a concrete description of the simple objects in our category, it is possible to explicitly implement the braiding and fusion operations defined abstractly in~\cite{Ogata2021}, and calculate those explicitly.
We then show that the category we constructed is indeed equivalent to the one defined abstractly in~\cite{Ogata2021}.

The key difference between the present work and the abelian case studied in~\cite{Naaijkens2011,Fiedler2014} is the use of \emph{amplimorphisms}, i.e. $*$-homomorphisms $\chi: \alg{A} \to M_n(\alg{A})$, instead of endomorphisms.\footnote{For technical reasons we will in fact need to consider amplimorphisms of some slightly bigger algebra $\alg{B}$ containing $\alg{A}$.}
This can be understood as follows: recall that in the quantum double models, we can define `ribbon operators' which create a pair of excitations from the ground state.
To obtain single-anyon states, one sends one of the excitations off to infinity.
For each irrep of $\caD(G)$, there is a corresponding multiplet of ribbon operators, transforming according to the irrep, with the total number of operators in the multiplet given by the dimension of the irrep. 
Hence for non-abelian representations, one has more than one ribbon operator, which combine naturally into an amplimorphism.

Although it is possible to pass from amplimorphisms to the endomorphisms used in~\cite{Fiedler2014,Ogata2021}, as we shall see later, doing so requires making some choices, and one loses the explicit description of the map.
Hence to identify the full superselection theory, we work mainly in the amplimorphism picture.
In particular, we show that the amplimorphisms constructed in~\cite{Naaijkens2015} can be endowed with a tensor product and a braiding, analogous to the tensor product and braiding of endomorphisms in the DHR theory.
More precisely, we construct a braided ${\rm C^*}$-tensor category $\Amp$ of localized and transportable amplimorphisms, which includes as objects the amplimorphisms constructed in~\cite{Naaijkens2015}.
We then consider the full subcategory $\Amp_f$ of $\Amp$ whose objects $\chi$ have finite dimensional Hom spaces $(\chi|\chi)$.
This category can be shown to be semi-simple and closed under the monoidal product on $\Amp$, and we study the fusion rules (how tensor products decompose into irreducible objects) and the braiding. The result is that the category $\Amp_f$ of such amplimorphisms is equivalent to $\Rep_f \caD(G)$ as braided tensor categories.
Using the classification result of anyon sectors in this model obtained by two of the authors~\cite{bols2023classificationanyonsectorskitaevs}, it then follows that the list of constructed anyon sectors is a complete list of representatives of irreducible anyon sectors. This then completes the classification. 

A similar approach using amplimorphisms was taken in~\cite{SzlachanyiV93,NillS97} to analyze topological defects of certain 1D quantum spin systems.
In their setting the anyon sectors are localized in finite intervals, with the corresponding algebra of observables localized in that region being finite dimensional.
This necessitated the use of amplimorphisms instead of endomorphisms.
In our case localization is in infinite cone regions, and the situation is different.
In particular, the unitary operators that can move the localization regions around no longer live in the quasi-local algebra $\alg{A}$.
From a technical point of view this means that we cannot restrict to a purely $\rm{C}^*$-algebraic approach with operators in the quasi-local algebra (or suitable amplifications) only, but have to consider von Neumann algebras as well, in particular the cone algebras $\pi_0(\cstar[\Lambda])''$.\footnote{This is already true for the abelian case, it is not specific to the non-abelian model.}
These cone algebras are ``big enough'' in the sense that they are properly infinite~\cite{Naaijkens2012,Fiedler2014,Tomba2023}.
This allows us to directly relate the localized and transportable amplimorphisms to localized and transportable endomorphisms of some suitably defined auxiliary algebra, making the connection with the usual DHR theory in terms of endomorphisms.

The paper is outlined as follows. In Section \ref{sec:setup and main result} we define the quantum double model and the associated categories of localized and tranportable amplimorphisms $\Amp$ and endomorphisms $\DHR$, as well as their `finite' versions $\Amp_f$ and $\DHR_f$.
We then state our main theorem, namely that the categories $\Amp_f$ and $\DHR_f$ are braided $\rm C^*$-tensor categories, equivalent to $\Rep_f \caD(G)$.
Section \ref{sec:braided tensor structure} is devoted to spelling out the braided $\rm C^*$-tensor structure of $\Amp$ and $\DHR$.
These two categories are then shown to be equivalent in Section \ref{sec:equivalence of Amp and DHR}.
Explicit localized and tranportable amplimorphisms corresonding to representations of $\caD(G)$ are constructed in Section \ref{sec:amplimorphism from ribbon operators} by taking limits of `ribbon multiplets'. These explicit amplimorphisms are organized into full subcategories $\Amp_{\rho}$ of $\Amp$ for a fixed half-infinite ribbon $\rho$, which are later shown to be equivalent to $\Amp_f$. This section also establishes the key properties of these ribbon multiplets that underlie the fusion and braiding structure of $\Amp_f$.
In Section \ref{sec:simples of Amp} we rephrase the main result of \cite{bols2023classificationanyonsectorskitaevs}, namely that the amplimorphisms corresponding to irreducible representations of $\caD(G)$ constructed in the previous section exhaust all simple objects of $\Amp$. Together with semi-simplicity of $\Amp_f$, this implies that the $\Amp_{\rho}$ are full and faithful subcategories of $\Amp_f$.
Finally, Section \ref{sec:proof of main THM} proves the main theorem. The appendices collect well-known facts about ribbon operators and some technical results related to taking their limits.

\textbf{Acknowledgements:}
We would like to thank Corey Jones, Boris Kj\ae r and David Penneys for helpful discussions.
MH was supported by EPSRC Doctoral Training Programme grant EP/T517951/1.
SV was funded by NSF grant number DMS-2108390.

\textbf{Copyright statement:} For the purpose of open access, the authors have applied a CC BY public copyright licence to any Author Accepted Manuscript version arising.

\textbf{Data availability:} We do not analyse or generate any datasets, because our work is entirely within a theoretical and mathematical approach.

\textbf{Conflict of interests:} The authors have no competing interests to declare that are relevant to the content of this article.



\section{Setup and main result} \label{sec:setup and main result}

\subsection{The quantum double model and its ground state}

\begin{figure}[t]
    \centering
    \includegraphics[width=0.5\linewidth]{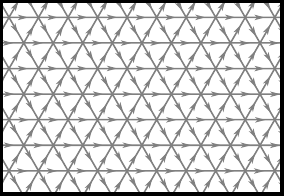}
    \caption{Snapshot of $\Gamma^E$. The edges are all oriented toward the right.}
    \label{fig:lattice snapshot}
\end{figure}

We first recall the definition of the quantum double model~\cite{kitaev2003fault} and introduce our notation.
Throughout the paper, we fix a finite group $G$.
Let $\Gamma$ be the triangular lattice in $\R^2$ and denote by $\Gamma^E$ the collection of oriented edges of $\Gamma$ which are oriented towards the right, see Figure~\ref{fig:lattice snapshot}.\footnote{We use the triangular lattice for simplicity, and to work in the same setting as~\cite{bols2023classificationanyonsectorskitaevs}, but believe the results hold for more general planar graphs as well.}
Denote by $\Gamma^V, \Gamma^F$ the set of vertices and faces of $\Gamma$ respectively. 
To each edge $e \in \Gamma^E$ we associate a degree of freedom $\caH_e \simeq \C[G]$ with basis $\{ |g\rangle_e \, : \, g \in G \}$ and corresponding algebra $\caA_e = \End(\caH_e) \cong M_{|G|}(\mathbb{C})$. We define in the usual way local algebras of observables $\caA^{\loc}_{X}$ supported on any $X \subset \Gamma^E$ and their norm closures $\caA_X := \overline{ \caA_X^{\loc}}^{\norm{\cdot}}$. We write $\caA = \caA_{\Gamma^E}$ and $\caA^{\loc} = \caA_{\Gamma^E}^{\loc}$.

The quantum double Hamiltonian is the commuting projector Hamiltonian given by the following formal sum
\begin{equation}\label{eq:Hamiltonian}
	H = \sum_{v \in \Gamma^V} \, (\1 - A_v) \, + \, \sum_{f \in \Gamma^F} \, ( \1 - B_f),
\end{equation}
where $A_v, B_f \in \cstar$ are the well-known \emph{star} and \emph{plaquette} operators of the quantum double model, which are mutually commuting projectors.
See Section \ref{subsubsec:gauge transformations and flux projections} in the appendix for precise definitions.

We say a state $\omega : \caA \rightarrow \C$ is a frustration free ground state of $H$ if
\begin{equation}
	\omega(A_v) = \omega(B_f) = 1
\end{equation}
for all $v \in \Gamma^V$ and all $f \in \Gamma^F$.
It is straightforward to verify that such a state $\omega$ indeed is a ground state for the dynamics generated by the Hamiltonian~\eqref{eq:Hamiltonian}.

The following theorem is proven in various sources~\cite{Fiedler2014, Cui2020kitaevsquantum, Tomba2023, bols2023classificationanyonsectorskitaevs}.
\begin{thm}
	The quantum double Hamiltonian $H$ has a unique frustration free ground state which we denote by $\omega_0$. The uniqueness implies in particular that $\omega_0$ is pure.
\end{thm}
We denote by $(\pi_0, \caH_0, \Omega)$ the GNS triple of the unique frustration free ground state $\omega_0$.
Note that $\pi_0$ is an irreducible representation since $\omega_0$ is pure.

\subsection{Cone algebras, Haag duality, and the allowed algebra}
\label{sec:cone_algebra}

The open cone with apex at $a \in \R^2$, axis $\hat v \in \R^2$, where $\hat v$ is a unit vector, and opening angle $\theta \in (0, 2\pi)$ is the subset of $\R^2$ given by
\begin{equation*}
	\Lambda_{a, \hat v, \theta} := \{ x \in \R^2 \, : \, (x - a) \cdot \hat v < \norm{x - a}_2 \cos(\theta/2)  \}.
\end{equation*}
We similarly define closed cones and call any subset of $\R^2$ that is either an open or a closed cone a \emph{cone}, so that the complement $\Lambda^c$ of any cone $\Lambda$ is again a cone. Note that a cone cannot be empty, nor can it equal the whole of $\R^2$.

For any $S \subset \R^2$ we denote by $\overline S$ the set of edges in $\Gamma^E$ whose midpoints lie in $S$. With slight abuse of notation we will simply write $S$ to mean the set of edges $\overline S$ unless otherwise stated.

To any cone $\Lambda$ we associate its \emph{cone algebra}
\begin{equation}
	\caR(\Lambda) := \pi_0(\caA_{\Lambda})'' \subset \caB(\caH_0).
\end{equation}
We remark that all these cone algebras are properly infinite factors~\cite{Naaijkens2012,Ogata2021}. We will moreover assume that Haag duality holds for cones.
\begin{asmp}[Haag duality for cones] \label{ass:Haag duality}
    For any cone $\Lambda$ we have
    $$\caR(\Lambda^c)' = \caR(\Lambda).$$
\end{asmp}

\begin{rem} \label{rem:Haag 2}
Haag duality for cones is proven in~\cite{Fiedler2014} in the case $G$ is an abelian group.
We believe the proof methods can be extended to the non-abelian case, however the analysis becomes considerably more technical since in the non-abelian case not all irreducible representations of the quantum double $\caD(G)$ are one-dimensional anymore.
In addition, a proof of Haag duality for a wide class of 2D quantum spin systems has been announced recently~\cite{HaagDuality}, including in particular for the non-abelian quantum double models considered here.

Finally, we comment on the role that Haag duality plays.
One can still construct the category of representations of superselection sectors, and show that the (equivalence classes of) irreducible representations are in one-to-one correspondence with the irreducible representations of $\caD(G)$~\cite{bols2023classificationanyonsectorskitaevs}.
By using this equivalence of categories the braided monoidal structure from $\Rep_f \caD(G)$ can be transported to the category of superselection sectors.
Haag duality is used to show that this in fact for example gives the natural braiding obtained from the Doplicher--Haag--Roberts approach.
That is, it has the correct physical interpretation.
Without Haag duality one can only do this for certain explicitly constructed representatives of each sector.\footnote{%
This is the category $\Amp_\rho$ that we will define later.
In this case, one can also explicitly construct the morphisms in the category as weak (or strong) operator limits of observables localized in some cone.
This gives enough control over the localization of these intertwiners, which requires Haag duality in general.
Using the explicit construction of the objects in the category, it can be directly checked that it is closed under the monoidal product of simple objects, and one can take finite direct sums.
However, this analysis only works for the amplimorphisms constructed explicitly, and does not extend to arbitrary ampli (or endo-)morphisms, even if they are in the same superselection sector.
}
For this reason, we prefer to assume (strict) Haag duality for cones to avoid making the analysis more technical than necessary.
\end{rem}

We fix a unit vector $\hat f \in \R^2$ and say a cone with axis $\hat v$ and opening angle $\theta$ is \emph{forbidden} if $\hat f \cdot \hat v < \cos(\theta /2)$. If a cone is not forbidden, then we say it is \emph{allowed}. The \emph{allowed algebra}
\begin{equation*}
	\caB = \caB_{\hat f} := \overline{\bigcup_{\Lambda \, \text{allowed}} \caR(\Lambda)}^{\norm{\cdot}} \subset \caB(\caH_0)
\end{equation*}
is the $\rm C^*$-algebra generated by the cone algebras of allowed cones.
Note that the set of allowed cones is a directed set for the inclusion relation.
Because we assume strict Haag duality, our algebra $\caB$ is the same as what is denoted by $\mathfrak{B}_{(\theta,\phi)}$ in~\cite[Eq. (2.5)]{Ogata2021} for suitable $(\theta,\phi)$.
If only approximate Haag duality holds, it can be replaced with the definition there.

Note that $\pi_0(\caA) \subset \caB$ as for any finite set $S \subset \Gamma^E$, we can find an allowed cone $\Lambda$ containing $S$. In addition, the allowed algebra will be seen to contain the intertwiners between the amplimorphisms we will consider. This will be crucial in defining the tensor product and the braiding.

It can be shown that the category of anyon sectors we define later does not depend on the choice of $\widehat{f}$.

\subsection{Categories of amplimorphisms and endomorphisms}

We largely follow the notation and terminology of~\cite{SzlachanyiV93}.
A *-homomorphism $\chi : \caB \rightarrow M_{n \times n}(\caB)$ is called an \emph{amplimorphism} of degree $n$.\footnote{One can take amplifications with infinite dimensional Hilbert spaces, but for our purposes it is enough to consider only the case where $n$ is finite.} We do not require such amplimorphisms to be unital. Given two amplimorphisms $\chi$ and $\chi'$ of degrees $n$ and $n'$ respectively, we let
\begin{equation} \label{eq:intertwiners of amplis defined}
    (\chi | \chi') := \{  T \in M_{n \times n'}(\caB(\caH_0)) \, : \, T \chi'(O) = \chi(O) T, \,\,\, O \in \caB, \,\,\, \chi(\1) T = T = T \chi'(\1) \}
\end{equation}
be the space of \emph{intertwiners} from $\chi'$ to $\chi$. The amplimorphisms $\chi$ and $\chi'$ are \emph{equivalent} if there is a partial isometry $U \in (\chi | \chi')$ such that $U^* U = \chi'(\1)$ and $U U^* = \chi(\1)$, in which case we write $\chi \sim \chi'$ and call $U$ an \emph{equivalence}.

An amplimorphism $\chi$ of degree $n$ is said to be \emph{localized} in a cone $\Lambda$ if for all $O \in \pi_0(\cstar[\Lambda^c])$, we have $\chi(O) = \chi(\1) (O \otimes \1_n)$. Such an amplimorphism is \emph{transportable} if for any cone $\Lambda'$ there is an amplimorphism $\chi'$ localized in $\Lambda'$ such that $\chi \sim \chi'$.

An amplimorphism $\chi$ is called \emph{finite} if the endomorphism space $(\chi | \chi)$ is finite dimensional. 
Note that $(\chi|\chi)$ is closed under taking adjoints. Hence if $\chi$ is finite, it follows that $(\chi|\chi)$ is isomorphic to a finite direct sum of full matrix algebras.

\begin{defn}
    We define $\Amp$ as the category whose objects are amplimorphisms that are localized in allowed cones, and are transportable.
    The morphisms between objects $\chi'$ and $\chi$ are given by $(\chi | \chi')$.
    The category $\Amp_f$ is the full subcategory of $\Amp$ whose objects are those amplimorphisms in $\Amp$ that are finite.
\end{defn}

In Section \ref{sec:braided tensor structure} we will show how the assumption of Haag duality allows us to endow $\Amp$ with the structure of a braided $\rm C^*$-tensor category. We will later see that the category $\Amp_f$ is closed under the monoidal product of $\Amp$ and therefore inherits the braided $\rm{C}^*$-tensor structure. The reduction to $\Amp_f$ is essential to establish equivalence with the category $\Rep_f \caD(G)$ of finite dimensional representations of the quantum double algebra. Indeed, $\Amp$ contains infinite direct sums, while $\Rep_f \caD(G)$ does not contain infinite direct sums by definition. We do not know if the infinite directs sums of objects of $\Amp_f$ exhaust all non-finite amplimorphisms of $\Amp$.

\begin{rem}
    In the algebraic description of anyons, it is commonly assumed that all anyons have a conjugate (see for example ~\cite[Sect. 6.3]{Wang2010}), meaning that each anyon type can fuse to the vacuum with some conjugate type.

    The assumption that an object in a $\rm C^*$-tensor category has a conjugate implies that it has a finite-dimensional endomorphism space~\cite[Lemma 3.2]{LongoRoberts97}. This is another way to see the necessity of restricting our attention to $\Amp_f$ if we want to show equivalence with $\Rep_f \caD(G)$. Indeed, all finite dimensional representations of $\caD(G)$ have conjugates. 
\end{rem}

\begin{defn}
    We denote by $\DHR$ the full subcategory of $\Amp$ whose objects are unital *-endomorphisms $\nu : \caB \rightarrow \caB$, \ie unital amplimorphisms of degree one. 
    Similarly, $\DHR_f$ is the full subcategory of $\DHR$ whose objects are finite endomorphisms.
\end{defn}
$\DHR$ is a braided $\rm C^*$-tensor subcategory of $\Amp$, see Section \ref{sec:braided tensor structure}. We show in Section \ref{sec:proof of main THM} that $\DHR_f$ is closed under the monoidal product of $\DHR$ and therefore inherits the braided $\rm C^*$-tensor structure.
The category $\DHR$ is equivalent to the category $\mathcal{O}_{\Lambda_0}$ defined in~\cite[Sect. 6]{Ogata2021}.
One can think of $\mathcal{O}_{\Lambda_0}$ as the subcategory of $\DHR$ restricted to endomorphisms localized in a specific cone $\Lambda_0$, however by the transportability requirement, one sees that this is equivalent to $\DHR$ (compare with Sect.~\ref{sec:proof of main THM} here).

\subsection{Main result}

We are now ready to give the main result of this paper, which states that the categories $\Amp_f$ and $\DHR_f$ introduced above are equivalent as braided $\rm C^*$-tensor categories to the category $\Rep_f \caD(G)$ of finite dimensional unitary representations of the quantum double $\caD(G)$ of the group $G$. See Appendix \ref{app:introduction to D(G)} for a brief review of $\caD(G)$ and its representation theory.

\begin{thm} \label{thm:main result}
    If Haag duality for cones (Assumption \ref{ass:Haag duality}) holds, then the categories $\Amp_f$ and $\DHR_f$ are braided $\rm C^*$-tensor categories with monoidal structure and braiding as described in Section \ref{sec:braided tensor structure}. Moreover, both of these categories are then equivalent to $\Rep_f \caD(G)$ as braided $\rm C^*$-tensor categories.
\end{thm}

Since $\Rep_f \caD(G)$ is a unitary modular tensor category (UMTC), it follows from this Theorem that $\Amp_f$ and $\DHR_f$ are also UMTCs. In particular, the anyon sectors are endowed with a duality which is inherited from the duality of finite dimensional representations of $\caD(G)$.




\section{Braided \texorpdfstring{$\rm C^*$}{}-tensor structure of \texorpdfstring{$\Amp$}{Amp} and \texorpdfstring{$\DHR$}{DHR}} \label{sec:braided tensor structure}

We spell out the $\rm C^*$-category structure of $\Amp, \Amp_f$, $\DHR$, and $\DHR_f$, as well as their finite direct sums and subobjects in Section~\ref{subsec:directsum_subojects}.

In Section \ref{subsec:braided monoidal} we use the assumption of Haag duality to endow $\Amp$ and $\DHR$ with braided ${\rm C}^*$-tensor structure. Most arguments in this section are straightforward adaptations of well-known constructions in the DHR superselection theory, see for example~\cite{SzlachanyiV93,NillS97,HalvorsonMueger,Ogata2021}. At this stage we do not know if the categories $\Amp_f$ and $\DHR_f$ are closed under the tensor product which we define for $\Amp$ and $\DHR$, a fact which will only be established in Proposition \ref{prop:equivalence of Amp_rho and Amp} and Lemma \ref{lem:DHR_f is braided monoidal} of Section \ref{sec:proof of main THM}.

\subsection{\texorpdfstring{$\rm C^*$}{C*}-structure, direct sums, and subobjects}
\label{subsec:directsum_subojects}

Let us first remark that the categories $\Amp$, $\Amp_f$, $\DHR$, and $\DHR_f$ are $\rm C^*$-categories (see~\cite{GhezLimaRoberts} or ~\cite[Definition 2.1.1]{neshveyev2013compact}). In this subsection we show that all these categories have finite direct sums and subobjects.

\subsubsection{Direct sums and subobjects of amplimorphisms} \label{subsec:direct sums and subobjects for Amp}

The direct sum of $\chi : \caB \rightarrow M_m(\caB)$ and $\psi : \caB \rightarrow M_n(\caB)$ is the amplimorphism $\chi \oplus \psi : \caB \rightarrow M_{m + n}(\caB)$ that maps $O \in \caB$ to the block diagonal matrix with blocks $\chi(O)$ and $\psi(O)$ with obvious projection and inclusion maps. If $\chi$ and $\psi$ are finite, then so is $\chi \oplus \psi$, so $\Amp_f$ is closed under this direct sum.

Before showing the existence of subobjects for $\Amp$ and $\Amp_f$, we state and prove two lemmas which will also be used to later to equip $\Amp$ and $\DHR$ with a tensor product.

\begin{lem} \label{lem:localized amplimorphisms amplify locally}
    Let $\Lambda$ be an allowed cone. If $\chi$ is a $\Lambda$-localized amplimorphism of degree $n$, then $\chi(\caR(\Lambda)) \subset M_{n}( \caR(\Lambda) )$.
\end{lem}

\begin{proof}
    If $O \in \caR(\Lambda^c)$ then the $\Lambda$-localization of $\chi$ implies that all components of $\chi(\1)$ commute with $O$, so $\chi(\I) \in M_n(\caR(\Lambda^c)') = M_n(\caR(\Lambda))$ by Haag duality.
    Now let $O \in \pi_0( \caA_{\Lambda^c})$ and $A \in \caR(\Lambda)$, then
    \begin{equation}
        \chi(O A) = \chi(O) \chi(A) = \chi(\1) (O \otimes \I_n) \chi(A) = (O \otimes \1_n) \chi(\I) \chi(A) = (O \otimes \I_n) \chi(A),
    \end{equation}
    but $\chi(O A) = \chi(A O)$ and by a similar computation we conclude that $(O \otimes \I_n) \chi(A) = \chi(A) (O \otimes \I_n)$. It follows that $\chi(A) \in M_{n}( \pi_0( \caA_{\Lambda^c})' ) = M_{n}( \pi_0( \caR(\Lambda^c)' ) = M_{n}( \caR(\Lambda) )$ by Haag duality.
\end{proof}

\begin{lem} \label{lem:localized morphisms for AMP}
    If $\chi_1, \chi_2$ are localized transportable amplimorphisms of degrees $n_1$ and $n_2$, and localized on cones $\Lambda_1, \Lambda_2$ respectively, and $\Lambda$ is a cone that contains $\Lambda_1 \cup \Lambda_2$, then $(\chi_1 | \chi_2) \subset M_{n_1 \times n_2}( \caR(\Lambda) )$.
\end{lem}

\begin{proof}
    If $T \in (\chi_1 | \chi_2)$ then for any $O \in \pi_0(\caA_{\Lambda^c})$ we have $\chi_1(O) T = T \chi_2(O)$. Since $O$ is supported outside of the cones $\Lambda_1, \Lambda_2$ on which $\chi_1$ and $\chi_2$ are localized, this implies $\chi_1(\1) (\pi_0(O) \otimes \I_{n_1}) T = T \chi_2(\1) (O \otimes \I_{n_2})$ for any $O \in \caA_{\Lambda^c}$. Using $\chi_1(\1) \in M_{n_1 \times n_2}( \caR(\Lambda_1) )$ (Lemma \ref{lem:localized amplimorphisms amplify locally}) and $\chi_1(\1) T = T = T \chi_2(\1)$ it follows that each component of $T$ belongs to $\pi_0(\caA_{\Lambda^c})' = \caR(\Lambda)$, where we used Haag duality. This proves the claim.
\end{proof}

We now establish the existence of subobjects.
Since at the moment we allow non-unital amplimorphisms, the construction is somewhat more elementary than the corresponding result for DHR endormorphisms (cf.~\cite[Lemma 5.8]{Ogata2021}).

\begin{prop} \label{prop:existence of subobjects for Amp}
    Let $\chi \in \Amp$ and $p \in (\chi | \chi)$ an orthogonal projector. Then there are localized and transportable amplimorphisms $\chi_1, \chi_2 \in \Amp$ and partial isometries $v \in (\chi | \chi_1)$, $w \in (\chi | \chi_2)$ such that $v v^* = p, w w^* = \chi(\1) - p$ ands $v v^* + w w^* = \chi(\1)$. In particular, $\chi$ is isomorphic to $\chi_1 \oplus \chi_2$. If $\chi$ is finite, then so are $\chi_1$ and $\chi_2$.
\end{prop}

\begin{proof}
    Consider the amplimorphism $\chi_1 : \caB \rightarrow M_n(\caB)$ given by $\chi_1(O) := p \chi(O) p$. By Lemma \ref{lem:localized morphisms for AMP} we have $p \in M_{n}( \caR(\Lambda) )$ where $n$ is the degree of $\chi$, so $\chi_{1}$ is localized on $\Lambda$. Moreover, $p \chi_1(O) = p \chi(O) p = \chi(O) p$ and $\chi(\1) p = p = p \chi_1(\1)$ which shows that $p \in (\chi | \chi_1)$.
    
    The amplimorphism $\chi_1$ is also transportable. Indeed, let $\Lambda'$ be some other cone. By transportability of $\chi$ there is an amplimorphism $\chi'$ of degree $n'$ localized on $\Lambda'$ and an equivalence $U \in (\chi|\chi')$. Consider the projection $q = U^* p U \in (\chi' | \chi') \subset M_{n'}( \caR(\Lambda') )$ and corresponding amplimorphism $\chi'_q(O) := q \chi'(O) q$ localized on $\Lambda'$. Then $$pU \chi'_q(O) = p U U^* p U \chi'(O) U^* p U = p \chi(\1) p \chi(O) p U = p \chi(O) p U = \chi_1(O) pU$$ and $pUU^*p = p \chi(\1) p = \chi_1(\1)$ while $$U^* p p U = U^* p \chi(\1) p U = q \chi'(\1) q = \chi'_q(\1),$$ so $pU$ is an equivalence of $\chi_1$ and $\chi'_q$.

    The same construction yields a localized transportable amplimorphism $\chi_{2}$ corresponding to the orthogonal projector $q = \chi(\1) - p \in (\chi | \chi)$. One easily checks that the claim of the proposition is satisfied with $v = p$ and $w = q$.

    Suppose $\chi_1$ were not finite, \ie $(\chi_1 | \chi_1)$ is infinite dimensional. Since $(\chi_1 | \chi_1)$ is isomorphic to $p (\chi | \chi) p$, this implies that $\chi$ is also not finite. With a similar argument for $\chi_2$, this shows that if $\chi$ is finite, then so are $\chi_1$ and $\chi_2$.
\end{proof}

\subsubsection{Direct sums and subobjects in \texorpdfstring{$\DHR$}{DHR}} \label{subsec:direct sums and subobjects for DHR}

The subcategory $\DHR$ is not closed under the direct sum described above, neither does the construction of subobjects stay in the $\DHR$ subcategory. However, $\DHR$ does have finite direct sums and subobjects, see \cite{Ogata2021}. The subcategory $\DHR_f$ is closed under these direct sums, and any subobject of a finite endomorphism must again be finite, so that $\DHR_f$ also has finite direct sums and subobjects.

\subsection{Braided \texorpdfstring{$\rm C^*$}{C*}-tensor structure of \texorpdfstring{$\Amp$}{Amp} and \texorpdfstring{$\DHR$}{DHR}} \label{subsec:braided monoidal}

Using the assumption of Haag duality for cones, we equip $\Amp$ and $\DHR$ with a monoidal product and a braiding, making them into braided $\rm{C}^*$-tensor categories (see Definition 2.1.1 of \cite{neshveyev2013compact}). At this point it is not clear that the tensor product of two finite amplimorphisms, as defined below, is again finite (and in fact one can construct examples of irreducible anyon sectors whose monoidal product decomposes into infinitely many irreducibles, see for example~\cite{Fredenhagen94}).
For this reason we can't yet equip $\Amp_f$ and $\DHR_f$ with the structure of braided $\rm{C}^*$-tensor categories. It will be shown in Proposition \ref{prop:equivalence of Amp_rho and Amp} and Lemma \ref{lem:DHR_f is braided monoidal} that $\Amp_f$ and $\DHR_f$ are in fact closed under the tensor product, and are therefore full braided $\rm C^*$-tensor subcategories of $\Amp$ and of $\DHR$ respectively.

\subsubsection{Monoidal structure}  \label{subsec:monoidal structure}

If $\chi : \caB \rightarrow M_n(\caB)$ is an amplimorphism of degree $n$ we denote by $\chi(O)^{i j}$ for $i, j = 1, \cdots, n$ the $\caB$-valued matrix components of $\chi(O)$. We endow $\Amp$ with a monoidal product $\times$ defined as follows. If $\chi_1$ and $\chi_2$ are amplimorphisms of degrees $n_1$ and $n_2$ respectively, then we define their tensor product $\chi_1 \times \chi_2 : \caB \rightarrow M_{n_1} \big( M_{n_2}(\caB) \big) \simeq M_{n_1 n_2}(\caB)$ to be the amplimorphism of degree $n_1 n_2$ with components
\begin{equation}
\label{eq:monoidal product}
	(\chi_1 \times \chi_2)^{u_1 u_2, v_1 v_2}(O) = \chi_1^{u_1 v_1} \big(  \chi_2^{u_2 v_2}(O)   \big) \quad \text{for all } \,  O \in \caB.
\end{equation}
Note that this is just $(\chi_1 \otimes \I_{n_2}) \circ \chi_2$ after identifying $\caB \otimes M_n(\mathbb{C})$ with $M_n(\caB)$.
For intertwiners $T \in (\chi | \chi')$ and $S \in (\psi | \psi')$ the tensor product $T \times S \in (\chi \times \psi | \chi' \times \psi')$ is defined by
\begin{equation} \label{eq:tensor of intertwiners}
	(T \times S)^{u_1 u_2, v_1 v_2} = \sum_{w_1, w_2} \chi^{u_1 w_1}( S^{u_2 w_2} ) T^{w_1 v_1} \delta^{w_2, v_2}
\end{equation}
which can also be written in matrix notation as $T \times S = \chi(S) (T \otimes I_{\psi'}) = (T \otimes I_{\psi}) \chi'(S)$.

The monoidal unit is the identity amplimorphism which is irreducible because $\caB'' = \caB(\caH_0)$ since $\pi_0$ is irreducible and $\pi_0(\alg{A}) \subset \alg{B}$.
Since the monoidal product is strict, it is trivially compatible with the ${\rm C}^*$-structure.
The subcategory $\DHR$ is closed under this monoidal product and contains the identity, it is therefore a monoidal subcategory of $\Amp$.

The monoidal product of objects is well defined thanks to Lemma \ref{lem:localized amplimorphisms amplify locally} and the monoidal product of intertwiners is well defined thanks to Lemma \ref{lem:localized morphisms for AMP}.
The monoidal product on $\DHR$ coincides with that defined in~\cite{Ogata2021} (see also the remarks around equations~(1.28)--(1.29) there).

\subsubsection{Braiding} \label{subsec:braiding}

It is well known that the category of localized endomorphisms for models in two spatial dimensions can be given a braiding \cite{frs1, frohlich1990braid, frohlich1988statistics}. Here we extend this to localized amplimorphisms.

\begin{figure}[!ht]
    \centering
    \includegraphics[width = 0.5\textwidth]{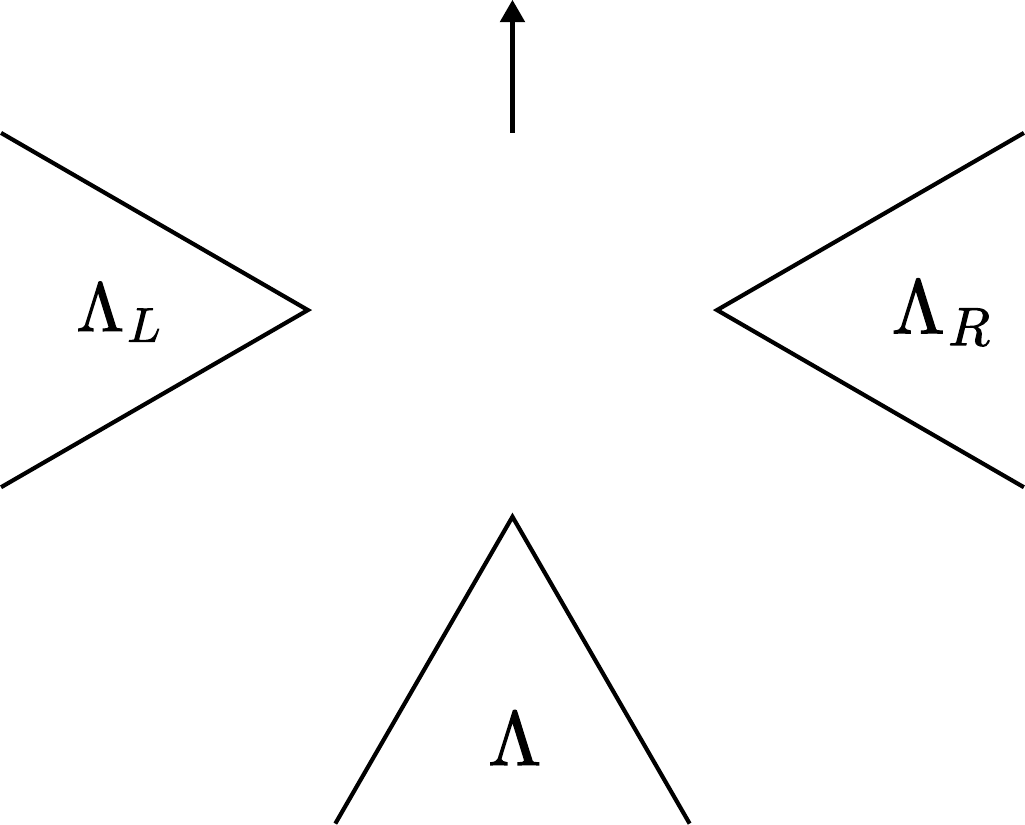}
    \caption{An example of the braiding setup. The arrow represents the forbidden direction.}
    \label{fig:braiding setup}
\end{figure}

The braiding on $\Amp$ is given by intertwiners $\ep(\chi, \psi) \in ( \psi \times \chi | \chi \times \psi)$ defined as follows. Since $\chi$ and $\psi$ are localized in allowed cones there is an allowed cone $\Lambda$ such that $\chi$ and $\psi$ are both localized in $\Lambda$. Let $\Lambda_L$ and $\Lambda_R$ be allowed cones `to the left and to the right' of $\Lambda$, \cf~Figure \ref{fig:braiding setup}. Let $\chi_R$ be a transportable amplimorphism localised in $\Lambda_R$ and fix an equivalence $U \in (\chi_R | \chi)$ with $U \in  M_m(\caR(\widetilde \Lambda_R))$ where $\widetilde \Lambda_R$ is an allowed cone that contains $\Lambda$ and $\Lambda_R$, but is disjoint from $\Lambda_L$. Similarly, pick a transportable amplimorphism $\psi_L$ localised in $\Lambda_L$ and a unitary $V \in (\psi_L | \psi)$ with $V \in M_n(\caR( \widetilde \Lambda_L))$. Such $\chi_R, U, \psi_L, V$ exist by transportability of $\chi$ and $\psi$. Now put
\begin{equation} \label{eq:braiding of amplimorphisms defined}
	\ep(\chi, \psi) := (V^* \times U^*) \cdot P_{12} \cdot (U \times V)
\end{equation}
where $P_{12} \in (\chi_R \times \psi_L | \psi_L \times \chi_R)$ is given by its components $P_{12}^{u_1 u_2, v_1 v_2} = \psi_L^{u_2 v_1}\left(\chi_R^{u_1 v_2}(\I)\right)$ (note the transposition of the indices compared to~\eqref{eq:monoidal product}).
That $P_{12}$ indeed is an intertwiner follows from a short calculation using that $\psi_L$ and $\chi_R$ are localized in disjoint cones, and hence $\psi_L^{u_1 u_2}(\chi_R^{v_1 v_2}(A)) = \chi_R^{v_1 v_2}(\psi_L^{u_1 u_2}(A))$ for all $A \in \caA$.
Alternatively,
\begin{equation}
    P_{12} = (\operatorname{id}_\caB \otimes P)((\psi_L \times \chi_R)(\I)), 
\end{equation}
where $P : M_n(\mathbb{C}) \otimes M_m(\mathbb{C}) \to M_m(\mathbb{C}) \otimes M_n(\mathbb{C})$ flips the tensor factors.
Using standard arguments, on can check that indeed $\ep(\chi, \psi) \in (\psi \times \chi | \chi \times \psi)$, that $\ep(\chi, \psi)$ is independent of the choices of $\chi_R, \psi_L, U, V$, and that $\ep$ is indeed a braiding for $\Amp$. See for example~\cite[Prop. 5.2]{SzlachanyiV93} for amplimorphisms, or \cite[Lemma 4.8]{Naaijkens2011},  \cite[Definition 4.10]{Ogata2021}, or \cite[Lemma 2.9]{bols2024double} for proofs of the analogous fact for the braiding of endomorphisms.\footnote{Note that in the case of approximate Haag duality (as in~\cite{Ogata2021}), one has to do some additional limiting procedure to define the braiding. This is because under the weaker localization properties, we do not necessarily have that $\rho \times \sigma = \sigma \times \rho$ if $\rho$ and $\sigma$ are approximately localized in disjoint cones.}
This braiding restricts to the $\rm C^*$-tensor subcategory $\DHR$, so $\DHR$ is a braided $\rm C^*$-tensor subcategory of $\Amp$.

\section{Equivalence of \texorpdfstring{$\Amp$}{Amp} and \texorpdfstring{$\DHR$}{DHR}}
\label{sec:equivalence of Amp and DHR}

\subsection{Reduction to unital amplimorphisms}

Our proof of the braided monoidal equivalence will rely on the fact that any amplimorphism of $\Amp$ is equivalent to a unital amplimorphism, a fact which we prove here. This fact will also be useful in Section \ref{sec:simples of Amp}, where the simple objects of $\Amp$ are characterized.

We say $\widetilde \Lambda$ is slightly larger than $\Lambda$, denoted $\Lambda \Subset \widetilde \Lambda$, if there exists another cone $\Lambda' \subset \widetilde \Lambda$ disjoint from $\Lambda$. 
That is, we can fit a cone in $\Lambda^c \cap \widetilde{\Lambda}$.
The following Lemma is proven in exactly the same way as \cite[Lemma 5.11]{Ogata2021}, and noting \cite[Corollary 6.3.5]{KadisonR97}.
We include it here for the convenience of the reader, as we will use this result repeatedly.

\begin{lem} \label{lem:projectors are infinite}
    Let $\Lambda \Subset \widetilde \Lambda$ and let $p \in M_n( \caR(\Lambda) )$ be an orthogonal projector. Then $p$ is infinite as a projector in $M_n(\caR(\widetilde \Lambda))$, and is Murray-von Neumann equivalent to $\1_n$.
\end{lem}
\begin{proof}
     By assumption, there is a cone $\Lambda' \subset \widetilde \Lambda$ that is disjoint from $\Lambda$. Since $\caR(\Lambda')$ is an infinite factor (see Sect.~\ref{sec:cone_algebra}), so is $M_n(\caR(\Lambda'))$ and we can apply the halving lemma~\cite[Lemma 6.3.3]{KadisonR97} to find an isometry $V \in M_n(\caR(\Lambda'))$ such that $V V^* < \1_n$. Note that $V$ and $V^*$ commute with $p$ since they have disjoint supports.
     The map $x \mapsto xp$ for $x \in M_n(\caR(\Lambda'))$ is a $*$-isomorphism from $M_n(\caR(\Lambda'))$ onto $M_n(\caR(\Lambda')p$ by~\cite[Prop. 5.5.5]{Kadison1983}.
     In particular, this implies that $V V^* p \neq p$, and hence is a proper subprojection of $p$.
     Put $\widetilde V = p V$, then
     \begin{equation}
         \widetilde V \widetilde V^* = p V V^* < p, \quad \widetilde V^* \widetilde V = p V^* V p = p.
     \end{equation}
     This shows that $p$ as a projection in $M_n(\caR(\widetilde \Lambda))$ is Murray von Neumann equivalent to its proper subprojection $p VV^*$ and thus $p$ is infinite in $M_n(\caR(\widetilde \Lambda))$. Murray-von Neumann equivalence to $\1_n$ now follows immediately from Corollary 6.3.5 of \cite{KadisonR97}.
\end{proof}

\begin{lem} \label{lem:localised amplis are equiv to unital amplis}
        Let $\chi$ be an amplimorphism of degree $n$ localized in a cone $\Lambda$, and $\widetilde \Lambda$ be another cone such that $\Lambda \Subset \widetilde \Lambda$. Then there exists a unital amplimorphism localized on  $\widetilde \Lambda$ that is equivalent to $\chi$.
\end{lem}

\begin{proof}
    By Lemma \ref{lem:localized amplimorphisms amplify locally} we have that the projector $\chi(\1)$ belongs to $M_n(\caR(\Lambda))$. By Lemma~\ref{lem:projectors are infinite}, it follows that $\chi(\1)$ is infinite as an element of $M_n(\caR(\widetilde\Lambda))$ and is Murray-von Neumann equivalent to $\1_n \in M_n(\caR(\widetilde{\Lambda}))$. Therefore there exists an isometry $V \in M_n(\caR(\widetilde\Lambda))$ such that $V V^* = \chi(\1)$ and $V^* V = \1_n$. 
    
    Let $\psi$ be given by $\psi(O) = V^* \chi(O) V$ for all $O \in \caB$, then $\psi(\1) = V^* \chi(\1) V = V^* V V^* V = \1_n$ so $\psi$ is indeed unital. In fact, we see that $V \in (\chi | \psi)$ is an equivalence.
    If $O \in \pi_0(\alg{A}_{\widetilde \Lambda^c})$ then 
    \begin{equation}
        \psi(O) = V^* \chi(O) V = V^* \chi(\1) (O \otimes \1_n) V = V^* (O \otimes \1_n) V = (O \otimes \1_n) V^* V = O \otimes \1_n,
    \end{equation}
    so $\psi$ is indeed localized on $\widetilde \Lambda$.
\end{proof}

If $\chi$ is in addition transportable, we can first transport to a smaller cone inside the localization region $\Lambda$, to make room for the `additional cone' needed in the proof.
The construction above does not affect transportability, so we immediately obtain the following corollary.

\begin{cor} \label{cor:ampli is equivalent to unital ampli in the same cone}
    Any localized and transportable amplimorphism $\chi$ is equivalent to a \emph{unital} transportable amplimorphism $\chi'$ localized in the same cone.
\end{cor}

\begin{proof}
    Let $\chi$ be localized in $\Lambda$. We have by transportability of $\chi$ that there exists an amplimorphism $\psi$ localized in a cone $\Lambda' \Subset \Lambda$ such that $\psi \sim \chi$. We have by Lemma \ref{lem:localised amplis are equiv to unital amplis} that there exists a unital amplimorphism $\chi'$ localized in $\Lambda$ such that $\chi' \sim \psi$, so we have $\chi' \sim \chi$. Transportability of $\chi'$ is immediate by the transportability of $\chi$.
\end{proof}

\subsection{Proof of equivalence}

We now show that instead of amplimorphisms, we can equivalently talk about endomorphisms.
For any cone $\Lambda$ and any $n \in \N$, fix a row vector $\Iso(\Lambda, n) := (V_1, \cdots, V_n)$ whose components are isometries $V_i \in \caR(\Lambda)$ satisfying $V_i^* V_j = \delta_{ij} \1$ and $\sum_{i = 1}^n V_i V_i^* = \1_n$.
(Since $\caR(\Lambda)$ is an infinite factor, we can repeatedly apply the halving lemma~\cite[Lemma 6.3.3]{KadisonR97} to obtain such isometries).
For any $\chi \in \Amp$ fix an allowed cone $\Lambda_{\chi}$ such that $\chi$ is localized on $\Lambda_{\chi}$ and write $\Iso_{\chi} = \Iso(\Lambda_{\chi}, n)$, where $n$ is the degree of $\chi$.

Now let $\chi \in \Amp$ be a unital amplimorphism of degree $n$. We define $\nu_{\chi} : \caB \rightarrow \caB$ to be the endomorphism given by
\begin{equation}
    \nu_{\chi}(O) := \Iso_{\chi} \, \chi(O) \Iso_{\chi}^*.
\end{equation}
Here we see $\Iso_{\chi}^*$ as a column vector with entries $V_i^*$.
One easily verifies that this indeed is an endomorphism and that $\nu_{\chi}$ is localized in $\Lambda_{\chi}$.

If $\chi, \chi' \in \Amp$ are unital amplimorphisms and $T \in (\chi | \chi')$, we define $t_T \in \caB(\caH_0)$ by $t_T = \Iso_{\chi} T \Iso_{\chi'}^*$. Then
\begin{equation}
    t_T \nu_{\chi'}(O) = \Iso_{\chi} \, T \, \Iso_{\chi'}^* \, \Iso_{\chi'} \, \chi'(O) \, \Iso_{\chi'}^* = \Iso_{\chi} \, T \, \chi'(O) \, \Iso_{\chi'}^* = \Iso_{\chi} \, \chi(O) \, T \, \Iso_{\chi'}^* = \nu_{\chi}(O) \, t_T
\end{equation}
so $t_T \in (\nu_{\chi} | \nu_{\chi'})$. The map $T \mapsto t_T$ defines a *-isomorphism of intertwiner spaces $(\chi | \chi')$ and $(\nu_{\chi} | \nu_{\chi'})$.

It follows in particular that the $\nu_{\chi}$ obtained in this way are transportable. Indeed, let $\Lambda'$ be some cone. By transportability of $\chi$ and Corollary \ref{cor:ampli is equivalent to unital ampli in the same cone} there is unital $\chi' \in \Amp$ localized on $\Lambda'$ and a unitary $U \in (\chi | \chi')$. Then $t_U \in (\nu_{\chi} | \nu_{\chi'})$ is also unitary.

Since $\Iso_{\chi} \in ( \nu_{\chi} | \chi )$ is an equivalence of amplimorphisms, we conclude in particular that every unital amplimorphsm in $\Amp$ is equivalent to an endomorphism in $\DHR$. Together with Corollary \ref{cor:ampli is equivalent to unital ampli in the same cone} we obtain the following lemma.

\begin{lem} \label{lem:ampli is equivalent to endo}
Every $\chi \in \Amp$ is equivalent to an endomorphism $\rho_\chi$ in the subcategory $\DHR$. 
\end{lem}

Even though we do not need it to prove Theorem \ref{thm:main result}, we can now easily obtain the following proposition which says that the localized and transportable amplimorphisms are equivalent to the endomorphisms studied in~\cite{Ogata2021}.
\begin{prop}
\label{prop:braided monoidal equivalence of Amp and DHR}
    $\DHR$ and $\Amp$ are equivalent as braided $\rm C^*$-tensor categories.
\end{prop}
\begin{proof}
    Let $F : \DHR \rightarrow \Amp$ be the embedding functor. Clearly $F$ is linear, fully faithful, braided monoidal, and respects the $*$-structure. It remains to check that $F$ is essentially surjective, but this is immediate from Lemma \ref{lem:ampli is equivalent to endo}.
\end{proof}




 \section{Amplimorphisms from ribbon operators} \label{sec:amplimorphism from ribbon operators}

In this section we construct for each half-infinite ribbon $\rho$ a full subcategory $\Amp_{\rho}$ of $\Amp$ whose objects are constructed as limits of certain `ribbon operators' taking unitary representations of $\caD(G)$ as input. (See Appendix \ref{app:ribbon operators} for the definition and basic properties of ribbons and ribbon operators).
From the equivalence of the localized and transportable amplimorphisms to DHR endomorphisms, this amounts to explicitly constructing examples of representations that satisfy the superselection criterion.
More importantly, we can also define the intertwiners as (weak operator) limits of elements in the quasi-local algebra.
In the notation of~\cite{Ogata2021}, this amounts to finding explicit examples of the maps $T$ defined there, as well as how they act on the intertwiners.

The very concrete description of $\Amp_{\rho}$ and its intertwiners will allow us to identify the braiding and fusion in this category.
We will use this to show in Section \ref{subsec:Amp_rho and Rep} that the categories $\Amp_{\rho}$ are equivalent to $\Rep_f\, \caD(G)$ as braided $\rm C^*$-tensor categories, and in Section \ref{subsec:Amp_rho and Amp_f} that they are equivalent to the whole of $\Amp_f$, thus establishing the equivalence of $\Amp_f$ and $\Rep_f\, \caD(G)$ as braided $\rm C^*$-tensor categories.

\subsection{Finite ribbon multiplets}

Throughout the rest of this manuscript the tensor product $\otimes$ of two matrices over $\caA$ will always mean the usual matrix tensor product, while the tensor product $\otimes$ of an element of $\caA$ with a matrix over $\C$ means the amplifying tensor product, yielding a matrix over $\caA$.

\begin{defn} \label{def:simple ribbon operators}
	For any $n$-dimensional unitary representation $D$ of $\caD(G)$ and any ribbon $\rho$ define $\bF^D_{\rho} \in M_n(\caA)$ by
	\begin{equation}
		\bF_{\rho}^D = \sum_{g, h} \, F_{\rho}^{g, h} \otimes D \big( g, h \big).
	\end{equation}
\end{defn}

\begin{figure}[!ht]
\centering
\includegraphics[width = 0.4\textwidth]{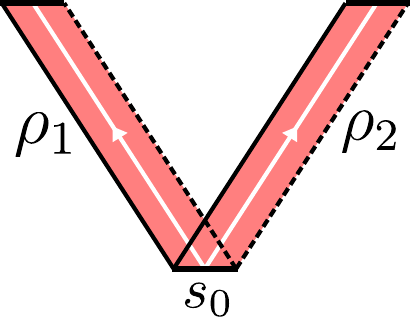}
\caption{Braiding positive ribbon operators, both having the same starting site $s_0$.}
\label{fig:braiding positive ribbons}
\end{figure}

\begin{prop} \label{prop:properties of simple ribbon operators}
	Let $\rho$ be a ribbon such that $s_i = \partial_i \rho$, $i = 1, 2$ have distinct vertices and faces, and let $D$ be an $n$-dimensional unital unitary representation of $\caD(G)$.
	\begin{enumerate}[label=(\roman*)]
		\item We have
            \begin{equation}
                \bF_{\rho}^{D} \cdot ( \bF_{\rho}^{D} )^* = (\bF_{\rho}^{D} )^* \cdot \bF_{\rho}^{D} = \1_n.
            \end{equation}
            In other words, $\bF_{\rho}^{D}$ is a unitary element of $M_n(\caA)$. \label{ribbon prop:unitarity}
		
            \item We have $\bF_{\bar \rho}^{D} = ( \bF_{\rho}^D )^*$. \label{ribbon prop:adjoint}
        
            \item \label{ribbon prop:sum and product} Let $D_1, D_2$ be unitary representations of $\caD(G)$. The direct sum and product of 
            ribbon operators $\bF_{\rho}^{D_1}$ and $\bF_{\rho}^{D_2}$ satisfy
    	\begin{equation}
    		\bF_{\rho}^{D_1} \oplus \bF_{\rho}^{D_2} =  \bF_{\rho}^{D_1 \oplus D_2}, \quad \bF_{\rho}^{D_1} \otimes \bF_{\rho}^{D_2} = \bF_{\rho}^{D_1 \times D_2}
            \end{equation}
            where the direct sum and tensor product on the left hand sides are the usual direct sum and tensor product of matrices (with $\caA$-valued components), and $D_1 \times D_2$ is the monoidal product of the two representations (see Appendix~\ref{app:introduction to D(G)}).
            
		\item If $\rho = \rho_1\rho_2$ then
            \begin{equation}
                \bF_{\rho}^D = \bF_{\rho_1}^D \cdot \bF_{\rho_2}^D.
            \end{equation} \label{ribbon prop:concatenation}
            
		\item If $t \in (D_1 | D_2)$ then
            \begin{equation}
                 \bF_{\rho}^{D_1} (\1 \otimes t) = (\1 \otimes t) \bF_{\rho}^{D_2}, \quad (\bF_{\rho}^{D_1})^* (\1 \otimes t) = (\1 \otimes t) (\bF_{\rho}^{D_2})^* .
            \end{equation} \label{ribbon prop:intertwiners}
		\item If $\rho_1$ and $\rho_2$ are positive ribbons with common initial site $s_0$ as in Figure \ref{fig:braiding positive ribbons}, then
            \begin{equation} \label{eq:braiding of positive ribbon multiplets}
                \bF_{\rho_2}^{D_2} \otimes \bF_{\rho_1}^{D_1} = (\I \otimes B(D_1, D_2)) \cdot ( \bF_{\rho_1}^{D_1} \otimes \bF_{\rho_2}^{D_2} ) \cdot (\I \otimes P_{12}).
            \end{equation} \label{ribbon prop:braiding}
            where $B(-, -)$ is the braiding on $\Rep_f \caD(G)$, and $P_{12}$ interchanges the factors in the tensor product of the representation spaces of $D_1$ and $D_2$ (see Appendix \ref{app:introduction to D(G)}).
	\end{enumerate}
\end{prop}

\begin{proof}
    By straightforward computations using Eqs. \eqref{eq:ribbon multiplication and adjoint}, \eqref{eq:ribbon reversal}, \eqref{eq:sum to identity}, and using the braid relation \eqref{eq:braiding positive ribbons} to obtain item \ref{ribbon prop:braiding}.
\end{proof}

\subsection{Amplimorphisms of the quasi-local algebra from ribbon multiplets}

\subsubsection{Construction}

For any finite ribbon $\rho$ and any $n$-dimensional unitary representation $D$ of $\caD(G)$, define linear maps $\mu_{\rho}^{D} : \caA \rightarrow M_n(\caA) \simeq \caA \otimes M_n(\C)$ by
\begin{equation}
	\mu_{\rho}^{D}(O) := \bF_{\rho}^{D} \cdot (O \otimes \I_n) \cdot (\bF_{\rho}^{D})^*.
\end{equation}
Note that by Proposition~\ref{prop:properties of simple ribbon operators} it follows directly that $\mu_\rho^D$ is a $*$-homomorphism.

A half-infinite ribbon $\rho = \{\tau_n\}_{n = 1}^{\infty}$ is a sequence of triangles labelled by $n \in \N$ such that $\partial_1 \tau_n = \partial_0 \tau_{n+1}$ for all $n\in \bbN$ and such that no edge of the lattice belongs to more than one of these triangles.

For any half-infinite ribbon $\rho = \{\tau_n\}$, denote by $\rho_n$ the ribbon consisting of the first $n$ triangles of $\rho$ and by $\rho_{>n} = \rho \setminus \rho_n$ the half-infinite ribbon obtained from $\rho$ by omitting the first $n$ triangles. Then a standard argument using Proposition~\ref{prop:properties of simple ribbon operators}\ref{ribbon prop:concatenation} shows the following limiting maps are well defined.
\begin{defn} \label{def:irreducible amplimorphism}
	For any half-infinite ribbon $\rho$ and any $n$-dimensional unitary representation $D$ of $\caD(G)$, define a linear map $\mu_{\rho}^{D} : \caA \rightarrow M_n(\caA)$ by
	\begin{equation}
		\mu_{\rho}^{D}(O) := \lim_{n \uparrow \infty} \, \mu_{\rho_n}^{D}(O).
	\end{equation}
\end{defn}

We have:
\begin{lem}[Lemma 5.2 of \cite{Naaijkens2015}] \label{lem:ampli properties}
	The map $\mu_{\rho}^{D}:\caA \rightarrow M_n(\caA)$ is a unital *-homomorphism. \ie it is an amplimorphism of $\caA$ of degree $n$. Moreover, if the support of $O \in \cstar$ is disjoint from the support of $\rho$ then $\mu_{\rho}^{D}(O) = O \otimes \I_{n}$. For any $O \in \caA^{\loc}$ we have $\mu_{\rho}^{D}(O) = \mu_{\rho_n}^{D}(O)$ for all $n$ large enough. 
\end{lem}

For each site in the model, it is possible to define an action $\gamma : \caD(G) \to \operatorname{Aut}(\alg{A})$ of the quantum double Hopf algebra.
The amplimorphisms constructed here transform covariantly with respect to this action.
These transformation properties (and of the ribbon multiplets themselves under this action) are essentially what connects these amplimorphisms to representations of $\caD(G)$.
For our purposes it is not necessary to spell out the details, and we refer the interested reader to~\cite{HamdanThesis}.

\subsubsection{Direct sum and tensor product}
The direct sum and tensor product of amplimorphisms of $\cstar$ are defined in the same way as amplimorphisms of $\caB$. We have for all $O \in \cstar$,
\begin{equation}
	(\mu_1 \times \mu_2)^{u_1 u_2, v_1 v_2}(O) = \mu_1^{u_1 v_1} \big(  \mu_2^{u_2 v_2}(O)   \big),
\end{equation}
and the direct sum of $\mu_1 : \cstar \rightarrow M_m(\cstar)$ and $\mu_2 : \cstar \rightarrow M_n(\cstar)$ is the amplimorphism $\mu_1 \oplus \mu_2 : \cstar \rightarrow M_{m + n}(\cstar)$ that maps $O \in \cstar$ to the block diagonal matrix with blocks $\mu_1(O)$ and $\mu_2(O)$.

\begin{lem} \label{lem:sum and product of ribbon amplimorphisms}
	If $\rho$ is a finite or half-infinite ribbon then
	\begin{equation}
		\mu_{\rho}^{D_1} \oplus \mu_{\rho}^{D_2} = \mu_{\rho}^{D_1 \oplus D_2}, \quad \mu_{\rho}^{D_1} \times \mu_{\rho}^{D_2} = \mu_{\rho}^{D_1 \times D_2}.
	\end{equation}
\end{lem}

\begin{proof}
	First consider the case where  $\rho$ is a finite ribbon.
    For ease of notation we omit the subscripts $\rho$ in the following.
    For any $O \in \caA$ we have
	\begin{align*}
		( \mu^{D_1} \oplus \mu^{D_2} )(O) &= \mu^{D_1}(O) \oplus \mu^{D_2}(O) = \bF^{D_1} (O \otimes \I_{n_1}) (\bF^{D_1})^* \oplus \bF^{D_2} (O \otimes \I_{n_2}) ( \bF^{D_2} )^* \\
						  &= \big( \bF^{D_1} \oplus \bF^{D_2} \big) \, (O \otimes \I_{n_1 + n_2} ) \, \big( \bF^{D_1} \oplus \bF^{D_2} \big)^* = \mu^{D_1 \oplus D_2}(O),
	\end{align*}
	where the last step uses item~\ref{ribbon prop:sum and product} of Proposition \ref{prop:properties of simple ribbon operators}.

	For the product, we compute componentwise
	\begin{align*}
		(\mu^{D_1} \times \mu^{D_2})(O)^{u_1 u_2; v_1 v_2} &= \mu^{D_1; u_1 v_1} \big(  \mu^{D_2; u_2 v_2}(O) \big) = \sum_{w_2} \, \mu^{D_1 ; u_1 v_1} \left( \bF^{D_2 ; u_2 w_2} \, O \, (\bF^{D_2 ; v_2 w_2})^* \right) \\
								   &= \sum_{w_1, w_2} \, \bF^{D_1; u_1 w_1} \, \bF^{D_2; u_2 w_2} \, O \, (\bF^{D_2; v_2 w_2})^* \, (\bF^{D_1;v_1 w_1})^* \\
								   &= \sum_{w_1, w_2} \, ( \bF^{D_1} \times \bF^{D_2} )^{u_1 u_2; w_1 w_2} \, O \, (( \bF^{D_1} \times \bF^{D_2} )^*)^{w_1 w_2 ; v_1 v_2} \\
								   &= \big( \bF^{D_1 \times D_2} \, (O \otimes \I_{n_1 n_2}) \, (\bF^{D_1 \times D_2})^* \big)^{u_1 u_2; v_1 v_2} = \mu^{D_1 \times D_2}(O)^{u_1 u_2; v_1 v_2}
	\end{align*}
	where the next to last step again uses item~\ref{ribbon prop:sum and product} of Proposition \ref{prop:properties of simple ribbon operators}.

    If $\rho$ is half-infinite, then the claim follows from the finite case by taking the limit of $\mu_{\rho_n}^D$.
\end{proof}

\subsubsection{Transportability}

We would like to extend the $\mu_{\rho}^{D}$ to amplimorphisms of the allowed algebra $\caB$. To this end, we must first establish their transportability.

We begin with a basic lemma which shows in particular that if $\rho$ and $\rho'$ coincide eventually, then $\mu_{\rho}^{D}$ and $\mu_{\rho'}^{D}$ are unitarily equivalent. Recall that if $\rho$ is a half-infinite ribbon, $\rho_n$ denotes the finite ribbon consisting of the first $n$ triangles of $\rho$, and $\rho_{> n}$ denotes the half-infinite ribbon obtained from $\rho$ by removing its first $n$ triangles. In particular, $\rho = \rho_n \rho_{>n}$.

\begin{lem} \label{lem:ribbon shortening intertwiner}
	Let $\rho$ be a half-infinite positive ribbon and let $D$ be an $n$-dimensional unitary representation of $\caD(G)$. Then
	\begin{equation}
		\mu_{\rho}^{D} = \Ad[ \bF_{\rho_n}^{D} ] \circ \mu_{\rho_{>n}}^{D}
	\end{equation}
	for any $n \in \N$.
\end{lem}

\begin{proof}
	This follows immediately from the definitions, Lemma \ref{lem:ampli properties}, and Proposition \ref{prop:properties of simple ribbon operators}.
\end{proof}

Since the $\bF_{\rho_n}^D$ are unitary operators, this establishes transportability over a finite distance.
To construct more general intertwiners, we need to use a limiting procedure.
\begin{defn} \label{def:bridge}
    Let $\rho$ and $\rho'$ be two half-infinite ribbons. A sequence of finite ribbons $\{ \xi_n \}_{n \in \N}$ is said to be a \emph{bridge} from $\rho$ to $\rho'$ if for each $n$ the concatenations $\sigma_n = \rho_n \xi_n \bar \rho'_n$ are finite ribbons and the bridges $\xi_n$ are eventually supported outside any ball. We call $\{ \sigma_n \}$ the intertwining sequence of the bridge $\{ \xi_n \}.$

    We say a half-infinite ribbon $\rho$ is `good' if it is supported in a cone $\Lambda$ and for any other cone $\Lambda'$ that is disjoint from $\Lambda$, there is a half-infinite ribbon $\rho'$ and a bridge from $\rho$ to $\rho'$. Note that any cone contains plenty of good half-infinite ribbons, both positive and negative ones.
\end{defn}

\begin{lem} \label{lem:construction of transporters}
	Let $\rho$ be a half-infinite positive ribbon and let $\rho'$ be half-infinite negative ribbon both supported in a cone $\Lambda$ and with initial sites $s, s'$ respectively. Suppose there is a bridge from $\rho$ to $\rho'$ with intertwining sequence $\{ \sigma_m = \rho_m \xi_m \overline \rho'_m \}$ all supported in $\Lambda$. Let $D$ be an $n$-dimensional unitary representation of $\caD(G)$. Then there is a unitary $U \in M_n(\caR(\Lambda))$ such that
	\begin{equation} \label{eq:transporter property}
		(\pi_0 \otimes \id_n) \circ \mu_{\rho'}^{D} = \Ad[ U ] \circ (\pi_0 \otimes \id_n) \circ \mu_{\rho}^{D}.
	\end{equation}
\end{lem}

\begin{proof}
	Consider the family of half-infinite ribbons $\rho^{(m)} = \rho'_m \overline{\xi_m} \rho_{>m}$, see Figure \ref{fig:bridge}. We first show that
	\begin{equation} \label{eq:finite intertwining}
		\mu_{\rho^{(m)}}^{D} = \Ad[ \bF_{\bar \sigma_m}^{D}  ] \circ \mu_{\rho}^{D}.
	\end{equation}
	Indeed, by Proposition \ref{prop:properties of simple ribbon operators} we have $\bF_{\bar \sigma_m}^{D} = \big( \bF_{\rho_m}^{D} \cdot \bF_{\xi_m}^{D} \cdot \bF_{\overline \rho'_m}^{D}  \big)^* = \bF_{\rho'_m}^{D} \cdot \bF_{\overline{\xi_m}}^{D} \cdot ( \bF_{\rho_m}^{D} )^*$ so for any $O \in \caA^{\loc}$ we have
	\begin{align*}
		\big( \Ad[ \bF_{\bar \sigma_m}^{D} ] \circ \mu_{\rho}^{D} \big)(O) &= \lim_{N \uparrow \infty} \, \Ad \left[ \bF_{\rho'_m}^{D} \cdot \bF_{\overline{\xi_m}}^{D} \cdot (\bF_{\rho_m}^{D})^* \cdot \bF_{\rho_m}^{D} \cdot \bF_{(\rho_{>m})_N}^{D} \right] (O \otimes \I_n). \\
    \intertext{Now we use unitarity to get}
         &= \lim_{N \uparrow \infty} \, \Ad \left[ \bF_{\rho'_m}^{D} \cdot \bF_{\overline{\xi_m}}^{D} \cdot \bF_{(\rho_{>m})_N}^{D} \right] (O \otimes \I_n) = \mu_{\rho^{(m)}}(O)
	\end{align*}
	as required. 
	
	\begin{figure}[!t]
	\centering
	\includegraphics[width = 0.4\textwidth]{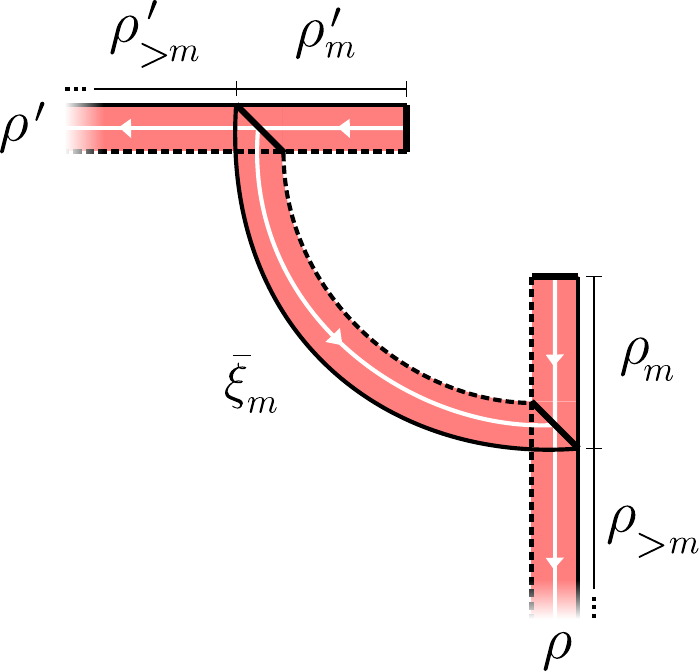}
	\caption{The finite ribbon $\overline{\xi_m}$ is a bridge from ribbon $\rho_m'$ to $\rho_m$.} 
    \label{fig:bridge}
	\end{figure}

	By Lemma \ref{lem:unitary transporters} the components of the image of  $\bF_{\bar \sigma_n}^{D}$ under $\pi_0 \otimes \id_n$ converge in the strong-* topology, and therefore so does the full image of $\bF_{\bar \sigma_n}^{D}$. Denote the limit by $U$. Since the $\bF_{\bar \sigma_n}^{D}$ are all unitary (Proposition \ref{prop:properties of simple ribbon operators}) it follows from Lemma \ref{lem:abstract unitarity} that $U$ is unitary. Since all the $\bF_{\bar \sigma_n}^{D}$ are supported in the cone $\Lambda$, it follows that $U \in M_n(\caR(\Lambda))$.
    
    Let $O \in \caA^{\loc}$. Then
    \begin{align*}
        U \cdot (\pi_0 \otimes \id_n) \big( \mu^D_{\rho}(O) \big) &= \lim_{n \uparrow \infty} \, (\pi_0 \otimes \id_n) \big( \bF_{\bar \sigma_n}^{D} \cdot \mu^D_{\rho}(O) \big) \\
        &= \lim_{n \uparrow \infty} \, (\pi_0 \otimes \id_n) \big(  \mu^D_{\rho^{(n)}}(O) \cdot \bF_{\bar \sigma_n}^{D}  \big) \\
        &= (\pi_0 \otimes \id_n) \big( \mu^D_{\rho'}(O) \big) \cdot U
    \end{align*}
    where we used componentwise continuity of multiplication in the strong operator topology in the first equality, Eq.~\eqref{eq:finite intertwining} to obtain the second equality, and the fact that $\mu^D_{\rho^{(n)}}(O) = \mu^D_{\rho'}(O)$ for $n$ large enough and again componentwise continuity of multiplication to obtain the last equality. Since $\caA^{\loc}$ is dense in $\caA$, we conclude that Eq. \eqref{eq:transporter property} holds, which completes the proof.
\end{proof}

\begin{rem}
This answers a question that was left open in~\cite{Naaijkens2015}, namely the construction of unitary charge transporters that transport charges between two cones, and not just over a finite distance.
Note that Lemma~\ref{lem:construction of transporters} implies that the representation $(\pi_0 \otimes \id_n) \circ \mu_{\rho}^D$ satisfies a variant of the superselection criterion, where we have $(\pi_0 \otimes \id_n) \circ \mu_{\rho}^D \upharpoonright \caA_{\Lambda^c} \cong n \cdot \pi_0 \upharpoonright \caA_{\Lambda^c}$.
That is, instead of unitary equivalence as in~\eqref{eq:sselect}, we have \emph{quasi-}equivalence.
As we shall see shortly, in the case at hand the two notions can be seen to coincide.
\end{rem}

\subsection{Amplimorphisms of the allowed algebra from ribbon multiplets}

The transportability of the $\mu_{\rho}^D$ established above in Lemma \ref{lem:construction of transporters} allows us to extend these amplimorphsisms to localized and transportable amplimorphisms of the allowed algebra $\caB$.

\begin{prop} \label{prop:Amp is nonempty}
	Let $\rho$ be a good half-infinite positive ribbon that is contained in an allowed cone $\Lambda$, then there exists a unique amplimorphism $\chi_{\rho}^{D} : \caB \rightarrow M_n(\caB)$ whose restriction to $\caR(\Lambda)$ is weakly continuous, and satisfies 
	\begin{equation}
		\chi_{\rho}^{D} \circ \pi_0(O) = (\pi_0 \otimes \id_n) \circ \mu_{\rho}^{D}(O).
	\end{equation}
    for all $O \in \cstar$. Moreover, $\chi_{\rho}^{D}$ is localized in $\Lambda$ and is transportable. It is therefore an object of $\Amp$.
\end{prop}

\begin{proof}
    Recall that $\caB$ is a direct limit of cone algebras $\caR(\Lambda)$.
    Note that $\mu_\rho^D$ restricts to an amplimorphism $\cstar[\Lambda] \to M_n(\cstar[\Lambda])$.
    We show that we can extend this (on both sides) to $\caR(\Lambda)$.
    This construction is compatible with the direct structure on the set of allowed cones, and hence defines an amplimorphism of $\caB$.

    To see that we can extend $\mu_\rho^D$ (restricted to $\cstar[\Lambda]$) to $\chi_\rho^D : \caR( \Lambda) \rightarrow M_n(\caR(\Lambda))$,
    note first that for every $\Lambda$ we have the existence of a forbidden cone $\widehat \Lambda$ disjoint from $\Lambda$.
    Since $\rho$ is good and by Lemma \ref{lem:construction of transporters}, we have that $\mu_\rho^D \simeq \mu_{\hat \rho}^D$ where $\hat \rho$ is localized in $\widehat \Lambda$.
    Let $U$ be the unitary implementing this equivalence.
    By locality we have that for all $O \in \cstar[ \Lambda]$, it holds that $\mu_{\widehat \rho}^D(O) = O \otimes \I_n$.
    
    Define $\chi_\rho^D(O) := \Ad[U](O \otimes \I_n)$ for all $O \in \caR(\Lambda)$.
    By construction, it follows that $\chi_\rho^D(O) = \mu_\rho^D(O)$ for all $O \in \cstar[\Lambda]$.
    Let $O \in \caR(\Lambda)$. 
    Then there exist $\cstar[\Lambda] \ni O_\lambda \to O$ weakly since $\cstar[\Lambda]$ is weak-operator dense in $\caR(\Lambda)$.
    Hence we have
    \[
        \lim_\lambda \mu_\rho^D(O_\lambda) = \lim_\lambda \Ad[U](O_\lambda \otimes \I_n) = \Ad[U](O \otimes \I_n) = \chi_\rho^D(O),
    \]
    where all limits are in the weak operator topology and we used that  $\Ad[U]$ is weakly continuous.
    Hence, $\chi_\rho^D$ is uniquely determined by $\mu_\rho^D$.
    This action on $\caR(\Lambda)$ is independent of the choice of forbidden cone $\widehat \Lambda$, so the extensions to $\caR(\widetilde \Lambda)$ for different cones are consistent with each other. These actions therefore define a *-homomorphism $\chi_\rho^D$ on all of $\caB$.

    Now consider some $O \in \cstar[\Lambda^c]^{\loc}$. Then there is a forbidden cone $\widehat \Lambda$, disjoint from $\Lambda$ and such that $O \in \caA_{\widehat \Lambda^c}$. Let $\mu_{\hat \rho}^D$ and $U$ be as above.
    We have 
    \begin{align}
        \chi_\rho^D(O) = U(O \otimes \I_n) U^* = U \mu_{\hat \rho}^D (O) U^* = \mu_\rho^D(O) = O \otimes \I_n.
    \end{align}
    Since this holds for any $O \in \caA_{\Lambda^c}^{\loc}$, we find that $\chi_\rho^D$ is localized in $\Lambda$.

    Now consider an allowed cone $\widetilde \Lambda$. Using transportability of $\mu_\rho^D$ (Lemma~\ref{lem:construction of transporters}) we have that there exists some $\mu_{\widetilde{\rho}}^D \simeq \mu_\rho^D$ localized in $\widetilde \Lambda$. Uniquely extend $\mu_{\widetilde\rho}^D$ to $\chi_{\widetilde \rho}^D$ as above. Then any unitary intertwiner from $\mu_{\widetilde \rho}^D$ to $\mu_{\rho}^D$ is an equivalence between $\chi_\rho^D$ and $\chi^D_{\widetilde\rho}$, showing that $\chi_\rho^D$ is indeed transportable. 
\end{proof}

This proposition allows the following definition.
\begin{defn} \label{def:amplis from ribbon operators}
    Let $\rho$ be a good half-infinite ribbon and $D$ a unitary representation of $\caD(G)$. Then we denote by $\chi_{\rho}^D$ the unique amplimorphism of $\caB$ that satsifies the properties of Proposition~\ref{prop:Amp is nonempty}.
\end{defn}

\begin{lem} \label{lem:sum and product of amplimorphism representations}
	For any good half-infinite ribbon $\rho$ supported in an allowed cone we have
	\begin{equation}
		\chi_{\rho}^{D_1} \oplus \chi_{\rho}^{D_2} = \chi_{\rho}^{D_1 \oplus D_2}, \quad \chi_{\rho}^{D_1} \times \chi_{\rho}^{D_2} = \chi_{\rho}^{D_1 \times D_2}. 
	\end{equation}
\end{lem}

\begin{proof}
	Follows immediately from Lemma \ref{lem:sum and product of ribbon amplimorphisms} and the uniqueness of the $\chi_{\rho}^D$ as extensions of the $\mu_{\rho}^D$.
\end{proof}

\subsection{Braided monoidal subcategory of \texorpdfstring{$\Amp$}{} on a fixed ribbon} \label{subsec:subcategory of amplimorphisms on a fixed ribbon}

We will call a half-infinite ribbon $\rho$ allowed if it is supported in some allowed cone. Let $\rho$ be a positive good allowed half-infinite ribbon and let $\Amp_{\rho}$ be the full subcategory of $\Amp$ whose objects are the localized and transportable amplimorphsisms $\chi_{\rho}^D$ for arbitrary unitary representations $D$. Lemma \ref{lem:sum and product of amplimorphism representations} shows that this subcategory is closed under direct sums and tensor products, so $\Amp_{\rho}$ is a full monoidal subcategory of $\Amp$. Being closed under the tensor product, the subcategory $\Amp_{\rho}$ inherits the braiding of $\Amp$ defined in Section \ref{subsec:braiding}. Finally, it follows from Proposition \ref{prop:T to t} below that $\Amp_{\rho}$ has subobjects, so it is in fact a full braided $\rm C^*$-tensor subcategory of $\Amp$.




\section{Simple objects of \texorpdfstring{$\Amp$}{Amp}} \label{sec:simples of Amp}

In the previous section we constructed full subcategories $\Amp_{\rho}$ of $\Amp$ whose objects are constructed from unitary representations of $\caD(G)$. These subcategories will play a crucial role in establishing the equivalence of $\Amp_f$ and $\Rep_f \caD(G)$. 

In order to do this we must first establish that the amplimorphisms $\chi_{\rho}^D$ are finite, so that they belong to $\Amp_f$. Then we must show that $\chi_{\rho}^D$ is a simple object whenever $D$ is an irreducible representation. Conversely, we must show that any simple object of $\Amp$ is equivalent to an amplimorphism $\chi^D_{\rho}$ for some irreducible representation $D$. In this section we prove these facts by appealing to the classification of irreducible anyon sectors of Kitaev's quantum double models achieved in \cite{bols2023classificationanyonsectorskitaevs}, which we first review.

\subsection{Classification of irreducible anyon sectors}

\begin{defn} \label{def:anyon representation}
    A *-representation $\pi : \caA \rightarrow \caB(\caH)$ is said to satisfy the superselection criterion with respect to the representation $\pi_0$ if for any cone $\Lambda$ there is a unitary $U : \caH_0 \rightarrow \caH$ such that
    \begin{equation*}
        \pi(O) = U \pi_0(O) U^*
    \end{equation*}
    for all $O \in \caA_{\Lambda^c}$. If $\pi$ is moreover irreducible, then we call $\pi$ an \emph{anyon representation}.
\end{defn}

The following theorem follows directly from Theorem 2.4 and Proposition 5.19 of \cite{bols2023classificationanyonsectorskitaevs}.
\begin{thm}[\cite{bols2023classificationanyonsectorskitaevs}] \label{thm:anyon classification}
    Let $\rho$ be a good half-infinite ribbon. The representations $\chi_{\rho}^D \circ \pi_0$ are anyon representations if and only if $D$ is irreducible. Two such anyon representations $\chi_{\rho}^{D_1} \circ \pi_0$ and $\chi_{\rho}^{D_2} \circ \pi_0$ are unitarily equivalent (disjoint) whenever the irreducible representations $D_1$ and $D_2$ are equivalent (disjoint).

    Moreover, any anyon representation $\pi$ is unitarily equivalent to $\chi_{\rho}^D \circ \pi_0$ for some irreducible representation $D$.
 \end{thm}

\subsection{Simple amplimorphisms}

Fix a good allowed half-infinite ribbon $\rho$.

\begin{prop} \label{prop:simple objects from irreducible representations}
    Let $D_1$ and $D_2$ be irreducible representation of $\caD(G)$. Then the amplimorphisms $\chi_{\rho}^{D_1}$ and $\chi_{\rho}^{D_2}$ are simple objects of $\Amp$. If they are equivalent, then the representations $D_1$ and $D_2$ must be equivalent.
\end{prop}

The converse to the second part, namely that $\chi_{\rho}^{D_1}$ and $\chi_{\rho}^{D_2}$ are equivalent if $D_1$ and $D_2$ are equivalent will be shown later in Proposition \ref{prop:t to T}. \\
 
\begin{proof}
    Suppose $\chi_{\rho}^{D_1}$ were not simple. Then there is a non-trivial orthogonal projector $p \in ( \chi_{\rho}^{D_1} | \chi_{\rho}^{D_1})$. Since $\chi_{\rho}^{D_1}$ is unital, this implies
    \begin{equation*}
        p \, \cdot \,  (\chi_{\rho}^{D_1} \circ \pi_0)(O) =  (\chi_{\rho}^{D_1} \circ \pi_0)(O) \, \cdot \,  p \quad \text{for all} \, O \in \caA.
    \end{equation*}
    But this shows that $p$ is in the commutant of the representation $\chi_{\rho}^{D_1} \circ \pi_0$. Since the latter representation is irreducible by Theorem \ref{thm:anyon classification}, $p$ cannot be a non-trivial projection. We conclude that $\chi_{\rho}^{D_1}$ is simple.

    Similarly, if $U \in (\chi_{\rho}^{D_2} | \chi_{\rho}^{D_1})$ is a unitary equivalence of unital amplimorphisms then $U$ is also a unitary intertwiner of representations $\chi_{\rho}^{D_1} \circ \pi_0$  and  $\chi_{\rho}^{D_2} \circ \pi_0$. By Theorem \ref{thm:anyon classification} such a $U$ can exists only if $D_1$ and $D_2$ are equivalent.
\end{proof}

\begin{prop} \label{prop:charcterization of simple amplis}
    Any simple object of $\Amp$ is equivalent to $\chi_{\rho}^D$ for some irreducible representation $D$.
\end{prop}

\begin{proof}
    Let $\chi$ be a simple amplimorphism of degree $n$. By Lemma \ref{lem:ampli is equivalent to endo} we can assume without loss of generality that $\chi$ is an endomorphism.
    
    Let us show that the *-representation $\chi \circ \pi_0 : \caA \rightarrow \caB(\caH_0)$ satisfies the superselection criterion, Definition \ref{def:anyon representation}. Let $\Lambda$ be a cone. By transportability there is an endomorphism $\chi' \in \DHR$ localized in $\Lambda^c$ such that $\chi \sim \chi'$. Let $U \in (\chi' | \chi)$ be a (necessarily unitary) equivalence. Then one has $(\chi \circ \pi_0)(O) = U^* \pi_0(O) U$ for any $O \in \caA_{\Lambda}$. Since $\Lambda$ was arbitrary, this shows that $\chi \circ \pi_0$ indeed statisfies the superselection criterion.

    We now use the assumption that $\chi$ is simple to show that $\chi \circ \pi_0$ is in fact an anyon representation. That is, we want to show that $\chi \circ \pi_0$ is irreducible. To obtain a contradiction, suppose $p \in \caB(\caH)$ is a non-trivial projection intertwining the representation $\chi \circ \pi_0$ with itself. Since commutation is preserved under weak limits, it follows that $p \in (\chi | \chi)$, contradicting simplicity of $\chi$. So $\chi \circ \pi_0$ is indeed an anyon representation.
    
    By Theorem \ref{thm:anyon classification} it follows that $\chi \circ \pi_0$ is unitarily equivalent as a $*$-representation of $\caA$ to $\chi_{\rho}^{D} \circ \pi_0$ for some irreducible representation $D$. Let $U$ be an intertwining unitary. It follows by continuity that in fact $U \in (\chi | \chi_{\rho}^D)$ is an equivalence of amplimorphisms, as required.
\end{proof}




\section{Equivalence of \texorpdfstring{$\Rep_f \caD(G)$}{RepD(G)}, \texorpdfstring{$\Amp_{\rho}$}{Amprho}, and \texorpdfstring{$\Amp_f$}{Ampf}} \label{sec:proof of main THM}

In this section we prove the remaining equivalences of categories needed to establish our main result, Theorem~\ref{thm:main result}.

\subsection{Equivalence of \texorpdfstring{$\Amp_{\rho}$}{} and \texorpdfstring{$\Rep_f \caD(G)$}{}} \label{subsec:Amp_rho and Rep}

Fix a good allowed half-infinite ribbon $\rho$. In this section we show that the category $\Amp_{\rho}$ introduced in Section \ref{subsec:subcategory of amplimorphisms on a fixed ribbon} is equivalent to $\Rep_f \caD(G)$, the category of finite dimensional unitary representations of $\caD(G)$.

\subsubsection{Monoidal equivalence} \label{subsec:monoidal equivalence}

Let us first show that for every intertwiner $t \in (D_1 | D_2 )$ of representations we can construct an intertwiner $T \in (\chi_{\rho}^{D_1} | \chi_{\rho}^{D_2})$ of amplimorphisms.

\begin{prop} \label{prop:t to T}
	If $t \in (D_1 | D_2)$ then $T := \1 \otimes t \in (\chi_{\rho}^{D_1} | \chi_{\rho}^{D_2})$.
\end{prop}
\begin{proof}
	For any $O \in \caA^{\loc}$ we have for all $n$ large enough (dropping $\pi_0$ from the notation)
    \[
    \begin{split}
		T \, \chi_{\rho}^{D_2}(O) &= T \, \mu_{\rho_n}^{D_2}(O) = (\1 \otimes t) \, \bF_{\rho_n}^{D_2} \, (O \otimes \I_n) \, (\bF_{\rho_n}^{D_2})^* \\
					  &= \bF_{\rho_n}^{D_1} \, (O \otimes \I_n) \, ( \bF_{\rho_n}^{D_1} )^* \, (\1 \otimes t) = \chi_{\rho}^{D_1}(O) \, T
    \end{split}
    \]
where we used item \ref{ribbon prop:intertwiners} of Proposition \ref{prop:properties of simple ribbon operators}. Let $\Lambda$ be an allowed cone containing $\rho$. Since $\cstar[\Lambda]^{\loc}$ is norm dense in $\cstar[\Lambda]$ which is in turn weakly dense in $\caR(\Lambda)$, using weak continuity of $\chi_\rho^{D_i}$ on cone algebras, this relation is true for all $O \in \caR(\Lambda)$. Since $\Lambda$ was an arbitrary allowed cone containing $\rho$, this relation holds for all $O \in \caB$. Thus $T \in (\chi_\rho^{D_1} | \chi_\rho^{D_2})$.
\end{proof}

Conversely, we want to show that all $T \in (\chi_{\rho}^{D_1} | \chi_{\rho}^{D_2})$ are of this form.
\begin{prop} \label{prop:T to t}
	If $T \in (\chi_{\rho}^{D_1} | \chi_{\rho}^{D_2})$ then $T = \1 \otimes t$ for some $t \in (D_1 | D_2)$. In particular, the amplimorphisms $\chi_{\rho}^{D}$ are finite so $\Amp_{\rho}$ is a full $\rm C^*$-subcategory of $\Amp_f$.
\end{prop}

\begin{proof}	
    Decompose $D_1$ and $D_2$ into direct sums of irreducibles (\cf Appendix \ref{app:introduction to D(G)}):
	\begin{equation}
		D_i \simeq \tilde D_i := \bigoplus_{r \in I} \, N_r^i \, \cdot \, D^{(r)},
	\end{equation}
	where $I$ is the finite set of equivalence classes of irreducible representations of $\caD(G)$ and $D^{(r)}$ is a representation in class $r$. Let $u_i \in ( D_i | \tilde D_i )$ be the unitaries implementing these equivalences. It follows from Proposition \ref{prop:t to T} that $U_i = (\1 \otimes u_i) \in (\chi_{\rho}^{D_i} | \chi_{\rho}^{\tilde D_i})$ and therefore $\tilde T := U_1^* T U_2 \in (\chi_{\rho}^{\tilde D_1} | \chi_{\rho}^{\tilde D_2})$.

    By Proposition \ref{prop:simple objects from irreducible representations},  $\{\chi_\rho^{D^{(r)}}\}_{r \in I}$ are disjoint simple objects of $\Amp_{\rho}$. Since the $\tilde D_i$ are direct sums of these it follows from Lemma \ref{lem:sum and product of amplimorphism representations} that the matrix blocks of $\tilde T$ mapping a $\chi_{\rho}^{D_r}$ subspace to a $\chi_{\rho}^{D_{r'}}$ are actually intertwiners of these amplimorphisms. It follows that the matrix blocks of $\tilde T$ corresponding to maps between copies of the same $\chi_{\rho}^{D_r}$ are multiples of the identity, and the other matrix blocks vanish, \ie $\tilde T = \1 \otimes \tilde t$ where
	\begin{equation}
		\tilde t = \bigoplus_{r} \tilde t_r \otimes \I_{n_r}
	\end{equation}
	with $\tilde t_r \in \Mat_{N^1_r \times N^2_r}(\C)$. Any such matrix $\tilde t$ belongs to $(\tilde D_1 | \tilde D_2)$. Since $u_i \in (D_i | \tilde D_i)$ it follows that $t = u_1 \tilde t u_2^* \in (D_1 | D_2)$. Now,
	\begin{equation}
		T = U_1 \tilde T U_2^* = (\1 \otimes u_1) (\1 \otimes \tilde t) (\1 \otimes u_2^*) = \1 \otimes t,
	\end{equation}
	which proves the claim.
\end{proof}

The two preceding propositions show that there is an isomorphsim between $(D_1 | D_2)$ and $( \chi_{\rho}^{D_1} | \chi_{\rho}^{D_2} )$ for all unitary representations $D_1, D_2$. We can use this isomorphisms to construct a monoidal equivalence between $\Rep_f \caD(G)$ and $\Amp_{\rho}$.

Consider the functor $F : \Rep_f \caD(G) \rightarrow \Amp_{\rho}$ which maps any unitary representation $D$ to the amplimorphism $\chi_{\rho}^{D}$, and maps any $t \in (D_1 | D_2)$ to $\1 \otimes t$. It follows from Proposition \ref{prop:t to T} that $F$ is indeed a functor. In fact, $F$ is linear and respects the $*$-structure. Moreover: 

\begin{prop} \label{prop:monoidal equivalence}
	The functor $F : \Rep_f \caD(G) \rightarrow \Amp_{\rho}$ is a monoidal equivalence. In particular, $\Rep_f \caD(G)$ and $\Amp_{\rho}$ are equivalent as $\rm C^*$-tensor categories.
\end{prop}

\begin{proof}
	Using Lemma \ref{lem:sum and product of amplimorphism representations} we find
	\begin{equation}
		F(D_1) \times F(D_2) = \chi_{\rho}^{D_1} \times \chi_{\rho}^{D_2} = \chi_{\rho}^{D_1 \times D_2} = F( D_1 \times D_2 ).
	\end{equation}
	Let $\id_{D_1, D_2} : F(D_1) \otimes F(D_2) \rightarrow F( D_1 \times D_2 )$ be the identity maps. Strict monoidality of $F$ means that the $\id_{D_1, D_2}$ form a natural transformation between functors $\times \circ (F, F) : \Rep_f \caD(G) \times \Rep_f \caD(G) \rightarrow \Amp_{\rho}$ and $F \circ \times :   \Rep_f \caD(G) \times \Rep_f \caD(G) \rightarrow \Amp_{\rho}$. Since  $\Amp_{\rho}$ is strict, this boils down to $F(t) \times F(t') = F(t \times t')$ for any $t \in (D_1 | D_2)$ and any $t' \in (D_1' | D_2')$, but this follows immediately from the definitions (recall in particular the definition in equation~\eqref{eq:tensor of intertwiners} of the tensor product of intertwiners of amplimorphisms).

	To see that $F$ is an equivalence of categories we note that $F$ is in fact an isomorphism, \ie $F$ is invertible with inverse $F^{-1}$ given on objects by $F^{-1}( \chi_{\rho}^D ) = D$ and on morphisms $T \in (\chi_{\rho}^{D_1} | \chi_{\rho}^{D_2})$ by $F^{-1}(T) = t$ with $t$ the unique intertwiner $t \in (D_1 | D_2)$ such that $T = \1 \otimes t$, \cf Proposition \ref{prop:T to t}.
\end{proof}

\subsubsection{Braided monoidal equivalence}

As remarked in Section \ref{subsec:subcategory of amplimorphisms on a fixed ribbon}, the subcategory $\Amp_{\rho}$ inherits the braiding of $\Amp$ defined in Section \ref{subsec:braiding}. Let us now compute the braiding between objects of $\Amp_{\rho}$ explicitly. 

\begin{figure}[!ht]
\centering
\includegraphics[width = 0.7\textwidth]{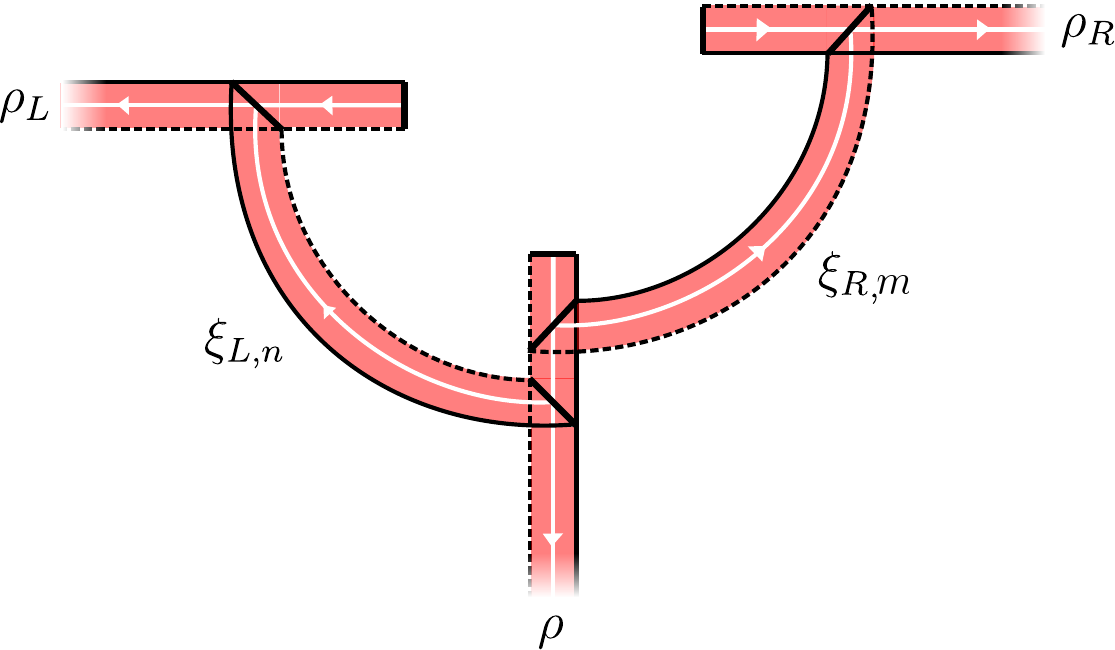}
\caption{The finite ribbon $\xi_{L,n}$ is a bridge from $\rho$ to $\rho_L$, and $\xi_{R,m}$ is a bridge from $\rho$ to $\rho_R$.}
\label{fig:explicit braiding}
\end{figure}

In order to compute $\ep( \chi_{\rho}^{D_1} , \chi_{\rho}^{D_2} )$ we fix good negative half-infinite ribbons $\rho_{L}$ and $\rho_R$ as in Figure \ref{fig:explicit braiding}. By the proof of Lemma \ref{lem:construction of transporters} there are unitaries $U \in (\chi_{\rho_R}^{D_1} | \chi_{\rho}^{D_1})$ and $V \in (\chi_{\rho_{L}}^{D_2} | \chi_{\rho}^{D_2})$ that are limits in the strong-* operator topology of unitary sequences $U_n = \bF_{\overline \sigma_{R, n}}^{D_1}$ and $V_n = \bF_{\overline \sigma_{L, n}}^{D_2}$ with ribbons $\sigma_{L, n} = \overline \rho_{L, n} \xi_{L, n} \rho_n$ and $\sigma_{R, n} = \overline \rho_{R, n} \xi_{R, n} \rho_n$ as in Figure \ref{fig:explicit braiding}, so the ribbons $\{ 
\xi_{L/R, n} \}$ are bridges from $\rho$ to $\rho_{L, R}$.

Let $\rho'_{L/R, n} = (\rho_{L/R})_{> n} \, \xi_{L/R, n} \, \rho_n$ and regard the unitaries $U_n$ and $V_n$ as intertwiners in $(\chi_{\rho_R}^{D_1} | \chi_{\rho'_{R, n}}^{D_1})$ and $(\chi_{\rho_L}^{D_2} | \chi_{\rho'_{L, n}}^{D_2})$ respectively, fix $l > 0$, and write $m = n + l$. Let $\zeta_{L/R, n} = \xi_{L/R, n} \overline \rho_{L/R, n}$ be such that $\sigma_{L, n} = \zeta_{L, n} \rho_n$ and $\sigma_{R, n} = \zeta_{R, n} \rho_n$.
Recall the braiding defined in equation~\eqref{eq:braiding of amplimorphisms defined}.
Noting that since all amplimorphisms are unital, the operator $P_{12}$ below used to define the braiding does not depend on $n$ or $m$, we have
\begin{align*}
	( V_{n+l}^* \times U_{n}^* )& \cdot P_{12} \cdot ( U_{n} \times V_{n+l} ) \\
         &= (V_m^* \otimes \1) \, \chi_{\rho_L}^{D_2}( U_n^* ) \cdot P_{12} \cdot  \chi_{\rho_R}^{D_1}( V_m ) \, (U_n \otimes \1) \\
             &= (V_m^* \otimes \1) (\1 \otimes U_n^*) \cdot P_{12} \cdot (\1 \otimes V_m) (U_n \otimes \1) \\
             &= ( \bF_{\rho_m}^{D_2} \otimes \1) ( \bF_{\zeta_{L, m}}^{D_2} \otimes \1 )  ( \1 \otimes \bF_{\rho_n}^{D_1}) ( \1 \otimes \bF_{\zeta_{R, n}}^{D_1} )  \cdot P_{12} \\
             & \quad\quad \cdot   ( \1 \otimes \bF_{\zeta_{L, m}}^{D_2} )^* ( \1 \otimes \bF_{\rho_m}^{D_2} )^* ( \bF_{\zeta_{R, n}}^{D_1} \otimes \1 )^* ( \bF_{\rho_n}^{D_1} \otimes \1 )^*\\
             \intertext{since $\rho_n$ is disjoint from the ribbons $\zeta_{L, n}$ and $\zeta_{R, n}$, and using item~\ref{ribbon prop:adjoint}  of Proposition \ref{prop:properties of simple ribbon operators} this becomes}
             &= ( \bF^{D_2}_{\rho_m} \otimes \bF^{D_1}_{\rho_n} ) ( \bF_{\bar \rho_m}^{D_2} \otimes \bF_{\bar \rho_n}^{D_1} ) \\
             & \quad \quad \cdot ( \bF_{\zeta_{L, m}}^{D_2} \otimes \bF_{\zeta_{R, n}}^{D_1} ) \cdot P_{12} \cdot ( \bF_{\zeta_{R, n}}^{D_1} \otimes \bF_{\zeta_{L, m}}^{D_2} )^* \\
             \intertext{using items \ref{ribbon prop:adjoint} and \ref{ribbon prop:sum and product} of Proposition \ref{prop:properties of simple ribbon operators} and unitarity, we get rid of the ribbon multiplets on $\rho_n, \rho_m$. The ribbons $\zeta_{R, n}$ and $\zeta_{L, m}$ are configured like the ribbons $\rho_1$ and $\rho_2$ of Figure \ref{fig:braiding positive ribbons} so we can apply item \ref{ribbon prop:braiding} of Proposition \ref{prop:properties of simple ribbon operators} to obtain}
			&= B(D_1, D_2).
\end{align*}

Since multiplication of operators is jointly continuous in the strong operator topology on bounded sets we have that
\begin{equation*}
    \ep( \chi_{\rho}^{D_1}, \chi_{\rho}^{D_2} ) = (V^* \times U^*) \cdot P_{12} \cdot (U \times V) = \lim_{n \uparrow \infty} ( V_{n+l}^* \times U_{n}^* ) \cdot P_{12} \cdot ( U_{n} \times V_{n+l} ).
\end{equation*}

We have thus shown
\begin{lem} \label{lem:explicit braiding}
	For any unitary representations $D_1$ and $D_2$ of $\caD(G)$ and any good positive half-infinite ribbon $\rho$ we have
	\begin{equation}
		\ep( \chi_{\rho}^{D_1}, \chi_{\rho}^{D_2} ) = B(D_1, D_2).
	\end{equation}
\end{lem}

The following proposition now follows immediately:

\begin{prop} \label{prop:braided equivalence of Amp_rho and Rep}
    The functor $F : \Rep_f \caD(G) \rightarrow \Amp_{\rho}$ is an equivalence of braided $\rm C^*$-tensor categories.
\end{prop}

\begin{proof}
    By Proposition \ref{prop:monoidal equivalence} it suffices to check
    \begin{equation}
	F( B(D_1, D_2) ) = \1 \otimes B(D_1, D_2) = \ep( \chi_{\rho}^{D_1}, \chi_{\rho}^{D_2}  ) = \ep( F(D_1), F(D_2) ).
    \end{equation}
    for any two unitary representations $D_1, D_2$, where we used Lemma \ref{lem:explicit braiding} in the second step.
\end{proof}

\subsection{Equivalence of \texorpdfstring{$\Amp_\rho$}{Amprho} and \texorpdfstring{$\Amp_f$}{Ampf}} \label{subsec:Amp_rho and Amp_f}

Let us first note that $\Amp_f$ is semisimple:
\begin{prop} \label{prop:amplimorphisms are direct sums of simples}
    Any amplimorphism $\chi \in \Amp_f$ is equivalent to a finite direct sum of irreducible amplimorphisms.
\end{prop}

\begin{proof}
    This follows immediately from Proposition \ref{prop:existence of subobjects for Amp} and the assumption that all objects of $\Amp_f$ are \emph{finite} amplimorphisms.
\end{proof}

\begin{prop}
\label{prop:equivalence of Amp_rho and Amp}
    The categories $\Amp_\rho$ and $\Amp_f$ are equivalent as $\rm C^*$-categories. In particular, $\Amp_f$ is closed under the tensor product of $\Amp$, so that $\Amp_f$ is a full braided $\rm C^*$-tensor subcategory of $\Amp$.
\end{prop}

\begin{proof}
    Recall Proposition \ref{prop:T to t} which shows that $\Amp_{\rho}$ is a full $\rm C^*$-subcategory of $\Amp_f$. Let us consider the functor $F : \Amp_{\rho} \rightarrow \Amp_f$ which embeds $\Amp_{\rho}$ into $\Amp_f$. We want to show that $F$ is an equivalence of $\rm C^*$-categories. Clearly, $F$ is linear, fully faithful, and respects the $*$-structure. The only thing that remains to be shown is that $F$ is essentially surjective, but this follows from Propositions \ref{prop:amplimorphisms are direct sums of simples} and \ref{prop:charcterization of simple amplis}.

    It follows that for any two amplimorphisms $\chi_1$ and $\chi_2$ of $\Amp_f$ there are representations $D_1$ and $D_2$ such that $\chi_1$ is equivalent to $\chi_{\rho}^{D_1}$ and $\chi_2$ is equivalent to $\chi_{\rho}^{D_2}$, and therefore $\chi_1 \times \chi_2$ is equivalent to $\chi_{\rho}^{D_1} \times \chi_{\rho}^{D_2} = \chi_{\rho}^{D_1 \times D_2}$ (see Lemma \ref{lem:sum and product of amplimorphism representations}). In particular, $\chi_1 \times \chi_2$ is finite (Proposition~\ref{prop:T to t}) and so $\Amp_f$ is closed under the tensor product. It is therefore a $\rm C^*$-tensor subcategory of $\Amp$, and inherits the braiding from $\Amp$.
\end{proof}

\begin{prop}
\label{prop:braided monoidal equivalence of Amp_rho and Amp}
    The categories $\Amp_\rho$ and $\Amp_f$ are equivalent as braided $\rm C^*$-tensor categories.
\end{prop}

\begin{proof}
    From Proposition \ref{prop:equivalence of Amp_rho and Amp} the embedding functor $F : \Amp_{\rho} \rightarrow \Amp_f$ is an equivalence of $\rm C^*$-categories, and $\Amp_{\rho}$ and $\Amp_f$ are braided $\rm C^*$-tensor subcateogries of $\Amp$. Clearly $F$ is monoidal and braided, which proves the claim.
\end{proof}

\subsection{Proof of Theorem \ref{thm:main result}}
 \label{subsec:proof of Theorem}

Before proving the main theorem, we must first establish that $\DHR_f$ is closed under the tensor product and therefore inherits the braided $\rm C^*$-tensor structure of $\DHR$.
\begin{lem} \label{lem:DHR_f is braided monoidal}
    The full subcategory $\DHR_f$ of $\DHR$ is closed under the tensor product. It is therefore a braided $\rm C^*$-tensor subcategory of $\DHR$ whith braiding inherited from $\DHR$.
\end{lem}

\begin{proof}
    Let $\nu_1$ and $\nu_2$ be endomorphisms belonging to $\DHR_f$. By Lemma \ref{lem:ampli is equivalent to endo} there are amplimorphisms $\chi_1$ and $\chi_2$ belonging to $\Amp$ such that $\nu_1$ is equivalent to $\chi_1$ and $\nu_2$ is equivalent to $\chi_2$. Moreover, since $\nu_1$ and $\nu_2$ are finite, so are $\chi_1$ and $\chi_2$. \ie $\chi_1$ and $\chi_2$ belong to $\Amp_f$. It follows that $\nu_1 \times \nu_2$ is equivalent to $\chi_1 \times \chi_2$, which is finite by Proposition \ref{prop:equivalence of Amp_rho and Amp}. This shows that $\nu_1 \times \nu_2$ is finite and so $\DHR_f$ is closed under the tensor product.
\end{proof}

We now proceed to prove our main result, Theorem \ref{thm:main result}, which we restate here for convenience.

\begin{thm}
    The categories $\Amp_f$, $\DHR_f$, and $\Rep_f \caD(G)$ are all equivalent as braided $\rm C^*$-tensor categories.
\end{thm}

\begin{proof}
    With Propositions \ref{prop:braided equivalence of Amp_rho and Rep} and \ref{prop:braided monoidal equivalence of Amp_rho and Amp} establishing the equivalence of $\Rep_f \caD(G)$ and $\Amp_f$, all that remains to be shown is the equivalence of $\DHR_f$ and $\Amp_f$ as braided $\rm C^*$-tensor categories.

    To see this, let $F : \DHR_f \rightarrow \Amp_f$ be the embedding functor. Clearly $F$ is linear, fully faithful, braided monoidal, and respects the $*$-structure. It remains to check that $F$ is essentially surjective, but this is immediate from Lemma \ref{lem:ampli is equivalent to endo}.
\end{proof}

\begin{rem}
As mentioned previously, we restrict to the category $\DHR_f$.
Since dualizable DHR endomorphisms are automatically finite (in our sense of the terminology) by~\cite{LongoRoberts97}, and all objects in the category $\Amp_f$ are dualizable, our results imply that the restriction of the category $\mathcal{O}_{\Lambda_0}$ (as defined by Ogata~\cite{Ogata2021}) to dualizable sectors (i.e., those who admit a conjugate) is precisely $\Rep_f \caD(G)$.
We do not expect that $\mathcal{O}_{\Lambda_0}$ has any objects which are not equivalent to (possibly infinite) direct sums of objects in $\Amp_f$.
For example, any simple direct summand of any such an object would be equivalent to a simple object in~$\Amp_f$.
\end{rem}




\section{Conclusions}

We explicitly characterized the category of anyon sectors for Kitaev's quantum double model for finite groups $G$.
As conjectured, the answer is that it is braided monoidally equivalent to $\Rep_f\,\caD(G)$.
This provides the first example where the category of anyon sectors is constructed explicitly for a model with non-abelian anyons.

The problem is tractable for the quantum double model largely because the Hamiltonian is of commuting projector type.
In general, we are interested in the whole quantum \emph{phase}.
The Hamiltonian of the quantum double model has a spectral gap in the thermodynamic limit, and roughly speaking another state is said to be in the same phase as the frustration free ground state $\omega_0$ of the quantum double model if they can be realised as ground states of a continuous path of gapped Hamiltonians.\footnote{Alternatively, it is possible to give a definition of a phase without referring to Hamiltonians at all, using e.g. finite depth quantum circuits or suitable locality preserving automorphisms.}
Using standard techniques (which we outline below) our results carry over to other states in the same gapped phase, which may no longer be ground states of a commuting projector Hamiltonian. 
One of the features of the quantum double model is that the physical features should be stable against small perturbations.
Indeed, the ground state has what is called local topological quantum order (LTQO)~\cite{Fiedler2014,Cui2020kitaevsquantum}.
This implies that sufficiently small local perturbations (even if applied throughout the system) do not close the spectral gap~\cite{michalakisz,brahami:2010}.

The result mentioned above implies that the ground states of the unperturbed and perturbed quantum double models can be related via an automorphism of $\alg{A}$ which is sufficiently local (meaning it satisfies a Lieb--Robinson type bound)~\cite{bachmannmns}.
Hence one can consider the \emph{phase} of a ground state as all states that can be connected via such a sufficiently local automorphism.
It turns out that the braided category of anyon sectors is an invariant of such a phase (that is, each state in the phase supports the same type of anyons).
This follows from the work of Ogata~\cite{Ogata2021} (see also~\cite{Ogata2021b} for a review), applied to the category $\DHR$ (or $\DHR_f$).
Alternatively, one can apply the approximation techniques developed there (necessary because one is forced to replace Haag duality by a weaker, approximate version) directly to the amplimorphisms constructed here.




\begin{subappendices}
\section{The quantum double of a finite group and its category of representations}
\label{app:introduction to D(G)}

Fix a finite group $G$. For any $g \in G$ we write $\bar g := g^{-1}$ for its inverse. We denote the unit of $G$ by $1 \in G$.
The quantum double algebra $\caD(G)$ of the finite group $G$ consists of formal $\C$-linear combinations of pairs of group elements $(g, h) \in G \times G$ equipped with product $\mu$, unit $\eta$, coproduct $\Delta$, counit $\ep$, and antipode $S$ defined by the linear extensions of
\begin{align*}
    \mu \big( (g_1, h_1), (g_2, h_2) \big) = \delta_{g_1, h_1 g_2 \bar h_1} (g_1, h_1 h_2), &\quad \Delta( g, h ) = \sum_{k \in G} (k, h) \otimes (\bar k g, h) \\
    \eta(1) = \sum_{k \in G} (k, 1), \quad \ep( g, h ) &= \delta_{g, 1} \quad S(g, h) = (\bar h \bar g h, \bar h),
\end{align*}
giving $\caD(G)$ the structure of a Hopf algebra. It is in fact a Hopf $*$-algebra with $(g, h)^* = (\bar h g h, \bar h)$, and is quasi-triangular with universal $R$-matrix
\begin{equation} \label{eq:R-matrix}
    R = \sum_{g, k \in G } (k, g) \otimes (g, 1).
\end{equation}

Let us recall some basic facts about the representation theory of $\caD(G)$ (see e.g.~\cite{Gould1993}) and establish notation. Denote by $\Rep_f \caD(G)$ the $\rm C^*$-category of finite dimensional unitary representations of $\caD(G)$, \ie representations $D$ such that $D(a^*) = D(a)^*$ for all $a \in \caD(G)$. We let $(D_2 | D_1)$ be the space of intertwiners from a representation $D_1$ to a representation $D_2$. We denote by $I$ the finite set of equivalence classes of irreducible representations and for each $i \in I$ we fix a representative $D^{(i)}$ from $i$. The algebra $\caD(G)$ is semisimple, from which it follows that any representation in $\Rep_f \caD(G)$ is equivalent to a direct sum of fintely many copies of the representatives $\{D^{(i)}\}_{i \in I}$.

The coproduct of $\caD(G)$ gives a monoidal product $\times$ of representations via
\[
(D_1 \times D_2)(a) := (D_1 \otimes D_2)(\Delta(a)),
\]
making $\Rep_f \caD(G)$ into a $\rm C^*$-tensor category. For $i, j \in I$ we have a unitary equivalence
\begin{equation*}
    D^{(i)} \times D^{(j)} \simeq \bigoplus_{k \in I} N_{ij}^k \cdot D^{(k)}
\end{equation*}
where the non-negative integers $N_{ij}^k$ stand for the multiplicity of $D^{(k)}$ in the direct sum.

The universal $R$-matrix of Eq. \eqref{eq:R-matrix} provides a braiding $B : \times \rightarrow \times^{\rm{op}}$ for $\Rep_f \caD(G)$ whose component maps are 
\begin{equation} \label{eq:braiding of Rep defined}
    B(D_1, D_2) := P_{12} \cdot (D_1 \times D_2) (R),
\end{equation}
where $P_{12}$ interchanges the factors in the tensor product of the representation spaces of $D_1$ and $D_2$. This makes $\Rep_f \caD(G)$ into a braided $\rm C^*$-tensor category.



\section{Ribbon operators}  \label{app:ribbon operators}
For the convenience of the reader, we recall the definition of ribbon operators and some of their properties, tailored to the triangular lattice we are using in this paper.
The material in this appendix is well-known, see e.g.~\cite{kitaev2003fault,Bombin2008,Yan2022} for more details.

\subsection{Triangles and ribbons}

Denote by $\Gamma^V, \Gamma^F$ the set of vertices and faces in $\Gamma$ respectively. An oriented edge $e \in \vec{\Gamma}^E$ may be identified with the pair of vertices $e = (v_0, v_1)$ where $v_0$ is the origin and $v_1$ the target of $e$. We write $\partial_0 e = v_0$ and $\partial_1 e = v_1$, and we have $\bar e = (v_1, v_0)$.

We say a vertex $v$ belongs to a face $f$ if $v$ sits on the boundary of $f$. A site $s = (v, f)$ is a pair of a vertex $v$ and a face $f$ such that $v$ belongs to $f$. We write $v(s) = v$ and $f(s) = f$.

Let $\bar\Gamma$ be the dual lattice to $\Gamma$. To each edge $e \in \Gamma^E$ we associate the oriented dual edge $e^*$ which crosses $e$ from right to left as follows : \includegraphics[width=0.6cm]{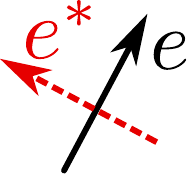}

A direct triangle $\tau = (s_0, s_1, e)$ consists of a pair of sites $s_0, s_1$ that share the same face, and an edge $e \in \Gamma^E$ connecting the vertices of $s_0$ and $s_1$. We write $\partial_0 \tau = s_0$ and $\partial_1 \tau = s_1$ for the initial and final sites of the direct triangle $\tau$, and $e_{\tau} = (v(s_0), v(s_1))$ for the oriented edge associated to $\tau$. The opposite triangle to $\tau$ is the direct triangle $\bar \tau = (s_1, s_0, e)$. Similarly, a dual triangle $\tau = (s_0, s_1, e)$ consists of a pair of sites $s_0, s_1$ that share the same vertex, and the edge $e$ whose associated dual edge $e^*$ connects the faces of $s_0$ and $s_1$. We write again $\partial_0 \tau = s_0$ and $\partial_1 \tau = s_1$, $e^*_{\tau} = (f(s_0), f(s_1))$ for the oriented dual edge associated to $\tau$, and define an opposite dual triangle $\bar \tau = (s_1, s_0, e)$.

A ribbon $\rho = \{ \tau_i \}_{i = 1}^l$ is a finite tuple of triangles such that $\partial_1 \tau_{i} = \partial_0 \tau_{i+1}$ for all $i = 1, \cdots, l-1$ and such that for each edge $e \in \Gamma^E$ there is at most one triangle $\tau_i$ for which $\tau_i = (\partial_0 \tau_i, \partial_1 \tau_i, e)$. We define $\partial_0 \rho = \partial_0 \tau_1$ and $\partial_1 \rho = \partial_1 \tau_l$. If $\rho$ consists of only direct triangles we say that $\rho$ is a direct ribbon, and if $\rho$ consists of only dual triangles we say ther $\rho$ is a dual ribbon. The empty ribbon is denoted by $\ep = \emptyset$.

\begin{figure}[!ht]
\centering
\includegraphics[width = 0.6\textwidth]{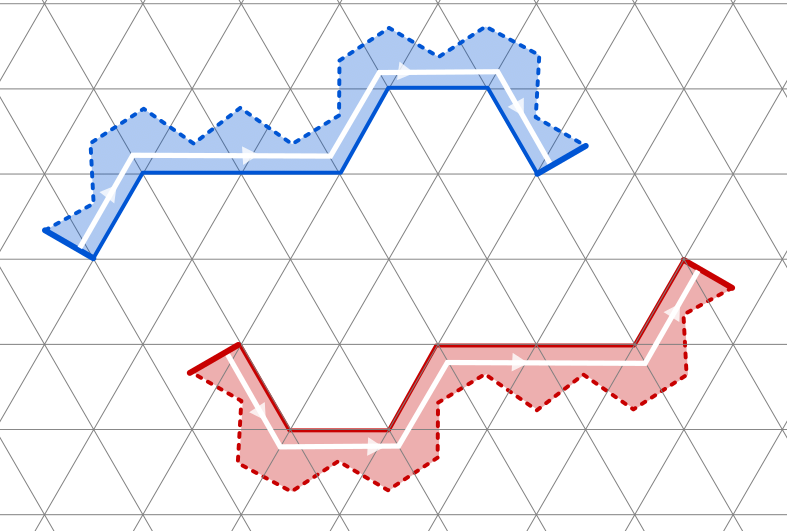}
\caption{An example of a positive ribbon (in red) and a negative ribbon (in blue).}
\label{fig:positive and negative ribbon}
\end{figure}

A ribbon is positively oriented (positive for short) if the sites of all its direct triangles lie to the right of their edges along the orientation of $\rho$ and vice versa for its dual triangles. The ribbon is negatively oriented (negative) otherwise, \cf Figure \ref{fig:positive and negative ribbon}. 

If we have two ribbons $\rho_1 = \{\tau_i\}_{i = 1}^{l_1}$ and $\rho_2 = \{ \tau_i \}_{i = l_1 + 1}^{l_1 + l_2}$ satisfying $\partial_1 \rho_1 = \partial_0 \rho_2$ then we can concatenate them to obtain a ribbon $\rho_1 \rho_2 = \{ \tau_i \}_{i = 1}^{l_1 + l_2}$. If $\rho_1$ and $\rho_2$ are non-empty then $\partial_0 \rho = \partial_0 \rho_1$ and $\partial_1 \rho = \partial_1 \rho_2$. The opposite ribbon to $\rho = \{\tau_i\}_{i = 1}^l$ is the ribbon $\bar \rho = \bar \tau_l \cdots \bar \tau_1$. If $\rho$ is positively oriented then $\bar \rho$ is negatively oriented and vice versa. The support of a ribbon $\rho = \{ \tau_i = (s_0^{(i)}, s_1^{(i)}, e_i)  \}_{i=1}^l$ is $\supp(\rho) := \{ e_i \}_{i = 1}^l.$

\subsection{Ribbon operators}

\subsubsection{Definitions and basic properties}

For each edge $e \in \Gamma^E$ we define the following operators on $\caH_e$:
\begin{equation}
	L_e^h := \sum_{g \in G} \, |hg \rangle \langle g|, \quad R_e^h := \sum_{g \in G} \, | g \bar h \rangle \langle g |, \quad T_e^g := | g \rangle \langle g|.
\end{equation}
The $L_e^h$ and $R_e^h$ are unitaries, implementing the left and right group action on $\caH_e$ respectively. The $T_e^g$ are projectors.

To each dual triangle $\tau = (s_0, s_1, e)$ we associate unitaries $L^h_{\tau}$ supported on the edge $e$ defined as follows. If $e^* = (f(s_0), f(s_1))$ and $v(s_0) = \partial_0 e$ then $L^h_{\tau} := L_e^h$. If $e^* = (f(s_0), f(s_1))$ and $v(s_0) = \partial_1 e$ then $L^h_{\tau} := R_e^{\bar h}$. If $e^* = (f(s_1), f(s_0))$ and $v(s_0) = \partial_0 e$ then $L^h_{\tau} := L_e^{\bar h}$. Finally, If $e^* = (f(s_1), f(s_0))$ and $v(s_0) = \partial_1 e$ then $L^h_{\tau} := R_e^h$. Similarly, to each direct triangle $\tau = (s_0, s_1, e)$ we associated projectors $T^g_\tau := T^g_{e}$ if $e = (v(s_0), v(s_1))$ and $T^g_{\tau} := T_e^{\bar g}$ if $e = (v(s_1), v(s_0))$.

To each finite ribbon $\rho$ we associate ribbon operators $F^{h, g}$ as follows. If $\ep$ is the trivial ribbon then $F_{\ep}^{h, g} := \delta_{g, 1} \I$. For ribbons composed of a single direct triangle $\tau$ we put $F_{\tau}^{h, g} = T_{\tau}^g$, and for ribbons composed of a single dual triangle $\tau$ we put $F_{\tau}^{h, g} = \delta_{g, 1} L_{\tau}^h$. The ribbon operators for general ribbons are defined inductively by the formula
\begin{equation}
\label{eq:ribbon_decomposition}
	F_{\rho_1 \rho_2}^{h, g} = \sum_{k \in G} \, F_{\rho_1}^{h, k} \, F_{\rho_2}^{\bar k h k, \bar k g}.
\end{equation}
The ribbon operator $F_{\rho}^{h, g}$ is supported on $\supp(\rho)$. Let us define
\begin{equation}
	T_{\rho}^g := F_{\rho}^{e, g}, \quad L_{\rho}^h := \sum_{g \in G} \, F_{\rho}^{h, g}.
\end{equation}
Then $F_{\rho}^{h, g} = L_{\rho}^h T_{\rho}^g = T_{\rho}^g L_{\rho}^h$. The multiplication and adjoint of ribbon operators is given by
\begin{equation} \label{eq:ribbon multiplication and adjoint}
	F_{\rho}^{h_1, g_1} \cdot F_{\rho}^{h_2, g_2} = \delta_{g_1, g_2}  F_{\rho}^{h_1 h_2, g_1}, \quad \big( F_{\rho}^{h, g} \big)^* = F_{\rho}^{\bar h, g},
\end{equation}
and reversing the orientation of a ribbon yields
\begin{equation} \label{eq:ribbon reversal}
	F_{\rho}^{h, g} = F_{\bar \rho}^{\bar g \bar h g, \bar g}.
\end{equation}
Note the natural appearance of the antipode of $\caD(G)$.

We also have the following property:
\begin{equation} \label{eq:sum to identity}
    \sum_k F_\rho^{e,k} = \I.
\end{equation}

If we have two positive ribbons $\rho_1$ and $\rho_2$ with common initial site as in Figure \ref{fig:braiding positive ribbons} then (\cf Eq.~(38) of \cite{kitaev2003fault}):
\begin{equation} \label{eq:braiding positive ribbons}
	F_{\rho_2}^{g_2, h_2} F_{\rho_1}^{g_1, h_1} = F_{\rho_1}^{g_1, h_1} F_{\rho_2}^{\bar g_1 g_2 g_1, \bar g_1 h_2}.
\end{equation}

\subsubsection{Gauge transformations and flux projectors} \label{subsubsec:gauge transformations and flux projections}

For any site $s$ there is a unique counterclockwise closed direct ribbon with end sites equal to $s$ which we denote by $\rho_{\triangle}(s)$. Similarly, there is a unique counterclockwise closed dual ribbon with end sites equal to $s$ which we denote by $\rho_{\star}(s)$. For any site $s$ we define gauge transformations $A_s^h$ and flux projectors $B_s^g$ by
\begin{equation} \label{eq:gauge transformations and flux projectors defined}
	A_s^h := L_{\rho_{\star}(s)}^{h, e}, \quad B_s^g := T_{\rho_{\triangle}(s)}^{e, g}.
\end{equation}

\begin{figure}[!ht]
    \centering
    \includegraphics[width=0.5\linewidth]{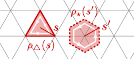}
    \caption{Example of $\rho_{\triangle}(s)$ and $\rho_\star(s')$.}
    \label{fig:enter-label}
\end{figure}

Let us define $U_s : \caD(G) \rightarrow \caA$ by
\begin{equation}
	U_s(g, h) := B_s^g A_s^h,
\end{equation}
extended linearly to $\caD(G)$. One easily checks that $U_s$ is an injective homomorphism of *-algebras, \ie the $B_s^{g} A_s^h$ span a representation of the quantum double algebra $\caD(G)$.

We note that the gauge transformations $A_s^h$ depend only on the vertex $v = v(s)$, so we may write $A_v^h := A_{s}^h$ for any site $s$ with $v = v(s)$. Moreover, the trivial flux projectors $B_s^e$ depend only on the face $f = f(s)$ so we write $B_f := B_s^e$ for any site $s$ with $f = f(s)$.

Finally, we define the projector onto states that are gauge invariant at the vertex $v$ by
\begin{equation}
	A_v := \frac{1}{\abs{G}} \sum_{h \in G} \, A_v^h.
\end{equation}
A straightforward calculation shows that this indeed is a projection.



\section{Convergence of transporters}

In this appendix we prove some technical lemmas needed to construct charge transporters.
The following Lemma, which we prove here for convenience, is well-known (c.f.~\cite[Prop. II.4.9]{TakesakiI}).

\begin{lem} \label{lem:abstract unitarity}
    Let $\alg{A} \subset \mathfrak{B}(\mathcal{H})$ be a $*$-algebra acting on some Hilbert space $\mathcal{H}$.
    Suppose that $\mathcal{H}_0 \subset \mathcal{H}$ is a dense subset of vectors.
    Let $U_\lambda \in \alg{A}$ be a uniformly bounded net such that for each $\xi \in \mathcal{H}_0$ both $U_\lambda \xi$ and $U_\lambda^* \xi$ converge in the norm topology of $\mathcal{H}$.
    Then $U_\lambda$ converges to some $U \in \alg{A}''$ in the strong-$*$ operator topology.
    If moreover each $U_\lambda$ is unitary, then the limit $U$ is unitary as well.
\end{lem}

\begin{proof}
    Choose $\epsilon > 0$ and $\xi \in \mathcal{H}$.
    Then there is $\xi_0 \in \mathcal{H}_0$ such that $\| \xi - \xi_0 \| < \epsilon$.
    By assumption, there is $M > 0$ such that $\| U_\lambda \| < M$ for all $\lambda$.
    From this we get
    \[
        \| (U_\lambda - U_\mu) \xi \| = \| (U_\lambda - U_\mu) (\xi-\xi_0)  + (U_\lambda-U_\mu) \xi_0\| \leq 2 M \epsilon + \| (U_\lambda-U_\mu) \xi_0 \|.
    \]
    Since $U_\lambda \xi_0$ converges by assumption, it follows that $U_\lambda \xi$ is Cauchy. We can therefore define $U \xi := \lim_\lambda U_\lambda \xi$.
    From the construction it is clear that $U$ is linear, and because $\| U_\lambda \|$ is uniformly bounded, it follows that $U$ is a bounded operator. A similar argument gives us an operator $\tilde{U}^*$, defined via $\tilde{U}^* \xi = \lim_\lambda U_\lambda^* \xi$.
    
    For all $\xi,\eta \in \mathcal{H}_0$ we have
    \[
    \begin{split}
        \left| \langle \eta, (U^* - \tilde{U}^*) \xi \rangle \right| 
        &= 
        \left| \langle \eta, (U^*-U_\lambda^*) \xi \rangle + \langle \eta, (U_\lambda^*-\tilde{U}^*) \xi \rangle \right| \\
        &\leq 
        \| (U-U_\lambda) \eta \| \|\xi\| + \| \eta\| \| (U_\lambda^*-\tilde{U}^*) \xi \|.
    \end{split}
    \]
    Since the right hand side tends to zero, it follows that $\tilde{U}^* = U^*$, and hence strong convergence of $U_\lambda \to U$ and $U_\lambda^* \to U^*$ gives that $U_\lambda \to U$ in the strong-$*$ operator topology.
    Since the ball of radius $M$ in $\alg{A}''$ is closed in the strong-$*$ operator topology, it follows that $u \in \alg{A}''$.
    
    Finally, suppose that the $U_\lambda$ are unitary.
    By strong-$*$ operator convergence, we have
    \[
        \left\| U \xi \right\| = \lim_\lambda \| U_\lambda \xi \| = \|\xi\|, \quad\quad\quad\quad  \| U^* \xi \| = \lim_\lambda \| U_\lambda^* \xi \| = \|\xi\|
    \]
    for all $\xi \in \mathcal{H}$.
    Hence both $U$ and $U^*$ are isometries, and it follows that $U$ is unitary.
\end{proof}
Note that we need to assume that \emph{both} $U_\lambda \xi$ and $U_\lambda^* \xi$ converge. Since the adjoint is not continuous with respect to the strong operator topology, one does not follow from the other.

\begin{lem} \label{lem:unitary transporters}
    Let $\rho_1$ be a half-infinite positive ribbon starting at the site $s_0$ and $\rho_2$ a half-infinite negative ribbon starting in $s_1$.
    Suppose that $\{\xi_n\}_{n \in \mathbb{N}}$ is a bridge from $\rho_1$ to $\rho_2$ in the sense of Definition~\ref{def:bridge}, and write $\sigma_n = \rho_{1,n} \xi_n \rho_{2,n}$ as in that definition.
    Finally, choose $g,h \in G$.
    Then $\pi_0(F_{\sigma_n}^{h,g})$ converges in the strong-* operator topology to an operator $F \in \pi_0(\alg{A})''$.
\end{lem}

\begin{proof}
    Recall that $(\pi_0, \mathcal{H}, \Omega)$ is the GNS representation for the frustration free ground state $\omega_0$ of the quantum double model. To ease the notation we omit $\pi_0$ on the operators.

    Let $A \in \caA^{\loc}$.
    Then there is some $k \in \mathbb{N}$ such that $\supp(A)^{+1} \cap \sigma_n$ is constant for all $n \geq k$, where the $+1$ superscript denotes a ``fattening'' of the set $\supp(A)$ by one site.
    For all $n \geq k$, write $\rho_{i,n\setminus k}$ for the (finite) ribbon $\rho_{i,n}$ with the first $k$ triangles removed, and define $\widehat{\xi}_n = \rho_{1,n \setminus k} \xi_n \rho_{2,n \setminus k}$.
    That is, $\sigma_n = \rho_{1,k} \widehat{\xi}_n \rho_{2,k}$.
    It follows from the choice of $k$ that $\supp(A)^{+1} \cap \widehat{\xi}_n = \emptyset$ for all $n \geq k$.
    Then, using the decomposition rule for ribbon operators, Eq.~\eqref{eq:ribbon_decomposition}, we have for all $n \geq k$ that
    \[
    \begin{split}
        F_{\sigma_n}^{h,g} A \Omega 
        &=
        \sum_{m_1, m_2 \in G} F_{\rho_{1,k}}^{h,m_1} F_{\widehat{\xi}_n}^{\bar m_1 h m_1, m_2}  F_{\rho_{2,k}}^{\overline{m_1 m_2} h (m_1 m_2), \overline{m_1 m_2} g} A \Omega \\
        &=
        \sum_{m_1, m_2 \in G} F_{\rho_{1,k}}^{h,m_1}   F_{\rho_{2,k}}^{\overline{m_1 m_2} h (m_1 m_2), \overline{m_1 m_2} g} A F_{\widehat{\xi}_n}^{ \bar m_1 h m_1, m_2} \Omega.
        \end{split}
    \]
    In the last step we used locality of the operators.
    Note that for $n,m \geq k$, the ribbons $\widehat{\xi}_n$ and $\widehat{\xi}_m$ have the same start and end points by construction.
    Since the action of ribbon operators on the ground state depends only on the endpoints of the ribbons (see e.g.~\cite{Bombin2008,HamdanThesis}) we have that $F_{\widehat{\xi_n}}^{\bar m_1 h m_1, m_2} \Omega = F_{\widehat{\xi_m}}^{\bar m_1 h m_1, m_2} \Omega$.
    It follows that the sequence $F_{\sigma_n}^{h,g} A \Omega$ converges in norm.
    Because the adjoint of a ribbon operator is again a ribbon operator (on the same ribbon, \cf \eqref{eq:ribbon multiplication and adjoint}), the argument above shows that $(F_{\sigma_n}^{h,g})^* A \Omega$ also converges in norm as $n \to \infty$.
    Note that for ribbon operators we have that $\| F_{\sigma_n}^{h,g} \| \leq 1$, regardless of $\sigma_n$.
    Hence by Lemma~\ref{lem:abstract unitarity}, the result follows.
\end{proof}

\end{subappendices}




\bibliographystyle{alpha}
\bibliography{category/category_refs}

%% file: main_symmetry_thesis.tex
\chapter{An Operator Algebraic Approach To Symmetry Defects And Fractionalization}
\label{chap:symmetry}
\chapterauthors{
\chapterauthor{Kyle Kawagoe}{Center for Quantum Information Science and Engineering, The Ohio State University, Columbus OH, 43210}
\chapterauthor{Siddharth Vadnerkar}{Department of Physics, University of California, Davis, Davis CA, 95616}
\chapterauthor{Daniel Wallick}{Department of Mathematics, The Ohio State University, Columbus OH, 43210}
}

This chapter is taken verbatim from \cite{kawagoe2024operator}, preprint available on ArXiv and under review in Communications in Mathematical Physics. Reprinted with permission from Kyle Kawagoe, Siddharth Vadnerkar, Daniel Wallick. Redistribution is allowed under the {copyright terms of this article} (\href{https://creativecommons.org/licenses/by-nc-sa/4.0/}{Creative Commons CC BY license}).  We first write a few words about the scope of this work. 

So far we've considered only systems without any global symmetry $G$. In the presence of an on-site symmetry $G$, there are many lattice systems where it's been demonstrated that one can ``break'' the symmetry along a path, leaving behind a symmetry domain wall. Domain walls, while interesting, still behave largely like topological phases, in that they have the same anyon category. However when one ``breaks'' the symmetry domain walls further, one can obtain somewhat arcane objects called symmetry defects. These objects are fairly commonplace in lattice systems and are sometimes called lattice disclinations. Symmetry defects act as ``sinks'' or ``sources'' for domain-walls, i.e, in that one can terminate domain-walls on symmetry defects. 

In the presence of symmetry defects (or simply defects), anyonic excitations can have much richer behaviour. The symmetry domain-walls can ``permute'' anyon labels and also act as a ``sink'' or ``source'' for individual anyons when ordinarily anyons can only be pair-created or annihilated. In addition, domain-walls are finitely transportable and thus have their own fusion and braiding structure. It is natural to ask whether there is a categorical structure behind the behaviour of defects.
Tthe correct algebraic language to describe symmetry defects is a $G$-crossed UMTC. The neutral component of this category recovers $\cC$, while the $g$-graded pieces describe symmetry defects and their crossed braiding with anyons. In parallel, fractionalization of symmetry on anyons appears as projective actions classified by group cohomology. This picture has been systematized in the category-theoretic literature \cite{PhysRevB.100.115147, drinfeld2010braided, MR3242743}.

Yet, importing this elegant description $G$-crossed BTC into the infinite-volume setting is highly nontrivial. Recall that in this setting, to recover the category of anyon sectors, one must propose a selection criterion. In fact, care must be taken to choose the right criterion that does not include spurious unphysical anyon sectors, and conversely does not miss legitimate ones. Some benefits of this criterion include (just like the case for anyon-selection criterion) stability under perturbations. And so it is a worthwhile question to ponder the existence of a defect-selection criterion in the infinite-volume setting.

At the time of publishing, the community was missing a suitable criterion that captures the story of symmetry defects. Since a symmetry enriched topological phase without any symmetry defects reduces to the usual anyon sector category, any proposed defect-selection criterion must subsume the anyon-seelction criterion.

Our paper has three main contributions. First is the proposal of a defect-selection criterion. We study the set of defect sectors (representations that satisfy the defect-selection criterion) and obtain that they form a $G$-crossed braided $\rmC^*$-tensor category. This approach heavily follows the path laid down by Ogata in the original derivation of the braided $\rmC^*$-tensor category of anyon sectors \cite{MR4362722}, and is also heavily inspired by the work of Müger in deriving a $G$-crossed braided $\rmC^*$-tensor category in $1+1D$ $G$-enriched rational CFTs \cite{MR2183964}.

Second, we work out many explicit lattice examples to demonstrate that our defect-selection criterion yields the correct category. In particular we work out the case of general $G$-SPTs and find that they form $\Gvec$, as well as a $\bbZ_2$-enriched Toric Code. To obtain these resuls, we also develop a practical algorithm for obtaining symmetry defects, which is of independent practical interest.

Thirdly, via the machinery developed here, we provide a route to compute the symmetry fractionalization data using purely bulk-manipulations, which at the time of publishing was an important open problem. Typically physicists rely on gauging the symmetry of a system, or the presence of a boundary to compute this data. So this work provides a practical toolkit to the working physicist.

\begin{chapterabstract}
We provide a superselection theory of symmetry defects in 2+1D symmetry enriched topological (SET) order in the infinite volume setting. For a finite symmetry group $G$ with a unitary on-site action, our formalism produces a $G$-crossed braided tensor category $\GSec$. This superselection theory is a direct generalization of the usual superselection theory of anyons, and thus is consistent with this standard analysis in the trivially graded component $\GSec_1$. This framework also gives us a completely rigorous understanding of symmetry fractionalization. To demonstrate the utility of our formalism, we compute $\GSec$ explicitly in both short-range and long-range entangled spin systems with symmetry and recover the relevant skeletal data.
\end{chapterabstract}

\minitoc

\section{Introduction}
Long range entangled topological orders in 2+1D are characterized by Unitary Modular Tensor Categories (UMTC) which arise from the superselection theory of their emergent anyons. In many cases, this physical proposition has been rigorously verified by using DHR theory from algebraic quantum field theory on infinite lattice models \cite{MR2804555, MR3426207, 2306.13762, 2310.19661}.
Interestingly, this story changes in the presence of a finite on-site symmetry group $G$. The landmark work \cite{PhysRevB.100.115147} gave a physical justification for why this classification is given by $G$-crossed braided categories for $G$-symmetry enriched topological (SET) order. SET models and symmetry fractionalization have been studied extensively in the physics literature \cite{CHEN20173,PhysRevB.65.165113,PhysRevB.74.174423, PhysRevB.94.235136}.
Despite the impact of this work, there is currently no rigorous  understanding of how these categories arise from a microscopic bulk analysis.
In particular, these SET models have not been studied before in the infinite volume setting. In this manuscript, we provide a complete formalism detailing how $G$-crossed braided fusion categories arise from a DHR style analysis of the symmetry defects of SET order. We also demonstrate our formalism in concrete examples.

The original DHR formalism comes from  \cite{MR297259, MR334742}, building on \cite{MR165864}.
It was constructed to describe continuous quantum field theory and uses finite regions of spacetime as its local regions.  
This work was later built on in \cite{MR660538} to describe states that are localized in spacelike cones instead of finite regions.  
This latter approach was then adopted to study topologically ordered quantum spin systems, starting with the Toric Code \cite{MR2804555, MR2956822, MR3135456}. 
These methods have been shown to be stable under perturbations \cite{MR4050095, MR4426734, MR4362722} and are thus an important step in understanding topological order in a model-independent way.
More recently, the DHR approach has been used to study anyons in the presence of a $U(1)$ symmetry \cite{2410.04736}. Our paper shares some aspects of their analysis, particularly in the construction of defects. However, many of their techniques and results are specific to $U(1)$ and thus not applicable to our setting since we focus on finite groups.
The DHR approach can also be generalized in the style of \cite{MR1231644}, as shown in \cite{2410.21454}.
Another DHR-inspired approach to topological order has been used in \cite{2304.00068, 2307.12552, MR4814524}.

Following in suit with these analyses, we consider a vacuum state $\omega_0$ of an SET and construct its GNS representation $\pi_0\colon\fA\rightarrow B(\cH)$ for the quasilocal algebra $\fA$. For each $g\in G$, we have a support preserving automorphism $\beta_g\in \Aut(\fA)$ which represents the symmetry action. We take our ground state to be symmetric in the sense that $\omega_0\circ\beta_g=\omega_0$. With such a ground state, we take inspiration from \cite{MR2183964} to define symmetry defect sectors. Physically, the analysis in \cite{MR2183964} should correspond to the boundary CFT at infinity surrounding the bulk SET which we study. We now state our main results. The main theoretical result is stated precisely in Theorem \ref{thm:main_result} and Corollary \ref{cor:anyon_equiv}.

\begin{thm_border}
The category of defect sectors with respect to $\pi_0$ is a $G$-crossed braided $\rmC^*$-tensor category whose trivially graded component is the braided tensor category of anyon sectors.
\end{thm_border}

This mirrors the prediction of \cite{PhysRevB.100.115147} in the operator algebraic setting. 
As this theorem suggests, their higher cohomological obstructions do not appear since we are considering strictly onsite symmetry in 2+1 dimensions.

We then demonstrate the utility of our formalism by computing this category in a variety of examples exhibiting both short-ranged and long-ranged order.

This applies to a broad class of SPTs. We show that under certain physically reasonable assumptions (Assumptions \ref{asmp:entangler commutes with symmetry}, \ref{asmp:strip_aut_is_an_FDQC}), the defect sectors of a $G$-SPT (Definition \ref{def:SPT_phase}) are $G$-graded monoidally equivalent to $\Vect{(G, \nu)}$ where $\nu$ is a 3-cocycle (precisely stated in Theorem \ref{thm:SPT theorem}). We comment that the analysis done by \cite{MR4354127} is similar in spirit to our defect construction technique, except their analysis is done with much weaker assumptions and is thus more general. However, our construction generalizes nicely to models with long-range order as we demonstrate later.

We then specialise to the case of the Levin-Gu SPT, which is an example of a non-trivial $\bbZ_2$-SPT. We compute the category of defect sectors for this model and its skeletal data in the bulk and obtain the following result (Theorem \ref{thm:category_of_Levin_Gu}).
Since SPT phases in 2+1D are conjectured to be completely determined by a 3-cocycle, we do not compute the braiding data.
However, it is certainly possible to compute the braiding data in these models with our formalism.

\begin{thm_border}
    The category of defect sectors of the Levin-Gu SPT is $\mathbb{Z}_2$-graded monoidally equivalent to $\Vect(\bbZ_2,\nu)$, where the 3-cocycle $\nu\colon \bbZ_2 \times \bbZ_2 \times \bbZ_2 \rightarrow U(1)$ represents the non-trivial element $[\nu]\in H^3(\mathbb{Z}_2,U(1))$.
\end{thm_border}

In addition to \cite{MR2804555, MR2956822, MR3135456}, there are several other works providing complete superselection analyses of infinite lattice models, such as \cite{MR3426207} which studies the abelian Quantum Double Model.  
More recently, these methods have been applied to the doubled semion model \cite{2306.13762} and the nonabelian Quantum Double Model \cite{2310.19661}.
A general treatment of this approach to topological order, using much weaker assumptions that we use in this paper, can be found in \cite{MR4362722}.

One of the main contributions of our work is a complete defect supeselection theory analysis of a symmetry enriched model of the Toric Code. This model is defined in Section \ref{sec:SET Toric Code}. We compute the category of defect sectors of this model and analyze the resulting skeletal data. The result below is the conclusion of our analysis and precisely stated in Theorem \ref{thm:SET_Toric_Code}.

\begin{thm_border}
    The defect sectors of the symmetry-enriched toric code model in Section \ref{sec:SET Toric Code} forms a $G$-crossed braided fusion category with trivial associators but non-trivial fractionalization data.
\end{thm_border}

After posting the preprint of this manuscript, we were made aware that another research group was nearing completion of a work covering many of the same topics. We encourage the interested reader to also check out \cite{2411.01210}.

This manuscript is organized as follows. We first propose a selection criterion to select relevant representations having symmetry defects, called defect sectors (Definition \ref{def:defect_sector}). In Section \ref{sec:symmetry_defects}, we then build a category of these defect sectors and show that it is a $G$-crossed braided tensor category. We then show in Section \ref{sec:coherance_data} that the coherence data of this category matches that of the algebraic theory of symmetry defects already discussed in the literature. We also show in Section \ref{sec:connection_to_anyon_sectors} that when the symmetry is trivial, this selection criterion reduces to that of anyon sectors and is thus a strict generalization.

In Section \ref{sec:general SPTs}, we treat the case of a general SPT built from a finite depth quantum circuit (FDQC), and show that there always exists a commuting projector Hamiltonian whose ground state houses a symmetry defect. As a capstone to this treatment, we verify that the category of defect sectors for any such SPT is equivalent to $\Vect(G,\nu)$ where $G$ is the underlying symmetry of the SPT state and $\nu$ is a 3-cocycle.

We specialize the treatment of Section \ref{sec:general SPTs} to $\bbZ_2$-SPTs in Sections \ref{sec:TrivialParamagnet} and \ref{sec:Levin Gu}.
In Section \ref{sec:TrivialParamagnet}, we analyze the trivial $\bbZ_2$-paramagnet and show that the category of defect sectors is equivalent to $\Vect(\bbZ_2)$.
Similarly, in Section \ref{sec:Levin Gu} we verify for the Levin-Gu SPT that the resulting category of defect sectors is equivalent to $\Vect(\bbZ_2,\nu)$ for a non-trivial cocycle $\nu$. 
In particular, we explicitly computing $\nu$ in the bulk using automorphisms that create a symmetry defect.

In Section \ref{sec:SET Toric Code}, we explore an SET commuting projector model that is obtained from the usual Toric Code using a FDQC. This model has non-trivial symmetry fractionalization data and is thus an excellent test chamber for our defect selection criterion. We find a completeness result for the defect sectors of this model and explicitly compute the $F,R$-symbols and the symmetry fractionalization data in the bulk.

We briefly comment the setting and assumptions for this paper in order to summarize the main results. If the reader is not already familiar with the operator algebra formalism, a brief introduction is provided in Appendix \ref{sec:background}.

\section{Setting and main results}
\begin{figure}[!htbp]
        \centering
\begin{subfigure}[tb]{0.4\textwidth}
\centering
    \includegraphics[width=0.5\linewidth]{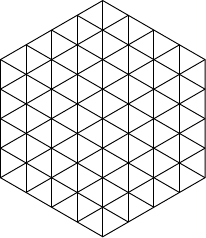}
        \caption{A portion of an infinite triangular lattice. Each vertex in the figure is a site.}
            \label{fig:example_triangular_lattice}
\end{subfigure}
\hfill
\begin{subfigure}[tb]{0.4\textwidth}
\centering
    \includegraphics[width=0.5\linewidth]{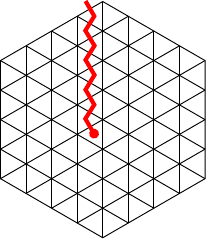}
    \caption{The chosen half-infinite dual-path $\bar \gamma_R$ on the triangular lattice. 
    }
    \label{fig:chosen_ray}
\end{subfigure}
\end{figure}

Let $\Gamma$ be a 2d cell complex consisting of vertices, edges, faces and equip the vertices in $\Gamma$ with the graph distance. For the examples we have in mind, we will often consider $\Gamma$ to be a regular lattice like the triangular lattice or the square lattice. An example is shown in Figure \ref{fig:example_triangular_lattice}.

Given a subset $\Sigma \subset \Gamma$, we denote by $\Sigma^c \subset \Gamma$ the complement of $\Sigma$, given by $\Sigma \cap \Sigma^c = \emptyset$ and $\Sigma \cup \Sigma^c = \Gamma$.

A \emph{cone} $\Delta \subseteq \bbR^2$ is a subset of the form 
\[
\Delta
\coloneqq
\{x \in \bbR^2 : (x - a) \cdot \hat{v}/2 > \|x - a\| \cos(\theta/2)\}.
\]
Here $a \in \bbR^2$ is the vertex of the cone, $\hat{v} \in \bbR^2$ is a unit vector specifying the axis of the cone, and $\theta \in (0, 2\pi)$ is the opening angle of the cone. We define a \emph{cone in $\Gamma$} to be a subset $\Lambda \subseteq \Gamma$ of the form $\Lambda = \Gamma \cap \Delta$, where $\Delta \subseteq \bbR^2$ is a cone.

We use the general term `site' to refer to a vertex, edge, or face. Associate a Hilbert space $\hilb_s = \bbC^{d_s}$ to each site $s \in \Gamma$, where $d_s \in \bbN$.
Let $\Gamma_f$ be the set of finite subsets of $\Gamma$. 
We can then define the tensor product over a finite set of sites $S \in \Gamma_f$ as $\hilb_{S} \coloneqq \bigotimes_{s \in S} \hilb_s$. Then $\cstar[S] \coloneqq B(\hilb_S)$ is a $C^*$-algebra.

Now let $S,S' \in \Gamma_f$ be such that $S \subset S'$. Then we can define the canonical inclusion $\cstar[S] \hookrightarrow \cstar[S']$ by tensoring with the identity element on all $s \in S' \setminus S$. With this we can define the algebra of local observables $\cstar[\loc]$ as $$\cstar[\loc] \coloneqq \bigcup_{S \in \Gamma_f} \cstar[S]$$ and its norm completion, $$\cstar \coloneqq \overline{\cstar[\loc]}^{||\cdot ||}$$ This algebra is known as the algebra of quasi-local observables, or simply, the \emph{quasi-local algebra}.

We assume that there is a symmetry action of a group $G$ on $\cstar$, i.e, a faithful homomorphism $\beta \colon G\to \Aut(\cstar)$ given by $g \mapsto \beta_g$ for all $g \in G$.
We call $\beta_g$ a \emph{symmetry automorphism}. In the cases we consider, the symmetry action is \emph{on-site}, i.e, for each $s \in \Gamma$, we assume that there is an action of $G$ on each $\hilb_s$ by unitaries $U^g_s$ acting on the site $s$. 

\begin{defn}
\label{def:GlobalSymmetryAutomorphism}
For each $A \in \cstar[S]$ with $S \in \Gamma_f$, we let $\beta_g\colon \cstar[S] \rightarrow \cstar[S]$ be the map defined by $$\beta_g (A) \coloneqq \left(\bigotimes_{s \in S} U_s^g\right) A \left(\bigotimes_{s \in S} U_s^{g}\right)^*.$$ 
We observe that this map can be uniquely extended in a norm continuous way to an automorphism $\beta_g$ acting on the whole of $\cstar$.
\end{defn}

We also sometimes consider situations where the symmetry only acts on a subset of the lattice. 

\begin{defn}
\label{def:SymmetryOnSubsets}
For any $S \subseteq \Gamma$, we let $\beta_g^S \colon \fA \to \fA$ be the map defined by 
\[
\beta_g^S(A)
\coloneqq
\left(\bigotimes_{s \in S} U_s^g\right) A \left(\bigotimes_{s \in S} U_s^g\right)^*,
\]
More precisely, one constructs $\beta_g^S \colon \fA \to \fA$ using the method used to construct $\beta_g$. An example symmetry action is shown in Figure \ref{fig:example_symmetry_action}.

\begin{figure}[!ht]
    \centering
    \includegraphics[width=0.3\linewidth]{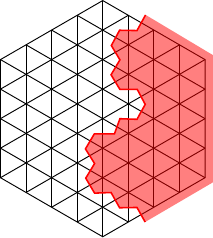}
    \caption{An example symmetry action $\beta_g^S$ on the triangular lattice, with $S$ being the region colored in red. On all sites $s$ in the red region, the symmetry acts as $U^g_s$, and $\mathds{1}_s$ otherwise.}
    \label{fig:example_symmetry_action}
\end{figure}
\end{defn}

\subsection{General Assumptions}
\label{sec:GCrossedAssumptions}

Fix a reference state $\omega_0$ and denote $\pi_0: \cstar \rightarrow B(\hilb_0)$ as its GNS representation. We now detail the assumptions that we will impose on the action by the group $G$ and on the state $\omega_0$ to ensure we obtain a $G$-crossed braided monoidal category.  

\begin{asmp}
\label{asmp:Faithfulness}
There is a fixed $n > 0$ such that for all balls $B \subseteq \Gamma$ of radius $n$, the representation of $G$ given by $g \mapsto \bigotimes_{s \in B} U^g_s$ is faithful.
\end{asmp}

Note that the faithfulness assumption implies that if $\beta_g|_{\cstar[B]} = \beta_h|_{\cstar[B]}$ for any large enough finite region $B \subseteq \Gamma$, then $g = h$.
Here $\beta_g \colon \fA \to \fA$ is the symmetry automorphism from Definition \ref{def:GlobalSymmetryAutomorphism}.

We now detail our assumptions on the chosen state $\omega_0 \colon \fA \to \bbC$.

\begin{asmp}
\label{asmp:GInvariance}
The state $\omega_0$ is \emph{$G$-invariant}, that is, $\omega_0 \circ \beta_g = \omega_0$ for all $g \in G$.
\end{asmp}

We observe that $G$-invariance of $\omega_0$ implies that the map $\beta_g \colon \fA \to \fA$ is implemented by a unitary in $B(\cH_0)$, where $\cH_0$ is the usual GNS Hilbert space for $\omega_0$.  
Hence $\beta_g$ extends to a WOT-continuous automorphism of $B(\cH_0)$.

We will also assume that the GNS representation $\pi_0$ for $\omega_0$ satisfies a generalization of Haag duality called \emph{bounded spread Haag duality} \cite[Def.~5.2]{2410.21454}.
This definition is analogous to the definition of weak algebraic Haag duality in \cite{2304.00068} and is stronger than the notion of approximate Haag duality used in \cite{MR4362722}.  

\begin{nota}
Let $\Lambda \subset \Gamma$. We denote by $\Lambda^{+r}$ is the set of points at most distance $r$ away from $\Lambda$.  
\end{nota}

\begin{defn}[{\cite[Def.~5.2]{2410.21454}}]
\label{def:BSHaagDuality}
Let $\pi \colon \fA \to B(\cH)$ be a representation.  
We say that $\pi$ satisfies \emph{bounded spread Haag duality} if there exists a global constant $r \geq 0$ such that for every cone $\Lambda \in \cL$, 
\[
\pi(\cstar[\Lambda^c])'
\subseteq
\pi(\cstar[\Lambda^{+r}])''.
\]
\end{defn}

\begin{asmp}
\label{asmp:BoundedSpreadHaagDuality}
The GNS representation $\pi_0$ for $\omega_0$ satisfies bounded spread Haag duality.
\end{asmp}

\begin{asmp}
\label{asmp:PureState}
The state $\omega_0$ is a pure state.
\end{asmp}

Note that $\omega_0$ being a pure state ensures that the cone algebras $\cR(\Lambda) \coloneqq \pi_0(\cstar[\Lambda])''$ are all factors.
We actually use a stronger assumption.  

\begin{asmp}
\label{asmp:InfiniteFactor}
For every cone $\Lambda$, the algebra $\cR(\Lambda)$ is an infinite factor.  
\end{asmp}

There are various reasonable assumptions on $\omega_0$ that ensure that the cone algebras are infinite, given that $\omega_0$ is pure.
For example, this holds when $\omega_0$ is translation invariant by using a standard argument \cite{MR2281418, MR2804555}.  
This also holds when $\omega_0$ is a gapped ground state of a Hamiltonian with uniformly bounded finite range interactions \cite[Lem.~5.3]{MR4362722}.

A (self-avoiding) \emph{finite  path} $\gamma \subset \Gamma$ is defined as a set of distinct edges $\{e_i \in \Gamma \}_{i=1}^N$ such that for all $i >1$, $e_i \cap e_{i-1}$ contains a single vertex. We call $\partial_0  \gamma \coloneqq \partial_0 e_1$ as the start of $ \gamma$ and $\partial_1  \gamma \coloneqq \partial_1 e_N$ as the end of $\gamma$. A \emph{half-infinite} path is a sequence of finite paths $\{\gamma_i\}$ such that $\gamma_i \subset \gamma_{i+1}$ and $\gamma_i$ all have a common start or end.

A dual path is a path on $\bar \Gamma$, the lattice dual to $\Gamma$ (c.f.~Section \ref{sec:cones}). We denote $e \in \bar \gamma$ for some edge $e \in \Gamma$ if $\bar e \in \bar \gamma$, where $\bar e$ is the dual edge to $e$.

Denote by $P (\Gamma)$ ( resp.~$\bar P(\Gamma)$) the collection of paths (resp.~dual paths) that are sufficiently nice, meaning roughly the path converges to a ray as it goes to infinity (cf.~discussion in Section \ref{sec:cones}).

Fix a half-infinite dual path $\bar \gamma_R \in \bar P(\Gamma)$ going straight up as shown in Figure \ref{fig:chosen_ray} for triangular lattices. An analogous ray can be drawn for square lattices. 
More general paths can be chosen as our reference path, c.f.~the discussion in Section \ref{sec:cones} for a general definition of allowed paths and how that modifies the definition of an allowed cone.
\begin{defn}
\label{def:allowed_cone}
We say that a cone $\Lambda \subseteq \Gamma$ is \emph{allowed} if for every translation $\Delta$ of $\Lambda$, $\bar \gamma_R \cap \Delta$ is finite. We take $\cL$ to be the set of allowed cones.
\end{defn}

\begin{defn}
\label{def:r(Lambda)}
For a cone $\Lambda \in \cL$, we define $r(\Lambda) \subseteq \Lambda^c$ to be the infinite region to the right of the path bounded by $\bar \gamma_R$ and the bounding rays of $\Lambda$. 
In the case where these two paths do not intersect, we connect them by the line segment connecting the endpoint of $\bar \gamma_R$ and the apex of $\Lambda$. 
By `to the right' of this path, we mean that $r(\Lambda)$ is a connected region just clockwise of $\bar \gamma_R$.
We define $\ell(\Lambda) \subseteq \Lambda^c$ to be $\Lambda^c \setminus r(\Lambda)$.  

An illustration of $r(\Lambda)$ and $\ell(\Lambda)$ for three cases is shown in Figure \ref{fig:defining_symmetry_action_to_the_right_of_cone}.
In each picture, $r(\Lambda)$ is the shaded region in $\Lambda^c$, and $\ell(\Lambda)$ is the unshaded region.
\begin{figure}[!ht]
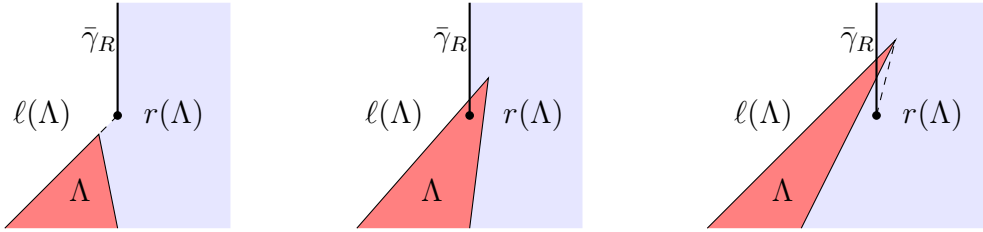

    \centering
    \begin{equation*}
\tikzmath{
\fill[fill=blue!10] (1.5, -1.5) -- (0, -1.5) -- (-.25, -.25) -- (0, 0) -- (0, 1.5) -- (1.5, 1.5);
\fill[fill=red!50] (-1.5, -1.5) -- (-0.25, -0.25) -- (0, -1.5);
\draw (-1.5, -1.5) -- (-0.25, -0.25) -- (0, -1.5);
\draw[thick] (0, 0) -- (0, 1.5);
\filldraw(0, 0) circle (0.05cm);
\draw[dashed] (0, 0) -- (-0.25, -0.25);
\node at (-0.5, -1) {$\Lambda$};
\node at (-.25, 1) {$\bar \gamma_R$};
\node at (.75, 0) {$r(\Lambda)$};
\node at (-1, 0) {$\ell(\Lambda)$};
}
\qquad\qquad
\tikzmath{
\fill[fill=blue!10] (1.5, -1.5) -- (0, -1.5) -- (-.25, -.25) -- (0, 0) -- (0, 1.5) -- (1.5, 1.5);
\fill[fill=red!50] (-1.5, -1.5) -- (0.25, 0.5) -- (0, -1.5);
\draw (-1.5, -1.5) -- (0.25, .5) -- (0, -1.5);
\draw[thick] (0, 0) -- (0, 1.5);
\filldraw(0, 0) circle (0.05cm);
\node at (-0.5, -1) {$\Lambda$};
\node at (-.25, 1) {$\bar \gamma_R$};
\node at (.85, 0) {$r(\Lambda)$};
\node at (-1, 0) {$\ell(\Lambda)$};
}
\qquad\qquad
\tikzmath{
\fill[fill=blue!10] (1.5, -1.5) -- (-1, -1.5)-- (.25, 1) -- (0, .75) -- (0, 1.5) -- (1.5, 1.5);
\fill[fill=red!50] (-2.25, -1.5) -- (0.25, 1) -- (-1, -1.5);
\draw (-2.25, -1.5) -- (0.25, 1) -- (-1, -1.5);
\draw[thick] (0, 0) -- (0, 1.5);
\filldraw(0, 0) circle (0.05cm);
\draw[dashed] (0, 0) -- (0.25, 1);
\node at (-1.25, -1) {$\Lambda$};
\node at (-.25, 1) {$\bar \gamma_R$};
\node at (.75, 0) {$r(\Lambda)$};
\node at (-1.5, 0) {$\ell(\Lambda)$};
}
    \end{equation*}
    \caption{Defining the symmetry action $\beta_g^{r(\Lambda)}$ for different cones.}
    \label{fig:defining_symmetry_action_to_the_right_of_cone}
\end{figure}
\end{defn}

\begin{defn}
\label{def:defect_sector}
    Let $\pi\colon \fA\rightarrow B(\cH_0)$ be a representation. We say that $\pi$ is \emph{$g$-localized} in a cone $\Lambda \in \cL$ if 
    $$
    \pi|_{\cstar[\Lambda^c]}=\pi_0 \circ \mu \circ \beta_g^{r(\Lambda)}|_{\cstar[\Lambda^c]},
    $$
    where $\mu = \Ad(\bigotimes_{s \in S} U_s^{g_s})$ for some $S \in \Gamma_f$.
    If $\mu = \id$, the identity automorphism, we say that $\pi$ is \emph{canonically $g$-localized}.
    We say that a $g$-localized representation $\pi$ is \emph{transportable} if for every cone $\Delta \in \cL$, there exists $\pi' \colon \fA \to B(\cH_0)$ such that $\pi' \simeq \pi$ and $\pi$ is $g$-localized in $\Delta$.
\end{defn}

\begin{rem}
    Note that if $\Lambda_1 \subseteq \Lambda_2$ and $\pi$ is $g$-localized in $\Lambda_1$, then $\pi$ is $g$-localized in $\Lambda_2$. 
    However, if $g \neq e$, then this does not hold if $g$-localized is replaced by canonically $g$-localized. 
    If $g = e$, then the definition of canonically $g$-localized recovers the definition of localized endomorphism used in \cite{MR297259, MR660538, MR2804555}, where it is true that if $\Lambda_1 \subseteq \Lambda_2$ and $\pi$ is localized in $\Lambda_1$, then $\pi$ is localized in $\Lambda_2$. 
\end{rem}

\begin{defn}
\label{def:g-defect_sector}
Let $\pi \colon \fA \to B(\cH_0)$ be a representation.
We say that $\pi$ is a \emph{$g$-defect sector} if it is $g$-localized and transportable.  
\end{defn}

\subsection{Main results}
We are now ready to state the main theoretical result of this paper. Recall Definitions \ref{def:G-crossed monoidal} and \ref{def:G-crossed braided} of a $G$-crossed braided tensor category\footnote{By \emph{tensor category} we mean a linear monoidal category that admits direct sums and subobjects.}.

\begin{thm}
\label{thm:main_result}
The category of defect sectors with respect to $\pi_0$ is a $G$-crossed braided $\rmC^*$-tensor category.
\end{thm}

We use $\rmC^*$-tensor category in the sense of \cite[Def.~2.1.1]{MR3204665}, following \cite{MR4362722}.
The dagger operation on morphisms is the usual adjoint in $B(\cH_0)$.
The fact that this category is a $G$-crossed braided tensor category is proved in parts using the results of Propositions \ref{prop:GCrossedMonoidal} and \ref{prop:GCrossedBraiding} and the discussion in Section \ref{sec:GSecCauchyComplete}.
It then follows that the category is a $\rmC^*$-tensor category by standard arguments; see for instance the proof of \cite[Thm.~5.1]{MR4362722}.

We now apply our defect selection criterion in a variety of models.

As stated in the introduction, we do not compute the braiding data for SPTs since SPT phases in 2+1D within the group cohomological classification are entirely determined by the 3-cocycle.
We give a procedure to compute the cocycle in the bulk, using specially constructed automorphisms that create a symmetry defect, which we explicitly carry out for the Levin-Gu SPT \cite{PhysRevB.86.115109}.

\subsubsection{General SPTs}
We show that our techniques can be used to construct defects for many SPT phases made using finite depth quantum circuits. Let $G$ be the symmetry group and $\beta_g$ the symmetry automorphism from Definition \ref{def:GlobalSymmetryAutomorphism}.

\begin{defn}
\label{def:FDQC}
    Let $\{\cU^d\}_{d = 1}^D$ be a family of sets of unitaries $U$ in $\cstar$ with $\supp(U)$ contained in a ball of diameter $N$ and having mutually disjoint supports. 
    An automorphism $\alpha$ is a \emph{finite depth (unitary) quantum circuit (FDQC)}\footnote{We use FDQC in the spirit of \cite{MR4544190}. 
    Some authors also consider non-unitary circuit elements, namely isometries and projections. 
    See \cite{2405.17379} for an instance where both definitions are discussed.} of the family $\{\cU^d\}_{d = 1}^D$ if for all $A \in \cstar[\loc]$ we have $$\alpha(A) = \alpha_D \circ \cdots \circ \alpha_1 (A), \qquad \qquad  \alpha_d(A) \coloneqq \Ad \! \left(\prod_{U \in \cU^d} U \right)\! (A).$$ We observe that $\alpha$ can be extended in a norm continuous way to all of $\cstar$.  
    We say $\cU^d$ is the set of \emph{entangling unitaries} of layer $d$ in the circuit. An example circuit in 1d with $N = 2, D=3$ is shown in Figure \ref{fig:FDQC_example}.
    \begin{figure}[!ht]
        \centering
        \includegraphics[width=0.5\linewidth]{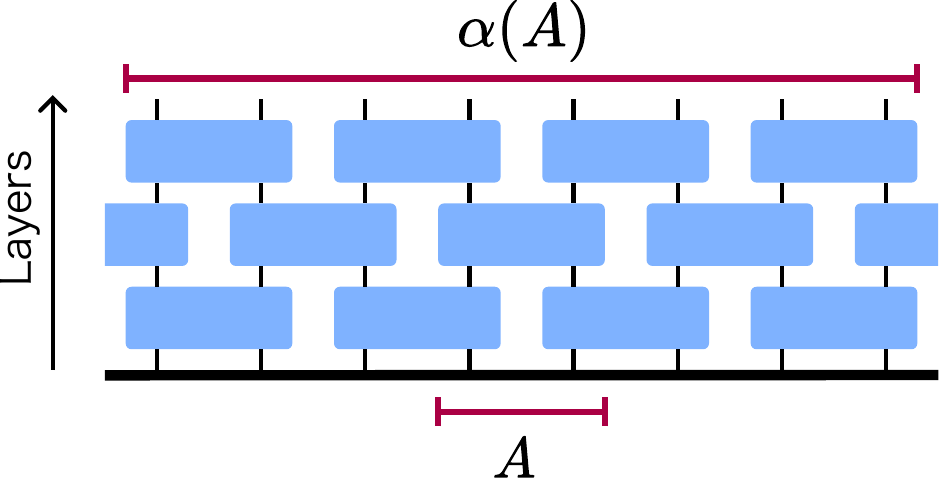}
        \caption{An example of a finite depth quantum circuit in 1 dimensions. Each block is an entangling unitary $U$ with support of 2 sites, so $N = 2$. The depth of this circuit is $D = 3$. We have $|\supp(A)| = 2$ and after the circuit, $|\supp(\alpha(A))| = 8$.}
        \label{fig:FDQC_example}
    \end{figure}
\end{defn}

\begin{defn}
\label{def:SPT_phase}
    We define a state $\omega$ to be a $G$-SPT state if there exists a product state $\omega_0$ satisfying $ \omega = \omega_0 \circ \alpha$, $\omega_0, \omega$ are both invariant under the action of $\beta_g$ for all $g \in G$, and $\alpha$ is an FDQC.
\end{defn}
\begin{rem}
    We note that this definition of an SPT is more strict than others appearing in the literature \cite{MR4354127}. In particular, it discounts locally generated automorphisms that are not FDQCs, and crystalline SPTs.
\end{rem}

Now fix $\omega_{SPT}$ to be a $G$-SPT, and let $\alpha$ be the automorphism implementing the FDQC.

\begin{asmp}
\label{asmp:entangler commutes with symmetry}
For every $g \in G$, $\alpha \circ \beta_g = \beta_g \circ \alpha$.
\end{asmp}

\begin{rem}
    We note that Assumption \ref{asmp:entangler commutes with symmetry} holds for a very general class of models \cite{PhysRevB.87.155114,PhysRevB.108.115144}.
\end{rem}

We also need a technical assumption which is physically reasonable and is satisfied for all known FDQC models. We elaborate the need for this assumption in Section \ref{sec:general SPTs}. Since $\beta_g$ is onsite, it can be restricted to any region (Definition \ref{def:SymmetryOnSubsets}). 
We let $r(L)$ denote the region to the right of the infinite dual path $L$.

\begin{asmp}
\label{asmp:strip_aut_is_an_FDQC}
    We assume that for any infinite dual path $L$, the automorphism $\alpha \circ  \beta_g^{r(L)} \circ \alpha^{-1} \circ (\beta_g^{r(L)})^{-1}$ is an FDQC built from unitaries of finite support and localized in $L^{+s}$. We remark that this condition is equivalent to the GNVW index \cite{MR2890305} of the aforementioned automorphism being trivial.
\end{asmp}

We mention the physical interpretation of our construction, as illustrated in Section \ref{sec:defect Hamiltonians SPT}.
Given a $G$-SPT (Definition \ref{def:SPT_phase}) satisfying Assumptions \ref{asmp:entangler commutes with symmetry}, \ref{asmp:strip_aut_is_an_FDQC}, for any chosen dual path $\gamma \in \bar P(\Gamma)$, there exists a commuting projector Hamiltonian $H_\gamma$ whose ground-state is a symmetry defect state, with the symmetry defect housed at the endpoints of $\gamma$ and a symmetry domain wall along $\gamma$.

We now fix $\pi_{SPT}$ as the GNS representation of $\omega_{SPT}$.

\begin{thm}
\label{thm:SPT theorem}
    Consider a $G$-SPT state $\omega_{SPT}$ constructed from a FDQC satisfying Assumptions \ref{asmp:entangler commutes with symmetry}, \ref{asmp:strip_aut_is_an_FDQC}. Then the category of defect sectors of a $G$-SPT (i.e, with respect to representation $\pi_{SPT}$) is monoidally equivalent to $\Vect(G, \nu)$ where $\nu: G \times G \times G \rightarrow U(1)$ is a $3$-cocycle. 
\end{thm}

\subsubsection{Levin-Gu SPT} The Levin-Gu SPT was first introduced in \cite{PhysRevB.86.115109} and serves as our first non-trivial test to the theory. In Section \ref{sec:Levin Gu} our main result is given by

\begin{thm}
\label{thm:category_of_Levin_Gu}
    The category of defect sectors of the Levin-Gu SPT is $G$-graded monoidally equivalent to $\Vect(\bbZ_2,\nu)$, where the 3-cocycle $\nu\colon \bbZ_2 \times \bbZ_2 \times \bbZ_2 \rightarrow U(1)$ represents the non-trivial element $[\nu]\in H^3(\mathbb{Z}_2,U(1))$.
\end{thm}

\subsubsection{An SET Toric Code}
We finally test the criterion for an SET related to the Toric Code, where the symmetry has non-trivial anyon data but does not permute anyon types.

Choose $\Gamma = \bbZ^2$ and let there be a qubit on each edge and on each vertex. As usual, we denote the resultant quasi-local algebra by $\cstar$. Let $\{\mathds{1}_v, \tau^x_v, \tau^y_v, \tau^z_v\}$ be the basis of $\cstar[v]$ where $\tau^x,\tau^y, \tau^z$ are the Pauli matrices. Let $\{\mathds{1}_e, \sigma^x_e, \sigma^y_e, \sigma^z_e\}$ be the basis of $\cstar[e]$ where $\sigma^x,\sigma^y, \sigma^z$ are also the Pauli matrices.

Define the star and plaquette operators of the usual Toric Code as $$A_v \coloneqq  \prod_{e \ni v} \sigma^x_e, \qquad \qquad B_f \coloneqq  \prod_{e \in f} \sigma^z_e.$$
We also define the following operators:
$$\tilde B_f \coloneqq  i^{- \sum_{e \in f}\sigma^x_e(\tau^z_{\partial_1 e} - \tau^z_{\partial_0 e})/2} B_f, \qquad Q_v \coloneqq  \tau^x_v i^{-\tau^z_v \sum_{e \ni v} f(e,v) \sigma^x_e/2}, \qquad \tilde Q_v \coloneqq  \frac{\mathds{1} + A_v}{2}Q_v,$$ where $f(e,v) = 1$ if $v = \partial_0 e$ and $f(e,v) = -1$ if $v = \partial_1 e$. 

The Hamiltonian for this model is then given by the formal sum
$$
H 
\coloneqq 
\sum_{v \in \Gamma} (\mathds{1} - A_v)/2 + (\mathds{1} - \tilde Q_v)/2 + \sum_{f \in \Gamma} (\mathds{1} - \tilde B_f)/2.
$$
We comment that that $H$ is a commuting projector Hamiltonian, and moreover is symmetric under the action of $\beta_g$ (recall Definition \ref{def:GlobalSymmetryAutomorphism}) where the unitaries are $U^g_e = \mathds{1}$ on each edge $e$ and $U^g_v = \tau^x_v$ on each vertex $v$. This Hamiltonian has a unique frustration-free ground-state $\tilde \omega$.

This model can be obtained from the usual Toric Code using a FDQC and thus has the same superselection theory \cite[Thm. ~6.1]{MR4362722}. 

We obtain a completeness result for the defect sectors of this model, and using that we construct the $G$-crossed braided monoidal category. Our final result for this paper is the following.

\begin{thm}
\label{thm:SET_Toric_Code}
    The category $\GSec^{ETC}$ of the defect sectors with respect to $\tilde \pi$, the GNS representation of $\tilde \omega$, is a $G$-crossed braided tensor category with trivial associators but non-trivial fractionalization data.
\end{thm}

\subsection{Constructing defect automorphisms}
\label{sec:DefectAutomorphismConstructionHeuristic}
Having stated the main results, in this section we give a sketch of the algorithm we used to construct symmetry defects. All of the examples discussed in the later sections apply this technique in order to construct defect endomorphisms and obtain defect representations.

Recalling our earlier notation, we denote $\beta_g$ to be the onsite symmetry action and $\omega_0$ to be a pure ground-state. We also have that interactions are uniformly bounded, and invariant under the symmetry action $\beta_g$. Since $\beta_g$ is onsite, for any dual path $ L \in \bar P(\Gamma)$ that divides $\Gamma$ into two halves, we have $$\beta_g =  \beta^{\ell(L)}_g \circ \beta^{r(L)}_g$$

\begin{figure}[!ht]
\centering
\begin{subfigure}[t]{0.4\linewidth}
    \includegraphics[width=\linewidth]{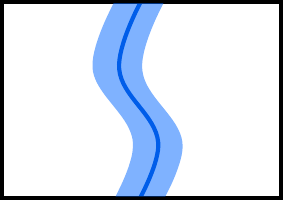}
    \caption{An infinite dual path $L$ shown in blue dividing the lattice into two halves. The strip $S$ is shown in light blue and centered at $L$.}
    \label{fig:strip_S}
\end{subfigure}
\hfill
\begin{subfigure}[t]{0.4\linewidth}
    \includegraphics[width=\linewidth]{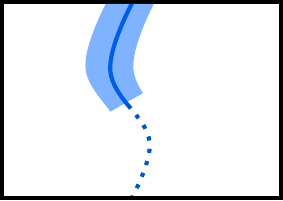}
    \caption{Half-strip $U$ shown in light blue. Using an automorphism $\alpha^D$ supported on half-strip $D = S \setminus U$, we erase the action of $\beta_g^{r(L)}$ such that the terms in the Hamiltonian supported outside $U$ are invariant under $\alpha^D \circ \beta_g^{r(L)}$.}
    \label{fig:partial_erasure}
\end{subfigure}
\caption{}
\end{figure}

Let us investigate the action of the restricted symmetry on the interaction terms in the Hamiltonian (which are invariant under $\beta_g$). If a term in the Hamiltonian has support disjoint from $L$, this term will remain invariant under the action of the restricted symmetry. We define a strip $S$ as the union of the supports of all interactions that are not invariant under $\beta_g^{r(L)}$(See Figure \ref{fig:strip_S}).

The goal then is to find an automorphism $\alpha$ localised in the strip $S$ that can `correct' the action of this restricted symmetry action on all the Hamiltonian terms. To do this, we seek to split $\alpha$ into two disjoint halves composed with some inner automorphism implemented by a local unitary where they meet. 
We expect that this step of the algorithm breaks down for more complicated models, in particular models exhibiting anyon permutation. 
Assuming that this can be done, we then cut $\alpha$ into $\Xi \circ (\alpha^U \otimes \alpha^D)$ where $\Xi$ is an inner automorphism implemented by a local unitary and $\alpha^U, \alpha^D$ are disjointly supported automorphisms both supported on $S$ and `erase' the restricted symmetry action on the terms in the Hamiltonian along their support, in the sense that $\alpha^D \circ \beta_g^{r(L)}$ leaves the terms in the Hamiltonian supported outside $U$ invariant (see Figure \ref{fig:partial_erasure}). 

\begin{figure}[!ht]
\centering
\begin{subfigure}[t]{0.4\linewidth}
    \includegraphics[width=\linewidth]{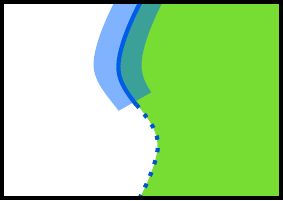}
    \caption{The symmetry defect is now given by $\kappa^U = \alpha^D \circ \beta_g^{r(L)}$ and acts trivially on Hamiltonian terms outside $U$ shown in light-blue.}
    \label{fig:symmetry_defect}
\end{subfigure}
\hfill
\begin{subfigure}[t]{0.4\linewidth}
    \includegraphics[width=\linewidth]{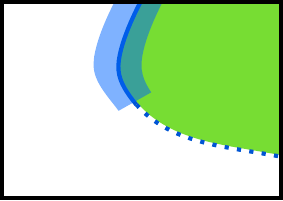}
    \caption{The dotted blue line can be freely transported while keeping the endpoint fixed, just like a string operator.}
    \label{fig:freely_transportable}
\end{subfigure}
\caption{Heuristic of a symmetry defect and its interpretation as being implemented by a string-operator.}
\end{figure}

A symmetry defect is then given by (see Figure \ref{fig:symmetry_defect}) $$\kappa^U = \alpha^D \circ \beta_g^{r(L)}.$$
We note that $\kappa$ depends on the entire path $L$ and not just the ray $U$, but we will always omit the dependence from the notation to prevent the notation from being too cluttered.

A symmetry defect can be interpreted as being obtained by adding an automorphism to partially `erase' the action of the restricted symmetry. We contrast this with the traditional paradigm of cutting a restricted symmetry action to obtain a symmetry defect.

We stress the word `partially' because even though the interaction terms supported away from the cut will not see the action of the restricted symmetry, there may still be observables in $\cstar$ that are transformed non-trivially under $\kappa$ along the erased symmetry action.

Written in this way, the symmetry defect can be interpreted as being implemented by something that behaves similarly to a string operator. More specifically, the ground state remains invariant under the action of $\kappa^U$ outside of some cone containing $U$. 
So the erased part of the symmetry defect can be freely transported outside of this cone (see Figure \ref{fig:freely_transportable}). A key difference between symmetry defects and anyons generated by string-operators is the presence of the $g$-action to the right of this string, so it is possible to detect the exact location of the defect line with local operators supported outside of the line, but not by the evaluation of these local operators in the ground state.

This also motivates our definition of a defect sector as a generalization of the anyon sector, in the sense that the `erased' part of the symmetry defect can always be moved into any allowed cone, as is typically done in anyon sector analysis. The key idea again is to account for the presence of the $g$-action to the right of this string.

To conclude this discussion, we summarize a simple algorithm to create symmetry defects, which we believe to be applicable to a wide variety of lattice models.

\vspace{5mm}
\centerline{%
\fbox{%
  \begin{minipage}{\dimexpr\linewidth-2\fboxsep-3\fboxrule\relax}
  \textbf{Creating a symmetry defect}
    \begin{enumerate}
    \item Observe the action of a restricted symmetry along a half-plane on the terms in the Hamiltonian.
    \item Devise an automorphism that `erases' this action on all such terms and is supported on some strip localised along the boundary of the restriction.
    \item If possible, cut this automorphism into 2 disjoint halves, possibly composed with some inner automorphism implemented by a local unitary.
    \item The symmetry defect is then given by composing the restricted symmetry action with the split automorphism.
    \end{enumerate}
  \end{minipage}%
}
}

\section{Symmetry Defects}
\label{sec:symmetry_defects}
Let $\omega_0 \colon \fA \to \bbC$ be a state.  
While in our examples, $\omega_0$ will usually be the unique frustration free ground state of a Hamiltonian (see Section \ref{sec:HamiltonianDynamics}), we do not assume that in this section.
We let $\pi_0 \colon \fA \to B(\cH_0)$ be the GNS representation for $\omega_0$.  
Note that $\pi_0$ is faithful since $\fA$ is UHF algebra and hence simple. 

\subsection{Category of \texorpdfstring{$G$}-defect sectors}
We recall the assumptions on the action by the group $G$ and on the state $\omega_0$ that we will impose to ensure we obtain a $G$-crossed braided monoidal category (Section \ref{sec:GCrossedAssumptions}). In this section, we construct the category of $G$-defect sectors.

Recall the definition of an allowed cone (Definition \ref{def:allowed_cone}). As before, we call $\cL$ the set of allowed cones with respect to a fixed path $\bar \gamma_R$. 

\begin{defn}
\label{def:sectorizable_rep} 
Let $\pi \colon \fA \to B(\cH)$ be an irreducible representation.  
We say that $\pi$ is \emph{$g$-sectorizable} with respect to $\pi_0$ if for every cone $\Lambda \in \cL$, 
\[
\pi|_{\cstar[\Lambda^c]}
\simeq
\pi_0 \circ \beta^{r(\Lambda)}_g|_{\cstar[\Lambda^c]}.
\]
\end{defn}

It may happen that $\pi \colon \fA \to B(\cH)$ is $g$-sectorizable and $h$-sectorizable for $g \neq h$.
Indeed, this happens for the trivial paramagnet model discussed in Section \ref{sec:TrivialParamagnet}.  
For the category we construct, we use a definition of $g$-defect sector which is a generalization of the usual notion of localized and transportable sector.
In particular, we modify the definition of localized and transportable analogously to \cite[Def.~2.6]{MR2183964}.

We now recall the definition of a defect sector (Definition \ref{def:defect_sector}). Let us set $\pi_0$ as the reference representation unless stated otherwise.

\begin{lem}
\label{lem:defect_sectorizable_reps_are_equivalent_to_a_defect_sector}
Let $\pi \colon \fA \to B(\cH)$ be a $g$-sectorizable irreducible representation.
Then $\pi \simeq \sigma$ for some $g$-defect sector $\sigma \colon \fA \to B(\cH_0)$.
In addition, if $\pi \colon \fA \to B(\cH_0)$ is an irreducible $g$-defect sector, then $\pi$ is $g$-sectorizable.
\end{lem}

\begin{proof}
This is an adaptation of the standard argument used for anyon sectors (see for instance \cite{MR3135456}).
Let $\Lambda \in \cL$.
Then by definition we have $\pi|_{\cstar[\Lambda^c]} \simeq \pi_0 \circ \beta_g^{r(\Lambda)}|_{\cstar[\Lambda^c]}$.  
Let $U \colon \cH \to \cH_0$ be a unitary implementing this equivalence, so for any $A \in\cstar[\Lambda^c]$,
\[
\pi_0 \circ \beta_g^{r(\Lambda)}(A)
=
U\pi(A)U^*.
\]
We then define $\sigma \colon \fA \to B(\cH_0)$ by $\sigma \coloneqq \Ad(U) \circ \pi$.  
Then by the above equation, $\sigma$ is canonically $g$-localized in $\Lambda$.  
Furthermore, since $\Lambda$ was arbitrary, the same procedure shows that $\sigma$ is transportable.
Hence $\sigma \simeq \pi$ is a $g$-defect sector. This shows the first result.

Now suppose $\pi \colon \fA \to B(\cH_0)$ is an irreducible $g$-defect sector and let $\Lambda \in \cL$.  
Then using transportability there exists $\pi' \colon \fA \to B(\cH_0)$ such that $\pi' \simeq \pi$ and $\pi'$ is $g$-localized in $\Lambda$.
Since $\pi'$ is $g$-localized in $\Lambda$, we have that 
\[
\pi'|_{\cstar[\Lambda^c]}=\pi_0 \circ \mu\circ \beta^{r(\Lambda)}_g|_{\cstar[\Lambda^c]},
\]
where $\mu$ is a symmetry action on finitely many sites.
Thus $\mu$ is an inner automorphism and we have,
\[
\pi_0 \circ \mu \circ \beta_g^{r(\Lambda)}|_{\cstar[\Lambda^c]}
\simeq
\pi_0 \circ \beta_g^{r(\Lambda)}|_{\cstar[\Lambda^c]}.
\]
We therefore have that 
\[
\pi|_{\cstar[\Lambda^c]}
\simeq
\pi'|_{\cstar[\Lambda^c]}
=
\pi_0 \circ \mu\circ \beta^{r(\Lambda)}_g|_{\cstar[\Lambda^c]}
\simeq
\pi_0 \circ \beta_g^{r(\Lambda)}|_{\cstar[\Lambda^c]}.
\qedhere
\]
\end{proof}

\begin{rem}
\label{rem:IrreducibilityAsAnAssumption}
In the definition for anyon sectors, it is commonly assumed that the anyon sectors $\pi$ are irreducible representations. This is because in examples, the category of anyon sectors is semisimple, meaning that every sector can be written as a direct sum of irreducible sectors.  
Hence irreducibility is a useful assumption in order to classify anyon sectors, as is done in \cite{2310.19661} for Kitaev's quantum double model.  
However, in our case, we wish to construct a category of defect sectors in a general setting, where the assumption of irreducibility is a hinderance since we want to take direct sums of defect sectors.
Additionally, we are able to state many of our results without assuming semisimplicity, so we do not assume that until it is necessary.
\end{rem}

We now show that we can extend every $g$-localized representation to an endomorphism of an auxiliary algebra (often called the allowed algebra), defined as in \cite{MR660538, MR2804555}.
Recall that $\cR(\Lambda) = \pi_0(\cstar[\Lambda])'' \subseteq B(\cH_0)$ for $\Lambda \subseteq \Gamma$. The \emph{auxiliary algebra} is defined to be
\[
\fA^a\coloneqq\overline{\bigcup_{\Lambda\in \cL} \cR(\Lambda)}^{||\cdot||}.
\]

\begin{lem}
    \label{lem:GDefectsDefinedOnAuxiliaryAlgebra}
    Let $\pi\colon\fA\rightarrow B(\cH_0)$ be a $g$-defect sector. 
    Then there is a unique extension $\pi^a$ of $\pi$ to $\fA^a$ such that $\pi^a|_{\cR(\Lambda)}$ is WOT-continuous for all $\Lambda \in \cL$.  
    Furthermore, $\pi^a(\fA^a) \subseteq \fA^a$, that is, $\pi^a \colon \fA^a \to \fA^a$ is an endomorphism.
\end{lem}
\begin{proof}
    We proceed as in the proofs of \cite[Lem.~4.1]{MR660538} and \cite[Prop.~4.2]{MR2804555}.
    Let $\Lambda\in\cL$. Then there exists some other $\Delta\in\cL$ such that $\Delta \subseteq r(\Lambda)$.
    Since $\pi$ is $g$-transportable, there exists some $U\in B(\cH_0)$ such that for all $A\in \cstar[\Delta^c]$, 
    $$
    U\pi(A)U^*=\pi_0 \circ \beta_g^{r(\Delta)}(A).
    $$
    Since $\Delta \subseteq r(\Lambda)$, $\Lambda \subseteq \Delta^c$ and $\Lambda$ intersects $r(\Delta)$ at most finitely many sites.  
    Therefore, there exists a unitary $V\in B(\cH_0)$ such that $V \pi_0(A) V^* = \pi_0 \circ \beta_g^{r(\Delta)}(A)$  for $A \in \cstar[\Lambda]$, so for all $A\in \cstar[\Lambda]$, we have that
    \[
    VU\pi(A)U^*V^*=\pi_0(A).
    \]
    Observe that we obtain a WOT-continuous formula for $\pi|_{\cstar[\Lambda]}$, namely $\pi|_{\cstar[\Lambda]} = \Ad(U^*V^*) \circ \pi_0|_{\cstar[\Lambda]}$, so $\pi|_{\cstar[\Lambda]}$ has a unique WOT-continuous extension to $\cR(\Lambda)$.
    (Note that we are implicitly identifying $\fA$ with $\pi_0(\fA)$, which we can do since $\pi_0 \colon \fA \to B(\cH_0)$ is faithful.)
    Since $\Lambda \in \cL$ was arbitrary, we obtain a unique extension of $\pi$ to $\bigcup_{\Lambda \in \cL} \cR(\Lambda)$ that is WOT-continuous on each $\cR(\Lambda)$.
    This extension is well-defined by continuity.
    It is also norm-continuous, so we obtain a unique extension $\pi^a$ of $\pi$ to $\fA^a$ with the desired properties.  

    It remains to show that $\pi^a(\fA^a) = \fA^a$.  
    To show this, it suffices to show that for all $\Lambda \in \cL$, $\pi^a(\pi_0(\cstar[\Lambda])) \subseteq \fA^a$.
    Let $\Lambda \in \cL$.
    Since $\pi$ is a $g$-defect sector, $\pi$ is $g$-localized in some $\widehat{\Lambda} \in \cL$.  
    Then there exists $\Delta \in \cL$ such that $\Lambda, \widehat{\Lambda} \subseteq \Delta$.
    In particular, we have that $\pi$ is $g$-localized in $\Delta$ and $\cstar[\Lambda] \subseteq \cstar[\Delta]$.
    By bounded spread Haag duality, we have that $\pi^a(\pi_0(\cstar[\Delta])) \subseteq \cR(\Delta^{+r})$ (see \cite[Lem.~2.12]{MR2183964}).  
    The result follows.
\end{proof}

In the remainder of the paper, we will abuse notation and identify the extension of $\pi$ to $\fA^a$ with $\pi$ for notational simplicity. The context should clarify any ambiguities.  

\begin{rem}
If $\pi \colon \fA \to B(\cH)$ is $g$-sectorizable, then it still holds that there is a unique extension of $\pi$ to $\fA^a$ such that $\pi|_{\cR(\Lambda)}$ is WOT-continuous for all $\Lambda \in \cL$.  
Indeed, the proof used in Lemma \ref{lem:GDefectsDefinedOnAuxiliaryAlgebra} still holds if $\pi$ is $g$-sectorizable.  
However, if $\pi$ is only $g$-sectorizable, then $\pi$ will in general not be an endomorphism of the auxiliary algebra.  
\end{rem}

\subsubsection{Category of homogeneous $G$-defect sectors}
\label{sec:Category_of_homogeneous_G-defect sectors}
We build a category $\GSec_{\hom}$ of homogeneous $G$-defect sectors as follows.  
The objects of our category are $g$-defect sectors for $g \in G$, and if $\pi, \sigma$ are $g$-defect sectors, a morphism $T \colon \pi \to \sigma$ is an intertwiner, i.e., an operator in $B(\cH_0)$ satisfying
\[
T\pi(-)
=
\sigma(-)T.
\]
We let $\GSec_g$ be the full subcategory of $g$-defect sectors for a fixed $g \in G$.  
Note that if $\pi$ and $\sigma$ are both canonically $g$-localized in $\Lambda \in \cL$, then an intertwiner $T \colon \pi \to \sigma$ satisfies that $T \in \cR(\Lambda^{+r})$ by bounded spread Haag duality (see \cite[Lem.~2.13]{MR2183964}).  
If $\pi$ and $\sigma$ are simply $g$-localized in $\Lambda$, then $T$ may not be in $\cR(\Lambda^{+r})$, where $r$ is specifically the spread for bounded spread Haag duality. 
However, we will have that $T \in \cR(\Lambda^{+R}) \subset \cstar^a$ for some $R \geq r$, since any $g$-localized map is unitarily equivalent to a canonically $g$-localized map by a unitary in $\fA_{\loc}$.  
We are now in a position to study $g$-defect sector endomorphisms that are $g$-localized to some $\Lambda\in\cL$ using the techniques of \cite{MR2183964}.\footnote{We thank David Penneys for the very helpful suggestion to apply the approach of \cite{MR2183964} to this problem.} 

Note that if $\pi$ is $g$-localized in $\Lambda \in \cL$ and $h$-localized in $\Lambda$, then $g = h$.  
Indeed, since $\pi$ is $g$-localized in $\Lambda$ we have that $\pi|_{\cstar[\Lambda^c]} = \pi_0 \circ \mu_1\circ \beta^{r(\Lambda)}_g$, where $\mu_1$ is a symmetry action on finitely many sites.  
Similarly, $\pi|_{\cstar[\Lambda^c]} = \pi_0 \circ \mu_2\circ \beta^{r(\Lambda)}_h$, where $\mu_2$ is a symmetry action on finitely many sites.  
But $\beta_g^{r(\Lambda)}$ and $\beta_h^{r(\Lambda)}$ differ on balls $B \subseteq \Gamma$ of arbitrarily large radius.  
Therefore, since the onsite symmetry is faithful on large enough balls, $g = h$. 
However, if $\pi$ is $g$-localized in $\Lambda \in \cL$ and $\sigma$ is $h$-localized in $\Lambda$, it may be the case that there exists a nonzero intertwiner $T \colon \pi \to \sigma$ even if $g \neq h$.  
However, by the lemma below, this intertwiner cannot be the cone algebra for any allowable cone. 
We therefore define the category $\GSec_{\hom} \coloneqq \bigsqcup_{g \in G} \GSec_g$, where $\bigsqcup$ denotes disjoint union.

\begin{lem}
Suppose $\pi$ is $g$-localized in $\Lambda \in \cL$ and $\sigma$ is $h$-localized in $\Lambda$, and suppose that $T \colon \pi \to \sigma$ satisfies that $T \in \cR(\Delta)$ for some $\Delta \in \cL$ and $T \neq 0$.  
Then $g = h$.
\end{lem}

\begin{proof}
Note that it suffices to consider the case where $\pi$ is canonically $g$-localized in $\Lambda$ and $\sigma$ is canonically $h$-localized in $\Lambda$, since a $g$-localized sector is unitarily equivalent to a canonically $g$-localized sector using a unitary in $\fA_{\loc}$.
(This assumption does not materially affect the argument, but it makes the notation easier.)
Since $\Lambda, \Delta \in \cL$, there exists a cone $\widetilde{\Lambda} \in \cL$ such that $\Lambda, \Delta \subseteq \widetilde{\Lambda}$.
Now, since $T \in \cR(\Delta)$, we have that for all $A \in\cstar[\Delta^c]$, $$T\pi_0(\beta_g^{r(\Lambda)}(A)) = \pi_0(\beta_g^{r(\Lambda)}(A))T$$

Similarly, since $\pi$ is $g$-localized in $\Lambda$, $\sigma$ is $h$-localized in $\Lambda$, and $T \colon \pi \to \sigma$, we have that for all $A \in\cstar[\Lambda^c]$, 
\[
T\pi_0(\beta_g^{r(\Lambda)}(A))
=
T\pi(A)
=
\sigma(A)T
=
\pi_0(\beta_h^{r(\Lambda)}(A))T.
\]
Using $\cstar[\Delta^c], \cstar[\Lambda^c] \subseteq \cstar[\widetilde{\Lambda}^c]$ and combining these equations, we get for all $A \in\cstar[\widetilde{\Lambda}^c]$, 
\[
\pi_0\big(\beta_g^{r(\Lambda)}(A) - \beta_h^{r(\Lambda)}(A)\big)T
=
0.
\]
Now we have $T \in \cR(\widetilde{\Lambda})$ and also for $A \in\cstar[\widetilde{\Lambda}^c]$ that,
\[
\pi_0\big(\beta_g^{r(\Lambda)}(A) - \beta_h^{r(\Lambda)}(A)\big)
\in\pi_0(\cstar[\widetilde{\Lambda}^c])
\subseteq
\cR(\widetilde{\Lambda})',
\]
Therefore, since $\cR(\widetilde{\Lambda})$ is a factor\footnote{We thank David Penneys for pointing out that this implies the desired result.} and $T \neq 0$, we obtain that for all $A \in\cstar[\widetilde{\Lambda}^c]$,
\[
\pi_0\big(\beta_g^{r(\Lambda)}(A) - \beta_h^{r(\Lambda)}(A)\big) = 0.
\]
Now there exists a ball $B \subseteq \widetilde{\Lambda}^c$ of arbitrarily large radius such that $\beta_g^{r(\Lambda)}|_B = \beta_g|_B$ and $\beta_h^{r(\Lambda)}|_B = \beta_h|_B$.
Since $\pi_0$ is faithful, we have that $\beta_g(A) = \beta_h(A)$ for all $A \in\cstar[B]$, so $g = h$.
\end{proof}

\subsubsection{Direct sums and subobjects of $G$-defect sectors}
\label{sec:GSecCauchyComplete}
Recall that we have assumed that the cone algebras are infinite factors.
Therefore, there exist isometries $V_1, \dots, V_n \in \cR(\Lambda)$ for all $\Lambda \in \cL$ such that $\sum_{i = 1}^n V_i V_i^* = \mathds{1}$ \cite[Halving Lemma 6.3.3]{MR1468230}.
We observe that the above conditions imply that $V_i^* V_j = \delta_{ij} \mathds{1}$.
For $\pi_1, \dots, \pi_n \in \GSec_{\hom}$, the map $\bigoplus_{i = 1}^n \pi_i \colon \fA^a \to \fA^a$ defined by $$\bigoplus_{i = 1}^n \pi_i(-) \coloneqq \sum_{i = 1}^n V_i \pi_i(-)V_i^*$$ satisfies the universal property of the direct sum.  

\begin{defn}
We define the category $\GSec \coloneqq \bigoplus_{g \in G} \GSec_g$, where the direct sums are taken using the above construction.
\end{defn}

Note that if $\Lambda \in \cL$ and $\pi_1, \dots, \pi_n \in \GSec$ are $g$-localized in $\Lambda$ and we choose $V_1, \dots, V_n \in \cR(\Lambda)$, then $\bigoplus_{i = 1}^n \pi_i$ is also $g$-localized in $\Lambda$  and  transportable.  
(This can be seen by adapting a standard argument; see for instance \cite[Lem.~6.1]{MR2804555}.)
Following \cite[Def.~2.8]{MR2183964}, we say that $\pi \in \GSec$ is \emph{$G$-localized} in $\Lambda \in \cL$ if 
\[
\pi(-)
=
\sum_{i = 1}^n V_i \pi_i(-)V_i^*,
\]
where each $\pi_i \in \GSec$ is $g_i$-localized in $\Lambda$ for some $g_i \in G$ and $V_1, \dots, V_n \in \cR(\Lambda)$.  
Additionally, if each $\pi_i$ is canonically $g_i$-localized, we say that $\pi$ is \emph{canonically $G$-localized}.  

We now show that our category admits subobjects using an adaptation of \cite[Lem.~5.8]{MR4362722}.  

\begin{lem}
Let $\pi \in \GSec_g$ $g$-localized in $\Lambda$ for some $g \in G$, and $p \colon \pi \to \pi$ be a projection.  
Then there exists an isometry $v \in \cR(\Lambda^{+r})$ such that $vv^* = p$.  
It follows that the map $\widehat{\pi} \colon \fA^a \to \fA^a$ given by $\widehat{\pi}(-) = v^* \pi(-)v$ is a $g$-defect sector localized in $\Lambda$ and that $v \colon \widehat{\pi} \to \pi$.
\end{lem}

\begin{proof}
This proof is a simplified version of the proof of \cite[Lem.~5.8]{MR4362722}.
Let $\widetilde{\Lambda}, \Delta \in \cL$ be disjoint cones such that $\widetilde{\Lambda}, \Delta \subseteq \Lambda$.
Let $\widetilde{\pi} \in \GSec_g$ be unitarily equivalent to $\pi$ and $g$-localized in $\widetilde{\Lambda}$, and $U \colon \pi \to \widetilde{\pi}$ be a unitary implementing the equivalence. 
Then $UpU^* \colon \widetilde{\pi} \to \widetilde{\pi}$ is an intertwiner. 
Since $\widetilde{\pi}$ is $g$-localized in $\widetilde{\Lambda}$, which is disjoint from $\Delta$, we have that $UpU^* \in \cR(\widetilde{\Lambda}^c)' \subseteq \cR(\Delta)'$.
Furthermore, by bounded spread Haag duality, we have that $UpU^* \in \cR(\Lambda^c)' \subseteq \cR(\Lambda^{+r})$, and additionally $\cR(\Delta) \subseteq \cR(\Lambda^{+r})$.
Thus, by \cite[Lem.~5.10]{MR4362722}, $UpU^*$ is Murray-von Neumann equivalent to $\mathds{1}$ in $\cR(\Lambda^{+r})$, as $\cR(\Delta), \cR(\Lambda^{+r})$ are infinite factors acting on a separable Hilbert space.  
Since $U \colon \pi \to \widetilde{\pi}$, we have that $U \in \cR(\Lambda^{+r})$, so $p$ is equivalent to $\mathds{1}$ in $\cR(\Lambda^{+r})$.  
Hence there exists an isometry $v \in \cR(\Lambda^{+r})$ such that $vv^* = p$.
One verifies that the map $\widehat{\pi} \colon \fA^a \to \fA^a$ given by $\widehat{\pi}(-) \coloneqq v^* \pi(-)v$ is $g$-localized in $\Lambda$ and transportable and that $v \colon \widehat{\pi} \to \pi$.
\end{proof}

\subsection{\texorpdfstring{$G$}{G}-crossed monoidal and braiding structure on \texorpdfstring{$\GSec$}{GSec}}
In this section, we show that $\GSec$ has the structure of a $G$-crossed braided monoidal category. 
To show this, we first construct the $G$-crossed monoidal structure and then construct the braiding.

\subsubsection{$G$-crossed monoidal structure on $\GSec$}
We henceforth identify $\cstar$ with $\pi_0(\cstar)$, since $\pi_0$ is a faithful representation.
We show that $\GSec$ has the structure of a strict $G$-crossed monoidal category.  
For $\pi, \sigma \in \GSec$, we define $\pi \otimes \sigma \coloneqq \pi \circ \sigma$ and for $T \colon \pi \to \pi'$ and $S \colon \sigma \to \sigma'$, we define $T \otimes S \coloneqq T\pi(S) = \pi'(S)T$.  
Note that for $\pi, \sigma \in \GSec$, $\pi \otimes \sigma \in \GSec$ by the following lemma.  
It follows that $\GSec$ is a strict monoidal category.  
In fact, this is the monoidal structure inherited from viewing $\GSec$ as a subcategory of $\End(\sB\fA^a)$, where $\sB\fA^a$ is the one-object category whose morphisms are elements of $\fA^a$.

\begin{lem}
\label{lem:TensorProductPreservesGGrading}
For $\pi \in \GSec_g$ and $\sigma \in \GSec_h$, we have that $\pi \otimes \sigma \in \GSec_{gh}$.
\end{lem}

\begin{proof}
Let $\pi \in \GSec_g$ and $\sigma \in \GSec_h$.  
Since $\pi \in \GSec_g$, there exists $\Lambda_1 \in \cL$ such that $\pi$ is $g$-localized in $\Lambda_1$.  
Similarly, since $\sigma \in \GSec_h$, there exists $\Lambda_2 \in \cL$ such that $\sigma$ is $h$-localized in $\Lambda_2$.  
Now, there exists $\Lambda \in \cL$ such that $\Lambda_1 \cup \Lambda_2 \subseteq \Lambda$, so $\pi$ is $g$-localized in $\Lambda$ and $\sigma$ is $h$-localized in $\Lambda$.  
We have that $\pi|_{\cstar[\Lambda^c]} = \mu_1\circ \beta^{r(\Lambda)}_g$ and $\sigma|_{\cstar[\Lambda^c]} = \mu_2\circ \beta^{r(\Lambda)}_h$, where $\mu_1$ and $\mu_2$ are symmetry actions on finitely many sites. 
We then have that $\mu_1 \circ \beta^{r(\Lambda)}_g\circ \mu_2\circ \beta^{r(\Lambda)}_h$ differs from $\beta_g^{r(\Lambda)} \circ \beta_h^{r(\Lambda)} = \beta_{gh}^{r(\Lambda)}$ respectively at finitely many sites, and 
\[
(\pi \otimes \sigma)|_{\cstar[\Lambda^c]}
=
(\pi \circ \sigma)|_{\cstar[\Lambda^c]}
=
\mu_1 \circ \beta^{r(\Lambda)}_g\circ \mu_2\circ \beta^{r(\Lambda)}_h|_{\cstar[\Lambda^c]}
=
\mu \circ \beta_{gh}^{r(\Lambda)}|_{\cstar[\Lambda^c]}
\]
Thus $\pi \otimes \sigma$ is $gh$-localized at $\Lambda$.
Now we show that $\pi \otimes \sigma$ is transportable. Choose $\Delta \in \cL$. 
Indeed, since $\pi, \sigma$ are transportable, so we have $\widehat{\pi} \simeq \pi$ and $\widehat{\sigma} \simeq \sigma$ where $\widehat{\pi}, \widehat{\sigma}$ are $g, h$-localized in $\Delta$ respectively. Let $U \colon \pi \to \widehat{\pi}$ and $V \colon \sigma \to \widehat{\sigma}$ be the unitaries implementing the equivalence. Then $U \otimes V \colon \pi \otimes \sigma \to \widehat{\pi} \otimes \widehat{\sigma}$ is a unitary, and $\widehat{\pi} \otimes \widehat{\sigma}$ is $gh$-localized in $\Delta$.
\end{proof}

It remains to show that $\GSec$ is $G$-crossed monoidal.  
We define $\partial \colon \GSec_{\hom} \to G$ by $\partial\pi \coloneqq g$ for $\pi \in \GSec_g$.  
Additionally, for $g \in G$, we define $\gamma_g \colon \GSec \to \GSec$ as follows. 
For $\pi \in \GSec$, we define $\gamma_g(\pi) \coloneqq \beta_g \circ \pi \circ \beta_g^{-1}$, and for $T \colon \pi \to \sigma$, we define $\gamma_g(T) = \beta_g(T)$.  
Observe that $\gamma_g(\pi)$ and $\gamma_g(T)$ are well-defined since $\pi$ and $\beta_g$ are endomorphisms of $\fA^a$.  
Additionally, with the above definitions, $\gamma_g(T) \colon \gamma_g(\pi) \to \gamma_g(\sigma)$, so $\gamma_g$ is a functor.

\begin{rem}
    According to physics literature \cite{PhysRevB.100.115147}, if there is a state $\omega_h$ housing a symmetry defect of type $h$, then under the action of the group symmetry $\beta_g$, we have $\omega_{h} \circ \beta_g = \omega_{ghg^{-1}}$. The physical significance of the functor $\gamma_g$ is thus the action of the symmetry $g$ on a symmetry defect. 
\end{rem}

\begin{prop}
\label{prop:GCrossedMonoidal}
The maps $\partial \colon \GSec_{\hom} \to G$ and $\gamma_g(\pi) = \beta_g \circ \pi \circ \beta_g^{-1}$ defined above equip $\GSec$ with the structure of a strict $G$-crossed monoidal category.  
\end{prop}

\begin{proof}
The proof proceeds analogously to that of \cite[Prop.~2.10]{MR2183964}.
Note that by construction, every object in $\GSec$ is a direct sum of objects in $\GSec_{\hom}$.  
Therefore, in order to show that $\GSec$ is strict $G$-crossed monoidal, we have to verify the following properties of $\partial$ and $\gamma$: 
\begin{enumerate}
    \item 
$\partial$ is constant on isomorphism classes, 
\item 
$\gamma_g \colon \GSec \to \GSec$ is a strict monoidal isomorphism, 
\item 
the map $g \mapsto \gamma_g$ is a group homomorphism, 
\item 
$\partial(\pi \otimes \sigma) = \partial \pi \partial \sigma$ for all $\pi, \sigma \in \GSec_{\hom}$
\item
$\gamma_g(\GSec_h) \subseteq \GSec_{ghg^{-1}}$
\end{enumerate}

\begin{enumerate}
    
\item[(1):] 
This follows from the fact that if $\pi \in \GSec_g$ and $\sigma \in \GSec_h$ for $g \neq h$, then there are no nonzero morphisms between $\pi$ and $\sigma$ in $\GSec$.  

\item[(2):]
Note that 
\(\gamma_g \circ \gamma_{g^{-1}} = \id_{\GSec} = \gamma_{g^{-1}} \circ \gamma_g,
\)
so $\gamma_g \colon \GSec \to \GSec$ is an isomorphism.  
It remains to show strict monoidality.  
Let $\pi, \sigma \in \GSec$.  
Then we have that 
\[
\gamma_g(\pi \otimes \sigma)
=
\beta_g \circ \pi \circ \sigma \circ \beta_g^{-1}
=
\beta_g \circ \pi \circ \beta_g^{-1} \circ \beta_g \circ \sigma \circ \beta_g^{-1}
=
\gamma_g(\pi) \otimes \gamma_g(\sigma).
\]
Similarly, for $T \colon \pi \to \pi'$ and $S \colon \sigma \to \sigma'$, we have that
\begin{align*}
\gamma_g(T \otimes S)
&=
\gamma_g(T\pi(S))
=
\beta_g(T)\beta_g(\pi(S))
=
\beta_g(T)\beta_g \circ \pi \circ \beta_g^{-1} (\beta_g(S))
\\&=
\gamma_g(T) \gamma_g(\pi)(\gamma_g(S))
=
\gamma_g(T) \otimes \gamma_g(S).
\end{align*}

\item[(3):]
Let $\pi \in \GSec$. We have, 
\[
\gamma_g \circ \gamma_h (\pi) = \beta_g \circ (\beta_h \circ \pi \circ \beta_h^{-1}) \circ \beta_g^{-1} = \beta_{gh} \circ \pi \circ \beta_{gh}^{-1}
=
\gamma_{gh}(\pi),
\]
and for $T \colon \pi \to \sigma$ in $\GSec$, we have that 
\[
\gamma_g \circ \gamma_h (T) = \beta_g \circ \beta_h (T) = \beta_{gh}(T)
=
\gamma_{gh}(T).
\]

\item[(4):]
This follows from Lemma \ref{lem:TensorProductPreservesGGrading}.

\item[(5):]
Let $\pi \in \GSec_h$. 
Then there exists $\Lambda \in \cL$ such that $\pi|_{\cstar[\Lambda^c]} = \mu\circ \beta^{r(\Lambda)}_h$, where is a symmetry action on finitely many sites.  
(Again, we are identifying $\cstar$ with $\pi_0(\cstar)$, since $\pi_0$ is a faithful representation.)
We then have that for $A \in\cstar[\Lambda^c]$, 
\[
\gamma_g(\pi)(A)
=
\beta_g \circ \pi\circ \beta_g^{-1}(A)
=
\beta_g \circ (\mu\circ \beta^{r(\Lambda)}_h) \circ \beta_g^{-1}(A), 
\]
and $\beta_g \circ \mu \circ \beta^{r(\Lambda)}_h\circ \beta_g^{-1}$ differs from $\beta_{ghg^{-1}}^{r(\Lambda)}$ at finitely many sites. 
Therefore, $\gamma_g(\pi)$ is $ghg^{-1}$-localized at $\Lambda$.
Now choose $\Delta \in \cL$ and $\widehat{\pi} \simeq \pi$, with $\widehat{\pi}$ being $g$-localized in $\Delta$, and $U \colon \pi \to \widehat{\pi}$ is a unitary. Then $\gamma_g(U) \colon \gamma_g(\pi) \to \gamma_g(\widehat{\pi})$ is a unitary, and $\gamma_g(\widehat{\pi})$ is $ghg^{-1}$-localized in $\Delta$. This shows $\gamma_g(\pi)$ is transportable.
\end{enumerate}
\end{proof}

\begin{rem}
Note that if $\pi \in \GSec_h$ is canonically $h$-localized in a cone $\Lambda \in \cL$, then it follows by the above argument that $\gamma_g(\pi)$ is canonically $ghg^{-1}$-localized in $\Lambda$.  
Thus, if $\pi \in \GSec$ is canonically $G$-localized in $\Lambda \in \cL$, then so is $\gamma_g(\pi)$.
\end{rem}

\begin{rem}
Since $\gamma_g \circ \gamma_h = \gamma_{gh}$, it may appear that we do not have any symmetry fractionalization data \cite{PhysRevB.100.115147}.
However in the case where our category is finitely semisimple, this data can be recovered from our construction in a manner similar to the one used in \cite{2306.13762} to recover $F$- and $R$-symbols.
We describe how to do this in Section \ref{sec:SymmetryFractionalization}.
\end{rem}

\subsubsection{$G$-crossed braided structure on $\GSec$}
\label{sec:GCrossedBraiding}
To simplify the construction, we now fix a cone $\Lambda \in \cL$ such that $\Lambda^{+r} \cap \bar \gamma_R = \emptyset$, where $\bar \gamma_R$ is the fixed half-infinite dual path from before.  
We let $\GSec(\Lambda)$ be the full subcategory of $\GSec$ consisting of sectors that are canonically $G$-localized in $\Lambda$.  
We use canonically $G$-localized here to simplify the computations.
We similarly define $\GSec(\Lambda)_{\hom}$ and $\GSec(\Lambda)_g$ for each $g \in G$. 
Note that if $T \colon \pi \to \sigma$ is an intertwiner in $\GSec(\Lambda)$, then $T \in \cR(\Lambda^{+r})$.
Indeed, if $\pi \in \GSec(\Lambda)$, then $\pi$ is of the form
\[
\pi(-)
=
\sum_{i = 1}^n V_i \pi_i(-)V_i^*,
\]
where each $\pi_i \in \GSec$ is canonically $g_i$-localized in $\Lambda$ for some $g_i \in G$ and $V_1, \dots, V_n \in \cR(\Lambda)$.
Now, for every $g \in G$, every intertwiner $T \colon \pi \to \sigma$ in $\GSec(\Lambda)_g$ lives in $\cR(\Lambda^{+r})$ by a standard argument (see for instance \cite[Lem.~2.13]{MR2183964}).  
Additionally, if $\pi \in \GSec(\Lambda)_g$, $\sigma \in \GSec(\Lambda)_h$, and $g \neq h$, then there are no nonzero morphisms $T:\pi \to \sigma$.
The desired result follows.

We now construct a $G$-crossed braiding on $\GSec(\Lambda)$. 

\begin{defn}
\label{def:sufficiently_to_the_left}
Let $\Delta \in \cL$.
We say that $\Delta$ is \emph{sufficiently to the left} of $\Lambda$ if $\Lambda^{+r}\cup r(\Lambda) \subseteq r(\Delta)$ and $\Delta^{+r} \subseteq \ell(\Lambda)$. The required geometry of cones is shown in figure \ref{fig:leftward_g_localised}.

\begin{figure}[!ht]
    \centering
    \includegraphics[width=0.5\linewidth]{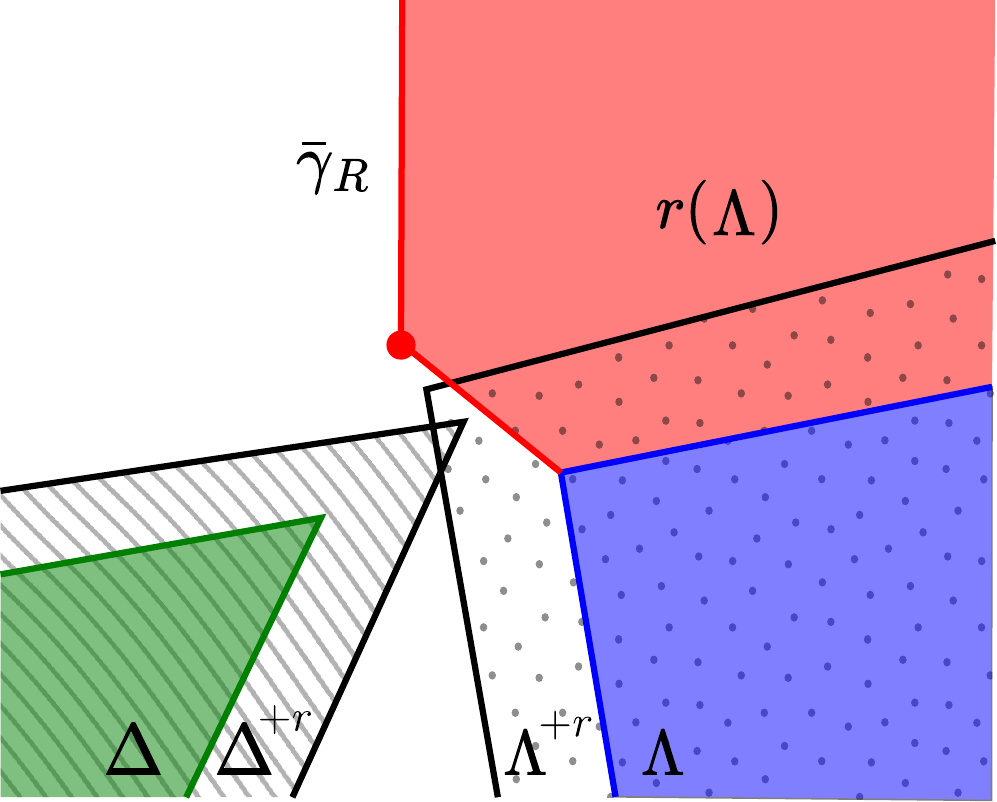}
    \caption{Example geometry of cones $\Lambda, \Delta$ needed for $\Delta$ to be sufficiently to the left of 
$\Lambda$ (see Def \ref{def:sufficiently_to_the_left}). The cone $\Lambda$ is shown in blue, $\Delta$ is shown in green, $r(\Lambda)$ is shown in red, $\Lambda^{+r}$ is shown with the dotted pattern, $\Delta^{+r}$ is shown with the striped pattern. In particular, $\Delta^{+r}$ and $\Lambda^{+r}$ are allowed to overlap, provided $\Delta$ does not overlap with $\Lambda^{+r}$ and $\Lambda$ does not overlap with $\Delta^{+r}$.}
    \label{fig:leftward_g_localised}
\end{figure}
\end{defn}

\begin{defn}
Let $\Delta \in \cL$ be sufficiently to the left of $\Lambda$.  
We say that $\pi \in \GSec_g$ is \emph{leftward $g$-localized} in $\Delta$ if $\pi$ is $g$-localized in $\Delta$ and $\pi|_{\cstar[\Lambda^{+r} \cup r(\Lambda)]} = \beta_g|_{\cstar[\Lambda^{+r} \cup r(\Lambda)]}$. 
\end{defn}

Let $\Delta \in \cL$ be sufficiently to the left of $\Lambda$ and $\pi, \sigma \in \GSec_g$ be leftward $g$-localized in $\Delta$. 
If $T \colon \pi \to \sigma$, then $T \in \cR(\ell(\Lambda))$.  
Indeed, if $\pi$ and $\sigma$ were canonically $g$-localized in $\Delta$, then $T \in \cR(\Delta^{+r}) \subseteq \cR(\ell(\Lambda))$.  
Since $\pi, \sigma \in \GSec_g$ are leftward $g$-localized in $\Delta$, they differ from being canonically $g$-localized in $\Delta$ by unitaries in $\pi_0(\cstar[\ell(\Lambda)])$, so $T \in \cR(\ell(\Lambda))$.  
Similarly, suppose $\Delta, \widehat{\Delta} \in \cL$ are both sufficiently to the left of $\Lambda$ and $\Delta \subseteq \widehat{\Delta}$. 
If $\pi$ is leftward $g$-localized in $\Delta$, then $\pi$ is leftward $g$-localized in $\widehat{\Delta}$.
Indeed, since $\pi$ is leftward $g$-localized in $\Delta$ and $\Delta \subseteq \widehat{\Delta}$, $\pi$ is $g$-localized in $\widehat{\Delta}$, and $\pi|_{\cstar[\Lambda^{+r} \cup r(\Lambda)]} = \beta_g|_{\cstar[\Lambda^{+r} \cup r(\Lambda)]}$.

The following lemma should be compared with \cite[Lem.~2.14]{MR2183964}.

\begin{lem}
\label{lem:GCrossedCommutation}
Let $\sigma \in \GSec(\Lambda)$ and $\pi \in \GSec_g$ be leftward $g$-localized in $\Delta$ for some $\Delta$ sufficiently to the left of $\Lambda$.  
Then $\pi \otimes \sigma = \gamma_g(\sigma) \otimes \pi$.
\end{lem}

\begin{proof}
We adapt the proof of \cite[Lem.~2.14]{MR2183964}. 
Note that $\sigma(-) = \sum_{i = 1}^n V_i \sigma_i(-)V_i^*$ where each $\sigma_i$ is canonically $h_i$-localized in $\Lambda$ for some $h_i \in G$ and $V_1, \dots, V_n \in \cR(\Lambda)$. 
This follows from the definition of $\GSec$ and that of $\GSec(\Lambda)$. 
Since $\Delta$ is sufficiently to the left of $\Lambda$ and $\pi$ is leftward $g$-localized in $\Delta$, $\pi(V_i) = \beta_g(V_i)$.  
Therefore, it suffices to show that $\pi \otimes \sigma = \gamma_g(\sigma) \otimes \pi$ for $\sigma$ canonically $h$-localized in $\Lambda$.  
We proceed by showing that $\pi \otimes \sigma(A) = \gamma_g(\sigma) \otimes \pi(A)$ for the following cases: 
\begin{enumerate}
    \item 
$A \in\cstar[r(\Lambda)]$,
\item 
$A \in\cstar[\Lambda]$, 
\item 
$A \in\cstar[\Delta]$, and
\item 
$A \in\cstar[\ell(\Lambda) \cap \Delta^c]$.
\end{enumerate}
\begin{enumerate}
\item[(1):]
In this case, $\sigma(A) = \beta_h(A)$, and since $\pi$ is leftward $g$-localized in $\Delta$, $\pi(A) = \beta_g(A)$.  
Therefore, we have that
\[
\pi \otimes \sigma(A)
=
\beta_g(\beta_h(A))
=
(\beta_g \circ \beta_h \circ \beta_g^{-1} \circ \beta_g(A))
=
(\beta_g \circ \sigma \circ \beta_g^{-1})(\beta_g(A))
=
\gamma_g(\sigma) \otimes \pi(A).
\]

\item[(2):]
In this case, $\sigma(A) \in \cR(\Lambda^{+r})$.  
Since $\Lambda^{+r} \subseteq r(\Delta)$ and $\pi$ is leftward $g$-localized in $\Delta$, we have that 
\[
\pi \otimes \sigma(A)
=
\beta_g(\sigma(A))
=
(\beta_g \circ \sigma \circ \beta_g^{-1} \circ \beta_g(A))
=
\gamma_g(\sigma) \otimes \pi(A).
\]

\item[(3):]
Since $\Delta \subseteq \Delta^{+r} \subseteq \ell(\Lambda)$, we have that $\sigma(A) = A$.  
Furthermore, since $\pi$ is $g$-localized in $\Delta$, $\pi(A) \in \cR(\Delta^{+r})$, and thus $\gamma_g(\sigma)(\pi(A)) = \pi(A)$.  
Therefore, we have that 
\[
\pi \otimes \sigma(A)
=
\pi(A)
=
\gamma_g(\sigma) \otimes \pi(A).
\]

\item[(4):]
Since $A \in\cstar[\ell(\Lambda)]$, $\sigma(A) = A$.  
In addition, since $A \in\cstar[\Delta^c]$ and $\pi$ is $g$-localized in $\Delta$, $\pi(A)$ has the same support as $A$.  
In particular, $\pi(A) \in \cstar[\ell(\Lambda)]$.  
Therefore, we have that 
\[
\pi \otimes \sigma(A)
=
\pi(A)
=
\gamma_g(\sigma) \otimes \pi(A).
\qedhere
\]
\end{enumerate}
\end{proof}

We now construct $c_{\pi, \sigma} \colon \pi \otimes \sigma \to \gamma_g(\sigma) \otimes \pi$ for $\pi \in \GSec(\Lambda)_g$ and $\sigma \in \GSec(\Lambda)$, using the approach of \cite[Prop.~2.17]{MR2183964}.

\begin{defn}
\label{def:GCrossedBraiding}
Let $\pi \in \GSec(\Lambda)_g$ and $\sigma \in \GSec(\Lambda)$ and choose $\Delta$ sufficiently to the left of $\Lambda$ and $\widetilde{\pi} \simeq \pi$ leftward $g$-localized in $\Delta$.  
Let $U \colon \pi \to \widetilde{\pi}$ be a unitary intertwiner.  
We then define the braiding isomorphism as 
\begin{equation*}
c_{\pi, \sigma}
\coloneqq
(\id_{\gamma_g(\sigma)} \otimes U^*)(U \otimes \id_\sigma)
=
\gamma_g(\sigma)(U^*)U.
\end{equation*}
Note that in defining $c_{\pi, \sigma}$ we are using that $\widetilde{\pi} \otimes \sigma = \gamma_g(\sigma) \otimes \widetilde{\pi}$ by Lemma \ref{lem:GCrossedCommutation}.

\begin{figure}[!ht]
\centering

\begin{subfigure}[t]{0.4\linewidth}
    \includegraphics[width=\linewidth]{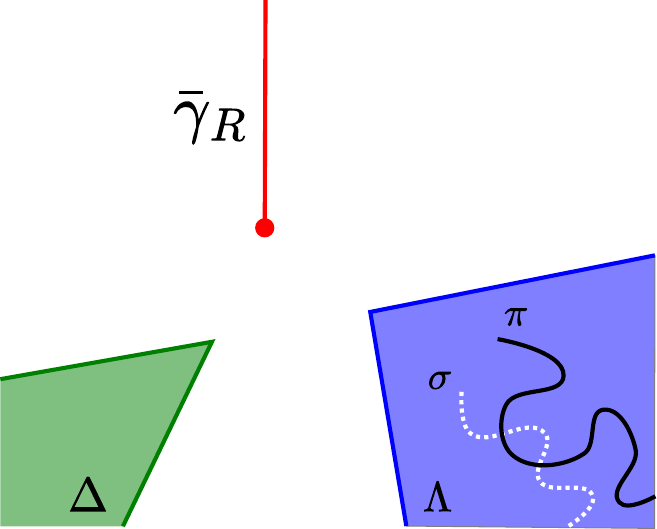}
    \caption{A cartoon of $\pi \otimes \sigma$, where the right component of the tensor is depicted by dotted lines.}
    \label{fig:defect_braid_1}
\end{subfigure}
\hfill
\begin{subfigure}[t]{0.4\linewidth}
    \includegraphics[width=\linewidth]{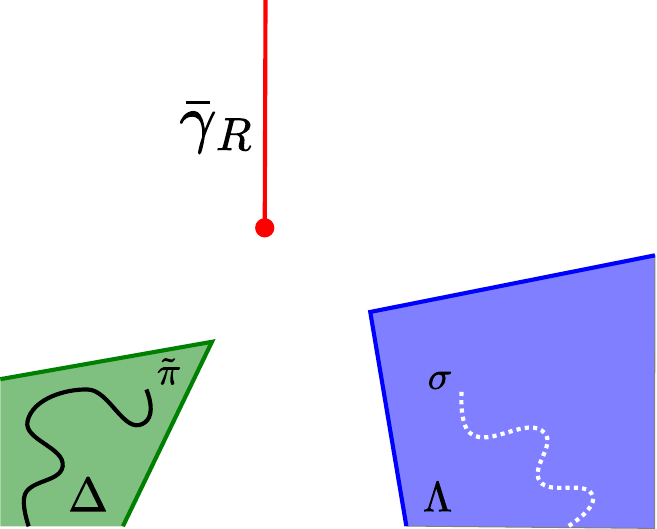}
    \caption{We conjugate by the unitary $U\colon \pi \rightarrow \tilde \pi$ to get $\tilde \pi \otimes \sigma = \Ad(U \otimes \Id_\sigma) (\pi \otimes \sigma)$ with $\tilde \pi$ leftward $g$-localised in $\Delta$.}
    \label{fig:defect_braid_2}
\end{subfigure}
\medskip
\begin{subfigure}[t]{0.4\linewidth}
    \includegraphics[width=\linewidth]{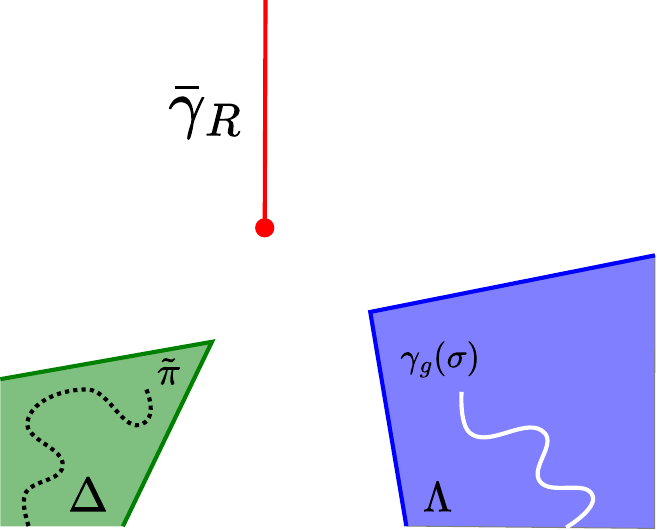}
    \caption{$\gamma_g(\sigma) \otimes \tilde \pi = \tilde \pi \otimes \sigma$ using Lemma \ref{lem:GCrossedCommutation} as $\tilde \pi$ is leftward $g$-localised in $\Delta$.}
    \label{fig:defect_braid_3}
\end{subfigure}
\hfill
\begin{subfigure}[t]{0.4\linewidth}
    \includegraphics[width=\linewidth]{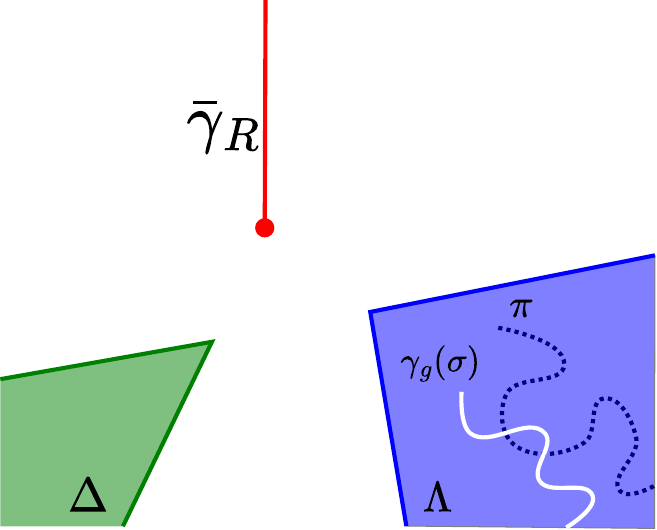}
    \caption{We can again use $U^*$ to get $\gamma_g(\sigma) \otimes \pi = \Ad(\Id_{\gamma_g(\sigma)}\otimes U^*) (\gamma_g(\sigma) \otimes \tilde \pi)$.}
    \label{fig:defect_braid_4}
\end{subfigure}
\caption{The procedure carried out by the \emph{braiding isomorphism} $c_{\pi, \sigma}$. Note that we require $\Delta, \Lambda$ to be such that $\Delta$ is sufficiently to the left of $\Lambda$ (Definition \ref{def:sufficiently_to_the_left}). We use dotted lines to represent the right component of the tensor.}
\label{fig:depiction_of_braiding_isomorphism}
\end{figure}
\end{defn}

\begin{rem}
    Fig \ref{fig:depiction_of_braiding_isomorphism} shows the procedure that is carried out by the braiding isomorphism $c_{\pi, \sigma}$ for defect sectors $\pi \in \GSec(\Lambda)_g, \sigma \in \GSec(\Lambda)$ (a cartoon is shown in Fig \ref{fig:defect_braid_1}). In order to apply Lemma \ref{lem:GCrossedCommutation}, we must first localise $\pi$ to $\Delta$ such that $\Delta$ is sufficiently to the left of $\Lambda$. To do so, we conjugate by the unitary $U\otimes \Id_\sigma$ to obtain $\tilde \pi \otimes \sigma$ (Fig \ref{fig:defect_braid_2}). We then use Lemma \ref{lem:GCrossedCommutation} to switch the tensor components and obtain $\gamma_g(\sigma) \otimes \tilde \pi$ (Fig \ref{fig:defect_braid_3}). Finally, we conjugate by the unitary $(\Id_{\gamma(g(\sigma)}\otimes U^*)$ to get $\gamma_g(\sigma) \otimes \pi$. Putting this sequence of operations together, we obtain $c_{\pi,\sigma} (\pi \otimes \sigma)(-) = (\gamma_g(\sigma) \otimes \pi)(-) c_{\pi, \sigma}$. 
\end{rem}

\begin{lem}
Let $\pi \in \GSec(\Lambda)_g$ and $\sigma \in \GSec(\Lambda)$.
The map $c_{\pi, \sigma}$ from Definition \ref{def:GCrossedBraiding} does not depend on the choices of $U$, $\widetilde{\pi}$, and $\Delta$.
\end{lem}

\begin{proof}
We adapt the proof in \cite[Prop.~2.17]{MR2183964}.
We first show independence of $\widetilde{\pi}$ and $U$.
Let $\widehat{\pi} \simeq \pi$ be another $g$-defect sector leftward $g$-localized in $\Delta$, and let $V \colon \pi \to \hat{\pi}$ be a unitary. 
We wish to show that 
\[
\gamma_g(\sigma)(U^*)U
=
\gamma_g(\sigma)(V^*)V.
\]
This is equivalent to showing that 
\[
\gamma_g(\sigma)(VU^*)
=
VU^*.
\]
Now, $VU^* \colon \widetilde{\pi} \to \widehat{\pi}$. 
Therefore, since $\Delta$ is sufficiently to the left of $\Lambda$ and $\widetilde{\pi}, \widehat{\pi}$ are both leftward $g$-localized in $\Delta$, $VU^* \in \cR(\ell(\Lambda))$.  
The desired result thus holds since $\sigma$ is canonically $G$-localized in $\Lambda$. 

It remains to show that $c_{\pi, \sigma}$ does not depend on $\Delta$.  
Suppose $\Delta, \widehat{\Delta} \in \cL$ are both sufficiently to the left of $\Lambda$. The existence of a single cone $\tilde \Delta$ satisfying both $\widetilde \Delta \subset \Delta, \widehat \Delta$ as well as $\tilde \Delta \subset \ell(\Lambda)$ is not necessary guaranteed.  
However, we are able to `zig-zag' between the two cones without leaving $\ell(\Lambda)$.\footnote{The paper \cite{2410.21454} uses zig-zags to show well-definedness of the braiding.
The idea to apply zig-zags here came from work on that paper, and more specifically from discussions with David Penneys.}  
More precisely, a \emph{zig-zag from $\Delta$ to $\widehat{\Delta}$} is a sequence of cones $(\Delta_1, \widetilde{\Delta}_1, \dots, \Delta_n, \widetilde{\Delta}_n, \Delta_{n + 1})$ such that $\Delta_1 = \Delta$, $\Delta_{n + 1} = \widehat{\Delta}$, and for each $i = 1, \dots, n$, $\Delta_i, \Delta_{i + 1} \subseteq \widetilde{\Delta}_i$ \cite[Sec~ 1.1]{2410.21454}. An example zig-zag with $n=2$ is shown in figure \ref{fig:zigzag}. 

Now, observe that given $\Delta, \widehat{\Delta} \in \cL$ sufficiently to the left of $\Lambda$, there exists a zig-zag from $\Delta$ to $\widehat{\Delta}$ where each cone in the zig-zag is sufficiently to the left of $\Lambda$.  
It therefore suffices to show that given $\Delta_i, \Delta_{i + 1} \subseteq \widetilde{\Delta}_i$, $g$-localizing in $\Delta_i$ and $\Delta_i$ give the same $c_{\pi, \sigma}$.  
But this follows since $g$-defect sectors leftward $g$-localized in $\Delta_i$/$\Delta_{i + 1}$ are leftward $g$-localized in $\widetilde{\Delta}_i$, and we already showed independence of $\widetilde{\pi}$ leftward $g$-localized in $\widetilde{\Delta}_i$.
\begin{figure}[!ht]
    \centering
    \includegraphics[width=0.5\linewidth]{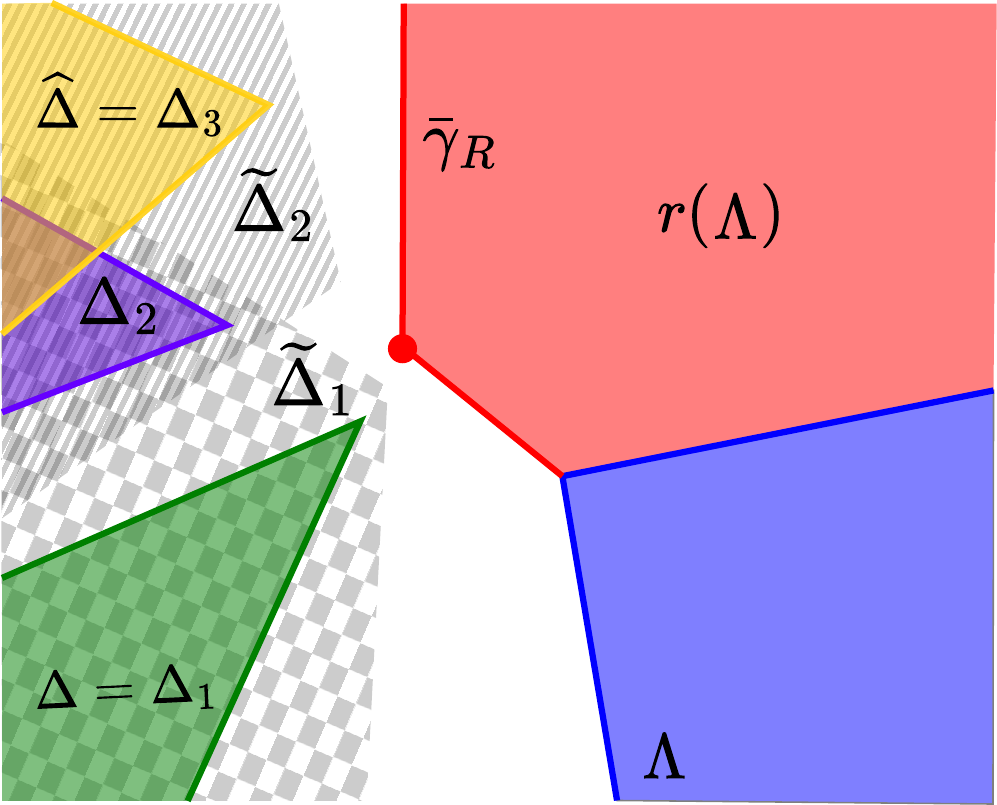}
    \caption{A zig-zag with $n=2$ between cones $\Delta$, $\widehat \Delta$. Notice that given the geometry of cones $\Delta, \widehat \Delta$, it is not possible to find a single cone $\widetilde \Delta \subset \ell(\Lambda)$ that contains both $\Delta, \widehat \Delta$. However we can zig-zag between them by choosing the zig-zag sequence ($\Delta_1 = \Delta, \widetilde \Delta_1, \Delta_2, \widetilde \Delta_2, \Delta_3 = \widehat \Delta)$ such that all cones in this sequence lie in $\ell(\Lambda)$.}
    \label{fig:zigzag}
\end{figure}
\end{proof}

\begin{prop}
\label{prop:GCrossedBraiding}
The category $\GSec(\Lambda)$ is $G$-crossed braided using the braid isomorphism from Definition \ref{def:GCrossedBraiding}.
\end{prop}

\begin{proof}
We proceed as in the proof in \cite[Prop.~2.17]{MR2183964}.
To show that Definition \ref{def:GCrossedBraiding} gives a $G$-crossed braiding on $\GSec(\Lambda)$, we must show that the following conditions of being a $G$-crossed braiding are satisfied:

\begin{enumerate}
    \item Naturality
    \item Monoidality
    \item Braiding is preserved by $\gamma_g$
\end{enumerate}

\begin{enumerate}
\item[{(1):}]
There are two naturality equations that must be verified; we verify each in turn.  
First, suppose $\pi \in \GSec(\Lambda)_g$, and $T \colon \sigma_1 \to \sigma_2$ is a morphism in $\GSec(\Lambda)$.  
We must show that 
\[
(\gamma_g(T) \otimes \id_\pi)c_{\pi, \sigma_1}
=
c_{\pi, \sigma_2} (\id_\pi \otimes T).
\]
Let $\Delta$ be sufficiently to the left of $\Lambda$, $\widetilde{\pi} \simeq \pi$ be leftward $g$-localized in $\Delta$, and $U \colon \pi \to \widetilde{\pi}$ be a unitary. 
The equation to verify then becomes
\[
\gamma_g(T)\gamma_g(\sigma_1)(U^*)U
=
\gamma_g(\sigma_2)(U^*)U\pi(T).
\]
We proceed starting with the right-hand side.  
We first observe that 
\[
\gamma_g(\sigma_2)(U^*)U\pi(T)=\gamma_g(\sigma_2)(U^*)\widetilde{\pi}(T)U=\gamma_g(\sigma_2)(U^*)\beta_g(T)U.
\]
The last equality follows since $\widetilde{\pi}$ is leftward $g$-localized in $\Delta$ and $T \in \cR(\Lambda^{+r})$. Now,
\[
\gamma_g(\sigma_2)(U^*)\beta_g(T)U
=
\beta_g(\sigma_2(\beta_g^{-1}(U^*))T)U=\beta_g(T\sigma_1(\beta_g^{-1}(U^*)))U=\gamma_g(T)\gamma_g(\sigma_1(U^*))U.
\]

Now, suppose $\sigma \in \GSec(\Lambda)$ and $T \colon \pi_1 \to \pi_2$ is a morphism in $\GSec(\Lambda)_g$.  
We must show that 
\[
(\id_{\gamma_g(\sigma)} \otimes T)c_{\pi_1, \sigma}
=
c_{\pi_2, \sigma}(T \otimes \id_\sigma).
\]
Let $\Delta$ be sufficiently to the left of $\Lambda$. 
For $i = 1, 2$, we let $\widetilde{\pi}_i \simeq \pi_i$ be leftward $g$-localized in $\Delta$, and $U_i \colon \pi_i \to \widetilde{\pi}_i$ be a unitary. 
The equation to verify then becomes 
\[
\gamma_g(\sigma)(T)\gamma_g(\sigma)(U_1^*)U_1
=
\gamma_g(\sigma)(U_2^*)U_2T.
\]
Note that the above equation is equivalent to 
\[
\gamma_g(\sigma)(U_2TU_1^*)
=
U_2TU_1^*.
\]
But since $U_2TU_1^* \colon \widetilde{\pi}_1 \to \widetilde{\pi}_2$ and $\widetilde{\pi}_1,\widetilde{\pi}_2$ are both leftward $g$-localized in $\Delta$, $U_2TU_1^* \in \cR(\Delta^{+r}) \subseteq \cR(\ell(\Lambda))$. 
Therefore, since $\sigma$ (and hence $\gamma_g(\sigma)$) is canonically $G$-localized in $\Lambda$, the desired equation holds.

\item[(2):]
Again, there are two equations that we must verify.  
First, let $\pi \in \GSec(\Lambda)_g$ and $\sigma, \tau \in \GSec(\Lambda)$.  We must show that
\[
c_{\pi, \sigma \otimes \tau}
=
(\id_{\gamma(\sigma)} \otimes c_{\pi, \tau})(c_{\pi, \sigma} \otimes \id_\tau).
\]
Let $\Delta$ be sufficiently to the left of $\Lambda$, $\widetilde{\pi} \simeq \pi$ be leftward $g$-localized in $\Delta$, and $U \colon \pi \to \widetilde{\pi}$ be a unitary. 
The equation to verify then becomes
\[
\gamma_g(\sigma \otimes \tau)(U^*)U
=
\gamma_g(\sigma)(\gamma_g(\tau)(U^*)U)\gamma_g(\sigma)(U^*)U.
\]
But this is easily seen to hold.  
Indeed, working from the right-hand side, we have that 
\[
\gamma_g(\sigma)(\gamma_g(\tau)(U^*)U)\gamma_g(\sigma)(U^*)U
=
(\gamma_g(\sigma) \circ \gamma_g(\tau))(U^*) \gamma_g(\sigma)(U)\gamma_g(\sigma)(U^*)U
=
\gamma_g(\sigma \otimes \tau)(U^*)U.
\]

Now, let $\pi \in \GSec(\Lambda)_g$, $\sigma \in \GSec(\Lambda)_h$, and $\tau \in \GSec(\Lambda)$.
We must show that 
\[
c_{\pi \otimes \sigma, \tau}
=
(c_{\pi, \gamma_h(\tau)} \otimes \id_\sigma)(\id_\pi \otimes c_{\sigma, \tau}).
\]
Let $\Delta$ be sufficiently to the left of $\Lambda$, $\widetilde{\pi} \simeq \pi$ be leftward $g$-localized in $\Delta$, $\widetilde{\sigma} \simeq \sigma$ be leftward $h$-localized in $\Delta$, and $U \colon \pi \to \widetilde{\pi}$ and $V \colon \sigma \to \widetilde{\sigma}$ be unitaries. 
Note that $\widetilde{\pi} \otimes \widetilde{\sigma}$ is leftward $gh$-localized in $\Delta$, and $U \otimes V = U\pi(V) = \widetilde{\pi}(V)U \colon \pi \otimes \sigma \to \widetilde{\pi} \otimes \widetilde{\sigma}$ is a unitary.
The desired equation therefore becomes
\[
\gamma_{gh}(\tau)(U^*\widetilde{\pi}(V^*))U\pi(V)
=
\gamma_g(\gamma_h(\tau))(U^*)U\pi(\gamma_h(\tau)(V^*)V)
=
\gamma_{gh}(\tau)(U^*)U\pi(\gamma_h(\tau)(V^*))\pi(V).
\]
Note that the above equation is equivalent to 
\[
\gamma_{gh}(\tau)(\widetilde{\pi}(V^*))U
=
U\pi(\gamma_h(\tau)(V^*))
=
\widetilde{\pi}(\gamma_h(\tau)(V^*))U,
\]
so it suffices to show that $\gamma_{gh}(\tau)(\widetilde{\pi}(V^*)) = \widetilde{\pi}(\gamma_h(\tau)(V^*))$.
But this holds by Lemma \ref{lem:GCrossedCommutation} since $\gamma_h(\tau) \in \GSec(\Lambda)$ and $\widetilde{\pi}$ is leftward $g$-localized in $\Delta$.  

\item[(3):]
We must show that for $\pi \in \GSec(\Lambda)_h$ and $\sigma \in \GSec(\Lambda)$, 
\[
\gamma_g(c_{\pi, \sigma})
=
c_{\gamma_g(\pi), \gamma_g(\sigma)}.
\]
Let $\Delta$ be sufficiently to the left of $\Delta$, $\widetilde{\pi} \simeq \pi$ be leftward $h$-localized in $\Delta$, and $U \colon \pi \to \widetilde{\pi}$ be a unitary. 
Note that $\gamma_g(\widetilde{\pi})$ is leftward $ghg^{-1}$-localized in $\Delta$, and $\gamma_g(U) \colon \gamma_g(\pi) \to \gamma_g(\widetilde{\pi})$ is a unitary. 
We therefore have that 
\begin{align*}
c_{\gamma_g(\pi), \gamma_g(\sigma)}
&=
\gamma_{ghg^{-1}}(\gamma_g(\sigma))(\gamma_g(U^*))\gamma_g(U)
=
\beta_{g}(\gamma_h(\sigma)(U^*))\beta_g(U)
=
\gamma_g(c_{\pi, \sigma}).
\qedhere
\end{align*}
\end{enumerate}
\end{proof}

\subsection{Connection to anyon sectors}
\label{sec:connection_to_anyon_sectors}

We now show that $\GSec(\Lambda)_e$ is precisely the braided $\rmC^*$-tensor category of superselection sectors with respect to $\pi_0$ \cite{MR4362722}.
This is to be expected since $\GSec(\Lambda)_e$ should correspond to the anyonic excitations, and anyonic excitations are described by superselection sectors.

The following definition of anyon sectors is identical to Definition \ref{def:AnyonSector} commonly found in the literature, with the sole exception that here we do not require irreducibility (c.f.~ Remark \ref{rem:IrreducibilityAsAnAssumption}).

\begin{defn}
Let $\pi_1 \colon \fA \to B(\cH_1)$ and $\pi_2 \colon \fA \to B(\cH_2)$ be representations of $\fA$.  
We say that $\pi_1$ satisfies the \emph{superselection criterion with respect to $\pi_2$} if for every cone $\Lambda$ (including $\Lambda \notin \cL$), 
\[
\pi_1|_{\cstar[\Lambda^c]}
\simeq
\pi_2|_{\cstar[\Lambda^c]}.
\]
If $\pi_2 = \pi_0$, we say that $\pi_1$ is an \emph{anyon sector}.  
\end{defn}

\begin{lem}
\label{lem:GDefectsRelationToAnyons}
The following statements are true:
\begin{enumerate}
    \item Let $\pi, \sigma \in \GSec_g$.  
Then $\pi$ satisfies the superselection criterion with respect to $\sigma$
\item Let $\sigma \in \GSec_g$, and let $\pi \colon \fA \to B(\cH)$ satisfies the superselection criterion with respect to $\sigma$. 
Then there exists $\widehat{\pi} \in \GSec_g$ such that $\pi \simeq \widehat{\pi}$
\end{enumerate}
\end{lem}

\begin{proof}
\begin{enumerate}
\item[(1):]  We must show that for every cone $\Lambda$ (including $\Lambda \notin \cL$), 
\[
\pi|_{\cstar[\Lambda^c]}
\simeq
\sigma|_{\cstar[\Lambda^c]}.
\]
Now we observe that for every cone $\Lambda$ (including $\Lambda \notin \cL$) there exists some cone $\Delta \subset \Lambda$ such that $\Delta \in \cL$.
Now since $\pi, \sigma \in \GSec_g$, we have that 
\[
\pi|_{\cstar[\Delta^c]}
\simeq
\pi_0 \circ \beta_g^{r(\Delta)}|_{\cstar[\Delta^c]}
\simeq
\sigma|_{\cstar[\Delta^c]},
\]

Noting $\cstar[\Lambda^c] \subset \cstar[\Delta^c]$, we have $\pi|_{\cstar[\Lambda^c]}
\simeq \sigma|_{\cstar[\Lambda^c]}$ as desired.

\item[(2):] 
Since $\sigma \in \GSec_g$, we have for some $\Lambda \in \cL$ that $\sigma|_{\cstar[\Lambda^c]} = \pi_0 \circ \mu\circ\beta^{r(\Lambda)}_g|_{\cstar[\Lambda^c]}$.
Noting that $\pi$ satisfies the superselection criterion with respect to $\sigma$, we have that 
\[
\pi|_{\cstar[\Lambda^c]}
\simeq
\sigma|_{\cstar[\Lambda^c]}
=
\pi_0 \circ \mu\circ \beta^{r(\Lambda)}_g|_{\cstar[\Lambda^c]}.
\]
Define $\widehat{\pi} \coloneqq \Ad(U) \circ \pi$, where $U$ is a unitary implementing the equivalence 
$\pi|_{\cstar[\Lambda^c]}
\simeq
\sigma|_{\cstar[\Lambda^c]}$.
Note that $\widehat{\pi}$ is $g$-localized in $\Lambda$ by definition.
Furthermore, for all $\Delta \in \cL$ there exists $\widetilde{\sigma} \simeq \sigma$ $g$-localized in $\Delta$.  
Since $\pi$ satisfies the superselection criterion with respect to $\sigma$, $\pi$ also satisfies the superselection criterion with respect to $\widetilde{\sigma}$.
Therefore, by the same argument that we used to find $\widehat{\pi}$, we can find $\widetilde{\pi} \simeq \pi \simeq \widehat{\pi}$ $g$-localized in $\Delta$, so $\widehat{\pi} \in \GSec_g$.
\end{enumerate}
\end{proof}

\begin{cor}\label{cor:anyon_equiv}
The braided tensor category $\GSec(\Lambda)_e$ is braided equivalent to the braided tensor category of superselection sectors localized in $\Lambda$.
\end{cor}

\begin{proof}
The category $\GSec(\Lambda)_e$ is equivalent to the category of superselection sectors localized in $\Lambda$ by Lemma \ref{lem:GDefectsRelationToAnyons} and the fact that $\pi_0 \in \GSec(\Lambda)_e$.  
Furthermore when $g=e$, the tensor product and braiding reduce to precisely those defined in \cite{MR2804555}.
The result follows.
\end{proof}

\subsection{Coherence data}
\label{sec:coherance_data}

In this section we discuss how to obtain the symmetry fractionalization and other cohrence data described in \cite{PhysRevB.100.115147}.
We proceed similarly to how \cite{2306.13762} obtain the $F$- and $R$-symbols in the case of anyon (superselection) sectors.
For this analysis, we introduce one more assumption. 

\begin{asmp}
The category $\GSec$ is finitely semisimple.
\end{asmp}

We let $\mathcal{K}_0(\GSec)$ be the fusion ring of $\GSec$ and let $I$ denote the basis of $\cK_0(\GSec)$. For each $i\in I$, we label the corresponding object in the category by $\pi_i$.
Note that $\pi_0$ is irreducible since $\omega_0$ is a pure state.
The assumption that $\GSec$ is finitely semisimple means that every object in $\GSec$ is isomorphic to finitely many direct sums of $\pi_i$'s.

\subsubsection{Symmetry fractionalization}
\label{sec:SymmetryFractionalization}
For every $g \in G$ and $i \in I$, we have that $\gamma_g(\pi_i)$ is irreducible, so we have that $\gamma_g(\pi_i) \simeq \pi_{i'}$ for a unique $i' \in I$.
We define $g(i) \coloneqq i'$ for notational clarity.
We let $V_g^i \colon \gamma_g(\pi_i) \to \pi_{g(i)}$ be a unitary. 
Now, for $g, h \in G$ and $i \in I$, we have that $V_g^{h(i)} \gamma_g(V_h^i) \colon \gamma_g(\gamma_h(\pi_i)) \to \pi_{g(h(i))}$, since $\gamma_g(V_h^i) \colon \gamma_g(\gamma_h(\pi_i)) \to \gamma_g(\pi_{h(i)})$.
Now, since $\gamma_g(\gamma_h(\pi_i)) = \gamma_{gh}(\pi_i)$, we also have that $V_{gh}^i \colon \gamma_{g}(\gamma_h(\pi_i)) \to \pi_{gh(i)}$.
This implies that $\pi_{gh(i)} \simeq \pi_{g(h(i))}$, so $gh(i) = g(h(i))$.
Furthermore, we have that
$$V_g^{h(i)}\gamma_g(V_h^i) (V_{gh}^i)^* \colon \pi_{gh(i)} \to \pi_{gh(i)}.$$
Therefore, since $\pi_{gh(i)}$ is irreducible, we have by Schur's lemma that 
\[
V_g^{h(i)}\gamma_g(V_h^i)
=
\eta(g, h)_i V_{gh}^i
\]
for some $\eta(g, h)_i \in U(1)$. 
This scalar is the symmetry fractionalization data described in \cite{PhysRevB.100.115147}.

The following lemma shows the desired coherence condition for the symmetry fractionalization which is analogous to \cite[Eq.~279]{PhysRevB.100.115147}. Note that these conditions are not identical because we have chosen different conventions.

\begin{lem}
\label{lem:SymmetryFractionalizationData}
Let $i \in I$ and $g, h, k \in G$.
We have that 
\[
\eta(g, h)_{k(i)} \eta(gh, k)_{i}
=
\eta(h, k)_i \eta(g, hk)_i.
\]
\end{lem}

\begin{proof}
We first observe that 
\[
V_g^{hk(i)} \gamma_g(V_h^{k(i)}) \gamma_{gh} (V_k^i)
\colon
\gamma_{ghk}(\pi_i) \to \pi_{ghk(i)}.
\]
We relate $V_g^{hk(i)} \gamma_g(V_h^{k(i)}) \gamma_{gh} (V_k^i)$ to $V_{ghk}^i$ in two different ways.  
Indeed, observe that 
\begin{gather*}
V_g^{hk(i)} \gamma_g(V_h^{k(i)}) \gamma_{gh} (V_k^i)
=
\eta(g, h)_{k(i)} V_{gh}^{k(i)} \gamma_{gh} (V_k^i)
=
\eta(g, h)_{k(i)} \eta(gh, k)_{i} V_{ghk}^i,
\\
\intertext{We may also proceed as follows,}
V_g^{hk(i)} \gamma_g(V_h^{k(i)}) \gamma_{gh} (V_k^i)
=
V_g^{hk(i)} \gamma_g((V_h^{k(i)}) \gamma_{h} (V_k^i))
=
V_g^{hk(i)} \gamma_g(\eta(h, k)_i V_{hk}^i)
=
\eta(h, k)_i \eta(g, hk)_i V_{ghk}^i.
\end{gather*}
Therefore, since $V_{ghk}^i$ is a unitary, we have that 
\[
\eta(g, h)_{k(i)} \eta(gh, k)_{i}
=
\eta(h, k)_i \eta(g, hk)_i.
\qedhere
\]
\end{proof}

We remark that different choices for $V_g^i$ are guaranteed to give equivalent $\eta(g, h)_i$, as the construction we used to obtain them is a formal categorical argument and the category $\GSec$ was already shown to be $G$-crossed monoidal (Proposition \ref{prop:GCrossedMonoidal}).

\subsubsection{Other coherence data}
\label{sec:OtherCoherenceData}
We now demonstrate how to obtain the rest of the coherence data discussed in \cite{PhysRevB.100.115147}.
For computational simplicity, for the remainder of the discussion we assume the category is \emph{pointed}, meaning that for every $i, j \in I$, $\pi_i \otimes \pi_j$ is irreducible.
The analysis can be done in more generality, but in that case more care must be taken.
Note that we are not constraining our general analysis with this assumption but are using it simply for demonstration purposes.

We first compute the $F$-symbols.
This proceeds exactly as done in \cite[Sec~2.3.1]{2306.13762}, but we repeat the discussion for convenience.  
For every $i, j \in I$, we have that $\pi_i \otimes \pi_j$ is irreducible, so $\pi_i \otimes \pi_j \simeq \pi_{ij}$ for some $ij \in I$.
Following \cite{2306.13762}, we let the tensorator $\Omega_{i, j} \colon \pi_i \otimes \pi_j \to \pi_{ij}$ be a unitary.
Now, for $i, j, k \in I$, we have that
\begin{gather*}
\Omega_{i, jk}(\id_{\pi_i} \otimes \Omega_{j,k})
=
\Omega_{i, jk}\pi_i(\Omega_{j,k})
\colon 
\pi_i \otimes \pi_j \otimes \pi_k \to \pi_{i(jk)},
\\
\Omega_{ij,k}(\Omega_{i, j} \otimes \id_{\pi_k})
=
\Omega_{ij,k}\Omega_{i, j}
\colon
\pi_i \otimes \pi_j \otimes \pi_k \to \pi_{(ij)k}.
\end{gather*}
Therefore, we have that $i(jk) = (ij)k \eqqcolon ijk$, and by Schur's lemma, we obtain that 
\[
\Omega_{ij,k}\Omega_{i, j}
=
F(i, j, k)
\Omega_{i, jk}\pi_i(\Omega_{j,k})
\]
for some $F(i, j, k) \in U(1)$.  
This is the $F$-symbol as defined in \cite{2306.13762}. We remark that the coherence condition for the $F$-symbols holds, omitting the proof as it is shown in \cite{2306.13762}.

\begin{lem}[{\cite[Prop.~2.11]{2306.13762}}]
For all $i, j, k, \ell \in \{0, 1, \dots, n\}$, 
\[
F(i, j, k) F(i, jk, \ell) F(j, k, \ell)
=
F(ij, k, \ell)F(i, j, k\ell).
\]
\end{lem}

We now compute the coherence data related to the tensorator of $\gamma_g$ in the skeletalization of the category that we are now working with; this corresponds to the data defined in \cite[Eq.~269]{PhysRevB.100.115147}.  
We let $i, j \in I$ and $g \in G$.  
We first observe that the following map is a unitary intertwiner: 
\[
V^{ij}_g \gamma_g(\Omega_{i, j})
\colon
\gamma_g(\pi_i \otimes \pi_j)
\to
\pi_{g(ij)}.
\]
In addition, since $\gamma_g(\pi_i \otimes \pi_j) = \gamma_g(\pi_i) \otimes \gamma_g(\pi_j)$, we also have that the following map is a unitary intertwiner: 
\[
\Omega_{g(i), g(j)}(V^i_g \otimes V^j_g)
=
\Omega_{g(i), g(j)}V^i_g\gamma_g(\pi_i)(V^j_g)
\colon
\gamma_g(\pi_i \otimes \pi_j)
\to
\pi_{g(i)g(j)}.
\]
We therefore have that $g(ij) = g(i)g(j)$, and by Schur's lemma, we obtain that 
\[
V^{ij}_g \gamma_g(\Omega_{i, j})
=
\mu_g(i, j) \Omega_{g(i), g(j)}V^i_g\gamma_g(\pi_i)(V^j_g)
\]
for some $\mu_g(i, j) \in U(1)$.
The following lemma follows from a straightforward computation.

\begin{lem}
Let $i, j, k \in I$ and $g \in G$.
We have that
\[
F(i, j, k) \mu_g(i, jk) \mu_g(j, k)
=
\mu_g(ij, k) \mu_g(i, j) F(g(i), g(j), g(k)).
\]
\end{lem}

Finally, we compute the $R$-symbols, analogously to \cite[Section 2.3.2]{2306.13762}.
We first assume that for all $i \in I$, $\pi_i \in \GSec(\Lambda)$, where $\Lambda \in \cL$ is a fixed cone such that $\Lambda^{+r} \cap R = \emptyset$.
(We do this since the braiding defined in Section \ref{sec:GCrossedBraiding} is defined on $\GSec(\Lambda)$ for such a $\Lambda \in \cL$.)
We now proceed as in \cite[Section 2.3.2]{2306.13762}.
Let $i, j \in I$.
Additionally, since $\pi_i \in \GSec(\Lambda)$ is irreducible, $\pi_i \in \GSec(\Lambda)_g$ for some $g \in G$.  
We then have that $c_{i, j} \colon \pi_i \otimes \pi_j \to \gamma_g(\pi_j) \otimes \pi_i$ is a unitary intertwiner, so we have that 
\[
\Omega_{g(j), i} (V^j_g \otimes \id_{\pi_i})c_{\pi_i, \pi_j}
=
\Omega_{g(j), i} V^j_g c_{\pi_i, \pi_j}
\colon
\pi_i \otimes \pi_j \to \pi_{g(j)i}.
\]
On the other hand, we have that $\Omega_{i, j} \colon \pi_i \otimes \pi_j \to \pi_{ij}$ is also a unitary intertwiner.  
Therefore, $ij = g(j)i$, and by Schur's lemma, 
\[
\Omega_{g(j), i} V^j_g c_{\pi_i, \pi_j}
=
R(i, j)\Omega_{i, j}
\]
for some $R(i, j) \in U(1)$.  
This defines the $R$-symbols for our category. The $R$-symbols satisfy several coherence relations. 
We single out the heptagon equations \cite[Eq.~286 \& 287]{PhysRevB.100.115147}.

\begin{lem}
Let $i, j, k \in \{0, 1, \dots, n\}$.
Suppose that $\pi_i \in \GSec(\Lambda)_g$ and $\pi_j \in \GSec(\Lambda)_h$. 
We then have that 
\begin{gather*}
R(i, k) F(g(j), i, k)^* R(i, j)
=
F(g(j), g(k), i)^* \mu_g(j, k)^*R(i, jk)F(i, j, k)^*,
\\
R(i, h(k))F(i, h(k), j) R(j, k)
=
F(gh(k), i, j) \eta(g, h)_k R(ij, k) F(i, j, k).
\end{gather*}
\end{lem}

\begin{proof}
We verify the second equation, which corresponds to \cite[Eq.~287]{PhysRevB.100.115147}. The computation can be graphically represented as in Figure \ref{fig:heptagon equation}. The other equation can be verified analogously.  
We consider the unitary 
\begin{equation}
\label{eq:UnitaryForHeptagonEquation}
\Omega_{gh(k)i, j} \Omega_{gh(k), i} V_g^{h(k)} c_{\pi_i, \pi_{h(k)}} \pi_i(V_h^k c_{\pi_j, \pi_k})
\colon
\pi_i \otimes \pi_j \otimes \pi_k
\to
\pi_{gh(k)ij}.
\end{equation}
We simplify this unitary in two different ways.
To make the simplification easier to follow, we color the terms that change at each step red or blue.
Specifically, the terms colored blue are the ones that were changed in the prior step, and the terms colored red are the ones that will change in the next step.  
If a term is colored purple, that means it is involved in the changes made at consecutive steps.  
For the first simplification, we have that 
\begin{align*}
\Omega_{gh(k)i, j} \textcolor{red}{\Omega_{gh(k), i} V_g^{h(k)} c_{\pi_i, \pi_{h(k)}}} \pi_i(V_h^k c_{\pi_j, \pi_k})
&=
\textcolor{blue}{R(i, hk)} \textcolor{red}{\Omega_{gh(k)i, j}} \textcolor{blue}{\Omega_{i, h(k)}} \pi_i(V_h^k c_{\pi_j, \pi_k})
\\&=
R(i, hk) \textcolor{violet}{\Omega_{ih(k), j}} \textcolor{red}{\Omega_{i, h(k)}} \pi_i(V_h^k c_{\pi_j, \pi_k})
\\&=
R(i, hk) \textcolor{blue}{F(i, h(k), j)}\textcolor{blue}{\Omega_{i, h(k)j}} \textcolor{violet}{\pi_i(\Omega_{h(k), j})} \textcolor{red}{\pi_i(V_h^k c_{\pi_j, \pi_k})}
\\&=
R(i, hk) F(i, h(k), j)\Omega_{i, h(k)j} \textcolor{blue}{\pi_i(}\textcolor{violet}{\Omega_{h(k), j} \pi_i(V_h^k c_{\pi_j, \pi_k})}\textcolor{blue}{)}
\\&=
R(i, hk) F(i, h(k), j)\textcolor{blue}{R(j, k)}\textcolor{red}{\Omega_{i, h(k)j}} \pi_i(\textcolor{blue}{\Omega_{j, k}})
\\&=
R(i, hk) F(i, h(k), j)R(j, k)\textcolor{blue}{\Omega_{i, jk}} \pi_i(\Omega_{j, k}).
\end{align*}

\begin{figure}[!ht]
    \centering
    \includegraphics[width=0.7\linewidth]{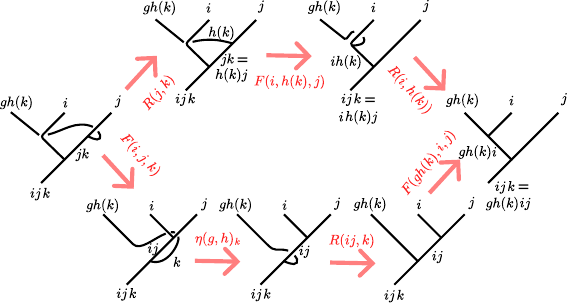}
    \caption{The graphical representation of the second heptagon equation of \cite[Eq. 287]{PhysRevB.100.115147}. We note that our definition of $R$ corresponds to $R^{-1}$ in their work. 
    For presentation purposes, some of the lines turn downwards, but one should interpret all of the lines as moving upward (i.e., we are not using any evaluation/coevaluation maps).}
    \label{fig:heptagon equation}
\end{figure}

For the other simplification, we recall the following naturality and monoidality equations for the braiding (proven in Proposition \ref{prop:GCrossedBraiding}).
\begin{facts}
\label{facts:braiding}
\leavevmode
\begin{itemize}
\item 
If $\pi \in \GSec(\Lambda)_g$, and $T \colon \sigma_1 \to \sigma_2$ is a morphism in $\GSec(\Lambda)$, then 
\[
\gamma_g(T)c_{\pi, \sigma_1}
=
c_{\pi, \sigma_2}\pi(T).
\]
\item 
If $\sigma \in \GSec(\Lambda)$ and $T \colon \pi_1 \to \pi_2$ is a morphism in $\GSec(\Lambda)_g$, then
\[
\gamma_g(\sigma)(T)c_{\pi_1, \sigma}
=
c_{\pi_2, \sigma}T.
\]
\item 
If $\pi \in \GSec(\Lambda)_g$, $\sigma \in \GSec(\Lambda)_h$, and $\tau \in \GSec(\Lambda)$, then
\[
c_{\pi \otimes \sigma, \tau}
=
c_{\pi, \gamma_h(\tau)}\pi(c_{\sigma, \tau}).
\]
\end{itemize}
\end{facts}
We now simplify the unitary in \eqref{eq:UnitaryForHeptagonEquation} using these facts, as well as the fact that $\pi_i \otimes \pi_j, \pi_{ij} \in \GSec(\Lambda)_{gh}$.
In particular, we have that 
\allowdisplaybreaks
\begin{align*}
\Omega_{gh(k)i, j} \Omega_{gh(k), i} V_g^{h(k)} c_{\pi_i, \pi_{h(k)}} \textcolor{red}{\pi_i(V_h^k c_{\pi_j, \pi_k})}
&=
\Omega_{gh(k)i, j} \Omega_{gh(k), i}V_g^{h(k)} \textcolor{red}{c_{\pi_i, \pi_{h(k)}}} \textcolor{violet}{\pi_i(V_h^k)}\textcolor{blue}{\pi_i(c_{\pi_j, \pi_k})}
\\&=
\Omega_{gh(k)i, j} \Omega_{gh(k), i} V_g^{h(k)}\textcolor{blue}{\gamma_g(V_h^k)}\textcolor{violet}{c_{\pi_i, \gamma_h(\pi_k)}} \textcolor{red}{\pi_i(c_{\pi_j, \pi_k})}
\\&=
\textcolor{red}{\Omega_{gh(k)i, j} \Omega_{gh(k), i}} V_g^{h(k)} \gamma_g(V_h^k)\textcolor{blue}{c_{\pi_i \otimes \pi_j, \pi_k}}
\\&=
\textcolor{blue}{F(gh(k), i, j) \Omega_{gh(k), ij} \pi_{gh(k)}(\Omega_{i, j})} \textcolor{red}{V_g^{h(k)} \gamma_g(V_h^k)}c_{\pi_i \otimes \pi_j, \pi_k}
\\&=
F(gh(k), i, j)\textcolor{blue}{\mu(g, h)_k}\Omega_{gh(k), ij} \textcolor{red}{\pi_{gh(k)}(\Omega_{i, j})}\textcolor{violet}{V_{gh}^k}c_{\pi_i \otimes \pi_j, \pi_k}
\\&=
F(gh(k), i, j)\mu(g, h)_k\Omega_{gh(k), ij} \textcolor{blue}{V_{gh}^k}\textcolor{violet}{\gamma_{gh}(\pi_k)(\Omega_{i, j})}\textcolor{red}{c_{\pi_i \otimes \pi_j, \pi_k}}
\\&=
F(gh(k), i, j)\mu(g, h)_k\textcolor{red}{\Omega_{gh(k), ij} V_{gh}^k}\textcolor{violet}{c_{\pi_{ij}, \pi_k}}\textcolor{blue}{\Omega(i, j)}
\\&=
F(gh(k), i, j)\mu(g, h)_k\textcolor{blue}{R(ij, k)}\textcolor{violet}{\Omega(ij, k)}\textcolor{red}{\Omega(i, j)}
\\&=
F(gh(k), i, j)\mu(g, h)_kR(ij, k)\textcolor{blue}{F(i, j, k)\Omega_{i, jk} \pi_i(\Omega_{j, k})}.
\end{align*}
Comparing the two simplifications of \eqref{eq:UnitaryForHeptagonEquation}, we obtain that 
\[
R(i, h(k))F(i, h(k), j) R(j, k)
=
F(gh(k), i, j) \eta(g, h)_k R(ij, k) F(i, j, k).
\qedhere
\]
\end{proof}

\section{General SPTs}
\label{sec:general SPTs}
In this section, given an SPT we will obtain states housing defects using defect automorphisms. We will then classify all possible $g$-sectors for this SPT.

Let $\hilb_v \simeq \bbC^{d_v}$ with $d_v \geq 2$ for each $v \in \Gamma$. For the sake of simplicity we take our lattice as the regular triangular lattice. We now define the symmetry action on $\cstar$. Let $G$ be the symmetry group and for every $g \in G$, let $g\mapsto U^g_v$ be its unitary representation onto the vertex $v$. 
We assume that this representation is faithful (Assumption \ref{asmp:Faithfulness}).
For each $A \in \cstar[V]$ with $V \in \Gamma_f$, we let $\beta_g$ be the map from Definition \ref{def:GlobalSymmetryAutomorphism}.

Recall the definition of a $G$-SPT (Definition \ref{def:SPT_phase}). We have for a $G$-SPT state $\tilde \omega$, the existence of a finite depth quantum circuit (FDQC) $\alpha$ such that $\omega_0 \circ \alpha = \tilde \omega$, where $\omega_0$ is some product state. 
For the entirety of this section, we let $s$ denote the spread of $\alpha$.
We impose one more assumption that allows us to apply the heuristic for defect automorphism construction discussed in Section \ref{sec:DefectAutomorphismConstructionHeuristic}.

\begin{asmp_recall}
(Assumption \ref{asmp:entangler commutes with symmetry})
For every $g \in G$, $\alpha \circ \beta_g = \beta_g \circ \alpha$.
\end{asmp_recall}

\begin{rem}
    We note that Assumption \ref{asmp:entangler commutes with symmetry} holds for a very general class of models like the ones constructed in \cite{PhysRevB.87.155114, PhysRevB.108.115144,PhysRevB.86.115109}, including the Levin-Gu SPT considered in Section \ref{sec:Levin Gu}. 
\end{rem}

\begin{rem}
\label{rem:GInvarianceTrick}
In the definition of $G$-SPT (Definition \ref{def:SPT_phase}), we assume that the both the product state $\omega_0$ and the state $\tilde \omega = \omega_0 \circ \alpha$ are invariant under $\beta_g$. 
However, if the FDQC $\alpha$ satisfies Assumption \ref{asmp:entangler commutes with symmetry}, then $\omega_0$ is $\beta_g$-invariant if and only if $\tilde \omega$ is. 
Indeed, suppose $\omega_0$ is $\beta_g$-invariant.
Then we have that 
\[
\tilde \omega \circ \beta_g
=
\omega_0 \circ \alpha \circ \beta_g
=
\omega_0 \circ \beta_g \circ \alpha
=
\omega_0 \circ \alpha
=
\tilde \omega.
\]
The other direction follows by the same argument.
\end{rem}

We first recall the following well-known Lemma.

\begin{lem}{\cite[Section 2.1.1]{MR2345476}}
\label{lem:can freely insert and remove P from the ground state.}
If $\omega(A) = 1$ for some $A \in \cstar$ satisfying that $A \leq \mathds{1}$, then we have for any $O \in \cstar$, $$\omega(O) = \omega(A O) = \omega(OA) = \omega(AOA)$$
\end{lem}

\begin{lem}
\label{lem:existence of gapped Hamiltonian}
    Let $\omega$ be a product state. Then for all $v \in \Gamma$, there exists a unique set $\{P_v^\omega\}_{v \in \Gamma}$ of rank-1 projections $P_v^\omega \in \cstar[v]$ such that $\omega$ is the unique state satisfying $\omega(P_v^\omega) = 1$ for all $v \in \Gamma$. 
    Moreover, the Hamiltonian given by $H_V^\omega \coloneqq \sum_{v \in V} \mathds{1} - P_v^\omega$ for $V \in \Gamma_f$ with derivation $\delta^\omega$ has $\omega$ as its unique ground state.
\end{lem}
\begin{proof}
    Since $\omega$ is a product state, $\omega^v \coloneqq \omega|_{\cstar[v]}$ is pure. Thus it represented by a vector $\ket{\psi_v} \in \hilb_v$. Let $P_v^\omega \in \hilb_v$ be the rank-$1$ projection to $\ket{\psi_v}$. Then the Hamiltonian $H^\omega_v \coloneqq  \mathds{1} - P_v^\omega$ has $\omega^v$ as its unique ground state. 
    Observe that $P^\omega_v$ is the unique state with this property since $\omega^v$ is represented by the vector $\ket{\psi_v}$.

    We can repeat this analysis to obtain the family of orthogonal rank-$1$ projections $\{P_v^\omega\}_{v \in \Gamma}$. Then $\omega$ satisfies $\omega(P_v^\omega) = 1$ for all $v \in \Gamma$, so $\omega(H_V^\omega) = 0$ for all $V \in \Gamma_f$. Since $\omega|_{\cstar[V]} = \bigotimes_{v \in V} \omega^v$ for all $V \in \Gamma_f$, the derivation $\delta^\omega$ corresponding to $H_V^\omega$ has $\omega$ as its unique ground state.

    Now let $\omega'$ be another state satisfying $\omega'(P_v^\omega) = 1$ for all $v \in \Gamma$. Then we have that $\omega'(H_V^\omega) = 0$ for all $v \in V$. By \cite[Lem ~3.8]{MR3764565}, $\omega'$ is a frustration free ground state of $\delta^\omega$. But the ground state of $\delta^\omega$ is unique, so $\omega' = \omega$.
\end{proof}

\begin{defn}
    For the product state $\omega_0$ satisfying $\omega_0 \circ \alpha = \tilde \omega$ for the $G-$SPT $\tilde \omega$, we define the corresponding unique rank-1 projections $P_v \coloneqq  P_v^{\omega_0}$, Hamiltonian $H_V^0 \coloneqq  H_V^{\omega_0}$ and corresponding derivation $\delta_0\coloneqq  \delta^{\omega_0}$ from Lemma \ref{lem:existence of gapped Hamiltonian}.
\end{defn}

\begin{lem}
\label{lem:Pv invariant under symmetry action}
    Let $\omega$ be a product state satisfying for all $g \in G$ that $\omega \circ \beta_g = \omega$. Then $\beta_g(P_v^\omega) = P_v^\omega$ for all $g \in G$ and $v \in \Gamma$, where $P_v^\omega$ are the projections defined in Lemma \ref{lem:existence of gapped Hamiltonian}. In particular, $\beta_g(H_V^\omega) = H_V^\omega$ for all $V \in \Gamma_f$ and $g \in G$.
\end{lem}
\begin{proof}
    Since $\omega$ is invariant under the symmetry, $\omega(\beta_g(P_v^\omega)) = 1$ for all $v \in \Gamma$. Since $\omega$ is a product state, applying Lemma \ref{lem:can freely insert and remove P from the ground state.}, we have $$1 = \omega (P_v^\omega) = \omega (\beta_g(P_v^\omega) P_v^\omega \beta_g(P_v^\omega)).$$ 
    But the condition $\omega (\beta_g(P_v^\omega) P_v^\omega \beta_g(P_v^\omega)) = 1$ only holds if $\beta_g(P_v^\omega) = P_v^\omega$ since $P_v^\omega, \beta_g(P_v^\omega) \in \cstar[v]$ are both rank-1 projections.
\end{proof}

\subsection{FDQC Hamiltonian}
We define for all $V \in \Gamma_f$ the Hamiltonian $H_V$ given by $$H_V \coloneqq \sum_{v \in V} \mathds{1} - Q_v \qquad \qquad Q_v \coloneqq \alpha^{-1}(P_v)$$ and define $\tilde \delta$ as the corresponding derivation.

\begin{lem}
\label{lem:unique GS of SPT}
    The state $\tilde \omega$ is the unique ground state of derivation $\tilde \delta$. 
    In addition, $\tilde \omega$ is the unique state satisfying that $\tilde \omega(Q_v) = 1$.
    In particular, $\tilde \omega$ is pure.
\end{lem}
\begin{proof}
We have by the definition of $\tilde \omega$ and by Lemma \ref{lem:existence of gapped Hamiltonian} that for all $v \in \Gamma$, $$\tilde \omega(Q_v) = \omega_0 \circ \alpha(\alpha^{-1}(P_v)) = \omega_0 (P_v) = 1.$$
Therefore, we have that $\tilde \omega(H_V) = 0$ for all $V \in \Gamma_f$.  
Thus, by \cite[Lem.~3.8]{MR3764565}, $\tilde\omega$ is a frustration free ground state for $\tilde \delta$.

We now show uniqueness for $\tilde\omega$.
First, suppose $\omega'$ is another ground state for $\tilde \delta$.  
Then by Lemma \ref{lem:QCAs preserve the ground state subspace}, $\omega' \circ \alpha^{-1}$ is a ground state for $\delta_0$.  
By Lemma \ref{lem:existence of gapped Hamiltonian}, $\omega' \circ \alpha^{-1} = \omega_0$.  
Therefore $\omega' = \omega_0 \circ \alpha = \tilde \omega$.

Now, suppose $\omega'$ is another state satisfying $\omega'(Q_v) = 1$ for all $v \in \Gamma$. Then we have that $\omega'(H_V) = 0$ for all $V \in \Gamma_f$, so $\omega'$ is a frustration free ground state (\cite[Lem ~3.8]{MR3764565}). But by the above argument there is a unique ground state of $\tilde \delta$. Thus $\omega' = \tilde \omega$ showing the required result.
\end{proof}

Note that the set $\{Q_v\}_{v \in \Gamma}$ is a set of commuting projections since they are the image of the projections $P_v$ under $\alpha^{-1}$. 
Therefore, $H_V$ is a commuting projector Hamiltonian for all $V \in \Gamma_f$.

\begin{lem}
    \label{lem:Qv are invariant under the symmetry action}
    For all $g \in G$ and $v \in \Gamma$ we have that $\beta_g(Q_v) = Q_v$.
    In particular, this implies $\beta_g(H_V) = H_V$ for all $V \in \Gamma_f$.
\end{lem}
\begin{proof}
By applying Lemma \ref{lem:Pv invariant under symmetry action} for $\omega_0$, we get $\beta_g(P_v) = P_v$ for all $v \in \Gamma$ and $g \in G$. By Assumption \ref{asmp:entangler commutes with symmetry}, we have that $\alpha^{-1} \circ \beta_g = \beta_g \circ \alpha^{-1}$, so for $v \in \Gamma$,
    \[
    \beta_g(Q_v)
    =
    \beta_g(\alpha^{-1}(P_v))
    =
    \alpha^{-1}(\beta_g(P_v))
    =
    \alpha^{-1}(P_v)
    =
    Q_v.
    \qedhere
    \]
\end{proof}

\subsection{Defects using automorphisms}
\label{sec:defects using auts}

\subsubsection{Paths and dual paths}
\label{sec:paths and dual paths}
We recall and elaborate on the definition of a path. Recall that a (self-avoiding) \emph{finite  path} $\gamma \subset \Gamma$ is defined as a set of distinct edges $\{e_i \in \Gamma \}_{i=1}^N$ such that for all $i >1$, $e_i \cap e_{i-1}$ contains a single vertex. We call $\partial_0  \gamma \coloneqq \partial_0 e_1$ as the start of $ \gamma$ and $\partial_N  \gamma \coloneqq \partial_1 e_N$ as the end of $\gamma$. When $\partial_0 \gamma \neq \partial_1 \gamma$ we call it an open path.

Two paths $\gamma_1, \gamma_2$ can be added, denoted $\gamma_1 + \gamma_2 \coloneqq  \gamma_1 \cup \gamma_2$ if the start of $\gamma_1$ is the end of $\gamma_2$ or vice versa. We can remove $\gamma_2$ from $\gamma_1$ if there exists a path $\gamma'$ such that $\gamma_1 = \gamma' + \gamma_2$; in that case, we write $\gamma' = \gamma_1 - \gamma_2$. Of course we may freely add or remove the empty path.

A \emph{positive half-infinite} path is defined as an ordered set $\gamma$ such that every finite interval $\gamma' \subset \gamma$ is a finite path, and that for every path $\gamma' \subset \gamma$ there exists another path $\gamma'' \subset \gamma$ such that $\gamma'' + \gamma' \subset \gamma$ is a path. For a positive half infinite path $\gamma$, we set $\partial_0 \gamma = \gamma_\emptyset$. Similarly, a \emph{negative half-infinite} is when $\gamma' + \gamma''\subset \gamma$ is a path instead. 
For a negative half infinite path $\gamma$, we set $\partial_1 \gamma = \gamma_\emptyset$.

An \emph{infinite path} $\gamma \subset \Gamma$ is defined as a subset $\gamma$ that is a positive half-infinite  path, as well as a negative half-infinite  path. For an infinite  path  $\gamma$ we set $\partial_0 \gamma = \partial_1 \gamma = \gamma_\emptyset$.

We use \emph{path} to refer to any of the above types of  paths when the distinction is unnecessary. We say that a  path is sufficiently nice if it has a well defined point at the boundary circle at infinity (cf.~ discussion in Section \ref{sec:cones}). The set of sufficiently nice  paths on $\Gamma$ is denoted by $P(\Gamma)$. For every sufficiently nice positive or negative half-infinite  path $\gamma$, there exists an infinite  path denoted by $L_\gamma$ such that $\gamma \subset L_\gamma$. We call this a \emph{completion} of $\gamma$.

Recall that a dual path is a path on $\bar \Gamma$, the lattice dual to $\Gamma$ (c.f.~Section \ref{sec:cones}). We denote $e \in \bar \gamma$ for some edge $e \in \Gamma$, if $\bar e \in \bar \gamma$ where $\bar e$ is the dual edge to $e$. We say that $\partial_i \bar \gamma = \triangle_{\partial_i \bar \gamma}$, where $\triangle_{\partial_i \bar \gamma}$ is the face corresponding to the dual vertex $\partial_i \bar \gamma$. All the above concepts can be imported for the definition of dual paths. 

\subsubsection{Construction of defect automorphisms} Since we consider only dual paths in this analysis, we drop $(\bar \cdot)$. We say the region $\Sigma \subset \Gamma$ is simply connected if it is a simply connect subcomplex of $\Gamma$. Note that $\Sigma^{+r}$ is then simply connected by definition for all $r \in \bbZ_{\geq 0}$ (except when $\Sigma^{+r}$ is an empty set).

We define entangled symmetry $\tilde \beta_g^\Sigma$ for a simply connected region $\Sigma \in \Gamma$ as $$\tilde \beta_g^\Sigma \coloneqq  \alpha^{-1} \circ \beta_g^\Sigma \circ \alpha$$

\begin{lem}
\label{lem:entangled symmetry looks like the regular symmetry in bulk}
    Let $\Sigma \subset \Gamma$ be a simply connected region, and let $s$ denote the spread of $\alpha$ as before. Then for all $A \in \cstar[\Sigma]$ and $A \in \cstar[(\Sigma^{+2s})^c]$, $$\tilde \beta_g^{\Sigma^{+s}}(A) = \beta_g^\Sigma(A) = \beta_g^{\Sigma^{+s}}(A).$$
\end{lem}
\begin{proof}
    First assume $A \in \cstar[\Sigma]$. 
    Then $\alpha(A) \in \cstar[\Sigma^{+s}]$, so we have have by Assumption \ref{asmp:entangler commutes with symmetry}, 
    \[
    \tilde \beta_g^{\Sigma^{+s}}(A)
    =
    \alpha^{-1} \circ \beta_g^{\Sigma^{+s}} \circ \alpha(A)
    =
    \alpha^{-1} \circ \beta_g \circ \alpha(A)
    =
    \alpha^{-1} \circ \alpha \circ \beta_g(A)
    =
    \beta_g^\Sigma(A).
    \]

    Now assume $A \in \cstar[(\Sigma^{+2s})^c]$.
    Then $\alpha(A) \in \cstar[(\Sigma^{+s})^c]$, so $\beta_g^{\Sigma^{+s}}(\alpha(A)) = \alpha(A)$. Therefore, we have that 
    \[
    \tilde \beta_g^{\Sigma^{+s}}(A)
    =
    \alpha^{-1} \circ \beta_g^{\Sigma^{+s}} \circ \alpha(A)
    =
    \alpha^{-1} \circ \alpha(A)
    =
    A
    =
    \beta_g^\Sigma(A).
    \]
    Combining these two results, we have shown the full result.
\end{proof}

\begin{rem}
    We note that Lemma \ref{lem:entangled symmetry looks like the regular symmetry in bulk} physically means that for all observables in the bulk of $\Sigma^{+s}$ or $(\Sigma^{+s})^c$, the FDQC commutes with the symmetry.
\end{rem}

Let us define $\epsilon_g^S := \tilde \beta_g^{\Sigma^{+s}} \circ (\beta_g^{\Sigma^{+s}})^{-1}$ to ease notation. Here $S := \Sigma^{+2s}\setminus \Sigma$ is the strip of width $2s$ localised on the boundary of $\Sigma^{+s}$. To motivate the use of $S$ we have the following Lemma.

\begin{lem}
\label{lem:localization along a strip}
    We have for any simply connected region $\Sigma \subset \Gamma$ that $\epsilon_g^S$ is an FDQC localized in S.
\end{lem}
\begin{proof}
    Observe that $\epsilon_g^S$ is an FDQC by construction.
    Now, consider $A \in \cstar[S^c]$. Then since $\beta_g$ is an onsite symmetry, we still have $\beta_g^{\Sigma^{+s}}(A) \in \cstar[S^c]$. By the result of Lemma \ref{lem:entangled symmetry looks like the regular symmetry in bulk} we then have,
    \[
        \epsilon_g^\Sigma (A) = \tilde \beta_g^{\Sigma^{+s}} \circ (\beta_g^{\Sigma^{+s}})^{-1}(A) = \beta_g^{\Sigma^{+s}} \circ (\beta_g^{\Sigma^{+s}})^{-1}(A) = A.
    \]

    We now show that $\epsilon_g^S(\cstar[S]) \subseteq \cstar[S]$.
    Let $A \in \cstar[S]$.
    Then for all $B \in \cstar[S^c]$, we have that 
    \[
    [\epsilon_g^S(A), B]
    =
    [\epsilon_g^S(A), \epsilon_g^S(B)]
    =
    \epsilon_g^S([A, B])
    =
    0.
    \]
    Hence $\epsilon_g^S(A) \in \cstar[S^c]' \cap \cstar = \cstar[S]$.
    Thus $\epsilon_g^S$ is localized in $S$.
\end{proof}

We now assume a technical condition that helps us prove that these automorphisms can be cut.

\begin{asmp_recall}
    (Assumption \ref{asmp:strip_aut_is_an_FDQC}) We assume that for any infinite dual path $L \in \bar P(\Gamma)$, the automorphism $\tilde \beta_g^{r(L)} \circ (\beta_g^{r(L)})^{-1}$ is an FDQC built from unitaries of finite support and localized in $L^{+s}$.
\end{asmp_recall}

\begin{rem}
    This assumption has been used mostly to ensure that the automorphism $\tilde \beta_g^{r(L)} \circ (\beta_g^{r(L)})^{-1}$ can be split into automorphisms $\alpha_\gamma, \alpha_{\eta}$ localized in $\gamma^{+s}$ and $\eta^{+s}$ respectively, where $L = \gamma \cup \eta \in \bar P(\Gamma)$ is a dual path.

    It is reasonable to assume this for FDQCs because this property seems to hold for all known SPT constructions in the literature with an onsite symmetry (for example \cite{PhysRevB.108.115144}). This assumptions also ensures that the index for QCAs as defined in \cite{MR2890305} is trivial, since there is no transfer along the cut.
\end{rem}

\begin{lem}
\label{lem:strip aut is split}
    Let $\gamma \in \bar P(\Gamma)$ be a half-infinite dual path and let $L_\gamma \in \bar P(\Gamma)$ be a completion of $\gamma$. Let $\xi \coloneqq L_\gamma - \gamma$. 
    Divide $L_\gamma^{+s}$ into disjoint halves $S^\gamma, S^{\xi}$ along $\gamma$, $\xi$, so that $S^\gamma \cup S^{\xi} = L_{\gamma}^{+s}$. We have
    $$\tilde \beta_g^{r(L_\gamma)} \circ (\beta_g^{r(L_\gamma)})^{-1} = \Xi \circ (\alpha_\gamma \otimes \alpha_{\xi}),$$
    where $\alpha_\gamma, \alpha_{\xi}$ are FDQCs localized in $S^\gamma$ and $S^{\xi}$ respectively and $\Xi$ is an inner automorphism implemented by a local unitary.
\end{lem}
\begin{proof}
    By Lemma \ref{lem:localization along a strip} $\tilde \beta_g^{r(L_\gamma)} \circ (\beta_g^{r(L_\gamma)})^{-1}$ is an FDQC localized in $L_\gamma^{+s}$. 
    Let the unitaries of the circuit be given by $\{\cB^d\}_{d=1}^D$. 
    By Assumption \ref{asmp:strip_aut_is_an_FDQC}, we may assume every $U \in \bigcup_{d = 1}^D \caB^d$ is localized in $L_\gamma^{+s}$. We can use this structure to define another automorphism implementing a FDQC localized around $\gamma$ as follows.
    For $d = 1, \dots, D$, we define
    $$\alpha^\gamma_d(A) \coloneqq \left(\prod_{U \in \cB^d, \supp(U) \subset S^\gamma} U\right) A \left(\prod_{U \in \cB^d, \supp(U) \subset S^\gamma} U\right)^*.$$
    Similarly, we define
    $$\alpha^{\xi}_d(A) \coloneqq \left(\prod_{U \in \cB^d, \supp(U) \subset S^{\xi} } U\right) A \left(\prod_{U \in \cB^d, \supp(U) \subset S^{\xi}} U\right)^*.$$
    Now we define $$\Xi_d(A) \coloneqq \left(\prod_{U \in \cB^d, \supp(U) \cap S^\gamma, S^{\xi} \neq \emptyset} U\right) A \left(\prod_{U \in \cB^d, \supp(U) \cap S^\gamma, S^{\xi} \neq \emptyset} U\right)^*.$$
    Since the unitaries in each depth have disjoint supports, we have $$\Xi_d \circ (\alpha^\gamma_d \otimes \alpha^{\xi}_d) (A) = \left(\prod_{U \in \cB^d} U\right) A \left(\prod_{U \in \cB^d} U\right)^*.$$
    Additionally, $\Xi_d$ is an inner automorphism implemented by a local unitary.

    We now define $\alpha^\gamma \coloneqq  \alpha^\gamma_1 \circ \cdots \circ \alpha^{\gamma}_D$ and $\alpha^{\xi} \coloneqq  \alpha^\xi_1 \circ \cdots \circ \alpha^{\xi}_D$.
    We then get $$\tilde \beta_g^{r(L_\gamma)} \circ (\beta_g^{r(L_\gamma)})^{-1} = \Xi_1 \circ (\alpha^\gamma_1 \otimes \alpha^{\xi}_1)  \circ \cdots \circ \Xi_D \circ (\alpha^\gamma_D \otimes \alpha^{\xi}_D) = \Xi \circ (\alpha^\gamma \otimes \alpha^{\xi}).$$
    Here we have used the fact that for any inner automorphism $\Xi$ implemented by a local unitary and FDQC $\eta$, there exists another inner automorphism $\Xi'$ also implemented by a local unitary such that $\Xi' \circ \eta = \eta \circ \Xi$.
    It is clear that $\alpha^\gamma, \alpha^{\xi}$ are FDQCs and that they are localized in $S^\gamma$ and $S^{\xi}$ respectively. The result follows.
\end{proof}

\begin{defn}
\label{def:defect automorphisms}
    Let $\gamma \in \bar P(\Gamma)$ be a half-infinite dual path.
    Let $L_\gamma \in \bar P(\Gamma)$ be a completion of $\gamma$, and let $\xi \coloneqq L_\gamma - \gamma$. 
    Then by Lemma \ref{lem:strip aut is split}, we have $\tilde \beta_g^{r(L_\gamma)} \circ (\beta_g^{r(L_\gamma)^{+s}})^{-1} = \Xi \circ (\eta^\gamma_g \otimes \eta_g^{\xi})$, where $S^\gamma, S^{\xi}$ are two halves of the strip $L_\gamma^{+s}$ along $\gamma, \xi$ respectively, $\Xi$ is an inner automorphism implemented by a local unitary, and $\eta_g^\gamma \in \Aut[\cstar[S^\gamma]], \eta_g^{\xi} \in \Aut[\cstar[S^{\xi}]]$.
    We define a $g$-defect automorphism to be
    \begin{equation*}
        \alpha_\gamma^g(A) \coloneqq  \eta_g^{\xi} \circ \beta_g^{r(L_\gamma)}.
    \end{equation*}
    Note that $\alpha_\gamma^g$ depends on the completion $L_\gamma$ of $\gamma$, but we suppress this dependence for ease of notation.
\end{defn}

In Section \ref{sec:SPTDefectSectorReps} below, we will show that these defect automorphisms can be used to define defect sectors according to Definition \ref{def:g-defect_sector}.

\begin{lem}
\label{lem:Qv relations with defect aut}
    Let $\gamma \in \bar P(\Gamma)$ be a half-infinite dual path and let $L_\gamma \in \bar P(\Gamma)$ be a completion of $\gamma$. 
    Then there exists a ball $V$ containing $\partial\gamma$ such that $\alpha_\gamma^g$ satisfies the following relations:
    \begin{equation*}
        \alpha^g_\gamma (Q_v) = \begin{cases}
            Q_v & v \in \Gamma \setminus (\gamma^{+2s} \cup V) \\
            \beta_g^{r(L_\gamma)}(Q_v) & v \in  \gamma^{+2s}\setminus V.
        \end{cases}
    \end{equation*}
\end{lem}
\begin{proof}
    We choose $V \subset \Gamma$ such that the local unitary implementing the automorphism $\Xi$ in Lemma \ref{lem:strip aut is split} is supported on $V$.
    As before, we let $\xi \coloneqq L_\gamma - \gamma$.
    
    First let $v \in (L_\gamma^{+2s})^c \setminus V$. We use Lemma \ref{lem:Qv are invariant under the symmetry action} to get $\beta_g(Q_v) = Q_v$. Now,
    \begin{align*}
        \alpha_\gamma^g(Q_v) = \eta_g^{\xi} \circ \beta_g^{r(L_\gamma)}(Q_v) = \beta_g^{r(L_\gamma)} (Q_v) = Q_v,
    \end{align*}
    where we have used that $\eta_g^{\xi}$ is localized around $\gamma^{+2s}$.
    Now let $v \in \xi^{+2s} \setminus V$. Then we have
    \begin{align*}
        \alpha_\gamma^g( Q_v) &= \eta_g^{\xi} \circ \beta_g^{r(L_\gamma)}( Q_v) = \tilde \beta_g^{r(L_\gamma)} \circ (\beta_g^{r(L_\gamma)})^{-1}\circ \beta_g^{r(L_\gamma)}( Q_v)\\
        &= \alpha^{-1} \circ \beta_g^{r(L_\gamma)} \circ \alpha (Q_v)
        = \alpha^{-1} \circ \beta_g^{r(L_\gamma)} (P_v) = \alpha^{-1}(P_v) = Q_v.
    \end{align*}
    Finally, let $v \in \gamma^{+2s} \setminus V $. Then since $Q_v \in \cstar[(\xi^{+2s})^c]$, we have
    \begin{align*}
        \alpha_\gamma^g( Q_v) &= \eta_g^{\xi} \circ \beta_g^{r(L_\gamma)}( Q_v) = \beta_g^{r(L_\gamma)}( Q_v).
    \end{align*}
    This completes all the cases and finishes the proof. 
\end{proof}

\subsection{Defect Hamiltonians}
\label{sec:defect Hamiltonians SPT}
Symmetry defects in topological order have been well explored in the literature (see for instance \cite{PhysRevLett.105.030403}, \cite{PhysRevB.100.115147}). Here we expand on the approach of \cite[Sec~ 5]{PhysRevB.100.115147} in order to explicitly construct a \emph{defect Hamiltonian}, whose ground state has a symmetry defect. 
In particular, we give the general procedure to construct a commuting projector Hamiltonian from $H_V$ that houses a symmetry defect at the end-points of $\gamma$, and a domain wall along $\gamma$. Choose a dual path $\gamma \in \bar P(\Gamma)$. 
We use the results of Lemma \ref{lem:Qv relations with defect aut} in the following definition.

\begin{defn}
    We define the \emph{defect Hamiltonian} to be $$H_V^{(g,\gamma)} \coloneqq  \sum_{v \in V} \mathds{1} - \widehat Q_v^g \qquad \qquad \widehat Q_v^g \coloneqq  (\alpha_\gamma^g)^{-1}(Q_v)$$ and let $\delta^{(g,\gamma)}$ be its corresponding derivation.
\end{defn}

For any chosen $g \in G$ and $\gamma \in \bar P(\Gamma)$, there always exists a commuting projector Hamiltonian $H_V^{(g,\gamma)}$ for all $V \in \Gamma_f$ with corresponding derivation $\delta^{(g,\gamma)}$ whose unique ground state houses a $g$-defect at the endpoints of $\gamma$ and is given by $\tilde \omega_\gamma^g \coloneqq  \tilde \omega \circ \alpha_{\gamma}^g$. Indeed, $H_V^{(g,\gamma)}$ is a commuting projector Hamiltonian since it is the image of the commuting projector Hamiltonian $H_V$ under the automorphism $(\alpha_\gamma^g)^{-1}$.
In addition, for all $v \in \Gamma$, $$\tilde \omega_\gamma^g(\widehat Q_v) = \tilde \omega \circ \alpha^g_\gamma(\widehat Q_v) = \tilde \omega(Q_v) = 1,$$ so $\tilde \omega^g_\gamma$ is a frustration free ground state of $\delta^{(g,\gamma)}$ by \cite[Lem ~3.8]{MR3764565}. By Lemmas \ref{lem:QCAs preserve the ground state subspace}, \ref{lem:unique GS of SPT} we have that $\tilde \omega^g_\gamma$ is the unique ground-state of $\delta^{(g,\gamma)}$.

\begin{rem}
    We note that the general idea of constructing a defect Hamiltonian $H_V^{(g, \gamma)}$ by `symmetry twisting'' the projections of the original Hamiltonian $H_V$ has already been discussed in \cite{PhysRevB.100.115147} and was the inspiration for this construction. The original construction does not specify how to handle the case when the defect lies in the support of the projections. Here we are able to design a commuting projector Hamiltonian while circumventing that issue. 
\end{rem}

\subsection{Defect sector representations}
\label{sec:SPTDefectSectorReps}
Recall the definition of a $g$-defect sector (Definition \ref{def:g-defect_sector}). Our reference representation will now be $\tilde \pi$, the GNS representation of $\tilde \omega$, unless stated otherwise. By Lemma \ref{lem:unique GS of SPT}, we have that $\tilde \pi$ is irreducible. 
We also note that $\pi_0 \circ \alpha \simeq \tilde \pi$ by uniqueness of the GNS representation.

\begin{lem}
\label{lem:trivial-para-Haag-duality}
    Let $\omega$ be a product state and $\pi$ its GNS representation. The representation $\pi$ satisfies strict Haag duality, i.e, we have for all cones $\Lambda$ that $$\pi(\cstar[\Lambda^c])' = \pi(\cstar[\Lambda])''.$$
\end{lem}
\begin{proof}
    Since $\omega$ is a product state we can apply \cite[Lem.~4.3]{MR4426734} to get that for any cone $\Lambda$, $\pi$ satisfies strict Haag duality.
\end{proof}

We now show that the assumptions in Section \ref{sec:GCrossedAssumptions} are satisfied.  
Assumptions \ref{asmp:Faithfulness} and \ref{asmp:GInvariance} are satisfied by assumption.
By Lemma \ref{lem:trivial-para-Haag-duality} we have that $\pi_0$ satisfies strict Haag duality. Using the fact that $\tilde \pi = \pi_0 \circ \alpha$ and Lemma \ref{lem:QCABSHaagDuality} we conclude that $\tilde \pi$ satisfies bounded spread Haag duality (Assumption \ref{asmp:BoundedSpreadHaagDuality}).
By Lemma \ref{lem:unique GS of SPT}, $\tilde \omega$ is pure, so Assumption \ref{asmp:PureState} is satisfied.  
Assumption \ref{asmp:InfiniteFactor} is satisfied by \cite[Lem.~5.3]{MR4362722} because $\tilde\omega$ is a gapped ground state of a Hamiltonian with uniformly bounded finite range interactions.

The following lemma shows that $\pi_0$ has trivial superselection theory since $\omega_0$ is a product state.

\begin{lem}
\label{lem:trivial para has trivial superselection theory}
    Let $\omega$ be a product state. The corresponding GNS representation $\pi\colon \cstar \rightarrow \caB(\hilb)$ has trivial superselection theory.
\end{lem}
\begin{proof}
    Since $\omega$ is a product state, for any chosen cone $\Lambda$ we have $\omega = \omega^{\Lambda} \otimes \omega^{\Lambda^c}$. Now let $\pi^\Lambda$ be the GNS representation of $\omega^{\Lambda}$ and $\pi^{\Lambda^c}$ the GNS representation of $\omega^{\Lambda^c}$. Using the uniqueness of the GNS representation, we have $\pi \simeq \pi^{\Lambda} \otimes \pi^{\Lambda^c}$. We now apply \cite[Thm.~4.5]{MR4426734} to get the required result.
\end{proof}

We now recall the automorphism $\beta_g^S \colon \cstar \to \cstar$ for $g \in G$ and $S \subset \Gamma$ (Definition \ref{def:SymmetryOnSubsets}).

\begin{lem}
\label{lem:triv para is sectorizable}
    Let $\omega$ be a product state such that $\omega \circ \beta_g = \omega$ and let $\pi$ be its GNS representation. For all $V \subseteq \Gamma$ and $h \in G$,
    $\pi \circ \beta_h^V$ is a $g$-sectorizable representation with respect to $\pi$ for all $g \in G$.
\end{lem}
\begin{proof}
    By Lemma \ref{lem:Pv invariant under symmetry action} we have $\beta_g(P_v^\omega) = P_v^\omega$ for all $v \in \Gamma$. Now, given $V \in \Gamma_f$ and $v \in \Gamma$, we have $$\omega \circ \beta_h^V (P_v^\omega) = \omega (P_v^\omega) = 1.$$ By Lemma \ref{lem:existence of gapped Hamiltonian} we thus have that $\omega \circ \beta_h^V = \omega$ for all $h$. So by uniqueness of the GNS representation, we get that $\pi \simeq \pi \circ \beta_h^V$. 
    Therefore, it suffices to show that $\pi$ is a $g$-sectorizable representation for all $g \in G$. 
    But this is true since for all $g \in G$, we have $\omega \circ \beta_g^{r(\Lambda)} = \omega$, implying $\pi \simeq \pi \circ \beta_g^{r(\Lambda)}$. 
    Thus $\pi$ and hence $\pi \circ \beta_h^V$ is $g$-sectorizable with respect to $\pi$.
\end{proof}

The previous Lemma implies in particular that $\pi_0$ is a $g$-sectorizable representation with respect to $\pi_0$ for all $g \in G$.

\begin{prop}
\label{prop:defect theory of trivial para}
    Let $\omega$ be a product state such that $\omega \circ \beta_g = \omega$ and let $\pi$ be its GNS representation. Then $\pi$ has trivial defect theory.
\end{prop}
\begin{proof}
    Let $\widehat\pi$ be a $g$-sectorizable representation with respect to $\pi$. Then from Lemma \ref{lem:defect_sectorizable_reps_are_equivalent_to_a_defect_sector}, $\widehat\pi \simeq \pi'$ where $\pi'$ is a $g$-defect sector.  
    Now, by Lemma \ref{lem:triv para is sectorizable}, $\pi$ is a $g$-sectorizable representation, so $\pi \simeq \sigma$, where $\sigma$ is another $g$-defect sector.
    Finally, by Lemma \ref{lem:GDefectsRelationToAnyons}, $\pi'$ is an anyon sector with respect to $\sigma$. Using the above results, $\widehat \pi \simeq \pi'$ and $\pi \simeq \sigma$, so $\widehat \pi$ must be an anyon sector with respect to $\pi$. From Lemma \ref{lem:trivial para has trivial superselection theory}, $\pi$ has trivial superselection theory, so $\widehat \pi \simeq \pi$.
\end{proof}

The following Lemma shows that $\tilde \pi$ has trivial superselection theory.

\begin{lem}
\label{lem:SPT has trivial superselection theory}
    Let $\pi$ be the GNS representation of a product state $\omega$, and let $\alpha$ be a quasi-factorizable automorphism. Then the representation $\widehat \pi\coloneqq \pi \circ \alpha$ has trivial superselection theory.
\end{lem}
\begin{proof}
    Let $\pi'$ be an anyon sector with respect to $\widehat \pi$. We apply \cite[Theorem ~4.7]{MR4426734} to get that $\pi' \circ \alpha^{-1}$ must be an anyon sector with respect to $\widehat \pi \circ \alpha^{-1} = \pi$. But from Lemma \ref{lem:trivial para has trivial superselection theory} we have that $\pi$ has trivial superselection theory, implying $\pi' \circ \alpha^{-1} \simeq \pi$. So we have $$\pi' \simeq \pi \circ \alpha \simeq \widehat \pi,$$ giving us the required result.
\end{proof}

We now show that the symmetry defect states $\tilde\omega_\gamma^g$ are finitely transportable. 
Let us first prepare an important lemma.

\begin{rem}
\label{rem:SubsetSpreadComplements}
If $V$ is a ball or a cone and $V \Subset_r V'$, then $V'^c \Subset_r V^c$.  
Indeed, we have that $((V^{+r})^c)^{+r} = V^c$, so we have that
\[
(V'^c)^{+r}
\subseteq
((V^{+r})^c)^{+r}
=
V^c.
\]
\end{rem}

\begin{lem}
\label{lem:restrictions of ground states are equal SPT generalization}
    Let $\gamma \in \bar P(\Gamma)$ be a half-infinite dual path, so that $\widehat Q_v^g = (\alpha^g_\gamma)^{-1}(Q_v)$.
    Let $V, V'$ be balls satisfying $V \Subset_{s^\gamma} V'$, where $s^\gamma$ is the spread of $(\alpha^g_\gamma)^{-1} \circ \alpha^{-1}$.
    Let $\omega_1, \omega_2 \in \cS(\cstar)$ be two states such that $\omega_1(\widehat Q_v^g) = \omega_2(\widehat Q_v^g) = 1$ for all $v \in V^c$. Then we have $$\omega_1 |_{\cstar[V'^c]} = \omega_2 |_{\cstar[V'^c]}.$$
\end{lem}
\begin{proof}
    Define the product state $\omega_0^{V^c} \coloneqq  \omega_0 |_{\cstar[V^c]} \in \cS(\cstar[V^c])$. 
    Suppose $\omega \in \cS(\cstar)$ satisfies that for every $v  \in V^c$, $\omega(P_v) = 1$.
    We claim that $\omega = \omega_0^{V^c}$.
    Indeed, let $A \in \cstar[V^c;\loc]$ be a simple tensor. 
    That is, $A = \bigotimes_{v \in W} A_v$ for some $W \in (V^c)_f$, where $A_v \in \cstar[v]$.
    We then have that 
    \[
    \omega(A)
    =
    \omega\!\left(\bigotimes_{v \in V} A_v\right)
    =
    \omega\!\left(\bigotimes_{v \in V} P_v A_v P_v \right).
    \]
    Now, $P_v A_v P_v \in \bbC P_v$ for all $v \in V$, so $\bigotimes_{v \in V} P_v A_v P_v = \lambda \bigotimes_{v \in V} P_v$ for some $\lambda \in \bbC$.
    Therefore, we have that 
    \[
    \omega(A)
    =
    \omega\!\left(\bigotimes_{v \in V} P_v A_v P_v \right)
    =
    \lambda \omega\!\left(\bigotimes_{v \in V} P_v\right)
    =
    \lambda.
    \]
    By the same argument, $\omega_0^{V^c}(A) = \lambda$.
    Since the simple tensors span a dense subset of $\cstar[V^c]$, we get that $\omega = \omega_0^{V^c}$.
    
    Now, suppose $\omega_1, \omega_2 \in \cS(\cstar)$ satisfy that $\omega_1(\widehat Q_v^g) = \omega_2(\widehat Q_v^g) = 1$ for all $v \in V^c$.
    In that case for $i = 1, 2$, we have that $\tilde \omega_i \coloneqq \omega_i  \circ (\alpha_\gamma^g)^{-1} \circ \alpha^{-1}$ satisfies that for all $v \in V^c$, 
    \[
    \tilde \omega_i(P_v)
    =
    \omega_i \circ (\alpha^g_\gamma)^{-1}(Q_v)
    =
    \omega_i(\widehat{Q}_v)
    =
    1.
    \]
    Thus by the previous paragraph, we have that for every $A \in \cstar[V^c]$, $\tilde \omega_1(A) = \tilde \omega_2(A)$.
    Now, since $V \Subset_{s^\gamma} V'$, we have that $V'^c \Subset_{s^\gamma} V^c$ by Remark \ref{rem:SubsetSpreadComplements}. 
    Thus, if $A \in \cstar[V'^c]$, we have that $ \alpha \circ \alpha^g_\gamma(A) \in \cstar[V^c]$, so we have that 
    \[
    \omega_1(A)
    =
    \tilde \omega_1( \alpha(\alpha^g_\gamma(A)))
    =\tilde \omega_2( \alpha(\alpha^g_\gamma(A)))
    =
    \omega_2(A).
    \qedhere
    \]
\end{proof}

\begin{lem}
\label{lem:defects are finitely transportable SPT case}
    Let $\gamma_1, \gamma_2 \in \bar P(\Gamma)$ be such that $\gamma_1 \cap \gamma_2 \in \bar P (\Gamma)$ (see Figure \ref{fig:paths that only differ finitely}). Then $\tilde \omega_{\gamma_1}^g \simeq \tilde \omega_{\gamma_2}^g$.
        \begin{figure}[!ht]
        \centering
        \includegraphics[width=0.2\linewidth]{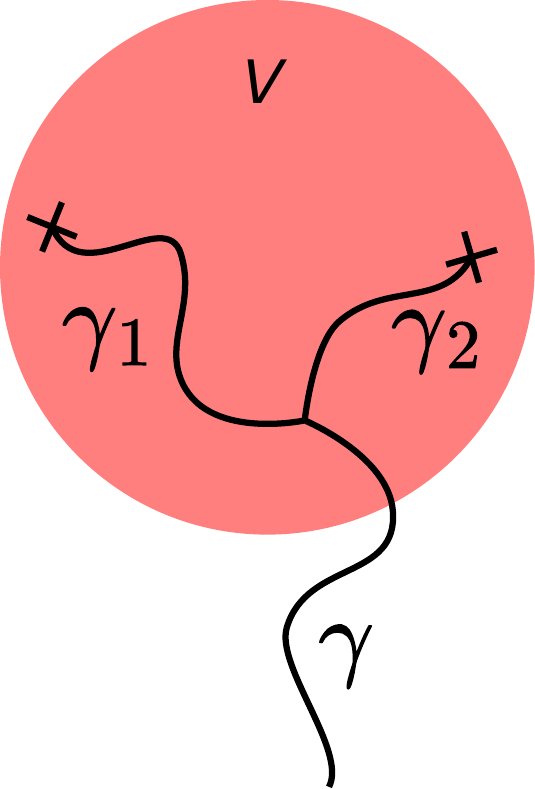}
        \caption{An example of two half-infinite dual paths $\gamma_1, \gamma_2 \in \bar P(\Gamma)$ such that $\gamma_1 \cap \gamma_2 = \gamma \in \bar P(\Gamma)$ is another half-infinite dual path, i.e, $\gamma_1, \gamma_2$ differ only in a finite region $V$. The region $V$ is designed such that $V^c \cap \gamma_1 \subset \gamma$ and $V^c \cap \gamma_2 \subset \gamma$}
        \label{fig:paths that only differ finitely}
    \end{figure}
\end{lem}
\begin{proof}
    Since $\tilde \omega_{\gamma_1}^g$ and $\tilde \omega_{\gamma_2}^g$ are pure states, $\tilde \omega_{\gamma_1}^g$ and $\tilde \omega_{\gamma_2}^g$ are equivalent if and only if they are quasi-equivalent \cite[Prop.~10.3.7]{MR1468230}.
    We can therefore apply \cite[Cor.~2.6.11]{MR887100}. 
    Since $\gamma_1 \cap \gamma_2 \in \bar P(\Gamma)$, by Lemma \ref{lem:Qv relations with defect aut}, there exists a ball $V \in \Gamma_f$ such that for every $v \in V^c$, $\alpha_{\gamma_1}^g(Q_v) = \alpha_{\gamma_2}^g(Q_v)$.  
    Therefore, by Lemma \ref{lem:restrictions of ground states are equal SPT generalization}, $\tilde \omega_{\gamma_1}^g|_{\cstar[(V^{+r})^c]} = \omega_{\gamma_2}^g|_{\cstar[(V^{+r})^c]}$ for some $r \geq 0$, so by \cite[Cor.~2.6.11]{MR887100}, $\tilde \omega_{\gamma_1}^g \simeq \tilde \omega_{\gamma_2}^g$.
\end{proof}

\begin{defn}
\label{def:set of half-infinite paths equivalent to R}
    We define for the fixed dual path ${\bar \gamma_R}$ (Figure \ref{fig:chosen_ray}) the set $\bar P_R(\Gamma)$ as follows. $$\bar P_R(\Gamma) \coloneqq  \{\gamma \in \bar P(\Gamma) : \gamma \text{ differs from ${\bar \gamma_R}$ on finitely many sites}\}$$
\end{defn}

We will now show that for $\gamma \in \bar P_R(\Gamma)$ and $g \in G$, $\tilde \pi_\gamma^g \coloneqq \tilde \pi \circ \alpha_\gamma^g$ is a $g$-defect sector with respect to $\tilde \pi$.
Note that $\tilde \pi_\gamma^g$ is a GNS representation of $\tilde\omega_\gamma^g = \tilde \omega \circ \alpha_\gamma^g$.

\begin{lem}
\label{lem:defect state GNS reps are sectorizable (SPT case)}
    For all $\gamma \in \bar P_R(\Gamma)$ and $g \in G$, $\tilde\pi^g_\gamma$ is an irreducible $g$-defect sector.
\end{lem}
\begin{proof}
    As before, we define $L_\gamma$ be a completion of $\gamma$ into a infinite dual path, and let $\xi \coloneqq L_\gamma - \gamma$.
    Let $\Lambda \in \cL$ be a cone such that $\xi^{+2s} \subset \Lambda$. 
    We show that $\tilde \pi_\gamma^g$ is $g$-localized in $\Lambda$.
    Recall that $\alpha^\gamma = \eta_g^{\xi} \circ \beta_g^{r(L_\gamma)}$, where $\eta_g^{\xi}$ only acts nontrivially on $\eta^{+2s}$.  
    Therefore, we have that $\eta_g^{\xi}|_{\cstar[\Lambda^c]} = \id$, so we have that 
    \[
    \tilde \pi_\gamma^g|_{\cstar[\Lambda^c]}
    =
    \tilde \pi \circ \alpha_\gamma^g|_{\cstar[\Lambda^c]}
    =
    \tilde \pi \circ \eta_g^{\xi} \circ \beta_g^{r(L_\gamma)}|_{\cstar[\Lambda^c]}
    =
    \tilde \pi \circ \beta_g^{r(L_\gamma)}|_{\cstar[\Lambda^c]}.
    \]
    Since $\gamma \in \bar P_R(\Gamma)$, $r(L_\gamma) \cap \Lambda^c$ differs from $r(\Lambda)$ by finitely many vertices.  
    Therefore, $\tilde \pi_\gamma^g$ is $g$-localized in $\Lambda$.

    It remains to show that $\tilde \pi_\gamma$ is transportable.  
    Let $\Lambda' \in \cL$ be another cone.  
    We choose a dual path $\widehat{\gamma} \in \bar P_R(\Gamma)$ and an extension $L_{\widehat\gamma}$ of $\widehat{\gamma}$ such that $(L_{\widehat\gamma} - \widehat{\gamma})^{+2s} \subset \Lambda'$.
    Then by the preceding argument, $\tilde \pi_{\widehat{\gamma}}^g$ is $g$-localized in $\Lambda'$.  
    Furthermore, by Lemma \ref{lem:defects are finitely transportable SPT case}, $\tilde \pi_\gamma^g \simeq \tilde \pi_{\widehat{\gamma}}^g$. 
    Thus $\tilde \pi_\gamma^g$ is transportable, so $\tilde \pi_\gamma$ is a $g$-defect sector.
\end{proof}

\begin{lem}
\label{lem:classification of defect sectorizable representations SPT}
    Let $g \in G$. Every $g$-sectorizable representation is unitarily equivalent to the representation $\tilde \pi_\gamma^g$ for some $\gamma \in \bar P_R(\Gamma)$.
\end{lem}
\begin{proof}
    Let $\pi$ be a $g$-sectorizable representation. We have from Lemma \ref{lem:defect_sectorizable_reps_are_equivalent_to_a_defect_sector} that $\pi \simeq \sigma$ for some $\sigma$ being a $g$-defect sector.
    From Lemma \ref{lem:defect state GNS reps are sectorizable (SPT case)} we have that $\tilde \pi_\gamma^g$ is a $g$-defect sector. 
    We have from Lemma \ref{lem:GDefectsRelationToAnyons} that $\sigma$ is an anyon sector with respect to $\tilde \pi_\gamma^g$. 
    Now, $\tilde \pi_\gamma^g = \tilde \pi \circ \alpha_\gamma^g$, and $\alpha_\gamma^g$ is an FDQC and thus quasi-factorizable by Lemma \ref{lem:FDQCQuasi-Factorizable}.
    Furthermore, by Lemma \ref{lem:SPT has trivial superselection theory} the superselection theory of $\tilde \pi$ is trivial.
    Therefore, the superselection theory of $\tilde \pi_\gamma^g$ is trivial by the proof of Lemma \ref{lem:SPT has trivial superselection theory}.
    Putting these results together, we have $$\tilde \pi^g_\gamma \simeq \sigma \simeq \pi,$$ which gives us the required result.
\end{proof}

We have shown the following classification result,

\begin{prop}
    \label{prop:general SPT defect sector classification}
    Let $\tilde \omega$ be a $G$-SPT (Definition \ref{def:SPT_phase}) satisfying Assumptions  \ref{asmp:entangler commutes with symmetry}, \ref{asmp:strip_aut_is_an_FDQC}. Let $\tilde \pi$ be the GNS representation of $\tilde \omega$ and define $\tilde \pi_{{\bar \gamma_R}}^g \coloneqq  \tilde \pi \circ \alpha_{\bar\gamma_R}^g$, where $\alpha_{\bar\gamma_R}^g$ is the defect automorphism in Definition \ref{def:defect automorphisms}.

    The representations $\{\tilde \pi_{{\bar \gamma_R}}^g\}_{g \in G}$ are a family of disjoint and irreducible defect sectors, and any defect sectorizable representation $\pi$ is unitarily equivalent to some $\pi_{{\bar \gamma_R}}^g$.
\end{prop}

\begin{cor}
The category of $G$-defect sectors with respect to $\tilde \pi$ is equivalent to $\Vect(G, \nu)$ for some 3-cocycle $\nu \colon G \times G \times G \to U(1)$.
\end{cor}

\begin{proof}
Let $\GSec$ denote the category of $G$-defect sectors with respect to $\tilde \pi$.
We show that $\GSec$ is a fusion category with the same fusion rules as $\Vect(G)$.
The result will then follow.
By Proposition \ref{prop:general SPT defect sector classification} and the discussion in Section \ref{sec:GSecCauchyComplete}, $\GSec$ is a semisimple category whose simple objects are given by $\{\tilde \pi_{{\bar \gamma_R}}^g\}_{g \in G}$.
For $g, h \in G$, $\tilde \pi_{{\bar \gamma_R}}^g \otimes \tilde \pi_{{\bar \gamma_R}}^h$ is a $gh$-defect sector by Lemma \ref{lem:TensorProductPreservesGGrading}. 
By Lemma \ref{lem:classification of defect sectorizable representations SPT}, $\tilde \pi_{{\bar \gamma_R}}^g \otimes \tilde \pi_{{\bar \gamma_R}}^h \simeq \tilde \pi_{{\bar \gamma_R}}^{gh}$.
Thus, $\GSec$ is a fusion category with the same fusion rules as $\Vect(G)$, as desired.
\end{proof}

\section{\texorpdfstring{$\bbZ_2$}{Z2} SPTs: Trivial \texorpdfstring{$\bbZ_2$}{Z2} paramagnet}

Assign a qubit to each vertex $v \in \Gamma$. For the sake of simplicity we take $\Gamma$ to be the regular triangular lattice. Let $\sigma^x_v, \sigma^y_v, \sigma^z_v$ denote the Pauli matrices in $\cstar[v]$. 
Then $\{\mathds{1}, \sigma^x_v, \sigma^y_v, \sigma^z_v\}$ is a basis of $\cstar[v]$ for each $v$.

We now define a symmetry action on $\cstar$. Let $G = \bbZ_2$ and let $g\mapsto U^g_v$ be its unitary representation onto the vertex $v$, with $U^g_v = \sigma^x_v$ for the non-trivial group element $g \in \bbZ_2$ and $U^1_v = \mathds{1}$ for the trivial group element $1 \in \bbZ_2$. We can then define $\beta_g$ as in Definition \ref{def:GlobalSymmetryAutomorphism}.

\label{sec:TrivialParamagnet}
\subsection{Hamiltonian and ground state}
We define for any $V \in \Gamma_f$ the Hamiltonian $$H^0_V \coloneqq \sum_{v \in V} (\mathds{1} - \sigma^x_v)/2.$$ 
It is easy to verify that $H_V^0$ is a commuting projector Hamiltonian, with the projections $(\mathds{1} - \sigma^x_v)/2$ trivially commuting since they have disjoint supports for different $v, v'$. Let $\delta_0$ be the generator of dynamics. 

We set up some notation which we will use in the next couple of subsections. Fix a vertex $v_0$ as the origin. Let $\Gamma^n \subset \Gamma_f$ be be the set of vertices that are a graph distance at most distance $n \in \bbN$ away from $v_0$. Define $\hilb^n \coloneqq \hilb_{\Gamma^n}$. 

\begin{lem}
\label{lem:GS of trivial para is unique}
    There is a unique state $\omega_0$ satisfying that $\omega_0(\sigma_v^x) = 1$ for all $v \in \Gamma$. Moreover, $\omega_0$ is pure and a product state.
\end{lem}
\begin{proof}
    Let $\ket{\Omega_0^v} \in \hilb_v$ be a unit vector satisfying $\ket{\Omega_0^v} = \sigma^x_v \ket{\Omega_0^v}$. Define a state $\omega_0^v$ on $\cstar[v]$ given by $\omega_0^v(A) \coloneqq  \inner{\Omega_0^v}{A \Omega_0^v}$ for all $A \in \cstar[v]$. We can then define a product state $\omega^V_0$ on $\cstar[V]$ for all $V \in \Gamma_f$ given by $\omega^V_0 \coloneqq  \bigotimes_{v \in V} \omega_0^v$. By continuity, we can extend $\omega^V_0$ to a state $\omega_0$ on $\cstar$ satisfying $\omega_0 (\sigma^x_v) = 1$ for all $v \in \Gamma$. This shows existence of $\omega_0$. Note that $\omega_0$ by construction is a product state.

    Uniqueness of $\omega_0$ (and hence purity, c.f.~ discussion at the end of section \ref{sec:HamiltonianDynamics}) is easily shown using operators $S_v \coloneqq  (\mathds{1} + \sigma^x_v)/2$, Lemma \ref{lem:can freely insert and remove P from the ground state.} with $A = S_v$, and standard continuity arguments.
\end{proof}

Note that the interactions for the Hamiltonian $H_V^0$ are translation invariant.
Additionally, observe that $\omega_0(H^0_V) = 0$ for all $V \in \Gamma_f$, and $\omega_0$ is translation invariant since for any translation $\tau$, $\omega_0 \circ \tau(\sigma^x_v) = 1$ for all $v \in \Gamma$.  
Thus by \cite[Thm.~6.2.58]{MR1441540}, $\omega_0$ is a ground state and hence a frustration free one. 

\begin{lem}
    \label{lem:GS of trivial para is unique actual lemma}
    The state $\omega_0$ is the unique ground state of $\delta_0$.
\end{lem}
\begin{proof}
    Let $\omega$ be a ground state of $\delta_0$. Then we have $-i \omega(A^* \delta_0(A)) \geq 0$. Restrict to observables $A \in \cstar[v]$. Then $\delta_0(A) \in \cstar[v]$ and thus $A^* \delta_0(A) \in \cstar[v]$. Now let $\omega^v$ be the restriction of $\omega$ onto $\cstar[v]$. On finite volume, the infinite volume ground state condition reduces to the finite volume ground state condition \cite[Lem.~3.4.2]{MR3617688}. Since $H_v$ is positive, a unit vector $\ket{\psi} \in \hilb_v$ is a ground state vector if and only if $H_v \ket{\psi} = 0$. This uniquely fixes $\ket{\psi} = \frac{1}{\sqrt{2}}(\ket{0} + \ket{1})$ in the eigenbasis of $\sigma^z_v$ (up to phase). Thus $\omega^v (A)= \inner{\psi}{A \psi}$ for all $A \in \cstar[v]$, and hence $\omega^v(\sigma^x_v) = 1$. 

    We then have that $\omega(\sigma^x_v) = \omega^v(\sigma^x_v) = 1$, showing that $\omega(\sigma^x_v) = 1$ for all $v \in \Gamma$. By Lemma \ref{lem:GS of trivial para is unique}, $\omega_0$ is the unique state satisfying $\omega_0(\sigma^x_v) = 1$ for all $v \in \Gamma$. Thus we have the result.
\end{proof}

Instead of starting with $H_V^0$ and obtaining $\omega_0$ as its unique ground-state, we could instead have proceeded in the opposite direction of constructing a product state $\omega_0$, then working out a commuting projector Hamiltonian whose ground-state is $\omega_0$. This will allow us to connect to Section \ref{sec:general SPTs}.

Let $\omega_0$ be the product state defined in Lemma \ref{lem:GS of trivial para is unique}. By Lemma \ref{lem:existence of gapped Hamiltonian} we have the existence of unique projections $P_v \coloneqq  P_v^{\omega_0}$, Hamiltonian $H_V^0 \coloneqq  H_V^{\omega_0}$, and corresponding derivation $\delta_0 \coloneqq  \delta^{\omega_0}$. Uniqueness of $P_v$ implies that $P_v = S_v = (\mathds{1}+ \sigma^x_v)/2$.

For $g \in G$ and $V \subseteq \Gamma$, we recall the symmetry automorphism $\beta_g^{V}\colon \cstar \rightarrow \cstar$ from Definition \ref{def:SymmetryOnSubsets}. 
We observe that $\beta_g(\sigma^x_v) = \sigma^x_v$ for every $v \in \Gamma$, so $\omega_0 \circ \beta_g = \omega_0$ by Lemma \ref{lem:GS of trivial para is unique}.
Hence the assumptions of Lemma \ref{lem:Pv invariant under symmetry action} hold.

\subsection{Defect sector category}
We define $\pi_0$ to be the GNS representation of the ground state $\omega_0$. Note that by Lemma \ref{lem:GS of trivial para is unique} we have that $\omega_0$ is pure. Thus $\pi_0$ is an irreducible representation. We note that the assumptions in Section \ref{sec:GCrossedAssumptions} are satisfied in this setup, as it is a special case of the discussion in Section \ref{sec:general SPTs}. 

Since $\omega_0$ is a product state, it follows by Proposition \ref{prop:defect theory of trivial para} that every defect sectorizable representation is unitarily equivalent to $\pi_0$.
We now compute the defect sectors with respect to $\pi_0$.
In particular, we will show that the category $\GSec$ of such sectors is equivalent to $\Vect(\bbZ_2)$.

\begin{figure}[!ht]
    \centering
    \includegraphics[width=0.2\linewidth]{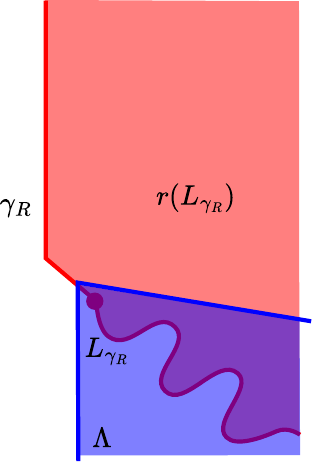}
    \caption{An example of $L_{{\bar \gamma_R}}$. We require the infinite dual path $L_{{\bar \gamma_R}} \in \bar P(\Gamma)$ to be the completion of ${\bar \gamma_R}$, such that $L_{{\bar \gamma_R}} - {\bar \gamma_R} \subset \Lambda$ for the chosen cone $\Lambda$ depicted in blue. The region $r(L_{{\bar \gamma_R}})$ is shown in red.}
    \label{fig:example of completion of gamma}
\end{figure}

Recall that $\bar \gamma_R$ is the fixed dual path as shown in Figure \ref{fig:chosen_ray}. 
Let $L_{\bar \gamma_R}$ be an appropriate extension of $\bar \gamma_R$ to form a infinite dual path and $r(L_{\bar \gamma_R}) \subseteq \Gamma$ be the region to the right of $L_{\bar \gamma_R}$. An example of $L_{\bar \gamma_R}$ is shown in Figure \ref{fig:example of completion of gamma}. We assume that the extension $L_{\bar \gamma_R}$ of $\bar \gamma_R$ is well-behaved in that there exists a cone $\Lambda \in \cL$ such that $L_{\bar \gamma_R} - \bar \gamma_R \subseteq \Lambda$.
We then have that $$\pi_0\circ \beta_g^{r(L_{\bar \gamma_R})}\big|_{\cstar[\Lambda^c]} = \pi_0 \circ \beta_g^{r(\Lambda)}\big|_{\cstar[\Lambda^c]},$$ so $\pi_0\circ \beta_g^{r(L_{{\bar \gamma_R}})}$ is $g$-localized in $\Lambda$.  
Furthermore, $\pi_0\circ \beta_g^{r(L_{{\bar \gamma_R}})}$ is $g$-sectorizable by Lemma \ref{lem:triv para is sectorizable}, so $\pi_0\circ \beta_g^{r(L_{{\bar \gamma_R}})}$ is unitarily equivalent to a $g$-defect sector by Lemma \ref{lem:defect_sectorizable_reps_are_equivalent_to_a_defect_sector} and hence transportable.
It follows that $\pi_0\circ \beta_g^{r(L_{{\bar \gamma_R}})}$ is in fact a $g$-defect sector. 
By Proposition \ref{prop:defect theory of trivial para}, $\pi_0$ is the only 1-defect sector and $\pi_0\circ \beta_g^{r(L_{{\bar \gamma_R}})}$ is the only $g$-defect sector, up to unitary equivalence.

We now compute the $F$-symbols for this category as done in Section \ref{sec:OtherCoherenceData}.
We observe that 
\[
\pi_0 \otimes \left(\pi_0\circ \beta_g^{r(L_{{\bar \gamma_R}})}\right) = \left(\pi_0\circ \beta_g^{r(L_{{\bar \gamma_R}})}\right) \otimes \pi_0 = \pi_0\circ \beta_g^{r(L_{{\bar \gamma_R}})}.
\]
Additionally, if $g \neq 1$, then 
\[
\left(\pi_0\circ \beta_g^{r(L_{{\bar \gamma_R}})}\right) \otimes \left(\pi_0\circ \beta_g^{r(L_{{\bar \gamma_R}})}\right)
=
\pi_0.
\]
Therefore, $\Omega_{1, 1} = \Omega_{g, 1} = \Omega_{1, g} = \Omega_{g, g} = \mathds{1}$.
Thus $\pi_0$ and $\pi_0\circ \beta_g^{r(L_{{\bar \gamma_R}})}$ generate a copy of $\Vect(\bbZ_2)$ with trivial cocycle inside $\GSec$.  
Since there is exactly one irreducible $g$-defect sector up to unitary equivalence for each $g \in \bbZ_2$, $\GSec \simeq \Vect(\bbZ_2)$.

\section{\texorpdfstring{$\bbZ_2$}{} SPTs: Levin-Gu SPT}
\label{sec:Levin Gu}
To define the non-trivial $\bbZ_2$-paramagnet, we follow \cite{PhysRevB.86.115109}. We import the setup from Section \ref{sec:TrivialParamagnet}. 
Given neighboring vertices $v,q,q' \in \Gamma$, we let $<vqq'>$ denote the (elementary) face formed by them. Let $\triangle_v$ denote the set of all triangles that vertex $v$ belongs to.

\subsection{Hamiltonian and ground state} We define for any $V \in \Gamma_f$ the Hamiltonian $H_V \in \cstar[V^{+1}]$ as follows $$H_{V} \coloneqq \sum_{v \in V} (\mathds{1} - B_v)/2 \qquad \qquad B_v \coloneqq - \sigma^x_v \prod_{<vqq'> \in \triangle_v} i^{\frac{1 - \sigma^z_q \sigma^z_{q'}}{2}}$$ 

Observe that $B_v, B_{v'}$ satisfy the following properties for all $v,v' \in \Gamma$:
\begin{align*}
    B_v^2 = 1, && B_v^* = B_v, && [B_v, B_{v'}] = 1.
\end{align*}
With the above properties, it is easily checked that $(\mathds{1} - B_v)/2$ is a projection, so the Hamiltonian $H_V$ is a commuting projector Hamiltonian for all $V \in \Gamma_f$. 
Let $\tilde \delta$ be the corresponding generator of dynamics.

We recall and rigorously define a useful `entangling' unitary given in \cite[Appendix A]{PhysRevB.86.115109}.  For each face in $\triangle \in \Gamma$, we define a unitary $U_{\triangle} \in \cstar[\triangle]$ given by $$U_{\triangle} \coloneqq e^{i \frac{\pi}{24} \left( 3 \prod_{v \in \triangle} \sigma^z_v - \sum_{v \in \triangle} \sigma^z_v \right)}.$$ 
Observe that we have $[U_{\triangle}, U_{\triangle'}] = 0$ for all faces $\triangle, \triangle' \in \Gamma$.

\begin{defn}
\label{def:LevinGuEntangler}
    For each $A \in \cstar[V']$ with $V' \in \Gamma_f$, let $V \in \Gamma_f$ be a sufficiently large supset of $V'$, i.e., for each site $v' \in V'$, $\triangle \subseteq V$ for all $\triangle \in \triangle_{v'}$. We define a map $\alpha \colon \cstar[V'] \rightarrow \cstar[V]$ given by 
    \begin{align*}
        \alpha (A) \coloneqq \left(\prod_{\triangle \subseteq V} U_{\triangle}\right) A \left(\prod_{\triangle' \subseteq V} U_{ \triangle'}\right)^*,
    \end{align*} 
    which can be uniquely extended in a norm continuous way to an automorphism $\alpha$ of $\cstar$.
\end{defn}

We remark that $\alpha$ is an FDQC of depth 3 with spread $s = 1$, as shown in detail in Lemma \ref{lem:alpha_theta is a bounded spread automorphism}.
In particular, $\alpha$ is a quasi-factorizable QCA.

Let $\omega_0$ be the unique ground-state of the trivial $\bbZ_2$ paramagnet as defined in Lemma \ref{lem:GS of trivial para is unique}. We define $\tilde\omega \coloneqq  \omega_0 \circ \alpha^{-1}.$

\begin{lem}
\label{lem:Levin-Gu ground-state is SPT}
    The state $\tilde \omega$ is a $\bbZ_2$-SPT satisfying Assumption \ref{asmp:entangler commutes with symmetry}.
\end{lem}
\begin{proof}
    Recall that $\omega_0$ is $\beta_g$-invariant, and by Lemma \ref{lem:alpha_theta is a bounded spread automorphism}, $\alpha$ is an FDQC.
    Therefore, by Remark \ref{rem:GInvarianceTrick}, $\tilde \omega$ is an $\bbZ_2$ provided that $\alpha$ satisfies Assumption \ref{asmp:entangler commutes with symmetry}. 
    We now show that Assumption \ref{asmp:entangler commutes with symmetry} is satisfied.
    By Lemma \ref{lem:automorphism connecting trivial para GS and Levin Gu GS}, we have that for every $v \in \Gamma$, 
    \begin{gather*}
    \alpha^{-1}(\beta_g(\sigma^x_v))
    =
    \alpha^{-1}(\sigma^x_v)
    =
    B_v
    =
    \beta_g(B_v)
    =
    \beta_g(\alpha^{-1}(\sigma^x_v)),
    \\
    \alpha^{-1}(\beta_g(\sigma^z_v))
    =
    \alpha^{-1}(-\sigma^z_v)
    =
    -\sigma^z_v
    =
    \beta_g(\sigma^z_v)
    =
    \beta_g(\alpha^{-1}(\sigma^z_v)).
    \end{gather*}
    Since $\{ \sigma^x_v, \sigma^z_v : v \in \Gamma\}$ generates $\cstar$, $\alpha^{-1} \circ \beta_g = \beta_g \circ \alpha^{-1}$, as desired.
\end{proof}

\begin{lem}
\label{lem:facts about Levin-Gu ground-state}
    We have the following facts about $\tilde \omega$:
    \begin{enumerate}
        \item $\tilde \omega$ is the unique state satisfying $\tilde \omega(B_v) = 1$ for all $v \in \Gamma$
        \item $\tilde \omega \circ \beta_g = \tilde \omega$ for all $g \in G$
        \item $\tilde \omega$ is the unique (hence pure) ground-state of $\tilde \delta$
        \item $\tilde \omega$ is translation invariant
    \end{enumerate}
\end{lem}
\begin{proof}
Since $\omega_0$ is a product state, from Lemma \ref{lem:existence of gapped Hamiltonian} we have corresponding projections $P_v \coloneqq P_v^{\omega_0}$ and Hamiltonians $H_V^0 \coloneqq H_V^{\omega_0}$. From Lemmas \ref{lem:existence of gapped Hamiltonian}, \ref{lem:GS of trivial para is unique} we get that $P_v = (\mathds{1} + \sigma^x_v)/2$ for all $v \in \Gamma$. 

Now, by Lemma \ref{lem:unique GS of SPT} we have that $\tilde\omega$ is the unique ground state of the Hamiltonian given by
$$H_V' = \sum_{v \in V} \mathds{1} - Q_v \qquad \qquad Q_v \coloneqq  \alpha(P_v)$$ 
and additionally that $\tilde \omega$ is the unique state satisfying $\tilde \omega(Q_v) = 1$ for all $v \in \Gamma$. 
By Lemma \ref{lem:automorphism connecting trivial para GS and Levin Gu GS}, $Q_v = \alpha(P_v) = (\mathds{1}+B_v)/2$, so $H_V' = H_V$.
Thus $\tilde \omega$ is the unique ground-state of $\tilde \delta$ and is the unique state satisfying $\tilde \omega(B_v) = 1$ for all $v \in \Gamma$. 
This proves (1) and (3) above. 
The statement in (2) is a direct consequence of Lemma \ref{lem:Levin-Gu ground-state is SPT}.
Finally, (4) follows from (1) since for any translation $\tau$, $\omega_0 \circ \tau(B_v) = 1$ for any $v \in \Gamma$.
\end{proof}

\subsection{Defect Hamiltonian}
\label{sec:Defect auts Hamiltonian Levin-Gu}

We construct a defect Hamiltonian by first constructing a defect automorphism that will give us a defect sector.
Recall the discussion in Section \ref{sec:paths and dual paths} about paths and dual paths, in particular the definition of $\bar P(\Gamma)$ and the definition of a completion $L_\gamma$ of a dual path $\gamma \in \bar P(\Gamma)$. 

We consider the defect automorphism $\alpha_\gamma^g$ from Definition \ref{def:defect automorphisms} with $g \in \bbZ_2$ being the non-trivial element. For the Levin-Gu SPT, it is possible to explicitly compute $\alpha_\gamma^g$. We do this computation in Appendix \ref{sec:LGDefectSectorComputationAppendix}.
In particular, this computation illustrates that Assumption \ref{asmp:strip_aut_is_an_FDQC} is satisfied for the Levin-Gu SPT.  

\begin{defn}
    Let $\gamma \in \bar P( \Gamma)$ be a dual path. Let $\alpha^\gamma \coloneqq  \alpha_\gamma^g$ be the defect automorphism (Definition \ref{def:defect automorphisms}) for the Levin-Gu SPT as constructed in Appendix \ref{sec:LGDefectSectorComputationAppendix}. 
    We the operator $\widehat B_v^\gamma$ as 
    \[
    \widehat B_v^\gamma
    \coloneqq
    (\alpha^\gamma)^{-1}(B_v).
    \]
    In cases where the dual path $\gamma$ is clear from context, we may simply write $\widehat B_v$ instead of $\widehat B_v^\gamma$.
\end{defn}

Specializing Lemma \ref{lem:Qv relations with defect aut} to the case of the Levin-Gu SPT, we have the following.
\begin{lem}
Let $\gamma \in \bar P(\Gamma)$.
If $v \notin \gamma$, then $\widehat{B}_v^\gamma = B_v$.
If $v \in \gamma - \partial_0\gamma - \partial_1\gamma$, then $\widehat{B}_v^\gamma = \beta_g^{r(L_\gamma)}(B_v)$, where $g \in \bbZ_2$ is the non-identity element.
\end{lem}

\begin{rem}
Let $\gamma \in \bar P(\Gamma)$.  
If $v \in \gamma - \partial_0\gamma - \partial_1\gamma$, then $\widehat{B}_v^\gamma = \beta_g^{r(L_\gamma)}(B_v)$ does not depend on the choice of $L_\gamma$.  
Indeed, if $L'_\gamma$ is another possible extension of $\gamma$, then $\beta_g^{r(L_\gamma)}(B_v) = \beta_g^{r(L'_\gamma)}(B_v)$ since $r(L_\gamma)$ and $r(L'_\gamma)$ only differ outside the support of $B_v$.
\end{rem}

\begin{defn}
\label{def:DefectHamiltonianLG}
For $\gamma \in \bar P(\Gamma)$, we define the \emph{defect Hamiltonian} as follows: for $V \in \Gamma_f$, 
$$H_V^\gamma \coloneqq \sum_{v \in V} (\mathds{1} - \widehat B_v^\gamma)/2 \in \cstar[V^{+1}].$$ 
We denote the corresponding derivation by $\tilde \delta^\gamma$. 
\end{defn}

Note that $H_V^\gamma$ is a commuting projector Hamiltonian for all $\gamma \in \bar P(\Gamma)$ and $V \in \Gamma_f$ since it is the image of the commuting projector Hamiltonian $H_V$ under $(\alpha^\gamma)^{-1}$.

The following lemma follows by specializing the discussion of Section \ref{sec:defect Hamiltonians SPT} to the Levin-Gu SPT.

\begin{lem}
\label{lem:construct defect GS using auts}
Let $\gamma \in \bar P(\Gamma)$.
The state $\tilde \omega^\gamma \coloneqq \tilde \omega \circ \alpha^\gamma$ is the unique state satisfying that $\tilde \omega^\gamma(\widehat{B}_v^\gamma) = 1$ for all $v \in \Gamma$.  
In addition, $\tilde\omega^\gamma$ is the unique ground state for the derivation $\tilde \delta^\gamma$ of the defect Hamiltonian in Definition \ref{def:DefectHamiltonianLG}.
\end{lem}

We note that $H_V^\gamma$ is not invariant under the action of $\beta_g$ for all $V \in \Gamma_f$. However we observe the following fact. Let $\Lambda$ be a cone such that $\gamma \subset \Lambda$, which is always guaranteed since $\gamma \in \bar P(\Gamma)$. Then for all $V \in \Gamma_f$ satisfying $V^{+1} \cap \Lambda = \emptyset$, we have that $H_V^\gamma = H_V$ and thus $\beta_g(H_V^\gamma) = H_V^\gamma$. 
Therefore, even though $\tilde \omega_\gamma$ is not a ground state of $\tilde \delta$, $\tilde \omega_\gamma$ still satisfies $\tilde \omega_\gamma(H_V) = \tilde \omega(H_V) = 0$ for all $V$ as above. In other words, $\tilde \omega_\gamma$ `looks like' $\tilde \omega$ outside of $\Lambda$.

\subsection{Defect sector category}
\label{sec:Defect sectors in the Levin-Gu SPT}
Recall the definition of a $g$-defect sector (Definition \ref{def:g-defect_sector}). We now set $\tilde \pi$, the GNS representation of the ground state $\tilde \omega$ of $\tilde \delta$ as our reference representation. 
Note that by the uniqueness of $\tilde \omega$, $\tilde \pi$ is an irreducible representation. 
In addition, the assumptions in Section \ref{sec:GCrossedAssumptions} are satisfied by the fact that Levin-Gu SPT is an SPT satisfying Assumptions \ref{asmp:entangler commutes with symmetry}, \ref{asmp:strip_aut_is_an_FDQC}.

Noting that $\alpha^{-1}$ is a quasi-factorizable automorphism (Lemma \ref{lem:FDQCQuasi-Factorizable}) and applying Lemma \ref{lem:SPT has trivial superselection theory} gives us that $\tilde \pi$ has trivial superselection theory. We observe that by definition, $\tilde \pi$ is a $1$-defect sector. By specializing Lemmas \ref{lem:defect state GNS reps are sectorizable (SPT case)}, \ref{lem:classification of defect sectorizable representations SPT} to the case of Levin-Gu SPT we get the following proposition.

\begin{prop}
    The representation $\tilde \pi_\gamma \coloneqq \pi_0 \circ \alpha^\gamma$ is an irreducible $g$-defect sector for all $\gamma \in \bar P_R(\Gamma)$, and every $g$-defect sector for $g \in \bbZ_2$, $g \neq 1$, is unitarily equivalent to $\tilde \pi_\gamma$.
\end{prop}

We now use the theory shown in Section \ref{sec:OtherCoherenceData} to compute the cocyle and show that the category of defect sectors is equivalent to $\Vect(\bbZ_2, \nu)$.
To do this, we pick out representative defect sectors $\tilde \pi$ and $\tilde \pi_{\bar \gamma_R}$ (recall $\bar \gamma_R$ is the fixed dual path given in Figure \ref{fig:chosen_ray}), and we compute the $F$-symbols using the procedure in Section \ref{sec:OtherCoherenceData}.
This computation will then imply that we have a tensor functor from $\Vect(\bbZ_2, \nu)$ to the category of defect sectors that is a tensor equivalence.

Observe that $\tilde \pi$ is a strict tensor unit for the category, so all $F$-symbols except $F(g, g, g)$ are guaranteed to be $1$.
We now compute $F(g, g, g)$.
To do so, we must compute $\Omega_{g, g}$, which we do by computing $\alpha^{{\bar \gamma_R}} \circ \alpha^{{\bar \gamma_R}}$.
We define $\xi \coloneqq L_{{\bar \gamma_R}} - {\bar \gamma_R}$ and $\partial r(\xi) \coloneqq r(L_{{\bar \gamma_R}}) \cap \xi$. 
Additionally, we define $N(\xi)$ to be the subgraph of $\Gamma$ consisting of all vertices in $\xi$ and edges between them (Figure \ref{fig:DecorationForLGDefectSector}).  
We calculate $\alpha^{\bar \gamma_R}$ in Appendix \ref{sec:LGDefectSectorComputationAppendix} to be of the form 
\[
\alpha^{{\bar \gamma_R}}(A)
=
\Ad\!\left(
\prod_{v \in \partial r(\xi)} \sigma^z_v
\prod_{qq' \in N(\xi)} i^{\frac{1 - (-1)^{\varepsilon_{qq'}}\sigma^z_q\sigma^z_{q'}}{2}} 
\right)\! \circ \beta_g^{r(L_{{\bar \gamma_R}})}(A)
\]
for $A \in \cstar[\loc]$.
Here $\varepsilon_{qq'} \in \{0, 1\}$ for each edge $qq'$, which is the edge between vertices $q$ and $q'$. 
The precise formula for $\varepsilon_{qq'}$ in terms of the edge $qq'$ is complicated and not necessary for our purposes, but  computed in Appendix \ref{sec:LGDefectSectorComputationAppendix}.

\begin{figure}[!ht]
    \centering
    \begin{tikzpicture}[scale=0.6]
    \draw[thick](5,5.5)--(4,5)--(4,4)--(5,3.5)--(5,2.5)--(4,2)--(4,1)--(5,.5)--(5,0);
    \draw[thick](5,5.5)--(5,4.5)--(4,4)--(4,3)--(5,2.5)--(5,1.5)--(4,1)--(4,0);
    \draw[thick](4,5)--(5,4.5)--(5,3.5)--(4,3)--(4,2)--(5,1.5)--(5,.5)--(4,0);
    \draw[thick,orange] (5, 5.5) -- (5, 0); 
    \draw[thick, cyan] (4, 5) -- (4, 0);
    \foreach \x in {0,...,4}{
    \foreach \y in {0,...,2}{
    \filldraw[thick] (2*\x,\y*3) circle(.1cm);
    \filldraw[thick] (2*\x,\y*3+1) circle(.1cm);
    \filldraw[thick] (2*\x,\y*3+2) circle(.1cm);
    \filldraw[thick] (2*\x+1,\y*3+.5) circle(.1cm);
    \filldraw[thick] (2*\x+1,\y*3+1.5) circle(.1cm);
    \filldraw[thick] (2*\x+1,\y*3+2.5) circle(.1cm);
    \draw[thick,dashed](4.5,8.5)--(4.5,5);
    \draw[thick, violet, dotted] (4.5, 5) -- (4.5, 0);
}}
\end{tikzpicture}
    \caption{An illustration of the notation used in defining $\alpha^{{\bar \gamma_R}}$.
    The black dashed dual path is ${\bar \gamma_R}$ and the purple, dotted dual path is $\xi$.
    The solid edges are those in $N(\xi)$, and the vertices along the orange edges are those in $\partial r(\xi)$.
    The orange edges are those in the dual path $\xi_{\mathrm{in}}$, and the blue edges are those in the dual path $\xi_{\mathrm{out}}$.}
    \label{fig:DecorationForLGDefectSector}
\end{figure}
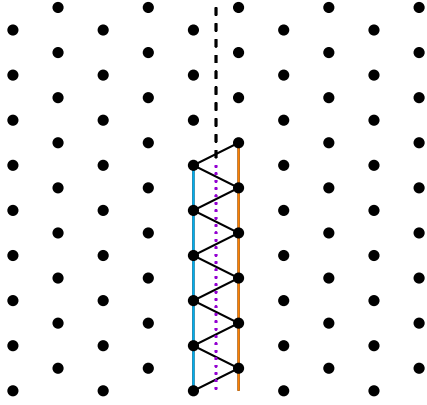

Note the vertices in $N(\xi)$ form two semi-infinite paths on the lattice (not the dual lattice!).
We let $\xi_{\mathrm{in}}$ denote the path of vertices in $N(\xi)$ that are in $r(L_{{\bar \gamma_R}})$ and $\xi_{\mathrm{out}}$ denote the path of vertices in $N(\xi)$ that are not in $r(L_{{\bar \gamma_R}})$.
We also let $\partial \xi_{\mathrm{in}}$ and $\partial \xi_{\mathrm{out}}$ denote the endpoints of $\xi_{\mathrm{in}}$ and $\xi_{\mathrm{out}}$ respectively.

It is shown in Lemma \ref{lem:LevinGuCocycleComputation} that 
\[
\alpha^{\bar \gamma_R} \circ \alpha^{\bar \gamma_R}
=
\Ad(\sigma^z_{\partial \xi_{\mathrm{in}}} \sigma^z_{\partial \xi_{\mathrm{out}}}).
\]
Using the notation of Section \ref{sec:OtherCoherenceData}, we have that $\Omega_{g, g} = \sigma^z_{\partial \xi_{\mathrm{in}}} \sigma^z_{\partial \xi_{\mathrm{out}}}$.
We now must find $F(g, g, g)$, which is determined by the equation
\[
\Omega_{1, g} \Omega_{g, g}
=
F(g, g, g) \Omega_{g, 1}\alpha^{{\bar \gamma_R}}(\Omega_{g, g}).
\]
Since $\tilde \pi$ is a strict tensor unity, $\Omega_{1, g} = \Omega_{g, 1} = 1$.  
Therefore, $F(g, g, g)$ is determined by 
\[
\Omega_{g, g}
=
F(g, g, g)\alpha^{{\bar \gamma_R}}(\Omega_{g, g}).
\]
We thus compute $\alpha^{{\bar \gamma_R}}(\Omega_{g, g})$.  
The key observation is that $\partial \xi_{\mathrm{in}} \in r(L_{{\bar \gamma_R}})$ but $\partial \xi_{\mathrm{out}} \notin r(L_{{\bar \gamma_R}})$, so $\beta_g^{r(L_{{\bar \gamma_R}})}(\sigma^z_{\partial \xi_{\mathrm{in}}} \sigma^z_{\partial \xi_{\mathrm{out}}}) = -\sigma^z_{\partial \xi_{\mathrm{in}}} \sigma^z_{\partial \xi_{\mathrm{out}}}$.
Therefore, we have that 
\begin{align*}
\alpha^{{\bar \gamma_R}}(\Omega_{g, g})
&=
\Ad\!\left(
\prod_{v \in \partial r(\xi)} \sigma^z_v
\prod_{qq' \in N(\xi)} i^{\frac{1 - (-1)^{\varepsilon_{qq'}}\sigma^z_q\sigma^z_{q'}}{2}} 
\right)\! \circ \beta_g^{r(L_{{\bar \gamma_R}})}
(\sigma^z_{\partial \xi_{\mathrm{in}}} \sigma^z_{\partial \xi_{\mathrm{out}}})
\\&=
\Ad\!\left(
\prod_{v \in \partial r(\xi)} \sigma^z_v
\prod_{qq' \in N(\xi)} i^{\frac{1 - (-1)^{\varepsilon_{qq'}}\sigma^z_q\sigma^z_{q'}}{2}} 
\right)\!
(-\sigma^z_{\partial \xi_{\mathrm{in}}} \sigma^z_{\partial \xi_{\mathrm{out}}})
=
-\sigma^z_{\partial \xi_{\mathrm{in}}} \sigma^z_{\partial \xi_{\mathrm{out}}}
=
-\Omega_{g, g}
\end{align*}
Hence $F(g, g, g) = -1$.  
We have thus shown that $F$ is the nontrivial cocycle on $\bbZ_2$.
The following result follows.

\begin{thm}
If the reference representation is the GNS representation for the ground state of the Levin-Gu SPT, then $\GSec \simeq \Vect(\bbZ_2, \nu)$, where $\nu$ is the nontrivial cocycle on $\bbZ_2$.  
\end{thm}
\begin{proof}
    Construct a functor that sends $\bbC_g \in \Vect(\bbZ_2, \nu)$ to $\tilde \pi \circ\alpha^{{\bar \gamma_R}}$. It is easily verified that this functor is a $G$-crossed monoidal equivalence.
\end{proof}

\begin{prop}\label{prop:LevinGuFrac}
    The symmetry fractionalization data is trivial for the Levin-Gu model.
\end{prop}
\begin{proof}
    Notice that $\beta_g\circ\alpha^\gamma \circ \beta_g^{-1}=\alpha^\gamma$. This implies that we may choose $V_g^\pi=1$ where $\pi$ is the non-trivial defect. We are also free to choose $V_g^1=V_1^1=V_1^\pi=1$. Furthermore, $g(\pi)=\pi$. This immediately implies that $\eta$ and $\mu$ are trivial.
\end{proof}

\section{A \texorpdfstring{$\bbZ_2$}{Z2}-Symmetry Enriched Toric Code}
\label{sec:SET Toric Code}
In this section, we will apply our general formalism to give a complete analysis of an infinite lattice model with $\mathbb{Z}_2$ onsite symmetry whose underlying topological order is that of the toric code. Our version of this model is closely related to the construction of \cite{PhysRevB.108.115144}.

Given the level of detail required in this type of analysis, we now give a brief outline of the following subsections. In Subsection \ref{sec:TC_review}, we quickly review the traditional toric code model.  This is followed by a presentation in Subsection \ref{sec:SET_model} of the $\mathbb{Z}_2$ symmetric toric code model and general analysis which shows that it has the same underlying topological order as the traditional toric code. We are then in a position to define the defects of this theory in Section \ref{sec:SET_defects} and provide a proof that they obey our selection criterion. This is followed by the calculation of the $F$-symbols, the symmetry fractionalization data, and the $G$-crossed braiding in sections \ref{sec:F_symbols}, \ref{sec:frac_data}, and \ref{sec:braid_data}.

\subsection{Review of Toric Code}
\label{sec:TC_review}


\subsubsection{Toric Code}

We first begin by recapitulating the construction and properties of the Toric Code model \cite{MR1951039} which is the string net generated by the unitary fusion category $\Vect(\mathbb{Z}_2)$ \cite{PhysRevB.71.045110, MR3204497, 10.22331/q-2024-03-28-1301}.
A thorough operator algebraic treatment can be found in \cite{MR2804555, MR2956822, MR3135456}.

Let our lattice be $\Gamma = \bbZ^2$ and place a qubit on each edge, i.e, $\hilb_e \simeq \bbC^2$. We can thus define the quasi-local algebra $\cstar$. A basis for $\cstar[e]$ is given by the Pauli matrices $\{\mathds{1}_v, \sigma^x_e, \sigma^z_e, \sigma^y_e\}$.

Let $v \in \Gamma$ be a vertex and $f \in \Gamma$ be a face. We will henceforth assume that $v$ refers to a vertex and $f$ to a face in $\Gamma$ whenever it is clear from the context. We can define the \emph{star} operator $A_v$ and \emph{plaquette} operator $B_f$ as follows: $$A_v \coloneqq  \prod_{e \ni v} \sigma^x_e \qquad \qquad B_f \coloneqq  \prod_{e \in f} \sigma^z_e.$$ It is easily checked for all $v,f \in \Gamma$ that $[A_v, B_f] = 0$.

Let $S \in \Gamma_f$ be a simply connected region.
Our finite volume Hamiltonian is given by $$H_S \coloneqq  \sum_{v \in S} (\mathds{1} - A_v)/2 + \sum_{f \in S} (\mathds{1} - B_f)/2.$$ 
This is a commuting projector Hamiltonian and thus has a frustration-free ground-state $\omega_0 \colon \cstar \to \bbC$ satisfying for all $v,f \in \Gamma$, $$\omega_0(A_v)= \omega_0(B_f) = 1.$$

\begin{lem}[\cite{MR2345476}]
    \label{lem:UniqueFrustrationFreeGSToricCode}
    The state $\omega_0$ is the unique state satisfying for all $v,f \in \Gamma$ $$\omega_0(A_v)= \omega_0(B_f) = 1$$
\end{lem}

We define $\pi_0$ to be the GNS representation of $\omega_0$ and $\cH_0$ to be the GNS Hilbert space. Note that $\pi_0$ is irreducible as $\omega_0$ is pure. 

\begin{lem}[{\cite{MR2956822}}]
\label{lem:TC Haag duality}
    The representation $\pi_0$ satisfies strict Haag duality. 
\end{lem}

Let $\bar \Gamma$ be the cell complex dual to $\Gamma$. Importantly, vectices in $\Gamma$ get mapped to faces in $\bar \Gamma$ and vice versa.
Edges get mapped to dual edges.
We recall the definitions of a path and dual path stated in Section \ref{sec:Defect auts Hamiltonian Levin-Gu}.
In order to remove ambiguity between paths and dual paths, we will denote dual paths by $(\bar \cdot )$. 

Closed paths $C \in \Gamma_f$ are paths such that $\partial_0 C = \partial_1 C$. We denote them as loops. Similarly, we define dual loops as closed dual paths $\bar C \in \Gamma_f$. Note that every loop/dual loop that is not empty divides $\Gamma$ into 2 simply connected regions $\Gamma_1, \Gamma_2 \subset \Gamma$, such that only one of them is finite. We call this region as the interior of the loop/dual loop respectively.

Some important objects in the study of the Toric Code are the \emph{string operators}. There are 2 different types of string operators, $F_\gamma, F_{\bar\gamma}$. 

\begin{defn}
    Let $\gamma$ be a finite path and $\bar \gamma$ a finite dual path. The \emph{string operators} are defined as $$F_\gamma \coloneqq  \prod_{e \in \gamma} \sigma^z_e \qquad \qquad F_{\bar \gamma} \coloneqq  \prod_{e \in \bar \gamma} \sigma^x_e.$$
\end{defn}

We have the relations $$F_\gamma F_{\bar \gamma} = (-1)^{c(\gamma, \bar \gamma)} F_{\bar \gamma} F_\gamma$$ where $c(\gamma, \bar \gamma)$ counts the number of crossings between $\gamma, \bar \gamma$. Using these string operators, we define the automorphisms which create the superselection sectors of the toric code.

\begin{defn}
\label{def:ToricCodeAnyonAutomorphisms}
    Let $\gamma$ be a half-infinite path and let $\gamma_i$ for $i\in\mathbb{N}$ be the path consisting of the first $i$ links of $\gamma$.
    Define $\bar{\gamma}_i$ with respect to a dual path $\bar{\gamma}$ similarly.
    We may then define the automorphisms $\alpha_\gamma^\epsilon, \alpha^m_{\bar \gamma}$ for all $A \in \cstar$ as  $$\alpha_\gamma^\epsilon(A) \coloneqq  \lim_{\gamma_i \uparrow \gamma} F_{\gamma_i} A F_{\gamma_i} \qquad \qquad \alpha_{\bar \gamma}^m(A) \coloneqq  \lim_{\bar \gamma_i \uparrow \bar \gamma} F_{\bar \gamma_i} A F_{\bar \gamma_i}.$$ as the charge/flux automorphisms respectively.
    Define also the following automorphism $$\alpha^\psi_{\gamma, \bar \gamma}(A) \coloneqq  \alpha_\gamma^\epsilon \circ \alpha_{\bar \gamma}^m(A).$$
\end{defn}

The following result is due to \cite{MR2804555}.

\begin{lem}[{\cite[Thm.~3.1]{MR2804555}}]
\label{lem:TC states are independent of paths}
    Let $\gamma, \gamma'$ be two arbitrary half-infinite paths and $\bar \gamma, \bar \gamma'$ be two arbitrary half-infinite dual paths. We have, $$\omega_0 \circ \alpha_\gamma^\epsilon \simeq \omega_0 \circ \alpha_{\gamma'}^\epsilon \qquad \omega_0 \circ \alpha_{\bar \gamma}^m \simeq \omega_0 \circ \alpha_{\bar \gamma'}^m \qquad \omega_0 \circ \alpha_{\gamma, \bar \gamma}^\psi \simeq \omega_0 \circ \alpha_{\gamma', \bar \gamma'}^\epsilon$$
\end{lem}

To prove the above result, \cite{MR2804555} uses the following lemma (\ref{lem:Naa31}), which is of independent interest.

\begin{lem}[{\cite[Lem.~3.1]{MR2804555}}]\label{lem:Naa31}
    Let $\gamma, \gamma'$ be two arbitrary half-infinite paths and $\bar \gamma, \bar \gamma'$ be two arbitrary half-infinite dual paths such that $\partial_0 \gamma = \partial_0 \gamma'$ and $\partial _0 \gamma' = \partial_0 \bar \gamma'$. Then we have, $$\omega_0 \circ \alpha_\gamma^\epsilon = \omega_0 \circ \alpha_{\gamma'}^\epsilon \qquad \omega_0 \circ \alpha_{\bar \gamma}^m = \omega_0 \circ \alpha_{\bar \gamma'}^m \qquad \omega_0 \circ \alpha_{\gamma, \bar \gamma}^\psi = \omega_0 \circ \alpha_{\gamma', \bar \gamma'}^\psi$$
\end{lem}

We use the above lemmas as motivation to define the following states: $$\omega^\epsilon_{\partial \gamma} \coloneqq  \omega_0 \circ \alpha_\gamma^\epsilon \qquad \omega^m_{\partial \bar \gamma} \coloneqq  \omega_0 \circ \alpha^m_{\bar \gamma} \qquad \omega^\psi_{\partial \gamma, \partial \bar \gamma} \coloneqq  \omega_0 \circ \alpha^\psi_{\gamma, \bar \gamma}$$

\begin{lem}
    We have $\pi_0 \circ \alpha_\gamma^\epsilon, \pi_0 \circ \alpha_{\bar \gamma}^m, \pi_0 \circ \alpha_{\gamma, \bar \gamma}^\psi$ are all localized in some cone and transportable for any chosen half-infinite paths/dual paths $\gamma/ \bar \gamma$. 
\end{lem}
\begin{proof}
    Straightforward from Lemma \ref{lem:TC states are independent of paths} and the definitions of the automorphisms.
\end{proof}

The following lemma has been discussed in several places in the literature. See for instance \cite{MR3764565}.

\begin{lem}[{\cite[Thm.~2.2]{MR3764565}}]
    Every ground state of the Toric Code model is equivalent to a convex combination of the states $\{\omega_0, \omega^\epsilon_v, \omega_f^m, \omega^\psi_{v,f}\}$ for some chosen $v,f \in \Gamma$.
    Furthermore, $\omega$ is a pure ground state of the Toric Code model if and only if $\omega$ is equivalent to one of these four states.
\end{lem}

We now fix a path $\gamma_0$ and dual path $\bar \gamma_0$ and define the representations $$\pi^\epsilon \coloneqq  \pi_0 \circ \alpha^\epsilon_{\gamma_0}, \qquad  \pi^m \coloneqq  \pi_0 \circ \alpha^m_{\bar \gamma_0}, \qquad \pi^\psi \coloneqq  \pi_0 \circ \alpha^\psi_{\gamma_0, \bar \gamma_0}$$

\begin{lem}[{\cite{MR2804555, MR3135456}}]
    \label{lem:ToricCodeAnyonSectors}
    The representations $\{\pi_0, \pi^\epsilon, \pi^m, \pi^\psi\}$ are anyon sectors, and any anyon sector is unitarily equivalent to one of these.
\end{lem}

Recall Definition \ref{def:allowed_cone} of the set of allowed cones $\cL$, and recall that the auxiliary algebra was defined to be 
\[
\fA^a
\coloneqq
\overline{\bigcup_{\Lambda \in \cL} \cR(\Lambda)}^{\| \cdot \|}
\subset
B(\cH_0).
\]

By \cite[Prop.~4.2]{MR2804555}, the maps $\pi_0 \circ \alpha_\gamma^\epsilon, \pi_0 \circ \alpha_{\bar \gamma}^m, \pi_0 \circ \alpha_{\gamma, \bar \gamma}^\psi$ all have a unique extension to $\cstar^a$ such that on any allowed cone $\Lambda$ the extension is weakly continuous. 
Furthermore, all of these extensions are endomorphisms of $\cstar^a$.

\begin{defn}
    An endomorphism $\rho$ of $\cstar^a$ is \emph{localized} in cone $\Lambda$ if for all $A \in \cR(\Lambda^c)$ we have $\rho(A) = A$. We say $\rho$ is \emph{transportable} if for any allowed cone $\Lambda'$ there exists an endomorphism $\rho'$ of $\cstar^a$ localized in $\Lambda'$ and satisfying such that $\rho \simeq \rho'$.
    We denote by $\DHR(\Lambda)$ the category of localized transportable endomorphisms of $\cstar^a$ that are localized in cone $\Lambda$, where the morphisms are intertwiners.
\end{defn}

\begin{rem}
    In \cite{MR2804555, MR3135456}, these localized and transportable endomorphisms are extended to the auxiliary algebra $\cstar^a$.
    One can then show that $\DHR(\Lambda)$ is a braided monoidal category.
\end{rem}

\begin{thm}[{\cite[Thm.~6.2]{MR2804555}}]
    The category $\DHR(\Lambda)$ is a braided monoidally equivalent to $\Rep(D(\bbZ_2))$.
\end{thm}

Specifically, if the simple objects in $\Rep (D(\bbZ_2))$ are denoted $1, e, m , \psi$, then we can make the identifications $$\id \mapsto 1 \qquad \pi_0 \circ \alpha_{\gamma_0}^\epsilon \mapsto e \qquad \pi_0 \circ \alpha^m_{\bar \gamma_0} \mapsto m \qquad \pi_0 \circ \alpha^\psi_{\gamma_0, \bar \gamma_0} \mapsto \psi$$

\subsection{SET toric code model}
\label{sec:SET_model}

In order to define the SET toric code mode, we include vertex spins in addition to the edge spins from the traditional toric code. This model is based on the models described in \cite{PhysRevB.108.115144}. For $i\in\{x,y,z\}$, the Pauli operators on the vertex spins will be denoted as $\tau^i_v$  and we will continue to call the $i$th Pauli operator on the edge $e$ $\sigma^i_e$. Since these operators have disjoint support, they must commute. 
On each finite region $S \in \Gamma_f$ with vertices $V(S)$ and edges $E(S)$ we have the Hilbert spaces $$\hilb^V_S \coloneqq  \bigotimes_{v \in V(S)} \hilb_v  \qquad\qquad \hilb^E_S \coloneqq  \bigotimes_{e \in E(S)} \hilb_e \qquad\qquad \hilb_S \coloneqq  \hilb^V_S \otimes \hilb^E_S.$$
We also define the following algebras for $S \in \Gamma_f$: $$\cstar[S]^V \coloneqq  B(\hilb^V_S) \qquad\qquad \cstar[S]^E \coloneqq  B(\hilb_S^E) \qquad\qquad \cstar[S] \coloneqq  B(\hilb_S).$$

We can define the following quasi-local algebras in the usual way as $$\cstar^V \coloneqq  \overline{\bigcup_{S \in \Gamma_f} \cstar[S]^V }^{||\cdot||} \qquad \cstar^E \coloneqq  \overline{\bigcup_{S \in \Gamma_f} \cstar[S]^E }^{||\cdot||} \qquad \cstar \coloneqq  \overline{\bigcup_{S \in \Gamma_f} \cstar[S] }^{||\cdot||}.$$
We note that in the previous section our operator algebra is what we are now calling $\cstar^E$. We also note that $\cstar = \cstar^E \otimes \cstar^V$.

On this new spin lattice, we are able to define the SET toric code Hamiltonian by finding an FDQC and applying it to a modified version of the traditional toric code Hamiltonian which accounts for the new vertex spins.
Recall that for any edge $e$, $\partial_0 e$ represents the source vertex whereas $\partial_1e$ represents the target vertex.
On each edge $e$, define 

$$W_e \coloneqq  i^{(1 - \sigma^x_e)(\tau^z_{\partial_1 e} - \tau^z_{\partial_0 e})/4} \in \cstar[e]\otimes \cstar[\partial_0 e] \otimes \cstar[\partial_1 e]$$ which gives rise the automorphism $\alpha\in \Aut(\fA_S)$ such that for all $A \in \cstar[S]$ $$\alpha_S(A) = \left(\bigotimes_{e \in S^{+1}} W_e\right) A \left(\bigotimes_{e \in S^{+1}} W_e^*\right).$$
This automorphism can be norm-continuously extended to an automorphism $\alpha$ of $\cstar$. This automorphism is an FDQC by the following lemma.
\begin{lem}
\label{lem:SETToricCodeFDQC}
    The automorphism $\alpha$ is an FDQC of depth 4.
\end{lem}
\begin{proof}
    Each $U_e$ acts only on an edge and its bounding vertices. The 4-coloring of the edges of the square lattice immediately tells us how to construct our FDQC.
\end{proof}

\subsubsection{Hamiltonian}
We can now define the Hamiltonian for all $S \in \Gamma_f$ as 
$$
H_S 
\coloneqq 
\sum_{v \in V(S)} \left(\dfrac{\mathds{1} - A_v}{2} + \dfrac{\mathds{1} - \tilde Q_v}{2}\right) + \sum_{f \in F(S)} \frac{\mathds{1} - \tilde B_f}{2}
$$
where we have denoted the vertices and faces in $S$ by $V(S)$ and $F(S)$, respectively, and we are defining 
$$
\tilde Q_v\coloneqq \dfrac{\mathds{1}+A_v}{2}\alpha(\tau_v^z)\quad\quad\text{and}\quad\quad \tilde{B}_f\coloneqq \alpha(B_f).
$$
We denote the corresponding derivation by $\tilde \delta$. We will first show that $H_S$ is a commuting projector Hamiltonian.
Then we show that there is a unique frustration-free ground state $\tilde \omega$ of $\tilde \delta$ by using our FDQC $\alpha$ to relate $\tilde{\omega}$ to the ground state of the traditional toric code. Finally, we will use this relationship to prove that the underlying braided fusion category of anyons in this theory is the equivalent to that of the toric code.

As is shown in the following lemma, we may alternatively define the operators in the Hamiltonian as follows:
$$\tilde B_f \coloneqq  i^{- \sum_{e \in f}\sigma^x_e(\tau^z_{\partial_1 e} - \tau^z_{\partial_0 e})/2} B_f \qquad\qquad \tilde Q_v\coloneqq \dfrac{\mathds{1}+A_v}{2} Q_v\qquad\qquad Q_v \coloneqq  \tau^x_v i^{-\tau^z_v \sum_{e \ni v} f(e,v) \sigma^x_e/2},$$ where $f(e,v) = 1$ if $v = \partial_0 e$ and $f(e,v) = -1$ if $v = \partial_1 e$. 

\begin{lem}
\label{lem:ImagesOfTCExtensionHamiltonianTermsUnderEntangler}
    Taking the above definition of $\tilde B_f$ and $Q_v$, we have $$\alpha(A_v) = A_v \qquad\qquad \alpha(B_f) = \tilde B_f \qquad\qquad \alpha(\tau^x_v) = Q_v.$$
\end{lem}

\begin{proof}
    
\begin{enumerate}
    \item[\underline{($\alpha(A_v) = A_v$):}] Since $W_e$ commutes with $A_v$ for all $e$, we straightforwardly have $\alpha(A_v) = A_v$.

    \item[\underline{($\alpha(B_f) = \tilde B_f$):}] We have for each $e \in f$, 
    \begin{align*}
    W_e B_f W_e^* 
    &= 
    i^{(1 - \sigma^x_e)(\tau^z_{\partial_1 e} - \tau^z_{\partial_0 e})/4} B_f i^{-(1 -\sigma^x_e)(\tau^z_{\partial_1 e} - \tau^z_{\partial_0 e})/4}
    \\&= i^{(1 - \sigma^x_e)(\tau^z_{\partial_1 e} - \tau^z_{\partial_0 e})/4} i^{-(1 + \sigma^x_e)(\tau^z_{\partial_1 e} - \tau^z_{\partial_0 e})/4} B_f  
    \\&= i^{- \sigma^x_e(\tau^z_{\partial_1 e} - \tau^z_{\partial_0 e})/2} B_f
    \end{align*}
    which gives us the required result after taking the product over all edges $e \in f$.

    \item[\underline{($\alpha(\tau^x_v) = Q_v$):}] Consider $e_1, e_2 \ni v$ such that $\partial_1 e_1 = \partial_0 e_2 = v$. Then we have,
    \begin{align*}
        W_{e_1} W_{e_2} \tau^x_v W_{e_2}^* W_{e_1}^* &= \Ad[W_{e_1}] (i^{(1 - \sigma^x_{e_2})(\tau^z_{\partial_1 e_2} - \tau^z_{v})/4} \tau_v^x i^{-(1 - \sigma^x_{e_2})(\tau^z_{\partial_1 e_2} - \tau^z_{v})/4})\\
        &= \Ad[W_{e_1}] ( \tau_v^x i^{(1 - \sigma^x_{e_2})(\tau^z_{\partial_1 e_2} + \tau^z_{v})/4} i^{-(1 - \sigma^x_{e_2})(\tau^z_{\partial_1 e_2} - \tau^z_{v})/4})\\
        &= \Ad[W_{e_1}] ( \tau_v^x i^{(1 - \sigma^x_{e_2}) \tau^z_{v}/2})\\
        &= \tau_v^x i ^{-(1 - \sigma^x_{e_1})\tau^z_v/2} i^{(1 - \sigma^x_{e_2}) \tau^z_{v}/2}\\
        &= \tau_v^x i^{- \tau_v^z(f(e_1,v)\sigma^x_{e_1}+ f(e_2)\sigma^x_{e_2,v})/2}
    \end{align*}
    Performing this conjugation on $\tau_v^x$ for both pairs of neighboring edges, we obtain the desired result.
\end{enumerate}
\end{proof}

We observe that Lemma \ref{lem:ImagesOfTCExtensionHamiltonianTermsUnderEntangler} implies that $H_S$ is a commuting projection Hamiltonian.
In the standard physics presentation, the Hamiltonian of an SET should be symmetric. Our analysis only directly requires that the ground state be symmetric. Nevertheless, it will be useful to prove the following lemma which implies the symmetry of our Hamiltonian.
Note that the symmetry action $\beta_g$ is given by Definition \ref{def:GlobalSymmetryAutomorphism} with $U_v^g = \tau^x_v$ and $U_e^g = \mathds{1}_v$.

\begin{lem}
\label{lem:SET Hamiltonian terms are symmetric}
    The operators $A_v, \tilde B_f, \tilde Q_{v}$ are invariant under the action of the symmetry $\beta_g$.
\end{lem}
\begin{proof}
    Using the definition of $A_v$, it is straightforward to verify that $\beta_g(A_v) = A_v$.

    We now show that $\tilde B_v$ is symmetric under $\beta_g$.
    \begin{align*}
        \beta_g(\tilde B_f) &= \left(\prod_{v \in f} \tau_v^x\right)\! i^{- \sum_{e \in f}\sigma^x_e(\tau^z_{\partial_1 e} - \tau^z_{\partial_0 e})/2} B_f \!\left(\prod_{v \in f} \tau_v^x\right)\\ &= i^{ \sum_{e \in f}\sigma^x_e(\tau^z_{\partial_1 e} - \tau^z_{\partial_0 e})/2} B_f\\
        &= i^{ \sum_{e \in f}\sigma^x_e(\tau^z_{\partial_1 e} - \tau^z_{\partial_0 e})} \tilde B_f.
        \intertext{We now use the fact that $(\tau^z_{\partial_1 e} - \tau^z_{\partial_0 e})$ always has eigenvalues $\pm 2,0$ and $\sigma^x_e$ has eigenvalues $\pm 1$, to observe that $i^{\sigma^x_e(\tau^z_{\partial_1 e} - \tau^z_{\partial_0 e})}$ has exactly the same spectral decomposition as $i^{(\tau^z_{\partial_1 e} - \tau^z_{\partial_0 e})}$. Therefore, using the fact that $i^{\pm\tau^z}=\pm i\tau^z$,}
        \beta_g(\tilde B_f)&= i^{\sum_{e \in f}(\tau^z_{\partial_1 e} - \tau^z_{\partial_0 e})} \tilde B_f = i^4(-i)^4\prod_{v \in f}(\tau^z_v)^2 \tilde B_f = \tilde B_f.
    \end{align*}
    
    Now we turn to $\tilde Q_v$. Consider the following calculation.
    \begin{align*}
        \beta_g(Q_v)&=\tau^x_v \tau^x_v i^{-\tau^z_v \sum_{e \ni v} f(e,v) \sigma^x_e/2} \tau^x_v = \left(i^{-\tau^z_v \sum_{e \ni v} f(e,v) \sigma^x_e}\right)\tau^x_v \left(i^{-\tau^z_v \sum_{e \ni v} f(e,v) \sigma^x_e/2}\right)\\
        &=  i^{\tau^z_v \sum_{e \ni v} f(e,v) \sigma^x_e} Q_v=  A_vQ_v
    \end{align*}
    where in the last step we have used that $\tau_v^z\sum_{e \ni v} f(e,v) \sigma^x_e \in \{\pm 4, 0\}$ on states where $A_v=1$ and $\tau^z_v\sum_{e \ni v} f(e,v) \sigma^x_e \in \{\pm 2\}$ on states where $A_v=-1$. Therefore, 
    \[
    \beta(\tilde{Q}_v)=\beta\!\left(\dfrac{\mathds{1}+A_v}{2}Q_v\right)=\dfrac{\mathds{1}+A_v}{2}A_vQ_v=\tilde{Q}_v.
    \qedhere
    \]
\end{proof}

By Lemma \ref{lem:ImagesOfTCExtensionHamiltonianTermsUnderEntangler} it follows that Hamiltonian $H_S$ is invariant under the $\bbZ_2$ symmetry.

\subsubsection{Relation to Toric Code}
We now define an augmented version of the traditional toric code Hamiltonian $H_S^0 \in \cstar$ to be $$H_S^0 \coloneqq  H_S^{TC} + \sum_{v \in V(S)} \dfrac{\mathds{1} - \tau_v^x}{2}, $$
where $H_S^{TC} \in \cstar^E$ is the Toric Code Hamiltonian on $S$.
We have simply added ancilla spins on the vertices and energetically enforced them to be in a product state.
Let $\delta^0$ be the derivation corresponding to this new Hamiltonian. It is easy to see that $H_S^0$ is still a commuting projector Hamiltonian. Let $\omega_0$ be a state on $\cstar$ defined by $$\omega_0 \coloneqq  \omega^E_{TC} \otimes \omega^V_0$$ where $\omega^E_{TC}$ (defined on $\cstar^E$) is the Toric Code frustration-free ground-state and $\omega^V_0$ is defined on $\cstar^V$ as a product state given by $\omega^V_0(A) \coloneqq  \bigotimes_{v \in \Gamma} \inner{\psi_v}{A \psi_v}$ and $\ket{\psi_v} \in \hilb_v$ satisfies $\ket{\psi_v} = \tau_v^x \ket{\psi_v}$.

Then it is easy to see that $\omega_0 $ is a frustration-free ground-state of $H_S^0$. We now list some useful facts about $\omega_0$, which have been shown in Appendix \ref{app:TC with ancillary vertex spins}.

\begin{prop}
    The state $\omega_0 = \omega^E_{TC}\otimes \omega_0^V$ satisfies the following:
    \begin{enumerate}
        \item $\omega_0$ is pure.
        \item $\omega_0$ is the unique frustration-free ground-state of $\delta^0$.
        \item $\omega_0$ is the unique state state satisfying for all $v,f$ $$\omega_0(A_v) = \omega_0(B_f) = \omega_0(\tau_v^x) = 1$$
    \end{enumerate} 
\end{prop}
This Proposition is proved in Lemma \ref{lem:TC extension FF GS is unique}.

We now define $\pi_0$ as the GNS representation of $\omega_0$. 

\begin{prop}
    The representation $\pi_0$ satisfies the following:
    \begin{enumerate}
        \item $\pi_0$ is irreducible.
        \item $\pi_0$ satisfies Haag Duality.
        \item Any anyon sector $\pi$ with respect to $\pi_0$ is unitarily equivalent to one of the mutually disjoint anyon sectors $\{\pi_0 \circ\zeta\}_\zeta$ where $\zeta \in \{\id, \alpha_\gamma^\epsilon, \alpha_{\bar \gamma}^m, \alpha_{\gamma, \bar \gamma}^\psi\}$ for fixed half-infinite $\gamma \in P(\Gamma), \bar \gamma \in \bar P(\Gamma)$.
    \end{enumerate}
\end{prop}
The fact that $\pi_0$ is irreducible follows from $\omega_0$ being pure. The rest of this proposition is proved in parts in Lemmas \ref{lem:TC extension Haag duality}, \ref{lem:TC extension anyon sectors bound}, \ref{lem:TC extension anyon sectors constructed} and Corollary \ref{cor:TC extension anyon sector classification}.

We now define the category $\DHR_{\pi_0}(\Lambda)$ as the braided $C^*$ tensor category of endomorphisms of $\cstar^a$ that are localized in the cone $\Lambda$ (with respect to $\pi_0$) and transportable. 
Note that we have $\DHR_{\pi_0}(\Lambda) \simeq \DHR_{\pi_0^{TC}}(\Lambda)$.

\subsubsection{Ground states}
Now we understand some facts about the ground state of $\tilde \delta$. We first define $\tilde \omega \coloneqq  \omega_0 \circ \alpha^{-1}$, then prove that it is the unique frustration-free ground-state of $\tilde \delta$.
\begin{lem}
\label{lem:SET toric code unique FF GS}
    The state $\tilde \omega$ is the unique state satisfying for all $v,f \in \Gamma$ 
    \begin{equation}
    \label{eq:SETTCFrustrationFreeGroundState-Condition1}
    \tilde \omega(A_v) = \tilde \omega(\tilde B_f) = \tilde \omega(Q_v) = 1.
    \end{equation}
    Additionally, $\tilde \omega = \omega_0 \circ \alpha^{-1}$ is the unique state satisfying for all $v,f \in \Gamma$ 
    \begin{equation}
    \label{eq:SETTCFrustrationFreeGroundState-Condition2}
    \tilde \omega(A_v) = \tilde \omega(\tilde B_f) = \tilde \omega(\tilde Q_v) = 1.
    \end{equation}
    In particular, $\tilde \omega$ is the unique frustration-free ground-state of $\tilde \delta$.
    Moreover, this state is symmetric under $\beta_g$, as are the dynamics generated by $\tilde \delta$.
\end{lem}
\begin{proof}
We first show that $\tilde \omega = \omega_0 \circ \alpha^{-1}$ satisfies \eqref{eq:SETTCFrustrationFreeGroundState-Condition1}.
Indeed, we have that for every $v,f$, we have by Lemma \ref{lem:ImagesOfTCExtensionHamiltonianTermsUnderEntangler} that
\begin{gather*}
\tilde \omega(A_v)
=
\omega_0 \circ \alpha^{-1} \circ \alpha(A_v)
=
\omega_0(A_v)
=
1,
\\
\tilde \omega(\tilde B_f)
=
\omega_0 \circ \alpha^{-1} \circ \alpha(B_f)
=
\omega_0(B_f)
=
1,
\\
\tilde \omega(Q_v)
=
\omega_0 \circ \alpha^{-1} \circ \alpha(\tau^x_v)
=
\omega_0(\tau^x_v)
=
1.
\end{gather*}
Now, suppose $\omega \in \cS(\cstar)$ is another state satsifying \eqref{eq:SETTCFrustrationFreeGroundState-Condition1}.
Then by Lemma \ref{lem:ImagesOfTCExtensionHamiltonianTermsUnderEntangler}, we have that 
\[
\omega \circ \alpha(A_v)
=
\omega \circ \alpha(B_f)
=
\omega \circ \alpha(\tau^x_v)
=
1.
\]
Therefore, by Lemma \ref{lem:TC extension FF GS is unique}, we have that $\omega \circ \alpha = \omega_0$, from which it follows that $\omega = \tilde \omega$.

We now show that $\tilde \omega$ is the unique state satisfying \eqref{eq:SETTCFrustrationFreeGroundState-Condition2}.
It suffices to show that a state $\omega$ satisfies \eqref{eq:SETTCFrustrationFreeGroundState-Condition1} if and only if it satisfies \eqref{eq:SETTCFrustrationFreeGroundState-Condition2}.
First, suppose that $\omega$ satisfies \eqref{eq:SETTCFrustrationFreeGroundState-Condition2}.
Then by Lemma \ref{lem:can freely insert and remove P from the ground state.}, we have that 
\[
\omega(Q_v)
=
\omega\!\left(\frac{\mathds{1} + A_v}{2}Q_v \right)
=
\omega(\tilde Q_v)
=
1,
\]
so $\omega$ satisfies \eqref{eq:SETTCFrustrationFreeGroundState-Condition1}.
Now, suppose $\omega$ satisfies \eqref{eq:SETTCFrustrationFreeGroundState-Condition1}.
We then have by Lemma \ref{lem:can freely insert and remove P from the ground state.} that 
\[
\omega(\tilde Q_v)
=
\omega\!\left(\frac{\mathds{1} + A_v}{2}Q_v \right)
=
\omega(Q_v)
=
1,
\]
so $\omega$ satisfies \eqref{eq:SETTCFrustrationFreeGroundState-Condition2}.

It follows that $\tilde \omega$ is the unique frustration-free ground state of $\tilde \delta$ by \cite[Lem.~3.8]{MR3764565}.

Also note that since $A_v,\tilde B_v,\tilde Q_v$ are symmetric by Lemma \ref{lem:SET Hamiltonian terms are symmetric}, $\tilde \omega\circ\beta_g$ satisfies \eqref{eq:SETTCFrustrationFreeGroundState-Condition2}.
By uniqueness, $\tilde \omega=\tilde\omega\circ\beta_g$, so $\tilde\omega$ is symmetric.
Lemma \ref{lem:SET Hamiltonian terms are symmetric} also directly implies that the dynamics generated by $\tilde\delta$ are symmetric.
\end{proof}

\begin{rem}
    The Hamiltonian $H_S$ was chosen specifically to be symmetric under the action of $\beta_g$, which we elaborate on below. However, it is also natural to consider the Hamiltonian $H_S' \coloneqq  \alpha(H_S^0)$ instead. Notice that $\tilde \omega$ is the unique frustration-free ground-state for both Hamiltonians, which follows from Lemma \ref{lem:SET toric code unique FF GS} as well as \cite[Lem.~3.8]{MR3764565}. This of course means that $\tilde \omega$ is $\beta_g$ invariant, since it is a ground-state of $H_S$, and can be obtained using a FDQC. These are the only properties required to completely determine the defect sectors with respect to $\tilde \omega$, and the choice of the dynamics is irrelevant to our story.
\end{rem}

Given that $\tilde \omega = \omega_0 \circ \alpha^{-1}$, we now let $\tilde \pi \coloneqq  \pi_0 \circ \alpha^{-1}$ be the GNS representation $\tilde \omega$, where $\pi_0$ is the GNS representation of $\omega_0$, the frustration-free ground state of $\delta^0$.

We now begin our analysis of the anyon sectors of the SET toric code.

\subsubsection{Anyon sectors}
We may easily obtain the new string operators by applying the entangling automorphism $\alpha$ to the old string operators. 

Recall the definition of string-operators on the original Toric Code. We have, $$F_\gamma \coloneqq  \prod_{e \in \gamma} \sigma^z_e \qquad \qquad F_{\bar \gamma} \coloneqq  \prod_{e \in \bar \gamma} \sigma^x_e$$
We now define the entangled string operators as $$\tilde F_\gamma \coloneqq  \alpha(F_\gamma) = F_\gamma \prod_{e \in \gamma} i^{\sigma^x_e(\tau^z_{\partial_1 e} - \tau_{\partial_0 e}^z)/2} \qquad \tilde F_{\bar \gamma} \coloneqq  \alpha(F_{\bar \gamma}) = F_{\bar \gamma}$$ 
The string operators still satisfy the identities of the Toric Code string operators, $$\tilde F_\gamma \tilde F_{\bar \gamma} = (-1)^{c(\gamma, \bar \gamma)} \tilde F_{\bar \gamma} \tilde F_\gamma$$ where $c(\gamma, \bar \gamma)$ counts the number of crossings between $\gamma, \bar \gamma$.

\begin{lem}
\label{lem:action of symmetry on string ops}
    We have the following identities: $$\beta_g(\tilde F_\gamma) = \tau^z_{\partial_1 \gamma} \tau^z_{\partial_0 \gamma} \tilde F_\gamma \qquad \qquad \beta_g(\tilde F_{\bar \gamma}) = \tilde F_{\bar \gamma}$$
\end{lem}
\begin{proof}
    The second identity is trivial. We prove the first identity. 
    \begin{align*}
        \beta_g(\tilde F_\gamma) &= \left( \prod_{v \in \gamma} \tau_v^x \right) F_\gamma \prod_{e \in \gamma} i^{\sigma^x_e(\tau^z_{\partial_1 e} - \tau^z_{\partial_0 e})/2} \left( \prod_{v \in \gamma} \tau_v^x \right)\\
        &= F_\gamma \prod_{e \in \gamma} i^{-\sigma^x_e(\tau^z_{\partial_1 e} - \tau^z_{\partial_0 e})/2}= \tilde F_\gamma \prod_{e \in \gamma} i^{-\sigma^x_e(\tau^z_{\partial_1 e} - \tau^z_{\partial_0 e})}\\
        \intertext{We now use the fact that $\sigma_e^x(\tau^z_{\partial_1 e} - \tau^z_{\partial_0 e})$ always has eigenvalues $\pm 2,0$ to observe that $i^{-\sigma^x_e(\tau^z_{\partial_1 e} - \tau^z_{\partial_0 e})}$ has the same spectral decomposition as $i^{(\tau^z_{\partial_1 e} - \tau^z_{\partial_0 e})}$. We then obtain}
        \beta_g(\tilde F_\gamma)&= \tilde F_\gamma \prod_{e \in \gamma} i^{(\tau^z_{\partial_1 e} - \tau^z_{\partial_0 e})}= \tilde F_\gamma i^{(\tau^z_{\partial_1 \gamma} - \tau^z_{\partial_0 \gamma})} = \tilde F_\gamma i{\tau^z_{\partial_1 \gamma} i(- \tau^z_{\partial_0 \gamma})} = \tilde F_\gamma \tau^z_{\partial_1 \gamma} \tau^z_{\partial_0 \gamma}.
    \end{align*}
    In the second to last equality we have used the fact that $i^{\pm\tau^z_v}=\pm i\tau^z_v$.
\end{proof}

We obtain new automorphisms for the anyon sectors as follows. Taking $\zeta_\gamma \in \{\alpha_\gamma^\epsilon, \alpha_{\bar \gamma}^m, \alpha_{\gamma, \bar \gamma}^\psi\}$, new automorphism $\tilde \zeta_\gamma$ is given by $$\tilde \zeta_\gamma \coloneqq  \alpha \circ \zeta_\gamma \circ \alpha^{-1}.$$
Here we note the dependence of $\zeta$ on $\gamma$ since the paths $\gamma, \bar \gamma$ are not fixed in this instance. 
Often, we will be considering fixed paths $\gamma, \bar\gamma$, and in that case we will drop the subscript $\gamma$ on $\zeta$.

Define the representations $\tilde \pi_{\zeta_\gamma} \coloneqq \pi_{\zeta_\gamma} \circ \alpha^{-1} = \tilde \pi \circ \tilde \zeta_\gamma$, which are irreducible since $\zeta_\gamma \circ \alpha^{-1}$ is an automorphism and $\pi_0$ is irreducible.

\begin{rem}\label{rem:SETTCAutomorphismDefn}
It is easy to verify that for $A \in \cstar$,
\[
\tilde \alpha^\epsilon_\gamma(A)
=
\lim_{n \to \infty}
\Ad(\tilde F_{\gamma_n})(A),
\qquad
\tilde \alpha^m_{\bar\gamma}(A)
=
\lim_{n \to \infty}
\Ad(\tilde F_{\bar\gamma_n})(A),
\qquad
\tilde \alpha^\psi_{\gamma, \bar\gamma}(A)
=
\lim_{n \to \infty}
\Ad(\tilde F_{\gamma_n}\tilde F_{\bar\gamma_n})(A),
\]
where the subscript $n$ denotes truncation after the first $n$ edges.
\end{rem}

\begin{rem}
    Let $\zeta_\gamma \in \{\alpha_\gamma^\epsilon, \alpha_{\bar \gamma}^m, \alpha_{\gamma, \bar \gamma}^\psi\}$. Recall that $\zeta_\gamma$ is an FDQC. Therefore, $$\tilde \omega_{\zeta_\gamma} \coloneqq  \tilde \omega \circ \tilde \zeta_\gamma = \omega_0 \circ \zeta_\gamma \circ \alpha^{-1} = \omega_{\zeta_\gamma} \circ \alpha^{-1}$$ are ground-states of $\tilde \delta$ using Lemma \ref{lem:QCAs preserve the ground state subspace}. 
    The physical relevance of the representations $\tilde \pi_{\zeta_\gamma}$ is that they are the GNS representations of ground-states $\tilde \omega_{\zeta_\gamma}$. 
\end{rem}

\begin{lem}\label{lem:TCSymmetry_permute}
    The action of the symmetry does not permute anyon types, i.e, $\gamma_g(\tilde \pi^{\zeta_\gamma}) \simeq \tilde \pi^{\zeta_\gamma}$.
\end{lem}

\begin{proof}
    We have for any observable $A \in \cstar[\loc]$,
    \[
        \gamma_g(\tilde \pi^{\zeta_\gamma})(A) = \beta_g \circ (\tilde \pi \circ \tilde\zeta_\gamma) \circ \beta_g(A) = \tilde \pi \circ \Ad[\beta_g(\tilde F_{\gamma'})](A),
    \]
    where $\gamma'$ is a truncation of $\gamma$ such that $\supp(A) \cap (\gamma - \gamma') = \emptyset$.
    Now, there exists $U \in \cstar[\loc]$ such that for any such truncation $\gamma'$ of $\gamma$, $\Ad[\beta_g(\tilde F_{\gamma'})](A) = \Ad[U\tilde F_{\gamma'}](A)$. 
    By continuity, we get that $\gamma_g(\tilde \pi^{\zeta_\gamma})(A) = \Ad[\tilde \pi(U)] \circ \tilde \pi^{\zeta_\gamma}(A)$ for all $A \in \cstar$.
\end{proof}

For the remainder of this subsection, we fix paths $\gamma, \bar \gamma$, so we drop subscripts on $\zeta$. 

\begin{lem}
\label{lem:anyon sectors of GS SET}
    The representations given by $\{\tilde \pi_\zeta \circ \alpha^{-1}\}_{\zeta}$ are mutually disjoint and anyon sectors with respect to $\tilde \pi$, and any anyon sector is unitarily equivalent to one of them.
\end{lem}
\begin{proof}
    Note that $\alpha$ is an FDQC. The result follows from Corollary \ref{cor:TC extension anyon sector classification}, Lemma \ref{lem:FDQCQuasi-Factorizable}, and \cite[Thm.~4.7]{MR4426734}.
\end{proof}

In fact, we can apply a theorem from \cite{MR4362722} to obtain a stronger result. 

\begin{prop}\label{prop:equiv_to_TC}
    Let $\Lambda$ be a cone. 
    The category $\DHR_{\pi_0}(\Lambda)$ is braided monoidally equivalent to $\DHR_{\tilde \pi}(\Lambda)$. In particular, $\DHR_{\tilde \pi}(\Lambda)$ is braided monoidally equivalent to $\DHR_{\pi_0^{TC}}(\Lambda)$. 
\end{prop}
\begin{proof}
    This follows from noting that $\alpha$ implements an FDQC and then applying Proposition \ref{prop:TC extension braided monoidal}, Lemma \ref{lem:FDQCQuasi-Factorizable}, and \cite[Thm.~6.1]{MR4362722}. 
\end{proof}

\begin{rem}
    The previous results strongly hint that we have found the full ground-state subspace for $H_S$. In fact, if we had chosen our Hamiltonian as $H'_S = \alpha(H_S^0)$, then the fact that $\alpha$ is a FDQC immediately implies that the ground-state subspace for the dynamics generated by $H_S'$ is the same as that of the dynamics generated by $H^0_S$ (which was in turn the same as that of the Toric Code).

    However, since we have chosen our Hamiltonian to be $H_S$ instead, which is not outright related to $H_S^0$ using an FDQC, we cannot guarantee that every pure ground state is equivalent to $\{\tilde \omega_\zeta\}_{\zeta}$. 
    
    We remark that the analysis of \cite{MR3764565} remains mostly applicable in our setting, and conjecture that this is the full subspace of $\tilde \delta$.
\end{rem}

\subsection{Symmetry defects}
\label{sec:SET_defects}
Recall that $ A_v, \tilde B_f, \tilde Q_v$, the terms of the SET toric code Hamiltonian, are invariant under the action of the symmetry (Lemma \ref{lem:SET Hamiltonian terms are symmetric}).
Our defect construction strategy will be similar to sections \ref{sec:Defect auts Hamiltonian Levin-Gu} and \ref{sec:defects using auts}. 
The idea is to observe the action of $\beta_g^{\bar L}$ along some dual path $\bar{L}$ on the terms of the SET toric code Hamiltonian. 
We then erase the action of the symmetry along some $\bar{\gamma} \subset \bar{L}$ using an explicit automorphism. 
However, we do not directly use these results concerning the symmetry action when showing that the representations we define are defect sectors. 
We therefore relegate this discussion to Appendix \ref{sec:SETToricCodeDefectHamiltonian}.

For a finite dual path $\bar \gamma \in \bar P(\Gamma)$, we define the \emph{symmetry erasing string operator} operator $$F_{\bar \gamma}^\sigma \coloneqq  \prod_{e \in \bar \gamma} e^{-i\pi p(e) \sigma^x_e/4}$$ where $p(e) = +1$ if $\partial_1 e$ is to the right of $\bar \gamma$ and $p(e) = -1$ otherwise. Note that `right' and `left' are considered with respect to the orientation of $\overline{\gamma}$. 
\begin{center}
    \begin{tikzpicture}
        \draw[thick,black,->](0,0)--(0,2);
        \draw[thick,black,->](4,1)--(6,1);
        \node at (-.8,1){Left};
        \node at (.8,1){Right};
        \node at (5,1.5){Left};
        \node at (5,.5){Right};
    \end{tikzpicture}
\end{center}

Let $A \in \cstar$ and consider a sequence of dual finite paths $\{\bar \gamma_n\}_{n \in \bbN}$ such that $\bar\gamma_n \subset \bar\gamma_{n+1}$ and $\partial_0\bar \gamma_n$ is constant for all $n$. Define the automorphism $\alpha^\sigma_{\bar \gamma}$ as $$\alpha^\sigma_{\bar \gamma} \coloneqq  \lim_{\bar \gamma_n \uparrow \bar \gamma} F_{\bar \gamma_n}^\sigma A (F_{\bar \gamma_n}^\sigma)^*.$$

Let $\overline{\xi}_1, \overline{\xi}_2$ be two dual paths with the same endpoints that do not intersect outside of the shared endpoints. 
In this setup, the two paths bound a surface $S(\overline{\xi}_1, \overline{\xi}_2)$. Note that by the way these paths are defined, $\overline{\xi}_1, \overline{\xi}_2$ are self-avoiding.
We can define the unitary $$\tilde F^\sigma_{\overline{\xi}_1, \overline{\xi}_2} \coloneqq  F^\sigma_{\overline{\xi}_1} (F^\sigma_{\overline{\xi}_2})^* \left(\bigotimes_{v \in S(\overline{\xi}_1, \overline{\xi}_2)} U_v^g\right).$$ 

\begin{lem}\label{lem:boundary_string}
    Let $S$ be a finite simply connected region. Then for the dual path $\tilde \gamma$ bounding $V(S)$, the collection of vertices in $S$, we have
    $$
    P_S\left(\prod\limits_{v\in V(S)} Q_v\right)=P_S F^\sigma_{\tilde{\gamma}}\left(\prod\limits_{v\in V(S)}\tau^x_v\right)
    $$
    regardless of the orientation of $\tilde{\gamma}$, where $P_S$ is the projection on to the $A_v=1$ subspace for all $v\in V(S)$.
\end{lem}
\begin{proof}
    We do this analysis in the case where $S$ is a single vertex spin $v$. The general case follows inductively by gluing smaller regions together and seeing that paths in opposite directions cancel.

    Starting with the edge directly above $v$ and going around clockwise, we label the edges neighboring $v$ as $1,2,3,4$. Then,
    $$
        \dfrac{1+A_v}{2}Q_v=\dfrac{1+A_v}{2}\tau^x_v i^{-\tau_v^z(\sigma^x_1+\sigma_2^x-\sigma^x_3-\sigma^x_4)/2}=\dfrac{1+A_v}{2} i^{\tau_v^z(\sigma^x_1+\sigma_2^x-\sigma^x_3-\sigma^x_4)/2}\tau^x_v\\
    $$
    Note that the spectrum of the exponent is always even, so we may remove $\tau^z_v$ from this expression. So we have that
    $$
    \dfrac{1+A_v}{2}Q_v=\dfrac{1+A_v}{2}e^{i\frac{\pi}{4}\sigma^x_1}e^{i\frac{\pi}{4}\sigma^x_2}e^{-i\frac{\pi}{4}\sigma^x_3}e^{-i\frac{\pi}{4}\sigma^x_4}\tau_v^x=\dfrac{1+A_v}{2}F^\sigma_{1\rightarrow 2\rightarrow 3\rightarrow 4\rightarrow 1}\tau_v^x.
    $$
    Note that we may related this path to the reverse path via the following:
    $$
    F^\sigma_{1\rightarrow 2\rightarrow 3\rightarrow 4\rightarrow 1}=A_vF^\sigma_{4\rightarrow 3\rightarrow 2\rightarrow 1\rightarrow 4}.
    $$
    Therefore, this result is independent of orientation.
\end{proof}

\begin{lem}
\label{lem:the pizza operators eval to 1 SET}
    Let $\overline{\xi}_1, \overline{\xi}_2$ be two dual paths with the same endpoints that do not intersect outside of the shared endpoints.
    Then
    $$\tilde \omega(\tilde F^\sigma_{\overline{\xi}_1, \overline{\xi}_2}) = 1.$$ 
\end{lem}
\begin{proof}
We prove this in the case where the region enclosed by these dual paths is some simply connected $S$. The non-simply connected case follows inductively. 

Let $\tilde{\gamma}$ be a dual path enclosing $S$ which runs parallel to $\overline{\xi}_1$ and anti-parallel to $\overline{\xi}_2$. If an edge is traversed in the same direction by the dual paths $\overline{\xi}_1$ and $\overline{\xi}_2$, then the operator $\tilde F^\sigma_{\overline{\xi}_1, \overline{\xi}_2}$ acts trivially on that edge. Therefore, we may write
$$
\tilde F^\sigma_{\overline{\xi}_1, \overline{\xi}_2}=F^\sigma_{\tilde{\gamma}}\prod\limits_{v\in V(S)}\tau_v^x.
$$
We may then use Lemma \ref{lem:boundary_string} to see that
$$
\tilde \omega(\tilde F^\sigma_{\overline{\xi}_1, \overline{\xi}_2}) =\tilde\omega\left(P_S F^\sigma_{\tilde{\gamma}}\prod\limits_{v\in V(S)}\tau_v^x\right)=\tilde\omega\left(P_S\prod\limits_{v\in V(S)}Q_v\right)=1
$$
where $P_S$ is the projection to the $A_v=1$ subspace for each $v\in S$.
\end{proof}

In what follows, we let $\tilde \cH$ denote the GNS Hilbert space corresponding to $\tilde \pi$ and let $\tilde \Omega$ denote the cyclic vector.

\begin{lem}
\label{lem:defects are cone transportable}
    Pick $\bar\gamma \in \bar P(\Gamma)$ and let $\bar L_1, \bar L_2$ be two different infinite extensions of $\bar \gamma$. 
    Let $\bar \eta_1 = \bar L_1 - \bar \gamma$ and $\bar \eta_2 = \bar L_2 - \bar \gamma$.
    Then $$\tilde \pi\circ \alpha_{\bar \eta_1}^\sigma \circ \beta^{r(L_1)}_g \simeq \tilde \pi \circ \alpha_{\bar \eta_2}^\sigma \circ \beta^{r(L_2)}_g.$$
    
    In fact, there is a unique unitary $V \in B(\tilde \cH)$ witnessing the above equivalence such that $V\tilde \Omega = \tilde \Omega$.  
    This unitary is the WOT-limit of the sequence $V_n \coloneqq \tilde \pi(\tilde F_{(\bar \eta_2)_n, \overline{\xi}_n}^\sigma)$, where $(\bar\eta_2)_n$ is the dual path consisting of the first $n$ steps of $\bar \eta_2$, and $\overline{\xi}_n$ is a dual path consisting of the first $n$ steps of $\bar \eta_1$ as well as a dual path $\varsigma_n$ connecting the $n$th step of $\bar \eta_1$ to that of $\bar \eta_2$ whose distance from $\partial \bar \gamma = \partial \bar \eta_1 = \partial \bar \eta_2$ goes to infinity as $n \to \infty$ (Figure \ref{fig:pizza}). 
    \begin{figure}[!ht]
\begin{center}
\begin{tikzpicture}
    \draw[thick,black](0,0)--(0,1)--(1,2);
    \draw[thick,blue](2,0)--(2,1)--(1,2);
    \draw[thick,red](1,2)--(1,3);
    \draw[thick,purple,snake it](0,.5)--(2,.5);
    \node at (1,0){$\varsigma_n$};
    \node at (0,-.2){$\overline{\eta}_1$};
    \node at (2,-.2){$\overline{\eta}_2$};
    \node at (1.2, 3) {$\overline{\gamma}$};
    \draw[thick,orange,dashed,<-](2,.65)--(.1,.65)--(.1,.92)--(1,1.8);
    \node at (.8,1.1){$\overline{\xi}_n$};
\end{tikzpicture}
\end{center}
\caption{The geometry of the paths in Lemma \ref{lem:defects are cone transportable}.}
\label{fig:pizza}
\end{figure}
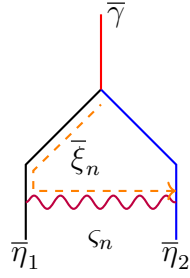
\end{lem}
\begin{proof}
    We consider states $\omega_{i} \coloneqq  \tilde \omega \circ \alpha_{\bar \eta_i}^\sigma \circ \beta^{r(L_\gamma)}_g$.
    Observe that for $i = 1, 2$, $\tilde \pi\circ \alpha_{\bar \eta_i}^\sigma \circ \beta^{r(L_i)}_g$ is a GNS representation for $\omega_i$. 
    Therefore, by uniqueness of the GNS representation, it suffices to show that $\omega_1 = \omega_2$.
    Let $A \in \cstar[\loc]$. Choose $n$ large enough such that $\supp(A)$ does not intersect $\varsigma_n$. We have,
    \begin{align*}
        \omega_{1}(A) &= \inner{\tilde \Omega}{\tilde \pi \circ  \alpha^\sigma_{\bar \eta_1} \circ \beta_g^{\bar L_1}(A)\tilde \Omega}= \inner{\tilde \Omega}{ \tilde \pi(\tilde F^\sigma_{(\bar \eta_2)_n, \overline{\xi}_n}) \tilde \pi \circ \alpha^\sigma_{\bar \eta_1} \circ \beta_g^{\bar L_1}(A) \tilde \pi(\tilde F^\sigma_{(\bar \eta_2)_n, \overline{\xi}_n})^*\tilde \Omega} \\
        &= \inner{\tilde \Omega}{ \tilde \pi(\tilde F^\sigma_{(\bar \eta_2)_n, \overline{\xi}_n}( \alpha^\sigma_{\bar \eta_1} \circ \beta_g^{\bar L_1}(A)) (\tilde F^\sigma_{(\bar \eta_2)_n, \overline{\xi}_n})^*)\tilde \Omega}= \inner{\tilde \Omega}{ \tilde \pi \circ \alpha^\sigma_{\bar \eta_2} \circ \beta_g^{r(\bar L_2)}(A)\tilde \Omega} =\omega_{2}(A).
    \end{align*}

    To show the second half of the lemma, we proceed as in \cite[Lem.~4.1]{MR2804555}.
    There exists a unitary $V \in B(\tilde\cH)$ satisfying intertwining $\tilde \pi\circ \alpha_{\bar \eta_1}^\sigma \circ \beta^{r(L_1)}_g $ and $ \tilde \pi \circ \alpha_{\bar \eta_2}^\sigma \circ \beta^{r(L_2)}_g$ satisfying that $V\tilde \Omega = \tilde \Omega$ by uniqueness of the GNS representation.  
    This unitary is unique by Schur's lemma, since $\tilde \pi\circ \alpha_{\bar \eta_1}^\sigma \circ \beta^{r(L_1)}_g $ and $ \tilde \pi \circ \alpha_{\bar \eta_2}^\sigma \circ \beta^{r(L_2)}_g$ are irreducible representations.
    We now show that the sequence $V_n = \tilde \pi(\tilde F_{(\bar \eta_2)_n, \overline{\xi}_n}^\sigma)$ converges WOT to $V$.
    Let $A, B \in \cstar[\loc]$.
    Let $n$ be large enough so that $\supp(B)$ does not intersect $\varsigma_n$.
    For ease of notation, we define $\tilde \pi_i \coloneqq \tilde \pi\circ \alpha_{\bar \eta_i}^\sigma \circ \beta^{r(L_i)}_g$ for $i = 1,2$.
    Then by the same argument as before, we have that
    \begin{align*}
    \langle \tilde \pi_1(A) \tilde \Omega, V_n \tilde \pi_1(B) \tilde \Omega \rangle
    &=
    \langle \tilde \pi_1(A) \tilde \Omega, \tilde \pi(\tilde F_{(\bar \eta_2)_n, \overline{\xi}_n}^\sigma) \tilde \pi_1(B) \tilde \Omega \rangle
    =
    \langle \tilde \pi_1(A) \tilde \Omega, \tilde \pi_2(B) \tilde \pi(\tilde F_{(\bar \eta_2)_n, \overline{\xi}_n}^\sigma)\tilde \Omega \rangle    
    \\&=
    \langle \tilde \pi_1(A) \tilde \Omega, \tilde \pi_2(B) \tilde \Omega \rangle 
    =
    \langle \tilde \pi_1(A) \tilde \Omega, \tilde \pi_2(B) V\tilde \Omega \rangle 
    =
    \langle \tilde \pi_1(A) \tilde \Omega, V\tilde \pi_1(B) \tilde \Omega \rangle.
    \end{align*}
    Now, since $\alpha_{\bar \eta_i}^\sigma \circ \beta^{r(L_i)}_g$ is an automorphism of $\cstar$, $\tilde \pi_i(\cstar[\loc])$ is dense in $\cstar$.
    Therefore, since $(V_n)$ is a uniformly bounded sequence, $V_n \to V$ WOT.
\end{proof}

Armed with these results, we now define the defect automorphisms.
\begin{defn}
\label{def:SETToricCodeDefectAutomorphism}
    Let $\bar \gamma \in \bar P(\Gamma)$ be a half-infinite path, $\bar L$ a completion of $\bar \gamma$ and $\bar \eta = \bar L - \bar \gamma$. Define the \emph{defect automorphism} $\tilde \alpha_{\bar \gamma}^\sigma$ to be $\tilde \alpha_{\bar \gamma}^\sigma \coloneqq \alpha_{\bar \eta}^\sigma \circ \beta^{r(\bar L)}_g.$
    Observe that $\tilde \alpha^\sigma_{\bar \gamma}$ depends on the completion $\bar L$ of $\bar \gamma$, but we suppress this dependence for notational convenience.
\end{defn}

\begin{rem}
    Note that by Lemma \ref{lem:defects are cone transportable}, we get that $\tilde \alpha^{\sigma}_{\bar \gamma}$ are all equivalent for different completions of $\bar \gamma$.
\end{rem}

\begin{rem}
    Physically, the defect automorphism $\tilde \alpha_{\bar \gamma}^\sigma$ creates a defect whose endpoint lives near $\partial_0 \bar \gamma$ and has a domain wall along the path $\bar \gamma$.
\end{rem}

\begin{rem}\label{rem:sigma2ism}

We have $$\tilde \alpha^\sigma_{\bar \gamma} \circ  \tilde \alpha^\sigma_{\bar \gamma} = \alpha_{\bar \gamma}^m,$$ which can be seen immediately by noting that $\beta_g^{r(\bar L)} \circ \alpha_{\bar \eta}^\sigma = \alpha_{\bar \eta}^\sigma \circ \beta_g^{r(\bar L)}$ and $\alpha^\sigma_{\bar \eta} \circ \alpha^\sigma_{\bar \eta} = \alpha_{\bar \eta}^m$.
\end{rem}

Now we define the following defect states as $$\tilde \omega_{\bar \gamma}^\sigma \coloneqq  \tilde \omega \circ \tilde \alpha_{\bar \gamma}^\sigma, \qquad \qquad \tilde \omega^{\sigma, \zeta_\gamma}_{\bar \gamma} \coloneqq  \tilde \omega^\sigma_{\bar \gamma} \circ \zeta_\gamma.$$

We recall the definition of $\bar P_R(\Gamma)$ (Definition \ref{def:set of half-infinite paths equivalent to R}).

\begin{lem}
\label{lem:SET toric code defects are finitely transportable}
    Pick $\bar \gamma_i\in \bar P_R(\Gamma)$ to be half-infinite dual paths for $i = 1,2$ such that $\gamma \coloneqq  \bar \gamma_1 \cap \bar \gamma_2 \in \bar P(\Gamma)$ is a half-infinite path.
    Then $\tilde \pi^\sigma_{\bar \gamma_1} \simeq \tilde \pi^\sigma_{\bar \gamma_2}.$
\end{lem}
\begin{proof}
    We consider pure states $\tilde \omega^\sigma_{\bar \gamma_i}$. Consider completions $\bar L_1$ of $\bar \gamma_1$ and $\bar L_2$ of $\bar \gamma_2$ such that $\bar L_1$ and $\bar L_2$ only differ in a finite region. Call that region $S$. Then for all observables $A \in \cstar[S^c]$ we have,

    $$\tilde \alpha_{\bar \gamma_1}^\sigma(A) = \alpha_{\bar L_1 - \bar \gamma_1} \circ \beta_g^{r(\bar L_1)}(A) = \alpha_{\bar L_2 - \bar \gamma_2} \circ \beta_g^{r(\bar L_2)}(A) = \tilde \alpha_{\bar \gamma_2}^\sigma(A)$$
    
    Note that since $\tilde \omega_{\bar \gamma_1}^\sigma$ and $\tilde \omega_{\bar \gamma_2}^\sigma$ are pure states, $\tilde \omega_{\bar \gamma_1}^\sigma$ and $\tilde \omega_{\bar \gamma_2}^\sigma$ are equivalent if and only if they are quasi-equivalent \cite[Prop.~10.3.7]{MR1468230}.
    We can therefore apply \cite[Cor.~2.6.11]{MR887100}.  Observe, for all $A \in \cstar[S^c]$ we have $$\tilde \omega_{\bar \gamma_1}^\sigma(A) = \tilde \omega \circ \tilde \alpha_{\bar \gamma_1}^\sigma(A) = \tilde \omega \circ \tilde \alpha_{\bar \gamma_2}^\sigma(A) =  \tilde \omega_{\bar \gamma_2}^\sigma(A),$$ so we have that $\tilde \pi^\sigma_{\bar \gamma_1} \simeq \tilde \pi^\sigma_{\bar \gamma_2}$.

    Now if we had chosen different completions $\bar L_i$ then let $\bar L_2'$ be a completion of $\bar \gamma_2$ such that $\bar L_2'$ only differs from $\bar L_1$ on a finite region. 
    Define $\pi' \coloneqq  \tilde \pi \circ \alpha_{\bar L_2' - \bar \gamma_2} \circ \beta_g^{r(\bar L_2')}$.
    By Lemma \ref{lem:defects are cone transportable}, $\pi' \simeq \tilde \pi^\sigma_{\bar \gamma_2}$. 
    But $\pi' \simeq \tilde \pi^\sigma_{\bar \gamma_1}$ by the prior analysis, so we still have $\tilde \pi^\sigma_{\bar \gamma_1} \simeq \tilde \pi^\sigma_{\bar \gamma_2}$ and have shown the required result.
\end{proof}

\subsection{Defect sectors}
Define
$\tilde \pi_{\bar \gamma}^\sigma \coloneqq  \tilde \pi \circ \tilde \alpha^\sigma_{\bar \gamma}$, and set $\tilde\pi$, the GNS representation of state $\tilde \omega$, as the reference representation. 
We first verify that all of the assumptions given in Section \ref{sec:GCrossedAssumptions} hold.
Since the representation of $G$ given by $g \mapsto U^g_v$ is faithful for every vertex $v \in \Gamma$, we have that Assumption \ref{asmp:Faithfulness} holds.
By Lemma \ref{lem:SET toric code unique FF GS}, Assumption \ref{asmp:GInvariance} holds.
By Lemmas \ref{lem:TC extension Haag duality}, \ref{lem:SETToricCodeFDQC}, and \ref{lem:QCABSHaagDuality}, Assumption \ref{asmp:BoundedSpreadHaagDuality} is satisfied using the fact that $\tilde \pi \simeq \pi_0 \circ \alpha^{-1}$.
By Lemma \ref{lem:SET toric code unique FF GS}, $\tilde \omega$ is pure, so Assumption \ref{asmp:PureState} holds.
Note that this implies that $\tilde \pi$ is irreducible.
In addition, $\tilde \omega$ is translation-invariant, so Assumption \ref{asmp:InfiniteFactor} holds by a standard argument \cite{MR2804555, MR2281418}.

\begin{lem}
\label{lem:defect sector TC case}
    For all paths $\bar \gamma \in \bar P_R(\Gamma)$, the representation $\tilde \pi^{\sigma}_{\bar \gamma}$ is a $g$-defect sector.
\end{lem}
\begin{proof}
    Let $\bar L$ be a completion of $\bar \gamma$ and let $\bar \eta = \bar L - \bar \gamma$. Choose a cone $\Lambda \in \cL$ such that $\bar \eta$ is contained in $\Lambda$ and let $A \in \cstar[\Lambda^c]$. We then have
    \begin{align*}
        \tilde \pi^\sigma(A) &= \tilde \pi \circ \tilde \alpha_{\bar \gamma}^{\sigma}(A) = \tilde \pi \circ \alpha_{\bar \eta}^{\sigma} \circ \beta_g^{r(\bar L)}(A) = \tilde \pi \circ \beta_g^{r(\bar L)}(A).
    \end{align*}
    In the last equality we used that $\beta_g(A) \in \cstar[\Lambda^c]$ since $\beta_g$ is an onsite symmetry and additionally that for any observable $A' \in \cstar[\Lambda^c]$, $\alpha^\sigma_{\bar \eta}(A) = A$ since $\alpha^\sigma_{\bar \eta}$ is localized inside $\Lambda$.
    Since $\bar \gamma \in \bar P_R(\Gamma)$, $r(\bar L) \cap \Lambda^c$ differs from $r(\Lambda)$ by finitely many vertices.
    This shows $\tilde \pi^\sigma_{\bar \gamma}$ is localized in $\Lambda$. 
    
    It remains to be shown that $\tilde \pi_{\bar \gamma}^\sigma$ is transportable. Choose another cone $\Lambda'$. We now choose another path $\bar \gamma' \in \bar P_R(\Gamma)$ and completion $\bar L'$ such that $\bar L' - \bar \gamma'$ lies entirely in $\Lambda'$. By the above argument, we have that $\tilde \pi^\sigma_{\bar \gamma'}$ is $g$-localized in $\Lambda'$. Using Lemma \ref{lem:SET toric code defects are finitely transportable} we get that $\tilde \pi^\sigma_{\bar \gamma} \simeq \tilde \pi^\sigma_{\bar \gamma'}$ giving us that $\tilde \pi^\sigma_{\bar \gamma}$ is transportable.
\end{proof}

We define some more representations given by $\tilde \pi^{\sigma, \zeta}_{\bar \gamma} \coloneqq  \tilde \pi_{\bar \gamma}^\sigma \circ \zeta$.
We omit subscripts on $\zeta$ since we are fixing the paths defining the automorphisms in this instance.

\begin{lem}
\label{lem:anyon sectors of defect SET}
    The representations given by $\{\tilde \pi^{\sigma, \zeta}_{\bar \gamma_R}\}_{\zeta}$ are mutually disjoint and anyon sectors with respect to $\tilde \pi^\sigma_{\bar \gamma_R}$, and any anyon sector is unitarily equivalent to one of them.
\end{lem}
\begin{proof}
    Since $\tilde \alpha^\sigma_{\bar \gamma_R}$ is an FDQC, the result follows from Lemmas \ref{lem:anyon sectors of GS SET}, \ref{lem:FDQCQuasi-Factorizable}, and \cite[Thm.~4.7]{MR4426734}.
\end{proof}

\begin{lem}
    Pick some $\bar \gamma_0 \in \bar P_R(\Gamma)$. The representations $\tilde \pi^{\sigma, \zeta}_{\bar \gamma_0}$ are $g$-defect sectors with respect to $\tilde \pi$ and the representations $\tilde \pi^{\zeta}$ are $1$-defect sectors for $\zeta \in \{\id, \alpha_\gamma^\epsilon, \alpha_{\bar \gamma}^m, \alpha_{\gamma, \bar \gamma}^\psi\}$, assuming that $\gamma \in P(\Gamma), \bar \gamma \in \bar P(\Gamma)$ are contained in some cone $\Lambda \in \cL$.
\end{lem}
\begin{proof}
    From Lemma \ref{lem:defect sector TC case} we know that $\tilde \pi_{\bar \gamma_0}^{\sigma}$ is a $g$-defect sector, and $\tilde \pi$ is obviously a $1$-defect sector. 
    Since $\tilde \pi^\zeta$ are anyon sectors with respect to $\tilde \pi$ (Lemma \ref{lem:anyon sectors of GS SET}) it follows that $\pi^\zeta$ are $1$-defect sectorizable.
    Since $\tilde \pi_{\bar \gamma_0}^{\sigma, \zeta}$ are anyon sectors with respect to $\tilde \pi_{\bar \gamma_0}^{\sigma}$, it follows that $\tilde \pi_{\bar \gamma_0}^{\sigma, \zeta}$ are $g$-defect sectorizable.

    Since $\gamma, \bar \gamma \subset \Lambda$, $\zeta$ is localized in $\Lambda$. 
    Thus for all observables $A \in \cstar[\Lambda^c]$ we have $\tilde \pi^{\zeta}(A)= \tilde \pi (A)$, so $\tilde \pi^{\zeta}$ is localized in $\Lambda$. By the above argument, $\tilde \pi^{\zeta}$ is a $1$-defect sector. 
    Similarly, choosing a completion $\bar L_0$ of $\bar \gamma_0$, we have $\tilde \pi_{\bar \gamma_0}^{\sigma, \zeta}$ is $g$-localized in some $\Lambda' \in \cL$ containing $\gamma$, $\bar \gamma$, and $\bar L_0 - \bar \gamma_0$, and therefore $\tilde \pi_{\bar \gamma_0}^{\sigma, \zeta}$ is a $g$-defect sector.
\end{proof}

We now fix the dual path to be $\bar \gamma_R$ as shown in Figure \ref{fig:chosen_ray}, and drop it from the notation. The defect will always be on this dual path. The new notation is $$\tilde \pi^\sigma \coloneqq  \tilde \pi^\sigma_{\bar \gamma_R}, \qquad \qquad \tilde \pi^{\sigma, \zeta} \coloneqq  \tilde \pi^\sigma \circ \zeta.$$

\begin{prop}
\label{lem:classification of defect sectorizable representations TC}
    Let $\zeta \in \{\id, \alpha_\gamma^\epsilon, \alpha_{\bar \gamma}^m, \alpha_{\gamma, \bar \gamma}^\psi\}$. The representations $\{\pi^\zeta\}_{\zeta}$ are irreducible and mutually disjoint defect sectors with respect to $\tilde \pi$ as well are $\{ \pi^{\sigma,\zeta}\}_{\zeta}$. Every defect-sectorizable representation is unitarily equivalent to one of them.
\end{prop}
\begin{proof}
    Note that the collections $\{\pi^\zeta\} _{\zeta}$ and $\{\pi^{\sigma,\zeta}\}_{\zeta}$ are mutually disjoint defect sectors by Lemmas \ref{lem:anyon sectors of GS SET} and \ref{lem:anyon sectors of defect SET}.
    Now, let $\pi$ be a $g$-sectorizable representation for $g$ being the non-trivial group element. We have from Lemma \ref{lem:defect_sectorizable_reps_are_equivalent_to_a_defect_sector} that $\pi \simeq \rho$ for some $\rho$ being a $g$-defect sector.
    From Lemma \ref{lem:defect state GNS reps are sectorizable (SPT case)} we have that $\tilde \pi_{}^{\sigma}$ is a $g$-defect sector. 
    We have from Lemma \ref{lem:GDefectsRelationToAnyons} that $\rho$ is an anyon sector with respect to $\tilde \pi^\sigma$. But by Lemma \ref{lem:anyon sectors of defect SET} $\rho \simeq \tilde \pi_{}^{\sigma, \zeta}$ for some $\zeta$ in the set. Repeating the same analysis for $g=1$ and using Lemmas \ref{lem:GDefectsRelationToAnyons}, \ref{lem:anyon sectors of GS SET} gives us the other case and hence the result.
\end{proof}

\begin{rem}
    It can be shown that the category $\DHR_{\tilde \pi^\sigma}(\Lambda)$ is braided monoidally equivalent to $\DHR_{\tilde \pi}(\Lambda)$. Since $\tilde \alpha_{\bar \gamma}^\sigma$ is an FDQC, the result immediately follows from \cite[Thm. ~6.1]{MR4362722}.

    Even though the above result is stronger than Proposition \ref{lem:classification of defect sectorizable representations TC}, we note that $\DHR_{\tilde \pi^\sigma}(\Lambda)$ is the category of anyon sectors with respect to $\tilde \pi^\sigma$ as the reference state. While the objects in this category are the objects we are ultimately interested in, we note that we do not want to inherit the fusion and braiding structure from this category as it disregards the presence of the defect. Below we construct the $G$-crossed braided monoidal category that we are interested in.
\end{rem}

\subsection{Defect tensor category}
\label{sec:F_symbols}

In this subsection, we describe the category of symmetry enriched toric code defects $\GSec^{ETC}$ as a $\mathbb{Z}_2$-graded tensor category. 
In the subsections which follow, we will give the rest of the $G$-crossed braiding data.

For the remainder of this manuscript, we fix a semi-infinite path $\cpath$ and dual path $\cdual$ by which we define our defect automorphisms.
Our specific choice for $\cpath$ and $\cdual$ will be explicated in Notation \ref{notation:defects}.
We denote $\DHR_{\tilde \pi^\sigma}(\Lambda)$ as the \textit{linear} category of localized transportable anyon sectors with respect to $\tilde \pi^\sigma_{\bar \gamma_R}$.
It is important to formally forget the usual tensor product in $\DHR_{\tilde \pi^\sigma}(\Lambda)$.
Recalling that $\DHR_{\tilde \pi}(\Lambda)$ is the category of localized transportable anyon sectors with respect to $\tilde \pi$, we have that
$$\GSec^{ETC}(\Lambda) = \DHR_{\tilde \pi}(\Lambda) \oplus \DHR_{\tilde \pi_{}^\sigma}(\Lambda)$$
where $\oplus$ is the direct sum of linear categories, rather than the direct sum of tensor or braided tensor categories.
We take the tensor product to be our usual tensor product of defect sectors.
It will later become apparent that the direct sum will be promoted to a $\mathbb{Z}_2$-grading under this tensor product.

We denote the simple objects in $\DHR_{\tilde \pi}(\Lambda)$ by $\{1, \epsilon, m , \psi\},$ where, for example, $\epsilon$ corresponds to extension of the automorphism $\alpha_{\cpath}^\epsilon$ to $\cstar^a$ (Lemma \ref{lem:GDefectsDefinedOnAuxiliaryAlgebra}). 
Likewise, the simple objects in $\DHR_{\tilde \pi_{}^\sigma}(\Lambda)$ are denoted by $\{1^\sigma, \epsilon^\sigma, m^\sigma , \psi^\sigma\}$.
By Lemma \ref{lem:anyon sectors of defect SET}, we may define these distinct defect sectors for each $a\in\{1, \epsilon, m , \psi\}$ by the unique extension of  $\tilde{\pi}^\sigma_{\cdual}\circ a|_{\cstar}$ to $\cstar^a$. 
In other words, we are defining
$a^\sigma:=1^\sigma\otimes a$ for each $a\in\{1, \epsilon, m , \psi\}$ where $1^\sigma$ is the extension of $\tilde{\pi}^\sigma_{\cdual}$ to $\cstar^a$.
We have chosen the basis $I:=\{1, \epsilon, m , \psi,1^\sigma, \epsilon^\sigma, m^\sigma , \psi^\sigma\}$ of $\mathcal{K}_0(\GSec^{ETC})$.

\begin{nota}
    To ease notation, from this point forward we omit the (dual) path subscript on automorphisms when the (dual) path is the canonical ($\cdual$) $\cpath$.
\end{nota}

\begin{nota}\label{notation:defects}
    We will use Figure \ref{fig:strings} to fix some geometric notation.
    First, we take the origin vertex $0$ to be the large yellow dot. 
    The orientation of each vertical edge is upward and the orientation of each horizontal edge is to the right.
    We take $\overline{L}$ to be the vertical gray line just to the right of the vertex $0$ so that $\beta^{r(\overline{L})}$ is the symmetry action on the vertices to the right of this line. The red ray is ${\bar \gamma_R}$. Take $\cpath$ to be the purple wiggling ray extending downward from $0$. From this, we have $\partial\cpath=0$. We take $\cdual$ to be the orange dual path which is just to the right of $\cpath$ and terminates at the edge neighboring the origin.
    The blue shading indicates the cone $\Lambda$.
    
    Note that since the half plane symmetry acts to the right of $\cpath$, we automatically have that $\alpha^\epsilon\circ \beta^{r(\overline{L})}_g =\beta^{r(\overline{L})}_g \circ\alpha^\epsilon$.

    \begin{figure}[!ht]
\begin{center}
\begin{tikzpicture}[scale=0.6]
    \filldraw[blue!30,thick,fill=blue!10](2.75,0)--(2.75,7.75)--(9,7.75)--(9,0);

    \foreach \x in {0,...,9}{
    \draw[thick,black](\x,0)--(\x,10);
    }
    \foreach \y in {0,...,10}{
    \draw[thick,black](0,\y)--(9,\y);
    }
    \draw[thick,black!50](4.5,10)--(4.5,0);
    
     \draw[very thick, purple,snake it](4,0)--(4,6);
    \foreach \x in {0,...,6}{
    \draw[very thick,orange](4,\x)--(5,\x);
    }
   \filldraw[purple,thick,fill=yellow] (4,6) circle(.3cm);
    \draw[thick,red](4.4,6.5)--(4.4,10);
    \filldraw[thick,red,fill=red](4.4,6.5) circle(.1cm);
\end{tikzpicture}
\end{center}
    \caption{The ray ${\bar \gamma_R}$ and $\epsilon, m, \psi$ strings.}
    \label{fig:strings}
\end{figure}
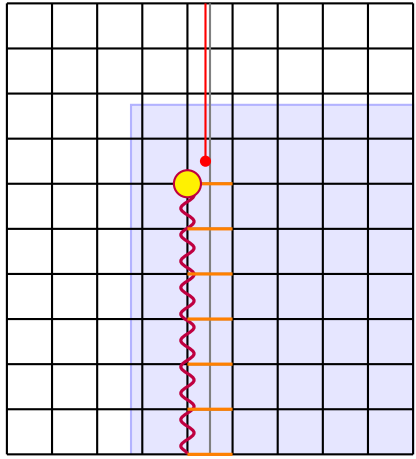
\end{nota}

\begin{lem}\label{lem:TC_Fusion}
The fusion rules of $\GSec^{ETC}$ are determined by the usual toric code fusion rules in $\DHR_{\tilde{\pi}}(\Lambda)$, as well as the equalities $a^{\sigma}\otimes b= (ab)^\sigma$ for $a,b\in\{1,\epsilon,m,\psi\}$ and $1^\sigma\otimes 1^{\sigma}= m$. 
\end{lem}
\begin{proof}
    We already proved in Proposition \ref{prop:equiv_to_TC} that $\DHR_{\tilde{\pi}}(\Lambda)$ obeys the usual fusion rules for the toric code.
    The second statement follows directly from the fact that $a^{\sigma}= 1^\sigma\otimes a$.
    Finally, we may notice that $\beta_g^{r(\overline{L})}\circ \alpha^\sigma=\alpha^\sigma\circ \beta_g^{r(\overline{L})}$ and by Remark \ref{rem:sigma2ism} $\alpha^\sigma\circ \alpha^\sigma=\alpha^m$. Therefore, 
    $$
    1^{\sigma}\circ 1^{\sigma}|_{\fA}=\alpha^\sigma\circ \beta_g^{r(\overline{L})}\circ \alpha^\sigma\circ \beta_g^{r(\overline{L})}=\alpha^\sigma\circ  \alpha^\sigma\circ \beta_g^{r(\overline{L})}\circ \beta_g^{r(\overline{L})}=\alpha^m.
    $$
    As endomorphisms of the auxiliary algebra, $1^\sigma \otimes 1^\sigma=m$.

    Note that the remaining fusion rules all follow from the existence of a $G$-crossed braiding and the fact that the symmetry acts trivially on the anyons, as was shown in Lemma \ref{lem:TCSymmetry_permute}.
\end{proof}

\begin{prop}\label{prop:triv_assoc}
    The tensorators $\Omega_{i,j}$ are trivial.
    Therefore the skeletalization of the tensor category $\GSec^{ETC}$ is strict.
\end{prop}
\begin{proof}
    To prove this, we simply need to show that we may pick representative endomorphisms for each isomorphism class so that the composition of any two is also a representative. After considering Lemma \ref{lem:TC_Fusion}, all that remains is to show that $a\otimes 1^{\sigma}=a^{\sigma}$ for $a\in\{1,\epsilon,m,\psi\}$. 
    The remaining equalities all easily follow from this fact.

    For $a\in \{1,\epsilon,m,\psi \}$, we have that 
    $$
    \alpha^a\circ\alpha^\sigma=\alpha^\sigma\circ\alpha^a\quad\quad \text{and}\quad\quad\alpha^a\circ\beta_g^{r(\overline{L})}=\beta_g^{r(\overline{L})}\circ\alpha^a.
    $$
    Therefore, we have that
    $$
    a\otimes 1^\sigma|_{\fA}=\alpha^a\circ\alpha^\sigma\circ\beta_g^{r(\overline{L})}=\alpha^\sigma\circ\beta_g^{r(\overline{L})}\circ \alpha^a=a^\sigma|_{\fA}.
    $$
    Extending this endomorphism to the auxiliary algebra $\cstar^a$ gives that $a\otimes 1^\sigma=a^\sigma$.
    Therefore, all of the tensorators are trivial.
\end{proof}

\subsection{Symmetry fractionalization}
\label{sec:frac_data}

We now compute the symmetry fractionalization data for $\GSec^{ETC}$ following our prescription in Section \ref{sec:SymmetryFractionalization}.
Recall that we have chosen our basis of $\mathcal{K}_0(\GSec^{ETC})$ to be $I=\{1, \epsilon, m , \psi,1^\sigma, \epsilon^\sigma, m^\sigma , \psi^\sigma\}$.
For each $h\in \mathbb{Z}_2$ and $i\in I$, we have the unitary intertwiner $V_h^i\colon\gamma_h(\pi_i)\rightarrow \pi_{h(i)}$.
We reserve $g\in \mathbb{Z}_2$ to be the non-trivial element.
When $h\neq g,$ $V^i_h=1$.

\begin{lem}\label{lem:symmetry_on_e}
    For any $x\in\cstar^a$,
    $$
    \tau^z_{0} \gamma_g(\alpha^\epsilon)(x)(\tau^z_{0})^*=\alpha^\epsilon(x).
    $$
\end{lem}
\begin{proof}
    By continuity, it is sufficient to prove this statement in the case where $x$ is a local operator.
    Using the definition $\gamma_g$, we have
    $$
    \tau^z_{0} \gamma_g(\alpha^\epsilon)(x)(\tau^z_{0})^*=\tau^z_{0} \beta_g\alpha^\epsilon(\beta_g^{-1}(x))(\tau^z_{0})^*.
    $$
    Since $x$ is local and $\beta_g^{-1}$ preserves the support of local operators, $\beta^{-1}_g(x)$ is also local.
    Take $\gamma'$ to be a finite subpath of the path $\cpath$ which defines $\alpha^\epsilon$ with $\partial_0\gamma'=\partial\cpath=0$.
    Using Remark \ref{rem:SETTCAutomorphismDefn}, since $\alpha^\epsilon$ is an FDQC, we may choose $\gamma'$ to be long enough such that
    $$
    \alpha^{\epsilon}(\beta_g^{-1}(x))=\alpha^{\epsilon}_{\gamma'}(\beta_g^{-1}(x))
    $$
    and such that the support of $\alpha^\epsilon(x)$ is disjoint from $\partial_1\gamma'$.
    Finally, this reasoning along with Lemma \ref{lem:action of symmetry on string ops} implies that
    \begin{align*}
    \tau^z_{0} \gamma_g(\alpha^\epsilon)(x)(\tau^z_{0})^*&=\tau^z_{0} \beta_g(\alpha^\epsilon_{\gamma'}(\beta^{-1}_g(x)))(\tau^z_{0})^*\\
    &=\tau^z_{0} \beta_g(\tilde{F}^\epsilon_{\gamma'})x\beta_g(\tilde{F}^\epsilon_{\gamma'})^*(\tau^z_{0})^*\\
    &=\tau^z_{\partial_1 \gamma'}\tilde{F}^\epsilon_{\gamma'}x(\tilde{F}^\epsilon_{\gamma'})^*(\tau^z_{\partial_1 \gamma'})^*\\
    &=\tau^z_{\partial_1 \gamma'}\alpha_{\gamma'}^\epsilon(x)(\tau^z_{\partial_1 \gamma'})^*\\
     &=\tau^z_{\partial_1 \gamma'}\alpha^\epsilon(x)(\tau^z_{\partial_1 \gamma'})^*\\
    &=\alpha^\epsilon(x).
    \qedhere
    \end{align*}
    
\end{proof}

This lemma shows that we may take $V^\epsilon_g=\tau^z_{0}$ where $g$ is the non-trivial element of $\mathbb{Z}_2$.
Similar computations reveal that $$V^1_g=V^m_g=V^{1^\sigma}_g=V^{m^\sigma}_g=\mathds{1}$$ and $$V^\psi_g=V^{\epsilon^\sigma}_g=V^{\psi^\sigma}_g=\tau^z_{\partial \gamma}.$$
Noting that $1^\sigma$ is invariant under the symmetry, we have that the only non-trivial values of $\eta$ (as operators corresponding to morphisms, rather than morphisms themselves) from Section \ref{sec:SymmetryFractionalization} are 
$$
\eta(g,g)_\epsilon=\eta(g,g)_\psi=\eta(g,g)_{\epsilon^\sigma}=\eta(g,g)_{\psi^\sigma}=-1.
$$

Since $V_1=\mathds{1}$, we have that for all basis elements $a,b\in \mathcal{K}_0(\GSec)$, $\mu_1(a,b)=1$.

Now we compute $\mu_g(a,b)$ where $g$ is the non-trivial element of $\mathbb{Z}_2$. Since Proposition \ref{prop:triv_assoc} tells us that all of the tensorators ($\Omega_{i,j}$) are trivial, we have that
$$
V_g^{ij}=\mu_g(i,j)V_g^i\gamma_g(\pi_i)(V_g^j).
$$
In this model, it can be easily checked that $V^{ij}_g=V^i_gV^j_g$.
Therefore,
$$\mu_g(i,j)\mathds{1}=\gamma_g(\pi_i)(V^j_g)^*V^j_g=V^i_g\pi_i(V^j_g)^*(V^i_g)^* V^j_g.$$
However, using the fact that $\pi_i(\tau^z_{0})=\tau^z_{0}$ and $[V_g^i,V_g^j]=0$ for all basis elements $i,j\in I$, we have that $\mu_g(i,j)=1$. Therefore, all of the skeletal data corresponding to $\mu$ is trivial.

\subsection{\texorpdfstring{$G$}{G}-crossed braiding}
\label{sec:braid_data}

We are now in a position to compute the $G$-crossed braiding data. Most importantly, we want to compute $c_{1^\sigma,1^\sigma}$, $c_{a,1^\sigma}$, and $c_{1^\sigma,a}$ for $a\in\{1,\epsilon,m,\psi\}$.

In Notation \ref{notation:braiding}, we will define the operators $U^\pi_N$ which limit to the operator $U^\pi$ which transports $\pi$ from $\Lambda$ to $\Delta$ for certain simple $\pi$ in $\GSec^{ETC}$.

\begin{nota}\label{notation:braiding}
    We will consider the two diagrams (Figures \ref{fig:sigma} and \ref{fig:U}) below to define $U_N^\pi$. 
    The figures depict the case where $N=4$. We start with the first diagram (Figure \ref{fig:sigma}). Refer to Notation \ref{notation:defects} for the definitions of the defect automorphisms in terms of this geometry. Recall that we take the vertex $0$ to be the large yellow dot. We take $\overline{L}$ to be the vertical gray line just to the right of the vertex $0$ (the origin) so that $\beta^{r(\overline{L})}_g$ is the symmetry action on the vertices to the right of this line. Just as before, we take $\cpath$ to be the purple wiggling ray and $\cdual$ to be the dual path just to the right of $\cpath$ terminating at the edge neighboring the origin. 
    The cone $\Lambda$ is given blue shading where as the cone $\Delta$ is in red.
    
    Take the path $\overline{\xi}_N$ to be the black wiggling dual path going clockwise around the large black dots. The edges in orange are the edges traversed by $\overline{\xi}_N$. This dual path is parameterized by $N$ so as to intersect $\gamma$ at the $N$th edge below the origin. Take the region $A_N$ to be the $(N+2)\times 5$ set of black dots bordered by $\overline{\xi}_N$.  We are now able to define
    $$
    U_N^\sigma
    \coloneqq 
    F^\sigma_{\overline{\xi}_N}\prod\limits_{v\in A_N}\tau_v^x.
    $$
    We also define
    $$
    U_N^m\coloneqq F^m_{\overline{\xi}_N}.
    $$
    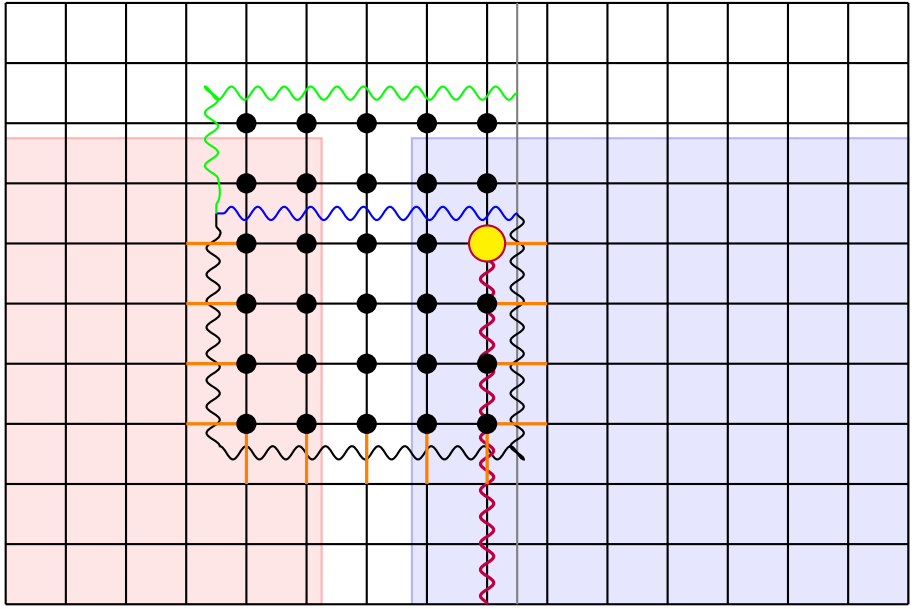
\begin{figure}[!ht]
    \begin{center}
\begin{tikzpicture}[scale=0.8]
    \filldraw[blue!30,thick,fill=blue!10](6.75,0)--(6.75,7.75)--(15,7.75)--(15,0);
    \filldraw[red!30,thick,fill=red!10](0,7.75)--(5.25,7.75)--(5.25,0)--(0,0);
    \foreach \x in {0,...,15}{
    \draw[thick,black](\x,0)--(\x,10);
    }
    \foreach \y in {0,...,10}{
    \draw[thick,black](0,\y)--(15,\y);
    }
    \draw[thick,black!50](8.5,10)--(8.5,0);
    \draw[thick,black,snake it](8.5,6.5)--(8.5,2.5)--(3.5,2.5)--(3.5,6.5);
    \draw[thick, blue, snake it] (8.5, 6.5) -- (3.5, 6.5);
    \draw[thick, green, snake it] (8.5, 8.5) -- (3.5, 8.5) -- (3.5, 6.5);
     \draw[very thick, purple,snake it](8,0)--(8,6);
    \foreach \x in {0,...,3}{
    \draw[very thick,orange](3,3+\x)--(4,3+\x);
    \draw[very thick,orange](8,3+\x)--(9,3+\x);
    \draw[very thick,orange](\x+4,2)--(\x+4,3);
    }
    \draw[very thick,orange](8,2)--(8,3);
    \foreach \x in {0,...,4}{
    \foreach \y in {0,...,5}{
    \filldraw[thick,black,fill=black](4+\x,3+\y)circle(.15cm);
    }
    }
   \filldraw[purple,thick,fill=yellow] (8,6) circle(.3cm);
    
\end{tikzpicture}
\end{center}
    \caption{Geometry of $1^\sigma$ strings and $U_N^m$ and $U_N^\sigma$ operators in the case where $N=4$.}
     \label{fig:sigma}
\end{figure}

    In the second diagram (Figure \ref{fig:U}), we have drawn the path $\zeta_N$ in purple in the case where $N=4$. It extends $N$ edges down from $0$, travels along $5$ edges toward $\Delta$, and then extends upward $N$ edges. Note that $\zeta_N$ and $\overline{\xi}_N$ share the $N$th edge below the origin.  From $\zeta_N$, we define
    $$
    U_N^\epsilon\coloneqq \tilde{F}^\epsilon_{\zeta_N}.
    $$
    We also define
    $$
    U_N^\psi\coloneqq U_N^\epsilon U_N^m.
    $$
    Note that we previously referred to $\tilde{F}^\epsilon_\gamma$ and $F^m_{\overline{\gamma}}$ simply as $\tilde{F}_\gamma$ and $F_{\overline{\gamma}}$, respectively. We will include these superscripts in the following discussion to avoid confusion.
    \begin{figure}[!ht]
\begin{center}
\begin{tikzpicture}[scale=0.8]
    \filldraw[blue!30,thick,fill=blue!10](6.75,0)--(6.75,7.75)--(15,7.75)--(15,0);
    \filldraw[red!30,thick,fill=red!10](0,7.75)--(5.25,7.75)--(5.25,0)--(0,0);
    \foreach \x in {0,...,15}{
    \draw[thick,black](\x,0)--(\x,10);
    }
    \foreach \y in {0,...,10}{
    \draw[thick,black](0,\y)--(15,\y);
    }
    \draw[thick,black!50](8.5,10)--(8.5,0);
    \draw[thick,black,snake it](8.5,6.5)--(8.5,2.5)--(3.5,2.5)--(3.5,6.5);

    \foreach \x in {0,...,4}{
    \foreach \y in {0,...,5}{
    \filldraw[thick,black,fill=black](4+\x,3+\y)circle(.15cm);
    }
    }
   \draw[very thick, purple](8,6)--(8,2)--(3,2)--(3,6);
   \filldraw[purple,thick,fill=yellow] (8,6) circle(.3cm);
       
\end{tikzpicture}
\end{center}
    \caption{Geometry of $U_N^\epsilon$ and $U_N^\psi$ operators in the case where $N=4$.}
     \label{fig:U}
\end{figure}
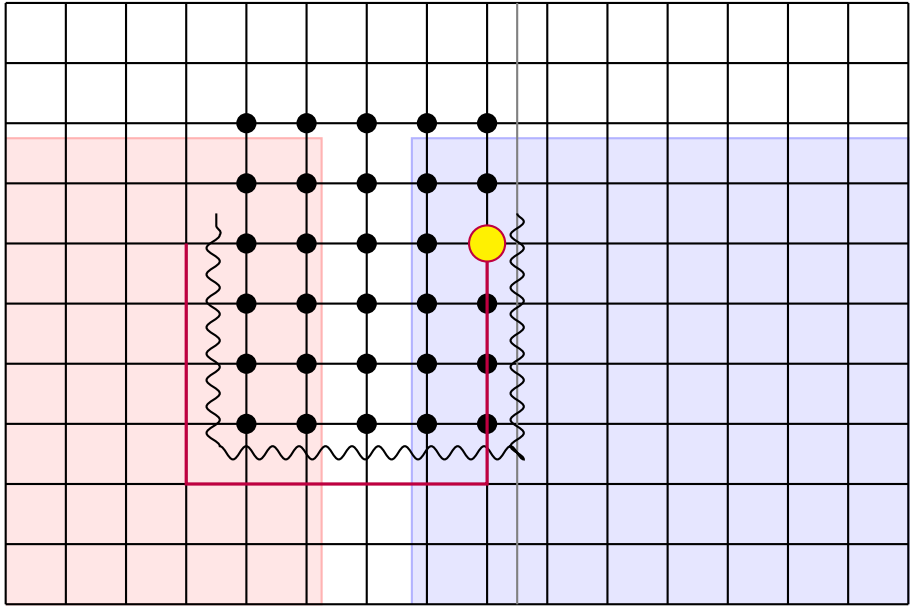

\end{nota}

\begin{prop}\label{prop:well-defined_transport}
    The sequences $(U_N^\sigma)$, $(U_N^m)$, $(U_N^\epsilon)$, and $(U_N^\psi)$ converge WOT to unitaries $U^\sigma,U^\epsilon,U^m,U^\psi$ which transport their corresponding defect from $\Lambda$ to $\Delta$.
\end{prop}
\begin{proof}
The statement for $U^\epsilon, U^m, U^\psi$ follows by appropriately modifying the proof of \cite[Lem.~4.1]{MR2804555} to take into account that the string operators for our model are not exactly the usual toric code string operators.
We now prove this statement for $U^\sigma$. 
Define $\overline{\xi}'$ to be the blue dual path which begins at $0$ in Figure \ref{fig:sigma}.
Take $S$ to be the $2\times 5$ grid of vertex spins directly above $\overline{\xi}'$ in Figure \ref{fig:sigma}.
Let $\overline{\gamma}'$ be the green dual path in Figure \ref{fig:sigma} oriented so that it has the same endpoint as $\overline{\xi}'$.
Let $\overline{\eta}_1$ be the vertical dual path starting at the endpoint of $\overline{\gamma}'$ and continuing downward, and let $\overline{\eta}_2 \coloneqq \overline{\xi}' + \overline{\eta}_1$. 
We let $\overline{L}_1$ be the infinite path that consists of the path $\overline{L}$ followed by $\overline{\gamma}'$ followed by $\overline{\eta}_1$.
Similarly, we let $\overline{L}_2$ be the infinite path that consists of the path $\overline{L}$ followed by $\overline{\eta}_2$.
Then we observe that $\tilde \pi \circ \alpha_{\overline\eta_1}^\sigma \circ \beta^{r(\overline{L}_1)}_g$ is a defect sector $g$-localized in $\Delta$.
Now, observe that the unitary $U \coloneqq F^\sigma_{\overline{\xi}'} \prod\limits_{v \in S} \tau^x_v$ intertwines the defect sector $g$-localized at $\Lambda$ with $\tilde \pi \circ \alpha_{\overline\eta_2}^\sigma \circ \beta^{r(\overline{L}_2)}_g$.
Then, applying Lemma \ref{lem:defects are cone transportable}, we obtain the desired result.
\end{proof}

In principle, we can define more unitary intertwiner to transport the other symmetry defects from $\Lambda$ to $\Delta$.
We omit the definition of these operators because they will not be used in the computation of the $G$-crossed braiding data.

In what follows, we will use the definition of the anyon automorphisms in terms of string operator adjunctions, as per Remark \ref{rem:SETTCAutomorphismDefn}.

\begin{prop}\label{prop:TCbraiding}
    The only non-trivial braiding isomorphisms of anyons are given by
    $$
    c_{m,\epsilon}=c_{m,\psi}=c_{\psi,\epsilon}=c_{\psi,\psi}=-1.
    $$
\end{prop}
\begin{proof}
    Based on the geometry of our set up, we have that whenever $a\in\{1,\epsilon\}$ or $b\in \{1,m\}$,
    $$
    c_{a,b}=b((U^a)^*)U^a=1.
    $$
    Let $N$ denote the $N$th edge below $0$. Using the definition $\alpha^m_{\cdual}$ and $U_N^\epsilon$, we have that 
    $$
    m((U^\epsilon_N)^*)U^\epsilon_N=\sigma^x_N (U^\epsilon_N)^* \sigma^x_N U^\epsilon_N=\sigma^x_N\sigma^z_N\sigma^x_N\sigma^z_N=-1.
    $$
    Therefore, $c_{m,\epsilon}=-1$.

    By continuity, we have
    $$
    c_{m,\psi}=\psi((U^m)^*)U^m=\epsilon(m((U^m)^*))U^m=\epsilon((U^m)^*)U^m=c_{m,\epsilon}=-1.
    $$
    Using Facts \ref{facts:braiding}, we have that
    $$
    c_{\psi,b}=c_{m,b}c_{\epsilon,b}=c_{m,b}.
    $$
    Therefore, we also have $c_{\psi,\epsilon}=c_{\psi,\psi}=-1$.
\end{proof}

\begin{prop}\label{prop:sigma_epsilon}
We have the braiding $c_{1^\sigma,\epsilon}=\tau^z_{0}$. 
\end{prop}
\begin{proof}
    For this proof, take the $\gamma_N$ to be the path made up of the top $N$ edges of $\cpath$. Define the self-adjoint operator $X_N\coloneqq \prod\limits_{v\in A_N}\tau^x_v$. Then
    $$
    U_N^\sigma=X_N F^\sigma_{\overline{\xi}_N}=F^\sigma_{\overline{\xi}_N}X_N.
    $$

    We label edge $1$ to be the topmost edge of $\gamma_N$ and $N$ to be the bottom most edge of $\gamma_N$. We label the $N+1$ vertices going from the top of edge $1$ to the bottom of edge $N$ as $0=\partial\cpath=\partial_0\gamma_N$ to $N=\partial_1\gamma_N$. Using this notation, we define
    $$
    S_N\coloneqq\prod\limits_{e=1}^N \sigma_e^z,
    \qquad\qquad
    T_n\coloneqq\prod\limits_{k=1}^{n}i^{\sigma^z_k(\tau^z_{k-1}-\tau^z_k)/2},
    $$
    so that $\tilde{F}^\epsilon_{\gamma_N}=S_N T_N$.
    A simple calculation reveals that $T_n^2=\tau_0^z\tau_n^z$.
    Also note that
    $$
    X_NT_N^*=X_NT_{N-1}^*i^{-\sigma^x_N(\tau_{N-1}^z-\tau_N^z)/2}=T_{N-1}X_N i^{-\sigma^x_N(\tau_{N-1}^z-\tau_N^z)/2}.
    $$
    We use these facts and notation to compute the following:
    \begin{align*}
    \epsilon((U^\sigma_N)^*)&=\tilde{F}_{\gamma_N}^\epsilon (U^\sigma_N)^* (\tilde{F}_{\gamma_N}^\epsilon)^*\\
    &=\left[\tilde{F}_{\gamma_N}^\epsilon (F^\sigma_{\overline{\xi}_N})^* (\tilde{F}_{\gamma_N}^\epsilon)^* \right]\left[\tilde{F}_{\gamma_N}^\epsilon X_N (\tilde{F}_{\gamma_N}^\epsilon)^*\right]\\
    &=\left[\sigma^z_N (F^\sigma_{\overline{\xi}_N})^* \sigma^z_N \right]\left[\tilde{F}_{\gamma_N}^\epsilon X_N (\tilde{F}_{\gamma_N}^\epsilon)^*\right]\\
    &=\left[e^{-i\frac{\pi}{2}\sigma_N^x}(F^\sigma_{\overline{\xi}_N})^*\right]\left[S_NT_N X_N T_N^* S_N\right]\\
    &=\left[e^{-i\frac{\pi}{2}\sigma_N^x}(F^\sigma_{\overline{\xi}_N})^*\right]\left[S_NT_{N-1}^2 i^{\sigma^x_N(\tau_{N-1}^z-\tau_N^z)/2} X_N  i^{-\sigma^x_N(\tau_{N-1}^z-\tau_N^z)/2}S_N\right]\\
    &=\left[-i\sigma_N^x(F^\sigma_{\overline{\xi}_N})^*\right]\left[S_NT_{N-1}^2 i^{\sigma^x_N\tau_{N-1}^z}  X_N S_N\right]\\
    &=\left[-i\sigma_N^x(F^\sigma_{\overline{\xi}_N})^*\right]\left[S_N\tau^z_0\tau^z_{N-1} (i\sigma^x_N\tau_{N-1}^z)  X_N S_N\right]\\
    &=\sigma_N^x\tau_0^z (F^\sigma_{\overline{\xi}_N})^* S_N\sigma^x_N S_N X_N\\
    &=-\tau^z_0(F^\sigma_{\overline{\xi}_N})^* X_N\\
    &=-\tau_0^z (U_N^\sigma)^*.
    \end{align*}
    
    Using Lemma \ref{lem:symmetry_on_e}, we then have
    $$
    \gamma_g(\epsilon)((U_N^\sigma)^*)=\tau_0^z(-\tau^z_0(U_N^\sigma)^*)\tau_0^z=\tau_0^z(U_N^\sigma)^*.
    $$
    Taking the appropriate limits and using continuity, we obtain
    \[
    c_{1^\sigma,\epsilon}=\gamma_g(\epsilon)((U^\sigma)^*) U^\sigma=\tau_0^z.
    \qedhere
    \]    
\end{proof}

\begin{lem}\label{lem:sigma_triv_braids}
    In $\GSec^{ETC}$, we have $c_{1^\sigma,1}=c_{1^\sigma,m}=c_{1^\sigma,1^\sigma}=1$.
\end{lem}
\begin{proof}
    Take $a\in\{1,m,1^\sigma\}$. Then $\gamma_g(a)=a$ and $a((U^\sigma_N)^*)=(U^\sigma_N)^*$. Therefore,
    \[
    c_{1^\sigma,a}=\gamma_g(a)((U^\sigma)^*)U^\sigma=1.
    \qedhere
    \]
\end{proof}

\begin{lem}\label{lem:sigma_psi}
    We have the braiding $c_{1^\sigma,\psi}=\tau_{0}^z$.
\end{lem}
\begin{proof}
    Following the proof of Lemma \ref{lem:symmetry_on_e}, one may show that $\gamma_g(\psi)(-)=\tau_0^z\psi(-)\tau_0^z$.
    We may then perform the following computation:
    \begin{align*}
        c_{1^\sigma,\psi}&=\gamma_g(\psi)((U^\sigma)^*)U^\sigma =\tau^z_{0}\epsilon(m((U^\sigma)^*))\tau^z_{0}U^\sigma =\tau^z_{0}\epsilon(c_{1^\sigma,m}(U^\sigma)^*)\tau^z_{0}U^\sigma\\
        &=\tau^z_{0}\epsilon((U^\sigma)^*)\tau^z_{0}U^\sigma =\gamma_g(\epsilon)((U^\sigma)^*)U^\sigma =c_{1^\sigma,\epsilon} =\tau^z_{0}.
    \end{align*}
    where the fourth equality follows from the fact that $c_{1^\sigma,m}=1$ from Lemma \ref{lem:sigma_triv_braids}, the fifth equality follows from Lemma \ref{lem:symmetry_on_e}, and the last equality follows from Proposition \ref{prop:sigma_epsilon}.
\end{proof}

\begin{lem}\label{lem:finish_fifth_row}
    For $b\in \{1,\epsilon,m,\psi\}$, $c_{1^\sigma,b^\sigma}=c_{1^\sigma,b}$.
\end{lem}
\begin{proof}
    Take $n=0$ if $b\in\{1,m\}$ and $n=1$ if $b\in\{\epsilon,\psi\}$. Using the fact that $c_{1^\sigma,1^\sigma}=1$ from Lemma \ref{lem:sigma_triv_braids} and generalizing Lemma \ref{lem:symmetry_on_e} to obtain $\gamma_g(\psi)(-)=\tau_0^z\psi(-)\tau_0^z$, we have that
    \begin{align*}
        c_{1^\sigma,b^\sigma}&=\gamma_g(b^\sigma)((U^\sigma)^*)U^\sigma =(\tau^z_0)^n b(1^\sigma((U^\sigma)^*))(\tau^z_0)^n U^\sigma\\
        &=(\tau^z_0)^n b(c_{1^\sigma,1^\sigma}(U^\sigma)^*)(\tau^z_0)^n U^\sigma =\gamma_g(b)(c_{1^\sigma,1^\sigma}(U^\sigma)^*) U^\sigma =c_{1^\sigma,b}.
        \qedhere
    \end{align*}
\end{proof}

\begin{lem}\label{lem:triv_a_sigma}
    We have $c_{a,1^\sigma}=1$ for all $a\in\{1,\epsilon,m,\psi\}$.
\end{lem}
\begin{proof}
    Let $a\in\{1,\epsilon,m,\psi\}$. 
    Then direct computation gives $1^\sigma((U^a_N)^*)=(U^a_N)^*$. Therefore,
    \[
    c_{a,1^\sigma}=1^\sigma((U^a)^*)U^a=1.
    \qedhere
    \]
\end{proof}

\begin{lem}\label{lem:braid_absigma}
    For $a,b\in\{1,\epsilon,m,\psi\}$, $c_{a,b^\sigma}=c_{a,b}$. 
\end{lem}
\begin{proof}
    Using the fact that $c_{a,b}$ is a scalar multiple of the identity from Proposition \ref{prop:TCbraiding},
    \begin{align*}
        c_{a,b^\sigma}&=b^\sigma((U^a)^*)U^a =1^\sigma(b((U^a)^*))U^a =1^\sigma(c_{a,b}(U^a)^*)U^a\\
        &=c_{a,b}1^\sigma((U^a)^*)U^a =c_{a,b}c_{a,1^\sigma} =c_{a,b}
        \qedhere
    \end{align*}
    where we have used and the fact that $c_{a,1^\sigma}=1$ from Lemma \ref{lem:triv_a_sigma}.
\end{proof}

\begin{lem}\label{lem:bottomleft}
    For $a,b\in\{1,\epsilon,m,\psi\}$, $c_{a^\sigma,b}=c_{a,b}c_{1^\sigma,b}$
\end{lem}
\begin{proof}
    This is a special case of Facts \ref{facts:braiding} by the following:
    \[
    c_{a^\sigma,b}=c_{1^\sigma,b}1^\sigma(c_{a,b})=c_{a,b}c_{1^\sigma,b}.
    \qedhere
    \]
\end{proof}

\begin{lem}\label{lem:bottomright}
    For $a,b\in\{1,\epsilon,m,\psi\}$, $c_{a^\sigma,b^\sigma}=c_{a,b}c_{1^\sigma,b}$.
\end{lem}
\begin{proof}
By Facts \ref{facts:braiding},
    $$
    c_{a^\sigma,b^\sigma}=c_{1^\sigma,b^\sigma}1^\sigma(c_{a,b^\sigma})=c_{1^\sigma,b^\sigma}c_{a,b^\sigma},
    $$
    where the second equality uses the fact that $c_{a,b^\sigma}$ is a scalar multiple of the identity operator by Lemma \ref{lem:braid_absigma} and Proposition \ref{prop:TCbraiding}. We then use Lemma \ref{lem:finish_fifth_row} to see that
    \[c_{a^\sigma,b^\sigma}=c_{1^\sigma,b}c_{a,b}.\qedhere\]
\end{proof}

\begin{thm}
    The $G$-crossed braiding of $\GSec^{ETC}$ is given in the following table:
\begin{center}
$$
\begin{array}{|c||c|c|c|c|c|c|c|c|}\hline
    c_{\pi_1,\pi_2} & \pi_2=1 &\epsilon & m &\psi & 1^\sigma & \epsilon^\sigma & m^\sigma & \psi^\sigma \\
    \hline\hline
  \pi_1=1   & 1&1 &1 &1 &1 &1 &1 &1 \\
  \hline
  \epsilon & 1&1 & 1 &1 &1 &1 &1 &1\\
  \hline
  m & 1&-1 & 1 &-1 &1 &-1 &1 &-1\\
  \hline
  \psi & 1&-1 & 1 &-1 &1 &-1 &1 &-1\\
  \hline
  1^\sigma   & 1& \tau^z_0 & 1&\tau^z_0&1 &\tau_0^z &1 &\tau^z_0 \\
  \hline
  \epsilon^\sigma & 1&\tau^z_0 &1 & \tau^z_0&1 &\tau^z_0 & 1&\tau^z_0 \\
  \hline
  m^\sigma & 1&-\tau^z_0 &1 &-\tau^z_0 &1 &-\tau^z_0 &1 &-\tau^z_0\\
  \hline
  \psi^\sigma & 1 &-\tau^z_0  &1 &-\tau^z_0 &1 &-\tau^z_0 &1 &-\tau^z_0 \\\hline
\end{array}
$$
\end{center}
\end{thm}
\begin{proof}
    The top left quadrant of this table is given in Proposition \ref{prop:TCbraiding}. The top right quadrant is then obtained from the top left quadrant and Lemma \ref{lem:braid_absigma}. The fifth row follows from Proposition \ref{prop:sigma_epsilon}, Lemma \ref{lem:sigma_triv_braids}, Lemma \ref{lem:sigma_psi}, and Lemma \ref{lem:finish_fifth_row}. The remainder of the bottom half is given by the top half, the fifth row, and Lemmas \ref{lem:bottomleft} and \ref{lem:bottomright}.
\end{proof}

\section{Discussion}

In this manuscript, we rigorously proved the expectation from \cite{PhysRevB.100.115147} that the symmetry defects of a 2+1D SET form a $G$-crossed braided tensor category. To do this, we defined symmetry defects in accord with the DHR paradigm. We demonstrated the utility of this definition by computing the defect category associated with SPTs and a lattice model of the $\mathbb{Z}_2$ symmetric toric code.

One potential direction for future work is to understand the role of antiunitary symmetries such as time reversal symmetry. A discussion of such SETs can be found in \cite{Barkeshli2020}, \cite{PhysRevLett.119.136801}, and \cite{PhysRevB.98.115129}. However, a detailed microscopic understanding of such bulk defects is missing, especially in the context of DHR theory.

We expect that there are many other lattice models which are amenable to our analysis. In particular, the models of SETs presented in \cite{PhysRevB.108.115144} give an extremely general class of models which are obtained by sequentially gauging abelian quotient groups of a global symmetry.
In addition to providing a large class of models to study, this research also suggests that it may be fruitful to understand the superselection theory in terms of gauging.

Finally, \cite{PhysRevB.94.235136} presents a model of the $\mathbb{Z}_2$-symmetric toric code where the symmetry swaps the anyons $\epsilon$ and $m$. 
In that example, the $\mathbb{Z}_2$-symmetry defects have non-integer quantum dimension, which provides an interesting challenge in terms of a DHR-style analysis.
This manuscript also presents a wide variety of other exactly solvable SETs which are related to string-nets by gauging the global symmetry.

\section*{Acknowledgments}

We would like to thank 
Juan Felipe Ariza Mej\'ia,  
Sven Bachmann, 
Tristen Brisky,
Chian Yeong Chuah, 
Martin Fraas,
Brett Hungar, 
Corey Jones, 
Michael Levin, 
Pieter Naaijkens, 
Bruno Nachtergaele, 
David Penneys, 
Sean Sanford, 
Wilbur Shirley, 
Daniel Spiegel, 
Dominic Williamson
for helpful conversations.  
SV was funded by the NSF grant number DMS-2108390.
DW was funded by the NSF grant number DMS-2154389.
KK was funded by NSF DMS 1654159 as well as the Center for Emergent Materials, an NSF-funded MRSEC, under Grant No. DMR-2011876.


\begin{subappendices}

\addtocontents{toc}{\protect\setcounter{tocdepth}{1}}
\section{Introduction to operators algebras and category theory}
\label{sec:background}
\subsection{General setting, cones, and the boundary at infinity}
\label{sec:cones}
We let $\Gamma$ be a 2d cell complex consisting of vertices, edges, faces and equip the vertices in $\Gamma$ with the graph distance. For the examples we have in mind, we often consider $\Gamma$ to be a regular lattice like the triangular lattice or the square lattice. An example is shown in Figure \ref{fig:example_triangular_lattice}.

Given a subset $\Sigma \subset \Gamma$, we denote by $\Sigma^c \subset \Gamma$ the complement of $\Sigma$, given by $\Sigma \cap \Sigma^c = \emptyset$ and $\Sigma \cup \Sigma^c = \Gamma$.

We now describe the `boundary circle at infinity' for $\bbR^2$ \cite[Sec~ A.2]{2410.21454}.  
More precisely, for some subsets $S \subseteq \bbR^2$, there is a corresponding subset $\partial_\infty(S) \subseteq S^1$ defined as 
\[
\partial_\infty(S)
\coloneqq
\lim_{r \to \infty} r^{-1} \cdot (C_r \cap S),
\]
where $C_r$ is the circle of radius $r$ centered at the origin.  

We consider two types of subsets $S \subseteq \bbR^2$ for which $\partial_\infty(S)$ is defined, in particular semi-infinite paths and cones.  
For many semi-infinite paths $p$, $\partial_\infty(p)$ consists of just a single point.  
We let $P(\bbR^2)$ denote the collection of such paths.
Often, a path in $\bbR^2$ determines a path on the dual lattice to $\Gamma$.  
More specifically, the dual lattice to $\Gamma$ is the lattice $\bar \Gamma$ whose vertices correspond to faces in $\Gamma$ and whose edges intersect those of $\Gamma$ transversely.
If a path $p \subset \bbR^2$ does not intersect any vertex in $\Gamma$, then it determines a path on the dual lattice, namely the path of dual edges corresponding to edges intersected by $p$.  
We denote by $\bar{P}(\Gamma)$ the collection of paths $\gamma \subset \bar\Gamma$ that correspond to a path $p \subset \bbR^2$ where $p \in P(\bbR^2)$.
In this paper, we fix a path $R \in P(\bbR^2)$ that corresponds to a dual path ${\bar \gamma_R} \in \bar{P}(\Gamma)$.
For simplicity, we will usually assume that $R$ is a ray, although one can consider more general paths $R$.  

We now describe the primary type of region we consider, namely cones.  
Specifically, a \emph{cone} $\Delta \subseteq \bbR^2$ is a subset of the form 
\[
\Delta
\coloneqq
\{x \in \bbR^2 : (x - a) \cdot \hat{v}/2 > \|x - a\| \cos(\theta/2)\}.
\]
Here $a \in \bbR^2$ is the vertex of the cone, $\hat{v} \in \bbR^2$ is a unit vector specifying the axis of the cone, and $\theta \in (0, 2\pi)$ is the opening angle of the cone.  
Note that if $\Delta \subseteq \bbR^2$ is a cone, $\partial_\infty(\Delta)$ is the interval in $S^1$ with midpoint $\hat{v}$ and length $\theta$.
We therefore term $\partial_\infty(\Delta)$ the \emph{boundary interval at infinity} for $\Delta$ \cite[Def.~A.5]{2410.21454}.

Finally, we define a \emph{cone in $\Gamma$} to be a subset $\Lambda \subseteq \Gamma$ of the form $\Lambda = \Gamma \cap \Delta$, where $\Delta \subseteq \bbR^2$ is a cone.  
Note that there are often many choices of $\Delta$ such that $\Lambda = \Gamma \cap \Delta$; however, all choices have the same boundary interval at infinity. 

\subsection{Operator algebras}
\label{sec:operator algebra basics}
In this section we provide a brief introduction to the operator algebraic approach to quantum spin systems on infinite lattices.  
For more detail, we refer the reader to \cite{MR3617688, MR887100, MR1441540}. 
In this section and the following ones that are model-independent, we use the word `site.'
In the examples we consider in this paper the sites will be the vertices of $\Gamma$, but the term `site' allows us to cover more general models (for instance those described in \cite{PhysRevB.94.235136}).  
Associate a Hilbert space $\hilb_s = \bbC^{d_s}$ to each site $s \in \Gamma$, where $d_s \in \bbN$.
Let $\Gamma_f$ be the set of finite subsets of $\Gamma$. 
We can then define the tensor product over a finite set of sites $S \in \Gamma_f$ as $\hilb_{S} \coloneqq \bigotimes_{s \in S} \hilb_s$. Then $\cstar[S] \coloneqq B(\hilb_S)$ is a $C^*$ algebra.

Now let $S,S' \in \Gamma_f$ be such that $S \subset S'$. Then we can define the canonical inclusion $\cstar[S] \hookrightarrow \cstar[S']$ by tensoring with the identity element on all $s \in S' \setminus S$. With this we can define the algebra of local observables $\cstar[\loc]$ as $$\cstar[\loc] \coloneqq \bigcup_{S \in \Gamma_f} \cstar[S]$$ and its norm completion, $$\cstar \coloneqq \overline{\cstar[\loc]}^{||\cdot ||}$$ This algebra is known as the algebra of quasi-local observables, or simply, the quasi-local algebra.

This algebra, as the name suggests, is the algebra whose elements can be approximated by strictly local observables, i.e, observables that act differently than the identity only on a finite subset $S \in \Gamma_f$. We say the \emph{support} of an observable $A \in \cstar$ is the smallest set $\Sigma \subset \Gamma$ such that $A \in \cstar[\Sigma]$, and we denote the support of $A$ by $\supp(A)$.

We note that we can define a quasi-local algebra $\cstar[\Sigma]$ on any (not necessarily finite) subset $\Sigma \subset \Gamma$ by first replacing $\Gamma$ with $\Sigma$ and then using the above procedure. We will use this fact primarily when talking about the quasi-local algebra $\cstar[\Lambda]$ on a cone $\Lambda$.

\subsubsection{States and representations} Let $\omega$ be a state on $\cstar$, meaning a positive linear functional of norm $1$. We denote by $\cS(\cstar)$ the space of all states on $\cstar$. 

Using a construction by Gelfand, Naimark, Segal (the GNS construction for short) one can associate to $(\omega, \cstar)$ a GNS triple $(\pi, \hilb, \ket{\Omega})$ where $\hilb$ is a Hilbert space, $\pi\colon \cstar \rightarrow B(\hilb)$ is a $^*$-representation onto $\hilb$ and $\ket{\Omega} \in \hilb$ is a cyclic vector, such that for all $A \in \cstar$ we have $\omega(A) = \inner{\Omega}{\pi(A) \Omega}$. The GNS triple for any state $\omega$ is unique up to unitary equivalence. 

We say that a state is \emph{pure} if for every $\phi \colon \fA \to \bbC$ satisfying that $0 \leq \phi \leq \omega$, $\phi = \phi(\mathds{1})\omega$. If $\omega$ is a pure state, then its GNS representation $\pi$ is irreducible.

If two representations $(\pi_1, \hilb_1)$ and $(\pi_2, \hilb_2)$ are unitarily equivalent, then we denote $\pi_1 \simeq \pi_2$. Two states $\omega_1, \omega_2$ of $\cstar$ are equivalent (denoted again by $\omega_1 \simeq \omega_2$) if their GNS representations are equivalent.

\subsubsection{Dynamics}
\label{sec:HamiltonianDynamics}
One can define a self-adjoint Hamiltonian $H_S \in \cstar[S]$ for any $S \in \Gamma_f$. 
In the infinite volume limit, $H_S$ is not convergent in norm, but remains meaningful as a generator of dynamics. For any observable $A \in \cstar[\loc]$, the limit $\delta(A) \coloneqq \lim_{S \rightarrow \Gamma} i [H_S, A]$ exists and extends to a densely defined unbounded $^*$-derivation on $\cstar$. A state $\omega_0$ is called a \emph{ground state} if for all $A \in \cstar[\loc]$ we have $$-i \omega_0(A^* \delta(A)) \geq 0,$$ and it is \emph{gapped} if there is some $g > 0$ such that for all $A \in \cstar[\loc]$ satisfying $\omega_0(A)=0$, we have $$-i \omega_0(A^* \delta(A)) \geq g \omega_0 (A^*A).$$

In our examples, our Hamiltonian $H_S$ will be of the form $H_S = \sum_{Z \subseteq S} \Phi(Z)$.  
Here $\Phi \colon \Gamma_f \to \fA_{\loc}$ is a map that satisfies the following conditions:
\begin{itemize}
\item 
$\Phi(Z) \in \cstar[Z]$ for $Z \in \Gamma_f$, and 
\item 
$\Phi(Z) \geq 0$ for all $Z \in \Gamma_f$.
\end{itemize}

We call the $\Phi(Z)$ \emph{interactions}.
We call the interactions \emph{finite range} if there exists $n > 0$ such that $\Phi(Z) = 0$ if $Z$ is not contained in a ball of radius $n$.
Note that in this case, we have that for $A \in \fA_{\loc}$ with $\supp(A) = S$, 
\[
\delta(A)
=
i\left[\sum_{Z \cap S \neq \emptyset} \Phi(Z), A\right],
\]
and the sum is finite since the interactions are finite range.
Similarly, we say that interactions are \emph{uniformly bounded} if there is some $N > 0$ such that for every $Z \in \Gamma_f$, $\|\Phi(Z)\| \leq N$.
In our examples, the interactions will be uniformly bounded and finite range.

A ground state $\omega_0$ is called \emph{frustration free} if for all $S \in \Gamma_f$ we have $\omega_0(H_S) = 0$. 
Note that by \cite[Lem.~3.8]{MR3764565}, a state $\omega_0 \colon \fA \to \bbC$ is a frustration free ground state if and only if $\omega_0(H_S) = 0$ for all $S \in \Gamma_f$. In our examples, we will have that there is a unique frustration free ground state $\omega_0$ for the derivation under consideration. 
By a standard argument (see for instance \cite[Cor.~2.24]{2307.12552}), $\omega_0$ must be a pure state.  
Indeed, suppose $\phi \colon \fA \to \bbC$ satisfies that $0 \leq \phi \leq \omega_0$.
Then for all $S \in \Gamma_f$ we have that
\[
0 
\leq
\phi(H_S)
\leq
\omega_0(H_S)
=
0,
\]
so $\phi(H_S) = 0$ for all $S \in \Gamma_f$.
Therefore, the map $\omega \colon \fA \to \bbC$ given by $\omega(A) = \frac{1}{\phi(\mathds{1})}\phi(A)$ for $A \in \fA$ is a state satisfying that $\omega(H_S) = 0$ for all $S \in \Gamma_f$, so $\omega$ is a frustration free ground state.  
Thus, $\omega = \omega_0$ and hence $\phi = \phi(\mathds{1})\omega_0$.

We let $(\pi_0, \hilb_0)$ be the GNS representation of $\omega_0$.

\subsubsection{von Neumann algebras} Let $(\pi_0, \hilb_0)$ be the GNS representation of the state $\omega_0 \colon \cstar \to \bbC$. For each set $S \subseteq \Gamma$ we can denote $\caR(S) \coloneqq \pi_0 (\cstar[S])'' \subseteq B(\hilb_0)$ where ($'$) denotes the commutant in $B(\hilb_0)$.
Equivalently, $\cR(S)$ is the closure of $\pi_0(\cstar[S])$ in the WOT-topology.  
In more detail, if $(A_i)$ is a net in $B(\cH_0)$, then $A_i \to A$ if for all $\overline{\xi}, \eta \in \cH_0$, $\langle \eta, A_i \overline{\xi} \rangle \to \langle \eta, A\overline{\xi}\rangle$.
In the case that the state $\omega_0$ is pure, the algebras $\cR(S)$ are factors, meaning that they have trivial center.  

There is a useful notion of two projections in a von Neumann algebra $M$ being equivalent.
If $p, q \in M$ are two projections, we say that $p, q$ are \emph{Murray von-Neumann equivalent}, denoted $p \sim q$, if there exists $v \in M$ such that $v^*v = P$ and $vv^*=Q$.
A von Neumann algebra $M$ is said to be \emph{infinite} if there exists $p \in M$ such that $p \neq \mathds{1}$ but $p \sim \mathds{1}$ in $M$.
There is a more specific notion of a von Neumann algebra $M$ being properly infinite; however, in the case that $M$ is a factor, this is equivalent to being infinite.
We will consider regions $\Lambda \subseteq \Gamma$ (specifically cones) such that the algebras $\cR(\Lambda)$ are infinite factors.

\subsubsection{Symmetry}
\label{sec:SymmetryGeneralSetup}
We assume that there is a symmetry action of a group $G$ onto $\cstar$, i.e, a faithful homomorphism $\beta \colon G\to \Aut(\cstar)$ given by $g \mapsto \beta_g$ for all $g \in G$.
We call $\beta_g$ a \emph{symmetry automorphism}. In the cases we consider, the symmetry action is \emph{on-site}, i.e, for each $s \in \Gamma$, we assume that there is an action of $G$ on each $\hilb_s$ by unitaries $U^g_s$ acting on the site $s$. 
In that case, $\beta_g$ is given by the formula in Definition \ref{def:GlobalSymmetryAutomorphism}.

Let $\alpha\colon \cstar \rightarrow \cstar$ be an automorphism. We say that $\alpha$ \emph{respects the symmetry} if we have $\alpha \circ \beta_g = \beta_g \circ \alpha$ for all $g \in G$.

\subsubsection{Anyon sectors} 
The following definition was first used by Doplicher-Haag-Roberts in axiomatic quantum field theory \cite{MR297259, MR334742}.
It was later adapted to the setting of lattice systems by Pieter Naaijkens in \cite{MR2804555}, using the framework developed by \cite{MR660538}. 
For this definition, we require $\pi_0$ to be an irreducible representation (equivalently, $\omega_0$ to be a pure state).
\begin{defn}
    \label{def:AnyonSector}
    An irreducible representation $(\pi, \cH)$ is said to satisfy the \emph{superselection criterion with respect to $(\pi_0,\hilb_0)$} if for any chosen cone $\Lambda$ we have the existence of a unitary $U\colon \cH \rightarrow \hilb_0$ such that for any chosen cone $\Lambda$ and $A \in \cstar[\Lambda^c]$ we have $$U \pi(A) U^* = \pi_0(A)$$
    We call such a representation $\pi$ an \emph{anyon sector}.
\end{defn}

\subsubsection{Automorphisms of the quasi-local algebra}
\label{sec:AutomorphismsOfQLAlgebra}

In this subsection, we discuss various types of automorphisms that preserve the structure of the quasi-local algebra. 
Often, we wish to consider automorphisms that preserve locality up to some spread; these are termed \emph{quantum cellular automata} \cite{quant-phys/0405174}. 

\begin{defn}
\label{def:QCA definition}
An automorphism $\alpha \colon \fA \to \fA$ is a \emph{quantum cellular automaton} (\emph{QCA} for short) if there exists $s > 0$ such that $\alpha(\cstar[S]) \subseteq \cstar[S^{+s}]$ and $\alpha^{-1}(\cstar[S]) \subseteq \cstar[S^{+s}]$, where $S^{+s}$ is the set of sites in $\Gamma$ that are distance at most $s$ from $S$.
We say that $s$ is the \emph{spread} of the QCA $\alpha$.
\end{defn}

\begin{lem}
\label{lem:QCAs preserve the ground state subspace}
    For any $S \in \Gamma_f$, let $H_{1,S} \coloneqq \sum_{Z \subseteq S} \Phi_1(Z)$ be a Hamiltonian with finite range interactions and $\delta_1$ the corresponding derivations. 
    
    Let $\alpha \colon \fA \to \fA$ be a QCA with spread $s$, and for $Z \in \Gamma_f$, define $\Phi_2(Z) \coloneqq \alpha(\Phi_1(Z))$.
    Let $H_{2,S} \coloneqq \sum_{Z \subseteq S} \Phi_2(Z)$ be the corresponding Hamiltonian and $\delta_2$ the corresponding derivation. 
    If $\omega_2$ is a ground state of derivation $\delta_2$, then $\omega_1 \coloneqq \omega_2 \circ \alpha$ is a ground state of $\delta_1$.
\end{lem}
\begin{proof}
    Since $\omega_2$ is a ground-state of $\delta_2$, we have for all $A \in \cstar[\loc]$ that $$-i \omega_2(A^* \delta_2(A)) \geq 0.$$
    Now let $A \in \cstar[\loc]$ with $\supp(A) = S$. Then we have,
    \begin{align*}
        -i \omega_1(A^* \delta_1(A)) &= -i \omega_1\left(A^*i\left[\sum_{Z \cap S \neq \emptyset} \Phi_1(Z), A\right]\right) = \omega_2 \circ \alpha \left(A^*\left[\sum_{Z \cap S \neq \emptyset} \Phi_1(Z), A\right]\right)\\
        &= \omega_2\left(\alpha(A)^*\left[\sum_{Z \cap S \neq \emptyset} \alpha(\Phi_1(Z)), \alpha(A)\right]\right) \\
        &= \omega_2\left(\alpha(A)^*\left[\sum_{Z \cap S \neq \emptyset} \Phi_2(Z), \alpha(A)\right]\right).
    \end{align*}
    Now, by how $\Phi_2(Z)$ is defined, 
    \[
    \delta_2(\alpha(A))
    =
    i \left[\sum_{Z \cap S \neq \emptyset} \Phi_2(Z), \alpha(A)\right]
    \]
    Therefore, we have that 
    \[
    -i \omega_1(A^* \delta_1(A))
    =
    -i\omega_2(\alpha(A)^*\delta_2(\alpha(A))
    \geq
    0.
    \]
    Thus $\omega_1$ is indeed a ground state of $\delta_1$.
\end{proof}

It is easy to see that the set of QCAs form a group.
A special type of QCA is the \emph{finite depth quantum circuit} (FDQC for short); see Definition \ref{def:FDQC}.

\begin{lem}
    \label{lem:FDQCBoundedSpread}
    Let $\alpha \colon \fA \to \fA$ be the FDQC built from $\{\cU^d\}_{d = 1}^D$, where each unitary in $\cU^d$ has support contained in a ball of diameter $N$. Then $\alpha$ is a QCA with spread $s = ND$.
\end{lem}

\begin{proof}
It suffices to show that for all $d = 1, \dots, D$, $\alpha_d$ is a QCA with spread $N$.  
Suppose that $S \subseteq \Gamma$ and $A \in \cstar[S, \loc]$
Then we have that 
\[
\alpha_d(A)
=
\Ad \! \left(\prod_{U \in \cU^d} U \right)\!(A).
\]
Since the support of each $U \in \cU^d$ is contained in a ball of diameter $N$, $\supp(\Ad \! \left(\prod_{U \in \cU^d} U \right)\!(A)) \subseteq \supp(A)^{+N}$.  
Thus $A \in \cstar[S^{+N}]$, so $\alpha_d$ is a QCA of at most $N$.
\end{proof}

Another useful notion is the notion of a \emph{quasi-factorizable} automorphism; these have been studied in \cite{MR4426734,MR4362722} as maps that preserve the anyon data when precomposed with the ground state.

\begin{defn}
    Let $\alpha$ be an automorphism of $\cstar$ and consider an inclusion of cones
    \[
    \Gamma_1' \subset \Lambda \subset \Gamma_2'
    \]
    We say that $\alpha$ is \emph{quasi-factorizable} with respect to this inclusion if there is a unitary $u \in \cstar$ and automorphisms $\alpha_\Lambda$ and $\alpha_{\Lambda^c}$ of $\cstar[\Lambda]$ and $\cstar[\Lambda^c]$ respectively, such that
    \[
    \alpha = \operatorname{Ad}(u) \circ \widetilde{\Xi} \circ (\alpha_\Lambda \otimes \alpha_{\Lambda^c}),
    \]
    where $\widetilde{\Xi}$ is an automorphism on $\cstar[\Gamma_2' \setminus \Gamma_1']$.
\end{defn}

\begin{lem}
\label{lem:FDQCQuasi-Factorizable}
If $\alpha \colon \fA \to \fA$ is a finite depth quantum circuit, then for every cone $\Lambda$, $\alpha$ is quasi-factorizable with respect to some inclusion of cones $\Gamma'_1 \subset \Lambda \subset \Gamma_2'$.
\end{lem}

\begin{proof}
We first observe that for each $d = 1, \dots, D$, we may assume that $\bigcup_{U \in \cU^d} \supp(U) = \Gamma$.  
Indeed, if this is not the case, we can always include $\mathds{1}_s$ for every $s \notin \bigcup_{U \in \cU^d} \supp(U)$ to $\cU^d$.
We now let $\Lambda$ be a cone.  
We define ${\cU}^1_{\mathrm{in}} \coloneqq \{U \in \cU^1 : U \in \cstar[\Lambda]\}$ and ${\cU}^1_{\mathrm{out}} \coloneqq \{U \in \cU^1 : U \in \cstar[\Lambda^c]\}$.
We also define $\Lambda_0 \coloneqq \Lambda$ and $\Lambda_0' \coloneqq \Lambda^c$.
For $d = 1, \dots, D - 1$, we inductively define 
\begin{align*}
\Lambda_d 
&\coloneqq 
\bigcup_{U \in {\cU}^d_{\mathrm{in}}} \supp(U), 
&
\Lambda_d' 
&\coloneqq 
\bigcup_{U \in {\cU}^d_{\mathrm{out}}} \supp(U),
\\
{\cU}^{d + 1}_{\mathrm{in}}
&\coloneqq
\{U \in \cU^{d + 1} : U \in \cstar[\Lambda_d]\},
&
{\cU}^{d + 1}_{\mathrm{out}}
&\coloneqq
\{U \in \cU^{d + 1} : U \in \cstar[\Lambda_d']\}.
\end{align*}
Observe that for all $d = 1, \dots, D - 1$, $\Lambda_{d - 1} \subseteq \Lambda_d$ and $\Lambda_{d - 1}' \subseteq \Lambda_d'$. 
In particular, for all for all $d = 0, 1, \dots, D - 1$, we have that $\Lambda_d \subseteq \Lambda$ and $\Lambda_d' \subseteq \Lambda^c$.

For $d = 1, \dots, D$, we define $\alpha_d^{\mathrm{in}} \colon \cstar \to \cstar$ and $\alpha_d^{\mathrm{out}} \colon \cstar \to \cstar$ by
\[
\alpha_d^{\mathrm{in}}(A)
\coloneqq
\Ad \! \left(\prod_{U \in {\cU}^d_{\mathrm{in}}} U \right)\!(A),
\qquad\qquad
\alpha_d^{\mathrm{out}}(A)
\coloneqq
\Ad \! \left(\prod_{U \in {\cU}^d_{\mathrm{out}}} U \right)\!(A).
\]
for $A \in \cstar[\loc]$.
Note that since $\Lambda_d \subseteq \Lambda$ and $\Lambda_d' \subseteq \Lambda^c$ for all $d = 0, 1, \dots, D - 1$, we have that $\alpha_d^{\mathrm{in}}$ is an automorphism of $\cstar[\Lambda]$ and $\alpha_d^{\mathrm{out}}$ is an automorphism of $\cstar[\Lambda^c]$ for all $d = 1, \dots, D$.
We therefore have that 
\[
\alpha_\Lambda 
\coloneqq
\alpha_D^{\mathrm{in}} \circ \dots \circ \alpha_1^{\mathrm{in}}, 
\qquad\qquad
\alpha_{\Lambda^c} 
\coloneqq
\alpha_D^{\mathrm{out}} \circ \dots \circ \alpha_1^{\mathrm{out}}
\]
are automorphisms of $\cstar[\Lambda]$ and $\cstar[\Lambda^c]$ respectively.  

We now consider the automorphism 
\[
\alpha_\Lambda \otimes \alpha_{\Lambda^c}
=
(\alpha_D^{\mathrm{in}} \otimes \alpha_D^{\mathrm{out}}) \circ \dots \circ (\alpha_1^{\mathrm{in}} \otimes \alpha_1^{\mathrm{out}}).
\]
We observe that for all $d = 1, \dots, D$, 
\[
\alpha_d^{\mathrm{in}} \otimes \alpha_d^{\mathrm{out}}(A)
=
\Ad \! \left(\prod_{U \in {\cU}^d_{\mathrm{in}} \cup {\cU}^d_{\mathrm{out}}} U \right)\!(A),
\]
for $A \in \cstar[\loc]$.
For $d = 1, \dots, D$, we define $\widehat{\cU}^d \coloneqq \cU^d \setminus ({\cU}^d_{\mathrm{in}} \cup {\cU}^d_{\mathrm{out}})$, and we define $\Xi_d \colon \cstar \to \cstar$ by 
\[
\Xi_d(A)
\coloneqq
\Ad \! \left(\prod_{U \in \widehat{\cU}^d} U \right)\!(A)
\]
for $A \in \cstar[\loc]$.
We similarly define $\Xi \colon \cstar\to\cstar$ by 
\(
\Xi
\coloneqq
\Xi_D \circ \dots \circ \Xi_1.
\)
Note that $\alpha_d = \Xi_d \circ (\alpha_d^{\mathrm{in}} \otimes \alpha_d^{\mathrm{out}})$.
By how $\alpha_d^{\mathrm{in}}$ and $\alpha_d^{\mathrm{out}}$ were defined, we have that $\Xi_d$ commutes with $\Xi_{d'}$ for all $d' \geq d$.  
Therefore, we have that 
\[
\Xi \circ (\alpha_\Lambda \otimes \alpha_{\Lambda^c})
=
(\Xi_D \circ (\alpha_D^{\mathrm{in}} \otimes \alpha_D^{\mathrm{out}})) \circ \dots \circ (\Xi_1 \circ (\alpha_1^{\mathrm{in}} \otimes \alpha_1^{\mathrm{out}}))
=
\alpha_D \circ \dots \circ \alpha_1
=
\alpha.
\]

It remains to show that there exists an inclusion of cones 
\(
\Gamma'_1 \subset \Lambda \subset \Gamma'_2
\)
such that $\Xi$ is an automorphism on $\cstar[\Gamma_2' \setminus \Gamma_1']$.
At this point, we use the assumption that for each $d = 1, \dots, D$, $\bigcup_{U \in \cU^d} \supp(U) = \Gamma$.
We also use the fact that every $U \in \bigcup_{d = 1}^D \cU^d$ has support at most $N$.
By these two facts, the unitaries in $\widehat{\cU}^1 = \cU^1 \setminus (\cU^1_{\mathrm{in}} \cup \cU^1_{\mathrm{out}})$ are all supported in the strip $\Delta_1 = \Lambda^{+N} \cap (\Lambda^c)^{+N}$.  
Similarly, for each $d = 2, \dots, D$, we have that the unitaries in $\widehat{\cU}^d$ are supported in the strip $\Delta_d \coloneqq \Delta_{d - 1}^{+N} = \Lambda^{+dN} \cap (\Lambda^c)^{+dN}$.
Therefore, we have that all unitaries in $\bigcup_{d = 1}^D \widehat{\cU}^d$ are supported in the strip $\Delta_D = \Lambda^{+DN} \cap (\Lambda^c)^{+DN}$.
In particular, $\Xi$ is an automorphism on $\cstar[\Delta_D]$.
Now, if we let $\Gamma'_1 \coloneqq \left((\Lambda^c)^{+DN}\right)^c$ and $\Gamma'_2 \coloneqq \Lambda^{+DN}$, then $\Gamma'_1 \subset \Lambda \subset \Gamma'_2$ and $\Delta_D = \Gamma_2' \setminus \Gamma_1'$.
The result follows.
\end{proof}

We recall the notion of bounded spread Haag duality (Definition \ref{def:BSHaagDuality}).
We then have the following result, which is a special case of \cite[Prop.~5.10]{2410.21454}. 

\begin{lem}
\label{lem:QCABSHaagDuality}
If $\pi \colon \cstar \to B(\cH)$ satisfies strict Haag duality and $\alpha \colon \cstar \to \cstar$ is a QCA with spread $s$, then $\pi \circ \alpha$ satisfies bounded spread Haag duality with spread $2s$.
\end{lem}

\begin{proof}
We consider the nets of von Neumann algebras given by $\pi(\cstar[\Lambda])''$ and $\pi \circ \alpha(\cstar[\Lambda])''$, where $\Lambda$ ranges over all cones.
Note that for every cone $\Lambda$, 
\begin{gather*}
\pi \circ \alpha(\cstar[\Lambda])''
\subseteq
\pi(\cstar[\Lambda^{+s}])'',
\\
\pi(\cstar[\Lambda])''
\subseteq
\pi\circ \alpha(\alpha^{-1}(\cstar[\Lambda]))''
\subseteq
\pi \circ \alpha(\cstar[\Lambda^{+s}])''.
\end{gather*}
By \cite[Prop.~5.10]{2410.21454}, $\pi \circ \alpha$ satisfies bounded spread Haag duality with spread $2s$.
\end{proof}

\subsection{Category theory}
In this section we define the primary category theoretic definitions that we will use in this paper. 
For more details, the reader can consult \cite{MR3242743} for the algebraic setting and \cite{MR808930, MR3687214} for the $\rmC^*$-/$\rmW^*$-setting.
In our examples, we will be working with a category $\cC$ that is a linear dagger category.  
A \emph{linear category} is a category $\cC$ such that for all $a, b \in \cC$, $\Hom{a \to b}$ is a vector space and composition is bilinear. 
A linear category $\cC$ is a \emph{dagger category} if for all $a, b \in \cC$, there is an anti-linear map $(-)^* \colon \Hom{a \to b} \to \Hom{b \to a}$ such that for all $f \colon a \to b$ and $g \colon b \to c$ in $\cC$, $(g \circ f)^* = f^* \circ g^*$.
Additionally, we have that our linear dagger categories are \emph{orthogonal Cauchy complete}, meaning that they admit all orthogonal direct sums and subobjects.  
Given, $a_1, \dots, a_n \in \cC$, the \emph{orthogonal direct sum} of $a_1, \dots, a_n$ is an object $\bigoplus_{i = 1}^n a_i$ along with morphisms $v_j \colon a_j \to \bigoplus_{i = 1}^n a_i$ for all $j \in \{1, \dots, n\}$ that satisfy the following properties: 
\begin{itemize}
\item 
$v_i^*v_i = \id_{a_i}$ for all $i \in \{1, \dots, n\}$, and
\item 
$\sum_{i = 1}^n v_i v_i^* = \id_{\bigoplus_{i = 1}^n a_i}$.
\end{itemize}
Note that the orthogonal direct sum $\bigoplus_{i = 1}^n a_i$ is unique up to unique isomorphism.  
We will also often drop the word `orthogonal' for simplicity.
Similarly, we say that our category \emph{admits all subobjects} if for every projection $p \colon a \to a$ in $\cC$ (that is, a morphism satisfying that $p^* = p = p^2$), there exists an object $b \in \cC$ (called a \emph{subobject}) and a map $v \colon b \to a$ such that $v^*v = \id_b$ and $vv^* = p$.
(The property of admitting subobjects is also called \emph{projection complete}, although we do not use this term in this paper.)
As with direct sums, given a projection $p \colon a \to a$ in $\cC$, any two subobjects corresponding to $p$ are isomorphic.

The categories we consider will also be strict monoidal categories. 
A category $\cC$ is a \emph{strict monoidal category} if there is a functor $- \otimes - \colon \cC \times \cC \to \cC$ such that $(a \otimes b) \otimes c = a \otimes (b \otimes c)$ for all $a, b, c \in \cC$, and such that there is an object $\mathds{1} \in \cC$ such that $\mathds{1} \otimes a = a = a \otimes \mathds{1}$.
(There is a more general notion of monoidal category that is not strict; however, our examples will be strict monoidal categories.)
A map between two (strict) monoidal categories is a monoidal functor.  
More specificaly, if $\cC$ and $\cD$ are (strict) monoidal categories, then we say that a functor $F \colon \cC \to \cD$ is \emph{monoidal} if there are natural tensorator isomorphisms $F^2_{a, b} \colon F(a) \otimes F(b) \to F(a \otimes b)$ and a unitor isomorphism $F^1 \colon F(\mathds{1}_\cC) \to \mathds{1}_\cD$ satisfying coherence conditions. 
We will usually consider monoidal functors that are \emph{strict}, meaning that for all $a, b \in \cC$, $F(a) \otimes F(b) = F(a \otimes b)$ and $F^2_{a, b} = \id_{F(a) \otimes F(b)}$, and additionally that $F^1 = \id_{\mathds{1}_\cD}$.

We now define strict $G$-crossed monoidal and $G$-crossed braided as done in \cite{MR2183964}.
If $\cA, \cB$ are subcategories of a category $\cC$, we say that $\cA$ and $\cB$ are \emph{disjoint} if $\Hom{a \to b} = \{0\}$ for every $a \in \cA$ and $b \in \cB$.
Now, let $G$ be a finite group.  
We say that a category $\cC$ is \emph{$G$-graded} if $\cC = \bigoplus_{g \in G} \cC_g$, where $\{\cC_g\}_{g \in G}$ is a collection of mutually disjoint subcategories of $\cC$.
In other words, we require that every object $a \in \cC$ is of the form $a = \bigoplus_{g \in G} a_g$, where $a_g \in \cC_g$.
We let $\cC_{\hom}$ denote the full subcategory of $\cC$ whose objects are those in $\bigcup_{g \in G} \cC_g$, and we say that $a \in \cC$ is \emph{homogeneous} if $a \in \cC_{\hom}$.
Following \cite{MR2183964}, we let $\partial \colon \cC_{\hom} \to G$ be the map defined by $\partial a \coloneqq g$ if $a \in \cC_g$.

\begin{defn}
\label{def:G-graded monoidal}
A category $\cC$ is a \emph{$G$-graded strict monoidal category} if $\cC$ is a $G$-graded category that is strict monoidal such that the grading $\partial\colon \cC_{\hom}\rightarrow G$ obeys $\partial(a \otimes b) = \partial a \partial b$ for all $a, b \in \cC_{\hom}$.
\end{defn}

\begin{defn}[{\cite[Def.~2.9]{MR2183964}}]
\label{def:G-crossed monoidal}
A \emph{strict $G$-crossed monoidal category} is a $G$-graded strict monoidal category $\cC$ along with strict monoidal isomorphisms $\gamma_g \colon \cC \to \cC$ such the following hold: 
\begin{itemize}
\item $g \mapsto \gamma_g$ is a group homomorphism, 
\item 
$\gamma_g(\cC_h) \subseteq \cC_{ghg^{-1}}$.
\end{itemize}
\end{defn}

\begin{defn}[{\cite[Def.~2.16]{MR2183964}}]
\label{def:G-crossed braided}
A \emph{braiding} on a strict $G$-crossed monoidal category is a collection of isomorphisms $c_{a, b} \colon a \otimes b \to \gamma_{\partial a}(b) \otimes a$ for $a \in \cC_{\hom}$ and $b \in \cC$ that satisfy the following coherence conditions: 
\begin{itemize}
\item 
(naturality)
for all $f_1 \colon a \to b$ in $\cC_{\hom}$ and $f_2 \colon c \to d$ in $\cC$, 
\[
(\gamma_{\partial a}(f_2) \otimes \id_a) \circ c_{a, c}
=
c_{a, d} \circ (\id_a \otimes f_2),
\qquad\qquad
(\id_{\gamma_{\partial a}(c)} \otimes f_1) \circ c_{a, c}
=
c_{b, c} \circ (f_1 \otimes \id_c),
\]
\item 
(monoidality)
for all $a, b \in \cC_{\hom}$ and $c, d \in \cC$, 
\[
c_{a, c \otimes d}
=
(\id_{\gamma_{\partial a}(c)} \otimes c_{a, d}) 
\circ
(c_{a, c} \otimes \id_d), 
\qquad\qquad
c_{a \otimes b, c}
=
(c_{a, \gamma_{\partial b}(c)} \otimes \id_b)
\circ
(\id_a \otimes c_{b, c}),
\]
\item 
($\gamma_g$ preserves braiding) for all $a \in \cC_{\hom}$, $b \in \cC$, and $g \in G$, $\gamma_g(c_{a, b}) = c_{\gamma_g(a), \gamma_g(b)}$.
\end{itemize}
\end{defn}

We remark that the above definition of strict $G$-crossed braided monoidal category is a strictified version of the definition of $G$-crossed braided monoidal category \cite[Def.~8.24.1]{MR3242743}.
However, the $G$-crossed braided monoidal categories we construct will be strict in this way.
The symmetry fractionalization data described in \cite{PhysRevB.100.115147} can nonetheless be recovered using an approach similar to the one used in \cite{2306.13762} to compute $F$- and $R$-symbols for anyon sectors. 
We illustrate this computation in Section \ref{sec:SymmetryFractionalization}.

We also remark that when the category is a dagger category, all of the coherence isomorphisms described in this section should be unitaries.

\section{Useful results for the Levin-Gu SPT}

We recall the automorphism $\alpha \colon \cstar \to \cstar$ (see Definition \ref{def:LevinGuEntangler}).

\begin{lem}
\label{lem:alpha_theta is a bounded spread automorphism}
    The automorphism $\alpha$ is a finite depth quantum circuit. 
    In particular, $\alpha$ is a quasi-factorizable QCA with spread $s = 1$.
\end{lem}
\begin{proof}

We group the triangles in $\Gamma$ into elementary hexagons that tile the entire plane.  
Note that this tiling of elementary hexagons can be colored using three colors, which we take to be red, blue, and green. An example is shown in Figure \ref{fig:example tiling of lattice into Hexagons}. We let $\scrR, \scrB, \scrG$ denote the collections of red, blue, and green hexagons, respectively.  
We now define $\cU^1$, $\cU^2$, and $\cU^3$ to be the following collections of unitaries: 
\[
\cU^1
\coloneqq
\left\{\prod_{\triangle \subseteq H} U_{\triangle} : H \in \scrR \right\},
\quad
\cU^2
\coloneqq
\left\{\prod_{\triangle \subseteq H} U_{\triangle} : H \in \scrB \right\},
\quad
\cU^3
\coloneqq
\left\{\prod_{\triangle \subseteq H} U_{\triangle} : H \in \scrG \right\}.
\]
Note that for $d = 1, 2, 3$, if $U_1, U_2 \in \cU^d$ with $U_1 \neq U_2$, then $\supp(U_1) \cap \supp(U_2) = \emptyset$.  
Furthermore, each $U \in \bigcup_{d = 1}^3 \cU^d$ acts only on a collection of seven vertices in an elementary hexagon.
Therefore, the collection $\{\cU^1, \cU^2, \cU^3\}$ defines a depth 3 quantum circuit.  
If we define $\widehat{\alpha}_d \colon \cstar\to \cstar$ by 
\[
\widehat{\alpha}_d(A)
=
\Ad\!\left(\prod_{U \in \cU^d}U \right)\!(A)
\]
for $A \in \cstar[\loc]$, then we have that for $A \in \cstar[\loc]$
\[
\alpha
=
\Ad\!\left(\prod_{\triangle \subseteq \Gamma} U_{\triangle}\right)\!(A)
=
\widehat{\alpha}_3 \circ \widehat{\alpha}_2 \circ \widehat{\alpha}_1(A).
\]
Thus, $\alpha$ is a finite depth quantum circuit.

\begin{figure}[!ht]
    \centering
    \includegraphics[width=0.2\linewidth]{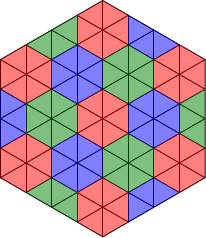}
    \caption{An example tiling of the triangular lattice into red, blue, green Hexagons. $\cU^1, \cU^2, \cU^3$ are supported on the red, blue, green colored triangles respectively. Each unitary $U$ on the hexagon $H$ centered at vertex $v$ is supported on $v^{+1}$.}
    \label{fig:example tiling of lattice into Hexagons}
\end{figure}

By Lemmas \ref{lem:FDQCBoundedSpread} and \ref{lem:FDQCQuasi-Factorizable}, $\alpha$ is a quasi-factorizable QCA.
To see that $\alpha$ has spread $1$, note that for $A \in \fA_{\loc}$, $\alpha(A)$ is the conjugation of $A$ by commuting unitaries that act on triangles.
Since every vertex on a face is distance 1 away from every other vertex, any vertex in the support $\alpha(A)$ is distance at most 1 from a vertex in $\supp(A)$.  
Thus $\supp(\alpha(A)) \subseteq \supp(A)^{+1}$.
\end{proof}

\begin{lem}
    \label{lem:automorphism connecting trivial para GS and Levin Gu GS}
    We have for all $v \in \Gamma$, $$\alpha^{-1}(\sigma^x_v) = B_v,$$
    In particular, this implies for all $V \in \Gamma_f$ that $$\alpha^{-1}(H_V^0) = H_V.$$ This is the result of \cite[Appendix A]{PhysRevB.86.115109} in the infinite volume setting.
\end{lem}
\begin{proof}
Let $\triangle_v$ be the set of triangles containing vertex $v \in \Gamma$. We denote $<vpq>$ to explicitly refer to a face $<vpq>\in \triangle_v$ with vertices $v,p,q \in \Gamma$.
    \begin{align*}
        \alpha^{-1}(\sigma^x_v) &= \prod_{\mathclap{\triangle \in \triangle_v}} e^{-\frac{i\pi}{24}(3 \prod_{v' \in \triangle} \sigma^z_{v'} - \sum_{v' \in \triangle} \sigma^z_{v'})} \sigma_v^x \prod_{\mathclap{\triangle \in \triangle_v}}e^{\frac{i\pi}{24}(3 \prod_{v \in \triangle} \sigma^z_v - \sum_{v \in \triangle} \sigma^z_v)}\\
        &= \sigma_v^x \prod_{\mathclap{<vpq> \in \triangle_v}} e^{\frac{i\pi}{8} \sigma^z_v \sigma^z_p \sigma^z_q + \frac{i\pi}{24}( -\sigma^z_v + \sigma^z_p + \sigma^z_q) + \frac{i\pi}{8} \sigma^z_v \sigma^z_p \sigma^z_q + \frac{i\pi}{24}( -\sigma^z_v - \sigma^z_p - \sigma^z_q)}\\
        &= \sigma_v^x \prod_{\mathclap{\triangle \in \triangle_v}} e^{\frac{i\pi}{4} \prod_{v' \in \triangle} \sigma^z_{v'} - \frac{i\pi}{12}\sigma^z_v} = \sigma_v^x  i^{(- \sigma^z_v)} \prod_{\mathclap{<vpq> \in \triangle_v}} i^{\frac{1}{2}(\sigma^z_v \sigma^z_p \sigma^z_q)}\\
        \intertext{In the previous equality we used the identity $e^{i\pi/2} = i$ and that there are 6 faces having $v$ as a vertex, so $\sum_{{\triangle \in \triangle_v}} \frac{i\pi}{12}\sigma^z_v = \frac{i\pi}{2}\sigma^z_v $. }
        &= \sigma_v^x  i^{3 \sigma^z_v} \prod_{\mathclap{<vpq> \in \triangle_v}} i^{\frac{1}{2}(\sigma^z_v \sigma^z_p \sigma^z_q)} = \sigma_v^x   \prod_{\mathclap{<vpq> \in \triangle_v}} i^{\frac{1}{2}(\sigma^z_v + \sigma^z_v \sigma^z_p \sigma^z_q)} = \sigma_v^x   \prod_{\mathclap{<vpq> \in \triangle_v}} i^{ \frac{1}{2}\sigma^z_v(1+ \sigma^z_p \sigma^z_q)} = \sigma^x_v M,
        \intertext{where $M =  \prod_{{<vpq> \in \triangle_v}} i^{ \frac{1}{2}\sigma^z_v(1+ \sigma^z_p \sigma^z_q)}$.
        We claim that $M = \prod_{{<vpq> \in \triangle_v}} i^{ \frac{1}{2}(1+ \sigma^z_p \sigma^z_q)}$.
        Indeed, since the eigenvalues of $\sigma^z_v, \sigma^z_p, \sigma^z_q$ are $\pm 1$, the eigenvalues of $i^{ \frac{1}{2}\sigma^z_v(1+ \sigma^z_p \sigma^z_q)}$ are exactly the same as the eigenvalues of $i^{ \frac{1}{2}(1+ \sigma^z_p \sigma^z_q)}$ (with exactly the same eigenvectors).
        Therefore, $i^{ \frac{1}{2}\sigma^z_v(1+ \sigma^z_p \sigma^z_q)} = i^{ \frac{1}{2}(1+ \sigma^z_p \sigma^z_q)}$, so $M = \prod_{{<vpq> \in \triangle_v}} i^{ \frac{1}{2}(1+ \sigma^z_p \sigma^z_q)}$. 
        Additionally, $M = M^{-1}$, since $i^{ \frac{1}{2}(1+ \sigma^z_p \sigma^z_q)}$ has eigenvalues $\pm 1$ and thus $M$ has eigenvalues $\pm 1$.
        This can be thought of as a gauge redundancy. 
        Using this fact we get}
        &= \sigma^x_v M^{-1} =  \sigma_v^x   \prod_{\mathclap{<vpq> \in \triangle_v}} i^{- \frac{1}{2}(1+ \sigma^z_p \sigma^z_q)} = \sigma^x_v i^{-3} \prod_{\mathclap{<vpq> \in \triangle_v}} i^{-\frac{1}{2}\sigma^z_p \sigma^z_q}\\
        &= -\sigma^x_v i^{3} \prod_{\mathclap{<vpq> \in \triangle_v}} i^{-\frac{1}{2}\sigma^z_p \sigma^z_r} = -\sigma^x_v \prod_{\mathclap{<vpq> \in \triangle_v}} i^{\frac{1}{2}(1-\sigma^z_p \sigma^z_r)} = B_v
    \end{align*}
    The statement of the lemma trivially follows from this result, since the Hamiltonians are a summation of these individual terms.

    The statement $\alpha^{-1}(H_V^0) = H_V$ now trivially follows for all $V \in \Gamma_f$.
\end{proof}

We recall the automorphism $\alpha^\gamma$ defined in Section \ref{sec:Defect auts Hamiltonian Levin-Gu}, as well as the representations $\tilde \pi$ and $\tilde \pi_\gamma$ defined in Section \ref{sec:Defect sectors in the Levin-Gu SPT}.

\begin{lem}
\label{lem:LGGSDefectStateInequivalent}
The representations $\tilde \pi$ and $\tilde \pi_\gamma$ are not equivalent.
(Equivalently, $\tilde \omega \not \simeq \tilde \omega_\gamma$.)
\end{lem}
\begin{proof}
Note that since $\tilde \omega$ and $\tilde \omega_{\gamma}$ are pure states, $\tilde \omega$ and $\tilde \omega_{\gamma}$ are equivalent if and only if they are quasi-equivalent \cite[Prop.~10.3.7]{MR1468230}.
We can therefore apply \cite[Cor.~2.6.11]{MR887100}. 
Let $V \in \Gamma_f$.  
Then since $\gamma$ is a half-infinite dual path, there exists $v \in \gamma - \partial_0 \gamma - \partial_1 \gamma$ such that every $\triangle \in \triangle_v$ satisfies that $\triangle \subseteq V^c$.  
Note that the last condition implies that $\supp(\widehat{B}_v) \subseteq V^c$.  
We let 
\[
\triangle_v^\gamma 
\coloneqq
\{<vqq'> \in \triangle_v : \text{$\gamma$ intersects the edge between $q$ and $q'$} \}.
\]
We therefore have that 
\begin{align*}
\widehat{B}_v
&=
-\sigma^x_v 
\prod_{<v q q'> \in \triangle_v^\gamma} i^{\frac{1 + \sigma^z_q \sigma^z_{q'}}{2}}
\prod_{<v q q'> \in \triangle_v \setminus \triangle_v^\gamma} i^{\frac{1 - \sigma^z_q \sigma^z_{q'}}{2}}
\\&=
-\sigma^x_v 
\prod_{<v q q'> \in \triangle_v^\gamma} i^{\sigma^z_q \sigma^z_{q'}} i^{\frac{1 - \sigma^z_q \sigma^z_{q'}}{2}}
\prod_{<v q q'> \in \triangle_v \setminus \triangle_v^\gamma} i^{\frac{1 - \sigma^z_q \sigma^z_{q'}}{2}}
\\&=
B_v \prod_{<v q q'> \in \triangle_v^\gamma}i^{\sigma^z_q \sigma^z_{q'}}.
\end{align*}
Therefore, since $\tilde \omega(B_v) = 1$ and $B_v \leq \mathds{1}$, we have by Lemma \ref{lem:can freely insert and remove P from the ground state.} that 
\[
\tilde\omega(\widehat{B}_v)
=
\tilde\omega(B_v\widehat{B}_v)
=
\tilde\omega\!\left(B_v^2 \prod_{<v q q'> \in \triangle_v^\gamma}i^{\sigma^z_q \sigma^z_{q'}}\right)
=
\tilde\omega\!\left(\prod_{<v q q'> \in \triangle_v^\gamma}i^{\sigma^z_q \sigma^z_{q'}}\right),
\]
where in the last step we used that $B_v^2 = 1$.  

Now, since $\gamma \in \bar{P}(\Gamma)$, we may assume that there exists $p \in \Gamma$ such that $p$ is contained in exactly one face $\triangle \coloneqq <vpp'> \in \triangle_v^\gamma$.  
(If such a $p \in \Gamma$ does not exist, then there is a different choice for $v$ for which such a $p$ does exist.)
Therefore, we have that 
\begin{align*}
\tilde\omega(\widehat{B}_v)
&=
\tilde\omega\!\left(\prod_{<v q q'> \in \triangle_v^\gamma}i^{\sigma^z_q \sigma^z_{q'}}\right)
=
\tilde\omega\!\left(B_p \prod_{<v q q'> \in \triangle_v^\gamma}i^{\sigma^z_q \sigma^z_{q'}}B_p\right)
\\&=
\tilde\omega\!\left(
\sigma_p^x 
\prod_{<p r r'> \in \triangle_p} i^{\frac{1 - \sigma^z_r \sigma^z_{r'}}{2}} i^{\sigma^z_p \sigma^z_{p'}}
\prod_{<v q q'> \in \triangle_v^\gamma \setminus \{\triangle\}}i^{\sigma^z_q \sigma^z_{q'}} 
\sigma_p^x 
\prod_{<p r r'> \in \triangle_p} i^{\frac{1 - \sigma^z_r \sigma^z_{r'}}{2}} 
\right)
\\&=
\tilde\omega\!\left(
i^{-\sigma^z_p \sigma^z_{p'}}
\prod_{<v q q'> \in \triangle_v^\gamma \setminus \{\triangle\}}i^{\sigma^z_q \sigma^z_{q'}} 
\sigma_p^x 
\prod_{<p r r'> \in \triangle_p} i^{\frac{1 - \sigma^z_r \sigma^z_{r'}}{2}}
\sigma_p^x 
\prod_{<p r r'> \in \triangle_p} i^{\frac{1 - \sigma^z_r \sigma^z_{r'}}{2}} 
\right)
\\&=
\tilde\omega\!\left(
i^{-\sigma^z_p \sigma^z_{p'}}
\prod_{<v q q'> \in \triangle_v^\gamma \setminus \{\triangle\}}i^{\sigma^z_q \sigma^z_{q'}} 
B_p^2
\right)
=
\tilde\omega\!\left(
i^{-\sigma^z_p \sigma^z_{p'}}
\prod_{<v q q'> \in \triangle_v^\gamma \setminus \{\triangle\}}i^{\sigma^z_q \sigma^z_{q'}} 
\right).
\end{align*}
Now, since the eigenvalues of $\sigma^z_p$ are $1$ and $-1$, we have that $i^{-\sigma^z_p \sigma^z_{p'}} = (i^{\sigma^z_p \sigma^z_{p'}})^{-1} = -i^{\sigma^z_p \sigma^z_{p'}}$.
Therefore, we have that 
\begin{align*}
\tilde\omega(\widehat{B}_v)
&=
\tilde\omega\!\left(
i^{-\sigma^z_p \sigma^z_{p'}}
\prod_{<v q q'> \in \triangle_v^\gamma \setminus \{\triangle\}}i^{\sigma^z_q \sigma^z_{q'}} 
\right)
=
\tilde\omega\!\left(
-i^{\sigma^z_p \sigma^z_{p'}}
\prod_{<v q q'> \in \triangle_v^\gamma \setminus \{\triangle\}}i^{\sigma^z_q \sigma^z_{q'}} 
\right)
\\&=
-\tilde\omega\!\left(\prod_{<v q q'> \in \triangle_v^\gamma}i^{\sigma^z_q \sigma^z_{q'}}\right)
=
-\tilde\omega(\widehat{B}_v).
\end{align*}
Thus, $\tilde\omega(\widehat{B}_v) = 0$.  
However, $\tilde\omega_\gamma(\widehat{B}_v) = 1$.  
Therefore, we have that 
\[
|\tilde\omega(\widehat{B}_v) - \tilde\omega_\gamma(\widehat{B}_v)|
=
1
=
\|\widehat{B}_v\|,
\]
so by \cite[Cor.~2.6.11]{MR887100}, $\tilde\omega \not \simeq \tilde \omega_\gamma$.
\end{proof}

\begin{lem}
    If $\gamma_1, \gamma_2 \in \bar P(\Gamma)$ such that $\gamma_1 \cap \gamma_2$ does not contain a half-infinite dual path, then we have that $\tilde \pi_{\gamma_1} \not\simeq \tilde \pi_{\gamma_2}$ (equivalently, $\tilde \omega_{\gamma_1} \not \simeq \tilde \omega_{\gamma_2}$).
\end{lem}
\begin{proof}
Note that since $\tilde \omega_{\gamma_1}$ and $\tilde \omega_{\gamma_2}$ are pure states, $\tilde \omega_{\gamma_1}$ and $\tilde \omega_{\gamma_2}$ are equivalent if and only if they are quasi-equivalent \cite[Prop.~10.3.7]{MR1468230}.
We can therefore apply \cite[Cor.~2.6.11]{MR887100}. 
Since $\gamma_1 \cap \gamma_2$ does not contain a half-infinite dual path, for every $V \in \Gamma_f$, there exists $v \in \Gamma$ such that $v \notin \gamma_1$, $v \in \gamma_2 - \partial_0 \gamma_2 - \partial_1 \gamma_2$, and every $\triangle \in \triangle_v$ satisfies that $\triangle \subseteq V^c$.  
Note that the last condition implies that $\supp(\widehat{B}_v^{\gamma_2}) \subseteq V^c$, where $\widehat{B}_v^{\gamma_2} = \beta_g^{r(L_{\gamma_2})}(B_v)$.  
Note that $\tilde\omega_{\gamma_1}(B_v) = 1$ since $v \notin \gamma_1$, so by the proof of Lemma \ref{lem:LGGSDefectStateInequivalent}, $\tilde \omega_{\gamma_1}(\widehat{B}_v^{\gamma_2}) = 0$.
On the other hand, $\tilde \omega_{\gamma_2}(\widehat{B}_v^{\gamma_2}) = 1$.
Therefore, we have that 
\[
|\tilde\omega_{\gamma_1}(\widehat{B}_v^{\gamma_2})
-
\tilde \omega_{\gamma_2}(\widehat{B}_v^{\gamma_2})|
=
1
=
\|\widehat{B}_v^{\gamma_2}\|,
\]
so by \cite[Cor.~2.6.11]{MR887100}, $\tilde \omega_{\gamma_1} \not \simeq \tilde \omega_{\gamma_2}$.
\end{proof}

\section{Toric Code with ancillary vertex spins}
\label{app:TC with ancillary vertex spins}

We now recall the Hamiltonian $H_S^0$ for this system to be to be $$H_S^0 \coloneqq  H_S^{TC} + \sum_{v \in V(S)} \dfrac{\mathds{1} - \tau_v^x}{2}, $$
where $H_S^{TC} \in \cstar^E$ is the Toric Code Hamiltonian on $S$. Let $\delta^0$ be the corresponding derivation. It is easy to see that $H_S^0$ is still a commuting projector Hamiltonian. Let $\omega_0$ be a state on $\cstar$ defined by $$\omega_0 \coloneqq  \omega^E_{TC} \otimes \omega^V_0$$ where $\omega^E_{TC}$ (defined on $\cstar^E$) is the Toric Code frustration-free ground-state and $\omega^V_0$ is defined on $\cstar^V$ as a product state given by $\omega^V_0(A) \coloneqq  \bigotimes_{v \in \Gamma} \inner{\psi_v}{A \psi_v}$ and $\ket{\psi_v} \in \hilb_v$ satisfies $\ket{\psi_v} = \tau_v^x \ket{\psi_v}$.

Then it is easy to see that $\omega_0 $ is a frustration-free ground-state of $H_S^0$. In fact, we have the following Lemma.

\begin{lem}
\label{lem:TC extension FF GS is unique}
    The state $\omega_0$ is the unique state satisfying for all $v,f$ $$\omega_0(A_v) = \omega_0(B_f) = \omega_0(\tau_v^x) = 1$$
    In particular, this means that $\omega_0$ is pure.
\end{lem}
\begin{proof}
We first observe that $\omega_0$ satisfies the above equation.  
Indeed, $\omega_0 = \omega_{TC}^E \otimes \omega_0^V$, and for every $v, f$, $\omega_{TC}^E(A_v) = \omega_{TC}^E(B_f) = 1$ and $\omega_0^V(\tau^x_v) = 1$.
Now, suppose $\omega \colon \cstar \to \bbC$ is a state satisfying that $\omega(A_v) = \omega(B_f) = \omega(\tau^x_v) = 1$ for all $v, f$.  
We first claim that $\omega = \omega^E \otimes \omega_0^V$ for some $\omega^E \colon \cstar^E \to \bbC$.  
Indeed, let $A \in \cstar[\loc]$ be a simple tensor.  
Then $A = A^E \otimes A^V$ for some $A^E \in \cstar^E$ and $A^V \in \cstar^V$.  
Furthermore, since $A$ is a simple tensor, we have that $A^V = \bigotimes_{v \in \supp(A^V)} A^V_v$ for some $A^V_v \in \cstar[v]$.
For each vertex $v$, we define $P_v \coloneqq (\mathds{1} + \tau^x_v)/2 \in \cstar[v]$.
Note that $P_v$ is a rank-1 projection, and $\omega(P_v) = 1$.
Using Lemma \ref{lem:can freely insert and remove P from the ground state.}, we then have that 
\[
\omega(A)
=
\omega(A^E \otimes A^V)
=
\omega\!\left(A^E \otimes \bigotimes_{v \in \supp(A^V)} P_v A^V_v P_v \right).
\]
Now, $P_vA^V_vP_v \in \bbC P_v$ for all $v \in \supp(A^V)$ since $P_v \in \cstar[v]$ is a rank-1 projection, so 
\[
\bigotimes_{v \in \supp(A^V)}P_vA^V_vP_v = \lambda \bigotimes_{v \in \supp(A^V)}P_v
\]
for some $\lambda \in \bbC$.
We therefore have that 
\[
\omega(A)
=
\omega\!\left(A^E \otimes \bigotimes_{v \in \supp(A^V)} P_v A^V_v P_v \right)
=
\lambda \omega\!\left(A^E \otimes \bigotimes_{v \in \supp(A^V)} P_v \right)
=
\lambda\omega(A^E)
=
\omega(A^E)\omega_0^V(A^V).
\]
Since the simple tensors span a dense subspace of $\cstar$, we get that $\omega = \omega^E \otimes \omega_0^V$ for some $\omega^E \in \cS(\cstar^E)$.  

It remains to show that $\omega^E = \omega^E_{TC}$.
However, this follows from Lemma \ref{lem:UniqueFrustrationFreeGSToricCode}.
\end{proof}

Define $\pi_0$ to be the GNS representation of $\omega_0$ and let $\pi_0^{TC}$ be the GNS representation of $\omega^E_{TC}$. Let also $\pi^V_0$ be the GNS representation of $\omega_0^V$.

Since $\omega^E_{TC}, \omega_0^V$ are both pure, it follows that $\omega_0$ is also pure. The corresponding GNS representations are all irreducible.

\begin{lem}
\label{lem:TC extension Haag duality}
    The representation $\pi_0$ satisfies Haag duality.
\end{lem}
\begin{proof}
    First, note that $\omega_0 = \omega^E_{TC} \otimes \omega_0^V$ and observe that $\pi_0 \simeq \pi_0^{TC} \otimes \pi_0^V$ by uniqueness of the GNS representation.
    In fact, without loss of generality, we may assume $\pi_0 = \pi_0^{TC} \otimes \pi_0^V$.
    Let $\Lambda$ be a cone.  
    Then since $\pi_0^{TC}$ and $\pi_0^V$ both satisfy Haag duality for cones, we have that
    \[
    \pi_0(\cstar[\Lambda])'
    =
    (\pi_0^{TC}(\cstar[\Lambda]^E) \otimes \pi_0^{V}(\cstar[\Lambda]^V))'
    =
    \pi_0^{TC}(\cstar[\Lambda]^E)' \otimes \pi_0^{V}(\cstar[\Lambda]^V)'
    =
    \pi_0^{TC}(\cstar[\Lambda^c]^E)'' \otimes \pi_0^{V}(\cstar[\Lambda^c]^V)''
    =
    \pi_0(\cstar[\Lambda^c])''.
    \qedhere
    \]
\end{proof}

\begin{lem}
\label{lem:TC extension anyon sectors bound}
There are at most four anyon sectors with respect to $\pi_0$.
\end{lem}
\begin{proof}
By Lemma \ref{lem:TC extension Haag duality}, $\pi_0$ satisfies Haag duality for cones.  
It is also easy to verify that $\pi_0$ satisfies the approximate split property for cones \cite[Def.~5.1]{MR2804555}.
This follows quickly from the fact that $\pi_0 \simeq \pi_0^{TC} \otimes \pi_0^V$, $\pi_0^{TC}$ satisfies the approximate split property, and $\pi_0^V$ satisfies the split property.
Therefore, by \cite[Thm.~3.6]{MR3135456}, the number of distinct anyon sectors can be bounded by computing the index of the following subfactor. 
Let $\Upsilon = \Lambda_1 \cup \Lambda_2$, where $\Lambda_1$ and $\Lambda_2$ are disjoint cones that are sufficiently far apart.  
Then the number of distinct anyon sectors is at most $[\pi_0(\cstar[\Upsilon^c])' : \pi_0(\cstar[\Upsilon])'']$.
Now, we have that 
\begin{gather*}
\pi_0(\cstar[\Upsilon])''
=
\pi_0^{TC}(\cstar[\Upsilon])'' \otimes \pi_0^{V}(\cstar[\Upsilon])'',
\\
\pi_0(\cstar[\Upsilon^c])'
=
\pi_0^{TC}(\cstar[\Upsilon^c])' \otimes \pi_0^{V}(\cstar[\Upsilon^c])'
=
\pi_0^{TC}(\cstar[\Upsilon^c])' \otimes \pi_0^{V}(\cstar[\Upsilon])'',
\end{gather*}
where the last step follows since $\pi_0^V$ is the GNS representation of a product state.
Therefore, applying \cite[Thm.~4.9]{MR3135456}, we have that 
\begin{align*}
[\pi_0(\cstar[\Upsilon^c])' : \pi_0(\cstar[\Upsilon])'']
&=
[\pi_0^{TC}(\cstar[\Upsilon^c])' \otimes \pi_0^{V}(\cstar[\Upsilon])'' : \pi_0^{TC}(\cstar[\Upsilon])'' \otimes \pi_0^{V}(\cstar[\Upsilon])'']
\\&= 
[\pi_0^{TC}(\cstar[\Upsilon^c])': \pi_0^{TC}(\cstar[\Upsilon])'']
=
4.
\qedhere
\end{align*}
\end{proof}

We now show that there are at least 4 anyon sectors with respect to $\pi_0$. 
To do this, we inherit the previously defined automorphisms of $\cstar^E$ given by $\alpha_\gamma^\epsilon, \alpha_{\bar \gamma}^m, \alpha^\psi_{\gamma, \bar \gamma}$ (Definition \ref{def:ToricCodeAnyonAutomorphisms}).

\begin{lem}
    \label{lem:TC extension anyon sectors constructed}
    Let $\zeta \in \{\id, \alpha_\gamma^\epsilon, \alpha_{\bar \gamma}^m, \alpha^\psi_{\gamma, \bar \gamma}\}$ be an automorphism of $\cstar^E$. Then the representations $\pi^\zeta \coloneqq  (\pi_0^{TC} \circ \zeta) \otimes \pi_0^V$ are mutually disjoint anyon sectors with respect to $\pi_0$.
\end{lem}
\begin{proof}
    The representation $\pi^\zeta$ is obviously irreducible, since $\pi_0^{TC}, \pi_0^V$ are both irreducible and $\zeta$ is an automorphism. We now check if it satisfies the superselection criterion. 
    To do so, we show that it is localized and transportable.
    
    We first check that it is localized in some cone $\Lambda$.
    Let $\Lambda$ be a cone containing $\gamma, \bar{\gamma}$. 
    We show that $\pi^\zeta$ is localized in $\Lambda$.
    It suffices to check that $\pi^\zeta(A) = \pi_0(A)$ for all simple tensors $A \in \cstar[\Lambda^c]$
    Let $A \in \cstar[\Lambda^c]$ be a simple tensor. We then have that $A = A^V \otimes A^E$ where $A^V \in \cstar[\Lambda^c]^V$ and $A^E \in \cstar[\Lambda^c]^E$, so we have that
    \begin{align*}
        \pi^\zeta(A) = (\pi_0^{TC} \circ \zeta) (A^E) \otimes \pi_0^V(A^V) = \pi_0^{TC} (A^E) \otimes \pi_0^V(A^V) = \pi_0 (A).
    \end{align*}
    Thus $\pi^\zeta$ is localized in $\Lambda$.

    We now check transportability.
    Let $\Lambda'$ be another cone. Since $\pi_0^{TC} \circ \zeta$ is transportable, there exists some automorphism $\zeta' \colon \cstar^E \to \cstar^E$ such that $\pi_0^{TC} \circ \zeta'$ is localized in $\Lambda'$ and $\pi_0^{TC} \circ \zeta' \simeq \pi_0^{TC} \circ \zeta$. 
    By the above argument, $(\pi_0^{TC} \circ \zeta') \otimes \pi_0^V$ is localized in $\Lambda'$, and 
    \[
    (\pi_0^{TC} \circ \zeta') \otimes \pi_0^V \simeq (\pi_0^{TC} \circ \zeta)  \otimes \pi_0^V = \pi^\zeta.
    \]
    Thus $\pi^\zeta$ is transportable.
    
    Finally, we show that the representations $\{\pi^\zeta\}_\zeta$ are mutually inequivalent. We consider the case $\zeta = \alpha_\gamma^\epsilon$ and $\zeta' = \alpha^\psi_{\gamma, \bar \gamma}$ as an example. The other cases proceed similarly. Let $\omega^\zeta \coloneqq  \omega_0 \circ \zeta$ and $\omega^{\zeta'} \coloneqq  \omega_0 \circ {\zeta'}$. We use corollary \cite[2.6.11]{MR887100} along with \cite[Prop.~10.3.7]{MR1468230}. Choose a finite simply connected region $S$. Then consider $F_C \in \cstar[S^c]$ for a big enough loop $C \in \Gamma_f$ such that $\partial \bar \gamma$ is in the interior of $C, \bar C$. We then have that $\omega^{\zeta'}(F_C) = 1$ while $\omega^\zeta(F_C) = -1$. Since for every chosen $S$, there exists a big enough loop $C$ such that $F_C \in \cstar[S^c]$, we have that $\pi^\zeta \not \simeq \pi^{\zeta'}$. 
    This shows the full result.
\end{proof}

\begin{cor}
    \label{cor:TC extension anyon sector classification}
    Any anyon sector $\pi$ with respect to $\pi_0$ is unitarily equivalent to one of the mutually disjoint anyon sectors $\{\pi^\zeta\}_\zeta$ defined in Lemma \ref{lem:TC extension anyon sectors constructed}.
\end{cor}

\begin{proof}
    This follows straightforwardly from Lemmas \ref{lem:TC extension anyon sectors bound} and \ref{lem:TC extension anyon sectors constructed}.
\end{proof}

In fact, we have the following stronger result.  

\begin{prop}
\label{prop:TC extension braided monoidal}
Given a cone $\Lambda$, the category $\DHR_{\pi_0}(\Lambda)$ of anyon sectors with respect to $\pi_0$ localized in $\Lambda$ is braided monoidally equivalent to the category $\DHR_{\pi_0^{TC}}(\Lambda)$ of anyon sectors with respect to $\pi_0^{TC}$ localized in $\Lambda$.
\end{prop}

\begin{proof}
By Lemma \ref{lem:TC extension Haag duality}, $\pi_0$ satisfies Haag duality, so anyon sectors localized in $\Lambda$ have a canonical extension to the auxiliary algebra $\cstar^a$ \cite{MR660538, MR2804555}.
Now, the automorphisms we use to construct the anyon sectors in Lemma \ref{lem:TC extension anyon sectors constructed} are exactly those used in \cite{MR2804555}.
Therefore, the category $\DHR_{\pi_0}(\Lambda)$ of anyon sectors with respect to $\pi_0$ localized in $\Lambda$ is the same as the category of anyon sectors with respect to $\pi_0^{TC}$ constructed in \cite{MR2804555}.
\end{proof}

\section{Automorphisms describing defect sectors for Levin-Gu SPT}
\label{sec:LGDefectSectorComputationAppendix}

Our aim now is to explicitly construction for the Levin-Gu SPT the defect automorphism $\alpha_\gamma^g$ from Definition \ref{def:defect automorphisms} with $g \in \bbZ_2$ being the non-trivial element.

Observe that an infinite dual path $\gamma \in \bar P(\Gamma)$ divides $\Gamma$ into two halves, denoted by $r(L_\gamma), \ell(L_\gamma)$. We first write down the automorphism $\tilde \beta_g^{r(L_\gamma)} = \alpha^{-1} \circ \beta_g^{r(L_\gamma)} \circ \alpha$. We can compute for all $A \in \cstar[loc]$ the following expression $\tilde \beta_g^{r(L_\gamma)}(A)$. Using the explicit form of $\alpha, \beta_g^{r(L_\gamma)}$ for the Levin-Gu SPT we get, 
\[
\tilde \beta_g^{r(L_\gamma)}(A) 
= 
\left(\prod_{v \in r(L_\gamma)} B_v\right)A\left(\prod_{v \in r(L_\gamma)} B_v\right)^*
\]
Observe that $\tilde \beta_g^{r(L_\gamma)}$ extends to a well-defined automorphism of $\cstar$. 

\begin{figure}[!ht]
\centering
\begin{tikzpicture}[scale=0.6]
    \foreach \x in {0,...,4}{
    \foreach \y in {0,...,5}{
    \filldraw[draw=red,thick,fill=red!100] (\x*3,2*\y) circle(.1cm);
    \filldraw[draw=cyan,thick,fill=cyan!100] (\x*3+1,2*\y) circle(.1cm);
     \filldraw[draw=orange,thick,fill=orange!100] (\x*3+2,2*\y) circle(.1cm);
     \filldraw[draw=red,thick,fill=red!100] (\x*3+1.5,2*\y+1) circle(.1cm);
     \filldraw[draw=cyan,thick,fill=cyan!100] (\x*3+2.5,2*\y+1) circle(.1cm);
     \filldraw[draw=orange,thick,fill=orange!100] (\x*3+.5,2*\y+1) circle(.1cm);
}}
\draw[thick,red](3.5,9)--(4,10)--(5,10)--(5.5,9)--(6.5,9)--(7,10)--(8,10)--(8.5,9)--(9.5,9)--(10,10)--(11,10)--(11.5,9)--(11,8)--(11.5,7)--(12.5,7)--(13,6)--(12.5,5)--(11.5,5)--(11,4)--(11.5,3)--(11,2)--(10,2)--(9.5,3)--(8.5,3)--(8,2)--(7,2)--(6.5,3)--(5.5,3)--(5,2)--(4,2)--(3.5,3)--(4,4)--(3.5,5)--(2.5,5)--(2,6)--(2.5,7)--(3.5,7)--(4,8)--(3.5,9);
\draw[thick,cyan](4.5,9)--(5,10)--(6,10)--(6.5,9)--(7.5,9)--(8,10)--(9,10)--(9.5,9)--(10.5,9)--(11,8)--(12,8)--(12.5,7)--(12,6)--(12.5,5)--(12,4)--(11,4)--(10.5,3)--(9.5,3)--(9,2)--(8,2)--(7.5,3)--(6.5,3)--(6,2)--(5,2)--(4.5,3)--(3.5,3)--(3,4)--(3.5,5)--(3,6)--(3.5,7)--(3,8)--(3.5,9)--(4.5,9);
\draw[thick,orange](4.5,9)--(5.5,9)--(6,10)--(7,10)--(7.5,9)--(8.5,9)--(9,10)--(10,10)--(10.5,9)--(11.5,9)--(12,8)--(11.5,7)--(12,6)--(11.5,5)--(12,4)--(11.5,3)--(10.5,3)--(10,2)--(9,2)--(8.5,3)--(7.5,3)--(7,2)--(6,2)--(5.5,3)--(4.5,3)--(4,4)--(3,4)--(2.5,5)--(3,6)--(2.5,7)--(3,8)--(4,8)--(4.5,9);
\draw[thick,black,dashed](2.5,6)--(4.25,9.5)--(10.75,9.5)--(12.5,6)--(10.75,2.5)--(4.25,2.5)--(2.5,6);
\draw[thick, violet, dashed](0, 9.5) -- (14.5, 9.5);
\node[violet] at (15, 9.5) {$L_\gamma$};
\draw[thick, dashed, brown] (6.75, 10.5) -- (6,8.5) -- (6.5, 8.5) -- (6.5, 7.5) -- (6, 7.5) -- (6, 6.5) -- (10.25, 6.5) -- (8.25, 10.5) -- (6.75, 10.5);
\node[brown] at (7.8, 8.5) {$\supp(A)$};
\end{tikzpicture}
\caption{A hexagon $V$ to the right of the line $L_\gamma$ over which we take the product of $B_v$ terms.
The hexagon is taken to be large enough that $\supp(A)$ does not intersect the corners of the hexagon.
The colors for the vertices and the edges correspond to the labels $a, b, c$.}
\label{fig:ProductOfBvTermsLevinGu}
\end{figure}
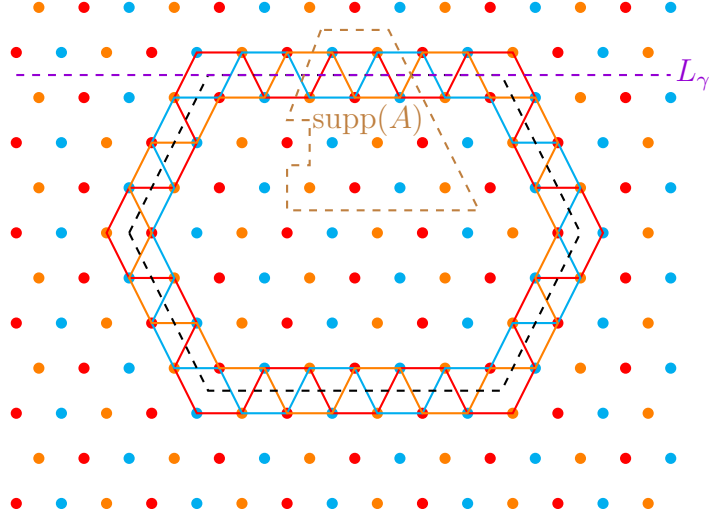

 Let $A \in \cstar[loc]$. We consider a hexagon $V\subseteq r(L_\gamma)$ such that one of the sides of the hexagon lies along $L_\gamma$ (see Figure \ref{fig:ProductOfBvTermsLevinGu}) and take $V$ to be large enough so that 
\[\tilde \beta_g^{r(L_\gamma)}(A)
=
\left(\prod_{v \in V} B_v\right)A\left(\prod_{v \in V} B_v\right)^*.
\]
We now compute the above expression. 
To do so, we notice that the lattice $\Gamma$ is tripartite.  
See Figure \ref{fig:ProductOfBvTermsLevinGu} to see the tripartite structure as well as the hexagon $V$ considered.
We let $a, b, c$ denote the labels of the vertices in $\Gamma$ according to the tripartite structure, and for $j = a, b, c$, we let $V_j \subseteq V$ be the collection of vertices labeled by $j$.  
Additionally, for $j = a, b, c$, we let $Q_j \coloneqq \prod_{v \in V_j} B_v$.  
Therefore, we obtain that 
\[
\tilde \beta_g^{r(L_\gamma)}(A)
=
(Q_aQ_bQ_c)A(Q_aQ_bQ_c)^*.
\]

We now compute $Q_j$ for each $j = a, b, c$.  
We observe that
\[
Q_j
=
\prod_{v \in V_j}B_v
=
\prod_{v \in V_j}\prod_{<vqq'> \in \triangle_v}-\sigma^x_v i^{\frac{1 - \sigma^z_q\sigma^z_{q'}}{2}}.
\]
Now, if $v \in V_j$ and $<vqq'> \in \triangle_v$, then $q, q'$ correspond to the other labels (not $j$).
Therefore, we have that 
\[
Q_j
=
\prod_{v \in V_j}\prod_{<vqq'> \in \triangle_v}-\sigma^x_v i^{\frac{1 - \sigma^z_q\sigma^z_{q'}}{2}}
=
Z_j X_j,
\]
where $X_j \coloneqq \prod_{v \in V_j} \sigma^x_v$ and $Z_j$ is some function of $\sigma^z_v$ for $v \in \Gamma$ corresponding to the non-$j$ labels.  
We compute $Z_j$ by considering every edge $qq'$ such that $<vqq'> \in \triangle_v$ for some $v \in V_j$.
There are two cases to consider. 
First, suppose $qq'$ is an edge labeled by the color corresponding to $j$ in Figure \ref{fig:ProductOfBvTermsLevinGu}.  
Then $i^{\frac{1 - \sigma^z_q\sigma^z_{q'}}{2}}$ shows up as a factor in $B_v$ for exactly one $v \in V_j$.
Now, suppose $qq'$ is an edge between two non-$j$ vertices in $V$ that does not lie along the boundary of $V$, so that $qq'$ is not a colored edge in Figure \ref{fig:ProductOfBvTermsLevinGu}.
Then $i^{\frac{1 - \sigma^z_q\sigma^z_{q'}}{2}}$ shows up in $B_v$ for two different $v \in V_j$.
We observe that 
\[
\left(i^{\frac{1 - \sigma^z_q\sigma^z_{q'}}{2}}\right)^2
=
i^{1 - \sigma^z_q\sigma^z_{q'}}
=
i(i^{-1})^{\sigma^z_q\sigma^z_{q'}}.
\]
Since the eigenvalues of $\sigma^z_q\sigma^z_{q'}$ are $\pm 1$, we have that 
\[
\left(i^{\frac{1 - \sigma^z_q\sigma^z_{q'}}{2}}\right)^2
=
i(i^{-1})^{\sigma^z_q\sigma^z_{q'}}
=
i(i^{-1})\sigma^z_q\sigma^z_{q'}
=
\sigma^z_q\sigma^z_{q'}.
\]
We therefore have that 
\[
Z_j 
=
\prod_{v \in V_{\widehat{j}}} \sigma^z_v \prod_{qq' \in {L_\gamma}_{\widehat{j}}} i^{\frac{1 - \sigma^z_q\sigma^z_{q'}}{2}},
\]
where ${L_\gamma}_{\widehat{j}}$ is the path corresponding to $j$ illustrated in Figure \ref{fig:ProductOfBvTermsLevinGu} and $V_{\widehat{j}}$ is the collection of non-$j$ vertices in $V$ that have an edge $e$ to another non-$j$ vertex in $V$ where $e \notin {L_\gamma}_{\widehat{j}}$.

Using this computation, we have that 
\[
\tilde \beta_g^{r(L_\gamma)}(A)
=
(Q_aQ_bQ_c)A(Q_aQ_bQ_c)^*
=
(Z_aX_aZ_bX_bZ_cX_c)A(Z_aX_aZ_bX_bZ_cX_c)^*.
\]
We compute the quantity on the right.  
We first observe that 
\[
X_a Z_b
=
\prod_{v \in V_a} \sigma^x_v \prod_{v \in V_{\widehat{b}}} \sigma^z_v \prod_{qq' \in {L_\gamma}_{\widehat{b}}} i^{\frac{1 - \sigma^z_q\sigma^z_{q'}}{2}}
=
(-1)^k \prod_{v \in V_{\widehat{b}}} \sigma^z_v \prod_{qq' \in {L_\gamma}_{\widehat{b}}} i^{\frac{1 - (-1)^{\varepsilon_a}\sigma^z_q\sigma^z_{q'}}{2}} \prod_{v \in V_a} \sigma^x_v,
\]
where $k \in \bbN$ and $\varepsilon_a = 1$ if one of $q, q' \in V_a$ and $\varepsilon_a = 0$ otherwise.  
Note that when we conjugate by the above operator, the factor of $(-1)^k$ cancels.  
Similarly, we have that
\[
X_aX_bZ_c
=
\prod_{v \in V_a \cup V_c} \sigma^x_v \prod_{v \in V_{\widehat{c}}} \sigma^z_v \prod_{qq' \in {L_\gamma}_{\widehat{c}}} i^{\frac{1 - \sigma^z_q\sigma^z_{q'}}{2}}
=
(-1)^k \prod_{v \in V_{\widehat{c}}} \sigma^z_v \prod_{qq' \in {L_\gamma}_{\widehat{c}}} i^{\frac{1 - (-1)^{\varepsilon_s}\sigma^z_q\sigma^z_{q'}}{2}}
\prod_{v \in V_a \cup V_c} \sigma^x_v,
\]
where again $k \in \bbN$ and $\varepsilon_s = 1$ if exactly one of $q, q' \in V$ and $\varepsilon_s = 0$ otherwise.
We then have that 
\begin{align*}
\tilde \beta_g^{r(L_\gamma)}(A)
&=
(Z_aX_aZ_bX_bZ_cX_c)A(Z_aX_aZ_bX_bZ_cX_c)^*
\\&=
\Ad\!\left(
\prod_{v \in V_{\widehat{a}}} \sigma^z_v \prod_{v \in V_{\widehat{b}}} \sigma^z_v \prod_{v \in V_{\widehat{c}}} \sigma^z_v 
\prod_{qq' \in {L_\gamma}_{\widehat{a}}} i^{\frac{1 - \sigma^z_q\sigma^z_{q'}}{2}} 
\prod_{qq' \in {L_\gamma}_{\widehat{b}}} i^{\frac{1 - (-1)^{\varepsilon_a}\sigma^z_q\sigma^z_{q'}}{2}} 
\prod_{qq' \in {L_\gamma}_{\widehat{c}}} i^{\frac{1 - (-1)^{\varepsilon_s}\sigma^z_q\sigma^z_{q'}}{2}}
\right)\! \circ \beta_g^{r({L_\gamma})}(A).
\end{align*}

Now, because $V$ is large relative to the support of $A$, we can ignore effects that occur at the corners of the hexagon $V$ and simplify the above expression.  
In particular, we have that 
\[
\Ad\!\left(\prod_{v \in V_{\widehat{a}}} \sigma^z_v \prod_{v \in V_{\widehat{b}}} \sigma^z_v \prod_{v \in V_{\widehat{c}}} \sigma^z_v \right)\!(A')
=
\Ad\!\left(\prod_{v \in \partial V} \sigma^z_v\right)\!(A'),
\]
where $\partial V$ is the collection of vertices along the boundary of $V$.  
Here $A'$ is defined by 
\[
A'
\coloneqq
\Ad\!\left(
\prod_{qq' \in {L_\gamma}_{\widehat{a}}} i^{\frac{1 - \sigma^z_q\sigma^z_{q'}}{2}} 
\prod_{qq' \in {L_\gamma}_{\widehat{b}}} i^{\frac{1 - (-1)^{\varepsilon_a}\sigma^z_q\sigma^z_{q'}}{2}} 
\prod_{qq' \in {L_\gamma}_{\widehat{c}}} i^{\frac{1 - (-1)^{\varepsilon_s}\sigma^z_q\sigma^z_{q'}}{2}}
\right)\! \circ \beta_g^{r({L_\gamma})}(A)
\]
Therefore, we have that 
\[
\tilde \beta_g^{r(L_\gamma)}(A)
=
\Ad\!\left(
\prod_{v \in \partial V} \sigma^z_v
\prod_{qq' \in {L_\gamma}_{\widehat{a}}} i^{\frac{1 - \sigma^z_q\sigma^z_{q'}}{2}} 
\prod_{qq' \in {L_\gamma}_{\widehat{b}}} i^{\frac{1 - (-1)^{\varepsilon_a}\sigma^z_q\sigma^z_{q'}}{2}} 
\prod_{qq' \in {L_\gamma}_{\widehat{c}}} i^{\frac{1 - (-1)^{\varepsilon_s}\sigma^z_q\sigma^z_{q'}}{2}}
\right)\! \circ \beta_g^{r(L_\gamma)}(A).
\]
Now, we can replace $V$ with $r({L_\gamma})$ in the above equation, and we can also redefine ${L_\gamma}_{\widehat{a}}, {L_\gamma}_{\widehat{b}}, {L_\gamma}_{\widehat{c}}$ to refer to their continuations along the path ${L_\gamma}$.
We then have that for any $A \in \cstar[\loc]$, 
\begin{align*}
    \tilde \beta_g^{r(L_\gamma)}(A)
&=
\Ad\!\left(
\prod_{v \in \partial r({L_\gamma})} \sigma^z_v
\prod_{qq' \in {L_\gamma}_{\widehat{a}}} i^{\frac{1 - \sigma^z_q\sigma^z_{q'}}{2}} 
\prod_{qq' \in {L_\gamma}_{\widehat{b}}} i^{\frac{1 - (-1)^{\varepsilon_a}\sigma^z_q\sigma^z_{q'}}{2}} 
\prod_{qq' \in {L_\gamma}_{\widehat{c}}} i^{\frac{1 - (-1)^{\varepsilon_s}\sigma^z_q\sigma^z_{q'}}{2}}
\right)\! \circ \beta_g^{r({L_\gamma})}(A)
\end{align*}

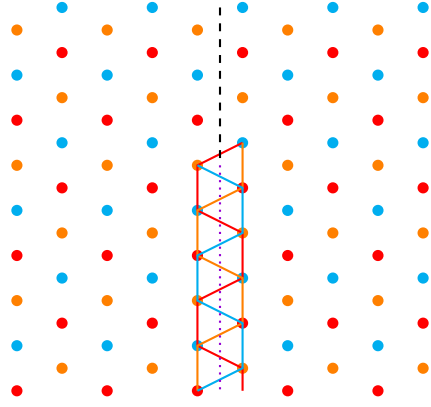
\begin{figure}[!ht]
    \centering
    \begin{tikzpicture}[scale=0.6]
    \foreach \x in {0,...,4}{
    \foreach \y in {0,...,2}{
    \filldraw[draw=red,thick,fill=red!100] (2*\x,\y*3) circle(.1cm);
    \filldraw[draw=cyan,thick,fill=cyan!100] (2*\x,\y*3+1) circle(.1cm);
    \filldraw[draw=orange,thick,fill=orange!100] (2*\x,\y*3+2) circle(.1cm);
    \filldraw[draw=orange,thick,fill=orange!100] (2*\x+1,\y*3+.5) circle(.1cm);
    \filldraw[draw=red,thick,fill=red!100] (2*\x+1,\y*3+1.5) circle(.1cm);
    \filldraw[draw=cyan,thick,fill=cyan!100] (2*\x+1,\y*3+2.5) circle(.1cm);
}}
\draw[thick,red](5,5.5)--(4,5)--(4,4)--(5,3.5)--(5,2.5)--(4,2)--(4,1)--(5,.5)--(5,0);
\draw[thick,orange](5,5.5)--(5,4.5)--(4,4)--(4,3)--(5,2.5)--(5,1.5)--(4,1)--(4,0);
\draw[thick,cyan](4,5)--(5,4.5)--(5,3.5)--(4,3)--(4,2)--(5,1.5)--(5,.5)--(4,0);
\draw[thick,dashed](4.5,8.5)--(4.5,5);
\draw[thick, violet, dotted] (4.5, 5) -- (4.5, 0);
\end{tikzpicture}
    \caption{An illustration of the notation used to define $\alpha^\gamma$.
    The dashed black path is $\gamma$, and the dotted purple path is $\xi$.
    The red, blue, and orange colors for the vertices correspond to the labels $a$, $b$, and $c$, and the red, blue, and orange edges denote the paths $\xi_{\widehat{a}}, \xi_{\widehat{b}}, \xi_{\widehat{c}}$ respectively. 
    The region $\partial r(\xi)$ consists of those edges to the right of the dashed/dotted line with an adjacent colored edge.}
    \label{fig:DecorationForLGDefectSectorTripartiteColor}
\end{figure}

Let $\xi = L_\gamma - \gamma$. For $j = a, b, c$, we define $\xi_{\widehat{j}}, \gamma_{\widehat{j}}$ analogously to ${L_\gamma}_{\widehat{j}}$ above (see Figure \ref{fig:DecorationForLGDefectSectorTripartiteColor}). Now we note that $$\tilde \beta_g^{r(L_\gamma)} \circ (\beta_g^{r(L_\gamma)})^{-1}(A) = \eta^\gamma \otimes \eta^\xi(A)$$ where

\begin{align*}
    \eta^\xi(A) &\coloneqq  \Ad\!\left(
\prod_{v \in \partial r(\xi)} \sigma^z_v
\prod_{qq' \in \xi_{\widehat{a}}} i^{\frac{1 - \sigma^z_q\sigma^z_{q'}}{2}} 
\prod_{qq' \in \xi_{\widehat{b}}} i^{\frac{1 - (-1)^{\varepsilon_a}\sigma^z_q\sigma^z_{q'}}{2}} 
\prod_{qq' \in \xi_{\widehat{c}}} i^{\frac{1 - (-1)^{\varepsilon_s}\sigma^z_q\sigma^z_{q'}}{2}}
\right)\!(A)\\
\eta^\gamma(A) &\coloneqq  \Ad\!\left(
\prod_{v \in \partial r(\gamma)} \sigma^z_v
\prod_{qq' \in \gamma_{\widehat{a}}} i^{\frac{1 - \sigma^z_q\sigma^z_{q'}}{2}} 
\prod_{qq' \in \gamma_{\widehat{b}}} i^{\frac{1 - (-1)^{\varepsilon_a}\sigma^z_q\sigma^z_{q'}}{2}} 
\prod_{qq' \in \gamma_{\widehat{c}}} i^{\frac{1 - (-1)^{\varepsilon_s}\sigma^z_q\sigma^z_{q'}}{2}}
\right)\!(A)
\end{align*}

So indeed, we note that Assumption \ref{asmp:strip_aut_is_an_FDQC} is satisfied by the Levin-Gu SPT.

Now we get the defect automorphism $\alpha^\gamma \coloneqq  \alpha^g_\gamma = \eta^\xi \circ \beta_g^{r(L_\gamma)}$ (Definition \ref{def:defect automorphisms}) as
\[
\alpha^{\gamma}(A)=
\Ad\!\left(
\prod_{v \in \partial r(\xi)} \sigma^z_v
\prod_{qq' \in \xi_{\widehat{a}}} i^{\frac{1 - \sigma^z_q\sigma^z_{q'}}{2}} 
\prod_{qq' \in \xi_{\widehat{b}}} i^{\frac{1 - (-1)^{\varepsilon_a}\sigma^z_q\sigma^z_{q'}}{2}} 
\prod_{qq' \in \xi_{\widehat{c}}} i^{\frac{1 - (-1)^{\varepsilon_s}\sigma^z_q\sigma^z_{q'}}{2}}
\right)\! \circ \beta_g^{r(L_{\gamma})}(A)
\]
for $A \in \cstar[\loc]$.
Here $\partial r(\xi)$ is the portion of $\partial r(L_{{\gamma}})$ along $\xi$.

We now write down the simplified form of the formula for $\alpha^\gamma$ that is used in Section \ref{sec:Defect sectors in the Levin-Gu SPT}.  
As done in that section, we define $N(\xi)$ to be the subgraph of $\Gamma$ consisting of all vertices in $\xi$ and edges between them. 
We let $\xi_{\mathrm{in}}$ denote the path of vertices in $N(\xi)$ that are in $r(L_{{\bar \gamma_R}})$ and $\xi_{\mathrm{out}}$ denote the path of vertices in $N(\xi)$ that are not in $r(L_{{\bar \gamma_R}})$ (Figure \ref{fig:DecorationForLGDefectSector}).
In that case, we have that 
\[
\alpha^{{\bar \gamma_R}}(A)
=
\Ad\!\left(
\prod_{v \in \partial r(\xi)} \sigma^z_v
\prod_{qq' \in N(\xi)} i^{\frac{1 - (-1)^{\varepsilon_{qq'}}\sigma^z_q\sigma^z_{q'}}{2}} 
\right)\! \circ \beta_g^{r(L_{{\bar \gamma_R}})}(A)
\]
for $A \in \cstar[\loc]$.
Note that $\varepsilon_{qq'} \in \{0, 1\}$, where the value of $\varepsilon_{qq'}$ is determined by the preceding discussion.

\begin{lem}
\label{lem:LevinGuCocycleComputation}
We have that $\alpha^{\bar\gamma_R} \circ \alpha^{\bar \gamma_R} = \Ad(\sigma^z_{\partial \xi_{\mathrm{in}}} \sigma^z_{\partial \xi_{\mathrm{out}}})$, where $\partial \xi_{\mathrm{in}}$ and $\partial \xi_{\mathrm{out}}$ are the endpoints of $\xi_{\mathrm{in}}$ and $\xi_{\mathrm{out}}$ respectively.
\end{lem}

\begin{proof}
It can be easily verified that 
\begin{align*}
\alpha^{{\bar \gamma_R}} \circ \alpha^{{\bar \gamma_R}}
&=
\Ad\!\left(
\prod_{v \in \partial r(\xi)} \sigma^z_v \beta_g^{r(L_{{\bar \gamma_R}})}(\sigma^z_v)
\prod_{qq' \in N(\xi)} i^{\frac{1 - (-1)^{\varepsilon_{qq'}}\sigma^z_q\sigma^z_{q'}}{2}} 
\beta_g^{r(L_{{\bar \gamma_R}})}\!\left(i^{\frac{1 - (-1)^{\varepsilon_{qq'}}\sigma^z_q\sigma^z_{q'}}{2}}\right)
\right)
\\&=
\Ad\!\left(
\prod_{v \in \partial r(\xi)} \sigma^z_v (-\sigma^z_v)
\prod_{qq' \in N(\xi)} i^{\frac{1 - (-1)^{\varepsilon_{qq'}}\sigma^z_q\sigma^z_{q'}}{2}} 
\cdot i^{\frac{1 - (-1)^{\varepsilon_{qq'}}(-1)^{\epsilon_s}\sigma^z_q\sigma^z_{q'}}{2}}
\right)
\\&=
\Ad\!\left(
\prod_{qq' \in N(\xi)} i^{\frac{1 - (-1)^{\varepsilon_{qq'}}\sigma^z_q\sigma^z_{q'}}{2}} 
\cdot i^{\frac{1 - (-1)^{\varepsilon_{qq'}}(-1)^{\epsilon_s}\sigma^z_q\sigma^z_{q'}}{2}}
\right).
\end{align*}
Here, $\epsilon_s = 1$ if exactly one of $q, q' \in r(L_\gamma)$ and $\epsilon_s = 0$ otherwise.
In particular, $\epsilon_s = 1$ if and only if $qq' \notin \xi_{\mathrm{in}}$ and $qq' \notin \xi_{\mathrm{out}}$.

Note that if $\epsilon_s = 1$, then 
\[
i^{\frac{1 - (-1)^{\varepsilon_{qq'}}\sigma^z_q\sigma^z_{q'}}{2}} 
\cdot i^{\frac{1 - (-1)^{\varepsilon_{qq'}}(-1)^{\epsilon_s}\sigma^z_q\sigma^z_{q'}}{2}}
=
i^{\frac{1 - (-1)^{\varepsilon_{qq'}}\sigma^z_q\sigma^z_{q'}}{2}} 
\cdot i^{\frac{1 + (-1)^{\varepsilon_{qq'}} \sigma^z_q\sigma^z_{q'}}{2}}
=
i^{\frac{1 - {(\sigma^z_q)}^2 {(\sigma^z_{q'})}^2}{4}}
=
1.
\]
On the other hand, if $\epsilon_s = 0$, then 
\[
i^{\frac{1 - (-1)^{\varepsilon_{qq'}}\sigma^z_q\sigma^z_{q'}}{2}} 
\cdot i^{\frac{1 - (-1)^{\varepsilon_{qq'}}(-1)^{\epsilon_s}\sigma^z_q\sigma^z_{q'}}{2}}
=
i^{\frac{1 - (-1)^{\varepsilon_{qq'}}\sigma^z_q\sigma^z_{q'}}{2}} 
\cdot i^{\frac{1 - (-1)^{\varepsilon_{qq'}}\sigma^z_q\sigma^z_{q'}}{2}}
=
i^{1 - (-1)^{\varepsilon_{qq'}}\sigma^z_q\sigma^z_{q'}}. 
\]
We therefore have that
\begin{align*}
\alpha^{{\bar \gamma_R}} \circ \alpha^{{\bar \gamma_R}}
&=
\Ad\!\left(
\prod_{qq' \in N(\xi)} i^{\frac{1 - (-1)^{\varepsilon_{qq'}}\sigma^z_q\sigma^z_{q'}}{2}} 
\cdot i^{\frac{1 - (-1)^{\varepsilon_{qq'}}(-1)^{\epsilon_s}\sigma^z_q\sigma^z_{q'}}{2}}
\right)
\\&=
\Ad\!\left(
\prod_{qq' \in \xi_{\mathrm{in}} \cup \xi_{\mathrm{out}}} i^{1 - (-1)^{\varepsilon_{qq'}}\sigma^z_q\sigma^z_{q'}} 
\right).
\end{align*}
Finally, we simplify $i^{1 - (-1)^{\varepsilon_{qq'}}\sigma^z_q\sigma^z_{q'}}$.
Observe that 
\[
i^{1 - (-1)^{\varepsilon_{qq'}}\sigma^z_q\sigma^z_{q'}} 
=
i\left( i^{(-1)^{\varepsilon_{qq'}+ 1}}\right)\sigma^z_q\sigma^z_{q'}
=
i\cdot i^{(-1)^{\varepsilon_{qq'}+ 1}} \sigma^z_q\sigma^z_{q'},
\]
where the last equality follows since the two operators have exactly the same eigenvalues and eigenvectors.  
Therefore, we have that 
\begin{align*}
\alpha^{{\bar \gamma_R}} \circ \alpha^{{\bar \gamma_R}}
&=
\Ad\!\left(
\prod_{qq' \in \xi_{\mathrm{in}} \cup \xi_{\mathrm{out}}} i^{1 - (-1)^{\varepsilon_{qq'}}\sigma^z_q\sigma^z_{q'}} 
\right)
=
\Ad\!\left(
\prod_{qq' \in \xi_{\mathrm{in}} \cup \xi_{\mathrm{out}}} i\cdot i^{(-1)^{\varepsilon_{qq'}+ 1}} \sigma^z_q\sigma^z_{q'}
\right)
\\&=
\Ad\!\left(
\prod_{qq' \in \xi_{\mathrm{in}} \cup \xi_{\mathrm{out}}} \sigma^z_q\sigma^z_{q'}
\right)
=
\Ad(\sigma^z_{\partial \xi_{\mathrm{in}}} \sigma^z_{\partial \xi_{\mathrm{out}}}).
\qedhere
\end{align*}
\end{proof}

Note that if $\gamma \in \bar P(\Gamma)$ is any semi-infinite path, we can still construct an automorphism $\alpha^\gamma \colon \cstar \to \cstar$ using the same procedure, although we do not do it here.  

\section{Relating SET toric code defect sectors to Hamiltonian terms}
\label{sec:SETToricCodeDefectHamiltonian}

Recall the SET toric code model discussed in Section \ref{sec:SET_model}.
The goal of this section is to prove some results concerning the symmetry action on the terms of the SET toric code Hamiltonian. 
This will relate the analysis for the SET toric code to our analysis of SPTs in sections \ref{sec:Defect auts Hamiltonian Levin-Gu} and \ref{sec:defects using auts}.

\begin{lem}
\label{lem:action of half plane sym on SET terms}
    Choose an infinite dual path $\bar L \in \bar P(\Gamma)$. We have, $$\beta_g^{r (\bar L)}(A_v) = A_v \qquad \beta_g^{r(\bar L)}(\tilde Q_v) = \tilde Q_v \qquad \beta_g^{r(\bar L)}(\tilde B_f) = \left(\prod_{v \in r(\bar L)\cap f}\prod_{e \ni v} i g(e,v) \sigma^x_e\right)\tilde B_f$$ where $g(e,v) = +1$ if $\partial_1 e = v$ and $g(e,v) = -1$ if $\partial_0 e = v$.
\end{lem}
\begin{proof}
    The first identity is obvious. For the second identity, we observe that,
    \begin{align*}
        \tilde Q_v &= \frac{\mathds{1}+A_v}{2}Q_v = \frac{\mathds{1}+A_v}{2} Q_v \frac{\mathds{1}+A_v}{2} = \frac{\mathds{1}+A_v}{2} \left(\tau_v^x i^{-\tau^z_v \sum_{e \ni v} f(e,v) \sigma^x_e/2}\right)\frac{\mathds{1}+A_v}{2}
        \intertext{Now we note that when $A_v = +1$, $\sum_{e \ni v} f(e,v) \sigma^x_e/2 $ has eigenvalues $\pm 2,0$ and therefore we can drop $-\tau^z_v$ from the exponent. We thus get,}
        &= \frac{\mathds{1}+A_v}{2}\left(\tau_v^x i^{ \sum_{e \ni v} f(e,v) \sigma^x_e/2}\right)\frac{\mathds{1}+A_v}{2}.
    \end{align*}
    Now it is obvious from the above explicit form of $\tilde Q_v$ that $\beta_g^{r(\bar L)}(\tilde Q_v) = \tilde Q_v$.

    For the third identity, we rewrite $\tilde B_f$ as follows:
    \[
    \tilde B_f = \prod_{e \in f}i^{-\sigma^x_e (\tau^z_{\partial_1 e} - \tau^z_{\partial_0 e})/2} B_f = \prod_{v \in f} i^{-\tau^z_v(\sum_{e \ni v} g(e,v)\sigma^x_e)/2} B_f.
    \]
    We now observe that
    \begin{align*}
        \beta_g^{r(\bar L)}(\tilde B_f)&=  \prod_{v \in f; v \in r(\bar L)} i^{\tau^z_v(\sum_{e \ni v}g(e,v) \sigma^x_e)/2}\prod_{v \in f; v \notin r(\bar L)} i^{-\tau^z_v(\sum_{e \ni v} g(e,v)\sigma^x_e)/2} B_f\\
        &= \prod_{v \in f; v \in r(\bar L)} i^{\tau^z_v(\sum_{e \ni v}g(e,v) \sigma^x_e)} \tilde B_f
        \intertext{Now we note that $\sum_{e \ni v}g(e,v) \sigma^x_e$ has eigenvalues in $\{\pm 2, 0\}$ and thus we can drop $\tau^z_v$. Now,}
        &= \prod_{v \in f; v \in r(\bar L)} i^{\sum_{e \ni v}g(e,v) \sigma^x_e} \tilde B_f = \prod_{v \in r(\bar L) \cap f} \prod_{e \ni v} i g(e,v) \sigma^x_e \tilde B_f,
    \end{align*}
    where in the last equality we've used that if $A^2= 1$, $i^A = iA$. This shows the result.
\end{proof}

Now recall the automorphisms $\alpha^\sigma_{\bar \gamma}$ defined in Section \ref{sec:SET_defects}.
The following lemma shows that the action of the symmetry along $r(\bar L)$ on $A_v, \tilde B_f, \tilde Q_v$ can be erased using the automorphism $\alpha^\sigma_{\bar \gamma}$ acting along a part of $\bar L$. 

\begin{lem}
\label{lem:erasure string erases SET}
    Let $\bar \gamma \in \bar P(\Gamma)$ be a half-infinite path, $\bar L$ a completion of $\bar \gamma$ and $\bar \eta = \bar L - \bar \gamma$. Choose a cone $\Lambda$ such that $\bar \gamma$ is contained in $\Lambda$. Then for all sites $s$ and $C_s \in \{A_v, \tilde B_f, \tilde Q_v\}$ such that $\supp(C_s) \subset \Gamma \cap \Lambda^c$, $$\alpha_{\bar \eta}^\sigma \circ \beta_g^{\bar L}(C_s) = C_s.$$
\end{lem}
\begin{proof}
    For all $s$ sufficiently far away from $\bar L$ this Lemma follows immediately from \ref{lem:SET Hamiltonian terms are symmetric}. 
    
    Now we note that for $A_v, \tilde Q_v$ having supports overlapping with $\bar L$, the result immediately follows from \ref{lem:action of half plane sym on SET terms} and the fact that $\alpha^\sigma_{\bar \eta}$ only consists of $\sigma^x_e$ terms. 
    
    All that remains is to check for the $\tilde B_f$ terms whose support overlaps with $\bar L$. We have from Lemma \ref{lem:action of half plane sym on SET terms} that $$\beta_g^{r(\bar L)}(\tilde B_f) = \left(\prod_{v \in r(\bar L)\cap f}\prod_{e \ni v,e\in f} i g(e,v) \sigma^x_e\right)\tilde B_f$$ where $g(e,v) = +1$ if $\partial_1 e = v$ and $g(e,v) = -1$ if $\partial_0 e = v$. Therefore,

    \begin{align*}
        \alpha_{\bar \eta}^\sigma \circ \beta_g^{\bar L}(\tilde B_f) &= \alpha^\sigma_{\bar \eta}\left(\left(\prod_{v \in r(\bar L)\cap f}\prod_{e \ni v,e\in f} i g(e,v) \sigma^x_e\right)\tilde B_f\right) = \left(\prod_{v \in r(\bar L)\cap f}\prod_{e \ni v,e\in f} i g(e,v) \sigma^x_e\right) \alpha^\sigma_{\bar \eta}(\tilde B_f)\\
        &= \left(\prod_{v \in r(\bar L)\cap f}\prod_{e \ni v,e\in f} i g(e,v) \sigma^x_e\right) \left(\prod_{e \in \bar L \cap f} e^{-i \pi p(e) \sigma^x_e/2}\right) \tilde B_f\\
        &= \left(\prod_{v \in r(\bar L)\cap f}\prod_{e \ni v,e\in f} i g(e,v) \sigma^x_e\right) \left(\prod_{e \in \bar L \cap f} (-i){ p(e) \sigma^x_e}\right) \tilde B_f = B_f,        
    \end{align*}
    where in the last equality comes from the fact that 
    $$\left(\prod_{v \in r(\bar L)\cap f}\prod_{e \ni v} i g(e,v) \sigma^x_e\right) = \left(\prod_{e \in \bar L \cap f} i{ p(e) \sigma^x_e}\right)$$ This identity may be verified by checking all cases of how the line $\bar L$ can intersect $f$. We have now shown the required result.
\end{proof}

Finally, we recall the defect state $\tilde \omega^\sigma_{\bar \gamma} = \tilde \omega \circ \tilde \alpha_{\bar \gamma}^\sigma$ for a dual path $\bar \gamma$, where $\tilde \alpha_{\bar \gamma}^\sigma$ is given in Definition \ref{def:SETToricCodeDefectAutomorphism}.

\begin{lem}
    Pick a dual path $\bar \gamma$. For all sites $s$ outside $\bar \gamma$, the state $\tilde \omega^\sigma_{\bar \gamma}$ looks like the ground-state. Specifically, pick $C_s \in \{A_v, B_f, Q_v\}$. Then for all sites $s$ outside $\bar \gamma$, $$\tilde \omega^\sigma_{\bar \gamma}(C_s) = 1$$
\end{lem}
\begin{proof}
    Follows immediately using the definition of $\tilde \alpha_{\bar \gamma}^\sigma$ and Lemmas \ref{lem:erasure string erases SET}, \ref{lem:SET toric code unique FF GS}.
\end{proof}

\end{subappendices}

\addtocontents{toc}{\protect\setcounter{tocdepth}{0}}  

\section*{Conflict of interest}

The authors have no conflicts of interest to disclose.

\section*{Data availability}

This manuscript has no associated data. 

\addtocontents{toc}{\protect\setcounter{tocdepth}{1}}  

\bibliographystyle{alpha}
\bibliography{symmetry/symmetry_bibliography}
    